Abstract

# Machine Learning Solutions for High Energy Physics:
## Applications to Electromagnetic Shower Generation, Flavor Tagging, and the Search for di-Higgs Production

Michela Paganini

2019

This thesis demonstrate the efficacy of designing and developing machine learning algorithms to selected use cases that encompass many of the outstanding challenges in the field of experimental high energy physics.

Although simple implementations of neural networks and boosted decision trees have been used in high energy physics for a long time (*e.g.* Ref. [1]), the field of machine learning has quickly evolved by devising more complex, fast and stable implementations of learning algorithms. The complexity and power of state-of-the-art deep learning far exceeds those of the learning algorithms implemented in the CERN-developed `ROOT` library.

All aspects of experimental high energy physics have been and will continue being revolutionized by the software- and hardware-based technological advances spearheaded by both academic and industrial research in other technical disciplines, and the emergent trend of increased interdisciplinarity will soon reframe many scientific domains. This thesis exemplifies this spirit of versatility and multidisciplinarity by bridging the gap between machine learning and particle physics, and exploring original lines of work to modernize the reconstruction, particle identification, simulation, and analysis workflows.

This contribution documents a collection of novel approaches to augment traditional domain-specific methods with modern, automated techniques based on industry-standard, open-source libraries. Specifically, it contributes to setting the state-of-the-art for impact parameter-based flavor tagging and di-Higgs searches in the $\gamma\gamma b\bar{b}$ channel with the ATLAS detector at the LHC, it introduces and lays the foundations for the use of generative adversarial networks for the simulation of particle showers in calorimeters.

These results substantiate the notion of machine learning powering particle physics in the upcoming years and establish baselines for future applications.

# Machine Learning Solutions for High Energy Physics:
## Applications to Electromagnetic Shower Generation, Flavor Tagging, and the Search for di-Higgs Production

A Dissertation
Presented to the Faculty of the Graduate School
of
Yale University
in Candidacy for the Degree of
Doctor of Philosophy

by
Michela Paganini

Dissertation Director: Professor Paul L. Tipton

May 2019

# Contents

# Listing of figures

To the future generation of graduate students.

# Acknowledgments

I would like to thank my parents for always giving me the best opportunities, even through the toughest of times. Thank you for being such inspiring, challenging, and educated role models, and for welcoming with enthusiasm my curiosity and dedication to academia.

Thanks to my life companion, who stood by me through the entirety of my graduate studies. You are my soul mate and best friend. You have witnessed the best and worst of me, you praised my achievements and held my hand through the difficult times when I considered giving up. You believed in me and turned me into a confident woman. I can't wait to spend the rest of my life with you.

Thanks to my best friends around the world. Your vision and intellect shaped me and the way I see the world. You are all successful, and I am so blessed our paths have crossed. Keep up the hard work and continue making the world a better place through your poems, resolutions, analyses, ideas, designs, and all the other priceless contributions that you bring to this world every day.

Thanks to my first post-docs, Johannes and Andrey, who are the sole reason I managed to navigate my way through CERN. Thanks for making it a much more bearable task and for coming to my rescue when I was lost. Thanks to the post-docs who followed me in my later years, Chase and Ben. Ben, your brilliant, creative mind will shape the future of our field. You always treated me as peer, valued my contributions, and had the patience and empathy to appreciate and encourage me through my unique path.

Thanks to the IML coordinators for being a light in the dark, and for building a welcoming and uplifting community.

Thanks to all the people who empowered me. You are the reason why this work is being completed. Your thoughtfulness and mentorship, through small or big gestures, was deeply appreciated and crucial throughout the years. Thank you Wahid, Mustafa, Paolo, Prabhat, and the rest of the LBNL and NERSC group. Thanks to all of those who used their stature to encourage and promote me throughout different stages of my career.

Thanks to the team at Facebook AI Research for believing in me and in my potential, and for welcoming me into your family. I am looking forward to my most productive research years, and to all the amazing work we will do together to advance the field on machine learning.

Thanks to all the inspiring women in STEM I have looked up to. You are strong, creative, supportive, intelligent, and empathetic researchers, colleagues, leaders, and friends.

Finally, thanks to those who never believed in me, who thought I would never make it because of my personality, gender, nationality, or lack of all the qualities and skills you thought I needed. Proving you wrong was my biggest source of inspiration.

# Preface

I never thought this thesis would be finished. Graduate school marked the lowest point of my life so far. Behind these colorful plots and contrived equations hides the solitary experience of a real human being.

The inspiration for completing this work came primarily from my younger peers. This thesis is dedicated to them. It warmed my heart to hear words of appreciation for the time spent over-engineering a tutorial, to receive enthusiastic and thankful feedback, to watch them thrive where I failed, and to empower them to take those steps many obstacles prevented me from taking. If only equal encouragement had come from those in my field who enjoy a stature that would allow them to truly promote ingenuity, collaboration, and initiative.

The portion of this thesis I enjoyed writing the most is the quick introduction to machine learning in Ch. 4. I hope for it to offer some essential pedagogical insight and intellectual inspiration to those intrepid souls embarking on the perilous quest of bringing innovative ideas to high energy physics experiments. May they encounter smoother sailing than I did throughout my odyssey.

My desire is for this thesis to bear witness to the change brought forward by the young generations of physicists. My naïveté gives me faith that one day ATLAS graduate students will be valued, not discounted, and receive the recognition they deserve.

For their sake and for my own curiosity, while writing this thesis, I set out to find out and compile an introductory documentation on the details of the implementation of ATLAS algorithms, and to clearly expose and effectively summarize the design philosophy and process that connects the several moving parts in the ATLAS codebase. To my dismay, but to no one's surprise, this turned out to be a titanic task. The lack of structure and accountability, and the failure to identify and address institutional shortcomings has resulted in dynamics in which only those who are (or have access to those who are) in the know will be allowed the advantage of contributing to the scientific program, while hiding behind a false sense of meritocracy. Meanwhile, others will be alienated and blamed for their perceived incapacity, when the truth is that preserving the status quo is easier than promoting ingenuity.

Leaving this chapter of my career behind, my personal intention is to recuperate from the emotional and physical burdens of this environment, and to reclaim a sense of agency, meaningfulness, and eagerness. The time for *mea culpa* is long gone, and I wish to turn to the next page with no regrets nor remorse, no bitterness, no guilt. Where there is shame, may there be dignity in knowing one's limits. Where there is injustice and prejudice and favoritism, may there be fairness, integrity, and companionship.

Ad maiora.

# 0
# Introduction

WHILE RESPECTING THE LONG-STANDING TRADITION OF EXCELLENCE in the field of particle physics, it is natural for every generation of researchers to want to infuse new ideas and methodologies into their field of research. The generational struggle that ensues may weed out many, but the paradigm shift brought about by novel thoughts and techniques is less easily eradicated.

The intellectual revolution partly documented in this thesis corresponds to the rapid advent and less rapid embrace of deep learning solutions in the field of high energy physics, and, more specifically, within the ATLAS collaboration at CERN. Certainly, machine learning solutions have recently been considered more favorably than in the past, also because of the desire to utilize heterogeneous computing architectures that have become available at computing centers around the globe.

In high energy physics, deep learning can be adopted for different intents and purposes. Not only can it directly be used to increase the sensitivity of physics analyses and reconstruction techniques, but it can also be useful to get a sense of where the Bayes optimum may approximately be located for a given task, thus providing a handle to evaluate the performance and potential for improvement of traditional physics algorithms.

The work described here has contributed to introducing modern deep learning models and libraries to the scientific community at CERN, and to advancing the overall field towards more efficient solutions for particle identification and related task. Unlike more traditional theses in this field, this work sits at the confluence of physics, statistics, computing, and machine learning. The results documented here prove that significant improvements in the reach of physics analyses can be achieved through more effective utilization of the current hardware resources and data collected by the ATLAS experiment, as long as proper statistical and computational software tools are developed and utilized. This work further demonstrates that, while machine learning has already positively impacted several areas in the experiment, many other aspects and projects can and should be making use of machine learning methods. It also attests that physics should look at other fields for inspiration: this work and my own academic experience bear witness to the fact that, while we stand on the shoulder of giants, giants no longer have to be found exclusively within our field.

In this thesis, the reader will find introductions to the fields of particle physics and machine learning, along with a series of examples of how the latter can be applied to the former. For a more detailed and reliable introduction to the topics of statistics and deep learning, one should consult one of the many excellent, well-established references, such as Ref. [47] and Ref. [48].

The rest of this chapter introduces a collection of useful notions, conventions, and terms that will repeatedly appear throughout this work.

## 0.1 Notation, Conventions, and Useful Notions

### 0.1.1 Units

The International System (SI) established benchmarks and units for physical measurements. However, the units of energy (joules), distance (m), etc., are often unsuitable to describe the smallest and most fundamental components of matter that particle physics aims to study. In physics, energy is measured in units of electronvolts (eV), defined as the amount of energy required to move an electron across a 1 Volt potential.

The natural unit system assumes that there is a natural and obvious way to measure certain quantities,

and this is to do so with respect to constants of nature. For example, the natural way of measuring velocities should be with respect of the speed of light $c$, meaning that it should be more convenient to say that a particle is moving at $\frac{1}{2}c$ rather than $1.499 \times 10^8$ m/s. Starting from the assumption that all velocities should be measured as multiples of the speed of light, one might then avoid explicitly writing factors of $c$, and just measure velocities in dimensionless units. Quantities such speed of light in vacuum $c$ and the Planck constant $\hbar$ are then rescaled to unity allowing for mass, energy, and momentum to be measured in the same units (omitting factors of $c$ and $\hbar$). These quantities are therefore all measured in eV or its SI-defined base-ten multiples and submultiples (keV, MeV, GeV, TeV, and so on). Length and time may be measured in $\text{eV}^{-1}$. Final results can be converted back into SI units by reintroducing the missing factors of $c$ and $\hbar$.

Cross-sections are measured in barns (b), units of area corresponding to $1 \times 10^{-28}$ m$^2$, or approximately the cross-sectional area of the uranium nucleus, as one would expect given the introduction of this unit during the Manhattan Project era. In natural units, it can be converted to $\text{GeV}^{-2}$ via the conversion 1 b / $\hbar^2 c^2$ = 2568.19 $\text{GeV}^{-2}$. A barn corresponds to a huge cross-section for high energy physicists, so fractional units such as femtobarns (1 fb = $10^{-15}$ b) are more commonly encountered.

In particle colliders, the instantaneous luminosity corresponds the number of collisions per cross-sectional area per unit time. Integrated luminosity expresses the number of collisions per cross-sectional area integrated over a period of time, and is often measured in inverse femtobarns ($\text{fb}^{-1}$). Therefore, an event with a cross-section of 1 fb is expected to have occurred approximately 50 times in 50 $\text{fb}^{-1}$ of data.

### 0.1.2 SPECIAL RELATIVITY

The kinematics of relativistic objects are expressed in terms of their four momentum $\mathbf{p}^\mu = (E, p_x, p_y, p_z) = (E, \vec{p})$ in Minkowski space. According to special relativity, it is possible to transform particle four-vectors from one inertial frame of reference to another traveling at a certain velocity with respect to it by applying a set of transformations that maintain four-vector inner products invariant. The invariant mass is therefore defined as $p^2 \equiv g_{\mu\nu} p^\mu p^\nu = E^2 - \vec{p}^2 = m^2$, where $g_{\mu\nu}$ is the Minkowski metric, and is conserved in particle interactions such that $\sum p_{\text{initial}} = \sum p_{\text{final}}$. Lorentz transformation satisfy the invariance condition of spacetime intervals and form the Lorentz group $O(1,3)$. Extending homogeneous Lorentz transformations with translations, which also preserve the interval, gives rise to the Poincaré group. Particles are irreducible representations of the Poincaré group.

Relativistic particles obey the equation $E = \sqrt{\vec{p}^2 + m^2} = \gamma m$, where $\gamma$ is the relativistic factor that appears in the equations for the Lorentz boost.

### 0.1.3 Particle Statistics

Particle decay is regulated by Poisson statistics. Given a particle's average lifetime $\tau$, its decay width is $\Gamma = 1/\tau$. Therefore, stable particles have infinitely narrow width, while particles with short lifetimes have wider widths and will appear as resonances around their nominal mass. The total decay width is computed as the sum of the partial decay widths due to the various available decay modes, where each mode's contribution is calculated as a matrix element using Fermi's golden rule. The Breit-Wigner distribution is used to statistically model the production rate of resonances in high energy experiments.

# 1
# Theoretical Background

ALL NATURAL PHENOMENA are governed and described by a few physical principles. The desire to explain and predict observations of physical events has driven physicists to formulate a mathematical description of the tiniest components of matter and their interactions. The most up-to-date formulation is known as the Standard Model of Particle Physics.

This chapter will introduce the particles and forces described by the Standard Model in Sec. 1.1, and then focus specifically on the pair-production of Higgs bosons, both in the context of the Standard Model and in new physics theories, in Sec. 1.2.

## 1.1 THE STANDARD MODEL OF PARTICLE PHYSICS

The Standard Model (SM) of Particle Physics is generally considered one of the most powerful and comprehensive scientific theories, because of its wide-ranging scope, coverage, strong predictive power and accurate description of observed interactions among elementary particles. From its debut in the second half of the 20$^{th}$ century, this elegant model has since grown into its current incarnation, described in Ref. [49] as "a triumph of the human intellect," thanks to the contributions of the international physics community, both from the theoretical and experimental side.

Its goal is to classify subatomic particles and accurately describe their mutual and self-interactions – a goal that can be attributed to the field of particle physics as a whole. Like a periodic table, the SM arranges known elementary particles based on their properties.

According to the model, relativistic particles arise as quantized excitations of fields, where a field is a function that associates a scalar, vector, or tensor to every point in space-time. In a quantum world, to allow for the creation and annihilation of particles and therefore account for a variable number of particles in the system, the fields introduced in the classical formalism are promoted to the status of operators, with suitable commutation or anticommutation relations among creation and annihilation operators.

The mathematical ground upon which the SM is rigorously built is provided by Quantum Field Theory (QFT), a framework that reconciles quantum mechanics and special relativity, in which particles can be described by quantum fields, and various theories, such as Quantum Electro-Dynamics (QED) or Quantum Chromo-Dynamics (QCD), govern their interactions. Simplistically, if we were to, in theory, write down all possible interactions among all types of fields, we would know how to compute probabilities for all physical phenomena observed in the universe. As experimentalists, in practice, make measurements of how particles truly interact, they effectively measure to what confidence level nature agrees with theory.

Over the past few decades, experimental evidence from colliders such as LEP and the Tevatron have placed the SM on firm footing, thanks to the astounding agreement between its predictions and empirical measurements. Modern experiments such as ATLAS at the LHC have reaffirmed its high compatibility with collected data, as shown in Fig. 1.1. Although experimental observations have largely confirmed the precision and validity of the SM predictions, there are reasons to believe it might not be a complete model. For example, it is yet to encompass a quantum theory of gravity, and is unable to provide a rationale for dark matter and dark energy in terms of known particle realizations.

This gave rise to an entire ecosystem of Beyond the Standard Model (BSM) theories that try to rectify problems with the SM by adopting mathematical solutions that, in turn, may make predictions for the existence of new particles that arise from the introduction of new fields and symmetries, or may introduce detectable changes in the properties of known SM particles.

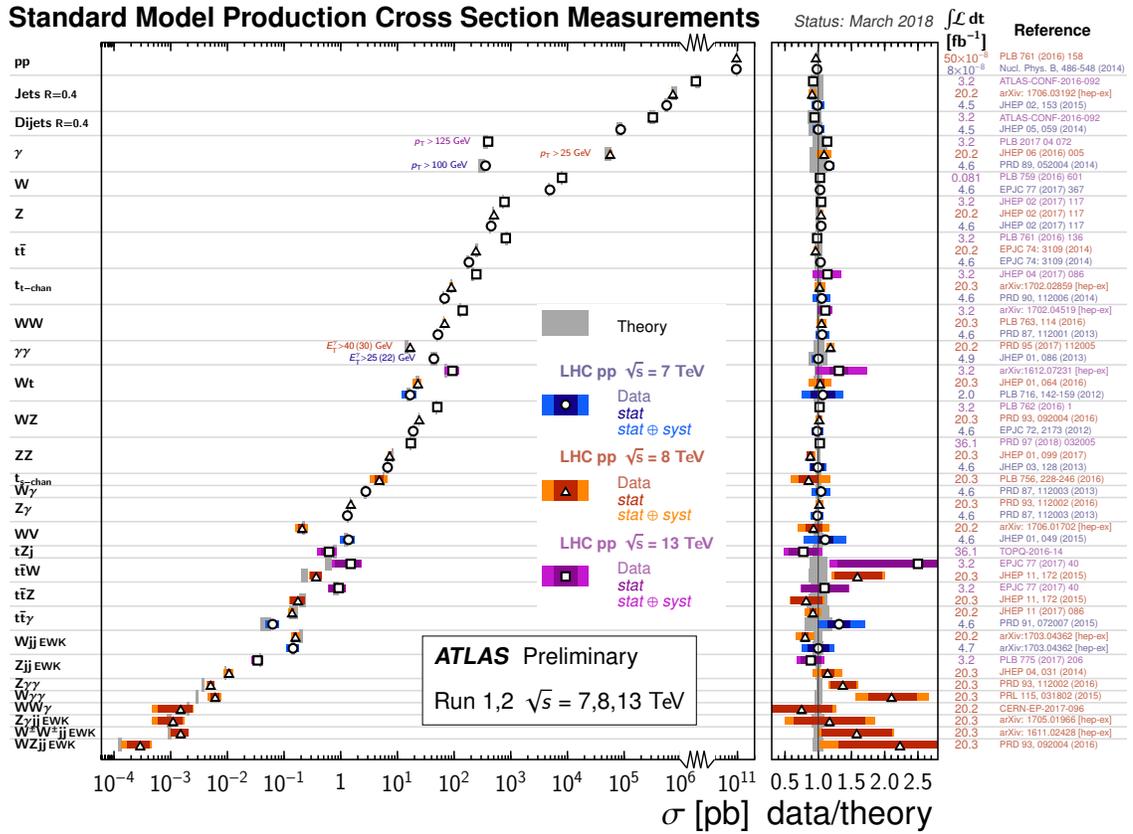

**Figure 1.1:** Summary of Standard Model production cross-sections as measured by the ATLAS detector at 7 TeV (blue), 8 TeV (orange), and 13 TeV (purple). Error bars show the statistical uncertainty in the darker shade, and the additional statistical uncertainty in the lighter shade. The theoretical predictions, computed at next-to-leading order (NLO) or higher order, are marked in gray. The right section of the plot shows the ratio between data measurement and theoretical prediction. The two columns report the integrated luminosity and the reference publication in which the measurement is first presented.

### 1.1.1 Elementary Particles

Elementary particles are fundamental particles that, as far as we currently know, have no internal structure. The Standard Model particles are arranged into the sets shown in Fig. 1.2. Fermions (spin 1/2 particles that obey Fermi-Dirac statistics) are subdivided into two distinct subsets: quarks, and leptons. The three increasingly heavier generations of quarks are $(u, d), (c, s), (t, b)$, and all possess electric charge in multiples of $e/3$. On the other hand, leptons, also arranged in three generations of increasing mass, have either unit charge $(e, \mu, \tau)$ or zero charge $(\nu_e, \nu_\mu, \nu_\tau)$. Bosons are integer-spin particles that obey Bose-Einstein statistics. Gauge bosons are spin-1 particles also known as force carriers, because they can be thought of as particles that are exchanged to mediate a particular type of fundamental interaction among the fermions. These include the massless photon $\gamma$ for electromagnetism, the massless gluons $g$ for the strong force, and the massive $Z^0$ and $W^\pm$ bosons for the weak force. Force mediators are represented in the mathematical language of QFT as vector fields. Neutrinos only interact via the weak interaction; charged leptons undergo both weak and electromagnetic interactions; quarks participate in all three fundamental interactions. The Higgs Boson, the remaining particle in the SM table in Fig. 1.2, is the spin-0 particle required for spontaneous electroweak symmetry breaking (see Sec. 1.1.2.1.4).

Every particle has a corresponding anti-particle with opposite charges, according to the CPT theorem.

Quarks are unique within the SM in that they never appear in nature at low energies as individual particles, but, with exception to the top quark, they are confined within composite particles known as hadrons (see Sec. 1.1.3). The top quark is the heaviest particle in the SM, and its short lifetime causes it to decay before it can hadronize. The mechanism that explains the nature and behavior of quarks is described in Sec. 1.1.2.2.

Spin statistic alone cannot account for the variety of elementary particles observed in nature. Other quantum numbers that govern the internal degrees of freedom must be built into a theory of interactions that progressively break symmetries to give rise to the diverse mass spectrum of elementary particles from initially degenerate states [50].

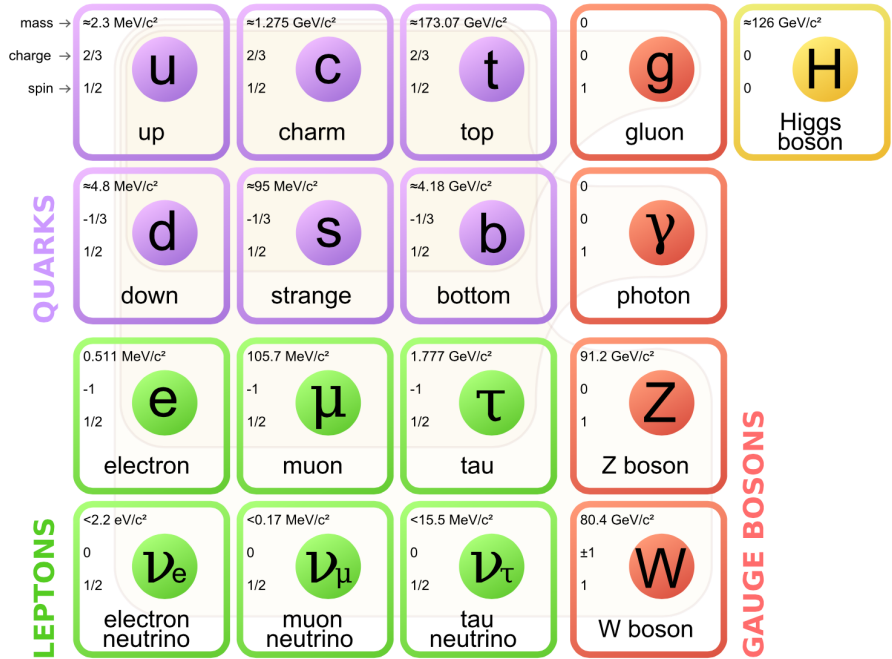

**Figure 1.2:** This Standard Model infographic lists the name and properties (mass, charge, spin) of all known SM particles. The types of particles are color coded. Quarks and leptons are arranged in three generations from left to right. The first row of quarks shows up-type quarks, the second down-type quarks. The shaded blobs in the background connect particles that couple to a given force mediator (gauge boson).

### 1.1.2 Forces and Interactions

The SM not only systematically catalogues elementary particles but also explains their interactions. Of the four fundamental interactions, only three (electromagnetism, weak force, and strong force) are described by the SM. No theory that attempts to include gravity in the picture has successfully been experimentally verified yet, but this is mostly inconsequential due to the weakness of gravity compared to the other forces.

Lagrangian formalism is used to enumerate kinetic and potenitial terms that regulate the possible interactions among fields and their relative strength.

In the language of group theory, the SM obeys an $SU(3) \times SU(2) \times U(1)$ symmetry. The SU(3) group arises from the theory of Quantum Chromo-Dynamics (see Sec. 1.1.2.2), the $SU(2)$ and $U(1)$ group from Electroweak Theory (see Sec. 1.1.2.1).

1.1.2.1   ELECTROWEAK THEORY

Electroweak (EW) theory unifies the theory of Quantum Electro-Dynamics (QED) and the Weak theory, and describes electromagnetism and the weak force as two low-energy manifestations of a more fundamendamental force.

1.1.2.1.1   QED   Quantum Electro-Dynamics is the quantum generalization of classical, relativistic electrodynamics, and can be obtained from the quantization of the electromagnetic field. In QED, given a spinor field $\Psi$ representing a fermionic field with mass $m$ as a solution to the Dirac equation, the free Lagrangian is simply $\mathcal{L} = i\bar{\Psi}\slashed{\partial}\Psi - m\bar{\Psi}\Psi$, where $\bar{\Psi}$ is the Dirac adjoint $\Psi^{\dagger}\gamma^0$ that makes the probability density $\bar{\Psi}\Psi$ and the fermion current $\bar{\Psi}\gamma^{\mu}\Psi$ a Lorentz scalar and a Lorentz vector respectively, $\Psi^{\dagger}$ is the hermitian conjugate, and, in Feynman slash notation, $\slashed{\partial} = \partial^{\mu}\gamma_{\mu}$, with $\gamma^{\mu} = (\gamma^0, \gamma^1, \gamma^2, \gamma^3)$ the Dirac gamma matrices that satisfy the anti-commutation relations $\{\gamma^{\mu}, \gamma^{\nu}\} = 2g^{\mu\nu}$. This defines the Clifford Algebra $\mathcal{C}\ell_{1,3}$. In the Weyl or chiral representation, the Dirac matrices are represented as:

$$\gamma^0 = \begin{pmatrix} 0 & \mathbb{I}_2 \\ \mathbb{I}_2 & 0 \end{pmatrix}, \quad \gamma^i = \begin{pmatrix} 0 & \sigma^i \\ -\sigma_i & 0 \end{pmatrix}, \tag{1.1}$$

where $\sigma^i$ are the Pauli matrices. In general, the $\gamma$ matrices can be used to form a set of bilinear covariants that, together with Dirac spinors, are useful to write down the sixteen quantities that we are allowed to use to construct terms in the Lagrangian. Scalars take the form $\bar{\Psi}\Psi$, pseudoscalars $\bar{\Psi}\gamma_5\Psi$, vectors $\bar{\Psi}\gamma_{\mu}\Psi$, axial vectors $\bar{\Psi}\gamma_5, \gamma_{\mu}\Psi$, antisymmetric tensors $\bar{\Psi}\frac{i}{2}[\gamma_{\mu}, \gamma_{\nu}]\Psi$.

If we require the Lagrangian to be invariant under a local gauge symmetry equivalent to a spinor phase transformation $\Psi \to \Psi' = e^{-ie\phi(x)}\Psi$, then the existence of a massless vector field is required in order to cancel the extra term in the Lagrangian involving the phase. We call the field $A_{\mu}$, and we know that for $\mathcal{L}$ to be invariant under gauge transformation, $A_{\mu}$ must transform as $A_{\mu} \to A'_{\mu} = A_{\mu} + \partial_{\mu}\phi$. The Lagrangian becomes $\mathcal{L} = -\frac{1}{4}F^{\mu\nu}F_{\mu\nu} + i\bar{\Psi}\slashed{D}\Psi - m\bar{\Psi}\Psi$, where $F^{\mu\nu} = \partial^{\mu}A^{\nu} - \partial^{\nu}A^{\mu}$ is the anti-symmetric gauge-invariant field strength tensor, $\slashed{D}$ is the covariant derivative needed to offset field transformations under U(1) gauge symmetry and recast the Lagrangian into gauge-invariant form. The first term is the kinetic term for $A_{\mu}$ that comes from classical EM theory, gives rise to the Maxwell equations, and involves the electric and magnetic fields: $-\frac{1}{4}F^{\mu\nu}F_{\mu\nu} = \frac{E^2 - B^2}{2}$. No mass term may appear in the Lagrangian for the gauge field $A_\mu$ because it would explicitly not be gauge invariant:
$\frac{1}{2}m^2 A^{\mu\prime} A'_\mu = \frac{1}{2}m^2(A^\mu + \partial^\mu\phi)(A_\mu + \partial_\mu\phi) \neq \frac{1}{2}m^2 A^\mu A_\mu$.

The constant $e$ can now be interpreted as being proportional the coupling strength of the vector field to fermions, known as the fine structure constant $\alpha = e^2/4\pi = 1/137$. The vector field $A_\mu$ describes the gauge boson of the theory, the photon.

In general, gauge transformations remove redundancies in the theory. In the case of QED, the redundancy was due the number of the degrees of freedom in the four-dimensional vector field $A_\mu$ being greater than the number of polarizations in EM radiation (2). Arbitrary, yet convenient, gauge conditions can be imposed on the components of $A_\mu$ to lower the degrees of freedom.

The spinor representation is reducible into two irreducible representations, so Dirac spinors can be decomposed into left-handed and right-handed Weyl spinors:

$$\Psi = \begin{pmatrix} \Psi_L \\ \Psi_R \end{pmatrix}. \tag{1.2}$$

The projections onto the right and left states are obtained, in terms of the matrix

$$\gamma^5 = i\gamma^0\gamma^1\gamma^2\gamma^3 = \begin{pmatrix} -\mathbb{I}_2 & 0 \\ 0 & \mathbb{I} \end{pmatrix}, \tag{1.3}$$

from the action of the projection operators $P_R = \frac{1+\gamma^5}{2}$ and $P_L = \frac{1-\gamma^5}{2}$:

$$P_R\Psi = \begin{pmatrix} 0 & 0 \\ 0 & \mathbb{I}_2 \end{pmatrix}\Psi = \Psi_R, \quad P_L\Psi = \begin{pmatrix} \mathbb{I}_2 & 0 \\ 0 & 0 \end{pmatrix}\Psi = \Psi_L. \tag{1.4}$$

The two Weyl spinors undergo the same transformation under rotations, but transform in opposite ways under Lorentz boosts.

Parity is the operation that transforms the sign of the space coordinates. Invariance under parity transformation is referred to as chiral symmetry. The complementary transformation that only flips the sign of the time component is called time reversal; both time reversal $T$ and parity $P$ are discrete symmetries of the Lorentz group. The two Weyl spinors are connected through a parity transformation:

$$P\Psi_{R/L}(t,\vec{x}) = \Psi_{L/R}(t,-\vec{x}).$$

Rewriting the Lagrangian explicitly in terms for right-handed and left-handed spinors, we obtain:

$$\mathcal{L} = -\frac{1}{4}F^{\mu\nu}F_{\mu\nu} + i\bar{\Psi}_L \slashed{D}\Psi_L + i\bar{\Psi}_R \slashed{D}\Psi_R - m\left[\bar{\Psi}_R\Psi_L + \bar{\Psi}_L\Psi_R\right]. \tag{1.5}$$

The mass term is responsible for the mixing of right and left-handed states. Conversely, massless fermions are stuck in their one intrinsic handedness state.

1.1.2.1.2    FERMI'S THEORY OF WEAK INTERACTIONS    It was initially strongly believed that parity would be conserved in all interactions. Experimental evidence from decays, however, showed no parity conservation in weak interactions, and that nature had a clear preference for left-handed neutrinos and right-handed anti-neutrinos. Fermi, who was particularly keen to solve the mystery of $\beta$ decay, in analogy to QED, proposed the description of the 3-body decay reaction $n \to p + e + \nu$ by writing down QED-like conserved vector current terms in the Legrangian of the type: $j_\mu^{np} = \bar{\Psi}_p\gamma_\mu\Psi_n$ and $j_{\nu e}^\mu = \bar{\Psi}_e\gamma^\mu\Psi_\nu$, replacing the electric charge with another coupling constant $G_F$ to parametrize the strength of the interaction. In practice, though, any bilinear covariant term can be used in the construction of valid terms in the Lagrangian. The best fit to experimental data provided evidence for the need of both vector and parity-violating axial vector interactions (V-A) in equal proportion (maximal parity violation) [51]. The interaction between two fields therefore becomes $\frac{1}{2}(\bar{\Psi}_A\gamma^\mu\Psi_B - \bar{\Psi}_A\gamma^\mu\gamma^5\Psi_B) = \bar{\Psi}_A\gamma^\mu\frac{1-\gamma^5}{2}\Psi_B$. The left projection operator appears in this formula, meaning that the charged weak force couples only to left-handed chiral component of fermions (and the right-handed chiral component of anti-fermions). In the SM, neutrinos are assumed to be massless particles, so their helicity and chirality are identical. Since neutrinos are always observed to have left-handed helicity, it means that their right-handed chiral component must be zero. The fact that only left-handed chiral states participate in the weak interaction as formulated thus far creates another issue: the mass term in Eq. 1.5 that allows left and right handed Weyl spinors to mix is no longer allowed.

Fermi's theory described $\beta$ decay and similar fermionic processes with remarkable accuracy, yet this four-fermion contact interaction picture would eventually be surpassed by a full theory of weak interactions mediated by charged and neutral vector bosons.

Although the weak Fermi interaction was eventually promoted to the status of universal interaction with its own coupling constant, similar to QED, its observed coupling strength in strange decays seemed to violate the principle of universality. Cabibbo put forward the hypothesis that the eigenstates the weak force acts on are linear combinations of the physical mass eigenstates that appear in QED and QCD. This allows the weak force to maintain its universal strength on the superposition object, while diminishing it on the individual mass eigenstates. The decrease was then parametrized in terms of the Cabibbo angle $\theta_C$, such that $d' = d \cos \theta_C + s \sin \theta_C$, where $d$ and $s$ are the down and strange mass eigenstates, respectively. This indicates the presence of flavor mixing as a rotation between mass and weak eigenstates, and allows for weak universality to be restored. This discovery was later extended into the Cabibbo-Kobayashi-Maskawa (CKM) matrix to account for the presence of three generations of quarks. Each weak eigenstate is expressed as a linear superposition of mass eigenstates $q'_i = \sum_{j=1}^{3} V_{ij} q_j$, where $V$ is the CKM matrix that connects different quark generations and allows for weak transitions among them.

The CKM matrix can be written as [2]:

$$V_{\text{CKM}} = \begin{pmatrix} V_{ud} & V_{us} & V_{ub} \\ V_{cd} & V_{cs} & V_{cb} \\ V_{td} & V_{ts} & V_{tb} \end{pmatrix} \tag{1.6}$$

$$= \begin{pmatrix} 0.97446 \pm 0.00010 & 0.22452 \pm 0.00044 & 0.00365 \pm 0.00012 \\ 0.22438 \pm 0.00044 & 0.97359^{+0.00010}_{-0.00011} & 0.04214 \pm 0.00076 \\ 0.00896^{+0.00024}_{-0.00023} & 0.04133 \pm 0.00074 & 0.999105 \pm 0.000032 \end{pmatrix}$$

$$= \begin{pmatrix} c_{12}c_{13} & s_{12}c_{13} & s_{13}e^{-i\delta} \\ -s_{12}c_{23} - c_{12}s_{23}s_{13}e^{i\delta} & c_{12}c_{23} - s_{12}s_{23}s_{13}e^{i\delta} & s_{23}c_{13} \\ s_{12}s_{23} - c_{12}c_{23}s_{13}e^{i\delta} & c_{12}s_{23} - s_{12}c_{23}s_{13}e^{i\delta} & c_{23}c_{13} \end{pmatrix} \tag{1.7}$$

$$= \begin{pmatrix} 1 - \lambda^2/2 & \lambda & A\lambda^3(\rho - i\eta) \\ -\lambda & 1 - \lambda^2/2 & A\lambda^2 \\ A\lambda^3(1 - \rho - i\eta) & -A\lambda^2 & 1 \end{pmatrix} + \mathcal{O}(\lambda^4) \tag{1.8}$$

In Eq. 1.7, $s_{ij} = \sin \theta_{ij}$ and $c_{ij} = \cos \theta_{ij}$. Three Euler angles and one complex phase $\delta$ are therefore sufficient to parametrize the matrix. Since this phase is the only one that appears in the SM, it is understood as the possible source of CP violation. These are free parameters of the SM that need to be measured experimentally.

In the last (Wolfenstein) parametrization, in Eq. 1.8:

$$\lambda = s_{12} \tag{1.9}$$

$$A\lambda^2 = s_{23} \tag{1.10}$$

$$A\lambda^3(\rho - i\eta) = s_{13}e^{-i\delta} \tag{1.11}$$

Here, as well as from the numerical magnitudes of the CKM elements, it is evident that the CKM matrix is almost diagonal.

The leptonic part of the Lagrangian may also exhibit flavor mixing in an analogous way to its quark counterpart. Lepton flavor violation is encoded into the unitary Pontecorvo-Maki-Nakagawa-Sakata (PMNS) matrix, introducing additional free parameters but providing a tool to parametrize the recently confirmed neutrino oscillations.

1.1.2.1.3  GLASHOW MODEL    QED and the Fermi theory of weak interactions described above are connected under the more general framework of EW theory. Fermi's formulation, in fact, diverges at high energies, and is therefore only acceptable as a first-order low-energy approximation of the true interaction potential. The Glashow model generalizes charges and currents, and dynamically mixes them in a unified theory of interactions.

Analogous to QED, the weak force, initially described as a non-renormalizable contact interaction by Fermi, can be formulated as proceeding via the exchange of parity-violating intermediate vector bosons in terms of a universal coupling of matter particles to these force carriers, as predicted by Schwinger [50]. The new dimensionless coupling constant to charged vector bosons $g$ is then related to the original Fermi constant $G_F$ by $\frac{g^2}{8m_W^2} = \frac{G_F}{\sqrt{2}}$, where $m_W$ is the mass of the weak force carrier.

The unification of electroweak theory stems from the application of Yang-Mills theory [52], *i.e.* the extension of gauge invariance to non-abelian theories, which leads to the prediction of the existence of massless charged vector bosons. In analogy to isospin, the three-component weak isospin $T$ connects quarks and leptons across generations, and is realized as a symmetry under SU(2)$_T$ transformations. The third component is identified at $T_3$ and is conserved in weak interactions.

Fermions are arranged into SU(2)$_T$ doublets and singlets:

$$\begin{pmatrix} \nu_{l,L} \\ l_L \end{pmatrix}, \nu_{l,R}, l_R \qquad \begin{pmatrix} u_{q,L} \\ d_{q,L} \end{pmatrix}, u_{q,R}, d_{q,R}, \tag{1.12}$$

with $l$ and $q$ running over the three lepton and quark generations.

The two components in left-handed doublets are labeled with quantum number $T_3 = \pm 1/2$, while right-handed singlets have $T_3 = 0$ and are fixed points of the rotation (they do not participate in the weak interaction). In other words, the singlets of the weak isospin group SU(2)$_T$ are invariant under SU(2)$_T$ rotations and transform in the trivial representation, while the doublets transform in the standard representation of SU(2)$_T$: $\Psi_L \to \Psi'_L = e^{i\alpha T} \Psi_L$. The free SU(2)-invariant Lagrangian can be written as $\mathcal{L} = \sum_L \bar{\Psi}_L \gamma^\mu \partial_\mu \Psi_L + \sum_R \bar{\Psi}_R \gamma^\mu \partial_\mu \Psi_R$, where the sums are over all the left-handed quark and lepton doublets ($L$), and all the right-handed quark and lepton singlets ($R$).

Expanding the symmetry to the larger group SU(2)$_T \times$ U(1)$_Y$, the eigenvalues of the two commuting hermitian operators $T_3$ and $Y$ generalize the concept of charge. $Y$ is known as hypercharge. The total electric charge is the conserved quantity $Q = T_3 + Y/2$. In 1961, Glashow [53] realized that extending the symmetry of the theory to SU(2)$_T \times$ U(1)$_Y$ accounted for an additional electrically neutral gauge boson, the $Z^0$ boson. However, this theory failed to explain how to generate masses without explicit symmetry violating terms.

In order for the Lagrangian to be invariant under local gauge transformations, one needs to introduce an isotriplet field $W_\mu$ and an isosinglet $B_\mu$ to offset extra symmetry-violating terms. For the triplet of vector fields, the field strength tensor becomes $W^{a\mu\nu} = \partial^\mu W^{a\nu} - \partial^\nu W^{a\mu} + gf^{abc}W^{b\mu}W^{c\nu}$, where $f$ is the totally anti-symmetric structure constant, and the covariant derivative picks up an extra gague field-dependent term: $\not{D}_L = \gamma^\mu(\partial_\mu - igt_a W^a_\mu)$, where the $t_a$'s are the Lie algebra generators with the following commutation relation: $[t^a, t^b] = if^{abc}t^c$. If, in addition, we add the extra weak isosinglet field $B_\mu$, which couples to the matter fields with strength $g'$, the partial derivative is now replaced by the covariant derivative

$$D_\mu = \partial_\mu + ig\frac{1}{2}\sigma^a W_{\mu a} + ig'\frac{1}{2}Y B_\mu \tag{1.13}$$

to preserve local gauge invariance. The Lagrangian becomes:

$$\mathcal{L}_G = -\frac{1}{4}B^{\mu\nu}B_{\mu\nu} - \frac{1}{4}W^{a\mu\nu}W_{a\mu\nu} +$$
$$+ \sum_L \bar{\Psi}_L \gamma^\mu (\partial_\mu - ig\frac{\sigma_a}{2}W^a_\mu + ig'\frac{1}{2}YB_\mu)\Psi_L + \sum_R \bar{\Psi}_R \gamma^\mu (\partial_\mu + ig'YB_\mu)\Psi_R \qquad (1.14)$$

Fermionic mass terms such as the ones in Eq. 1.5 are no longer allowed as the two Weyl spinors transform differently under isospin rotation: $\Psi_R$ is an isospin singlet ($\Psi_R \to \Psi'_R = e^{i\beta Y}\Psi_R$), while $\Psi_L$ is an isospin doublet ($\Psi_L \to \Psi'_L = e^{i\alpha T + i\beta Y}\Psi_L$).

Although the Glashow model introduced the four EW gauge bosons, it did not account for a mechanism for them to acquire mass without spoiling gauge invariance.

1.1.2.1.4 ELECTROWEAK SYMMETRY BREAKING  Several lecture series, such as Ref. [54], are available on this subject. This section aims to summarize the key aspects of the Brout-Englert-Higgs mechanism in the context of EW theory.

It became evident from the short range of weak interactions that the force had to be mediated by massive vector bosons. However, such mass terms were forbidden (as in the QED Lagrangian) because they would spoil gauge invariance. The forbidden fermion and gauge boson mass terms and the unphysical presence of Goldstone modes in electroweak theory point to the need for spontaneous symmetry breaking in the vacuum state of the gauge-invariant Lagrangian. Weinberg and Salam realised that a unified SU(2)$_T$ × U(1)$_Y$ gauge theory with spontaneously broken symmetry down to U(1)$_{\text{EM}}$ would generate one massless and three massive gauge fields, the photon and the $W^\pm, Z^0$ bosons. Therefore, problems with masslessness and gauge invariance can be solved by introducing spontaneous symmetry breaking. The underlying mechanism was first proposed in 1964 by three different groups [55, 56, 57].

Electroweak symmetry breaking (EWSB) requires the introduction of an isospin doublet of complex scalar fields that preserves SU(2)$_T$ × U(1)$_Y$ symmetry, and transforms in a two-dimensional representation:

$$\Phi = \begin{pmatrix} \Phi^+ \\ \Phi^0 \end{pmatrix} = \frac{1}{\sqrt{2}} \begin{pmatrix} \Phi_1 + i\Phi_2 \\ \Phi_3 + i\Phi_4 \end{pmatrix}. \qquad (1.15)$$

This is a left-handed isospin doublet with hypercharge $Y = 2(Q - T_3) = 1$.

The Lagrangian becomes:

$$\mathcal{L} = \mathcal{L}_G + (D^\mu \Phi)^\dagger (D_\mu \Phi) - \mu^2 \Phi^\dagger \Phi - \lambda (\Phi^\dagger \Phi)^2 \tag{1.16}$$

where the $D_\mu$ is the covariant derivative in Eq. 1.13, and $\mathcal{L}_G$ is the Lagrangian in Eq. 1.14.

With $\lambda > 0$ and $\mu^2 < 0$, the vacua are shifted with respect to the origin, so investigating the particle spectrum by studying small perturbations around the minimum becomes more intuitive with an appropriate change of variables. Without loss of generality, out of the infinite number of degenerate minima, pick the vacuum location with $\Phi_1 = \Phi_2 = \Phi_4 = 0$ and $\Phi_3 = v = \sqrt{-\frac{\mu^2}{\lambda}}$ and study small perturbations around it:

$$\Phi_{\text{vacuum}} = \frac{1}{\sqrt{2}} \begin{pmatrix} 0 \\ v+h \end{pmatrix} \tag{1.17}$$

in the unitary gauge, where gauge fixing removes the degree of freedom associated with rotations along the locus of minima. In other words, the massless Goldstone bosons that arise as perturbations of the field $\Phi$ along the parallel direction through the minima can be gauged away.

One can check that the vacuum breaks $SU(2)_T$ and $U(1)_Y$, while $U(1)_{\text{EM}}$ remains unbroken. The four EW gauge bosons $(W_1, W_2, W_3, B)$ that represent an orthogonal basis of weak eigenstates become massive, while the EM gauge boson, the photon, remains massless. In other words, a suitable change of basis that diagonalizes the bilinear, symmetric form representing the mass yields a set of broken generators and an unbroken generator corresponding to the three massive and one massless mass eigenstates $W^\pm, Z^0, \gamma$ [58]. To verify this, take the kinetic term for the vacuum scalar field $(D^\mu \Phi)^\dagger (D_\mu \Phi)$:

$$D_\mu \Phi = \frac{1}{\sqrt{2}} \left[ ig \frac{1}{2} \sigma^a W_{\mu a} + ig' \frac{1}{2} Y B_\mu \right] \begin{pmatrix} 0 \\ v \end{pmatrix} \tag{1.18}$$

$$= \frac{i}{2\sqrt{2}} \left[ g \left( \begin{pmatrix} 0 & W_1 \\ W_1 & 0 \end{pmatrix} + \begin{pmatrix} 0 & -iW_2 \\ iW_2 & 0 \end{pmatrix} + \begin{pmatrix} W_3 & 0 \\ 0 & -W_3 \end{pmatrix} \right) + g' \begin{pmatrix} YB_\mu & 0 \\ 0 & YB_\mu \end{pmatrix} \right] \begin{pmatrix} 0 \\ v \end{pmatrix} \tag{1.19}$$

$$= \frac{iv}{2\sqrt{2}} \begin{pmatrix} g(W_1 - iW_2) \\ -gW_3 + g'YB_\mu \end{pmatrix} \tag{1.20}$$

so that

$$(D^\mu \Phi)^\dagger (D_\mu \Phi) = \frac{v^2}{8} \left[ g^2(W_1^2 + W_2^2) + (-gW_3 + g'YB_\mu)^2 \right]. \tag{1.21}$$

In this basis, the bilinear form $M$ is clearly not diagonal:

$$M = \frac{v^2}{4} \begin{pmatrix} g^2 & 0 & 0 & 0 \\ 0 & g^2 & 0 & 0 \\ 0 & 0 & g^2 & -gg' \\ 0 & 0 & -gg' & g'^2 \end{pmatrix}. \tag{1.22}$$

Since $M$ is a symmetric matrix, a new orthonormal basis can be defined via orthogonal diagonalization $M = Q\Lambda Q^T$ by solving for the eigenvalues and eigenvectors of $M$. The new basis becomes:

$$W^\pm = \frac{1}{\sqrt{2}}(W_1 \mp iW_2) \tag{1.23}$$

$$Z_\mu = \frac{1}{\sqrt{g^2 + g'^2}}(gW_3 - g'B_\mu) \tag{1.24}$$

$$A_\mu = \frac{1}{\sqrt{g^2 + g'^2}}(g'W_3 + gB_\mu) \tag{1.25}$$

and one can verify that in this basis, by construction, the mass matrix $\Lambda$ is diagonal, so that Eq. 1.21 becomes:

$$(D^\mu \Phi)^\dagger (D_\mu \Phi) = \frac{v^2}{8} \left[ g^2(W^+)^2 + g^2(W^-)^2 + (g^2 + g'^2)Z_\mu^2 + 0 \cdot A_\mu^2 \right]. \tag{1.26}$$

This is exactly the form of four mass terms for the $W^\pm$, $Z^0$ and $\gamma$ force carriers. The last term confirms that the linear combination of weak gauge fields that gives rise to the photons preserves its expected masslessness. On the other hand:

$$\frac{1}{2} m_{W^\pm}^2 (W^\pm)^2 = \frac{v^2}{8} g^2 (W^\pm)^2 \rightarrow m_{W^\pm} = \frac{v}{2} g \tag{1.27}$$

and

$$\frac{1}{2} m_Z^2 Z^2 = \frac{v^2}{8}(g^2 + g'^2)Z^2 \rightarrow m_Z = \frac{v}{2}\sqrt{g^2 + g'^2}. \tag{1.28}$$

The ratio of the two is expressed in terms of the Weinberg angle, or weak mixing angle, a free parameter in the SM, as:

$$\frac{m_W}{m_Z} = \frac{g}{\sqrt{g^2 + g'^2}} = \cos\theta_W \tag{1.29}$$

Investigating physics at the electroweak symmetry breaking scale is one of the original physics objective of the ATLAS experiment at the LHC. On July 4$^{th}$ 2012, both the ATLAS and CMS collaborations at CERN announced the discovery of a new scalar boson whose properties are in agreement with the $h$ boson predicted by the Standard Model [59, 60]. This result solidifies the basis of electroweak theory and so far confirms our understanding of how fermions and gauge bosons acquire mass in a local gauge invariant way.

Further measurements of the Higgs potential beyond EWSB and the study of the properties of the recently-discovered particle, such as its trilinear self-coupling, are necessary to confirm its role within the Standard Model.

1.1.2.1.5  YUKAWA INTERACTION   The fermionic fields do not acquire mass through the same mechanism as the gauge bosons, but their mass terms are introduced through Yukawa coupling to the Higgs field. The Yukawa interaction describes a system with interacting Dirac spinor $\Psi$ and a scalar field $\Phi$. This coupling is used in the SM to add mass terms for chiral fermions through interactions with the Higgs field while preserving gauge invariance.

The fermion-boson interaction is expressed as a cubic interaction term of the form $g_Y \bar{\Psi} \Phi \Psi$, where $g_Y$ is the Yukawa coupling constant. The Lagrangian takes the following form:

$$\mathcal{L} = \mathcal{L}(\Phi) + \mathcal{L}(\Psi) + \mathcal{L}(\Phi, \Psi) = \frac{1}{2} D^\mu \Phi D_\mu \Phi - V(\Phi) + \bar{\Psi}(i\slashed{D} - m)\Psi - g_Y \bar{\Psi} \Phi \Psi. \quad (1.30)$$

Specifically, in the case in which $\Phi$ is the Higgs field, this interaction assigns mass to quarks and leptons. Expanding the last term explicitly using the chiral spinors in Eq. 1.12 and the Higgs field representation in Eq. 1.15,

$$g_Y \bar{\Psi} \Phi \Psi = \sum_{l=1}^{3} g_{Y,l}(\bar{\nu}_{l,L}\Phi^+ + \bar{l}_L \Phi^0)(\nu_{l,R} + l_R) + \sum_{q=1}^{3} g_{Y,q}(\bar{u}_{q,L}\Phi^+ + \bar{d}_L \Phi^0)(u_{q,R} + d_R) + h.c. \quad (1.31)$$

where $l$ and $q$ label the three generations of leptons and quarks.

Under spontaneous symmetry breaking, the scalar field acquires non-zero vacuum expectation value and can be written as in Eq. 1.17. Fermionic mass terms emerge naturally from the Yukawa Lagrangian above in the form: $m_i \bar{\Psi}_i \Psi_i = g_{Y,i} v/\sqrt{2}$, where the Yukawa couplings $g_{Y,i}$ are free parameters of the

SM.

### 1.1.2.2 QCD

Quantum Chromo-Dynamics (QCD) is the fundamental gauge theory of strong interactions, which describes the coupling of quarks and gluons according to the coupling $\alpha_S$. Collectively, these particles are known as partons.

Historically, the conjecture that quarks existed came from the interpretation of hadron spectroscopy experiments. Further experimental evidence from lepton-nucleon scattering, and in $e^+e^-$ annihilation resulting in hadrons provided irrefutable verification for the existence of quarks. The TASSO collaboration at the Positron-Electron Tandem Ring Accelerator (PETRA) at DESY is credited for finding the first direct evidence for the existence of the gluon in 3-jet events, where one of the jets was identified as the signature of a hard gluon radiation [61], as predicted in prior theoretical work [62].

Given the observational evidence, for quarks to be spin 1/2 particles that obey Fermi-Dirac statistics, a new quantum number ought to exist for the total wave function to be anti-symmetric. To address this, color was introduced as an internal SU(3) symmetry. Experimentally, measurements of the $R$-ratio

$$R_{\text{had}} = \frac{\sigma(e^+e^- \to \text{hadrons})}{\sigma(e^+e^- \to \mu^+\mu^-)} = N_c \sum_f q_f^2 \tag{1.32}$$

where $q_f$ is the EM charge of each quark flavor $f$, demonstrated the presence of $N_c = 3$ colors. The fundamental SU(3) representation is given by the color triplet (*red*, *blue*, *green*). Quarks can be found in one of the three color-charge states, and the color-state covariant vectors can be rotated by 3 × 3 unitary matrices. Specifically, a valid SU(3) rotation is a transformation of the form $U = e^{-i\sum_{j=1}^{8} \theta_j \lambda_j / 2}$ with det$U = 1$, where $\theta_i$ are arbitrary parameters and $\lambda_i$ are conventionally chosen to be the Gell-Mann matrices. On the other hand, antiquarks are described by contravariant vectors. In other words, quarks transform as a 3 representation of the SU(3) symmetry and antiquarks transform as $\bar{3}$ representation, such that the inner product of the two representations is invariant under SU(3). Taking direct products of these two fundamental representations of SU(3) gives linear combinations of higher dimensional representations of SU(3) that can be arranged in multiplets. These represent composite quark bound states (see Sec. 1.1.3).

The QCD Lagrangian must therefore be invariant under local SU(3) gauge transformations. The Dirac spinor $\Psi_{fa}$ that represents the quark field is a solution to the Dirac equation. It is labeled by its weak eigenvalue (flavor) $f$ and strong eigenvalue (color) $a$. Requiring gauge invariance mandates the introduction of the covariant derivative $\partial_\mu \to D_\mu = \partial_\mu + i\alpha_S t^i A^i_\mu$ in analogy with QED, where $A^i_\mu$ are the 8 massless spin-1 gluon fields with field strength tensor $F_{\mu\nu i} = \partial_\mu A_{\nu i} - \partial_\nu A_{\mu i} + \alpha_S f_{ijk} A_{\mu j} A_{\nu k}$, with $f_{ijk}$ the SU(3) structure constants, and $t_i = \lambda_i/2$.

The gauge invariant lagrangian of QCD is a sum of the gluon and quark terms:

$$\mathcal{L} = -\frac{1}{4} F^i_{\mu\nu} F^{i\mu\nu} + \bar{\Psi}_{fa}(i\gamma^\mu \partial_\mu \delta_{ab} - \alpha_S \gamma^\mu t^i_{ab} A^i_\mu - m_f \delta_{ab}) \Psi_{fb} \tag{1.33}$$

where $f$ labels the quark flavor, $i \in [1, 8]$ labels the gluons, and the $t^i_{ab}$'s are the SU(3) group generators. Specifically, from the standpoint of group theory, in SU($n$), the initial $2n^2$ parameters are reduced to $n^2 - 1$ by the unitarity and determinant conditions, meaning that, for SU(3), there will be 8 generators $t$. The generators of the group are $3 \times 3$ hermitian (due to unitarity), traceless (because det$U = 1$) matrices $t_a = \lambda_a/2$.

SU(3) is a non-abelian group, meaning that the elements of the group do not commute: $[t_a, t_b] = i \sum_{c=1}^{8} f_{abc} t_c$. A consequence of the non-abelian nature of the symmetry is that the gauge bosons of the theory, the gluons, also carry color charge and can therefore couple to themselves. The self-coupling is evident in the last term of the field strength tensor which couples gluon fields.

The first term in the Lagrangian describes free gluons; the third term is the color current that couples quarks and gluons with strength $\alpha_S$.

The label $f$ runs over the $N_f = 6$ quark flavors and introduces 6 free parameters for the quark masses $m_f$. Since the mass of the $u$ and $d$ quarks are similar but not identical, isospin is only an approximate symmetry of the Lagrangian.

1.1.2.2.1  THE ROLE OF $\alpha_S$ IN QCD    The phenomenology of strong interactions varies at different distance scales, due to the presence of color-charged force carriers that may appear in loops. The fundamental coupling of QCD, known as $\alpha_S$, is responsible for regulating the interaction strength of quarks and gluons. It is the free parameter that enters the QCD Lagrangian, and, as such, it represents a cornerstone for our understanding of strong interactions. Its effective magnitude depends on the momentum transfer involved in the interaction, and this dependence controls the quark-gluon dynamics at different scales. Each process is characterized by a momentum-transfer scale $Q^2 = -q^2 (> 0$ for space-like processes) at which the value of the coupling $\alpha_S$ is evaluated. It is, therefore, imperative to fully understand the dependence of the coupling on the momentum transfer to be able to describe phenomena at both long and short distances. Uncertainties in the modeling of $\alpha_S(Q^2)$ influence results of experiments at the Large Hadron Collider in the form of theoretical uncertainties. The study of grand unification theories also relies on precise measurements of coupling constants as a way of validating predictions of their values.

QCD exhibits different interaction strengths in the short and long distance scale regimes. At short distances, or high momentum transfer, partons are asymptotically free, and the interaction strength decreases. At large distance scales, the strong force turns on, forcing partons to hadronize. The two regimes are described separately.

At high $Q^2$, perturbative QCD (pQCD) can be used for the description of short-distance phenomena where asymptotic freedom holds. These are relevant, for example, for the continued probing of high-energy unification theories. Regularization of UV-divergent terms in the Lagrangian, due to phenomena like vacuum polarization and quark self-energy, causes the coupling constant to acquire a dependence on the renormalization scale used to ground the value of observables to their measured values. The renormalization mass scale $\mu$ is introduced to represent the arbitrary momentum scale for the subtraction of UV-divergent terms. All observables must be independent of this subtraction point, while the scale dependence is folded into the definition of the scale-dependent coupling $\alpha_S$. Choosing the renormalization scale to equal the momentum transfer of the process $Q$, and introducing the renormalization scheme-dependent QCD scale parameter $\Lambda$, pQCD relates the nature of interactions at one length scale to that at the next scale via the equation

$$\alpha_S(Q^2) = \frac{4\pi}{\beta_0 \ln(Q^2/\Lambda^2)}, \tag{1.34}$$

where $\beta_0$ is the 0[th] order term in the perturbative expansion of the $\beta$-function that appears in the renormalization group equation, and, at $\beta_0$ order, $\Lambda^2 = \mu^2 \exp\left(-\frac{4\pi}{\beta_0 \alpha_S(\mu^2)}\right)$. Substituting, to first order:

$$\alpha_S(\mu^2) = \frac{\alpha_S(Q^2)}{1 + \frac{\beta_0}{4\pi}\alpha_S(Q^2)\ln(\mu^2/Q^2)} \quad (1.35)$$

Expanding this gives the approximation:

$$\alpha_S(\mu^2) \approx \alpha_S(Q^2)\left(1 - \frac{\beta_0}{4\pi}\alpha_S(Q^2)\ln(\frac{\mu^2}{Q^2}) + \mathcal{O}(\alpha_S^2)\right) \quad (1.36)$$

Therefore, given a renormalization scheme, the value of the coupling $\alpha_S(Q^2)$ at any order can be expressed in terms of the numerical value of $\alpha_S$ at a known scale, such as $M_Z^2$. Experimental measurements of $\alpha_S$ are usually compared at the fixed energy scale corresponding to the $Z$ boson mass. pQCD is therefore able to model the energy dependence of $\alpha_S$, but its value at any given energy must be measured experimentally [63].

Since the terms in the $\beta$-function depend on $n_f$, the number of quark flavors for which $m_q^2 \ll Q^2$ at a given scale $Q$, so does $\alpha_S$, and threshold matching is necessary to allow it to vary smoothly at the transition points. At 1-loop order, the term $\beta_0 = 11 - \frac{2}{3}n_f$ receives contributions from two diagrams: the second term is negative and proportional to the number of active quarks, and, just like in QED, it strengthens the interaction at short distances; the first term, instead, is positive and typical of QCD, and it represents the antiscreening effect due to gluon self-coupling that decreases the strength of the force at short distances [64].

Eq. 1.36 implies that at leading order the effective coupling is approximately constant, while the running behavior emerges at next-to-leading order. The dependence only fully disappears if all terms are included; otherwise, the cancellation will not be perfect. Varying the (unphysical) renormalization scale $\mu$ then modifies the value of $\alpha_S$ and of $\alpha_S$-dependent cross-sections. These variations in the renormalization scale are used to estimate the uncertainty associated to the approximation of $\alpha_S$ to leading orders, while ignoring the effects of higher-order terms. $\mu$ is habitually varied up and down by a factor of 2 to quantify this uncertainty.

At low $Q^2$, the value of $\alpha_S$ is expected to "freeze out" to a scale-independent point, so that models of non-perturbative QCD must be adopted to describe physical processes at the infrared (IR) limit, such as color confinement and hadronization. Lattice QCD is one such framework in which space is discretized into finite-size hypercubes.

To establish a connection between the two regimes, it is possible to identify a scale at which the crossover between the perturbative and non-perturbative regimes occurs. By evolving the pQCD calculation to $Q = \Lambda$, we encounter a divergence in the $\alpha_S$ equation, known as the Landau Pole. It is at this (unphysical) pole that the perturbative approach breaks down. Perturbative QCD is therefore only valid down to energies $\approx 1$ GeV, or distance scales of the order of hadron sizes.

Non-perturbative effects enter, for example, the calculation of hadronic cross-sections in the form of scale-dependent parton distribution functions $f(x, Q^2)$ that describe the collinear emission of quarks and gluons carrying a certain fraction $x$ of the hadron momentum that are seen as modifying the internal structure of the hadron. Parton distribution functions (PDFs) encode the probability of resolving a parton carrying a fraction $x$ of the hadron momentum at the energy scale $Q^2$. The factorization theorem is what allows us to fold the scale dependence of the hadronic cross-section into the PDF, while decoupling it from the perturbative calculation of the parton-level cross-section. The Dokshitzer–Gribov–Lipatov–Altarelli–Parisi (DGLAP) equations are a set of integro-differential equations used to describe the leading-order coupled scale evolution of the running parton densities $f$ [4]:

$$\frac{d}{d\log Q} f_g(x, Q) = \frac{\alpha_S(Q^2)}{\pi} \int_x^1 \frac{dz}{z} \{P_{g\leftarrow q}(z) \sum_f \left[ f_f\left(\frac{x}{z}, Q\right) + f_{\bar{f}}\left(\frac{x}{z}, Q\right) \right] + P_{g\leftarrow g}(z) f_g\left(\frac{x}{z}, Q\right)\} \tag{1.37}$$

$$\frac{d}{d\log Q} f_f(x, Q) = \frac{\alpha_S(Q^2)}{\pi} \int_x^1 \frac{dz}{z} \{P_{q\leftarrow q}(z) f_f\left(\frac{x}{z}, Q\right) + f_{\bar{f}}\left(\frac{x}{z}, Q\right) + P_{q\leftarrow g}(z) f_g\left(\frac{x}{z}, Q\right)\} \tag{1.38}$$

$$\frac{d}{d\log Q} f_{\bar{f}}(x, Q) = \frac{\alpha_S(Q^2)}{\pi} \int_x^1 \frac{dz}{z} \{P_{q\leftarrow q}(z) f_{\bar{f}}\left(\frac{x}{z}, Q\right) + f_{\bar{f}}\left(\frac{x}{z}, Q\right) + P_{q\leftarrow g}(z) f_g\left(\frac{x}{z}, Q\right)\} \tag{1.39}$$

where $x$ is the fraction of momentum carried by the parton, and the leading order splitting functions (or

kernels) are:

$$P_{q \leftarrow q}(z) = \frac{4}{3}\left[\frac{1+z^2}{(1-z)_+} + \frac{3}{2}\delta(1-z)\right] \tag{1.40}$$

$$P_{g \leftarrow q}(z) = \frac{4}{3}\left[\frac{1+(1-z)^2}{z}\right] \tag{1.41}$$

$$P_{q \leftarrow g}(z) = \frac{1}{2}[z^2 + (1-z)^2] \tag{1.42}$$

$$P_{g \leftarrow g}(z) = 6\left[\frac{1-z}{z} + \frac{z}{(1-z)_+} + z(1-z) + \left(\frac{11}{12} - \frac{n_f}{18}\right)\delta(1-z)\right] \tag{1.43}$$

with $n_f$ the number of quark flavors that can be treated in the massless limit at the scale $Q$, and $\int_0^1 \frac{f(x)}{(1-x)_+}dx = \int_0^1 \frac{f(x)-f(1)}{1-x}dx$ [65]. The DGLAP equations equate the change in parton density to the convolutions of the parton densities at $x/z$ with the probabilities of parton emission with momentum fraction $z$.

Fig. 1.3 shows parton distribution functions at two different energy scale. As the scale increases and the inner structure is probed at finer scales, the number of resolved constituents within the hadron increases, though the majority of them will carry only a small fraction of the hadron's momentum. The full evolution of the fractions of proton momentum carried by different partons as a function of the probing energy is displayed in Fig. 1.4. Incidentally, this shows that quark contributions are not sufficient to add up to the corresponding hadronic masses. About half of the total comes from the non-perturbative emission and absorption of soft gluons and the presence within each hadron of a 'sea' of partons, in addition to the valence quarks.

Accurate determination of the coupling constant at a conventional scale such as the one defined by the mass of the $Z$ boson, as well as the running of $\alpha_S$ at different energy scales have been the subject of various measurements across experiments using a variety of observables from many physics processes. Combined results for $\alpha_S(M_Z^2)$ and $\alpha_S(Q)$ are provided, respectively, in Fig. 1.5(a) and 1.5(b).

Since $\alpha_S$ is not itself an observable quantity, performing a measurement of its value at some energy scale requires examining many physical processes with incoming or outgoing quarks and gluons, and measuring the value of a variety of observables that can be expanded in a perturbation series.

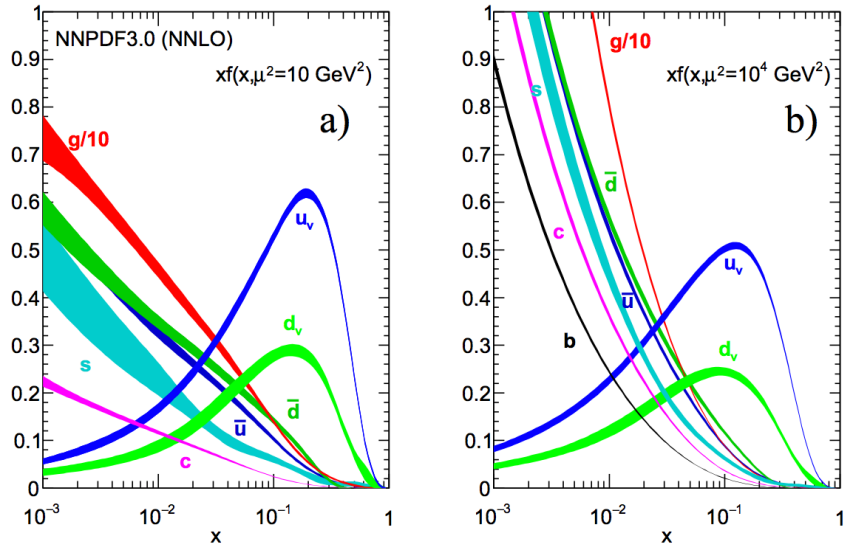

**Figure 1.3:** Structure function $xf(x,\mu^2)$ as a function of the momentum fraction $x$ at low ($\mu^2 = 10$ GeV$^2$ in a)) and high ($\mu^2 = 10^4$ GeV$^2$ in b)) energy scale, where $f(x,\mu^2)$ is the unpolarized parton distribution function. Results are obtained from NNLO NNPDF3.0 with a default value of the strong coupling constant $\alpha_S(M_Z^2) = 0.118$. Reproduced from Ref. [2].

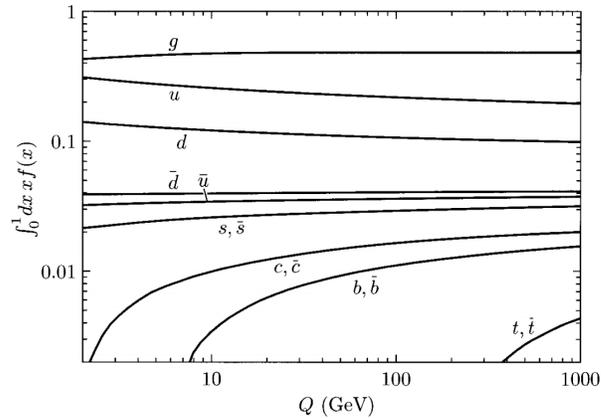

**Figure 1.4:** Evolution, as a function of energy $Q$ at which the internal structure of the proton is probed, of the fraction of the hadron's momentum that is carried by partons of each kind. The contribution of the valence $u$ and $d$ quarks is progressively diminished as more and more partons from the 'sea' are resolved. The gluons carry approximately half of the momentum/energy. The data comes from results from deep inelastic scattering from the CTEQ Collaboration [3]. Reproduced from Ref. [4].

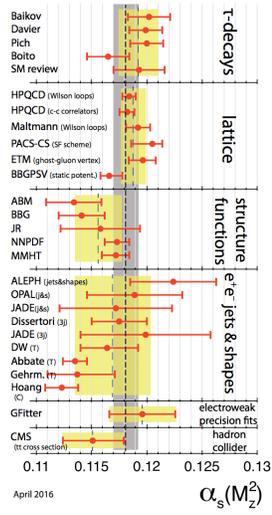
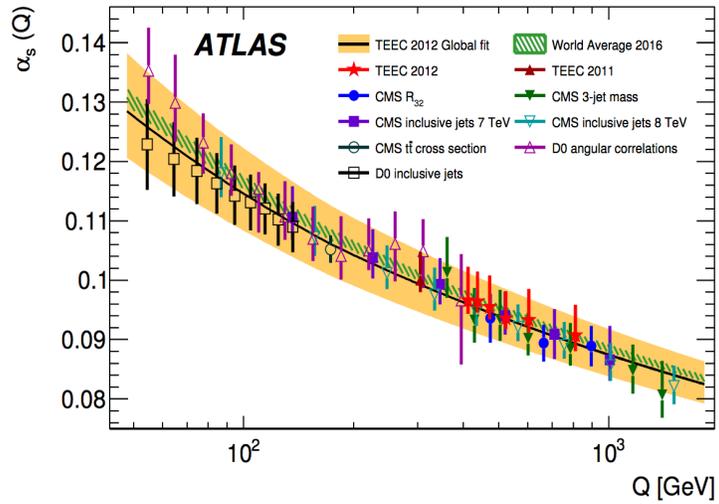

(a) $\alpha_S(M_Z^2)$

(b) $\alpha_S$ as a function of $Q$

**Figure 1.5:** Historical measurements of $\alpha_S$ at $Q = M_Z^2$ and as a function of $Q$, from a variety of different experiments and physical processes.

### 1.1.3 Composite Particles and the Quark Model

Unlike elementary particles, composite particles are bound states of fundamental particles. A consequence of confinement is the empirical observation that only color-neutral systems of quarks and gluons exist in nature in a stable state at ordinary energies. An example of such particle systems are protons and neutrons, the building blocks of atomic nuclei. Historically, the quark model emerged as a formalism to categorize the abundance of quark bound states observed in nature. First theorized in the early 1960's by Gell-Mann and Zweig, the popularity of the quark model can be traced back to its early success in predicting the existence of yet to be observed hadronic states, and analytically inferring their masses and quantum properties.

Quark-antiquark pairs form *mesons*; three quark systems are called *baryons*; tetraquarks and pentaquarks are rarer bound states of four and five quarks respectively. All particles observed in nature are colorless, or, rather, color singlets. Quark-antiquark bound states in mesons achieve color neutrality by pairing a same color quark and antiquark. Colorless 3-quark bound states, instead, are groups of quarks of three different colors, with every color appearing in the same amount.

Hadrons composed of quarks from higher generations decay into more stable hadrons from the early generations. Their lifetime depends on the strength of the interaction that mediates the reaction, with weak flavor-changing phenomena encoded in the CKM matrix (see Sec. 1.1.2.1.2). Lifetimes impose restrictions on detector technologies and effectively limit our ability to detect them if not through the detection of their decay products. The average hadron decay length, in its own rest frame is given by $\beta c\tau$, where $\tau$ is the average lifetime, while in the lab frame, where time is dilated by $\gamma$, the particle travels a distance of $\gamma\beta c\tau$, where $\gamma\beta = p/m$. Jet flavor identification techniques make use of this fact (see Sec. 6.1).

Composite particles can be characterized in terms of their quantum numbers. The quark number $N_q$ is the (number of quarks - number of antiquarks) in a particle. The baryon number is then computed as $B = N_q/3$, so that baryons, bound states of three quarks, have baryon number of 1. Similarly, lepton number $L$ is the (number of leptons - number of antileptons) in a particle. Each family of leptons has its associated lepton number: $L_e$, $L_\mu$, $L_\tau$.

## 1.2 Di-Higgs Production

To investigate the role of the newly found particle at the LHC and to evaluate different theoretical hypotheses for the origin of EWSB, it is important to probe the full Higgs potential, including self-coupling terms. Although the SM trilinear coupling will only become accessible with sufficient sensitivity in future runs of the LHC, BSM physics may provide enhancements to the production rate through resonant and non-resonant effects [66].

### 1.2.1 Standard Model Di-Higgs Production

In the Standard Model potential, the trilinear and quartic Higgs self-interaction terms arise from the expansion of the Higgs potential around the electroweak symmetry breaking vacuum expectation

$$V(h) = \frac{1}{2}m_h^2 h^2 + \lambda_{hhh} v h^3 + \frac{1}{4}\lambda_{hhhh} h^4 \tag{1.44}$$

In the Standard Model, the self-coupling constants are uniquely fixed values, which can be expressed in terms of the Higgs mass $m_h$ and vacuum expectation value $v$ as

$$\lambda_{hhh} = \lambda_{hhhh} = \frac{m_h^2}{2v^2} \tag{1.45}$$

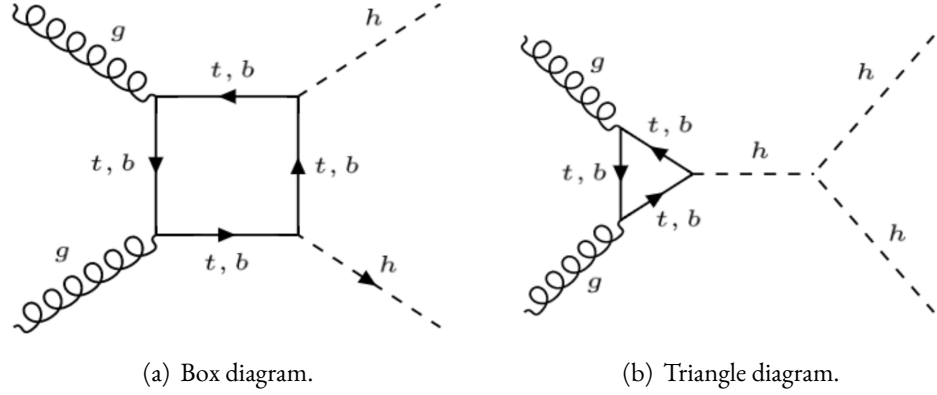

(a) Box diagram.  (b) Triangle diagram.

**Figure 1.6:** Feynman diagrams of the two leading di-Higgs production processes in the gluon fusion production channel.

which is approximately equal to 0.13 for $m_h$ = 125 GeV, $v = \frac{2}{g}m_W = \frac{2}{g}\sqrt{\frac{\sqrt{2}g^2}{8G_F}} = 246$ GeV if we plug in the value of $G_F$ = 1.166 $\times 10^{-5}$ GeV$^{-2}$ measured from muon decay.

The trilinear Higgs self-coupling is accessible via double Higgs production. The production mechanisms are the same as for single Higgs production, with the requirement of the intermediate Higgs state be off-shell in order to split up into two real Higgs bosons [67]. At the LHC, gluon fusion ($gg \to hh$) is the dominant production channel, whereas vector boson fusion ($qq' \to V^*V^*qq' \to hhqq'$ ($V = W, Z$)), associated production with $t$ quark ($pp \to t\bar{t}hh$) and Higgs-strahlung ($q\bar{q}' \to V^* \to Vhh$ ($V = W, Z$)) represent secondary production mechanisms [66].

At leading order (LO), two Feynman diagrams contribute to the di-Higgs production in the gluon fusion mechanism. Figure 1.6(a) shows the so-called 'Box Diagram', in which the $hh$ production is mediated by an internal fermionic loop, with $t$, or secondarily $b$, quarks circulating in the loop. Figure 1.6(b), instead, depicts the 'Triangle Diagram', where a triple Higgs interaction gives rise to the $hh$ signal from a $t\bar{t}$ (or $b\bar{b}$) generated Higgs boson. In general, in the SM, it is safe to neglect the tiny (0.2%) contribution coming from the $b$-mediated diagrams. The two diagrams destructively interfere, hence reducing the total cross-section. Only the triangle diagram is sensitive to the trilinear self-coupling. Similarly, in the other production channels, only one type of Feynman diagram involves the trilinear self-coupling, while the others dilute its contribution to the di-Higgs cross-section because of the appearance of the couplings of the Higgs to gauge bosons and fermions.

At next-to-learning order (NLO) and next-to-next-to-leading order (NNLO), more complex internal loops contribute to the overall cross-section. An Effective Lagrangian can be derived using the Low En-ergy Theorem (LET) in the limit of infinite quark mass, and used to obtain QCD corrections as a perturbative approximation to the full NLO result [68]. Recent updated calculations [69] provide the most advanced approximations currently available for the di-Higgs production cross-section in the $gg$ channel. Taking into account the various uncertainty sources, the total Higgs pair production cross-section in the gluon fusion process at the LHC at 13 TeV for $m_h \approx 125$ GeV, calculated at NNLO FT$_{\text{approx}}$, is [69]:

$$31.05 \text{ fb}^{+2.2\%}_{-5.0\%} \text{ (QCD scale)} \pm 2.6\% \, (m_t) \pm 3.0\% \, (\text{PDF}+\alpha_s).$$

### 1.2.2 Beyond the Standard Model Higgs

The observation of a di-Higgs event is very challenging because of the small expected SM cross-section. In fact, at 14 TeV, we only expect a production cross-section about 3 orders of magnitude smaller than that of single Higgs SM production. Extensions of the SM, with an expansion of the Higgs sector, can cause a significant variation in the Higgs self-coupling compared to the expected SM predictions [70, 71]. The rate of Higgs pair production can be altered, even with enhancements larger than one order of magnitude [72], as proposed by some theoretical models, and the variations can be expressed in terms of free parameters currently constrained by single Higgs measurements. Di-Higgs searches provide a model-independent probe to physics beyond the Standard Model that may interact with the Higgs sector and create a clear signature [72].

#### 1.2.2.1 $hh$ in 2HDM

Given the relatively low mass of the SM Higgs boson, many BSM theories have invoked the presence of new symmetries to protect the Higgs mass from receiving large contributions. Among these are supersymmetric models that require two Higgs doublets [73, 74, 75, 76].

The Two Higgs Doublet Model (2HDM) requires the existence of five different physical scalar particles after EWSB: two neutral CP-even scalars ($h, H$), one neutral CP-odd pseudoscalar ($A$), and two charged scalars ($H^\pm$) [77]. The Lagrangian has up to 11 independent physical parameters that determine the strength of the new interactions and the masses of the new states. Mixing between scalars can alter the couplings of the SM-like Higgs $h$ [71]. When kinematically allowed, the heavy scalar Higgs $H$ may decay to two light CP-even Higgses as $H \to hh$. This decay mode might vanish in the exact alignment limit, *i.e.* when the $h$ couplings are SM-like, but it still dominates when moving even slightly away from the limit.

Assuming the lightest CP-even Higgs boson $h$ is the SM-like one with mass of ∼125 GeV, and the remaining scalars have $m_H \sim m_A \sim m_{H^\pm} >> v$, then observable deviations in the production and decay rates of the SM-like Higgs from the SM expectations can be parametrized in terms of two free parameters, the angles $\alpha$, the mixing angle between the two CP-even Higgs bosons $h$ and $H$, and $\beta$, the angle whose tangent is the ratio of the vacuum expectation values of the two doublets [71, 77]. Given the observed agreement of the properties of the SM-like Higgs – as measured by the ATLAS and CMS collaborations – with the SM predictions, the 2HDM EWSB sector is constrained to lie close to the alignment sector where $\sin(\beta - \alpha) = 1$ [71, 77].

### 1.2.2.2 $hh$ IN MSSM

The advent of the Minimal Supersymmetric extension of the Standard Model (MSSM), a subgroup of Type-II 2HDM [71], enhanced the relevance of the 2HDM. In this model, the existence of two isodoublets of complex scalar fields of opposite hypercharge is required for two main reasons: first, to give mass to both isospin-type fermions without introducing conjugate fields in the SUSY superpotential; second, to preserve renormalizability and avoid chiral anomalies caused by the higgsino, the spin 1/2 superpartner of the scalar field [78].

Electroweak symmetry breaking gives rise to the SM gauge bosons ($W_L^+, W_L^-, Z_L$), leaving 5 physical states ($h, H, A, H^+, H^-$), as we saw before [79].

In the decoupling limit (where $\alpha \sim \beta - \pi/2$), the deviations in parameters such as the partial width of the SM-like scalar can be parametrized in terms of $\tan\beta$, $m_A$ and a factor that captures SUSY radiative corrections [71].

This theory has the advantage of being characterized by a smaller number of free parameters. Searching for di-Higgs events, in the limit of low $\tan\beta$, is the most promising strategy to reconstruct these parameters: measurements of the light Higgs couplings under different $\tan\beta$ hypotheses will help estimate the mass of the CP-odd state. The assumption $2m_h < m_H < 2m_t$ causes a large branching ratio for $H \to hh$, with possible further enhancements from squarks in the loops [70].

1.2.2.3 $hh$ in the Higgs Portal Scenario

The Higgs Portal Scenario provides a viable candidate for dark matter by introducing a hidden-sector Higgs field $\Phi_H$ that extends the Higgs potential in a mirrored way and is coupled to the SM-field $\Phi_S$ via a renormalizable quartic interaction of strength $\eta_\chi$ [71]:

$$V = \mu_S^2 |\Phi_S|^2 + \lambda_S |\Phi_S|^4 + \mu_H^2 |\Phi_H|^2 + \lambda_H |\Phi_H|^4 + \eta_\chi |\Phi_S|^2 |\Phi_H|^2. \tag{1.46}$$

The new field is otherwise a singlet under SM gauge symmetries. After symmetry breaking in the SM and hidden sector, both fields develops vevs, and the physical eigenstates can be obtained from a two-dimensional isometry which mixes the visible and hidden Higgs states [70]:

$$\begin{cases} h = \cos\chi\, \Phi_S + \sin\chi\, \Phi_H \\ H = -\sin\chi\, \Phi_S + \cos\chi\, \Phi_H \end{cases} \tag{1.47}$$

with a mixing angle [71]:

$$\tan(2\chi) \simeq -\frac{\eta_\chi v_S}{\lambda_H v_H} \tag{1.48}$$

where $v_S$ and $v_H$ are respectively the SM and hidden sector vevs.

This model predicts visible enhancements in both resonant and non-resonant di-Higgs productions. All Higgs couplings, including the trilinear one, will be universally affected by the mixing [71] according to

$$g_h = \cos\chi\, g_h^{\text{SM}}, \tag{1.49}$$

and the light Higgs cross-section will be modified to [80]

$$\sigma = \cos^2\chi\, \sigma^{\text{SM}} \tag{1.50}$$

Available constraints prefer a minimal mixing solution with $\cos\chi^2 \sim 1$ [70]. Ref. [70] provides analytical formulas for the trilinear couplings strengths expanded around the vevs.

The phenomenology of this model is similar to that of di-Higgs production in the low $\tan\beta$ regime in MSSM [70]. The measurement of the $H \to hh$ decay and of the other trilinear couplings will con-strain the currently large available parameter space that leaves room for phenomenologically interesting scenarios at the LHC.

### 1.2.2.4 OTHER BSM

Another possibility is to interpret a new heavy resonance at the TeV scale as a spin-2 Kaluza-Klein graviton $G$ in a Randall-Sundrum model with warped extra-dimensions [81]. Since experimental data seems to disfavor low mass gravitons, boosted analyses, such as the $hh \to b\bar{b}b\bar{b}$ and $hh \to b\bar{b}\tau\bar{\tau}$, are the best candidates to test this interpretation [82].

Composite Higgs scenarios, where electroweak symmetry is explicitly broken and the EW group is a subgroup of a larger spontaneously broken symmetry group, new fermionic interaction vertices of the form $f\bar{f}hh$ become relevant for di-Higgs production. In these models, the strongly-interacting light Higgs is interpreted as a pseudo-Goldstone boson that can therefore bear much lighter mass than its partners [83]. The modification of the SM Higgs couplings are expressed in terms of $\xi = (v/f)^2$ where $v$ is the SM Higgs vev and $f$ is the Goldstone scale, which goes to infinity to recover the SM scenario, and equals $v$, for example, in Technicolor models [80].

Several other models, such as di-Higgs production through light colored scalars [84], have been put forward over the years with clear di-Higgs enhancement predictions and suggested phenomenological signatures.

# 2

# The ATLAS Experiment at the LHC

THIS IS A THESIS ON EXPERIMENTAL PARTICLE PHYSICS. The strength of the fundamental forces predicted by the Standard Model and by other theories that attempt to supersede it manifests itself in the phenomenology of particle interactions in nature. Therefore, the theoretical constructs examined in the previous chapter are to be empirically validated against experimental results, to verify the agreement of their predictions with observations from data.

Investigating physics at high energies requires the construction of pieces of apparatus able to explore such energy regimes. One such machine is the Large Hadron Collider (LHC) at CERN, the European Organization for Nuclear Research. From the early days of collider physics, pioneered by Ernest Rutherford with his gold foil experiment, perfected by Ernest Orlando Lawrence and generations of physicists after him, experimentalists have mastered the art of constructing larger and larger machines to accelerate, collide, and detect particles. Today, over 13,000 users from 112 countries work together at CERN to advance the limits of human knowledge [85].

In a collider, particle beams are accelerated to ultra-relativistic speeds and brought to collision in or-der to create heavier particles. Although the Holy Grail of particle physics is to probe energies close to the grand unification energy, the current LHC design specifications only allow for collisions in the TeV regime. However, at this scale, located just beyond the electroweak symmetry breaking point, *new physics* is expected to manifest itself, according to many popular theories.

## 2.1 Properties of Colliders

The center of mass energy $E_{CM}$ is an important parameter to describe the discovery potential of a high-energy experiment. To resolve the tiniest structures of matter, high energies and momenta are necessary, according to the De Broglie relation $\lambda = h/p$, where $\lambda$ is the De Broglie wavelength that indicates the distance scales that can be probed, $h$ is the Planck constant, and $p$ is the incoming particle momentum. In addition, to produce (potentially new) heavy particles, the interaction energy must be greater than the particle's rest mass. Thus, maximizing the momentum, or energy, of colliding particles is a priority in experimental design, and provides a straightforward explanation for the popularity of particle colliders. To visualize the dependence on the center-of-mass energy of the cross-section for the production of a series of SM processes at proton-proton colliders, refer, for example, to Fig. 2.1.

Compared to fixed-target experiments, colliders are capable of achieving higher center-of-mass energies. In the collision of two particles $p_1^\mu$ and $p_2^\mu$, the center of mass energy, also defined as the square root of the Mandelstam variable $s$, is a Lorentz invariant, and is calculated as:

$$E_{CM} = \sqrt{s} = \sqrt{(p_1^\mu + p_2^\mu)^2} = \sqrt{(p_1^\mu)^2 + (p_2^\mu)^2 + 2g_{\mu\nu}p_1^\mu p_2^\nu} \qquad (2.1)$$

In colliders, where the relativistic limit $E \approx |\vec{p}| \gg m$ holds and the particles collide head-on:

$$E_{CM} \approx \sqrt{2g_{\mu\nu}p_1^\mu p_2^\nu} \qquad (2.2)$$
$$= \sqrt{(E_1 + E_2)^2 - (\vec{p}_1 + \vec{p}_2)^2} \qquad (2.3)$$
$$= \sqrt{\cancel{E_1^2} + \cancel{E_2^2} + 2E_1 E_2 - \cancel{\vec{p}_1^{\,2}} - \cancel{\vec{p}_2^{\,2}} - 2\vec{p}_1 \cdot \vec{p}_2} \qquad (2.4)$$
$$= \sqrt{2E_1 E_2 - 2E_1 E_2 \cos(\theta)} = \sqrt{2E_1 E_2 + 2E_1 E_2} = 2\sqrt{E_1 E_2}. \qquad (2.5)$$

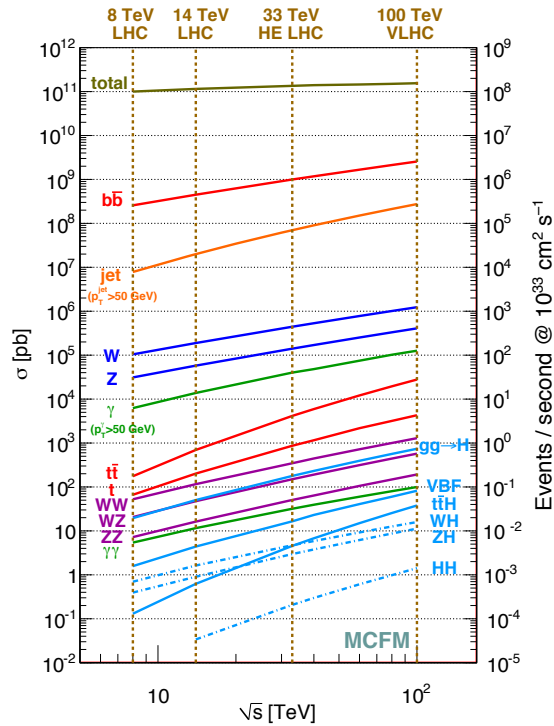

**Figure 2.1:** Simulated cross-section for selected SM events as a function of the center-of-mass energy at past, current, and potential future proton-proton colliders. The left vertical axis converts cross-sections into the event rate at a fixed instantaneous luminosity of $10^{33}$ cm$^2$s$^{-1}$. This plot was obtained using the MCFM software package [5, 6].

On the other hand, for a fixed-target experiment, $E_1 \approx |\vec{p}_1| \gg m_1$ and $E_2 = m_2$, $|\vec{p}_2| = 0$, so:

$$E_{CM} \approx \sqrt{m_2^2 + (E_1 + m_2)^2 - \vec{p}_1^2} \tag{2.6}$$

$$= \sqrt{m_2^2 + \cancel{E_1^2} + m_2^2 + 2E_1 m_2 - \cancel{E_1^2}} \tag{2.7}$$

$$= \sqrt{2m_2^2 + 2E_1 m_2} \approx \sqrt{2E_1 m_2} \tag{2.8}$$

assuming the incoming energy of the particle beam $E_1 \gg m_2$, the target particle's rest mass. Therefore, colliders are more energy efficient than fixed-target experiments. However, the ease of colliding a beam of particles with a fixed target leads to higher luminosity and cross-section.

Several design choices go into the construction of a particle collider. Two major classes of colliders differ in the geometry of the collider structure: *linear* colliders are composed of two longitudinal, symmetric halves that accelerate beams in a straight line and collide them at their junction; *circular* colliders, on the other hand, use bending magnets to accelerate beams in both directions along a circular path before bringing them to collision at designated points along the ring.

The choice of particles to collide determines the cross-section and luminosity machines can deliver. Hadron colliders are best suited for probing high energy regimes, while precision measurements are the focus of lepton colliders. At higher energies, more massive colliding particles are necessary to avoid excessive energy losses through synchrotron radiation, which scales as $\Delta E \propto 1/m^4$ [86]. Protons are not elementary particles, so, upon collision, interactions will instead occur among their partons, thus activating multiple production mechanisms involving quarks and gluons. Hadronic collisions lack in cleanliness inasmuch as all sorts of soft collisions, plus initial and final state radiation, add to the complexity of the event. A slew of possible interactions may take place, and the precise energy of the hard parton collision is unknown. To account for the parton energy distribution, cross-sections calculated as a function of the momentum fraction of the partons are weighted by the Parton Distribution Function (PDF, see Sec. 1.1.2.2.1) and integrated over the unknown momentum fraction. PDFs are experimentally measured distribution functions that model the probability of finding a parton within a colliding hadron with a given momentum. One advantage of the parton dynamics is that actual parton interactions may occur along an energy band around the nominal center-of-mass energy.

In hadron colliders, the choice between colliding protons or proton-antiproton beams is informed by both practical and phenomenological considerations. Antiprotons are more difficult to produce and store, which would limit the overall luminosity, but their quark content allows for different final states, and their opposite charge allows for the use of the same set of magnets and vacuum chamber for counter rotation with respect to the positively charged protons.

The event rate $R$, or number of events per unit time, is solely determined by the cross-section of the event $\sigma$ and the collider instantaneous luminosity $\mathcal{L}$:

$$R = \frac{dN}{dt} = \sigma \times \mathcal{L}. \qquad (2.9)$$

In particle colliders, where particles circulate in bunches, the luminosity is proportional to the bunch crossing frequency $f$, the number of bunches $n_b$, the bunch sizes $n_1$ and $n_2$, and inversely proportional to the beam rms widths $\sigma_x$ and $\sigma_y$ in the bending and transverse directions. Factoring degradation effects, such as the dispersion along the beam direction $\sigma_z$ and the crossing angle between the colliding bunches $\theta_c$, into the geometric luminosity reduction factor

$$\mathcal{F} = \left(1 + \left(\frac{\theta_c \sigma_z}{2\sigma_x}\right)^2\right)^{-1/2} \leq 1, \qquad (2.10)$$

and assuming identical Gaussian transverse beam parameters,

$$\mathcal{L} = f \frac{n_1 n_2}{4\pi \sigma_x \sigma_y} \mathcal{F} = f \frac{n_1 n_2}{4\pi \sqrt{\epsilon_x \beta_x \epsilon_y \beta_y}} \mathcal{F}, \qquad (2.11)$$

where $\beta$ is the amplitude function, designed to be optimally narrow at the interaction points, and $\epsilon$ is the emittance, defined as the beam area that contains one standard deviation of the particles [2]. Given this relation, it is clear that beam optimization consists in maintaining large numbers of particles, $n_1$ and $n_2$, in circulation with high frequency, while lowering the emittance and the amplitude function.

### 2.1.1 The Large Hadron Collider (LHC)

Housed at CERN (the European Organization for Nuclear Research), the Large Hadron Collider (LHC) is a proton-proton circular synchrotron located 100 m under ground, in the old LEP [87] tunnel, which traverses the French-Swiss border outside of Geneva, Switzerland. It is the most powerful particle accel-erator ever built, and, as the flagship machine at CERN, it carries out the laboratory's mission of peaceful scientific discovery, technological innovation, international collaboration, and education, in continuity with its storied predecessors.

The LHC is a discovery machine, built with the objective of probing physics at untapped energy regimes. Its unprecedented center-of-mass energy has provided ample opportunities to perform high-precision measurements of Standard Model processes and search for rich new physics phenomena. The high luminosity is required to compensate the tiny cross-sections of many interesting physics processes that physicists wish to investigate. Evidence for gravitons, black holes, extra dimensions, supersymmetric particles, etc., may become available within the lifetime of the LHC, although the window for many Beyond the Standard Model theories is quickly shrinking.

With its 26.7 km ring of superconducting magnets, the LHC accelerates and focuses beams of protons traveling at a top speed of 0.999999991 times the speed of light which are injected from the Super Proton Synchrotron (SPS) after previous stages of acceleration depicted in Fig. 2.2 [8]. Two beamlines, enclosed in the same cryo-system and magnetic field, house the two magnetically coupled, counter-rotating proton beams. Beams can continue circulating for hours before being dumped and replaced, due to degradation and luminosity decay.

The ring is in fact composed of eight straight sections, interleaved with bending areas. The LHC's 1232 NbTi dipoles operating at a temperature of 1.9 K are responsible for maintaining the protons along their circular path by creating a magnetic field of up to 8.3 T. 392 quadrupoles are used to focus the beam, and radio-frequency (RF) cavities – 8 per beam – keep the protons tightly bunched and accelerate them to provide an extra energy boost [88, 8]. Each cavity delivers 2 MV at a frequency of 400 MHz.

The RF cavities cause the protons to be bundled in bunches of approximately 120 billion particles (at the start), with a maximum value of 2808 bunches per beam, separated by approximately 7.5 m (or 25 ns), producing about a billion collisions per second [8]. The size of the bunches varies as they circulate around the LHC ring: they are squeezed to a width of about 20 $\mu$m at the collision points to increase their density and, therefore, the probability of collision.

Among other factors, the beam energy is limited by the radius of the accelerator as well as the mag-netic field produced by its bending magnets. Equating the magnetic force to the centripetal force,

$$evB = \gamma m \frac{v^2}{R} \rightarrow E \sim p = \frac{e}{c}RB \approx 0.3 \frac{\text{GeV}}{\text{m} \cdot \text{T}} RB. \tag{2.12}$$

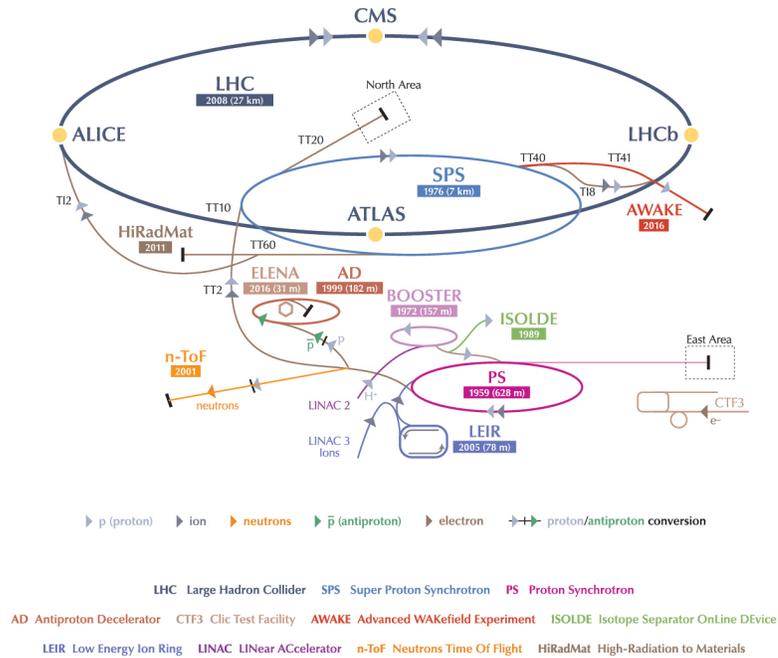

**Figure 2.2:** Diagram of the stages of acceleration of protons in the CERN accelerator complex [7]. This series of machines accelerates protons extracted from the ionization of a tank of pure hydrogen gas. The LINear ACcelerator (LINAC) 2 receives the protons and accelerates them to 50 MeV, or $\sim 0.3c$. The protons are subsequently inserted into a series of circular accelerators: first, the Proton Synchrotron Booster, that brings their energy up to 1.4 GeV; second the Proton Synchrotron (PS), which increases the beam energy to 25 GeV; third the Super Proton Synchrotron (SPS), that accelerates them to 450 GeV. At this point, protons are already traveling at 99.9998% of the speed of light. Finally, the protons are injected into the two LHC beam pipes, until the design number of bunches is reached. The process of filling each LHC ring takes over four minutes, while the final acceleration stage, which brings each beam energy up to 6.5 TeV, takes approximately 20 minutes [8, 9].

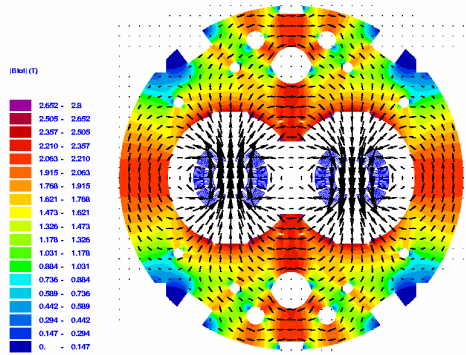

**Figure 2.3:** Magnetic flux in a typical twin-aperture LHC dipole. Each dipole surrounds a beam line. Image reproduced from Ref. [10]

Since the radius is fixed due to the constraints of inheriting the LEP tunnel – which aimed at limiting curvature due to the significantly higher levels of synchrotron radiation in $e^+e^-$ colliders –, the magnetic field becomes a key component in any attempt to increase the center-of-mass energy. The twin-bore design of the LHC magnets is itself also a compromise due to the limited tunnel diameter. The magnetic flux produced by the LHC dipole magnets is represented in Fig. 2.3. For future runs of the LHC, new magnet technologies based on Nb$_3$Sn and High-Temperature Superconductivity (HTS) are currently under investigation [89].

More details on the LHC machine, magnets, RF cavities, cryogenics, vacuum system, shielding, operation, and performance, can be found in Ref. [90, 10].

The protons are brought to collision in four designated points, where the four major LHC experiments (ATLAS [91], CMS [92], LHCb [93], and ALICE [94]) are located. Three smaller detectors have been installed close to other experiments' caverns: LHCf [95] (near ATLAS), TOTEM [96] (near CMS), and MoEDAL [97] (near LHCb).

The LHC project was first approved by the CERN Council in 1994. After the decommissioning of LEP in 2000, construction of the LHC continued for almost a decade [90]. The LHC was initially designed to deliver $\sqrt{s}$ = 14 TeV collisions and to begin operations in 2008. However, on 19 September 2008, during the commissioning phase, an electrical fault between two magnets punctured the liquid helium enclosure, causing the release of a 6t helium leak which damaged several dipole and quadrupole magnets in three adjacent subsectors. Operations were resumed in 2009, with the first $\sqrt{s}$ = 7 TeV collisions viable for physics produced in early 2010. In 2012, the energy was ramped up to $\sqrt{s}$ = 8 TeV. This data collection phase is identified as Run I. The 2013-2015 shutdown enabled repairs and upgrades in both the LHC and the detector systems. Run II at the LHC began in 2015 and is still underway, with $\sqrt{s}$ = 13 TeV. The tentative long-term schedule for the LHC is shown in Fig. 2.4, along with the projections for peak instantaneous and integrate luminosities in Fig 2.5.

The size of the ATLAS collected dataset was 0.047 fb$^{-1}$ in 2010, and 5.5 fb$^{-1}$ in 2011. During the 2012

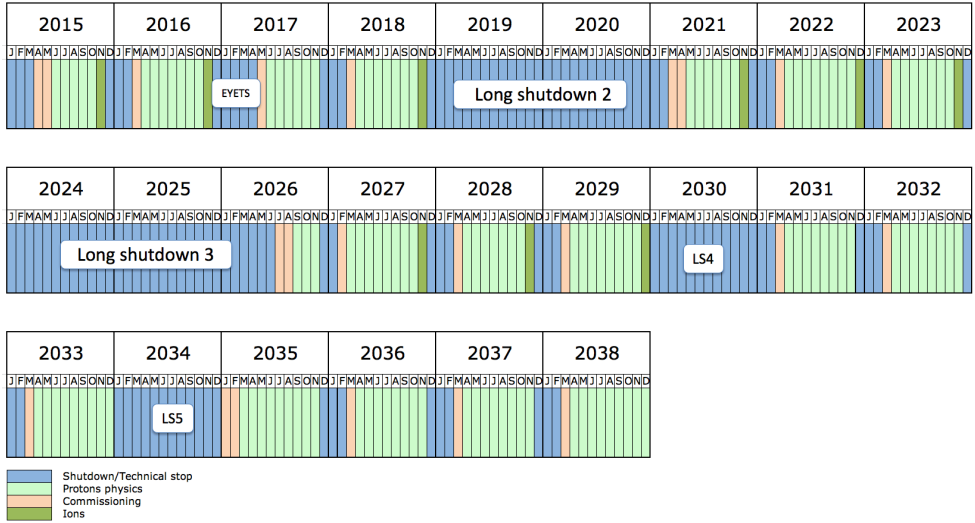

**Figure 2.4:** Tentative schedule for LHC operations to 2038. At the end of the current year (2018), after the heavy ions run, the LHC will enter the Long Shutdown 2 phase, which will extend until 2021, at the beginning of Run III.

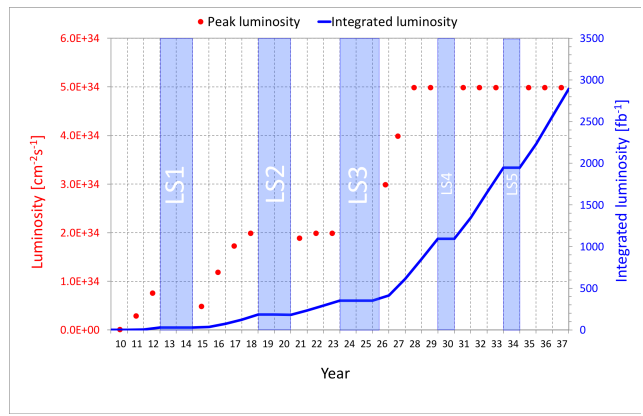

**Figure 2.5:** Prediction of integrated and peak instantaneous luminosity growth over the lifetime of the LHC. The blue bands indicate Long Shutdown periods.

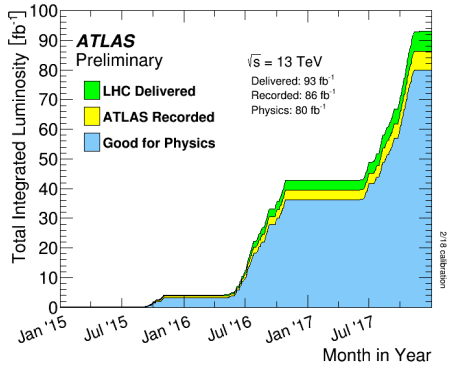
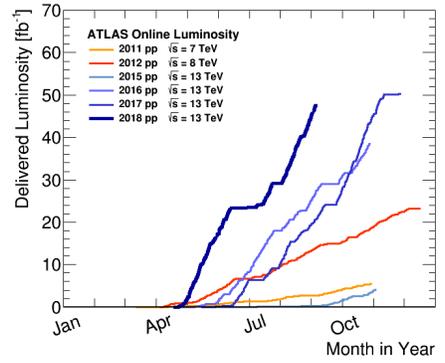

(a) Total integrated luminosity delivered by the LHC (green), recorder by ATLAS (yellow) and passing the quality requirements for physics analysis (blue) in Run II, up to the beginning of 2018, with corresponding numerical values in inverse femtobarns.

(b) Per-year total integrated luminosity delivered to ATLAS by the LHC over the twelve month period. The 2011 and 2012 data-taking campaigns took place at lower center-of-mass energies, as specified in the legend.

**Figure 2.6:** The impressive performance of the LHC machine has allowed experiments like ATLAS to record record numbers of collisions over its years of operation.

data taking period, the LHC delivered 22.8 fb$^{-1}$ of total integrated luminosity, of which 21.3 fb$^{-1}$ were recorded by the ATLAS detector, and 20.3 fb$^{-1}$ were labeled as 'good' for physics analysis, according to a variety of quality criteria to guarantee acceptable detector conditions. Summary of luminosity and LHC performance measurements for Run I are available in Ref. [98, 99]. The LHC continued to deliver record-breaking luminosity levels during Run II, including 4.2 fb$^{-1}$ in 2015, and 38.5 fb$^{-1}$ in 2016. In 2017, the LHC produced approximately 50 fb$^{-1}$ of collisions over a run period that saw the previous instantaneous luminosity records crushed by the new achievement of $2.09 \times 10^{34}$ cm$^2$s$^{-1}$. 2018 is off to a great start with 47.5 fb$^{-1}$ delivered as of September, and peak luminosity of $2.14 \times 10^{34}$ cm$^2$s$^{-1}$. Live monitoring of the LHC status is available at [100].

Precise measurements of the luminosity delivered by the LHC to each experiment are crucial to estimate the expected number of signal and background events, and are carried out by dedicated subdetectors. The total instantaneous luminosity can be rewritten in terms of the single bunch luminosity $\mathcal{L}_b$ as

$$\mathcal{L} = \sum_{b=1}^{n_b} \mathcal{L}_b = n_b \langle \mathcal{L}_b \rangle = n_b f \frac{\langle \mu \rangle}{\sigma_{\text{inel}}} \tag{2.13}$$

where $\langle \mu \rangle$ is the average number of inelastic interactions per bunch crossing, and $\sigma_{\text{inel}}$ is the proton-proton inelastic cross-section. In fact, on top of the hard-scatter event, at each bunch crossing, multiple interactions among the partons in each beam are expected to occur. These are primarily soft QCD scatterings, resulting in hadronic activity at low momentum transfer. The radiation from these inelastic particle interactions overlays the principal event. The additional proton-proton interactions are referred to as *pile-up*. We distinguish between *in-time* and *out-of-time* pile-up as the additional interactions originating from the same bunch crossing as the primary event, versus those coming from bunch crossings immediately before or after, which may crowd the detector and complicate reconstruction.

At each collision point along the LHC, efficiencies of detector and identification algorithms enter as identical multiplicative factors in the scaling of the registered number of interactions and visible cross-section: $\mu_{\text{vis}} = \varepsilon \mu$ and $\sigma_{\text{vis}} = \varepsilon \sigma_{\text{inel}}$. By monitoring the bunch luminosity and the in-situ registered $\mu_{\text{vis}}$ under controlled conditions during van der Meer scans, in which the horizontal and vertical beam separations are varied in steps, the value of $\sigma_{\text{vis}}$ for various detectors can be calibrated [98]. The cross-section $\sigma_{\text{inel}}$ for inelastic $pp$ interactions at $\sqrt{s} = 13$ TeV in ATLAS has been measured to be $78.1 \pm 2.9$ mb [101] and increases with center-of-mass energy [2].

The instantaneous luminosity is then integrated over time assuming that the instantaneous luminosity remains approximately constant over the unit time of approximately one minute identified as a luminosity block (LB), as defined, in the case of ATLAS, by its central trigger processor [98]. The integrated luminosity delivered to the ATLAS detector over the years, as a function of the average pile-up level is shown in Fig. 2.7. Several LBs within the same beam injection form a run. Runs that share detector and beam conditions are grouped into periods.

## 2.2 Particle Detection via Matter Interactions

After a proton collision, a spray of particles encoded with the information about the physical process that occurred at the interaction point will leave the collision area in various directions. Those particles need to be detected, measured, and identified.

Particle detection occurs via the interaction with detector material. Instruments for particle detection are therefore designed to optimize these interactions and to create unique patterns per particle type. Once particles are produced at colliders, their precise detection and identification becomes paramount.

Detector development is tightly coupled to projected physics outcomes and phenomenological expectations, binding the design to the optimization for particular physics signatures while attempting to preserve the generality of the apparatus. What drives the optimization procedure is a series of figures of merit associated with particular physics processes of interest. As explained in Ref. [91], physics goals can be translated into detector requirements. These include: fast, radiation-hard sensors and electronics, high resolution detectors, nearly full coverage in both the azimuthal and polar direction, and high distinguishability among particle signatures upon interaction with the detector. The signatures of various physics objects in the finalized ATLAS detector and the way in which they can be exploited for particle identification are addressed in Ch. 5.

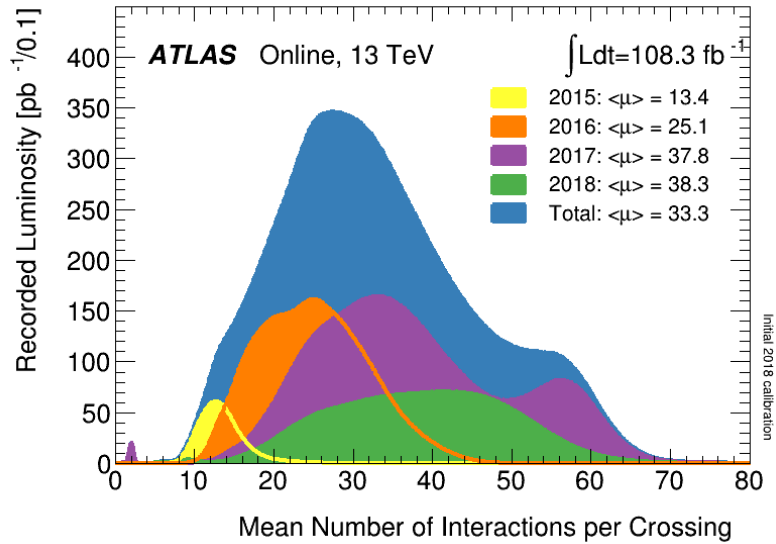

**Figure 2.7:** Luminosity-weighted distribution of the mean number of interaction registered by the ATLAS experiment in stable-beam proton-proton collision data over the years, during Run II. 2018 data corresponds to the partial dataset collected as of 12 June 2018. The average pile-up value per year is provided in the legend.

Despite the identical constraints and purpose, the ATLAS [91] and CMS [92] collaborations settled on rather different design choices. Designing and constructing a particle detector is a venture fraught with technical, theoretical, and sociological hurdles whose nature demands compromises and resolutions often based on predicted experimental outcomes and future technology developments.

### 2.2.1 THE ATLAS DETECTOR

ATLAS (A Toroidal LHC ApparatuS) is a multi-purpose detector situated in the experimental cavern at point 1 along the LHC at CERN [91]. Initially proposed in 1994 [102], its design emerges from its main physics targets.

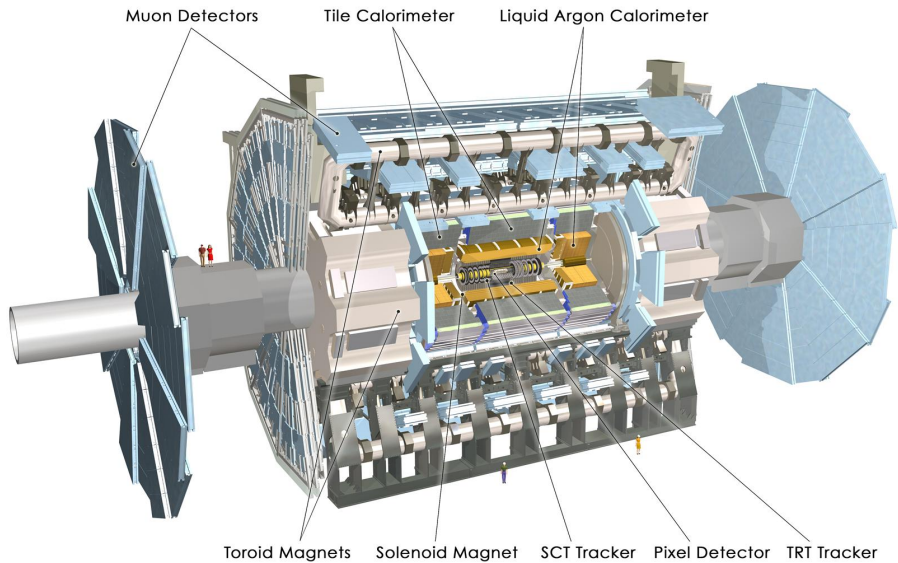

**Figure 2.8:** The ATLAS detector and its sub-detectors [11].

When the two beams collide, the ATLAS detector records the collision products across almost the full $4\pi$ solid angle as they interact with its various layers. The ATLAS detector, pictured in Fig.2.8 is centered around the collision point and extends with cylindrical symmetry around the beam line. Thanks to its length of 46 m and diameter of 25 m, this 7000 tonnes particle detector is the largest ever constructed. The cavern that houses the detector is located approximately 100 m under ground.

Along the beam axis, detector layers are arranged as concentric cylinders with endcap regions perpendicular to the beam line. The layered geometry is composed of a series of subdetectors described in this section. Additional, smaller, forward detectors, such as LUCID and ALFA, contribute to monitoring the luminosity delivered to ATLAS [91].

Designed to prioritize its discovery potential at the TeV scale, the development and performance of its various sub-detectors have been optimized to search for the Higgs boson. Much of the design was dictated by the choice of opting for its unique magnet structure, with the inner detector immersed in the uniform field generated by a thin solenoid, and the signature superconducting toroids with eight-fold azimuthal symmetry encapsulating the calorimeters [91].

The right-handed coordinate system used to describe position within the detector volume is defined as follows. With the origin located at the designated collision point, the $z$ direction extends along the beam-line, the positive $y$ direction points upwards towards the sky, and the positive $x$ direction points towards the center of the LHC ring. The $x$ and $y$ axes form the so-called the transverse plane; vector components in the $x - y$ plane are called transverse components (e.g. the transverse momentum $p_T$ is the momentum component perpendicular to the beam line). Observables in the transverse plane play a critical role in ATLAS analyses because they are invariant to unknown boosts along the beam axis due to colliding partons' initial velocities with respect to the center-of-mass frame. These unknown non-zero longitudinal components of the net initial momentum cause collision products to travel in a preferred direction, informing the decision of constructing a cylindrical, rather than spherical, detector. In addition, in head-on collisions, any resulting transverse momentum component is due to the underlying physics process. We identify particles with high $p_T$ as *hard*, and particles with low $p_T$ as *soft*.

Due to the cylindrical symmetry in the detector geometry, it is more suitable to define angular coordinates: $\phi$, the azimuthal angle, revolves around the beam direction in a counter-clockwise direction from the $x$ axis upwards, while $\theta$, the polar angle, is the angle that defines the aperture with respect to the positive beam axis. $\phi$ is zero towards the center of the ring, while $\theta$ is zero along the beam line. The polar angle is reparametrized in terms of the more commonly used pseudorapidity $\eta = -\ln\tan(\theta/2)$, so that particles traveling along the beam direction (forward particles) have $\eta = \infty$, and particles perpendicular to it (central particles) have $\eta = 0$. The usefulness of the definition of the $\eta$ coordinate relies on the invariance of distances in $\eta$ under unknown boosts along the beam direction, which results in a nearly uniform particle distribution in $\eta$. In the massless limit $E \approx |\vec{p}| \gg m$, the rapidity, defined as $y = \frac{1}{2}\ln[(E + p_z)/(E - p_z)]$, approaches the pseudorapidity. Angular distances in $\eta - \phi$ space are usually measured in units of $\Delta R = \sqrt{(\Delta\eta)^2 + (\Delta\phi)^2}$.

#### 2.2.1.1 The Inner Detector

The inner detector, shown in Fig. 2.10, consists of four sub-detectors equipped with different hardware technologies that trace out the trajectories of charged particles close to the interaction point with minimal interference on the particle's energy.

These particles are deflected and bent into helical motion by a 2 T magnetic field parallel to the beam line produced by a superconducting central solenoid magnet that encloses the inner detector, and facilitates the measurement of their direction, momentum, and charge. The central solenoid is a 5.3 m long coil of doped Al-stabilized NbTi/Cu that operates at 7600 A. It is housed in the same cryostat as the electromagnetic calorimeter to reduce radiative loss [103]. Eq. 2.12 relates the momentum of a particle with charge $e$ to the radius $R$ of its trajectory in a magnetic field $B$. This implies that only particles with $p_T \gtrsim 0.3$ GeV can ever be expected to leave hits in all inner detector layers, given its thickness of $\sim 1$m. On the other hand, large $p_T$ particles will escape the inner detector almost unbent.

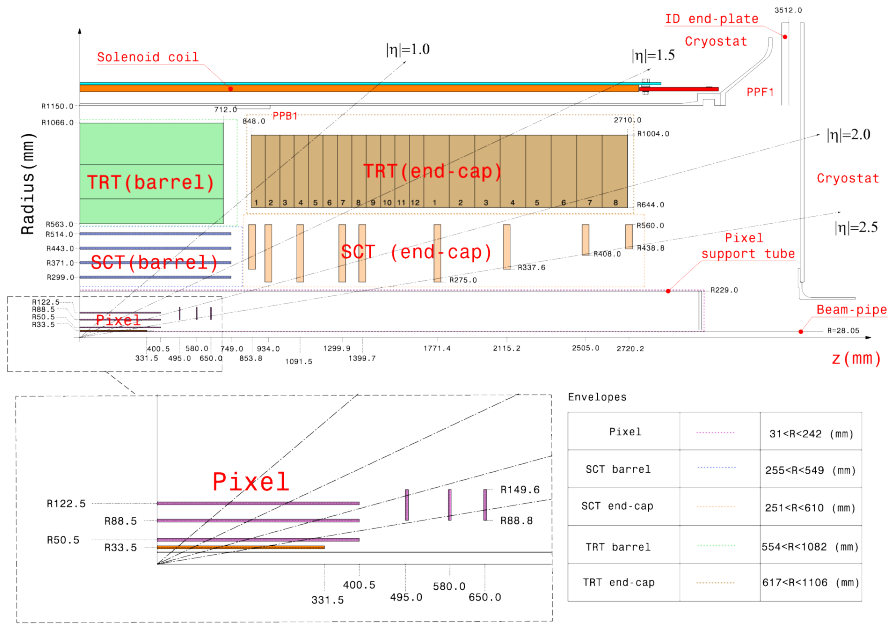

**Figure 2.9:** Cross-section view in the $r-z$ plane of one quadrant of the ATLAS inner detector, including the end-cap portion, which extends the acceptance up to $|\eta| = 2.5$. The bottom section zooms into the pixel detector region. The interaction point is located in the bottom left corner. Image reproduced from Ref. [12].

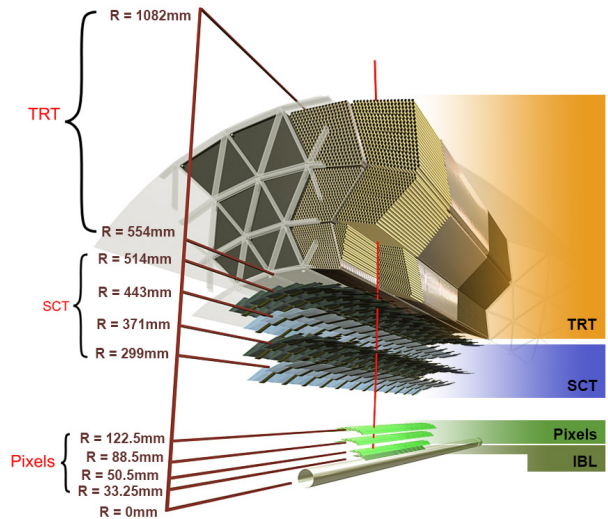

**Figure 2.10:** Three dimensional diagram of the barrel region of the he ATLAS inner detector [13].

Because of its vicinity to the interaction region, the inner detector requires the highest granularity among all ATLAS sub-detectors. This provides high-definition track reconstruction in $(r, \phi, z)$ from *hits* in the various layers (see Sec. 5.1). As particles travel outward from the collision point, they encounter, in order, the Pixel Detector, the Semi-Conductor Tracker (SCT) and the Transition Radiation Tracker (TRT). These sub-detectors are instrumented with different technologies to provide complementarity and fault tolerance in the measurements. A cross-sectional view of the inner detector region is provided in Fig. 2.9.

The innermost instrument is the silicon pixel detector, which, with its 92 million channels, develops over 3 endcap disks and 4 concentric barrel layers to provide tracking capabilities up to $|\eta| < 2.5$, and is characterized by spatial resolutions of 10 $\mu$m in the $R - \phi$ direction and 115 $\mu$m in $z$. The latest addition to the system, known as Insertable B-Layer (IBL) [104], was installed closest to the beam line between Run I and II to improve tracking resolution and mitigate the radiation damage suffered by the B-Layer. Located at an average distance from the beam line of 33.25 mm, because of its proximity to the beam pipe – which underwent a reduction in diameter to make room for the new detector layer –, the IBL has to fulfill stringent radiation hardness and resolution requirements, which demand a smaller pixel surface (reduced from the $50 \times 400$ $\mu$m$^2$ of the remaining pixel layers to $50 \times 250$ $\mu$m$^2$) for planar sensors, the introduction of new 3D sensors, and a dedicated $CO_2$ cooling system. Along with the other inner detector layers, the IBL improves the ability of ATLAS to identify particles, such as bottom quarks, by providing up to four precise measurements to improve impact parameter resolution and vertex reconstruction efficiency [105]. In particular, the upgrade improves trajectory extrapolation in the low and medium $p_T$ regime by producing an additional space point close to the interaction region [37].

Concentric to the inner detector, the SCT barrel consists of 4 double layers of 80 $\mu$m strips of length 126 mm, located across 4088 modules, and arranged either parallel to the beam axis or at an angle of 40 mrad. The design achieves resolutions of 17 $\mu$m in the $\phi$-direction and 580 $\mu$m in the $z$-direction. In the endcap region, this technology is extended to another set of strips distributed over 9 disks that brings the total number of SCT channels to about 6.3 million [91]. In the pixel detector and in the SCT, hits are measured as electron-hole drifts in the applied electric field after the electrons have been knocked out of orbit by the incoming particle.

Finally, the TRT extends ATLAS' tracking capability out to a radial distance of approximately 1 m. Its geometry and functionality is explained, for example, in Ref. [106] and is summarized here. The TRT is constructed from 4 mm diameter straw tubes that detect particles via the generation and collection of transition radiation and ionization charge. The drift tubes are filled with a Xe, $CO_2$, $O_2$ gaseous mixture and contain a grounded gold-plated tungsten wire at their center, creating a potential difference of 1.5 kV with the walls. When a charged particle ionizes the gas, the electrons drift towards the wire, producing a two-dimensional hit measurement. The measured drift time is proportional to the distance between the hit location and the wire. In addition, transition radiation is generated due to the inhomogeneity of the medium. At the interface between the straws and the polymer fiber or foil-filled gaps, transition radiation is given off with probability proportional to $\gamma = E/m$, offering a handle for particle identification and discrimination. Transition radiation photons can be used, for example, to discriminate electrons from pions.

In the barrel region, the 1.5 m long tubes are arranged in 70 layers parallel to the beam axis to reach $|\eta| < 1$ coverage. In the endcap region, straws are positioned along disks perpendicular to the beam line to extend coverage to the $1 < |\eta| < 2$ region. Measurements take advantage of drift time corrections to improve the spatial resolution to 130 $\mu$m in the $\phi$ direction, which is further compensated by the thickness of the detector, which improves the momentum resolution and increases the average number of hits to about 36 per particle.

2.2.1.2 THE CALORIMETER

Calorimeters are energy detectors. Placed after the tracking detectors, they aim at stopping particles by absorbing their energy. They are vital components of detectors at the LHC for their ability to measure particle properties that yield distinctive traces of phenomena of interest, which can be used for offline identification as well as event triggering. Calorimetry is a detection method that provides fast information about both charged and neutral particles. The precise understanding of the means through which particles of different kinds and energies deposit energy in a medium is fundamental to inform the design choices of detectors such as the ATLAS calorimeter system. This section first digresses in a discussion of the physical processes involved in shower creation and evolution, which will be useful to understand the physical constraints that guided the constructions of the ATLAS electromagnetic and hadronic calorimeters, described later in the section.

Electromagnetic showers develop from particles such as electrons and photons, whereas particles such as pions and protons are subject to strong nuclear interactions as well.

Electromagnetic (EM) calorimeters are designed to measure the energy deposited by the evolution of electromagnetic showers. As they traverse the calorimeter, photons and charged particles experience well-understood electromagnetic interactions with the dense detector medium. The nature of photon interactions are physically different from those of electrons and positrons. In the vicinity of an atomic nucleus with charge number $Z$, the Coulomb field $Ze$ affects the incoming charged particles via several different processes. For example, it may induce pair production $\gamma \to e^+e^-$ from incoming high energy photons, and photon emission via bremsstrahlung from incoming high energy electrons (with energy spectrum falling off as 1/energy [107]). The latter may leave behind a multitude of primarily soft photons. Particles may ionize the material, especially at lower energies, or induce excitations in the atoms or molecules that may result in scintillation signal upon de-excitation [107]. The average energy dissipation through ionization per unit length $\langle dE/dx \rangle$, also known as specific ionization or ionization density, is well modeled by the Bethe-Bloch formula [107]. Minimum ionizing particles (MIPs) are defined in terms of $\langle dE/dx \rangle$ as unit charge particles with energy equal to the position of the minimum in the $\langle dE/dx \rangle$ curve. At low energy, photons dissipate energy through photoelectric effect and Compton scattering. The availability of electrons in the medium, hence the charge number $Z$ of the material, heavily affects the cross-section for photoelectric processes, which, however, scales cubically with inverse incoming photon energy, thus losing overall relevance to competing processes very rapidly [107]. Repeated Compton scatterings by atomic electrons lead the energy deposition of photons via energy and momentum transfer. With a cross-section not typically exceeding a few percent of all other processes, photonuclear reactions may also contribute to photon absorption whenever the photon energy matches the binding energy of the last nucleon. Positrons annihilate with electrons, giving rise to photon pairs. Coulomb and Rayleigh scattering also contribute to altering the trajectory of incoming particles. Čerenkov light is also emitted by particles traveling faster than the speed of light in a given medium.

The radiative process is the dominant energy-loss process for high energy electrons and positrons with mass $m_e$, but it begins to significantly contribute to the absorption of heavier charged particles with mass $m$ at an energy value that scales as $(m/m_e)^2$ [107]. We can define a critical energy $E_c$ for a material to equal the crossover energy between the ionization and the radiation regimes [49]. For comparison, the second lightest charged particle, the muon, has mass $m$ about 200 times larger than that of the electron, requiring a critical energy about 40,000 times higher [107].

These interactions cause an exponentially growing cascade (*shower*) of thousands of interacting, secondary particles to form along the direction of the incoming particle, each with increasingly lower energy. High energy particles are likely to produce more particles via a multiplication mechanism that acts primarily through pair production and bremsstrahlung. Therefore, as a function of depth in the calorimeter, the energy deposited per unit length initially increases as the particle multiplicity increases, up to the so-called shower maximum. After this point, most shower components will have low enough energy to preferentially lose energy through processes that do no increase the particle count. Therefore, higher energy showers will, on average, penetrate deeper into the calorimeter.

The composition of the material chosen for the EM calorimeter determines the detector's radiation length $X_0$, defined as the average distance traveled by an electron before its energy is reduced by a factor of $1/e$. The radiation length can be approximated, in terms of the atomic weight $A$ and atomic number $Z$ as [2]:

$$X_0 = \frac{716.4\ A}{Z(Z+1)\ln(287/\sqrt{Z})} \text{g cm}^{-2} \tag{2.14}$$

Unlike electrons, with some probability, photons may traverse portions of the material without undergoing any interaction. Therefore, for photons, the concept of radiation length is instead replace by that of mean free path, where the average photon mean free path is approximately $(9/7 X_0)$, resulting in discernible differences in the shower evolution between $e$'s and $\gamma$'s. The radiation length, along with the Molière radius $\rho_M$, defined as

$$\rho_M = m_e c^2 \sqrt{4\pi/\alpha} \frac{X_0}{E_c}, \tag{2.15}$$

shift the conversation from $Z$-dependent calculations to allow for a material-independent description of both the longitudinal and lateral development of showers, respectively. Lateral shower spread is due to trajectory changes caused by multiple Coulomb scatterings, as well as isotropic processes such as Comp-ton scattering and photoelectric production [107].

Calorimeters are designed to contain the majority of shower developments in the energy ranges of interest. Shower length is proportional to the log of the energy. Therefore, to optimize shower containment, detector design requires mindful accounting for the expected longitudinal and transversal shower size as a function of incoming particle energy and $Z$ of the detector medium.

Hadronic calorimeters are equipped to measure physical interactions involving hadronic particles subject to additional strong interactions. The number of physical processes that proceed via the strong force, their complexity, and cross-section cause hadronic showers to display higher degree of diversity. On average, hadronic showers are expected to reach larger longitudinal and lateral displacements than electromagnetic showers. While nuclear interactions are the only viable energy loss mechanism for neutral hadrons, a charged hadron that travels through a detector might undergo several interactions that ionize the medium and finally interact strongly with a nucleus, fundamentally altering its nature and that of the nucleus, and creating multiple secondary particles. Similar to EM shower, hadronic showers begin with a multiplicative chain reaction that increases the number of particles in the shower, until a maximum is reached; after that point, absorption processes take over. A percentage of particles in hadronic showers decays electromagnetically, adding an EM component to the shower that is usually identifiable as a narrower central cone surrounded by a larger halo formed by the non-EM shower component. The EM fraction includes $\pi^0$'s, which make up roughly a third of the particles produced during the first nuclear interaction in the chain, when the decay is energetically favorable and allowed by baryon number conservation. At equal energy, proton showers have a shorter longitudinal profile than pion showers.

Hadronic showers may comprise an invisible energy component that is undetectable by the calorimeter.

Given the unique signatures of particles in calorimeters, recognizing their presence becomes an inverse problem of accurately measuring shower properties by making use of appropriate calorimetry technologies. The ATLAS detector contains two types of calorimeters: the electromagnetic calorimeter and the hadronic calorimeter. The whole complex is depicted in Fig. 2.11.

The electromagnetic calorimeter is a 0.5 m thick, longitudinally segmented, sampling structure of alternating active and passive media. The active material, in this case liquid argon (LAr), is the signal-generating component, while steel and lead are chosen as the high density absorbers. Argon was chosen for its purity and affordability. It is characterized by an accordion geometry, pictured in Fig. 2.12(c), that reduces the signal collection and transfer time, and minimizes cable length and dead space compared to the standard parallel-plate geometry [108]. In the barrel, the width of each accordion module increases with depth in order to subtend a constant angular distance $\phi$. 4 mm LAr gaps are sandwiched between millimeter thin lead layers in an alternating fashion, and the whole structure is sub-divided into three consecutive sublayers along the longitudinal direction, preceded by a presampler designed to estimate the energy lost prior to the particle reaching the calorimeter volume. Each compartment has different granularity in the $\eta$ and $\phi$ directions: the first layer extends for a depth of $4.3X_0$ and is segmented into 4 mm wide strips which provide fine $\eta$ resolution; the second layer extends for $16X_0$ with squared $4 \times 4$ cm$^2$ towers; and the third and final layer has a depth of $2X_0$ with twice the tower length in the $\eta$ direction compared to the previous layer. A diagram of its geometry is available in Fig. 2.12(a). The geometry and granularity of the first layer is particularly suited for the identification of electrons and photons, while the total thickness is sufficient to absorb nearly all electrons and photons and minimize particle punch-through events in the muon system [91]. The calorimeter system has a much coarser granularity than the inner detector.

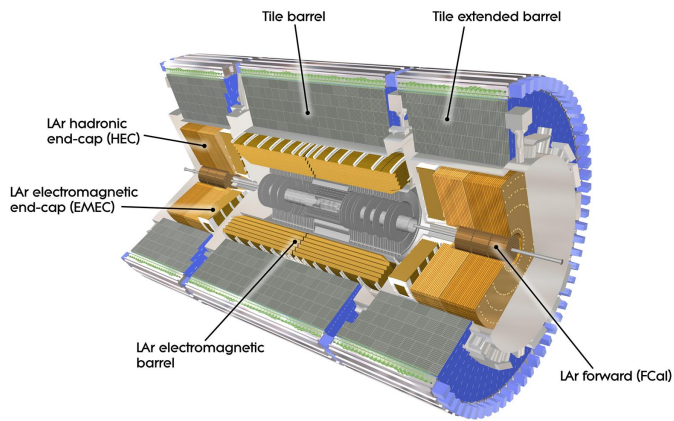

**Figure 2.11:** Diagram of the ATLAS calorimeter region.

The ATLAS hadronic calorimeter [109, 110, 111] surrounds the electromagnetic calorimeter, and is instrumented with 3 mm alternating plastic scintillating tiles as the active material and iron plates as the absorber. The role of the absorber in this sampling calorimeter is to advance the shower generation process, while the active material is responsible for transforming the ionization signal into light. 64 modules

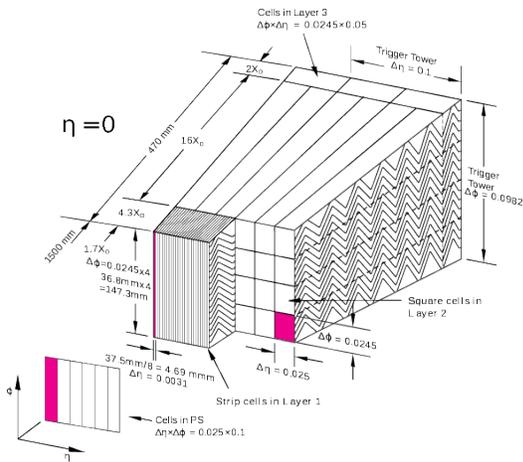
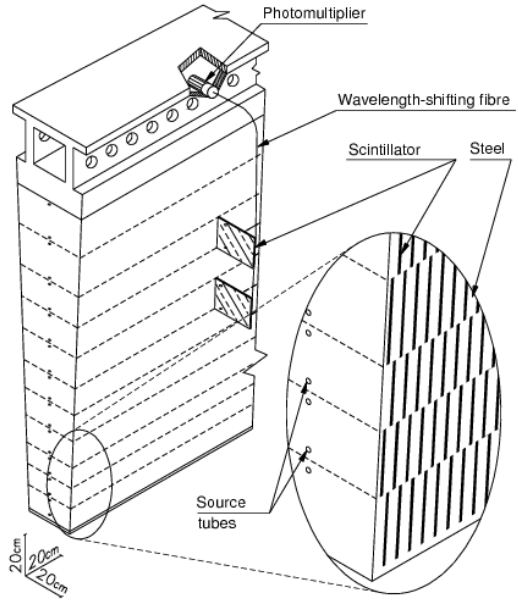

(a) Schematic representation of a wedge of the ATLAS electromagnetic calorimeter.

(b) Schematic representation of a wedge of the ATLAS hadronic calorimeter.

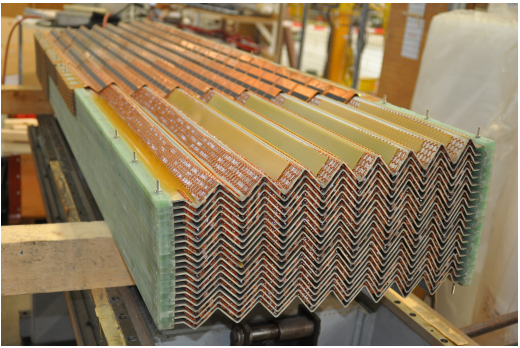
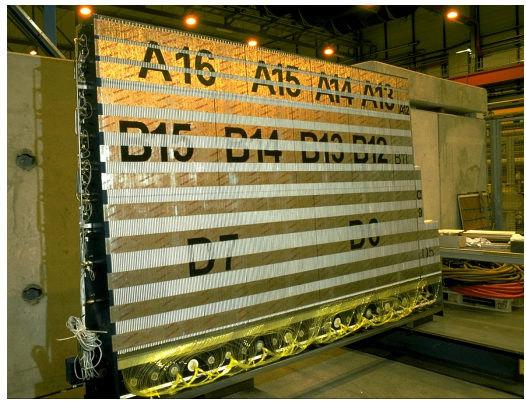

(c) Photo of the accordion geometry that characterizes the ATLAS electromagnetic calorimeter.

(d) Photo of a module of the ATLAS tile calorimeter.

**Figure 2.12:** These pictures shed light on the geometry, materials, and technologies adopted for the construction of the ATLAS elecromagnetic and hadronic calorimeters.

are arranged in each cylinder, each containing a series of radially staggered tiles arranged in the plane perpendicular to the beam axis. A diagram of the structure of this detector is show in Fig. 2.12(b), and module is pictured in Fig. 2.12(d).

About 10,000 photomultiplier tubes (PMTs), two for each cell, are placed on the outer surface of the calorimeter, collect the scintillation light transported from the tiles through wavelength shifting fiber optics mounted at the extremities of the modules, and convert it to electrical signals that are subsequently read out, amplified, and digitized every 25 ns.

The hadronic calorimeter is assembled in three consecutive components located along the beam line: a central barrel region ($|\eta| < 1.0$) is placed between two side (or extended) barrel portions ($0.8 < |\eta| < 1.7$). The choice of a 4.7 to 1 ratio of iron to scintillator material in the ATLAS hadronic calorimeter results in a nuclear interaction length of $\lambda$ = 20.7 cm. The Molière radius is 20.5 mm,

To factorize the material dependence, the depth of hadronic calorimeters is usually parameterized in terms of the nuclear interaction length $\lambda$, in anomaly to $X_0$ for electromagnetic interactions.

The physical hardware segmentation in the calorimeter corresponds to read-out cells of size $0.1 \times 0.1$ (or $0.2 \times 0.1$ in the outer D layer) in $\eta$-$\phi$ space, that connect to adjacent ones to form three main longitudinal layers. The segmentation is important to extract detailed kinematic information and identify particles based on their shower profile. The layers have thickness $1.5\lambda$, $4.1\lambda$ and $1.8\lambda$ in the barrel region and $1.5\lambda$, $2.6\lambda$, $3.3\lambda$ in the extended barrel region. Instrumentation gaps due to the presence of service electronics from other detector layers are filled with crack scintillators that fill holes in the calorimeter coverage and improve its hermeticity.

LAr technology interleaved with copper plates is adopted in the hadronic end-cap (HEC) and forward (FCAL) portions to cover a pseudorapidity range of $|\eta| < 4.9$ [91]. These "wheels" are located behind the electromagnetic end-cap detectors, perpendicular to the beam line.

The HEC covers the $1.5 < |\eta| < 3.2$ region and is segmented longitudinally into four sampling section, with granularity $0.1 \times 0.1$ in $\eta$-$\phi$ up to $|\eta| < 2.5$, and $0.2 \times 0.2$ beyond. The absorber material in the HEC is chosen to be copper, and the plates are arranged perpendicular to the beam to form two wheels.

The hadronic FCAL presents two sampling structures of thickness 3.6 $\lambda$ each, granularity $0.2 \times 0.2$ in $\eta$-$\phi$, and with tungsten as the absorber. Copper is chosen as the absorber in the EM FCAL wheel, instead. The technology used in this detector corresponds to long (45 cm), thin (5 mm in diameter) tungsten rods placed in a tungsten matrix and arranged parallel to the beam axis. These electrodes are held at 250 V, and separated from a concentric, grounded tube by a very narrow LAr gap (with width that increases up to ~500 μm across the layers). The hadronic FCAL is designed for reconstruction of forward jets and for operation in dense environments, where the high rate of pile-up events may cause large noise levels, detector saturation, and radiation damage. Pile-up reduction requires adequately fast response, guaranteed by the extremely narrow LAr gaps. A shield is located behind this detector to reduce punch-through effects in the muon system.

Good calorimeters are compact structures with excellent energy resolutions and little to no particle shower leakage. Energy resolution is limited by energy dependent factors, such as statistical event-by-event fluctuations in the shower development, noise, and inhomogeneous detector response, whose relative importance varies as a function of the energy scale [112].

The statistical nature of energy deposition results in energy resolution $\sigma_E \propto 1/\sqrt{N} \propto 1/\sqrt{E}$, where $N$ is the number of particles in a shower, and $E$ is the energy of the shower-initiating particle. The stochastic term is a fundamentally limiting factor in sampling detectors, with higher sampling frequency playing a role in ameliorating the resolution. The sampling structure makes it so that the resolution improves as more particles in a shower are sampled, but can also be improved by increasing the sampling frequency by reducing the thickness of each absorber plate, for example. Electronic noise in the readout components can instead be decreased by increasing the sampling fraction, thus increasing the signal-to-noise ratio in each readout channel. Another factor in the noise term is due to pile-up, which contributes to worsening the performance at low $p_T$. *Leakage*, shower particles escaping the calorimeter, contribute to the uncertainty of the measurement.

In hadronic calorimeters, random event-by-event fluctuations in invisible energy affect the energy resolution. Non-gaussian fluctuations also alter the EM-shower fraction distribution in hadronic showers. The calorimeter response to the EM and hadronic components becomes uneven. It is possible, however, to tune the response to achieve a response ratio $e/\pi = 1$).

The nominal energy resolution for the hadronic calorimeter is $\sigma_E/E \sim 50\%/\sqrt{E} \oplus 3\%$, while the electron resolution for the electromagnetic calorimeter is $\sigma_E/E \sim 10\%/\sqrt{E} \oplus 0.7\%$, though this approximation ignores the $\eta$ dependence of these terms [113]. The operator $\oplus$ signifies addition in quadrature.

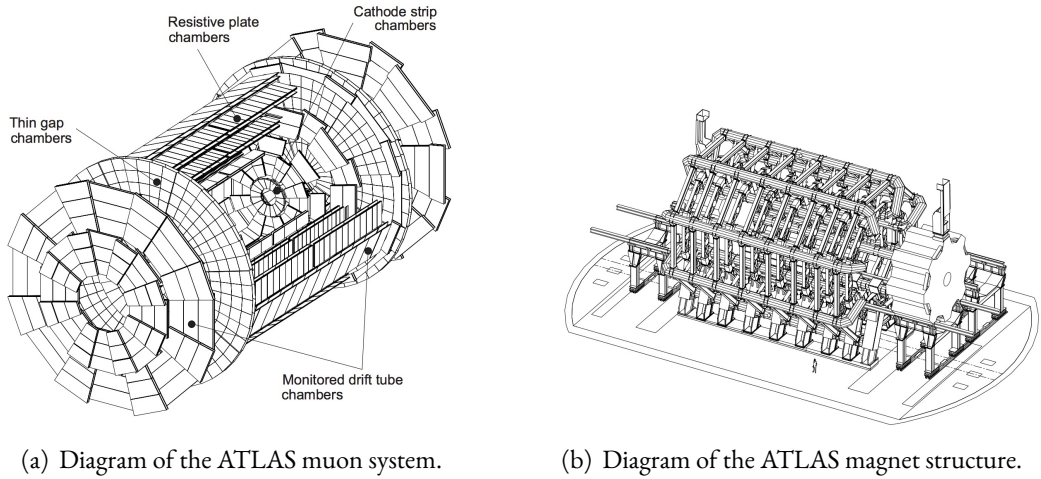

(a) Diagram of the ATLAS muon system.  (b) Diagram of the ATLAS magnet structure.

**Figure 2.13:** The unique magnet structure and extensive muon chambers in the ATLAS detector are optimized to bend and detect muons.

#### 2.2.1.3 The Muon Spectrometer

Muons are relatively long lived particles that belong to the second lepton family and are often described as the heavy siblings of the electrons. Muons may appear in the final states of Higgs events, or provide a signature for the presence of $b$-quarks. The phenomenology of the interactions of high energy muons significantly differs from that of lighter EM interacting particles, with radiation processes becoming comparable in cross-section with respect to ionization processes only at very high energies.

The layout of the ATLAS muon spectrometer is described in [114] and is shown in Fig. 2.13(a). The eight coils in the air-core toroidal magnets and two end-cap toroids provide the strong bending capabilities required to achieve excellent momentum resolution (Fig. 2.13(b)). These operate at temperatures of 4.7K. The ATLAS experiment owes its name to its unique toroidal magnet structure.

The muon spectrometer is the outermost layer of the ATLAS detector, and extends from a radius of about 5 m to about 10 m. Its goal is to detect muons that escape the previous layers of the detector and accurately measure their momentum. The barrel region is composed of three concentric cylinders that cover a pseudorapidity range of $|\eta| < 1$, while the four end-cap disks extend between $1 < |\eta| < 2.7$. This sub-detector is a combination of 4000 individual muon tracking chambers arranged along slightly overlapping large and small sectors, and built using four different chamber technologies. Around the azimuthal direction, to match the symmetry of the toroidal magnets, the muon design displays a 16-fold segmentation in which a large and a small sector alternate within each octant. In the barrel region, muon chambers form three concentric cylindrical layers, while, in the end-cap region, the large wheels perpendicular to the beam line are placed beyond the toroid magnets at distances of $|z| \sim 7.4, 10.8, 14,$ and 21.5 m from the interaction point [115].

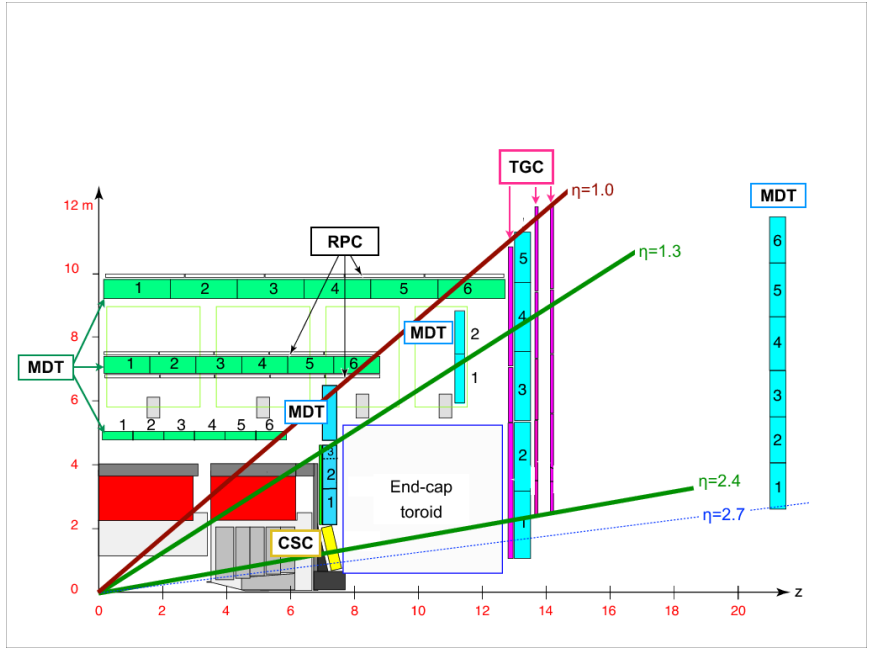

**Figure 2.14:** Placement of the muon chambers withing the volume of one quadrant of the ATLAS detector. This slice is taken in the positive $y$-$z$ plane, with the inetraction point at the origin, in the bottom left corner. The MDTs are colored in cyan if perpendicular to the beam line, and lime green if parallel to it. The CSCs are filled in yellow, and the TGCs in magenta. The RPCs are the white colored segments mounted along the barrel MDTs. The thick lines that traverse the detector roughly indicate the trajectory of a particle with the corresponding value of $\eta$. Image reproduced from Ref. [14].

Three sets of Resistive Plate Chambers (RPCs) in the barrel region and three sets of Thin Gap Chambers (TGCs) in the end-cap region are used for fast triggering on muon tracks. Precision chambers with Monitored Drift Tube (MDT) technology are used everywhere else in the detector for high accuracy momentum measurements, except for the inner circle of chambers in the innermost end-cap disk, where the design choice favored higher granularity Cathode Strip Chambers (CSCs). The layout of the various muon chambers is displayed in Fig. 2.14.

In the RPCs [115, 114], parallel electrode plates, 2 mm in thickness and separated by a narrow 2 mm wide gas-filled gap (for fast readout and triggering), exert a strong uniform electric field to collect ionization charges built up from the passage of muons in the barrel region $|\eta| < 1.05$. Their structure is shown in Fig. 2.15(a). The bakelite plates are held in place by a set of spacers located every 10 cm. The electric field is induced by applying a 9.8 kV voltage, and is used to produce an avalanche effect that multiplies the ionization charges before collection. To guarantee operations in avalanche mode and reduce steamer effect, in ATLAS, the gas mixture is chosen to be 94.7% $C_2H_2F_4$, 5% $C_4H_{10}$, and 0.3% $SF_6$. Readout strips in each chamber provide a track measurement in the $\eta$ and $\phi$ directions, with sufficient time resolution to enable triggering.

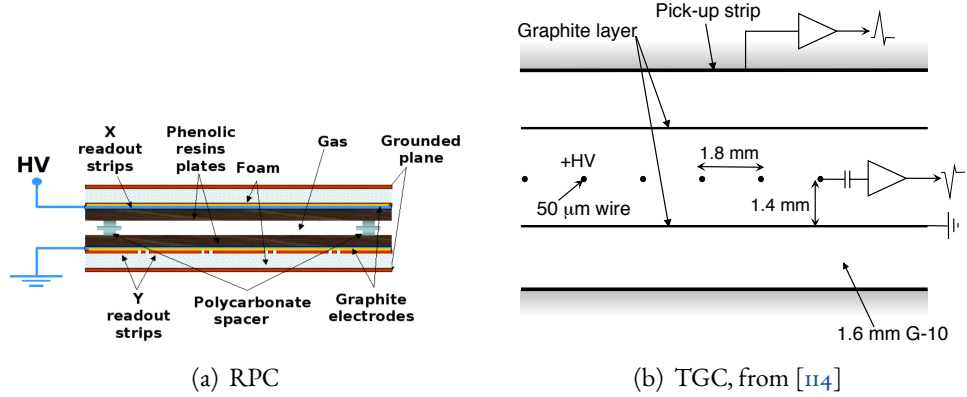

(a) RPC  (b) TGC, from [114]

**Figure 2.15:** Triggering muon chambers.

The TGCs [116, 114] take over in the $1.05 < |\eta| < 2.4$ region for triggering, and up to $|\eta| < 2.7$ for tracking. As the name might suggest, TGCs are proportional chambers in which a thin gap filled with parallel wire tubes held at a high voltage connects two grounded cathode planes and collects ionization charges with excellent spatial resolution. The gas is chosen to be a highly quenching mixture of 45% n-$C_5H_{12}$ and 55% $CO_2$ for a total gas volume of 16 m$^3$ [117]. These units are arranged over seven layers in either triplets or doublets of wire planes. They are characterized by an anode-to-anode distance greater than the cathode-to-anode distance, as shown in Fig. 2.15(b). They provide second coordinate measurements in the end-cap region for high $p_T$ muons, where they need to operate in saturated proportional mode and withstand increased track density and noise rates.

MDTs [117, 114] are high precision detectors that provide muon tracking and reconstruction capabilities by measuring hits in the bending direction of the magnetic field. Their coverage extends to the $|\eta| < 2.7$ region. Each unit is composed of a multilayer grid of gas-filled, 30 mm in diameter, cylindrical Al drift tubes with a 50 $\mu$m in diameter tungsten-rhenium gold-plated anode wire suspended at their center with individual resolution of 80 $\mu$m. The tubes are filled with a gas mixture of 91% Ar, 7% $CO_2$

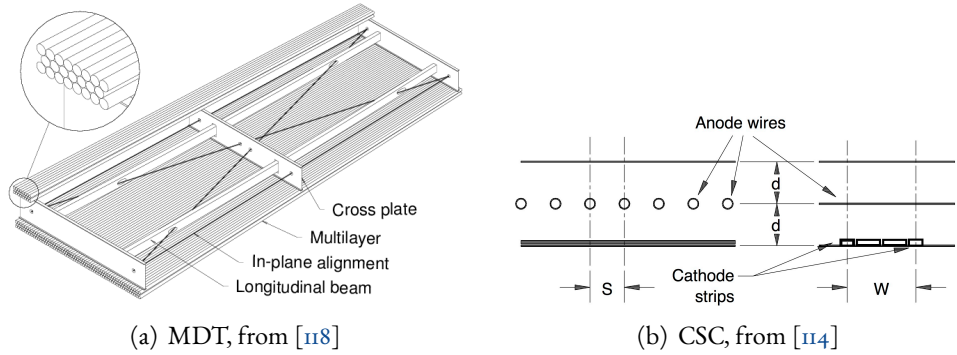

(a) MDT, from [118]     (b) CSC, from [114]

**Figure 2.16:** Tracking muon chambers.

held at a pressure of 3 bar. The layout of a chamber is shown in Fig. 2.16(a)

Finally, the fine granularity CSCs [117, 114], shown in Fig. 2.16(b), augment the end-cap coverage of the MDTs in the range $2.0 < |\eta| < 2.7$. These are multi-wire proportional chambers designed for higher occupancy conditions and arranged at an 11.59° angle from the normal to the beam axis, in a small wheel in front of the end-cap toroid magnets. The inclination is chosen so that the average track from the collision point would be perpendicular to this detector component, thus reducing the spacial resolution degradation due to the wire length. They are characterized by identical anode-to-anode and cathode-to-anode distance. Avalanche formation is favored by the choice of gas mixture (30% Ar, 50% $CO_2$, and 20% $CF_4$, for a total volume of 1.1 m$^3$). Charges are then collected by segmented cathode strips, and read out by a series of amplifiers and discriminators.

The design choices were driven by the objective of achieving a position resolution of 50 $\mu$m, which requires the uncertainty on each chamber position to be lower than 30 $\mu$m. Dedicated mechanical and optical alignment strategies have been devised to position and monitor and chambers [91].

## 2.3  Data Acquisition and Storage

The data acquisition system is responsible for collecting the signal registered by the various detector components and channel it to storage locations.

Specialized readout and front-end electronics dedicated to each detector unit are designed to collect, amplify, and digitize the signal collected in that portion of the ATLAS detector, and are located in the proximity of the corresponding hardware component.

### 2.3.1 Trigger

Given the number of readout channels in the ATLAS detector (hundreds of millions) and the number of collisions per second (about a billion), recording and storing every event (typical individual event size $\approx$ 1 MB) would require amounts of storage well beyond current availability. In addition, the cross-section for events of interest to the ATLAS physics program is orders of magnitude lower than the total event cross-section, so the majority of events only constitute a distraction, a haystack among which we look for the needle. Most events that occur within the ATLAS detector are therefore quickly discarded, and only interesting collisions are stored for further analysis. The method for automated real-time event selection is known as triggering. As data percolates through multiple trigger levels, its volume is reduced to manageable sizes. Trigger systems include both hardware and software based solutions, and require extremely low latency (or decision delay) to match the collision frequency.

Triggers are designed to isolate rarer events from the abundant background of soft or otherwise uninteresting collisions. Unusual events may contain large transverse momentum objects, or large transverse momentum imbalance. The production of energetic photons or leptons is another useful signature of electroweak physics at hadron colliders. The objective of trigger systems is therefore to rapidly identify these events and communicate the decision downstream. A flowchart of the ATLAS TDAQ system is provided in Fig. 2.17.

Similar to many institutions of higher education in the United States, the ATLAS trigger prides itself with an extremely low admission rate. A series of algorithms enforces selective cuts to identify worthy candidates. The level 1 (L1) is a hardware-level trigger implemented on custom ASICs and FPGAs located in close proximity to the detector to minimize latency (the maximum allowed latency is 2.5 $\mu$s). It operates at 40 MHz to reduces the event rate to $\mathcal{O}(100,000)$ per second. For decision making, the Central Trigger Processor (CTP) makes use of energy deposits in the calorimeters and track segments in the trigger-dedicated muon chambers, as well as signals from other dedicated subsystems. Regions of interest are found according to a set of decision rules.

The event rate is further reduced to a few thousands events per second by the High Level Trigger (HLT) that more carefully inspects the regions of interest flagged by the L1 trigger and buffered in the read-out system, and selects them for permanent storage on site and at Tier-0 computing facilities [119].

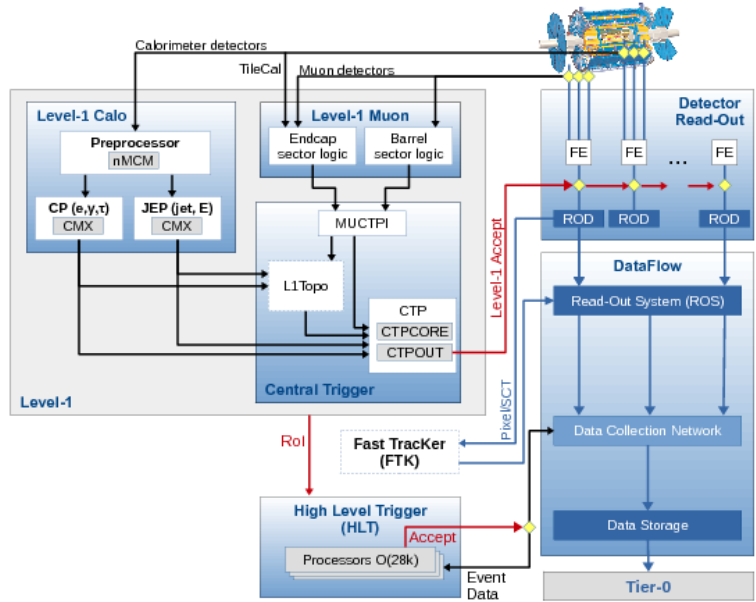

**Figure 2.17:** Schematic representation of the ATLAS Trigger and Data Acquisition system (TDAQ) for Run II. The FTK system is still in commissioning phase. Image reproduced from Ref. [15].

The Run II trigger upgrade combined the previously separated L2 and Event Filter triggers into a single event processing farm [120].

A new hardware-based trigger component, the FastTracKer (FTK), is currently in commissioning phase [121, 122]. It consists of more than 8,000 ASIC chips and 2,000 FPGAs. This massively parallel system is designed to integrate full track reconstruction for tracks with $p_T > 1$ GeV from hits in the ATLAS inner detector with latency $< 100$ μs. It does so by partitioning the detector in $4 \times 16$ towers in $\eta - \phi$ space that are processed in parallel by pattern matching algorithms. A linear approximation is then used for track fitting, as opposed to the full helical fit. The high-quality tracks serve as inputs to the HLT for early utilization of track-level information in the trigger system, with expected benefits for $\tau$ and $b$ selection in particular.

A trigger menu provides analyzers with the ability to retrieve events with desired signatures. Trigger items are, by design, generic enough to serve multiple analyses. Object-based triggers flag, for example, the presence of muons, photons, or high transverse momentum jets, while event-level triggers focus on global properties of the event, such as high missing transverse energy. The table in Fig. 2.18 reports the observed trigger rates for the trigger menu items optimized for the 2017 ATLAS data-taking campaign.

The ATLAS trigger underwent significant upgrades during the first long shutdown to ready it for the higher luminosity and collision rate of Run 2 [15]. Future challenges in online selection will continue to arise with increased pile-up.

| Trigger | Typical offline selection | Trigger Selection | | Level-1 Peak Rate (kHz) | HLT Peak Rate (Hz) |
|---|---|---|---|---|---|
| | | Level-1 (GeV) | HLT (GeV) | $L = 1.7 \times 10^{34}$ cm$^{-2}$s$^{-1}$ | |
| Single leptons | Single isolated $\mu$, $p_T > 27$ GeV | 20 | 26 (i) | 15 | 180 |
| | Single isolated tight $e$, $p_T > 27$ GeV | 22 (i) | 26 (i) | 28 | 180 |
| | Single $\mu$, $p_T > 52$ GeV | 20 | 50 | 15 | 61 |
| | Single $e$, $p_T > 61$ GeV | 22 (i) | 60 | 28 | 18 |
| | Single $\tau$, $p_T > 170$ GeV | 100 | 160 | 1.2 | 47 |
| Two leptons | Two $\mu$, each $p_T > 15$ GeV | $2 \times 10$ | $2 \times 14$ | 1.8 | 26 |
| | Two $\mu$, $p_T > 23, 9$ GeV | 20 | 22, 8 | 15 | 42 |
| | Two very loose $e$, each $p_T > 18$ GeV | $2 \times 15$ (i) | $2 \times 17$ | 1.7 | 12 |
| | One $e$ & one $\mu$, $p_T > 8, 25$ GeV | 20 ($\mu$) | 7, 24 | 15 | 5 |
| | One $e$ & one $\mu$, $p_T > 18, 15$ GeV | 15, 10 | 17, 14 | 2.0 | 4 |
| | One $e$ & one $\mu$, $p_T > 27, 9$ GeV | 22 (e, i) | 26, 8 | 28 | 3 |
| | Two $\tau$, $p_T > 40, 30$ GeV | 20 (i), 12 (i) (+jets, topo) | 35, 25 | 5 | 61 |
| | One $\tau$ & one isolated $\mu$, $p_T > 30, 15$ GeV | 12 (i), 10 (+jets) | 25, 14 (i) | 2.1 | 10 |
| | One $\tau$ & one isolated $e$, $p_T > 30, 18$ GeV | 12 (i), 15 (i) (+jets) | 25, 17 (i) | 4 | 15 |
| Three leptons | Three loose $e$, $p_T > 25, 13, 13$ GeV | $20, 2 \times 10$ | $24, 2 \times 12$ | 1.3 | < 0.1 |
| | Three $\mu$, each $p_T > 7$ GeV | $3 \times 6$ | $3 \times 6$ | 0.2 | 6 |
| | Three $\mu$, $p_T > 21, 2 \times 5$ GeV | 20 | $20, 2 \times 4$ | 15 | 8 |
| | Two $\mu$ & one loose $e$, $p_T > 2 \times 11, 13$ GeV | $2 \times 10$ ($\mu$) | $2 \times 10, 12$ | 1.8 | 0.3 |
| | Two loose $e$ & one $\mu$, $p_T > 2 \times 13, 11$ GeV | $2 \times 8, 10$ | $2 \times 12, 10$ | 1.7 | 0.1 |
| One photon | One loose $\gamma$, $p_T > 145$ GeV | 22 (i) | 140 | 28 | 43 |
| Two photons | Two loose $\gamma$, $p_T > 55, 55$ GeV | $2 \times 20$ | 50, 50 | 2.6 | 6 |
| | Two medium $\gamma$, $p_T > 40, 30$ GeV | $2 \times 20$ | 35, 25 | 2.6 | 17 |
| Single jet | Jet ($R = 0.4$), $p_T > 435$ GeV | 100 | 420 | 3.3 | 33 |
| | Jet ($R = 1.0$), $p_T > 480$ GeV | 100 | 460 | 3.3 | 24 |
| | Jet ($R = 1.0$), $p_T > 450$ GeV, $m_{\rm jet} > 50$ GeV | 100 | $420, m_{\rm jet} > 40$ | 3.3 | 29 |
| $E_T^{\rm miss}$ | $E_T^{\rm miss} > 200$ GeV | 50 | 110 | 5 | 110 |
| Multi-jets | Four jets, each $p_T > 125$ GeV | $3 \times 50$ | $4 \times 115$ | 0.5 | 16 |
| | Five jets, each $p_T > 95$ GeV | $4 \times 15$ | $5 \times 85$ | 5 | 10 |
| | Six jets, each $p_T > 80$ GeV | $4 \times 15$ | $6 \times 70$ | 5 | 4 |
| | Six jets, each $p_T > 60$ GeV, $|\eta| < 2.0$ | $4 \times 15$ | $6 \times 55, |\eta| < 2.4$ | 5 | 15 |
| $b$–jets | One $b$ ($\epsilon = 40\%$), $p_T > 235$ GeV | 100 | 225 | 3.3 | 15 |
| | Two $b$ ($\epsilon = 60\%$), $p_T > 185, 70$ GeV | 100 | 175, 60 | 3.3 | 12 |
| | One $b$ ($\epsilon = 40\%$) & three jets, each $p_T > 85$ GeV | $4 \times 15$ | $4 \times 75$ | 5 | 15 |
| | Two $b$ ($\epsilon = 70\%$) & one jet, $p_T > 65, 65, 160$ GeV | $2 \times 30, 85$ | $2 \times 55, 150$ | 1.2 | 15 |
| | Two $b$ ($\epsilon = 60\%$) & two jets, each $p_T > 65$ GeV | $4 \times 15, |\eta| < 2.5$ | $4 \times 55$ | 3.2 | 13 |
| $B$-Physics | Two $\mu$, $p_T > 11, 6$ GeV | 11, 6 | 11, 6 (di-$\mu$) | 2.5 | 47 |
| | Two $\mu$, $p_T > 6, 6$ GeV, $2.5 < {\rm m}(\mu, \mu) < 4.0$ GeV | $2 \times 6$ ($J/\psi$, topo) | $2 \times 6$ ($J/\psi$) | 1.6 | 48 |
| | Two $\mu$, $p_T > 6, 6$ GeV, $4.7 < {\rm m}(\mu, \mu) < 5.9$ GeV | $2 \times 6$ ($B$, topo) | $2 \times 6$ ($B$) | 1.6 | 5 |
| | Two $\mu$, $p_T > 6, 6$ GeV, $7 < {\rm m}(\mu, \mu) < 12$ GeV | $2 \times 6$ ($\Upsilon$, topo) | $2 \times 6$ ($\Upsilon$) | 1.4 | 10 |
| Total Rate | | | | 85 | 1550 |

**Figure 2.18:** List of main ATLAS triggers on the 2017 trigger menu, as reported in Ref. [16], which tabulates the observed event rates for each menu item.

2.3.2  STORAGE

The Worldwide LHC Computing Grid (WLCG) is a global network of computing centers located across 42 countries which harnesses computing resources for data analysis, storage, and simulation. The centers in this hierarchical network are organized in tiers. The CERN computing center, also known as Tier-0, is the common entry point into the computing network for the data collected by the experiments Data Acquisition systems and represents approximately 20% of the net computing resources.

Upon delivery of RAW data [123, 124, 125] from the online ATLAS facility, the CERN Tier-0 resources are enlisted for prompt data reconstruction, calibration, and archival tasks. Over 200 petabytes of data are stored on tape in CERN Advanced STORage system (CASTOR) at the CERN Data Center, and approximately 1 petabyte is processed every day [126].

The ATLAS Computing Resources Management established replication policies to distribute the data collected on site to the various grid location. Federation of data storage resources allows for efficient computing usage. The XROOT protocol [127] with the Xcache proxy cache [128] integrated with the Rucio data management service [129] are all used for remote data access by collaborators around the world.

# 3

# ATLAS Computing, Simulation, and Data Analysis

The outcomes of collisions generated by the LHC and detected by the various modules in the ATLAS detector produce petabytes of data per year. This data is analyzed by comparing it to expected results obtained from the computer simulation of outcomes under Standard Model and new physics scenarios.

The road from detection to final result requires the development of sophisticated software, and the availability of worldwide computing resources to perform the required operations of data collection, simulation, digitization, reconstruction, and analysis. The initial steps of collecting or simulating data occur separately, involve different computing resources, and require the development of specialized software. The output of the two streams is then processed through identical analysis steps. The analysis strategy proceeds through a series of derivations, in which information from a previous step is diluted and compressed into more compact yet more specialized formats. A stylized diagram of the ATLAS sim-ulation and data processing pipeline is provided in Fig. 3.1. From left to right, we see a reduction in file size as the information becomes more heavily processed to suit individual analysis groups.

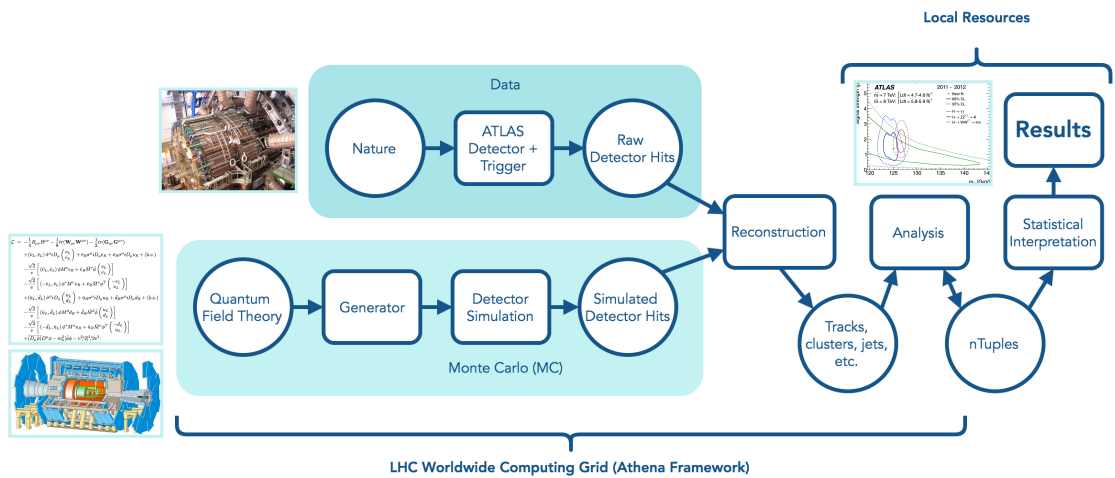

**Figure 3.1:** The data collections and simulation pipelines articulate themselves over several specialized steps and converge to produce identical output formats that can be processed by downstream reconstruction, identification and analysis steps. The data flow defined by this pipeline consists in the production of more and more specialized and compressed derived datasets from the initial collected or simulated detector readouts. Most of the data processing occurs on the Grid, where ATLAS software is deployed across hundreds of sites and machines. The final steps of data analysis and statistical interpretation instead, often, take place on individual users' laptops.

A description of the simulation portion of the pipeline will follow in Sec. 3.1, and an overview of the ATLAS computing environment is presented in Sec. 3.2.

## 3.1 Simulation

Models of particle collision and detection are central to the field of high energy physics. Virtual experiments are conducted on a large scale to inform the construction of particle detectors, gain insight on the response of the instrumentation upon interaction with elementary particles, make predictions for ob-servation, and interpret results in light of competing theories. Randomized methods for Monte Carlo sampling are at the base of modern particle physics simulators.

Experiments' outcomes are governed by physical processes spanning length scales over 20 orders of magnitude, all of which need to be accurately modeled in simulation. The net effect factorizes into contributions from processes at different energy regimes, allowing for the sequential treatment of simulated processes at different energy scales. The simulation pipeline can then be envisioned as a series of steps, starting from the smallest length scales (highest energy scales) at the interaction point, and ending at the largest length scales (lowest energy scales). Therefore, simulation requires many sophisticated packages for its many components, from generating particle collisions and their immediate decays, to constructing detector geometries and simulating the interaction of individual particles with hundreds of thousands of readout components. The ATLAS simulation infrastructure [130] connects and interfaces the various simulation routines.

### 3.1.1 Event Generators and Parton Shower Algorithms

The Monte Carlo simulation workflow starts with a physics generator that simulates the user-requested, hard-scatter subprocess – a high-momentum-transfer parton interaction from proton collisions and prompt decay of the interaction products. Empirical parton distribution functions (PDFs) are needed at this stage to model the non-perturbative behavior of colliding partons inside the protons. The evolution of the momentum fraction carried by partons is regulated by the DGLAP equations (Eq. 1.37, 1.38, 1.39). The LHAPDF library [131, 132] offers a universal interface to access several PDF sets, such as the ones provided by the CTEQ [133] and NNPDF [134] collaborations. For a recent report on uncertainty determination on parton distribution functions, see Ref. [135].

To calculate the cross-section of a subprocess, event generators [136], such as PYTHIA and SHERPA, allow to directly select a PDF set from the built-in list or to inject one from LHAPDF. Generators first simulate the physics event to idealistic precision, without detector effects. Monte Carlo techniques are used to describe the probabilistic nature of physics phenomena encoded in process amplitudes. Generators allow users to restrict the simulation to (possibly rare) selected subprocesses of interest within the SM or BSM frameworks. These tools handle the calculation of approximate matrix elements, up to a certain order in perturbation theory, and their integration over the large final-state phase space using Monte

Carlo integration techniques. The matrix elements can be obtained by summing over the contribution of all Feynman diagrams up to the specified order. Multipurpose event generators can be interfaced with dedicated software for leading order (LO) or next-to-leading order (NLO) cross-section computation, such as MadGraph5_aMC@NLO [137], to obtain higher-precision predictions. Matrix element methods are reliable and efficient in handling hard partons, but they present shortcomings in accounting for soft, collinear parton radiation or multi-parton diagrams.

Parton shower descriptions are combined and merged with matrix element calculations. Perturbative evolutionary processes take us from one end of the momentum transfer spectrum to the other via a Markov chain, where each scale solely depends on a stochastic transition from the output of the previous scale. These parton shower algorithms evolve color-charged QCD partons through the probabilistic radiation of gluons and quark-antiquark pair production, leading to a cascade, or *shower* of particles.

Given a process with final state partons $i$ with cross-section $\sigma_0$, the cross-section for the same event to be accompanied by the collinear radiation of an extra parton $j$ with momentum fraction $z$ is [136]:

$$d\sigma \approx \sigma_0 \sum_{\text{partons},i} \frac{\alpha_S}{2\pi} \frac{d\theta^2}{\theta^2} dz P_{ji}(z,\phi) d\phi, \qquad (3.1)$$

where $\theta$ is the opening angle between partons $i$ and $j$, $\phi$ is the azimuthal angle between partons $i$ and $j$, and $P_{ji}(z,\phi)$ are the spin-dependent splitting functions that encode the probability of radiating a parton. Divergences are handled by selecting a cutoff that establishes a criterion for the emission to be resolvable, and invoking unitarity to easily compute the probability of non-emission. This gives rise to an iterative algorithm, where, at each factorization step, radiation may or may not take place, and the final state of that process is used as the initial state of the subsequent process. Radiation that takes place after the hard scattering process is known as *final-state radiation* (FSR). If incoming partons radiate before the primary interaction takes place, the process is referred to as *initial-state radiation* (ISR). ISR is accounted for by back-tracing the hard-scatter through the DGLAP equations in order to compute the probability that a participating parton had evolved from another original incoming parton.

Electromagnetic charge gives rise of similar QED-regulated radiation effects that contribute to the shower evolution.

Jet formation is the phenomenon that describes the presence of a collimated bundle of partons, origi-nating from either soft or approximately collinear emission with respect to the original outgoing parton direction. The radiation pattern within a jet carries useful information to infer the nature of the original parton. This information can be used in the jet classification process called flavor tagging. More information on jet reconstruction and identification is available in Sec. 5.2.1.

At confinement scales of order 1 GeV, perturbation theory breaks down and non-perturbative interactions take over to form final-state, colorless hadrons from outgoing, colored partons, thus requiring tunable, non-perturbative hadronization models, such as the Lund string model [138]. Unstable hadrons are then decayed into lighter long-lived hadrons.

In addition, the underlying event, *i.e.* the products of soft interactions involving secondary partons of colliding hadrons, also require non-perturbative modeling with tunable parameters. This background radiation does not originate from the hard scatter vertex, but is instead likely to come from secondary interactions among incoming partons. For this reason, the tunable models developed to describe these components are called *multi-parton interaction* (MPI) models. Similar phenomenological models are used to describe most other interactions at the LHC that, contrary to the majority of interesting signal processes, can be categorized as soft processes, yielding low transverse momentum particles [136].

Monte Carlo generators store the truth-level information they produce in a C++ object representation stored in an EVNT file [139].

The diagram in Fig. 3.2 shows the complexity of the physical processes occurring at each hadron collision, all parts of which need to be carefully simulated by event generators to match empirical observations.

### 3.1.1.1 Pythia

Pythia is a general-purpose event generator that will be used extensively for the majority of studies in this thesis. It is therefore worth spending a few lines describing some of its unique properties.

The current Pythia 8 C++ library is the product of decades of development and of the overhaul of the previous, Fortran-based, Pythia 6 and Jetset (1978) versions. Now one of the most widely adopted generators in high energy collision physics, Pythia manages the hard-process generation, as well as the perturbative shower evolution phase, the hadronization, and the simulation of underlying soft processes. It also provides hooks to other specialized libraries to employ specific enhanced features.

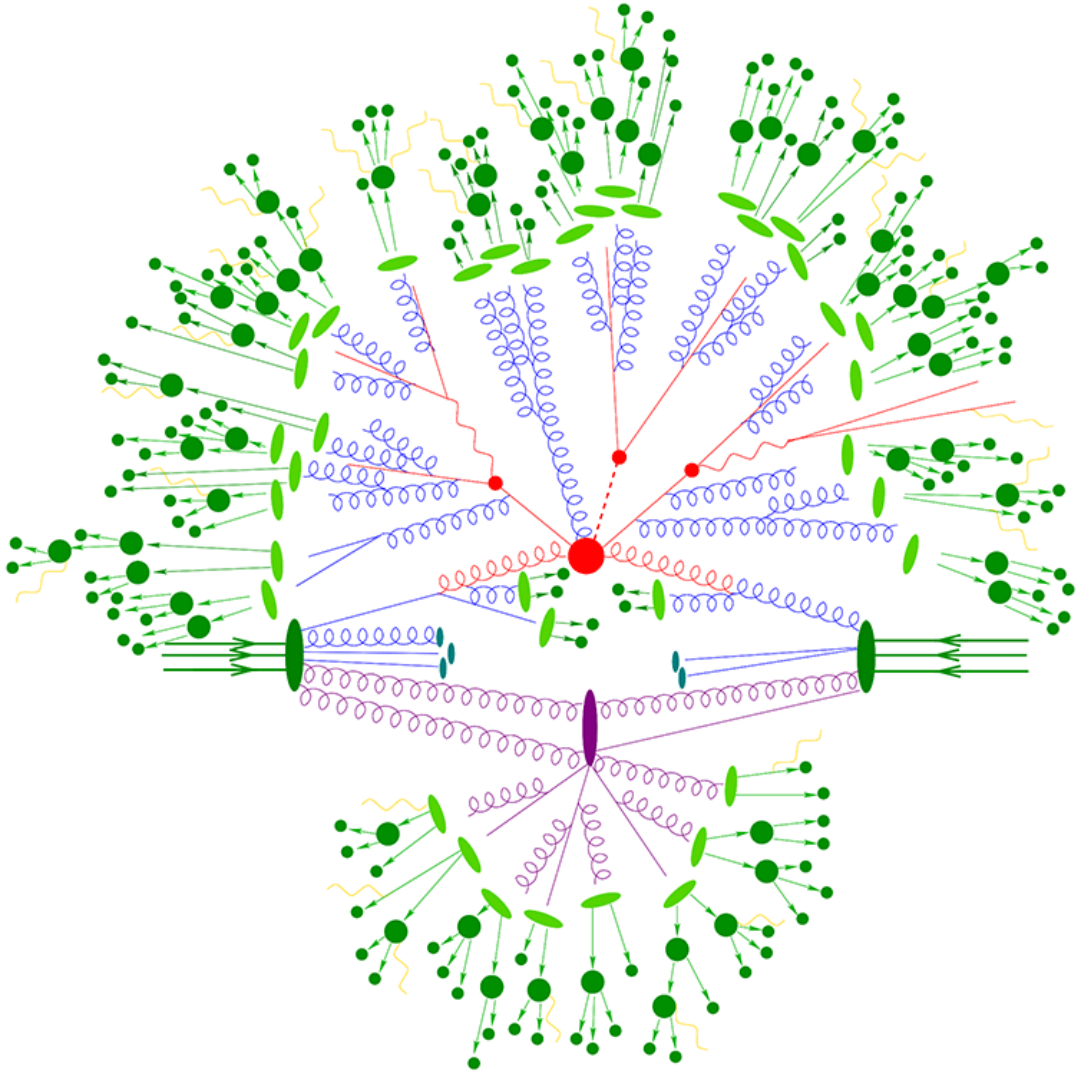

**Figure 3.2:** This diagram of a $t\bar{t}h$ event presents all the components that need to be simulated by event generators. The hard scatter event is highlighted in red, and matrix element calculations are used to compute its contribution. Initial and final state radiation is highlighted in blue and handled by parton shower algorithms. The purple lines in the bottom portion of the figure represent secondary interactions modeled by multi-parton interaction (MPI) models. The light green shapes indicate hadronization processes, while the dark green portions show the decay of unstable hadrons into lighter hadrons. QED radiation is shown in yellow. A version of this picture appeared in Ref. [17].

It includes both theory driven calculations and phenomenological models of non-perturbative effects.

Among the internally available hard processes, we primarily find $2 \to 1$ and $2 \to 2$ with some $2 \to 3$ processes, which include both SM and BSM options, such as QCD and EW events, Higgs processes, as well as SUSY, extra-dimension events, and more. Pythia also ingests LHA/LHEF inputs for additional hard process specification. MadGraph can be used to auto-generate Pythia-style classes for ad-hoc matrix elements.

Pythia's distinctive point is that it adopts an ordering in momentum perpendicular to the parton direction ($p_\perp$) in the dipole-style shower evolution [140]. Intuitively, this gives precedence to harder processes over softer radiation, regardless of their MPI, ISR, or FSR origin. The evolution equation that expresses the total differential branching probability that regulates probabilistic evolution in Pythia is given by:

$$\frac{d\mathcal{P}}{dp_\perp} = \left( \frac{d\mathcal{P}_{\text{MPI}}}{dp_\perp} + \sum \frac{d\mathcal{P}_{\text{ISR}}}{dp_\perp} + \sum \frac{d\mathcal{P}_{\text{FSR}}}{dp_\perp} \right) \\ \times \exp\left( - \int_{p_\perp}^{p_{\perp i-1}} \left( \frac{d\mathcal{P}_{\text{MPI}}}{dp'_\perp} + \sum \frac{d\mathcal{P}_{\text{ISR}}}{dp'_\perp} + \sum \frac{d\mathcal{P}_{\text{FSR}}}{dp'_\perp} \right) dp'_\perp \right), \quad (3.2)$$

where the subscript $i - 1$ identifies the momentum scale at the previous step in the chain, and the sums run over all incoming partons for ISR, and all outgoing partons for FSR [136, 140]. The analytical expression for each term, as well as much more information about Pythia 8, can be found in Ref. [140] [*].

Once energy scales $\mathcal{O}(1)$ GeV have been reached, pQCD breaks down, shower evolution slows down, and the hadronization phase begins. Hadronization is modeled in Pythia using the Lund string model [138], in which the energy build-up between two outgoing partons is modeled as a color flux tube of transverse size $\sim 1$ fm that behaves as a string with no transverse degrees of freedom. As the particles separate, the string may break, giving rise to a new quark-antiquark pair out of the potential energy previously stored in the string.

### 3.1.2 Detector Simulation

Particles produced or simulated in $pp$ collisions traverse the detector volumes and interact with the various media that compose the different detector layers. Detector effects must be accounted for in simula-tion to match reality with high precision. This is the goal of software libraries known as detector simulators.

---

[*]and in the online manual at http://home.thep.lu.se/~torbjorn/pythia82html/Welcome.html

Unlike event generation, which is experiment agnostic, detector simulation is highly coupled to detector geometry. Common tools exist in the community, but they need to be optimized and configured for final utilization by each experiment.

This portion of the simulation workflow is as important as it is time consuming. High precision, yet computationally demanding, routines are adopted in detector simulators like GEANT4 to accurately reproduce the behavior of particle interactions with detector material. The complexity of the geometry of the ATLAS detector, the largest particle detector ever built, poses a significant challenge to the scaling of the simulation effort. In practice, large quantities of high-quality simulated events cannot always be made available for all applications.

Full Simulation strategies are presented in Sec. 3.1.2.1, while Fast Simulation techniques are presented in Sec. 3.1.2.2. All simulation approaches are available in the ATLAS Integrated Simulation Framework (ISF), allowing for full and fast simulation strategies of different detector components to be interleaved in the same job. For example, in ATLAS, the combination of fast calorimeter simulation and full simulation of the tracking detectors is known as ATLFAST-II [141].

From the computing standpoint, the simulated energy depositions obtained from the detector interaction simulation step are saved in HITS format [123, 124, 125].

At the end of the simulation workflow, to simplify the comparison, the digitization process, which models all the sensor and electronics responses not captured by detector simulation packages, also ensures that synthetic events are stored in the same format as real collisions detected by the ATLAS detector. The Raw Data Object (RDO) file format is used for simulated events to match the RAW bytestream format from the data acquisition stream. Thanks to this standardized format, both real and simulated events can subsequently be processed through identical trigger and reconstruction steps, to ultimately produce common Analysis Object Data (AOD) files. ATLAS adopted a new common analysis format for Run II known as xAOD, which simultaneously supports ROOT I/O and Athena processing [139].

A database containing individual data collection run conditions (such as instrumentation faults and beam properties) is used to ensure maximal agreement between real and simulated datasets.

Validation is carried out in dedicated test-beam runs, in which relevant parameters, such as the energy dependence of calorimeter response, are measured and compared in data and simulation [142].

3.1.2.1 FULL SIMULATION

Full simulation is CPU and memory intensive, but achieves unparalleled levels of accuracy because of its careful accounting for individual detector cells and tracing of every particle involved in an event along their passage through matter. Particle propagation and interactions are developed step by step using the GEANT4 toolkit [143].

GEANT is an international collaboration that provides software for simulation of physical interactions to a large community of scientific and industry partners across fields such as high energy physics and medicine. The release of GEANT4 surpassed the previous Fortran-based GEANT3 version by employing modern object-oriented programming paradigms in C++.

In general, the first step in the full simulation pipeline involves the construction of a high-fidelity detector geometry, engineered from geometric primitives and elementary geometrical objects, and assembled into more and more complex detector volumes. Enormous and complex detectors like ATLAS can be rendered with unprecedented precision. These volumes are then embedded into magnetic field maps, that bend the trajectory of charged particles. Alignment transforms keep track of the movement of volumes over time.

Finally, one by one, particles are added to a stack and tracked as they traverse the detector material. Iterative Monte Carlo methods are used to model the inherently probabilistic nature of particle motion through matter. A series of steps in a chain describes the interaction history of a particle which may deposit energy throughout the detector and cause the creation of additional particles. Physics lists encode physics models used to describe interactions of particles with matter in different energy regimes and with different degrees of accuracy.

As mentioned, accurately modeling complex detector geometries, physical processes, and interactions happening at distance scales as small as $10^{-20}$m and as large as tens of meters requires extensive computing resources. Bulk production of simulated events is primarily generated using Tier-2 computing resources on the WLCG which support the ATLAS simulation software stack.

Improvements in the computational efficiency of full simulators are under investigation. For exam-ple, static GEANT4 builds versus dynamic linking of libraries with large address spaces has been shown to significantly improve the speed of full simulation [144, 145, 146].

### 3.1.2.2 Fast Simulation

Compromises between computational complexity and simulation accuracy need to be made to reduce CPU costs in scenarios in which full accuracy is not required. In general, fast simulation strategies provide an operational trade-off between precision and computational cost. Substantial speedups can be achieved at the price of higher levels of simplification in the detector and interaction description.

Simple, idealized solutions include Monte Carlo smearing to emulate the average detector response. Involved parametrization techniques lie a step above in terms of fidelity and detail.

In ATLAS, fast simulation solutions target detector components separately. All solutions are designed to be seamlessly interfaced with ATLAS digitization and reconstruction techniques.

Of all layers, calorimeters take up about 75 % of the total simulation computing time. The main fast calorimeter simulation program is known as FASTCALOSIM and has successfully been deployed since Run I [147]. Recent improvements, which make use of machine learning techniques for optimized I/O and memory usage, have been documented in Ref. [148, 149]. This tool aims at reproducing calorimeter cell readout responses by parametrizing shower evolution while taking into account shower shape properties. Single particles such as photons and electrons (for EM showers) and charged pions (for hadronic showers) are used as inputs to the simulator, binned in slices of energy and pseudorapidity [150]. The output of the fully simulated showers is parametrized, first by using principal component analysis to decorrelate longitudinal energy depositions across successive sampling layers and transform them into orthogonal components. The linearly independent directions uncovered this way are, however, perhaps highly correlated with conditioning factors such as incoming particle and shower properties, thus requiring several parametrizations in bins of quantities of interest (in this case, particle type, energy and $\eta$ of the original particle). At simulation time, random values are drawn and assigned to each latent component, and then rotated back into the detector readout space. For each PCA bin, a parametrization of the lateral shower profile is obtained by building template histograms in $\eta$-$\phi$ space, and approximating them with a neural network regressor [149]. A simplified geometry replaces the accordion setup, and a wiggle function is used to correct cell assignment.

In the FCAL, libraries of binned, pre-simulated, *frozen showers* are available to quickly load in electromagnetic cascades at simulation time. This solution primarily targets low energy showers ($< 1$ GeV) by substituting them with information-compressed, stored templates divided into learned pseudorapidity and distance bins [151].

In ATLAS, calorimeter-dedicated fast simulation performs roughly one to two orders of magnitude faster than full detector simulation [139, 148, 152].

Tracking detectors, such as the ID and MS, have their own dedicated fast simulation program known as FATRAS, born from the idea that detector simulation can be sped up by only simulating information needed as input to track reconstruction algorithms [153]. The complex detector geometry is then simplified to match the stripped-down geometry used by reconstruction software. Reconstruction efficiencies for different EM particle type and track parameter resolution comparisons between FATRAS and full simulations are documented in Ref. [152].

Commissioning of new ATLAS fast simulation procedures is constantly underway to test their applicability across a range of physics analysis scenarios. New solutions that employ generative adversarial networks are proposed in this thesis (see Sec. 8).

For very fast simulation techniques, the bottleneck in the pipeline shifts to the digitization and reconstruction steps. The ATLAS Fast Chain project [154] aims at streamlining all these components to optimize resource consumption with an eye towards Run III.

## 3.2 ATLAS Computing Infrastructure

Analyses consist of measurements of properties of known particles and interactions, and searches for new physics phenomena. Traditionally, analyses are performed using a *cutflow*, a series of cuts, or decisions, to select events in order to exclude background processes and identify signal-rich regions of phase space. The ATLAS computing infrastructure needs to support tooling for event processing that operates identically on simulated and collected data and allow users to define event and object selection cuts. It further ought to allow the flexibility to integrate shared, central collaboration services and functionalities into analysis-specific code developed by individual analysis groups. In particular, it needs to provide users with reproducible environments and common builds of supported libraries for data analysis.

ATLAS software is usually categorized in online (if it runs in real time during the data taking phase)

and offline (if it runs any time after the data has been collected and stored), though significant overlap exists between the two.

Simulation, as any other part of the ATLAS offline workflow as well as the online High Level Trigger, exists within the ATLAS Athena framework [†] [155]. Athena is itself based on the Gaudi framework that ATLAS shares with the LHCb collaboration. Several C++ projects exist within the Athena codebase and can be built independently from a fraction of the code. Projects are built daily (or "*nightly*"). Production-ready releases are tagged and versioned, and released on the ATLAS CVMFS server. Within the Athena framework, Python is used as a configuration language to control and specify settings and arguments in scripts called `jobOptions`. Tools and algorithms are implemented within fixed abstract interfaces, and are designed to be modular enough to be composed, shared, and reused for similar purposes across analyses. ATLAS software can be deployed everywhere thanks to Docker containers that mount CVMFS volumes.

Most of the codebase consists of research quality, experiment specific code accumulated over the decades, which makes little use of parallel programming techniques. The Long Shutdown 1 was instrumental in migrating from the CLHEP [156] to the Eigen [157] library for linear algebra operations [158]. Recent efforts to migrate ATLAS software to a multithread-friendly version, known as AthenaMT, are documented, for example, in Ref. [159, 160]. A highly vectorized upgrade to the GEANT library, known as GEANTV [161] is also in the making.

As shown in Fig. 3.1, beside the very final nTuple processing and statistical analysis steps, which are often run on local resources, much of the ATLAS workflow takes place on the nodes in the network of computing centers around the world known as the WLCG, or simply as 'The Grid' (see Sec. 2.3.2 for a brief introduction in the context of data storage). ATLAS, along with the other LHC experiments at CERN, subscribes to a highly decentralized computing paradigm that relies on large amounts of global shared resources located across hundreds of off-site locations to counter the demands and challenges of its data analysis workflows [162]. The computing centers that participate in this global partnership are divided into four tiers, and differ in the type of tasks they perform (Fig. 3.3): the Tier-0 facilities immediately store and begin to process data from the nearby experiments; RAW data is replicated at the thirteen

---

[†]The ATLAS Athena software can be found at the following link: http://acode-browser.usatlas.bnl.gov/lxr/source/athena

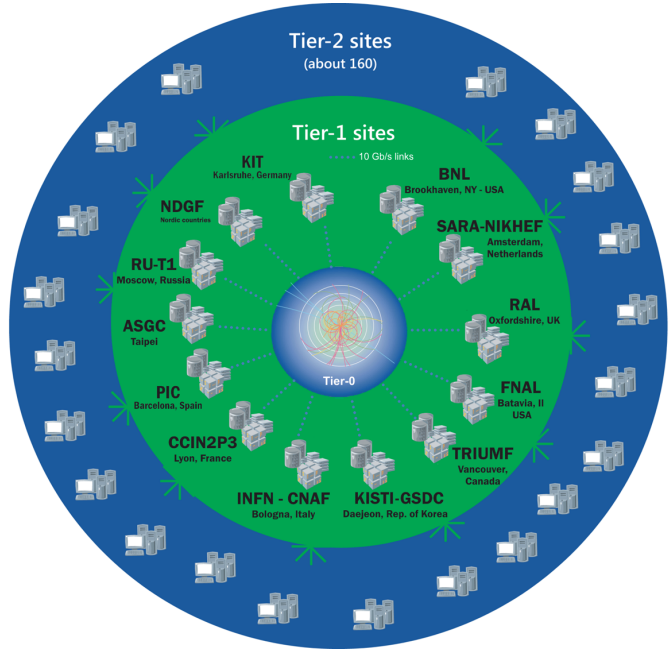

**Figure 3.3:** The Worldwide LHC Computing Grid (WLCG) relies on shared computing resources located at research centers and universities across the globe, and organized in tiers that fulfill diverse roles.

Tier-1 large computer centers for reprocessing and centralized analysis tasks; the output of derivation tasks that take place at Tier-1 facilities is then copied over to the ∼ 160 Tier-2 facilities at universities and research centers, where users can submit specialized analysis jobs as well as simulation requests. Local resources and smaller department-level clusters are informally identified as Tier-3 centers. CERN offers its users a Linux cluster called LXPLUS.

The WLCG provides impressive connectivity and networking capabilities [163], with total transfer rates up to 100 Gbps across the three data links that connect the Tier-0 facilities at CERN and in Budapest, Hungary. The LHC Optical Private Network (LHCOPN) handles Tier-0 to Tier-1 and inter-Tier-1 data transfer and communication across the globe at speeds up to 100 Gbps. Increased flexibility, fault tolerance, redundancy, and cost effectiveness are all advantages of the Grid concept over a centralized solution.

Moving towards the HL-LHC, ATLAS will need to continue looking for ways to include opportunistic computing resources and new technologies to confront the growing volumes and demands of heterogeneous workloads and computing needs [164, 165].

# 4

# Machine Learning for High Energy Physics

## 4.1 Statistical Learning Theory

Physics, statistical learning, and machine learning all share the ambitious goal of devising mathematical models that best fit a set of observations to then possibly be able to make principled predictions of similar outcomes. Statistical learning theory, the subject of this section, is concerned with finding parameters that allow models to behave optimally.

Given a model parametrized by weights $\theta$, and a set of random variables $(x, y) \sim P_{\text{data}}(x, y)$ where $P_{\text{data}}(x, y)$ is the true but unknown data distribution, training on a finite sample drawn from this distribution can be seen as the process of finding the optimal values for $\theta$ that best approximate the true data distribution. Generative modeling aims at learning an approximation to the true underlying data distribution $P_{\text{model}} \approx P_{\text{data}}$, while discriminative modeling restricts itself to the goal of learning the conditional distribution $P_{\text{model}}(y|x; \theta) \approx P_{\text{data}}(y|x)$.

4.1.1 MAXIMUM LIKELIHOOD

A common approach to optimizing the parameters $\theta$ is known as Maximum Likelihood Estimation (MLE). Given a finite sample from the data distribution $X = \{x | x \sim P_{\text{data}}(x)\}, |X| = n$, we construct a parametric model $P_{\text{model}}(x; \theta)$ and build the likelihood

$$\mathscr{L}(\theta; X) = \prod_{x \in X} P_{\text{model}}(x; \theta). \tag{4.1}$$

The goal of this optimization procedure is to maximize the likelihood $\mathscr{L}(\theta; X)$. Since the product of several terms that are $< 1$ tends to be numerically unstable to compute, it is instead preferred to minimize the negative log-likelihood:

$$\theta^* = \arg\min_\theta(-\ln \mathscr{L}(\theta; X)) = \arg\min_\theta \left( -\sum_{x \in X} \ln P_{\text{model}}(x; \theta) \right). \tag{4.2}$$

The solution can sometimes be found analytically, and often numerically, but the simplicity of the concept conceals the complexity of the implementation.

4.1.2 THE EM ALGORITHM

In partially observed models, parameter estimation becomes more complex. Expectation Maximization (EM) [166] is an iterative strategy for maximum likelihood estimation in the presence of hidden factors of variation. The iterative nature of the algorithm manifests itself in a two-step cycle: a value assignment step for the hidden variables (*E-step*), and a MLE step given the assignment in the previous step (*M-step*).

Given a set of observations $X = \{x | x \sim P(x)\}, |X| = n$ and a set of unobserved variables $Z = \{z | z \sim Q(z)\}, |Z| = n$, the likelihood of a probabilistic model parametrized by $\theta$ is $\mathscr{L}(\theta) = P(X|\theta) = \sum_Z P(X, Z|\theta)$. The goal of the optimization procedure is to find the optimal weights that maximize the incomplete log-likelihood according to $\theta^* = \arg\max_\theta \ln P(X|\theta)$. But $\ln P(X|\theta)$ can be

expressed as:

$$\begin{aligned}
\ln P(X|\theta) &= \sum_Z Q(Z) \ln P(X|\theta) \\
&= \sum_Z Q(Z) \ln \left[ \frac{P(X,Z|\theta)}{P(Z|X,\theta)} \frac{Q(Z)}{Q(Z)} \right] \\
&= \sum_Z Q(Z) \ln P(X,Z|\theta) - \sum_Z Q(Z) \ln Q(Z) + \sum_Z Q(Z) \ln \frac{Q(Z)}{P(Z|X,\theta)}
\end{aligned}$$

where the central term can be identified as the entropy of the distribution over the latent variables, which, combined with the first term, can be interpreted as a free energy, while the last term is the Kullback-Lieber divergence $KL(Q(Z)||P(Z|X,\theta))$. More information on divergences can be found in Sec. 4.6.2.2.

In both steps of the EM algorithm, the free energy is maximized. In the E-step, the distribution of hidden variables is updated by estimating the posterior and solving $Q(Z)_{(t)} = \arg\max_Q F\left(Q(Z), \theta_{(t-1)}\right)$; in the M-step, the weights are updated by maximizing the joint distribution of observed and latent variables, *i.e.* by solving $\theta_{(t)} = \arg\max_\theta F(Q(Z)_{(t)}, \theta)$. From a variational standpoint, the E-step maximizes a lower bound $F$ on the likelihood, while the M-step updates the model parameters such that $F$ matches the likelihood.

EM is guaranteed to converge to a local optimum of the likelihood, and the likelihood monotonically increases throughout the procedure: $P(X|\theta_{(t)}) \leq P(X|\theta_{(t+1)})$.

### 4.1.3 Empirical Risk Minimization

An alternative method for model training is the Empirical Risk Minimization (ERM) procedure. It relies on the idea that if a loss function $\mathcal{L}$ can be defined to measure the cost of a model's decision, then the objective is to minimize the expected loss over all possible observable events. In the context of supervised learning (for simplicity), define the risk $R$ as the expectation of the loss function

$$R(\theta) = \mathbb{E}_{(x,y) \sim P(x,y)}[\mathcal{L}((x,y); \theta)], \tag{4.3}$$

where $\theta$ are the parameters of the model, $P(x,y)$ is the ground truth distribution of observable properties that describe the examples in the dataset, and the expectation (or expected value) of a random variable is the probability-weighted sum over all possible outcomes. For a continuous random variable with probability density function $p(x)$, $\mathbb{E}[f(X)] = \int_{-\infty}^{+\infty} f(x)p(x)dx$; for a discrete random variable, $\mathbb{E}[f(X)] = \sum_{\text{possible } i\text{'s}} f(x_i)p_i$.

The ideal goal would be to minimize the risk $R$. The Bayes decision rule is the optimal classifier with parameters $\theta^*$ that minimize the risk. The value of the risk for the Bayes decision rule is called the Bayes risk, or optimal risk.

However, since the joint data distribution $P(x, y)$ is often unknown, the risk cannot be computed nor minimized exactly. It can be approximated by the *empirical risk*, the average loss over the empirical distribution defined by a dataset

$$\hat{R}(\theta) = \frac{1}{n} \sum_{i=1}^{n} \mathcal{L}((x_i, y_i); \theta), \tag{4.4}$$

where $n$ is the number of examples in the discrete dataset.

It can be shown that the empirical risk is an unbiased estimator of the risk:

$$\begin{aligned}
\mathbb{E}[\hat{R}(\theta)] &= \mathbb{E}\left[\frac{1}{n} \sum_{i=1}^{n} \mathcal{L}((x_i, y_i); \theta)\right] \\
&= \frac{1}{n} \sum_{i=1}^{n} \mathbb{E}\left[\mathcal{L}((x_i, y_i); \theta)\right] \\
&= \frac{1}{n} \sum_{i=1}^{n} \mathbb{E}\left[\mathcal{L}((x, y); \theta)\right] \\
&= \frac{1}{n} n \mathbb{E}\left[\mathcal{L}((x, y); \theta)\right] \\
&= R(\theta)
\end{aligned}$$

Learning consists of searching through a set of allowed functions parametrized by weights $\theta$. When learning proceeds by empirical risk minimization, the objective of the learning algorithm is to find the optimal parameters that minimize the empirical risk over the training dataset, *i.e.* the average training loss:

$$\hat{\theta}^* = \arg\min_{\theta} \sum_{i=1}^{n_{\text{train}}} \mathcal{L}((x_i, y_i); \theta). \tag{4.5}$$

However, the empirical training risk is a biased estimate of the risk:

$$\hat{R}\left(\{x_i, y_i\}_{i=1}^{n_{\text{train}}}; \hat{\theta}^*\right) \leq R(\hat{\theta}^*), \tag{4.6}$$

meaning that the loss calculated on the training set will be lower than the loss calculated on any other dataset drawn from $P(x, y)$, because, intuitively, it accounts for bias in the model, but not for variance (see Sec. 4.1.3.1 for details). The difference between the two can be upper bounded.

Doubts have recently been raised [167] on the suitability of using ERM for neural network training, due to concerns around generalization and robustness to adversarial examples [168].

### 4.1.3.1 Bias-Variance Decomposition

The risk function can be decomposed into different sources of error. The expected error is the sum of the estimation error and the approximation error. The estimation error can be conceptualized as the difference between the risk of the current classifier and the risk of the best classifier that can be trained given the family of functions that can be approximated using the chosen model. The generalization error is the difference between the best classifier that can be trained given the allowed family of functions and the Bayes risk. Formally, the difference between the risk $R(\theta)$ of the classifier parametrized by weights $\theta$ and the Bayes risk $R(\theta^*)$ can be rewritten as:

$$R(\theta) - R(\theta^*) = R(\theta) - \inf_{\theta \in \Theta} R(\theta) + \inf_{\theta \in \Theta} R(\theta) - R(\theta^*). \tag{4.7}$$

The first term can be interpreted as the variance of the model, the second as its bias. In practice, if the set of functions defined by $\Theta$ does not contain the Bayes optimum, then the model will consistently underfit. Increasing the complexity of the model corresponds to increasing the set of functions that the model can successfully approximate, to the point that, if the Bayes optimum were to be included in this set, the second term could be reduced to zero. This can be achieved by increasing the capacity of the model. The first term can also be reduced to zero, as it is theoretically possible, in the limit of infinite data and infinite time, to train the classifier to converge to the best possible classifier in the family of allowed ones. However, in practice, as the family grows, and with it the capacity of the model, the number of possible solutions grows and the probability of converging to the best allowed one decreases. In fact, models with higher capacity are likely to overfit to the training set and fail to generalize. Convergence is also inherently limited by the optimization procedure (in particular, the lack of convexity of the optimization problem and, thus, any guarantees of convergence) and by susceptibility to initial conditions.

Evidently, the family of allowed functions should neither be too large nor too small. Regularization is a set of strategies to impose a prior on the set of functions to search through, and will be the subject of Sec. 4.1.5.

In addition to the two sources of reducible error mentioned above, there may also be an irreducible error, coming from the possible dependence of the label on both observed and unobserved variables, as well as from sources of error in the measurement itself. If a set of unknown variables affects the label beyond the explanatory power of the model's input variables, then the label will appear to be noisy. Even assuming complete describability of the true labels from the observations, there may still be statistical flactuations that contaminate the collected labels. In either case, for an observation $x_i$, the label can then be expressed as $y_i = f(x_i) + \epsilon_i$, where $\epsilon_i$ represents the label noise from the data collection procedure.

#### 4.1.3.2 Cross-Validation

To achieve the goals of model selection and expected risk estimation, methods such as cross-validation become useful. The full dataset should be partitioned into a set used for the training procedure, and a test set for final performance estimation. Given a set of tunable parameters $\phi$ and trainable parameters $\theta$, the cross-validation procedure proceeds as follows. The training set is further divided into $K$ distinct partitions of equal size $\{\mathcal{D}_k\}_{k=1}^{K}$; the following iteration is repeated for each configuration of hyper-parameters $\phi$: for each partition $k$, the remaining $K-1$ are used to select the model with the best set of weights that offer a solution to $\theta^*_{\phi,k} = \arg\min_\theta \hat{R}(\mathcal{D}_{i \neq k}; \theta)$; the risk $R(\theta^*_{\phi,k})$ associated with the model trained on all but the $k^{\text{th}}$ subset and with hyper-parameters $\phi$ can be approximated by the empirical risk computed on the $k^{\text{th}}$ dataset:

$$\hat{R}(\mathcal{D}_k; \theta^*_{\phi,k}) = \frac{1}{n/K} \sum_{(x,y) \in \mathcal{D}_k} \mathcal{L}((x,y); \theta^*_{\phi,k}). \tag{4.8}$$

Finally, this risk is used in the estimation of the total risk of the model by computing $R(\theta^*_\phi) = \frac{1}{K} \sum_{k=1}^{K} R(\theta^*_{\phi,k})$. The best hyper-parameters configuration is selected by risk minimization:

$\phi^* = \arg\min_\phi R(\theta^*_\phi)$. Once the hyper-parameters have been selected, the model is re-trained to optimize the trainable weights $\theta^* = \arg\min_{\theta_{\phi^*}} \hat{R}(\theta_{\phi^*})$. A better estimate of the risk is finally computed as the empirical risk on the independent test test $\frac{1}{|\mathcal{D}^{\text{test}}|} \sum_{(x,y) \in \mathcal{D}^{\text{test}}} \mathcal{L}((x,y);\theta^*)$.

### 4.1.4 Optimization

As illustrated, for example, in Sections 4.1.1, 4.1.2, and 4.1.3, learning often refers to the task of minimizing or maximizing a function with respect to some of its parameters. Optimality is defined as the achievement of the set of values for the model parameters that incurs in the lowest possible value of the cost function, or, conversely, in the highest possible value of the utility function associated to the problem.

For the interested reader, a thorough and rigorous presentation of optimization methods for machine learning can be found in Ref. [169]. This section merely summarizes introductory notions that can be useful to machine learning practitioners in applied fields such as high energy physics.

The typical formulation of an optimization problem can be phrased as follows:

$$\text{Minimize } f(\theta) \tag{4.9}$$

$$\text{subject to } \theta \in \Theta$$

where $f$ is a loss function parametrized by a real vector of parameters $\theta$ that can take on values in the feasible set $\Theta$. If the feasible set $\Theta$ is a subset of $\mathbb{R}^d$, where $d$ is the dimensionality of the parameter vector $\theta$, then the optimization problem is constrained.

Although a few such problems can be solved exactly (perhaps even by hand), in general, optimization problems are not analytically solvable, but optimization methods often offer approximate solutions.

The complexity of the optimization process depends on the formulation of the objective function (see Sec. 4.1.4.1) and of the mapping from inputs to outputs. Convex functions on convex sets define optimization problems such that all local minima are global minima. On the other hand, complex, multidimensional learning tasks often result in intractable optimization problems. In most cases, it is therefore acceptable (if not even desirable, for generalization purposes) to converge to a good, approximate, local minimum, instead of the global minimum.

Once the optimization problem has been codified into mathematical form, the type of algorithms that can be used to solve it are known as optimizers. These optimization methods can be divided into groups according to the level of information they use and the complexity to calculate it. Zero-order methods only use the value of the function $f(\theta)$; first-order methods also require the value of the first derivative $f'(\theta)$; second-order methods need access to the Hessian $f''(\theta)$. Sec. 4.1.4.2 is dedicated to the description of gradient-based optimization methods.

#### 4.1.4.1 Loss Function

The loss function, also known as objective function or cost function, is a criterion that can be used in an ERM-like optimization procedure. It is, in other words, the function to be minimized in the problem. The loss function must be a real-valued function that encodes a notion of disadvantage of a solution compared to other feasible solutions.

The choice of cost function is tied to a deep understanding of the problem itself and different choices may result in different optimal solutions. Many complex, real-world problem are difficult to phrase in terms of a unique, unequivocal cost function, but optimizing the wrong function can introduce issues that fall under the broader umbrella of *alignment* problems. If a proxy is used for an unquantifiable objective, undesirable behavior may occur.

The loss can be expressed in terms of target labels in supervised learning settings, or as functions of the input data in unsupervised settings. Depending on the domain of the target values, loss formulations that enforce output constraints may be found to be more suitable than others. Brief lists of common loss functions used in classification and regression problems can be found, respectively in Sec. 4.3.1 and 4.3.2.

There exist many ways of optimizing the parameters of a model by minimizing the loss function with respect to the parameters. A family of such procedures falls in the category of gradient-based methods, and is further discussed in Sec. 4.1.4.2.

#### 4.1.4.2 Gradient-Based Optimization

Given a loss function that is subdifferentiable with respect to the model parameters, the optimization can proceed by gradient descent (GD). It is known, in fact, that the gradient indicates the direction of fastest local ascent in the value of a function. If the function is twice subdifferentiable, the second order condition for a minimum requires that $f''(\theta^*) \geq 0$.

For the problem description in Eq. 4.9 with subdifferentiable $f$, a solution can be found iteratively by initializing the weights to $\theta_0$ and updating them at each iteration $i$ according to the rule

$$\theta_{i+1} = \theta_i - \lambda_i \nabla_\theta f(\theta_i) \tag{4.10}$$

where $\lambda_i$ is the tunable step size, or learning rate – a problem-dependent parameter that determines how big a step to make in the direction opposite to that indicated by the gradient.

The learning rate can prevent convergence if too large, or slow it down if too small. Its magnitude can be held constant during the optimization procedure, or adjusted at each iteration using various methods. If held constant, it is often suggested to set it to a relatively small value compared to the scale of the problem, because the optimality of the direction indicated by the gradient only holds for infinitesimally small steps. However, in the presence of long flat regions along the optimization surface (*plateaus*), small step sizes might result in a large number of iterations needed to escape a potentially suboptimal region. There exist data driven strategies, based on the computation of the eigenvalues of the Hessian, that can help determine an ideal learning rate. If the learning rate is to be adjusted over time, a simple strategy is to decay it at each iteration, according to schema such as $\lambda_{i+1} = \lambda_i/(i+1)$, or other powers of $i+1$. Other popular options include finite step decay, in which the decay rate is decreased by a constant factor every certain number of epochs, and exponential annealing. Alternatively, one could use line search by solving for $\lambda_i^* = \arg, \min_{\lambda_i} \theta_i - \lambda_i \nabla_\theta f(\theta_i)$ for a fixed set of values $\lambda_i$. Recent contributions have derived empirical proportionality laws between the effects of decaying the learning rate to encourage convergence within a particular region and increasing the batch size [170, 171, 172, 173, 174]. As should now be evident, selecting an appropriate learning rate is a difficult problem, complicated by the additional intricate relationship with batch size. The issue is, therefore, not completely settled, with much of the recent debate and scholarly literature [175] centering around warmup strategies, adaptive scheduling, cyclical learning rates, and the emergence of adaptive learning methods such as RMSProp [176], Adam [177], Nadam [178], Adagrad [179], Adadelta [180], ESGD [181], Adasecant [182], and more [183]. Some scholars, however, challenge the preconceived notions that adaptive methods are unequivocally superior to gradient descent or stochastic gradient descent, due to the poor understanding of their gen-eralization properties and their apparent preference for sharp minima, and therefore warn against their indiscriminate use in neural network trainings [184].

For a convex and differentiable function $f$ with gradients with Lipschitz constant [185, 186, 187] $L > 0$, as long as the learning rate is set to $\lambda \leq 1/L$, after $k$ iterations, gradient descent is guaranteed to converge to a solution $\theta_k$ such that, compared to the optimal solution $\theta^*$, the following convergence relation holds, showing a $\mathcal{O}(1/k)$ convergence rate:

$$f(\theta_k) - f(\theta^*) \leq \frac{||\theta_0 - \theta^*||_2^2}{2\lambda k} \tag{4.11}$$

When utilizing gradient descent to learn an optimal configuration of model parameters, the user can choose how often to update the parameters $\theta$, *i.e.* how many input data points to use to estimate the local gradient direction. This choice partitions gradient-based techniques into separate, related algorithms with unique convergence properties. For example, computing the gradient from a single training example (*online*, or *stochastic gradient descent* (SGD)) is feasible and fast, but results in a very noisy gradient approximation with high variance updates. Conversely, computing the gradient using the entirety of the training dataset yields a very reliable estimate of the gradient, but is computationally expensive and data inefficient.

*Minibatch stochastic gradient descent* [188], often itself referred to as SGD, strikes a compromise between these two extrema, by making use of a limited collection of randomly sampled examples to compute a fast yet robust update:

$$\theta_{i+1} = \theta_i - \lambda_i \frac{1}{|\mathcal{B}|} \sum_{x \in \mathcal{B}} \nabla_\theta f(x, \theta_i), \tag{4.12}$$

where $\mathcal{B}$ identifies the set of randomly selected examples in a minibatch, $|\mathcal{B}|$ is its size, and $x$ is one such example. In practice, minibatch SGD computes an unbiased, variable estimate of the full gradient, while reducing the variance compared to the single-sample approximation. SGD also encourages exploration of the parameter landscape, avoids costly computations of an only locally optimal descent direction, and acts as a source of regularization through the injected noise.

Stochastic gradient approximation provides tangible computational advantages, which result in faster rate of convergence compared to the whole-dataset-level gradient method, due to its efficient use of the available training examples and its quicker initial improvement that is accomplished without waiting for all events to be processed at least once. More theoretically grounded reasons to prefer stochastic methods are summarized in Ref. [169].

If the optimization is not convex, as in the case of deep learning, SGD is not guaranteed to converge to the global minimum. In practice, though, it has been repeatedly observed that, when applied to the training of deep neural networks, it often converges to acceptable solutions with comparably low loss values. A more detailed discussion of the mathematical properties of critical points to which SGD is found to converge is presented in Sec. 4.4.1.2.

Improvements upon the vanilla minibatch SGD include the addition of momentum, which modifies the update rule by adding a weighted term that considers previously computed update steps in formulating the next parameters update:

$$\Delta\theta_{i+1} = -(1-\mu)\lambda_i \frac{1}{|\mathcal{B}|} \sum_{x \in \mathcal{B}} \nabla_\theta f(x, \theta_i) + \mu \Delta\theta_i, \qquad (4.13)$$

where $\mu$ rescales the fractional contributions from the current gradient calculation and from the previous updates. Intuitively, momentum prevents individual high-variance updates from dominating the optimization, and instead collects consensus from previous gradient calculations to inform the parameter update. This can be beneficial to speed up the learning process in zones of low but persistent curvature.

An increasingly popular alternative to the momentum formulation described above is to use Nesterov momentum, which suggests computing the local gradient contribution at the weight configuration obtained after first applying the momentum-weighted update from the previous iteration [189]. It follows that the final update can benefit from peaking ahead towards the configuration suggested by the momentum-based strategy and correct it or refine it to improve convergence. In fact, given certain conditions on the range of the learning rate, Nesterov's accelerated gradient (NAG) is guaranteed to converge faster than regular gradient descent in the case of convex optimization – $\mathcal{O}(1/k^2)$ versus the $\mathcal{O}(1/k)$ behavior reported in Eq. 4.11.

In addition to first-order methods, various second-order techniques are available to improve the convergence properties towards critical solutions to the optimization problem. Second-order methods make use of the Hessian, the matrix of second partial derivatives of $f$ with respect to its arguments.

Newton's methods is a common root-finding algorithm that can be applied to optimization when the function of interest is $g(\theta) = f'(\theta)$. A critical point of $f(x)$, indeed, would occur wherever its first derivative vanishes. This is equivalent to finding the roots of $g(x)$, which can in fact be achieved through Newton's method. Assume the optimization starts out at a value $\theta_0$ close to the optimal $\theta^*$. Then, by Taylor expansion, a better guess for the location of the minimum can be found as follows:

$$0 = g(\theta_1) \approx g(\theta_0) + g'(\theta_0) \cdot (\theta_1 - \theta_0) + \mathcal{O}\left((\theta_1 - \theta_0)^2\right), \quad (4.14)$$

with solution

$$\theta_1 = \theta_0 - \frac{g(\theta_0)}{g'(\theta_0)}. \quad (4.15)$$

Therefore, recalling the relation between $g$ and $f$, the location of the critical points of $f$ can be approximated by iteratively stepping in the direction of the gradient with step size scaled by the inverse of the Hessian.

Calculating and storing the Hessian matrix is not always computationally feasible. Inverting it might also be a challenge. Therefore, the original formulation of this method becomes unfeasible beyond a few dimensions. Quasi-Newton methods present computationally advantageous alternatives that allow to approximate the Hessian using first order information only. Several approximation techniques are available, including, for the example, BFGS [190].

In high dimensional non-convex scenarios, saddle points in the loss landscape, which occur whenever the Hessian is not a positive definite matrix, may become attractors under Newton or quasi-Newton formulations, as pointed out in Ref. [191], which makes them ineffective in light of the recent observations on the proliferation of saddle points in deep learning optimization (see Sec. 4.4.1.2).

### 4.1.4.3 Backpropagation

For any learning algorithm to take advantage of gradient-based optimization, the loss function must be differentiable with respect to the inputs and parameters that appear in the operations that compose the model. For certain models such as linear regression (in the form of Ordinary Least Squares) and logistic regression, such differentiation can easily be handled analytically. However, since deep neural net-work models have complex structure, control flow, and lack continuity in portions of the computational graph, analytically computing gradients is often not feasible.

The backpropagation algorithm presents a solution by providing the ability to recursively calculate gradients of composable functions in an efficient manner that maximizes sub-expression reuse.

As early as the 1960s [192, 193, 194], early versions of backpropagation-like algorithms existed, often inspired by dynamic programming. In 1970, Ref. [195] introduced *Reverse-Mode Automatic Differentiation*, which would later be applied to neural networks in Ref. [196]. Backpropagation was finally popularized in Ref. [197], which empirically showed that this specific version of Reverse-Mode Automatic Differentiation is effective for learning representations in neural networks with hidden layers. A more detailed history of automatic differentiation and its simultaneous rediscovery in the context of machine learning in relation to backpropagation can be found in Ref. [198].

Simply stated, backpropagation is designed to ascend a computational graph and compute nested gradients with respect to all the parameters in the composite functions that connect outputs and inputs within a model. Its effect is equivalent to recursively applying the chain rule along the graph of operations that define the network structure while making use of precomputed quantities. The goal is to track functional dependencies and efficiently assign credit or blame for observed shifts in the cost. Backpropagation is simply an intelligent way of keeping track of parameter dependencies for smart derivative taking using the chain rule. The way in which the derivatives are then used depends on the choice of specific optimizer (SGD or others).

Efficient backpropagation is often implemented as a shadow graph that augments the forward operations with corresponding backward methods. It is a form of book-keeping and memoization of differentiation results that makes use of caching of reusable quantities computed in both the forward and upstream backward pass, in order to avoid repeating the same expensive calculations for each individual weight update. In practice, backpropagation allows to relate and connect operations for gradient computation so that the chain rule does not have to be recomputed from the root node each time, but its previous results can instead be reused as information is tracked down the graph structure.

4.1.5 Regularization

Regularization techniques are introduced to limit the capacity of the model in order to avoid overfitting to the fluctuations or to the noise level displayed in the training dataset, to reduce the complexity of the family of functions a model may to converge to, or to impose other forms of constraints on the model structure or parameters that reflect prior domain knowledge. These are mathematical forms of encouragement towards simpler, more robust models that penalize solutions that fail to adequately generalize. Regularization methods generally fall into two main categories: norm-based and stochastic regularization.

Norm-based regularization generally imposes explicit penalties on the magnitude of certain model parameters in the form of additional terms $\lambda l(\theta)$ to the loss function, where $\theta$ identifies a set of parameters, $\lambda$ is the regularization coefficient, and $l$ is a measure of the norm of the parameters. The Frobenius norm (or Euclidean norm) encourages small weights; the element-wise 1-norm (the sum of the absolute values of all entries) induces sparsity in the parameters. Spectral regularization, which expresses the penalty in terms of an induced norm, appears less frequently in the literature [199], but has recently enjoyed considerable success when applied to the context of generative adversarial networks for discriminator stabilization [200]. A much more detailed, intuitive, and pedagogical explanation of the effects of various types of regularization is available in Chapter 7 of Ref. [48]. In the deep learning use case, different regularization coefficients and strategies can be adopted for individual layers or sub-networks within a larger model, at the expense of an increased number of hyper-parameters to tune.

Stochastic regularization is comprised of a variety of techniques, including dropout [201], batch normalization [202], and early stopping [203], that implicitly act as capacity constraints on the learning algorithm. Theoretical work on the investigation of the properties and induced effects of stochastic regularization is still underway. Meanwhile, though, thanks to their empirical success and ease of implementation, these methods have surpassed their norm-based counterparts in terms of popularity as forms of regularization in deep learning models.

Dropout, in particular, may currently be one of the most frequently utilized components in neural networks, surely in part because of its computational and conceptual simplicity. As Ref. [204] describes it, dropout is the mathematical equivalent to "the concept that if you can learn how to do a task repeat-edly whilst drunk, you should be able to do the task even better when sober". In practice, at every iteration of the training procedure, dropout consists of randomly sampling a binary mask over the network units and apply it to drop a percentage of node connections among adjacent layers, thus discouraging neuron co-adaptation [205, 206]. In fact, the random dropping of selected connections increases the likelihood that all units learn independently useful and robust features, without strongly relying on the computation performed by any other individual neuron. Weight scaling by the probability of dropping a given unit must be applied at inference time to correct for the training-time shift in magnitude in the weights due to the dropping of units [205]. The practical effect of dropout on the learned model has been interpreted as being approximately equivalent to learning an ensemble of networks that share the same underlying structure [206]. Its probabilistic nature is consistent with the Bayesian interpretation of mathematically approximating a deep Gaussian process, and can be further used to capture and model the uncertainty associated with the network's prediction without the need for extraneous probes that hamper its performance [207].

Stochastic regularization can also be implemented within specific neural network operations, an example of which is stochastic pooling among convolutional layers [208] or recurrent dropout in RNNs [209]. Several other variants and expansions upon the dropout algorithms are available in the literature [210, 211, 212].

## 4.2 Unsupervised Learning

Unsupervised learning captures a broad set of diverse tasks that share the generic objective of uncovering latent structure in unlabeled data. In unsupervised learning, the objective functions are fully parametrized by the data, without the need for external labels.

While some tasks are purely unsupervised in nature, others may benefit from a first unsupervised pass that transforms the data representation in order to speed up and simplify downstream computation.

Several interesting classes of machine learning problems can be phrased as unsupervised learning tasks. Clustering, density estimation, dimensionality reduction, and probability distribution modeling, for example, are among some of the most frequently encountered tasks that fit this paradigm.

Given the significance of clustering tasks in high energy physics, and the utility of this technique as a data exploration and interpretation tool, the following sections are specifically dedicated to brief re-marks on a few of the different machine learning approaches to clustering. Excellent overviews and tutorials on the several other topics that fall within the purview of unsupervised learning can be found elsewhere [213, 214].

### 4.2.1 Clustering

Clustering is the process of partitioning data points into related sub-groups with shared features, and can be used, for example, as a data exploration tool, or to construct fast lookup structures and speed up nearest neighbor retrieval. In general, the clustering problem definition is the following: given a set of points $\{x_i\}_{i=1}^N$ with $x_i \in \mathbb{R}^n$, the goals of a clustering algorithm is to assign each sample to one or more clusters, and to determine properties of the clusters.

The many types of clustering algorithms can be subdivided based on the input parameters they require, such as cluster size or cluster number. They can also be classified as probabilistic or soft (each element belongs to each cluster with some probability), hard (elements either belong or do not belong to each cluster), exclusive (each element belongs to exactly one cluster), hierarchical (clusters are ordered in tiers and structured like a tree), etc.

Cluster assignment usually requires the definition of a distance metric to be used to measure compatibility between objects. A typical metric between $a, b \in \mathbb{R}^n$ is the Euclidean distance $d(a, b) = ||a - b||_2^2$. However, domain knowledge might inform the use of alternative measures of similarity for specific problems.

Jet clustering algorithms used in high energy physics and described in Sec. 5.2.1.2 are application specific methods that combine machine learning clustering strategies with domain specific constraints to inform the algorithm design to best match the physical properties of jets.

This section explores a selection of clustering algorithms to highlight the diversity of strategies that exist in the field.

#### 4.2.1.1 $k$-Means

One of the most popular clustering algorithms is $k$-means. It begins by initializing the centroid locations of the $k$ clusters to $\{\mu_j\}_{j=1}^k$ with $\mu_j \in \mathbb{R}^n$. These can be initialized at random, for example by selecting $k$ random data points, obtaining seeds from another algorithm, or through other heuristics. Data points are assigned to the closest cluster according to the distance metric $d$. In the first step, then, the cluster assignment $c_i$ for the example $x_i$ is determined as:

$$c_i \leftarrow \arg\min_j d(x_i, \mu_j). \tag{4.16}$$

After each point $x_i$ has been assigned to the nearest cluster, the second step in the algorithm consists of optimizing the position of each cluster's centroid by moving it to the average position of of all points assigned to said cluster, or:

$$\mu_j \leftarrow \frac{1}{|C_j|} \sum_{i \in C_j} x_i, \tag{4.17}$$

where $C_j$ is the set of points assigned to cluster $j$. The two steps constitute an iteration, which is to be repeated until satisfactory convergence. Convergence is often expressed in terms of a stopping criterion, which could correspond to the cluster assignment step not modifying previous assignments, or the change in position of the centroids being smaller than a certain threshold, etc.

The global optimization problem can then be summarized as:

$$\min_{\{\mu_j\}} \min_{\{C_j\}} \sum_{j=1}^{k} \sum_{i \in C_j} d(x_i, \mu_j). \tag{4.18}$$

The objective function of the optimization is guaranteed to decrease or remain constant at each iteration, as the two steps in the EM algorithm contribute to performing coordinate descent by alternating the minimization along one dimension while keeping the other model parameters fixed. However, given the non-convex nature of the objective, the algorithm is only guaranteed to converge to a local minimum, where bad local minima can arise from pathological initializations.

The time complexity of each iteration is $\mathcal{O}(kn)$ for the first step, $\mathcal{O}(n)$ for the second step.

#### 4.2.1.2 Mixture Models

Mixture models represent a set of strategies to incorporate prior knowledge, in the form of distributional assumptions, in the formation of clusters of elements within the sampled dataset. They further provide the advantage of being capable of yielding soft probabilistic cluster assignment, which can then be resolved into hard assignment by selecting the highest probability cluster association.

Mixture models are usually defined in terms of a number of clusters (or components) $k$ and a functional form $f$ parametrized by different $\theta$ parameters for each component, which need to be optimized to best fit the $N$ observed data points $\{x_i | x_i \in \mathbb{R}^n\}_{i=1}^{N}$. Heuristics to infer an appropriate number of clusters $k$ are often employed in conjunction with the optimization of the model, but can be considered as logically distinct steps that augment the algorithm. Formally, the density is modeled as:

$$p(x|\theta) = \prod_{i=1}^{N} \prod_{c=1}^{k} w_c \, f(x_i|\theta_c) \tag{4.19}$$

where $w_c$ is the mixture weight assigned to component $c$, with $\sum_{c=1}^{k} w_c = 1$, and $f(x_i|\theta_c)$ is the probability distribution of event $x_i$, computed using parameter values $\theta_c$. If $f$ is chosen to be a Gaussian distribution parametrized by $\theta_c = (\mu_c, \Sigma_c)$, where $\mu_c \in \mathbb{R}^n$ identifies the components' means and $\Sigma_c \in \mathbb{R}^{n \times n}$ represents the covariance, the model is referred to as a Gaussian Mixture Model.

Expectation maximization is often used to estimate the optimal parameters of the model.

Given the optimized model and its parameters, one can obtain probabilistic cluster assignments for each sample $x_i$ by computing:

$$p(c = C|x_i) = \frac{w_C \, f(x_i|\theta_C)}{\sum_{c=1}^{k} w_c \, f(x_i|\theta_c)}. \tag{4.20}$$

### 4.2.1.3 Hierarchical Clustering

Clusters with hierarchical structure imply that an object that belongs to a certain cluster will also automatically belong to all upstream clusters in the cluster hierarchy.

The construction of clusters trees can take place using a top-down or bottom-up approach. In the bottom-up (*agglomerative*) case, the $N$ clusters all contain one of the $N$ examples in $\{x_i\}_{i=1}^{N}$. The pairwise cluster distances are then computed and used to identify the closest cluster pair, according to a pre-defined distance metric $d$. The two clusters that form the pair are merged into a new cluster, which is added to the set of clusters, while the two children are removed. The algorithm repeats by computing the new cluster's pairwise distance to all remaining clusters, finding the new closest pair, etc. The algorithm terminates if a minimum number of clusters is reached, or if the closest cluster pair distance is greater than a specified maximum threshold.

The top-down (*divisive*) method, instead, applies reverse logic, with all examples starting out in one cluster. A distance metric $d$ is used to identify the largest outlier, and two new sub-clusters can be formed by iteratively transitioning objects out of the original cluster if they are more similar to the new cluster formed around the outlier than to the rest.

Cluster distance is a somewhat ill-defined quantity that can be calculated in several ways. It can be defined as the distance between cluster centroids, the distance between the two closest elements across the two clusters, the maximum distance between elements in the two clusters, or in any other consistent manner, according to any available domain knowledge about the problem statement.

## 4.3 Supervised Learning

Supervised learning identifies the set of learning tasks aimed at optimizing models using labeled data, in order to predict targets on future unseen, unlabeled data.

Mathematical models are then built to predict outcomes from a set of predictors. The properties of the family of functions of the type $f : \mathbb{R}^d \to \mathbb{R}^t$ that map inputs from the input domain $\mathbb{R}^d$ to outputs from the target space $\mathbb{R}^t$ are determined by the expressivity and flexibility of the chosen learning algorithm. As in many other scenarios, the choice of function family is often informed by prior knowledge of the problem and domain, and by the use of inductive bias to design models that incorporate it (see Sec. 4.4.3 for a discussion of inductive bias, primarily in the context of neural networks). The goal of the learned function $f$, in general, is to approximate the true relationship between inputs and outputs – instances of which can be observed from the training examples – to then successfully predict $f(x) \sim y$ for unseen input-output pair $(x, y) \in \mathbb{R}^d \times \mathbb{R}^t$, *i.e.* to generalize the mapping without memorizing.

If the target output $y$ is a discrete, categorical quantity, the exercise of predicting its value is known as classification (see Sec. 4.3.1). In the case of continuous outputs, the prediction task is known as regression (see Sec. 4.3.2).

### 4.3.1 Classification

Classifying samples consists of assigning them to the correct, exclusive category our of a list of options $C$. The deterministic class assignment problem can be phrased in terms of the empirical risk (Eq. 4.4) of misclassification computed over the dataset $\{(x_i, y_i)\}_{i=1}^n$ as:

$$\hat{R}(f) = \frac{1}{n} \sum_{i=1}^{n} [f(x_i) \neq y_i], \qquad (4.21)$$

where the decision $f(x_i)$ returns the most likely class for object $x_i$ under the model, and is penalized if it does not correspond to the true class $y_i$.

Binary classification is the simplest incarnation of the generic classification problem. It consists of partitioning data points into two classes, that can be labeled, for example, as class 0 and class 1. Assuming class membership is exclusive, $p(y = 1|x) = 1 - p(y = 0|x)$. By construction, $p(y|x)$ must be bounded between 0 and 1.

While directly regressing the quantity $p(y|x)$ cannot ensure the preservation of its logical bounds $[0, 1]$, applying a logit function $\log \frac{p}{1-p}$ to it results in an unbounded quantity that can be described by a linearly activated model $g(x; \theta)$ with output in $(-\infty, +\infty)$. In this case, the quantity of interest can be extracted from the output of the model by applying the logistic transformation:

$$p(y|x; \theta) = \frac{1}{1 + e^{-g(x;\theta)}} \qquad (4.22)$$

With $p(y|x; \theta) \in [0, 1]$, the class assignment can be determined by drawing a decision boundary along the distribution of $p(y|x; \theta)$ itself: for example, $\hat{y} = 1$ if $p(y|x; \theta) \geq 0.5$, $\hat{y} = 0$ if $p(y|x; \theta) < 0.5$. For this reason, the logistic function is often employed as the activation function for the output layer of neural networks that perform binary classification.

The probabilistic output models the class assignment as a Bernoulli random variable, for which the likelihood can be explicitly expressed as: $\mathcal{L}(p) = p^y(1-p)^{1-y}$. Maximizing the likelihood, or, better, minimizing the negative log-likelihood

$$\theta^* = \arg\min_{\theta} \left( -\ln \mathcal{L}(p(y|X; \theta)) \right) \qquad (4.23)$$

$$= \arg\min_{\theta} \left( -\sum_{x \in X} \ln(p^y(1-p)^{1-y}) \right) \qquad (4.24)$$

$$= \arg\min_{\theta} \left( -\sum_{x \in X} y \ln(p) + (1-y) \ln(1-p) \right) \qquad (4.25)$$

corresponds to minimizing the cross entropy.

This can be generalized to the case of multi-class classification, in which the optimization problem formulation replaces the two possible outcomes in binary classification with a sum over all possible classes $c \in C$:

$$\theta^* = \arg \min_\theta \left( -\sum_{c \in C} \sum_{x \in X} y_c \ln p(c|x; \theta) \right), \tag{4.26}$$

where $y_c$ is a binary indicator of whether the true class of the example $x$ is indeed $c$, and $p(c|x; \theta)$ is the output of the model associated with class $c$ for input $x$. The latter is calculated using the generalization of the logistic function to multi-class classification scenarios, known as *softmax*:

$$p(c|x; \theta) = \frac{e^{g(c|x;\theta)}}{\sum_{i \in C} e^{g(i|x;\theta)}}, \tag{4.27}$$

where $g(i|x; \theta)$ is the pre-activation output corresponding to class $i \in C$ computed by the model. The partition function at the denominator is used to ensure that the sum of all softmax-activated outputs sums to $1$.

Other loss formulations used in the context of classification include the hinge loss and its various modifications, which emerge from the application of support vector machines. In the context of binary classification, for labels $y = \{-1, +1\}$, the hinge loss takes on the following form:

$$\mathcal{L} = \max(0, 1 - y \cdot \hat{y}), \tag{4.28}$$

where $\hat{y}$ is the neural network output. This cost function is best understood in terms of margin separation, where the classification goal is conceptualized as finding the optimal hyperplane to separate data points belonging to different classes. If $y \cdot \hat{y} < 0$, the point lies on the wrong side of the margin and is being misclassified, thus incurring in a penalty; if $0 < y \cdot \hat{y} < 1$, the point lies on the correct side of the margin and is being correctly classified, but the distance from the margin is not sufficiently large and should therefore be increased; in all other cases, as long as points are located on the correct side of the margin and far enough away from it, no extra penalty is applied, regardless of the hierarchy of scores one would expect the model to assign to a set of examples.

The hinge loss is mathematically appealing because it represents a strict convex upper bound on the empirical risk. Other types of available classification losses share this property, as shown in Fig. 4.1, so the choice must be informed by knowledge of the relative importance of different types of error in the specific application under consideration.

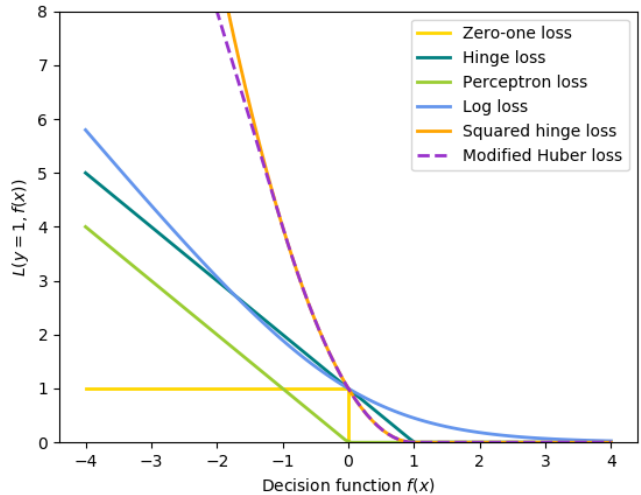

**Figure 4.1:** Comparison of loss functions for classification supported by `sklearn.linear_model.SGDClassifier`. The curves quantify the different penalties associated with a model's output value $f(x)$, given truth label $y = 1$. In all cases, a prediction $f(x) \geq 0$ results in small or no cost. Opposite sign predictions may be more or less strongly penalized depending on the choice of loss function. Image reproduced from http://scikit-learn.org/stable/modules/sgd.html.

#### 4.3.1.1 Metrics

In the context of binary classification, in which examples can be divided into two classes called signal and background, or positive and negative, or $\{0, 1\}$, several metrics can be defined to interpret the performance of a classifier.

The number of positive examples that the model correctly identifies as positives is called true positives (TP). False positives (FP) are the negative events incorrectly labeled as positives. Similarly, the positive examples incorrectly labeled as negatives are known as false negatives (FN). Finally, the negative examples correctly labeled as negatives are called true negatives (TN).

Several functions of these four quantities can be computed to highlight problematic behaviors in a classifier. Only the ones that are commonly used in high energy physics will be mentioned here. For example, the true positive rate (TPR), also known as recall, and false positive rate (FPR) are defined as:

$$\text{TPR} = \frac{\text{true positives}}{(\text{true positives} + \text{false negatives})} \tag{4.29}$$

$$\text{FPR} = \frac{\text{false positives}}{(\text{false positives} + \text{true negatives})}. \tag{4.30}$$

In high energy physics, these are often called signal and background efficiencies. Despite the name, the background efficiency is not the probability of correctly classifying background samples in the background category (also known as the true negative rate, TNR), but, instead, it is the frequency with which background samples are misidentified as signal samples. Background rejection, on the other hand, indicates the reciprocal 1/FPR, and should therefore be maximized, as per intuition. The precision, also known in physics as purity, is the percentage of positively labeled examples that are actually positive, and is calculated as:

$$\text{precision} = \frac{\text{true positives}}{(\text{true positives} + \text{false positives})}. \tag{4.31}$$

The accuracy, instead, describes the percentage of total test samples that are correctly classified, and is computed as:

$$\text{accuracy} = \frac{\text{true positives} + \text{true negative}}{(\text{allevents})}. \tag{4.32}$$

Maximizing the TPR while minimizing the FPR requires a trade-off between the two objectives. This is best illustrated using a receiver operating characteristic (ROC) curve, which is a graphical representation of the discrimination power of a model over a binary classification task, as the threshold cut on the classifier output that defines the boundary between the two classes is moved along the output distribution. Classically, the ROC curve plots the FPR on the $x$-axis and the TPR on the $y$-axis, but the ATLAS collaboration prefers the TPR on the $x$-axis and $1/$FPR on the $y$-axis. Working points are the defined along the curve at a handful of fixed TPR values.

The area under the ROC curve (AUC) can be used as a single metric to compare classifiers by summarizing the information displayed by the ROC curve itself. Intuitively, the AUC can be interpreted as the probability of the model assigning a higher output value to a positive example than it would to a negative example. If, however, only a specific range of efficiency values is relevant for a task, or if two ROC curves cross, the AUC may not be appropriate to fully capture the relationship among classifiers.

The work presented in Ref. [215] argues that different problems and disciplines present different challenges and therefore demand different evaluation metrics, and explains why the oft-quoted AUC may not be a relevant metric for searches in HEP.

### 4.3.2 REGRESSION

Regression tasks differ from classification tasks in the nature of the output they attempt to predict. Instead of assigning events to distinct categories, the goal of a regression task is to approximate one or more real valued targets. Regression models encode the way in which the varying of these response variables is informed by changes in the explaining variables.

Depending on the application task, the target's domain may be bounded or unbounded. It is important, then, to transform the model's output to fall within the physical range observed over the training samples. If, for example, the quantity to regress is strictly positive, passing the output through a rectified linear unit [216, 217] is one possible method to enforce a sensible output range. Similarly, a bounded activation function can ensure that predicted outputs do not exceed the bounds of the training target range.

Regression tasks are often optimized using loss functions such as the mean squared error (Eq. 4.33) or the mean absolute error (Eq. 4.34):

$$\mathcal{L}_{\text{MSE}} = ||y - \hat{y}||_2^2 \qquad (4.33)$$

$$\mathcal{L}_{\text{MAE}} = ||y - \hat{y}||_1 \qquad (4.34)$$

where $y$ is the ground truth and $\hat{y}$ is the model prediction. A hybrid between the two is know as Huber loss, which combines the quadratic and absolute value components in a piecewise manner, with the quadratic portion occurring within $\pm\delta$ of the origin. These three objectives are symmetric around the origin, so they may not always reflect the true cost of making a mistake in realistic applications. Often times, in certain application domains, the penalty for overshooting might be higher (or lower) than the penalty associated with undershooting. This can be handled by defining a piecewise loss function to target different output regimes, or, equivalently, by transforming the output space so that a symmetric objective function becomes suitable. A commonly encountered instance of loss modification towards special treatment of different output ranges is to combine an absolute error on one side of the origin with a squared error on the other.

For practical reasons, binned regression tasks are sometimes phrased as classification or ranking problems, in which the model is tasked with assigning each example to the correct bin in the output distribution.

Expanding on this practice, a simple alternative problem formulation that encourages the learning of well-behaved functional approximations in continuous value regression problems consists of learning a softmax activation over an over-complete basis representation. This is perhaps best elucidated in the context of neural networks, an introduction of which is postponed to Sec. 4.5. Let us consider an input vector $x$ with $p$ features that is the result of forward propagation of a minibatch of examples, and the problem of predicting $q$ regression outputs. Choosing some over-complete basis of size $(q \times r)$, with $r > 1$, one can replace the traditional approach of learning a fully-connected layer with $p$ inputs and $q$ outputs, with a fully-connected layer with $p$ inputs and $(q \times r)$ outputs. Reshaping this to a (minibatch size $\times q \times r$) tensor, each of the $q$ rows can be thought of as a histogram representing the unnormalized, binned distribution for each random variable to regress, while the $r$ rows index the histogram bins. Taking a softmax over the $r$ bins in each distribution is conceptually equivalent to scaling the intermediate learned histogram representation to having bin heights that fall in the range $[0, 1]$ and sum to 1, in order to acquire the meaning of a set of probability density functions, which can ultimately be used to compute the expected value of each of the $q$ random variables. In order to do so, it suffices to take the column-wise dot product of the $q$ softmax outputs with a final trainable matrix $B$, also of dimensions $(q \times r)$, which represents the learned numerical location of each of the $r$ bins across each of the $q$ spectra. Its entries can either be fixed to predetermined values (given a strong prior on an effective binning strategy) or allowed to float as an additional set of learnable parameters in the algorithm. Note that this is essentially a change of variables: instead of expressing a continuous output as a volatile linear map, one can instead map the continuous value problem onto a softmax-weighting problem. More formally, let $x \in \mathbb{R}^p$ be the input representation for a single example in a minibatch. The simplex computes $z = \mathsf{softmax}(Wx + b) \in \mathbb{R}^{q \times r}$, where $W$ and $b$ represent the usual fully-connected weight and bias tensors, respectively. The final output then becomes $\hat{y} = \mathrm{diag}(B^T z) \in \mathbb{R}^q$, which corresponds of the prediction of $q$ regression outputs per example, as defined in the problem statement.

## 4.4 Neural Networks and Deep Learning

Neural Networks are among the most powerful systems that are currently being employed in some of the most advanced artificial intelligence projects, with applications ranging from Computer Vision to Natural Language Processing. Thanks to algorithmic breakthroughs, data availability, and recent hardware improvements, in particular the advent and diffusion of GPUs, deep neural networks have finally become a viable modeling choice, trainable even on personal laptops.

Despite the complexity of some of the recently developed neural architectures, at their core, neural networks simply consist of compositions non-linear transformations. In fact, they can be described as series of linear and non-linear tensor operations used to approximate one or more output targets as non-linear functions of the input variables. They can be symbolically represented by Directed Acyclic Graphs (DAGs), where a directed graph $G = (V, E)$ has edges $(u, v) \in E$ such that $u \to v$ but $v \not\to u$, and the acyclic property demands that $\forall v \in V$, $\nexists$ a path $v \to v$ [218]. For neural networks, vertices are tensor operations (i.e. additions, multiplications, etc.) while edges represent the use of the output of one operation as the input to another operation.

For a generic feed-forward neural network, forward propagation consists of traversing the DAG of operations. Given $x$, the vector of physical quantities that define the properties of the physics object under investigation, forward propagation describes a flow of operations that transforms $x$ through the layers on the neural network to obtain the final output $\hat{y}$, the array of predictions for each example. The final output vector $\hat{y}$ will depend not only on the input vector $x$ but also on the current values of the weights in each layer of the network.

Formally, a sequential model with $L$ layers is simply computing

$$\hat{y} = f^{(L)}(z_{L-1}) \tag{4.35}$$

where

$$z_k = f^{(k)}(z_{k-1}), \tag{4.36}$$

$$z_0 = x \tag{4.37}$$

and each $f^{(k)}$ is a non-linear tensor map. The more flexible this mapping, the broader the class of functions we can hope to learn. The functions $f$ are themselves parametrized by the weights associated with the units in the corresponding network layers. Therefore, $\hat{y}$ can be interpreted as a function of both the inputs $x$ and the weights of the model.

Training a neural networks is then equivalent to learning the most appropriate superposition of activations to construct a desirable response function for a given input. The hierarchical nature of the model allows to build nested representations of the data, enhancing predictive features and suppressing less relevant information.

### 4.4.1 Theoretical Underpinning

Despite the availability of a number of mathematical results that justify some of the emergent properties in neural networks from a principled, mathematically rigorous standpoint, a general, unified theory of deep learning is still absent. The limited ability to solve the exact learning dynamics in deep neural networks does not preclude the fact that their success across a variety of complex applications, including the many encountered in HEP and described in this thesis, lies on mathematical foundations that guarantee, at least in idealized scenarios, their universality, *i.e.* the ability to approximate arbitrary continuous functions with arbitrary precision.

The space of learnable functions $f_\phi^L$ by a feed forward network of $L$ layers with activation function $\phi$ can be investigated in terms of the richness of the output space it admits, or, in other words, the complexity of decision boundaries it can implement. Although the problem is intractable in the general case, the simple case of the space of functions $f_\sigma^1$ representable by one-layer neural networks continuous sigmoidal activation functions $\phi$ can shed light onto the expressive power of neural networks.

Ref. [219] shows that a one-layer neural network with sigmoidal activation is able to represent an output function space that is dense in the space $C(I_n)$ of continuous functions on the $n$-dimensional unit cube, $I_n = [0,1]^n$, meaning that the network is able to construct a function that is only $\epsilon$-away from any target function in $C(I_n)$. In other words, a one-layer neural network with sigmoidal activation can separate arbitrarily well any set of compact, disjoint subsets in $I_n$, granted no limitations on the number of nodes in the layer and on the size of the weights.

While the proof of universality of single-layer neural networks as function approximators provides a solid theoretical result for the development of the field, it has also been empirically shown that the same level of expressivity can be achieved, with major performance improvements, by deeper networks with a finite and reasonable number of units per layer.

#### 4.4.1.1 Depth

The power of modern deep networks, compared to their shallow predecessors, is largely to be attributed to the benefits that arise in connection to their depth. Training deep nets, with several layers of nodes that transform the outputs of adjacent ones, had for long looked like an impractical task due to problems such as vanishing and exploding gradients [220]. The advent of practical solutions such as rectified linear units [216, 217], better weight initialization schemes, and auto-differentiation frameworks contributed to the rapid growth in feasibility and popularity of deep learning.

Deep nets often contain millions of parameters, dozens of layers, convolutional and recurrent units to process structured inputs, non-sequential connections among nodes, as well as non-linear activations that allow NNs to represent any function by learning complex, non-linear transformations of the inputs.

The power of deep learning resides in its flexibility to rely on several nested mappings from the inputs to the target output, which allow the algorithm to progress in the learning task through several different stages of complexity.

The superiority of deeper networks may be ascribable to their ability to perform metric learning, *i.e.* to learn advantageous, gradual embeddings of the input space that deform it so that similar data points are brought closer to one another, while points belonging to different classes are further separated. Networks increase angular separation between inputs that map to different outputs, and contract the space between similar inputs. However, no mature theoretical framework exists that demonstrates the optimality, in the information-theoretic sense, of the representation that deep networks internally learn. Intuitively, in fact, much of the power of deep networks is thought to come from their reliance on highly redundant information, without an attempt to reduce the stored information to a minimal set of sufficient bits.

In addition, deep networks may also attribute their power to the aptitude for learning invariance relations among inputs, so as to encode the symmetries of the system observed in the training data.

4.4.1.2   Geometry of the Loss Surface

Modern neural network training can be interpreted as a form of empirical risk minimization (see Sec. 4.1.3) which relies on the formulation and minimization of a loss function (see Sec. 4.1.4.1).

The geometric and topological structure of the loss surface is tightly coupled to the generalizability of the solution learned by the neural network. However, its properties and shape in a general setting are still elusive.

Given the non-convexity and high dimensionality of typical loss functions constructed by the hierarchical, non-linear transformations encoded in neural network structures, loss landscapes end up being characterized by an abundance of saddle points [191], with competitive local minima being exponentially more likely to be localized in a relatively flat neighborhood of the global minimum [221]. Empirical studies have found that using ReLUs as activation functions, which induce an abundance of non-firing (zero-weight) units in large neural networks, results in several different random initializations to converge to very different network configurations with similar generalization errors, leading to the conclusion that many local minima might be global minima [222, 221, 223, 224, 225]. Ref. [223] provides an initial theoretical conjecture as to how seemingly ad-hoc techniques (such as ReLUs [226], Max-Pooling, and dropout [201]), which have achieved impressive empirical results in convergence rate and generalization power of the solution [227, 228, 229, 230, 231, 232, 233, 210], satisfy the positive homogeneity property that the authors suggest might be key for convergence and local minima disappearance in deep networks.

Complications from the optimization standpoint emerge from the presence of plateaus in the vicinity of saddle points, which inhibit the learning process driven by standard gradient descent algorithms. Intuitively, the presence of saddle points in a low-dimensional problem is unlikely, while, moving to high dimensions, local minima become so rare that only the ones that are also global minima, or are very close to an error value compatible with that of the global minimum, are likely to appear [234, 191, 221]. On the other hand, the majority of critical points with higher associated error have high probability of being saddle points surrounded by flat regions that slow down training and simulate the presence of local minima in the training dynamics. Indeed, many observations of the trends of loss curves for deep networks point to learning dynamics characterized by long periods of slow improvement followed by sudden drops in error value. In general, gradient-based optimization algorithms on non-convex problems can only guarantee the convergence towards a critical point, but the quality of the point reached by the optimization procedure also depends on exogenous factors such as the initial conditions. While a variety of second order methods have been devised to take into consideration the curvature of the loss landscape, their practicality is hampered by their computational cost.

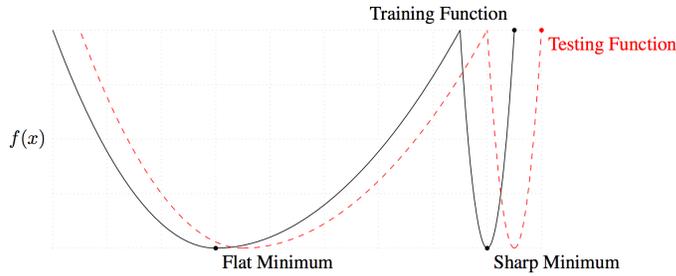

**Figure 4.2:** Diagram representing hypothetical loss landscapes for the training (black) and test (red) datasets. This illustrates that converging to flat minima is preferable for generalization purposes, because these regions are intuitively more likely to also correspond to low-cost configurations in parameter space when the loss is evaluated on the test set. Conversely, even small shifts in the train and test distributions may cause sharp minima in the training loss function to occur in high-cost locations in the test loss function. Image reproduced from Ref. [18].

Another property of critical points used in the literature to assess their quality is *flatness*, i.e., the extent of the neighborhood around the point of interest over which the value of the loss function does not change significantly [235, 236]. Since the goal of many learning algorithms is not simply to find the global minimum of the training loss, but, instead, to converge to solutions that generalize and reduce the error rate on unseen examples, it is preferable to settle for wide flat minima than sharper ones, because of their higher likelihood to represent configurations in parameter space with large overlap with low expected loss regions, as shown in Fig. 4.2. The flatness (or, conversely, sharpness) of minima has itself been linked to the choice of other hyper-parameters in the learning algorithms, such as batch size [18].

The choice of hyper-parameters, including learning rate, batch size, and regularization, all affect the shape of the loss landscape in ways that have not entirely been analytically decoded yet. The desire to exploit knowledge of local geometry has given rise to the introduction of several adaptive learning algorithms, such as ESGD [181] – which explicitly makes use of flatness information to seek out more generalizable minima – or other methods mentioned in Sec. 4.1.4.2. The success of particular types of network architectures, such as those employing residual connections [237], can be linked to the way in which they modify the shape of the loss landscape. Very deep neural networks without skip connections often suffer from ill-conditioning and lack of appropriate feedback, which cause failures in the optimization procedure and may result in decreased performance with depth.

Interesting work towards visualizing the geometry of the loss surface in neural networks, given different architectures and optimization strategies, can be found in Ref. [238].

### 4.4.2 Practical Enhancements

Similar to physics, the machine learning community is composed of two somewhat overlapping and interacting sub-groups of researchers, theorists and experimentalists, with the prevalence of theoretical and practical results alternating throughout the history of the field. Several ad-hoc tricks and theoretically founded improvements have been suggested over time to improve the convergence rate of deep neural networks. While not all commonplace heuristics have found a clear theoretical justification, not all theory-grounded suggestions have enjoyed the practical success one could hope for, either.

This section highlights a few useful empirical tools that facilitate the training of deep networks. Further tips and tricks are documented in Ref. [239].

#### 4.4.2.1 Activation Functions

A key feature of neural networks is the contribution of point-wise non-linearities to their learning capacity. Commonly called *activation functions*, these usually simple functions of the units' outputs add expressiveness as well as both theoretically and experimentally motivated boosts to performance.

Early activation functions drew inspiration from biological neural networks, often times relying on simplified models of neural activity. One such example, commonly referred to as the McCullough-Pitts neuron [240], simply used a Heaviside step function to compute the output of a given neuron, as shown in Eq. 4.38, where $x$ is an input to a neuron, $w$ is a set of "synaptic weights" of the neuron, and $\theta$ is the activation threshold:

$$f(x; w, \theta) = \begin{cases} 1 & w^T x > \theta \\ 0 & \text{otherwise} \end{cases} \quad (4.38)$$

Until the past decade, the most commonly used activation functions in neural networks were the logistic unit (Eq. 4.39) and the hyperbolic tangent (Eq. 4.40):

$$f(x) = \frac{1}{1 + e^{-x}} \tag{4.39}$$

$$f(x) = \tanh(x) = \frac{e^x - e^{-x}}{e^x + e^{-x}} \tag{4.40}$$

These are monotonically increasing functions with similar asymptotic behavior known as sigmoids. While the logistic curve is bound between 0 and 1, others, such as the hyperbolic tangent, are bound between $-1$ and 1 and are symmetric with respect to the origin.

Though these activation functions still find use in special use cases such as the internal transformations that define Highway layers [241], LSTMs [242], and gated convolutions [243], their usage in vanilla multi-layer perceptrons has faded due to their issues with saturation. In both Eq. 4.39 and Eq. 4.40, as $|x| \longrightarrow \infty$, $f'(x) \longrightarrow 0$, which, when compounded with the multiplicative nature of the chain rule present in the definition of backpropagation (Sec. 4.1.4.3), can lead to rapidly vanishing gradients [244].

A more commonly used activation function that does not suffer from such issues is the Rectified Linear Unit (ReLU) [216, 217]:

$$f(x) = \begin{cases} x & x > 0 \\ 0 & \text{otherwise} \end{cases} \tag{4.41}$$

In addition to added benefits from lack of saturation, ReLUs offer computational speedups, as their gradients can only ever be 0 or 1, and can be computed via a single comparison. Other related activation functions with various associated enhancements include the Leaky Rectified Linear Unit (LeakyReLU) [230], the Scaled Exponential Linear Unit (SELU) [245], the Parameterized Rectified Linear Unit (PReLU) [246], and the SoftPlus function [217].

A useful lens with which to view activation functions and understand the circumstances in which their use is appropriate is that of inductive bias [247] (see Sec. 4.4.3). Consider the softmax activation

(Eq. 4.42) applied to a $D$ dimensional vector $x$:

$$f(x) = \frac{e^x}{\sum_{i=1}^{D} e^{x_i}} \tag{4.42}$$

This type of activation is used in cases where one discrete object needs to be selected out of many, and the choice is mutually exclusive. The formulation in Eq. 4.42 allows a model to learn to select such an object in a probabilistic manner, roughly learning a posterior for each given item. This notion has been extended in recent literature on the topic of attention [248], where a softmax is used to structurally impose that a model learn a mutually exclusive selection process.

Another example of inductive bias in the selection of activation function is the modern usage of the logistic function, defined in Eq. 4.39, at the output layer. If a neural network is tasked with selecting between two options, $a, b \in \mathbb{R}$, the choice can be reparametrized as selecting some $s \in \{0, 1\}$, to then compute the final output as $sa + (1-s)b$, where $s$ comes to symbolize the relative preference for $a$ over $b$. Letting the value of $s$ be learnable via a logistic activation function (which is bounded within $(0, 1)$) allows to train a neural network to (continuously) select between the two discrete values. This intuition is the basis for many modern models that involve a notion of gating [241, 243, 242].

#### 4.4.2.2　Weight Initialization

Despite the recent growth in exploration and understanding of training dynamics and loss landscapes, in the majority of deep learning applications, the initial conditions for the optimization procedure still play an important role in determining the critical points towards which the parameter configuration will converge.

Starting from a suitable initialization of network weights is important to increase the likelihood of finding a minimum with low generalization error. A good start is half the battle.

Random weight initialization equips the network with an inherent source of stochasticity, which can be quantified by repeating the training numerous times with different initial conditions, in order to registered the expected change in outcome. The extent of the delta in generalization error depends on the properties of the loss surface, which is itself determined by the properties of the architecture and by the hyper-parameters of the training procedure (see Sec. 4.4.1.2).

For symmetry reasons, the network weights should not be all identically initialized to zero, otherwise each unit only behaves as a copy of the other units, without differentiation. Weights should instead be initialized at random from a distribution that coacts with the chosen activation function to produce outputs within sensible ranges and with desirable distributional properties.

The nature of random process from which the initial weights are sampled matters. Not all initializations are good initializations. Extensive research in the conditions that the weight initialization distribution must satisfy has culminated in the community's adoption of a few popular initialization strategies. A common thread among them is the control of the variance of each layer's outputs as information flows through the layers and their activations. Assuming, for example, inputs to a unit and its weights drawn from Gaussian distributions, the distribution of the output will itself be Gaussian, with variance rescaled by the number of inputs to the unit. Therefore, to constrain the variance and avoid excessive growing or vanishing of parameters, the variance of the weights should intuitively be rescaled by a factor of $1/n$, where $n$ is the length of the input vector [239]. In practice, initialization schemes such as those presented in Ref. [249] recommend drawing weights from either uniform or normal distributions with variance that depends on both the number of incoming and outgoing connections at each unit. An improved initialization strategy for ReLU-activated layers is presented in Ref. [246]. Bias units are commonly initialized to zero. Other popular options include those proposed in Ref. [250].

#### 4.4.2.3 Input Normalization

With the exclusion of tree-based methods, most machine learning algorithms benefit from feature scaling as a preprocessing step. Generally speaking, normalizing the input features to have zero mean and common (preferably 1) standard deviation has been observed to speed up the training procedure. Subtracting the mean of each input training feature has the advantage of symmetrizing the distribution with respect to the origin, with similar pulls in the positive and negative directions. In addition, rescaling all variables to the same range evens out the steepness of the loss landscape along each direction, removing the necessity of learning network weights that span several orders of magnitude to achieve similar rates of change across all dimensions. In fact, the conditioning of the Hessian is related to the speed of convergence.

The mean and standard deviation computed on the input data need to be stored for future input transformation at test time.

Other types of input transformations that involve the application of monotonic transformations (such as taking the log) of the input space do not modify the amount of information contained in the data but empirically improve the numerical stability of the optimization process.

#### 4.4.2.4 BATCH NORMALIZATION

Similar to input normalization, batch normalization (BN) [202] performs a transformation that standardizes distributions to zero mean and unit standard deviation at every layer in a neural network. Each layer is seen as the input layer of a new sub-network, which, like the proper input layer, would benefit from normalized inputs, as discussed in Sec. 4.4.2.3.

At each training iteration, weights in each layer are optimized to improve that layer's processing of the information it receives. However, the quantitative properties of the inputs to each layer are entirely dependent on the upstream processing that occurs in previous network layers. Therefore, as weights are globally updated after a training step, the new distribution of inputs a layer receives at the following forward pass will be different from the previous one it was just optimized to handle, forcing the weights to continuously try to adapt to the new input distributions. In other words, weight updates can be seen as always being a step behind, optimizing their configuration for inputs they received in the past, as opposed to the ones they will receive next. These distributional mismatches create analogous problems to those that one can expect from out-of-domain evaluation of a fully trained neural network, when the distribution of inputs used at test time does not match the distribution fed into the network at training time (covariate shift [251]). Issues arising from distributional shifts at the input layer are well-known and are being addressed in the domain adaptation literature. In analogy, though, unexpected variations in the input distributions to each layer can also regularly occur in the training process, intuitively causing major learning retardations, due to the weights being optimized to process a different distribution of inputs. This challenge of coordinating weight updates across layers is known as *internal covariate shift*: as a set of weights is updated, other sets of codependent weights are also simultaneously updated, so that the expected input distributions will no longer be present at the next iteration. In fact, in practice, given the relation between the magnitudes of gradients and input values, to avoid divergence, researchers are forced to rely on finely tuned initialization strategies and conservative learning rates.

The unstable dynamics of this learning process can be ameliorated with the introduction of batch normalization. This improvement can be interpreted as a reparametrization of the network's internal computations that factors out the distributional shifts, in order to preserve the network's capacity and information processing power while modifying the learning dynamics to a more favorable state. The two parametrizations are equivalent in terms of family of functions the network is able to represent, but the latter addresses difficulties introduced by the internal covariate shift, making it especially suited for learning very deep models with gradient-based optimization strategies. This is achieved by replacing the complex interactions with upstream layers, after first normalizing out their contribution to the distributional distortions, with an independent set of learnable parameters $(\gamma, \beta)$ in the BN layer (placed before the regular neural network layer under consideration), which decouple the empirical mean and standard deviations of the inputs at the given layer from the neural activity in previous layers.

The reparametrization in batch normalization is adaptive: the normalization parameters are recomputed at the minibatch level for each minibatch of training examples that participates in the optimization. It follows that, at training time, the output of the network given a specific input is no longer deterministic, but instead depends on the statistics of the minibatch in which it appears. Each feature in a minibatch input matrix is normalized independently to zero mean and unit variance, without computing the full covariance. This has been shown to improve convergence even in the presence of correlated features.

At training time, the computation performed by the BN layer over a minibatch of training examples $\mathcal{B} = \{x_i\}_{i=1}^{b}$ is the following:

$$\mu_{\mathcal{B}} = \frac{1}{b} \sum_{i=1}^{b} x_i \tag{4.43}$$

$$\sigma_{\mathcal{B}}^2 = \frac{1}{b} \sum_{i=1}^{b} (x_i - \mu_{\mathcal{B}})^2 \tag{4.44}$$

$$\hat{x} = \frac{x - \mu_{\mathcal{B}}}{\sqrt{\sigma_{\mathcal{B}}^2}} \tag{4.45}$$

$$y = \gamma \hat{x} + \beta \tag{4.46}$$

$$\equiv \text{BN}_{\gamma, \beta}(x)$$

where $\gamma$ and $\beta$ are the scale and shift parameters that need to be learned to restore the network's representation power after the feature normalization step. In practice, Eq. 4.43 computes the mean of the features over the minibatch, Eq. 4.44 computes the standard deviation, Eq. 4.45 makes use of them to standardize the input feature, and Eq. 4.46 transform the normalized minibatch using the learned weights $\gamma$ and $\beta$.

Batch normalization is devised to stabilize the learning dynamics at training time. At test time, however, deterministic per-example network outcomes must be restored. To do so, population-level (instead of minibatch-level) statistics are utilized to normalize the features, modifying the training-time equations above as follows:

$$\hat{x} = \frac{x - \mathbb{E}[x]}{\sqrt{\mathrm{Var}[x]}} \tag{4.47}$$

where

$$\mathbb{E}[x] = \mathbb{E}_{\{\mathcal{B}\}}[\mu_\mathcal{B}] \tag{4.48}$$

$$\mathrm{Var}[x] = \frac{b}{b-1}\mathbb{E}_{\{\mathcal{B}\}}[\sigma_\mathcal{B}^2] \tag{4.49}$$

These expectations are computed over the set of training minibatches $\{\mathcal{B}\}$ – a running average of these population statistics is available at any point at training time for validation purposes. The entire transformation performed at test time by the batch normalization layer can therefore be rewritten as:

$$y = \gamma\hat{x} + \beta = \gamma\left(\frac{x - \mathbb{E}[x]}{\sqrt{\mathrm{Var}[x]}}\right) + \beta = \frac{\gamma}{\sqrt{\mathrm{Var}[x]}} \cdot x + \left(\beta - \frac{\gamma\mathbb{E}[x]}{\sqrt{\mathrm{Var}[x]}}\right) \tag{4.50}$$

It is important to note that BN is also observed to possess indirect regularization properties, thus diminishing the need for aggressive dropout or other common regularization methods. This is due to the stochasticity of the minibatch sampling strategy, which percolates down into the non-deterministic nature of the output associated with each input at training time.

Recent results [252] have begun to challenge the notion that the success of batch normalization is related to the phenomenon of internal covariate shift mitigation, and have argued, instead, in favor of an explanation based on the positive effects of batch normalization towards smoothing the loss landscape.

4.4.2.5 Gradient Clipping

To avoid wild weight updates in the presence of a cliff or wall in the loss landscape, excessive gradient magnitudes can be artificially capped by imposing a hard threshold on them [253]. The component of the natural gradient in the direction parallel to the wall would otherwise overpower all other gradient components, resulting in a large jump in parameter space, which may push the optimization away from a promising valley in the loss surface, undo much of the previous learning, and ultimately slow down the convergence process. Without changing the direction of the gradient vector, the clipped weight update preserves its directionality but shrinks its extend to match the value of the threshold. The hard limit is generally imposed on the $L2$-norm of the gradient, but other gradient norms could be used, if desired. This method has successfully been employed in the training of recurrent neural networks, which are very prone to issues arising from exploding gradients in long sequence learning [254].

4.4.3 Inductive Bias

Many recent advances in the deep learning literature claim so-called "end-to-end" learning, that is, learning complex, multi-step mappings from raw, unstructured inputs without a priori construction of useful features for a learning algorithm, which is often placed at odds with domain specific variable engineering. Although the latter has enjoyed great success in limited, niche tasks, much of the intelligence of those algorithms is still to be attributed to the user, with little automatic understanding coming from the model alone. In contrast, the promise of deep learning is to shift towards more easily generalizable architectures that require fewer direct inputs from the user. However, recent position work [247] has argued for a reinterpretation of the role of human knowledge in setting up of machine learning problems, pointing out that inductive bias has shifted from the construction of features to the design of the learning algorithm or network architecture themselves. The role of deep learning architects has then become that of equipping neural networks with building blocks that are capable of handling multiple tasks with similar characteristics. Inductive bias, in fact, allows to build nets from intuition, and construct useful structure *into* the learning model, based on prior assumptions about features and properties that are likely to appear in the data and to help inform the learning procedure [255, 256]. Imposing penalties on complex models, as well as constructing loss functions that codify the intolerance towards certain types of errors, are all ways to explicitly express a preference over the hypothesis space. These choices result in the model reducing the search space and prioritizing families of predictive functions that are expected to be most suitable and powerful for the task. Inductive bias can thus be thought of as a strict prior on how information is allowed to flow from input to output.

| Architectural Component | Imposed Inductive Bias |
|---|---|
| Convolutional Layer | Spatial structure / translation invariance |
| Recurrent Layer | Temporal dependence / invariance |
| Attention Mechanism | Specific components of input are important |
| Locally Connected Layer | Location-specific spatial structure |

Table 4.1: Deep learning concepts and associated inductive bias.

Being able to successfully skew the learning towards more favorable configurations requires prior knowledge not only of the problem, but also of the availability of methods, their strengths, and their domain of applicability. In the case of computer vision, for example, the question of whether an object exists in an image can be answered irrespective of where the object is located; thus, it is natural to restrict a model to have components that are invariant to translations and other spatial deformations. A few examples of inductive bias, as it pertains to deep learning components, are illustrated in Table 4.1.

This idea is powerful for any applications to high energy physics, to improve interpretability and understanding. Deep learning blocks lend themselves well to domain-specific inductive bias injection; many examples of recent work at the intersection of machine learning and HEP successfully incorporate physics inductive bias, such as utilizing tree structures from physics-driven clustering algorithms [257], enforcing insensitivity to nuisance parameters [258, 259], or explicitly favoring physical energy depositions in generative models for physics simulation (see Sec. 4.6). The extension of machine learning methods to the field of geometric deep learning in non-Euclidean spaces has the potential of further impacting the physical sciences and similar application domains [260].

The emergent field of meta-learning, still vastly under development in the machine learning literature and almost entirely unexplored in HEP, promises to shift the bias yet again, removing some of the current architecture decisions, and aiming for automatic neural architecture search and parameter initialization, with the goal of learning optimal learning algorithm [261, 262, 263, 264, 265, 266, 267].

### 4.4.4 Dataset Size

Learning from limited samples is a challenging task, but its constant recurrence in many application domains has given rise to extensive research fields tackling the questions of zero, one, and few shot learning. On the other side of the spectrum, even with infinite data one can only expect the error to asymptotically approach the Bayes optimum. Analyzing the behavior of the generalization error in the regime that connects the two extremes can help quantify the added value of increasing dataset size.

Both theoretical bounds and empirical observations on the generalization error as a function of dataset size have become available.

From the theoretical standpoint, Ref. [169] explains that the bound on the gap between expected and empirical risk is a function of both dataset size and model capacity, the latter of which can be measured in terms of Vapnik–Chervonenkis (VC) dimensions of the family of functions represented by the model [268, 269]. Fixing the VC dimension leads to uniform convergence with an increase in training examples, but the two quantities need identical rates of change to maintain the gap constant.

On the empirical front, instead, although the previous two sections make informed suggestions aimed at speeding up the discovery of learning models that perform well on a given problem, Ref. [270] shows that the performance improvements due to increasing training dataset size are often predictable and, across a variety of tasks in machine translation, language modeling, image processing, and speech recognition, the accuracy scales as a power law of training set size. On this particular set of tasks, sensible architectural or hyper-parameter modifications only add a vertical offset to the accuracy scaling, but are unlikely to change the power at the exponent in the power law behavior. As a consequence, it should be possible to perform reliable model exploration and optimization with relatively low amounts of data, thanks to the possibility of full performance extrapolation to larger dataset sizes.

Therefore, in general, given sufficient computational power, performance improvements can often be expected from augmenting the training dataset. The exact functional dependence of relevant metrics on the training set size is likely to vary across tasks and application domains. However, a cost-benefit analysis of collecting or simulating more training examples might highlight more affordable ways of improving upon the state of the art.

4.5   Hardware and Implementations

Much of the recent popularity explosion in artificial intelligence can be attributed to seemingly ever-increasing computational capabilities on modern hardware. Novelties and improvements at the hardware level, coupled with a shift towards open source software in the machine learning community, has allowed for massive growth in both the number and depth of research projects in artificial intelligence.

4.5.1   Open Source Software

One of the main catalysts of the pace of modern deep learning research has been the prioritization of open source software by both university research groups and corporate research labs alike. Although many libraries, such as Torch [271] and Theano [272], have been used for over a decade, the 2015-2017 period saw a rapid increase in interest and subsequent releases of new open source libraries, such as Chainer [273], TensorFlow [274], Keras [275], and PyTorch [276].

These libraries allow users to define symbolical mathematical functions and compose them into neural network structures using as high or as low a level of abstraction as desired. They translate neural network specifications into highly optimized low-level instructions with fused computation, when possible, and improved implementations for numerical stability. The breadth of available deep learning libraries exposes the variety of design choices and target audiences that are present along the machine learning spectrum. Different use cases across various domains might benefit from different levels of complexity, interface minimalism, modularity, or platform compatibility, all of which need to be carefully considered before engaging in model design. In particular, in the case of high energy physics, while most applications thus far have fit into the design paradigms defined by some of the high-level packages that provide common neural network building blocks, it is important not to write off more complex and ambitious projects for which a low-level library might be more adept.

Two common threads across the open source landscape that have helped accelerate research are the availability automatic differentiation (see Sec. 4.5.3) and GPU implementation (see Sec. 4.5.4) via bindings to NVIDIA's CUDA and CuDNN libraries. In fact, much of the convenience of modern deep learning libraries lies in their seamless cross-platform compatibility, and in the implementation of efficient versions of automatic and symbolic differentiation, when suitable, which handle the automatic computation and hierarchical dependence of the many partial derivatives that arise in gradient-based neural network optimization. While symbolic manipulation allows for the analytical treatment of variable replacement and functional composition, automatic differentiation offers more scalable tools to numerically compose terms in chain rule computations across graphs (see Sec. 4.5.3).

### 4.5.2 Static versus Dynamic Graphs

The dependence of the objective function and its derivatives on each model parameter and input is best organized in the form of an acyclic graph, in which each node represents a basic scalar, vector, or tensor operation (addition, multiplication, etc.), in terms of which more complex, arbitrary functions can be expressed.

Modern open-source deep learning libraries implement these graphs of operations using symbolic representation, in which functions are defined in terms of other parameters, without the need to immediately associate a numerical value to the computation. Users can define these relations through configuration files or programmatically, depending on the software.

The details of graph implementation separate the two major schools of thought that define the landscape of available deep learning frameworks. On one hand, the define-*and*-run scheme adopted by libraries such as Theano [272] and TensorFlow [274] renders these tools highly optimized compilation engines that interpret the graph definition, fix and optimize its structure into a static representation, but limit runtime flexibility. On the other hand, frameworks such as PyTorch [276] and Chainer [273] that follow the define-*by*-run scheme dynamically execute every iteration as independent program, allowing for data-driven and user-specified modifications to the forward pass, which are reflected in the corresponding backward pass through the use of automatic differentiation. This enables greater flexibility and debugging capabilities.

### 4.5.3 Automatic Differentiation

Underpinning the modern accelerated research pace and sped-up experimentation cycle around designing and testing new architectures is the integration of automatic differentiation (AD) into the machine learning toolbox.

Automatic differentiation techniques generalize backpropagation and use the knowledge that com-plex computer programs can be broken down into a series of elementary operations to automatically infer the set of partial derivatives to compute, in order to differentiate an entire program in a scalable manner.

Unlike pure symbolic differentiation, AD does not try to analytically compose a sequence of functions into one single equation that solely depends on the parameters with respect to which the target function is being differentiated. Instead, AD refers to the association of a gradient op to each elementary component of a computer program, thus operating at an atomic level and leaving the composition of the chain rule to the sequential gradient value computation for a given set of numerical inputs. This makes AD more scalable and suitable for the differentiation of large scale programs.

AD implementations are divided into two main modes of accumulating information at runtime and subsequently using that information to calculate a gradient: forward-mode, and reverse-mode. Their respective advantages depend on the logic of the application. In forward-mode AD, partial contributions to the gradients are computed along with the program execution, starting from the inputs and arriving at any connected step in the computational graph; in reverse-mode, intermediate quantities computed during the program execution must be stored and reused for subsequent backward gradient calculation, which begins at the output and can stop at any node in the graph. In other words, at any node, the forward mode computes the derivative of the output of that node with respect to a program's input, passing through all intermediate upstream nodes and chain rule contributions, while the reverse mode computes the gradient of the program's output with respect to the inputs at that node, passing through all intermediate downstream nodes and chain rule contributions.

More information on automatic differentiation and its use in machine learning is available, for example, in Ref. [198, 277, 278, 279].

### 4.5.4 GPU versus CPU

One of the earliest criticisms of early neural network models was computation time. Naively, a forward pass of a $d$ training examples through an $L$-layer feed forward neural network with $n$ inputs takes $\mathcal{O}(d \cdot L \cdot n^2)$ operations. Although parallelism speedups can be obtained on traditional CPU architectures, such hardware architectures do not lend themselves well to the `axpy`- and inner-product-heavy nature of neural network computations, which tend to require fine-grained parallelism. In addition, many promising models such as LeNet [280], which rely on convolution operators, struggle to find adequate speed on traditional CPUs.

Beginning with Ref. [281], Graphics Processing Units, or GPUs, have shown immense promise in speeding up neural network primitives for deep learning. GPUs were conceived by NVIDIA for the purpose of accelerating the heavy computational requirements of graphics programmers building shaders and 3D renderers. One of the key architectural considerations of the GPU is to maximize the trade-off between on-chip transistors devoted to data processing and transistors dedicated to caching and control flow in favor of data processing capabilities. GPUs are able to handle and schedule across hundreds or thousands of cores, albeit with a reduced instruction set compared to CPUs. This makes GPUs extremely performant for tasks such as large matrix multiplications and inner products.

Modern deep learning libraries succeed at abstracting away many of the implementation details necessary to port an application from CPU to GPU, enabling machine learning practitioners to focus on higher level design choices in their neural network architectures.

## 4.6  Deep Generative Modeling

A major component of this thesis focuses on the presentation of research work in which deep generative modeling is employed as a method to speed up physics simulation at the LHC. This section is therefore aimed at providing the necessary theoretical background that sets the stage for the applications described in Chapter 8. It opens with a generic introduction to generative modeling techniques in machine learning, arranged in the taxonomy proposed by Ref. [20] (see Sec. 4.6.1), continues with the presentation of the Generative Adversarial Network (GAN) framework (see Sec. 4.6.2), and concludes with the exposition of selected improvements on the original GAN formulation, intended to improve the empirical or theoretical understanding of GANs' behavior (see Sec. 4.6.2.1 and 4.6.2.2).

The utility of generative modeling in the sciences surpasses the pure and fundamental research endeavor that has been driving much of the effort within the machine learning academic research community. Despite also finding applicability in domain adaptation and latent representation learning tasks (among others), generative models have sparked the interest of the scientific community because of their potential as pragmatic and flexible data augmentation tools with the promise of emulating compute intensive parts of scientific simulation pipelines.

Simply stated, the goal of generative modeling is to take a finite number of observations drawn form a ground truth, unknown distribution, and learn a model that estimates that distribution function for future controlled reuse.

Formally, assume events of interest $x$ are produced in nature according to a generative process modeled by a data distribution $p_{\text{data}}(x)$, from which we are allowed to draw a finite set of $n$ samples $X = \{x | x \sim p_{\text{data}}(x)\}, |X| = n$. In addition, assume that $p_{\text{data}}$ is *opaque*, that is, it is only possible to sample from it and there is no access to the ground truth analytical distribution. The task of generative modeling, then, is to learn a model $p_{\text{model}}$ and optimal parameters $\theta^*$ such that, according to some metric of interest $\rho$,

$$\theta^* = \arg\max_{\theta} \rho(p_{\text{model}}(x;\theta), p_{\text{data}}(x)). \tag{4.51}$$

Quantitatively measuring distributional proximity can be achieved through many distance measures, resulting in a similar variety of learning algorithms to train generative models. A large portion of deep learning approaches to generative modeling can be thought of as explicitly or implicitly minimizing either an $f$-divergence, such as the Kullback-Liebler divergence, the Hellenger distance, or the Total Variation distance, or an Integral Probability Metric (IPM), such as the Wasserstein distance. Other distance metrics, such as the neural network distance introduced in Ref. [282], can be used as a measure of generalization in GANs. Sec. 4.6.2.2 connects the choice of distance metrics used to measure distributional distance to the emergence of distinct GAN formulations and training strategies.

### 4.6.1 Generative Modeling with Maximum Likelihood

In ordinary settings, distributional fits and generative modeling in science and engineering are performed through direct Maximum Likelihood Estimation (MLE). In most scientific applications, a small number of interpretable parameters are specified for $p_{\text{model}}(x;\theta)$, and the likelihood of the data under the parameterized model,

$$\mathscr{L}(\theta; X) = \prod_{x \in X} p_{\text{model}}(x;\theta), \tag{4.52}$$

is directly maximized to find optimal values for the parameters.

As Ref. [20] suggests, the abundance of generative modeling formulations has given rise to an entire taxonomy of approaches that can be phrased to fit the MLE framework. The several approaches can be compared to one another by contrasting the ways in which they compute or approximate the likelihood and its gradients. These formulations either define an explicit functional form for the density, whether tractable or intractable, or only make use of it in an implicit fashion.

For explicit density models, evaluation or approximation of the density function $p_{\text{model}}(x; \theta)$ is possible and optimization can be straightforward, but capturing the right level of complexity displayed by the data in a hand-designed functional form can be challenging. Trading in tractability for scalability, one encounters approximation techniques, such as variational methods, that make use of either deterministic or stochastic approximations in their formulation. Canonical examples of explicit density generative modeling algorithms include PixelRNN [283], which admits a tractable autoregressive density, and Variational Auto-encoders [284], which build an approximate density using a variational approximation.

Implicit models, on the other hand, only admit a means to interact with the unknown generative distribution, usually through the generation of new samples $x \sim p_{\text{model}}(x; \theta)$, without any explicit formalization of the density function itself. Generative Adversarial Networks are among the most common implicit generative models and are particularly suited for applications in which the user's objective is to sample from a complex high-dimensional training distribution without explicitly modeling it. GANs enable this by reparametrizing the problem: one can first sample from simple (normal or uniform) noise distributions, and then train a neural network to learn a complex transformation that maps the simple distribution to the target training distribution.

This framework has been popularized because of its clever formulation and its higher quality, more visually compelling samples. GANs also offer the added advantages of being computationally cheap to evaluate in forward mode, and relatively easy to modify to include application-specific constraints.

The following section provides a more in depth introduction to the GAN framework.

### 4.6.2 Generative Adversarial Networks

Generative Adversarial Networks (GAN) [285] are a method to learn a generative model by recasting the generation procedure as a minimax game between two actors that are parameterized by deep neural networks. A *generator* network is tasked with mapping a $d$-dimensional latent prior $z \sim p_z(z), z \in \mathcal{Z} \subseteq \mathbb{R}^d$, which acts as the source of stochasticity in the generative process, to a reputable synthetic sample that a *discriminator* network (the adversary) is tasked with classifying as either real (from $p_{\text{data}}(x)$) or synthetic (from $p_{\text{model}}$). $p_z(z)$ is generally a multivariate normal distribution, $\mathcal{N}(0, I)$, which is often recommended over a uniform prior [286].

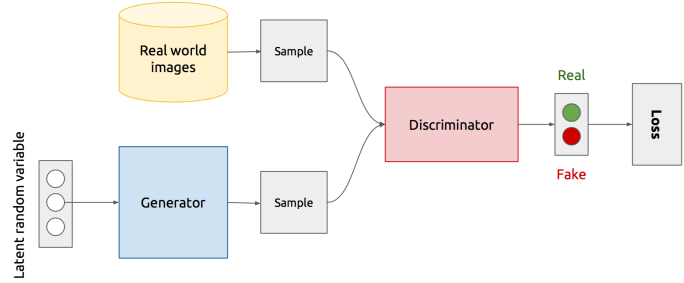

**Figure 4.3:** Schematic representation of the GAN setup, showing the two data streams that the discriminator is tasked to distinguish. Image reproduced from Ref. [19].

A visual representation of a typical GAN setup is provided in Fig. 4.3. In practice, the discriminator examines batches of data coming from two streams – either from the lot of real examples or from the output of the generator – and tries to tell the two apart. The intuitive goal of the generator is to deceive the adversary by producing credible samples that could be mistaken for real.

Formally, the generator network defines a map $G(\cdot; \theta_G) : \mathcal{Z} \longrightarrow \mathcal{X}$ and the discriminator network defines a map $D(\cdot; \theta_D) : \mathcal{X} \longrightarrow [0, 1]$, where 0 is associated with very synthetic-looking samples, and 1 very realistic ones. $G$ and $D$ are trained in an alternating fashion in order to learn a parametric approximation to $p_{\text{data}}(x)$ via a minimax game defined on a value function:

$$V(D, G) = \underbrace{\mathbb{E}_{x \sim p_{\text{data}}(x)}[\log D(x; \theta_D)]}_{\substack{\text{term associated with the discriminator}\\\text{perceiving a real sample as real}}} + \underbrace{\mathbb{E}_{z \sim p_z(z)}[\log(1 - D(G(z; \theta_G); \theta_D))]}_{\substack{\text{term associated with the discriminator}\\\text{perceiving a generated sample as fake}}}. \quad (4.53)$$

As per the minimax formulation, the generator and discriminator are required to play the game defined by the objective:

$$\min_G \max_D V(D, G). \quad (4.54)$$

Convergence and equilibrium are characterized as follows. For a given generator parametrized by fixed weights $\theta_G$, the optimal discriminator $D^*$ is given by $D^*(x) = \frac{p_{\text{data}}(x)}{p_{\text{data}}(x) + p_{\text{model}}(x; \theta_G)}$. From this, optimality conditions for the generator $G$ can be determined, and, as a direct result, a convergence result for the game defined in Eq. 4.54 can also be derived. Under Eq. 4.53 and Eq. 4.54, and when fixing $D^*$, the op-timal generator minimizes the Jensen-Shannon divergence and admits a unique global minimum where $p_{\text{model}} = p_{\text{data}}$ and $D^* = 1/2$. This can be shown by first defining $C(G) = V(D^*, G)$. The generator seeks to minimize $C(G)$ subject to $D^*$. Substituting optimality condition into the formulation in Eq. 4.53 yields:

$$\begin{aligned} C(G) &= \mathbb{E}_{x \sim p_{\text{data}}} \left[ \log \frac{p_{\text{data}}}{p_{\text{data}} + p_{\text{model}}} \right] + \mathbb{E}_{x \sim p_{\text{model}}} \left[ \log \frac{p_{\text{model}}}{p_{\text{data}} + p_{\text{model}}} \right] \\ &= \mathbb{E}_{x \sim p_{\text{data}}} \left[ \log \frac{p_{\text{data}}}{p_{\text{data}}/2 + p_{\text{model}}/2} \right] + \mathbb{E}_{x \sim p_{\text{model}}} \left[ \log \frac{p_{\text{model}}}{p_{\text{data}}/2 + p_{\text{model}}/2} \right] - \log(4). \end{aligned}$$
(4.55)

This is the summation of two Kullback-Leibler divergences, which can be combined into the definition of the Jensen-Shannon divergence:

$$\begin{aligned} C(G) &= D_{\text{KL}} \left( p_{\text{data}} \middle\| \frac{p_{\text{data}} + p_{\text{model}}}{2} \right) + D_{\text{KL}} \left( p_{\text{model}} \middle\| \frac{p_{\text{data}} + p_{\text{model}}}{2} \right) - \log(4) \\ &= 2 \cdot \text{JSD}\left(p_{\text{data}} \middle\| p_{\text{model}}\right) - \log(4) \end{aligned}$$
(4.56)

Therefore, given an optimal discriminator, the sample-based cross-entropy formulation of GANs performing adversarial classification corresponds to the more intuitive scenario of the generator optimizing a quantity $C(G)$, which is directly proportional to the Jensen-Shannon divergence between the learned and ground truth distributions. The unique minimum and equilibrium of $C(G^*) = -\log(4)$ is obtained when $p_{\text{model}} = p_{\text{data}}$, which, when substituted back into $D^*$, results in $D^* = 1/2$ for an optimal $G^*$. However, these results rely on infinite capacity $G$ and $D$, which are clearly not realistic assumptions. Since the players are parametrized as deep neural networks, they have highly expressive yet finite capacity.

A natural question that arises from the preceding discussion is the extent to which the notion of infinite capacity for a discriminator impacts such a system, since the generators ability to obtain meaningful gradient signal depends upon the fidelity and utility of the discriminator. As an example of how the system may be impacted by finite capacity, if a discriminator $D$ is limited to contain a total of $n$ trainable parameters, $D$ will be unable to distinguish (at an $\varepsilon$-level) between a true data distribution $p_{\text{data}}$ and a set of samples from the data distribution $\hat{p}_{\text{data}}$ with $|\hat{p}_{\text{data}}| = \mathcal{O}(\frac{n \log(n/\varepsilon)}{\varepsilon^2})$ [282]. This implies that a finite capacity discriminator is unable to distinguish between a small subset of the support of $p_\text{data}$ and $p_\text{data}$ itself, leading to the conclusion that gradients of $D$ may be unable to assist in moving towards a solution where $p_\text{model} \longrightarrow p_\text{data}$. This result merely suggests that nothing intrinsic about the GAN setup gives satisfying theoretical conditions for a generator to obtain equilibrium when paired with a finite discriminator. This does not, however, preclude such a setup from yielding positive results in practice.

GANs are trained by alternating optimization steps that independently update the weights of $D$ and $G$, while keeping the other set of weights fixed. However, by looking at the magnitude of the gradient as a function of the discriminator's output (displayed in Fig. 4.4), it becomes immediately apparent that the original GAN objective formulation causes the majority of the gradient signal to be dominated by the error associated with relatively high quality (but not yet perfect) samples, while no strong mathematical incentive is provided to recover the data distribution in badly mismodeled regions. Similar issues still arise if the GAN learning procedure is instead phrased as a maximum likelihood problem, in which maximizing the likelihood of the training data under the model is equivalent to minimizing the KL divergence between the true and estimated data distributions. Both formulations of the value function saturate the learning when the generator poorly recovers $p_\text{data}$, as clearly visible in Fig. 4.4.

To overcome this ineffective learning behavior, Ref. [285] proposes a non-saturating heuristic: instead of minimizing the value function in Eq. 4.53, the generator maximizes

$$V_G(D, G) = \mathbb{E}_{z \sim p_z}[\log D(G(z; \theta_G); \theta_D)], \qquad (4.57)$$

which does not suffer from saturation problems.

Though not yet fully developed from a theoretical perspective, the literature around convergence [287], generalization [282], and dynamics of GAN systems [288] is maturing and quickly evolving. The following presents a snapshot in time of current widely adopted empirical tricks and theoretical results towards improving the performance of GANs.

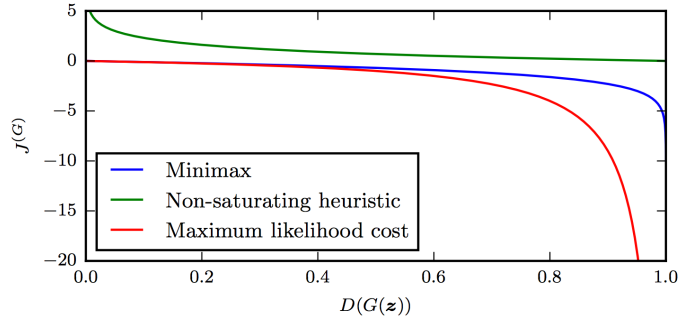

**Figure 4.4:** The value function for the generator under three different cost formulations. Note that when the discriminator correctly classifies a generated sample as very fake (towards zero), the gradient of the value function for the MLE and the original minimax formulation saturate, providing no valuable gradient. This is ameliorated in the non-saturating heuristic. Image reproduced from Ref. [20].

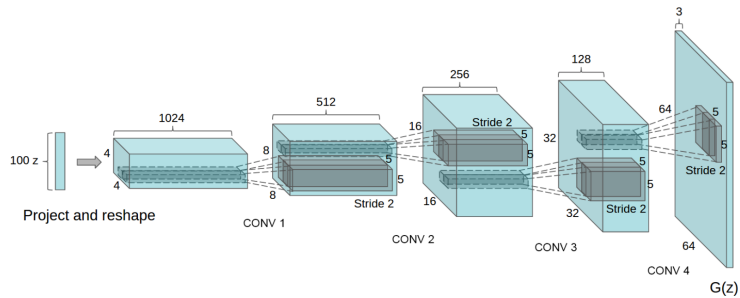

**Figure 4.5:** An example DCGAN-style architecture. Note the usage of transposed convolutions to increase the dimensionality away from the latent prior. Image reproduced from Ref. [21].

#### 4.6.2.1 Experimental Improvements and Modifications of the GAN Architecture and Training Procedure

The first major improvement in generative model quality for image production is attributable to the introduction of Deep Convolutional Generative Adversarial Networks (DCGANs) [21]. The DCGAN architecture, displayed in Fig. 4.5, places architectural constraints on the generator and discriminator networks, and provides well-tuned suggestions for optimization parameters and architectural decisions. Specifically, DCGAN uses strided convolutional discriminators [289] with LeakyReLU activations and batch normalization in order to achieve stable gradients. In addition, the generator network is comprised of strided transposed convolutions [289] to boost the dimensionality of the latent prior up to match those of the sample space. To this day, DCGAN represents a powerful baseline architecture for image generation tasks, and its initialization and optimization parameters provide stable defaults for the development of GANs in the computer vision domain.

Another early realization in the GAN literature is the utility of side information in creating clearer, controllable samples from complex, diverse data distributions. For a given sample $x_i \in X$, consider the case where side information $y_i$ is available. Examples of such information in the context of natural images include class, caption, or any number of attributes such as style, lighting, or pose. In a particle physics application, these may instead include a particle's energy, position, or flavor, or a detector's configuration, for example.

Conditional GANs [290] introduce a simple conditioning mechanism by which side information is fed into both the generator and discriminator, leading to a game in which $G(z, y; \theta_G)$ implicitly defines a conditional distribution $p_{\mathsf{model}}(x|y; \theta_G)$ which is minimized by way of $D(x, y; \theta_D)$. The idea of showing side information is also explored in Ref. [291], where structural key points are directly incorporated into image channels in both generator and discriminator, allowing the system to learn to generate structurally guided natural images.

Alternatively, Auxiliary Classifier GANs [292] require the discriminator to jointly maximize Eq. 4.54 and minimize a loss $\mathcal{L}_{\mathsf{aux}}$ associated with the discriminator's ability to reconstruct the side information $y_i$ for a given $x_i$. Though Ref. [292] explicitly tests this formulation to control the class of a generated image, extensions allow for this solution to work on continuous valued side information.

InfoGAN [293], instead, introduces an unsupervised method for learning disentangled representations within the latent prior $z$. Specifically, through information theoretic means, a portion of the entries in the latent space $z$ are encouraged to take on the meaning of a set of conditioning random variables $c_i$ that represent latent attributes of the distribution. Without direct intervention, InfoGAN is designed to maximize the mutual information between $c \sim P$ and $p_{\mathsf{model}}(x|c; \theta_G)$, the distribution implicitly defined by $G(z, c; \theta_G)$. The intuitive meaning of the attributes $c_i$ emerges automatically and can be investigated post-facto, without directly guiding the network towards discovering these variables.

Other empirical improvements developed to ameliorate the training of GANs move beyond attribute conditioning to develop ad-hoc tricks that address specific shortcomings observed across the majority of GAN formulations and applications. For instance, a common failure mode in GANs is their inability to fully explore a target distribution $p_{\mathsf{data}}$. This issue commonly manifests in *mode collapse*, where a GAN learns to only represent a limited subset of the support, and ignores large sections that are more difficult to learn. To overcome this suboptimal behavior, Ref. [294] suggests the use of Minibatch

Discrimination (MD) to allow the discriminator to directly access a measure related to overall diversity of produced samples. Specifically, this method relies on adding transformed intra-batch distances, computed across examples in a minibatch, to the set of internally calculated features used by the discriminator. This directly allows $D$ to access and make use of approximate distributional information when determining whether a batch of samples is synthetic or real. Formally, let $h(x_i) \in \mathbb{R}^A$ be the vector of hidden features produced as output of the neural network layer preceding MD for the single input example $x_i$. The MD layer is parametrized by a learnable tensor $T \in \mathbb{R}^{A \times B \times C}$ such that $h(x_i) \times T = M(x_i) \in \mathbb{R}^{B \times C}$. For each $x_i$, then, the entries in the matrix $M(x_i)$ can be used to compute the distances $d_b(x_i, x_j) = e^{-|M(x_i)_b - M(x_j)_b|}$, where $x_j$ is another example that belong to the same minibatch as $x_i$, and $b$ indexes the rows of $M$. For each row $b$, each sample $x_i$ will have a number of distances associated with it that equals the minibatch size $n$. To then compute a single quantity per sample per row, a sum is performed over all $j$'s in the minibatch: $o(x_i)_b = \sum_{j=1}^{n} d_b(x_i, x_j)$. Each sample $x_i$ will then end up with $B$ of these features, which contain information about the intra-minibatch diversity from the sample source from which the minibatch originates:

$$o(x_i) = [o(x_i)_0, o(x_i)_1, ..., o(x_i)_{B-1}] \in \mathbb{R}^B. \tag{4.58}$$

Intuitively, if the generated samples show a lesser degree of diversity then the training samples, which could be caused by a poor exploration of the full data distribution, the MD features computed from batches from the two sources should be quantitatively different, and can therefore be used by the discriminator to improve its ability to classify real and fake minibatches.

Another established heuristic introduced in Ref. [294] is the usage of feature matching to modify the objective of the generator. In lieu of optimizing Eq. 4.57, the generator instead minimizes

$$V_G(D, G) = \|\mathbb{E}_{x \sim p_{\text{data}}}[f_D(x)] - \mathbb{E}_{z \sim p_z}[f_D(G(z))]\|_2^2, \tag{4.59}$$

*i.e.*, the norm of the difference between a set of intermediate features $f_D(\cdot)$ computed within the discriminator for real and fake samples. The intuition behind this technique is to prevent the generator from overfitting to the output of the generator, by tasking it instead with matching some statistics of the data distribution. In this formulation, to reduce inductive bias, the features to be matched are not explicitly specified in terms of known, interpretable variables, but are taken from the set of useful hidden features automatically learned and built within the discriminator to optimally perform its classification task. The opposite approach can however be more suitable for applications of GANs to scientific domains where strong priors on specific feature matching requirements exists.

Label flipping and label smoothing have also been observed to be beneficial to GAN training [294]. These respectively refer to random hard bit flips in the labeling scheme, which turn the label associated with training samples to fake and vice versa, and to the smooth perturbation of labels to the set of real numbers $\in [0, 1]$ [295]. Both are intended to add beneficial noise to the training procedure, which fosters exploration, reduces the likelihood of the training getting stuck in suboptimal configurations, and artificially increases the overlap of the supports of $p_{\text{data}}$ and $p_{\text{model}}$.

Progressively fixing, in a supervised manner, the stages of magnification of images produced by GANs in computer vision applications has also empirically demonstrated to help stabilize and speed up training, resulting in the production of images of outstanding quality [296, 297]. While producing high resolution images from scratch seems hopeless – due, among others, to the complexity of learning the transformation from noise to detailed image all at once – progressively increasing the resolution allows the network to tackle increasingly more difficult generative tasks while retaining the prior knowledge acquired at earlier stages when learning to produce lower resolution samples.

Although appealing from a purely experimental standpoint, not all ad-hoc improvements always offer fully satisfying theoretical arguments that justify and support their utilization to improve sample quality and stability. In parallel, though, other theoretical work has focused on systematically analyzing and remedying the mathematical deficiencies that arise in GANs. Towards understanding and designing theoretically driven improvements, researchers have investigated the effects of using better game parameters, and the feasibility and conditions of convergence to the Nash equilibrium.

Whether an equilibrium for the game exists, and whether the GAN formulation converges and, if so, how fast, are all open research questions currently under investigation. When the optimization takes place in function space, in the idealized scenario of infinite capacity models, GANs, which often display non-linear dynamics, can be shown to converge to an equilibrium point via simultaneous stochastic gradient descent [285, 298, 288]. The case of GANs parametrized by realistic, finite-capacity neural networks, is more complex, and theoretical guarantees are not always available. Equilibrium may still exist and convergence can be achieved under certain conditions, such as appropriate learning rate decay schedules between generator and discriminator [299], in the case of mixtures of models [282], or for suitable initialization conditions that start the learning near the equilibrium [288]. In practice, acceptable convergence levels are empirically observed from the samples produced by several GAN models, and the heuristics described above have contributed to speeding up the rate of convergence, which is otherwise observed to be relatively slow.

### 4.6.2.2 Improved Distance Metrics

As demonstrated in Sec. 4.6.2, the original GAN formulation can be seen as minimizing the JS divergence between the data and the model distributions. This distance metric belongs to a broader family of non-negative measures of similarity between distributions, called $f$-divergences [300, 301]. For two distributions $P$ and $Q$ with densities $p$ and $q$, respectively, the $f$-divergence, $D_f(P\|Q)$ is defined as

$$D_f(P\|Q) = \int_\Omega f\left(\frac{p(x)}{q(x)}\right) q(x) d\mu(x), \qquad (4.60)$$

with the requirement that $p$ and $q$ be absolutely continuous with respect to a measure $\mu$ on the space $\Omega$, and that $f : \mathbb{R}_{>0} \longrightarrow \mathbb{R}$ be convex and lower-semicontinuous, and satisfy $f(1) = 0$ [302]. Various functional choices for $f$ realize different known divergences.

In an attempt to unify the rapidly expanding number of of GAN variants, Ref. [302] provides an appealing generalization of many GAN formulations by introducing the $f$-GAN framework, which connects various related GAN variants that are found to minimize different types of $f$-divergences. Examples of these divergence functions are listed in Table 4.2.

Since the actual distributions $p$ and $q$ may not be accessible in functional form – indeed, for GANs, they never are, and one can only implicitly sample from them – a variational method is needed to estimate the $f$-divergences given only empirical distributions of samples from $p$ and $q$.

The notion of an $f$-GAN allows the examination of some assumptions that $f$-divergence-based GANs implicitly make about $p_{\text{data}}$ and $p_{\text{model}}$ at any point in time during training. In particular, the requirement of absolute continuity of $p$ and $q$ with respect to a reference distribution $\mu$ is strong, and often times not true in practice. Specifically, a result from Ref. [303] allows to state that when $p_{\text{model}}$ and $p_{\text{data}}$ lie in two manifolds $\mathcal{A}$ and $\mathcal{B}$ that do not have perfect alignment, the optimal discriminator will be perfect, and will admit a fully zero gradient. Practically, this means that if $\exists x : (p_{\text{data}}(x) > 0 \land p_{\text{model}}(x) = 0) \lor (p_{\text{model}}(x) > 0 \land p_{\text{data}}(x) = 0)$, or, in other words, if $\text{supp}(p_{\text{model}}) \neq \text{supp}(p_{\text{data}})$, then a GAN system defines invalid divergences that do not provide a gradient signal to the generator.

| Name | $D_f(P\|Q)$ | Generator $f(u)$ |
|---|---|---|
| Kullback-Leibler | $\int p(x) \log \frac{p(x)}{q(x)} \, dx$ | $u \log u$ |
| Reverse KL | $\int q(x) \log \frac{q(x)}{p(x)} \, dx$ | $-\log u$ |
| Pearson $\chi^2$ | $\int \frac{(q(x)-p(x))^2}{p(x)} \, dx$ | $(u-1)^2$ |
| Squared Hellinger | $\int \left(\sqrt{p(x)} - \sqrt{q(x)}\right)^2 dx$ | $(\sqrt{u}-1)^2$ |
| Jensen-Shannon | $\frac{1}{2}\int p(x) \log \frac{2p(x)}{p(x)+q(x)} + q(x) \log \frac{2q(x)}{p(x)+q(x)} \, dx$ | $-(u+1)\log \frac{1+u}{2} + u \log u$ |
| GAN | $\int p(x) \log \frac{2p(x)}{p(x)+q(x)} + q(x) \log \frac{2q(x)}{p(x)+q(x)} \, dx - \log(4)$ | $u \log u - (u+1)\log(u+1)$ |

**Table 4.2:** A selection of popular $f$-divergences with their associated generator functions. Table adapted from Ref. [302].

To ameliorate the issue of divergences being ill-defined with respect to distinct supports between generator and data, other distance metrics have been proposed and tested. For example, the Wasserstein GAN (WGAN)[304] makes use of the Wasserstein distance in lieu of an $f$-divergence. In this scenario, the distance is well-defined over non-overlapping supports and explicitly able to account for the encounter of these pathological regions in training. Ref. [304], in particular, suggests the use of the Wasserstein-1 distance, often called the Earth Mover's distance, which intuitively specifies the cost of optimally shifting probability mass between two distributions to make them identical:

$$W(\mathcal{P}_{\text{model}}, \mathcal{P}_{\text{data}}) = \inf_{\gamma \in \Pi(\mathcal{P}_{\text{model}}, \mathcal{P}_{\text{data}})} \mathbb{E}_{(x,y) \sim \gamma} \|x - y\|. \tag{4.61}$$

The intractable version of the Earth's Movers distance in Eq. 4.61 can be expressed, using the Kantorovich-Rubinstein duality, as

$$K \cdot W(\mathcal{P}_{\text{model}}, \mathcal{P}_{\text{data}}) = \sup_{f : \|f\|_L \leq K} (\mathbb{E}_{x \sim p_{\text{model}}}[f(x)] - \mathbb{E}_{x \sim p_{\text{data}}}[f(x)]), \tag{4.62}$$

where $f$ must be chosen to be $K$-Lipschitz [185, 186, 187], and the dual expectations lend themselves to minibatch stochastic optimization. To ensure such a restriction on Lipschitz continuity, the original formulation clips weight parameters in the critic to be in a compact space (*i.e.*, $[-0.01, 0.01]^d$), implic-itly setting a value for $K$ to systematically reduce computational capacity. In fact, $K$ is a function of the network capacity and clamping threshold value.

An alternative solution that avoids such stringent and counterproductive capacity reduction consequences is to make use of a gradient penalty, which directly penalizes the critic for being far from a 1-Lipschitz functional [305]. This formulation, known as WGAN-GP, expresses the discriminator's cost as

$$\mathcal{L}_{\mathsf{GP}} = (\|\nabla_{\tilde{x}} D(\tilde{x}; \theta_d)\|_2 - 1)^2, \qquad (4.63)$$

where $\hat{x} = \varepsilon x + (1-\varepsilon)\tilde{x}$, with $x \sim p_{\mathsf{data}}$, $\tilde{x} \sim p_{\mathsf{model}}$, and $\varepsilon \sim U(0,1)$. The $\mathcal{L}_{\mathsf{GP}}$ term forces $D$ to be approximately 1-Lipschitz.

WGAN and WGAN-GP are some of the most widely used GAN formulations across many application domains, because of the simultaneous empirical and theoretical appeal.

# 5

# Physics Objects and Where to Find Them

As particles traverse the detector volume, they typically interact with them and dissipate a fraction of their energy in the process, until they are fully absorbed or escape the detector. The type of particle and energy scale that characterizes the interaction determine the nature and phenomenology of the physical processes that takes place.

This chapter discusses reconstruction and identification techniques adopted by the ATLAS collaboration to recognize the presence and characteristics of physical objects that make up the composition of a complex event recorded by the detector's sensors.

*Reconstruction* refers to the task of analyzing raw sensor signals to infer the presence of particles, and translating detector readouts into physical objects, proto-particles, and particle candidates that may have generated the signal registered by the detector, while extracting their physical properties. Because of the vast diversity of detector technologies employed in detectors like ATLAS, each sub-detector component is instrumented with dedicated readout electronics, and targeted algorithms are implemented to process each signal type and convert it into particle features.

Reconstruction algorithms are usually multi-stage, sequential software routines that iteratively simplify the signal representation from raw detector hits, to intermediate object definition, and finally to concise, tabular, particle-level properties for subsequent analysis. Similar, but not always identical, reconstruction algorithms are used in the two stages of online triggering and in offline reconstruction, with the online analysis often making use of a simplified detector geometry and uncalibrated inputs for real-time decision making. The technical description of on-device electronic readout strategies is omitted, and, in the rest of this section, the reconstruction procedure is described starting from the minimally processed collection of individual position and energy measurements in each detector sensor.

Particle *identification*, on the other hand, refers to the task of classifying the nature of particles in the event, starting from the reconstructed properties of a series of objects obtained in the reconstruction phase.

The performance of reconstruction and identification algorithms is expressed in terms of figures of merit such as the selection efficiency, the resolution with which a quantity is measured, the false positive rate that contaminates a selection, the robustness to perturbations in the inputs, the computational cost, and the correlation with physical quantities, among others.

Due to the LHC being a "jet factory", most reconstruction and identification methods need to be extraordinarily robust to the copious amounts of QCD jets from quark and gluon fragmentation that constitute the dominant background to many tasks.

## 5.1 Reconstruction from Detector Hits

The first level of reconstruction often consists of forming clusters of detector hits and connecting them to other causally-related clusters. This section describes the techniques adopted in the reconstruction of objects from hits in the detector layers that a particle would encounter while moving outward through the ATLAS detector from the collision point.

### 5.1.1 Tracking

A *track* is a collection of connected clusters of hits in the inner detector (and, in the case of muons, in the muon spectrometer as well), which represents the trajectory of a charged particle, and from which it is possible to infer the particle energy, momentum, vertex position, and charge. Charged particles, such as $e^{\pm}$ and $\mu^{\pm}$, can travel for measurable distances, deposit energy along the way, and their trajectories can be precisely traced out by tracking detectors. The charge sign determines the direction in which the track bends, according to the Lorentz force.

The ATLAS inner detector (ID) is the main instrument tasked with reconstructing charged particle tracks from the energy deposited within its volume. The severe detector occupancy conditions, due to the several hundreds of tracks emerging from each event, and the short timing separation between bunch crossings require excellent spatial resolution and fast response from the inner detector sensors and readout electronics. The intrinsic resolutions of the ID components are: 8 $\mu$m ($r$-$\phi$) × 40 $\mu$m ($z$) in the IBL layer, 10 $\mu$m ($r$-$\phi$) × 115 $\mu$m ($z$) in the Pixel layers, 17 $\mu$m ($r - \phi$) in the SCT, and 130 $\mu$m ($r - \phi$) in the TRT [306]. The lower precision TRT measurements are compensated by the large number of expected hits per particle ($> 30$) in that portion of the detector, which, along with the large lever arm created by its distance from the interaction point, contributes to improving the momentum resolution. On the other hand, high sensor density in the silicon layers is fundamental, for example, to reconstruct secondary vertices in heavy flavor decays. Augmenting high-precision hit measurements from the silicon layers close to the interaction point with the addition of several TRT hits in the outer portion of the ID produces robust and reliable inputs for tracking algorithms in $r$-$\phi$-$z$ space. To get a sense of the occupancy levels and required spatial resolution for track reconstruction at the LHC, hits collected by the ID layers and corresponding reconstructed tracks from one of the first Run II events observed by ATLAS at 13 TeV are shown in Fig. 5.1.

Unlike other particles, muons are usually not fully absorbed in the calorimeters, but instead traverse the entire detector volume and arrive to the muon spectrometer, where they are likely to leave hits guided by the magnetic field induced by the toroidal coils. The hits in the muon chambers can be matched to those in the inner detector layers for full track reconstruction (see Sec. 5.2.5).

The ATLAS tracking procedure is summarized in this section from the information in Ref. [105]. The pixel detector measures the charge collected by its sensors and relates it to the corresponding ionizing particle's energy loss $dE/dx$ [307]. Hits are registered using the time-over-threshold (ToT) technique, which indicates the duration for which the signal exceeds a certain threshold, and is proportional to the magnitude of the energy deposit [308]. As an ionizing particle traverses a silicon detector layer, depending on its angle of incidence, it may deposit energy across a large enough volume to results in charge collection over multiple pixels or strips. Connected component analysis is first used to cluster sensor components that register energy measurements above a given threshold. A neural network then outputs a space point for the location where each particle trajectory intersects each detector component. Three space points are sufficient to form a track seed. Seeds that pass quality requirements are fed into a combinatorial Kalman filter to build track candidates by selecting further compatible space points. A track score is assigned to each track candidate based on the goodness of the fit, the track momentum, the cluster weight, the presence of holes (which reduces the score), and other track quality criteria. In the presence of shared space points, an iterative ambiguity solver is called on the track-score-ordered track candidates collection to limit the number of shared, but not merged, clusters. A neural network is used to classify pixel clusters into single or merged [309, 310]. A cluster that is identified by the neural network as a single-particle cluster but is assigned to two or more track candidates is labeled as shared, and its assignment must be resolved. Techniques for splitting SCT clusters based on expected and observed cluster width, in the absence of charge information, have also recently been developed in ATLAS [311]. Tracks can be extended to include TRT hits, and individual TRT segments can be promoted to the state of tracks from conversion or material interactions.

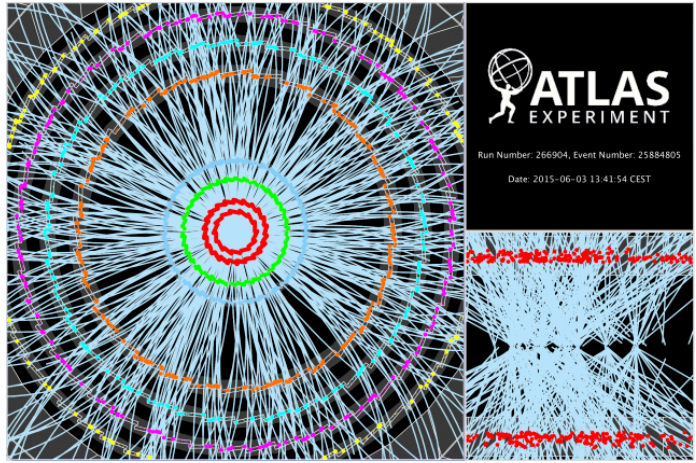

**Figure 5.1:** ATLAS event display of an early Run II $pp$ collision event at $\sqrt{s} = 13$ TeV [22]. Seventeen individual vertices are reconstructed in this event. Several charged particles originate at these vertices and travel outward, leaving *hits* in the inner detector layers. These are shown as small dots, color-coded to represent the detector sub-layers. Tracks are fitted by connecting these dots in order to reconstruct particle trajectories. The higher occupancy levels in the innermost layers demand finer granularity and higher radiation resistance.

The final track fit is parametrized by five perigee parameters and their uncertainties.

In offline tracking, the Kalman filter is replaced by a generalized Gaussian Sum Filter, which results in up to 6% improvement in downstream electron reconstruction efficiency [312, 313].

Track curvature, and hence momentum, are usually obtained from sagitta measurements, so the momentum resolution is intrinsically related to the resolution in the measurement of the sagitta $s = R(1 - \cos\theta)$, assuming the circular arc spans an angle of $2\theta$ and the radius of the circle is $R$. At high $p_T$, $s \approx R\theta^2/2 \propto R\frac{L^2}{R^2} \propto L^2B/p_T$, where $B$ is the magnetic field, and $L$ is the chord of the circle connecting the two hits that define the extrema of the arc over which the measurement is made. Therefore, the fractional momentum resolution $\sigma_{p_T}/p_T$ becomes proportional to $\sigma_s p_T/(BL^2)$, where $\sigma_s$ is the uncertainty on the sagitta. As expected, the momentum resolution degrades as $p_T$ increases. A larger tracking detector and stronger magnetic field would improve the momentum resolution.

Two track selection recommendations and working points are provided by ATLAS, according to the purity and track quality requirements of individual use cases. The selection criteria are primarily based on kinematics and on number of hits in each detector layer that are used to fit the track. The *Loose* selection is the standard one applied at reconstruction time, which provides a compromise between excellent tracking efficiency and the inclusion of a fraction of fake tracks. The *Tight Primary* selection includes additional, stricter requirements on the number of hits per track, thus lowering the contribution of fake tracks to negligible amounts, for tasks such as vertexing (see Sec. 5.1.2) where track quality is paramount [23].

Although tracking precision is high [314, 23, 315], specific physics scenarios appear particularly problematic and may contribute to a decrease in track reconstruction performance. The failure rate for track reconstruction is quantified in the tracking efficiency measurement, which varies as a function of several parameters, including charged component multiplicity, location in the detector, and transverse momentum of the incoming particle.

Detector geometry and particle kinematics limit the range for high quality tracking to $|\eta| < 2.5$ and $p_T > 400$ MeV. Nonetheless, even within these constraints, particles in the high $|\eta|$ and low $p_T$ ranges are still likely to fail the track quality selection. Tracking efficiency as a function of track $\eta$ and $p_T$, measured in ATLAS simulation at $\sqrt{s} = 13$ TeV, is shown in Fig. 5.2. The more detector material the particle has to traverse (which increases with $|\eta|$), the higher the probability for multiple scattering processes that may alter the particle trajectory [316].

At the same time, improved material modeling in the inner detector is necessary to reduce the large

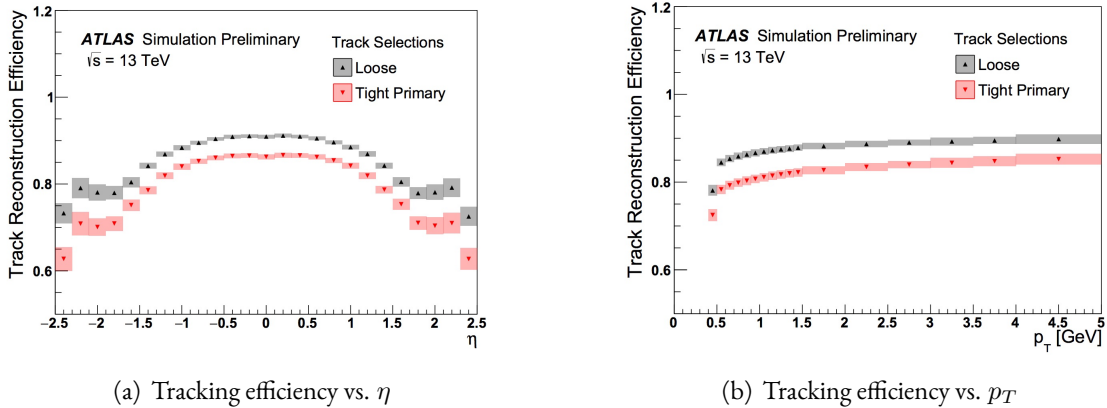

(a) Tracking efficiency vs. $\eta$

(b) Tracking efficiency vs. $p_T$

**Figure 5.2:** Track reconstruction efficiency as a function of the kinematic properties of the track for two sets of track identification selection criteria (Loose and Tight Primary) defined in Ref. [23]. The plots are produced from Monte Carlo simulated minimum bias events at $\sqrt{s} = 13$ TeV. The Loose selection accepts $\sim 5\%$ more tracks than the Tight Primary selection across the entire $p_T$ spectrum, with enhancements at high $|\eta|$, but, at the same time, also incurs in increased fake rates. The shaded regions indicate the systematic uncertainty. Images reproduced from Ref. [23].

(dominant) contribution to the tracking efficiency uncertainty in simulation. Unaccounted misalignments of detector subcomponents are also sources of error that degrade tracking resolution.

Correctly accounting for all particles in an event becomes an increasingly onerous task in dense environments, such as those defined by collimated jets, due to track overlap and large detector occupancy. Higher momentum translates into more collimated decay products, with higher probability of overlap within a detector module. As particle density grows, clusters tend to merge because connected component analysis may identify only one cluster out of the energy deposited by multiple nearby particles, which in turns worsens the track reconstruction performance. The ability to single out individual charged particle trajectories is limited by the inner detector resolution, especially in dense environments when the distance among tracks is similar to the size of a detector element. This hardware limitation compromises the performance of software-based identification algorithms for track reconstruction and particle identification, and requires the development of ambiguity solving strategies to resolve individual tracks and mitigate performance losses [317, 105, 311].

A recent, dedicated ATLAS optimization study of tracking in dense environments provides state-of-the-art performance results for tracking efficiency, and enhances previous optimization campaigns with a novel in-situ tracking efficiency estimation technique [105].

Concerns regarding the ability of the current tracking algorithms to sustain the increased luminosity conditions of the HL-LHC has awoken new interest in machine learning methods to revamp the ATLAS tracking software. The ongoing "TrackML Particle Tracking Challenge" on Kaggle [318] is likely to bring about novel strategies to cope with the expected worsening of tracking conditions, while meeting the requirements for low fake rates, good resolution, and low computational demands.

### 5.1.2 Vertexing

*Vertices* are the locations in the proximity of which multiple tracks intersect. Tracking information is used to infer the position of the primary interaction vertex (PV). Among all vertices identified in an event, unless an analysis requests otherwise, the point where the associated tracks (requiring at least two good tracks) converge and add up to the largest sum of squared transverse momenta $\sum p_T^2$ is called the primary vertex.

For tracks to be used in the vertex reconstruction procedure, they need to satisfy quality requirements: $p_T > 400$ MeV, $|\eta| < 2.5$, number of hits in silicon $\geq 9$ if $|\eta| \leq 1.65$ or $\geq 11$ if $|\eta| > 1.65$, IBL hits + B-layer hits $\geq 1$, at most 1 shared pixel hit or 2 shared SCT hits, 0 pixel holes, SCT holes $\leq 1$ [24].

An iterative approach to vertex finding and fitting has been adopted by the ATLAS collaboration since Run I [319]. First, the longitudinal location of the vertex seed is estimated by iteratively updating the value of the mode of a growing collection of tracks, whose impact parameter $z_0$ is computed with respect to the beam spot. All tracks compatible with the seed position are then selected and used for a second iterative procedure for vertex $\chi^2$-fitting and covariance matrix estimation. Once a vertex is reconstructed, its associated tracks are refitted to be constrained to originate from the vertex, and removed from the initial collection, so that the procedure can be repeated with the remaining candidates in order to find more vertices.

After the vertex fitting operation, track association to vertices is revisited, leading to the definition of three track-vertex-association working points (*Loose*, *Nominal*, *Tight*) that provide different trade-off levels between including all tracks that truly originate from a given vertex and rejecting background tracks that do not belong to the vertex. The selection criteria are based on track impact parameter properties.

In general, vertex reconstruction efficiency increases with track multiplicity at the vertex, growing quickly to above 95 % with three tracks, as shown in Fig. 5.3 for simulated and early Run II ATLAS data [24]. As intuitively expected, however, increased pile-up levels negatively affect tracking and vertexing performance, by increasing the number of spurious tracks and deteriorating parameter resolution. In particular, the probability of reconstructing two independent, nearby vertices as one increases with pile-up. A series of working points for track-to-vertex association based on track impact parameter quantities are derived, resulting in different levels of impurity, defined as the percentage of pile-up tracks that end up being associated to the hard-scattering vertex.

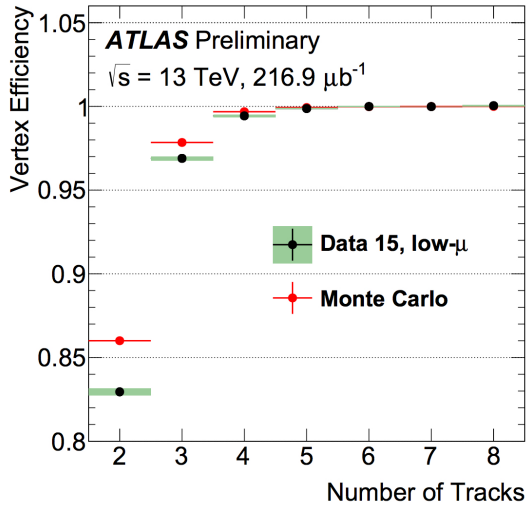

**Figure 5.3:** Comparison of vertex reconstruction efficiency as a function of the track multiplicity in the event, in low pile-up data collected during the ATLAS run no.267359 and in Monte Carlo simulation. The efficiency is quantified as the ratio of events in which a vertex is reconstructed to the events with at least two reconstructed tracks that pass the quality criteria enumerated in the text. Image reproduced from Ref. [24].

To estimate the vertex position uncertainties, the Split-Vertex method separates the tracks that define the vertex into two subgroups of comparable $\sum p_T^2$, resulting in the formation of two new temporary vertices. Their separation gives an estimate of the intrinsic vertex position resolution [319].

Computer vision techniques can be adopted for simultaneous, robust vertex finding. An image-based technique had been explored in ATLAS before, with the goal of improving the computational efficiency of seed and vertex finding, and mitigating the effects of vertex merging at high pile-up [320, 321, 322]. It makes use of hand-designed image transformations, coupled with an additional clustering algorithm to find peaks in the three-dimensional image representation. This method flags potential vertex positions, removing the outer iteration loop and the hierarchy of vertices in the track assignment procedure previously described.

Automated, deep-learning-based computer vision techniques represent the next logical solution to build upon currently available vertexing methods, and to remove much of the clumsy, ad-hoc algorithmic design. Exploratory studies of the application of convolutional neural networks to vertex finding have been proposed by the HEP.TrkX collaboration [323].

### 5.1.3 Energy Depositions in the Calorimeter

Connecting calorimeter units that register plausibly related signals from both charged and neutral particles is an important step in trying to make sense of the high-dimensional detector readout. Pattern recognition algorithms are employed to identify signal over the large amounts of electronics and pile-up noise.

Calorimeter segmentation into cells is exploited to measure the deposited energy as a function of the position of each voxel. Nearby cells that register the passage of an incoming particle are clustered using one of the several available clustering algorithms [324].

In the L1 trigger, the geometry of the calorimeters is simplified into coarser $0.1 \times 0.1$ towers in $\eta$-$\phi$ space. Under the assumption that showers are characterized by narrow width, which causes related energy depositions to occur within a small neighborhood of cells, localized energy deposits discovered within the calorimeter volume are promoted to Regions Of Interest (ROI) for further processing [325]. Specifically, object identification from energy depositions in the calorimeter begins with the construction of $2 \times 2$ ROIs segmented into $0.1 \times 0.1$ sub-regions in $\eta$-$\phi$ space around cluster seeds (trigger towers). The total energy is computed by adding the analog signals collected by the calorimeter cells in these sub-regions; a transverse energy threshold is then applied to remove spurious upward fluctuations. The local maximum is pinpointed using a sliding-window algorithm in the core of the region. In the case of EM showers, the surrounding cells, known as the isolation ring, as well as the adjacent volumes in the hadronic calorimeter, provide handles to isolate the electromagnetic core from hadronic contributions [313]. A visual representation of the definition of a RoI with associated isolation rings is available in Fig. 5.4.

On the other hand, in the HLT trigger and offline, in order to identify three-dimensional structure of related energy deposits across the longitudinally and laterally segmented volume of the ATLAS calorimeters while implicitly suppressing noise, neighboring topologically-connected calorimeter cells that collect energy likely to have originated from a common ancestor are grouped into calorimeter clusters (*topoclusters*) [25]. This algorithm implements the logic of merging locally-connected energy deposits that have high probability of describing different branchings of the same particle shower. Recognizing and exploiting the relation among spatially proximal clusters improves fidelity in the reconstruction of primary physics objects. In the topocluster formation algorithm, cluster growth is governed by a per-cell signal significance quantity $\varsigma_{\text{cell}}^{\text{EM}}$ that rescales the collected energy by the expected noise level, both measured at the EM scale, in order to reduce false positive signal detection. Proto-clusters are initialized from cells that pass a $|\varsigma_{\text{cell}}^{\text{EM}}| > 4$ requirement, then iteratively grown to include neighboring cells that satisfy $|\varsigma_{\text{cell}}^{\text{EM}}| > 2$ and $|\varsigma_{\text{cell}}^{\text{EM}}| \geq 0$, and merged with nearby proto-clusters in case of overlap. The presence of more than one non-isolated local maximum is ground for splitting the resulting cluster into two daughters, with fractional per-cell contributions alloted among the two. Clusters may extend across multiple calorimeter sub-detector layers, in which case neighboring cells are defined as cells with at least partial overlap in $\eta$-$\phi$. An example of the result of topological clustering is provided in Fig. 5.5.

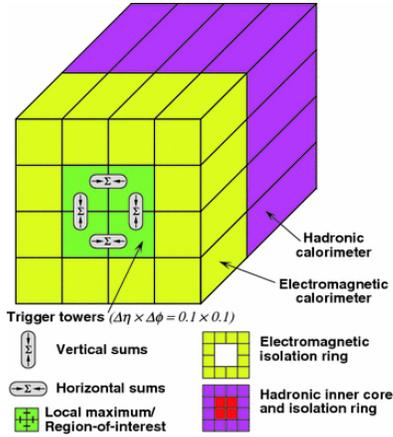
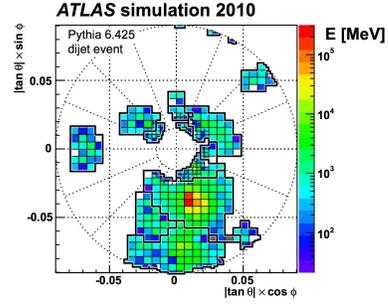

Figure 5.4: Schematic representation of the formation of regions of interest in the calorimeter region using the sliding window algorithm, used in the L1 trigger.

Figure 5.5: Topoclusters formed in the first module of the ATLAS FCAL in a MC dijet event with at least one jet entering this region of the detector. The event is produced using Pythia 6 and emulates the condition of early 2010 ATLAS runs in the absence of pile-up. Image reproduced from Ref. [25].

Negative-energy cells that may be included in clusters (with net positive or negative cluster energy) are primarily due to fluctuations induced by noise and residual radiation from out-of-time pile-up. Reconstructing negative-energy clusters provides a useful handle to estimate the amount of noise in an event, and including negative-energy cells in the clustering algorithm helps suppress random contributions from upward fluctuations, thus reducing the overall noise level.

Although the main objective of the topological clustering algorithm is to provide dynamic cluster re-shaping and resizing to adapt to the large variability in single shower development topologies, topoclusters are not expected to always map to one physical entity in its entirety. Hadronic showers, in particular, because of their innate complexity and extent, may develop sub-showers in the non-immediate neighborhood of the main hadronic energy deposition and are often spread over multiple topoclusters.

The pile-up fit (*pufit*) method is used to estimate and remove the transverse energy contributions of secondary interactions: topoclusters are aggregated into $0.7 \times 0.7$ high- or low-$E_T$ towers; a function is fitted to the low-$E_T$ tower heights to estimate the pile-up contribution; the deduced pile-up level in high-$E_T$ towers is subtracted from the total tower height to obtain pile-up-corrected hard towers [326].

In addition to contributing to the faithful reconstruction of EM and hadronic showers, precise energy measurements in the calorimeter can also help infer the presence of neutrinos in the event. These elusive particles travel undetected, without interacting, through the entire detector apparatus. However, their presence can be deduced from the difference in initial transverse energy compared to the same quantity measured in the calorimeters. To reconstruct non-interacting neutrinos and potential BSM particles, precise accounting of all other momenta in the event is necessary. This is fundamentally limited by energy resolution.

Topoclusters are often inputs to particle identification algorithms discussed in the following section. As such, quantitative descriptions of the properties of topoclusters are provided in terms of several cluster moments and features, defined, for example, in Ref. [25].

Topoclusters can be calibrated using the Local Cell Weighting (LCW) method, which applies different corrections to hadronic and electromagnetic showers after calculating the probability that each topocluster belongs to one of the two categories [25].

## 5.2 Physics Objects

Once spikes in detector activity have been investigated and clustered as described in the previous section, the next step is to map them to the nature of the physical objects that might have generated them. Different types of particles interact in unique ways with the multitude of sub-detector media that are present in the ATLAS detector, thus leaving unique signatures in the various detector layers. The diagram in Fig. 5.6 represents a cross-sectional view of a wedge of the ATLAS detector, along with the average signature of different particle types as they traverse and are stopped by the ATLAS detector.

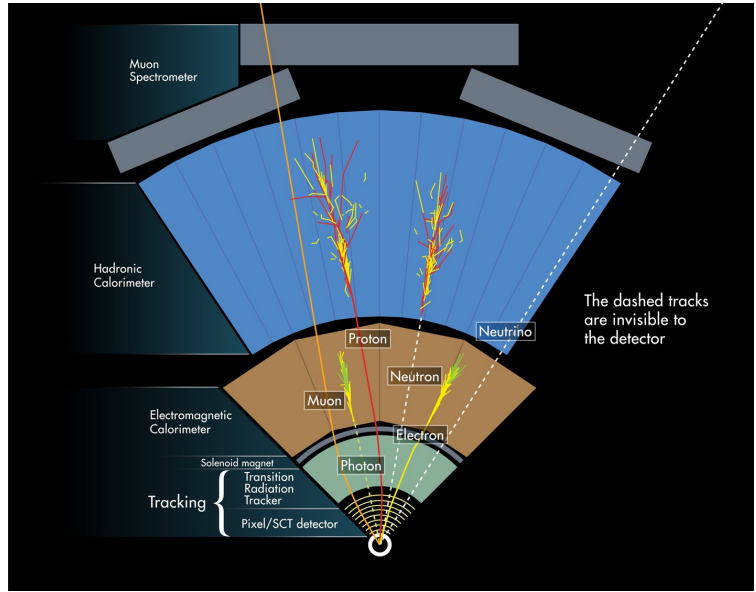

**Figure 5.6:** A diagram representing how the ATLAS detector geometry and material cause each elementary particle type to appear in the detector. The peculiar characteristics of each signature are exploited to identify the presence of a specific particle in an event.

This section describes the various physics objects that can be identified by the ATLAS detector, and addresses how they are individually reconstructed and identified.

### 5.2.1 Jets

Jets are sprays of final-state hadrons originating from collinear or soft splittings from the initiating parton during the parton showering process. Jet properties are therefore correlated with the properties of the original parton, which need to be recovered to infer the nature of the hard-scattering process.

Both incoming and outgoing partons give rise to radiation, known, respectively, as initial-state and final-state radiation. Jets are therefore ubiquitous in high momentum transfer events dominated by hadronic activity.

By construction, jets are not proper physics objects with a unique definition. The choice of algorithm used to quantify locality and to bundle the energy depositions from hadrons and from their decay products determines the nature of the jet itself. Jet-defining, or jet-clustering, algorithms specify routines that associate energy clusters from final-state objects to jets. A review of jet-clustering algorithms can be found in Ref. [27].

Geometric cone strategies or mainstream machine learning clustering algorithms may not be readily adopted for jet clustering. In fact, a viable jet-clustering algorithm has to satisfy two important physical requirements: infrared (IR) and collinear (C) safety. The former requires invariance of jet properties under the addition of infinitely soft radiation, while the latter requires invariance under the substitution of any parton with a series of collinear partons with identical total momentum. Therefore, ad-hoc physics-driven hierarchical clustering algorithms are preferred in HEP.

| Algorithm | $d_{ij}$ | $d_{iB}$ | References |
|---|---|---|---|
| $k_t$ | $\min\left[p_{T,i}^2, p_{T,j}^2\right] \frac{\Delta R_{ij}^2}{R^2}$ | $p_{T,i}^2$ | [328, 329] |
| C/A | $\frac{\Delta R_{ij}^2}{R^2}$ | 1 | [330, 331] |
| anti-$k_t$ | $\min\left[\frac{1}{p_{T,i}^2}, \frac{1}{p_{T,j}^2}\right] \frac{\Delta R_{ij}^2}{R^2}$ | $\frac{1}{p_{T,i}^2}$ | [332] |
| VR | $\min\left[\frac{1}{p_{T,i}^2}, \frac{1}{p_{T,j}^2}\right] \Delta R_{ij}^2$ | $\frac{\rho^2}{p_{T,i}^4}$ | [333] |

**Table 5.1:** Popular jet clustering algorithms currently in use in HEP. $R$ is the user-specified jet radius passed to the algorithm at invocation time. The variable radius (VR) formulation refers specifically to what Ref. [333] denotes as "AKT-VR"; here $\rho$ is a dimensionless constant, such that $R_{\text{eff}}(p_T) = \rho/p_T$.

These sequential jet-clustering algorithms primarily differ among each other in the transverse-momentum-based weighting scheme used to prioritize constituents in the iterative clustering process. Starting from topocluster or track four-momenta, clustering algorithms define the jet-distance metric $d_{ij}$ for constituent $i$ and jet $j$ and the beam-distance metric $d_{iB}$ as shown in Table 5.1 for four prominent algorithms. At each iteration the smallest entry in the set of $\{\{d_{ij}\}, \{d_{iB}\}\}$ is selected. If it is a constituent-to-jet distance, the jet four-momentum is corrected by the vectorial addition of the constituent's four-momentum; this correction to the jet kinematics forces all constituent-to-jet distances to have be recomputed. If, instead, the smallest entry is a constituent-to-beam distance, the constituent is removed until the clustering of the next jet begins [327].

Typical jet-clustering algorithms require the specification of a radius parameter $R$, but no indication of the number of clusters to look for. In ATLAS, for example, *fat jets* typically have radius $R = 1.0$, while small sub-jets within large-radius jets are generally defined with $R = 0.2$. The ideal value for the $R$ parameter varies with $p_T$.

As an alternative to sequential algorithms, a popular cone clustering techniques is the seedless infrared-safe cone (SIScone) algorithm [26]. Cone clustering algorithms have, however, mostly fallen out of favor compared to sequential clustering algorithms.

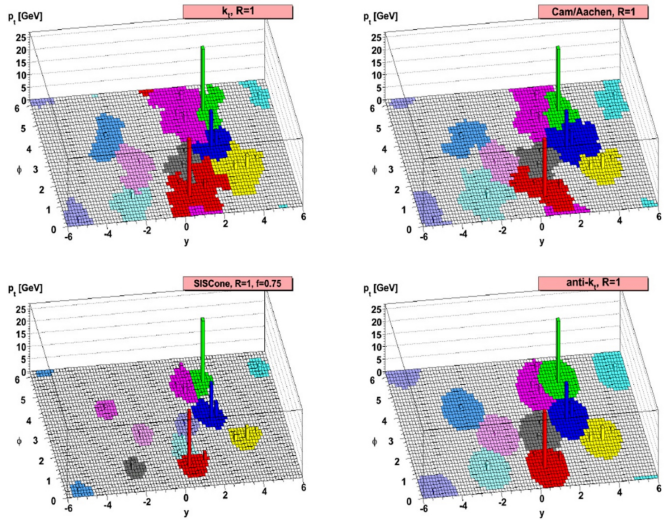

**Figure 5.7:** The three sequential jet-clustering algorithms listed in Table 5.1 and the SIScone algorithm [26] are applied, using the same $R$ parameter value of $1.0$, on a jet-clustering task in an event that features multiple jets. The performance differences among the algorithms are noticeable in terms of the shape, area, and number of jets reconstructed in the event. The C/A and $k_t$ algorithms produce irregular shapes, while the anti-$k_t$ algorithm outputs the most regular circular shapes. Image reproduced from Ref. [27].

The outcome of four jet clustering algorithms can be visualized in Fig. 5.7. Notice how the shape, size, and quantity of jets in the event vary with the algorithm selection, despite all methods receiving the same $R$ parameter value as input. One of the factors to which the anti-$k_t$ algorithm owes its popularity is the circular shape of jets it produces, which simplifies interpretability.

Perhaps the most popular implementation of jet clustering algorithms can be found in the FASTJET package [334].

Jet-calibration corrections are applied to rescale the jet kinematics to account for pile-up contributions, detector response, data-MC discrepancies, and to more closely recover the properties of the jet-initiating particle. At the trigger level, the calibration procedure varies between large ($R = 1.0$) and small ($R = 0.4$) radius jets. All jets are subject to jet energy scale (JES) calibration, which rescales towards unity the jet energy response, *i.e.* the ratio of the reconstructed jet energy to the true amount of deposited energy from simulation, which is highly dependent on jet energy and position in the detector. The correction is derived from simulated dijet samples by matching reconstructed and "truth" jets (jets obtained from running clustering algorithms on truth-level particles produced and catalogued by the simulator), binning their distributions, and computing the per-bin average correction needed to offset the measured-to-true energy discrepancy.

Small-$R$ jets undergo pile-up subtraction, JES calibration, a global sequential correction that modifies jets based on their associated track properties (if available) and on their longitudinal shower shape profile in order to remove flavor-dependent differences. A final, residual, data-driven calibration further removes data-MC differences. Large-$R$ jets are subject to jet trimming [327], during which smaller radius ($R = 0.2$) sub-jets that contribute to less than a fraction $f_\text{cut}$ of the jet $p_T$ are removed from the jet, JES calibration, and mass calibration to more accurately recover the mass of the original particle. Besides trimming, other jet grooming and pile-up mitigation techniques have been developed in order to reduce the contributions of soft radiation and pile-up to the substructure definition of jets [335, 336, 337, 338, 339, 340, 341, 342].

#### 5.2.1.1 Track Jets

Neutral hadrons traverse the tracker without leaving hits and deposit their energy into the hadronic (and in smaller percentage, electromagnetic) calorimeter, while charged hadrons differ due to the presence of tracks in the inner detector. As a consequence, building jet-like objects out of tracks in the inner detector (and associating them to calorimeter jets) can enhance the performance of tagging algorithms, especially in boosted environments, where the decay products become too collimated for standard jet reconstruction and flavor identification algorithms [343]. Track jets are therefore primarily born out of the necessity of incorporating flavor tagging and jet substructure techniques.

Track jets are physics objects developed from the clustering of high quality tracks that are compatible with the primary vertex hypothesis. These track selection criteria are enforced by applying the requirements defined by the loose track selection coupled with the tight track-vertex-association working point with respect to the PV, in order to reduce the fractional contributions of tracks from pile-up vertices.

The formation of track jets relies on the definition of locality provided by the same clustering algorithms used for calorimeter jet clustering. Because of the higher resolution achieved in the tracking layers, track jets can be reconstructed with smaller $R$ parameter than calorimeter jets; values up to $R = 0.4$ have been investigated in ATLAS, with general preference for $R = 0.2$ track jets, due to performance improvements in flavor tagging scenarios. Variable-radius track jets have also become a popular alterna-tive thanks to their flexibility of smoothly adjusting the radius parameter as a function of $p_T$.

A further track-to-jet association step can be used to assign unmatched tracks to track jets, using a $p_T$-dependent cone (see Sec. 6.1.1).

Once formed, track jets can be associated to corresponding jets from calorimetric activity. The method that connects the two types of physics objects in flavor tagging applications is known as *ghost association* [344, 345, 346], and has been found to outperform geometric cone association procedures and reduce ambiguities in the matching phase. Ghost association consists of reclustering calorimeter jets with additional track jet constituents after setting the track jet four-momenta to zero in order not to alter the kinematic properties of the calorimeter jet. To improve the match between calorimeter and track jets and increase the mass resolution, the track jet mass can be rescaled by the ratio of the matched calorimeter jet $p_T$ to the track jet $p_T$, to obtain the so-called track-assisted mass.

Track jets provide independent probes for jet flavor identification by complementing the flavor tagging program with substructure-level tagging abilities within large-$R$ jets in boosted scenarios. Track jets have also been observed to provide higher momentum and angular resolution with respect to the properties of the heavy hadron within their radius. The benefits of track jets include a more natural connection with tracking information used in many low-level flavor tagging algorithms (see Sec. 6.1), while strengthening pile-up independence due to the strong PV origin requirements enforced in the track jet definition [343].

### 5.2.1.2   Jet Clustering and Representation with Machine Learning

While the jet clustering ecosystem offers a variety of IRC-safe algorithms to formally define the boundaries of a jet, one can alternatively use machine learning to learn an optimal representation. The format of this representation has been the subject of much of the recent literature at the intersection of machine learning and jet physics. The jet representation options can be positioned along two main axes: fixed or variable size, and learned or pre-determined representation using physics domain knowledge. A recent review of this rapidly evolving field can also be found in Ref. [347].

5.2.1.2.1   Fuzzy Jets   One possibility is to use gaussian mixture models to describe a multi-jet event as a superposition of gaussian components, each approximately defining a jet in $\eta$-$\phi$ space. This provides the user with an explicit parametric likelihood that can be maximized to learn the optimal values of the parameters that describe the objects in the event. The objects thus defined, known as *Fuzzy Jets* [348], have soft boundaries and are probabilistic in nature, in that, by design, the clustering algorithm returns a probabilistic assignment of each constituent to each jet. In other words, this method adjusts the parameters of the gaussians so that each region of space will be assigned a probability of membership to any gaussian component. The learned parameters, such as the width of the gaussians, which is variable by construction, have discriminative power that can be exploited for downstream jet classification tasks.

5.2.1.2.2 SEQUENCE REPRESENTATION OF JETS  Among the most popular and widely explored variable-length jet representations is that of a sequence of tracks, towers, or other jet constituents, including vertices. This intuitive format describes the jet as an ordered list of objects containing the properties of each jet constituent. Constituents can be ordered by $p_T$, though alternative ordering schemes can be selected for specific tasks based on performance arguments. This jet representation closely resembles the one used to encode sentences in the field of natural language processing (NLP). Therefore, adopting a sequence-based jet representation has the advantage of enabling the use of methods that have been developed and tested for NLP tasks.

The ATLAS collaboration has adopted this jet representation for the application of a subset of its $b$-tagging algorithms, the details of which are the topic of Sec. 6.1 [36, 37].

The absence of a natural ordering scheme to sort constituents within the sequence has led to recent attempts of learning an optimal, task-specific ordering, and to the exploration of machine learning methods that are able to learn from unordered sets [349].

5.2.1.2.3 TREE AND GRAPH REPRESENTATION OF JETS  Another possibility is to make use of the decision history of any iterative clustering algorithm used to define the jet to build a tree structure that represents the jet itself [257]. The sequential combination of energy clusters into larger, more abstract objects creates a natural hierarchy among these entities that can be visualized as a tree whose structure depends on the unique assumptions of each jet clustering algorithm. Each jet has a unique topology based on its clustering history, but the structure of all jets is logically related by the application of the same criteria for iterative assignment. In this case, the jet representation is fixed, and an identical learning unit (in the case of Ref. [257], a GRU) is applied to each node in the tree to recover a final, fixed-length vec-

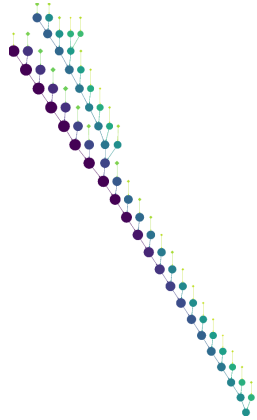
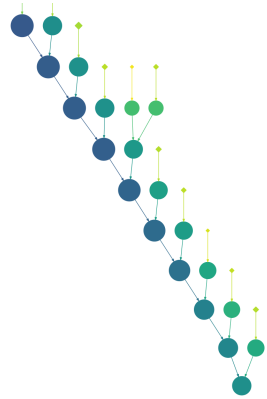

(a) A gluon jet tree representation, where the tree structure represents the clustering history of jet constituents using the anti-$k_t$ algorithm.

(b) A quark jet tree representation, where the tree structure represents the clustering history of jet constituents using the anti-$k_t$ algorithm.

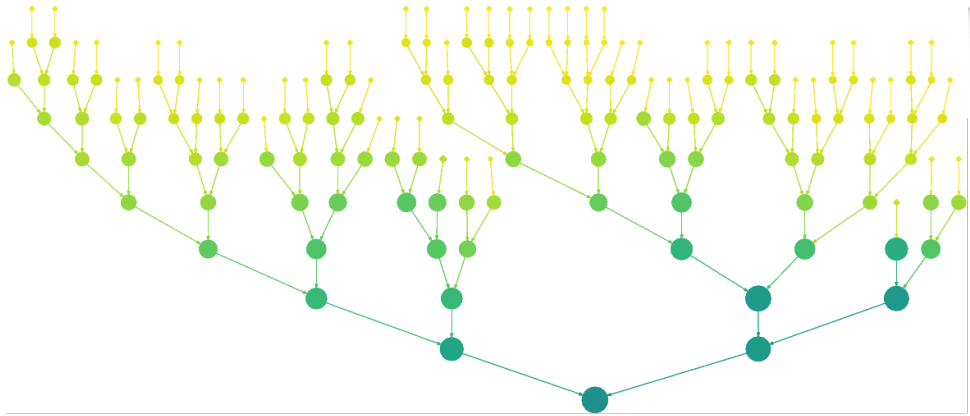

(c) A jet tree representation, where the tree structure represents the clustering history of jet constituents using the $k_t$ algorithm.

**Figure 5.8:** These jet tree representations, reproduced from Ref. [28], highlight the differences in jet topologies determined by the sequential algorithm used for jet clustering ($k_t$ versus anti-$k_t$), as well as the structural differences between quark and gluon initiated jets, which can be exploited for jet classification purposes. Each leaf represents an elementary jet constituent, used as input to the clustering algorithm, and usually described in terms of its four-vector. Each node in the tree represents a clustering step. The root of the tree represents the final jet embedding.

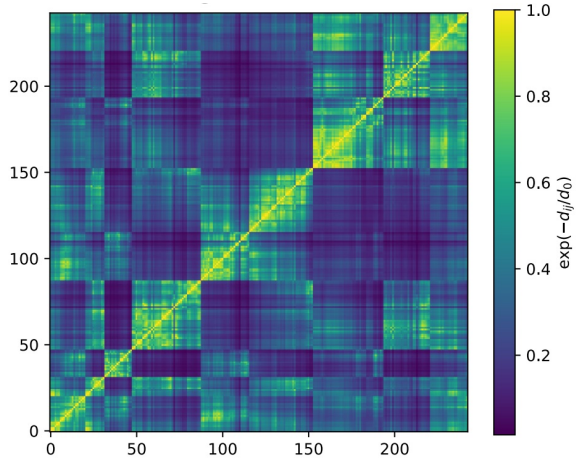

**Figure 5.9:** Adjacency matrix that defines the graph representation of a jet clustered with the C/A algorithm. The $x$ and $y$ axes represent individual jet constituents. The matrix encodes their distances.

torial representation of the jet, to be used downstream by identification or analysis tasks. The advantage of using cells like GRUs lies in the ability of learning a gating mechanism to weight the relative importance of inputs from daughter objects. Other orderings of the proto-particles, such as simple descending $p_T$, were found to be equally competitive. This method has also applied to quark-gluon tagging [28], by exploiting the topological differences between gluon and quark initiated jets, which are evident in Fig. 5.8(a) and 5.8(b), respectively. As a comparison, Fig 5.8(c) shows the topology of a tree defined by the $k_t$ algorithm.

A further evolution of this approach is to encode the jet structure into a physics-aware graph, by defining the adjacency matrix in terms of the $d_{ij}$ distance between each pair of jet constituents [350]. Any distance metric can be used, but selecting one of the $d_{ij}$ formulations from Table 5.1 allows to inject physics domain knowledge into the learning process. Alternatively, a parametric distance metric can be learned automatically to simultaneously yield a new jet clustering algorithm. An example of jet adjacency matrix obtained using the C/A clustering algorithm is presented in Fig. 5.9. Graph convolutional neural networks can then be adopted to learn a series of abstract jet characteristics.

5.2.1.2.4  JET IMAGES   A more popular option is to take the final state energy depositions in the calorimeter, discretize them according to the natural detector granularity, and produce a representation of finite and constant dimensionality. The idea of using a fix-length jet representation with the added benefit of easy visual interpretability is not new [351, 352, 353, 354, 355]. While image representation of jets have

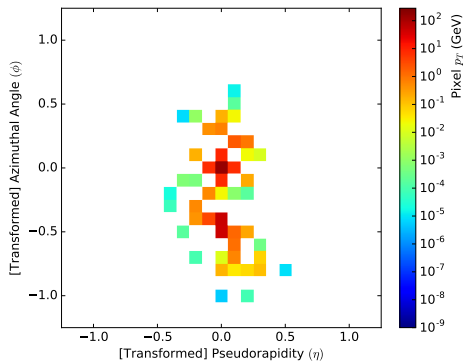
(a) A single jet image.

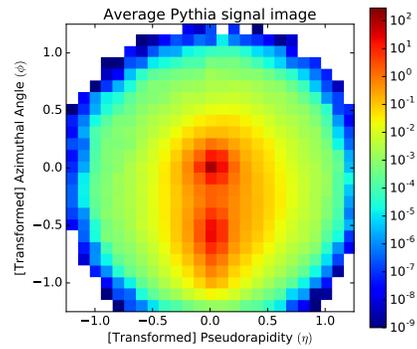
(b) Average of approximately 10,000 jet images.

**Figure 5.10:** Pythia generated jet images from the two-prong decay of a boosted $W$ boson from the publicly available dataset [29] used in Ref. [30, 31].

appeared in the literature in the past, only recently has the use of *jet images* [356] become mainstream as a mean to test computer vision solutions on HEP data. Projecting the calorimeter layers onto a single cylindrical surface and then unrolling it into a two-dimensional plane is a natural choice, given the pixel-like calorimeter segmentation, which allows us to visualize high energy collisions in an analogous way to natural images.

The traditional jet image formulation equates the intensity of each pixel in the image to the sum of the energies deposited by all particles incident to that $\eta$-$\phi$ region of the calorimeter. In other words, jet images are simply two-dimensional histogram projected onto the $x$-$y$ plane (or, in this case, the $\eta$-$\phi$ plane) where the value along the $z$-axis, *i.e.* the height in each bin, corresponds to the energy of the particles that fall in that bin.

A typical boosted $W$ boson jet image from the public dataset available in Ref. [29] is shown in Fig. 5.10, along with the average over approximately 10,000 jet images.

Jet images do not encompass the whole event but rather focus on the region of the detector where a jet is located. Their extent in $\eta$ and $\phi$ is determined by the size of jets selected for the analysis. Since jets may appear with any orientation within the detector but this rotational degeneracy carries no physical information, rotational symmetry is invoked as a pre-processing tool to align all jets along the same axis. The most energetic local maximum is normally centered in the middle of the image. If other local maxima are present (typically identified with sub-jets that define the broader jet's substructure), a rotation convention is adopted, which often consists of aligning the sub-leading sub-jet or dominant principal component either straight above or straight below the leading sub-jet. Finally, a parity transformation can be applied to flip images so that one half of the image is always more energetic. To simplify the image transformations, jet images are often constructed using the transverse momentum $p_{T,\text{ cell}} = E_{\text{cell}}/\cosh(\eta_{\text{cell}})$, instead of the energy, as pixel intensity value, because of its invariance to longitudinal Lorentz boosts (translations along $\eta$). After applying all transformations, images are re-pixelated to match the granularity of the detector using cubic-spline interpolation to redistribute the energy among the cells in the new grid. More procedural details on jet image preprocessing can be found in Ref. [356, 30, 357, 31].

Jet images are inherently sparse images, with only $\approx 10\%$ non-zero pixels in each image. In addition, their unnormalized pixel intensities may span several orders of magnitude, and an RGB channel augmentation technique has yet to be codified as a standard procedure. Finally, the heavy preprocessing removes much of the location invariance of the images, making the exact position of their features very physically meaningful. Therefore, imperceptible variations in the pixel intensities result in jets with significantly altered physical properties. This requires extra care when applying of out-of-the-box traditional machine learning solutions to jet images analysis.

Exploratory work has shown promising results when combining different sources of local information into different channels of the image, such as the charged particles' momentum, the neutral particles' momentum, and the charged particle multiplicity [358], or total and pile-up-subtracted energy depositions [341].

After the first modern revisitation of jet images [356] and the first demonstrations of the power of deep learning techniques in extracting useful information from them [359, 30], the same methods have been recycled for application to several classification tasks, including boosted $W$ boson tagging [356, 30, 360, 361], quark-gluon tagging [358, 40], and top tagging [359, 362, 363]. The successful use of jet images has also been demonstrated for pile-up mitigation [341] and jet simulation [31].

Representing jets in image format has the advantage of making visual inspection of the inner workings of a learning algorithm easier to perform and more intuitive to interpret. Some commonly adopted diagnostics are summarized in Ref. [347].

Novel physics-driven jet representations that can also be visualized in image format include the Lund

Jet Plane [364].

Whole-detector events have also been represented in image format, in order to investigate the power of deep learning techniques to find patterns in high-dimensional sensor-level data, and bypass object reconstruction and identification altogether, in line with the desire of avoiding unnecessary and lossy information reduction steps [365, 366, 367].

### 5.2.2 Photons

Photons are neutral particles, so they cannot be directly tracked in the inner detector. Their typical signature is an electromagnetic energy deposition in the calorimeter caused by their shower evolution. Properties that describe the shower shape and development are used to distinguish photons from other particles. Photon reconstruction is seeded using energy clusters reconstructed in the EM calorimeter using the sliding-window algorithm.

However, between 20-50% of photons (depending on $\eta$) convert to an electron-positron pair within the inner detector [314]. This occurrence is signaled by the presence of a vertex with at least one track pointing towards the calorimeter cluster. Dedicated software tools look for the presence of conversion vertices in the events. Converted photons are distinguished into different types depending on the number of tracks, and the position and quantity of their hits. Improved reconstruction of TRT tracks highly impacts the reconstruction performance for converted photons. Unconverted photons have no such vertex or track, and need to be reconstructed exclusively from energy clusters in the EM calorimeter.

Photon (or electron) candidates, therefore, consist of clusters with potential associated tracks pointing to the corresponding calorimeter region and collections of conversion vertices. Each object's reconstructed four-vector is recovered from the kinematics associated with the calorimeter cluster or with the tracks, in case of conversion. For example, transverse energy $E_T$ can be computed as a function of the cluster energy $E$ and the pseudorapidity $\eta$ of the cluster's barycenter in the second layer of the EM calorimeter.

An ambiguity tool categorizes candidate objects into photons and electrons, or flags them as ambiguous.

Quality selection criteria exclude poorly reconstructed photon candidates, such as the ones that enter the cracks of the calorimeter region, by vetoing the regions $1.37 < |\eta| < 1.52$, as well as those in the high pseudorapidity region $|\eta| > 2.37$. A categorical quality flag is used to decorate objects in the presence of detector malfunctionings.

Isolation working points are designed for a range of physics analyses that require a corresponding range of signal efficiency and fake rejection rates. These are implemented by applying overlap cuts in the calorimeter and/or the tracker region. For example, the tight selection only recovers very high quality photons with clear signatures, and provides the highest background rejection of all working points. Isolation cuts are often expressed as fractional requirements with respect to the transverse energy $E_T$.

Photon identification is driven by the design of MC-optimized identification strategies based on shower shape variables that describe the longitudinal and transverse geometry of the EM shower. This strategy is primarily limited by the ability to accurately describe shower evolution in simulation. Fudge factors can be applied to remedy the data-MC disagreement in shower shapes description [368]. Cut optimization is performed separately for converted and unconverted photons, to account for different width profiles, and in pseudorapidity bins, to account for different detector material depth.

Recent improvements in ATLAS reconstruction algorithms have shown the benefits of adopting topological cell clustering techniques [25] with variable radius clusters for the reconstruction of electromagnetically interacting objects, especially in specific topologies such as photon conversion and low-energy photon bremsstrahlung from an electron that may benefit from the formation of shared superclusters [369]. Topoclusters that are suitable for the reconstruction of EM showers are identified by a cut on the fraction of cluster energy deposited in the EM calorimeter layers. Dynamic resizing of clusters provides superior flexibility, compared to fixed-size clusters reconstructed by the sliding-window algorithm, to connect and combine logically related nearby sub-clusters, in a way that reflects the natural fluctuations in shower size as it develops along the depth of the detector. Jointly reconstructing primary seed clusters along with the radiative portions coming from in-detector bremsstrahlung or conversion bridges the gap between reconstructed and true deposited energy from objects with common origin, thus improving the total energy resolution.

### 5.2.3 Electrons

$e^{\pm}$ reconstruction and identification is crucial in order to investigate events with leptonic decays with high efficiencies while reducing to a minimum the trigger rates for these types of signatures.

The typical measurement of the presence of electrons involves tracking in the ID as well as EM calorimetry. In addition, signal from the hadronic calorimeter can support electron identification through the measurement of the tails of the shower distribution, and through veto of electron-like signatures with high associated hadronic activity.

Calorimeter clusters are used as starting points for electron reconstruction. Electrons, however, also deposit 30-60% of their energy in the inner detector [314]. Tracking is therefore vital for improving the resolution of track parameters used to match tracks to calorimeter clusters and to identify electrons. Better description of electron radiative losses by bremsstrahlung is provided by substituting traditional Kalman Filters with Gaussian Sum Filters for non-linear track fitting, which characterize noisy trajectories as sums of gaussian components [312]. Tracks are extrapolated to the central layer of the EM calorimeter to be matched to nearby clusters. Crucial information for the identification of electrons, especially at low energies, is provided by the TRT [314].

The combined object's kinematics are extracted from the energy measured in the EM calorimeter clustered, rescaled by calibration factors [370]. The precise $\eta$-$\phi$ direction is obtained from tracking and vertexing information.

Central electron selection criteria exclude objects with $|\eta| > 2.47$ and $p_{T,\text{ cluster}} < 4.5$ GeV, and candidates falling in the calorimeter cracks. Forward electrons lack tracking information, so they are entirely reconstructed from topological clusters with high transverse energy and hadronic veto.

The performance of electron reconstruction algorithms is evaluated using tag-and-probe methods in $W$, $Z$, and $J/\psi$ decays [32].

Electrons and positrons from photon conversion often need to be separated from those originating at the PV. PV consistency is measured in terms of the track impact parameters.

Calorimetric shower shape variables have been designed to facilitate electron identification. For example, $e^{\pm}$ can be distinguished from charged pions based on their shower depth and width, with pions being more penetrating and characterized by a larger lateral development. The fraction of energy deposited by each type of shower in each of the sublayers of a longitudinally segmented EM calorimeter, as well as the lateral width of the shower measured by a laterally segmented calorimeter, are both indicative of the nature of particle that have rise to the shower. The ratio of their energy $E$ deposited in the EM calorimeter to their momentum $p$ registered in the tracker is another discriminating feature for classification, with $e^{\pm}$ depositing nearly all of its energy in the EM calorimeter, as opposed to pions.

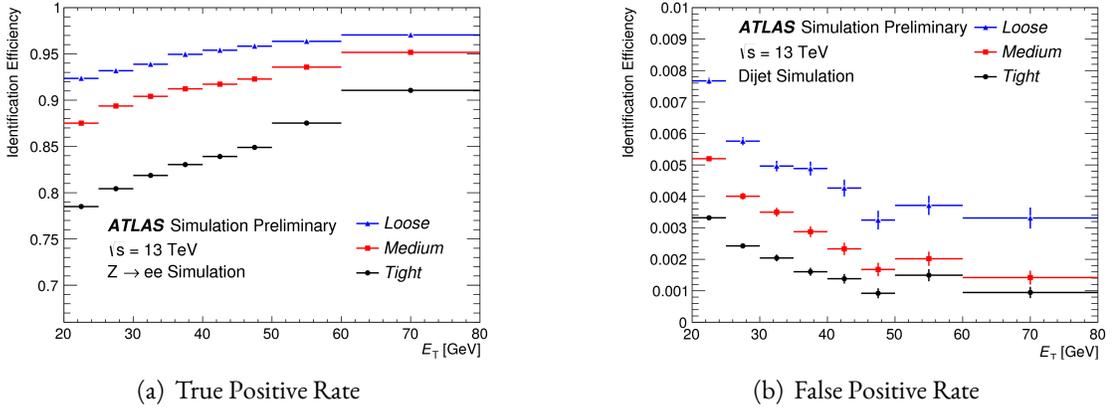

(a) True Positive Rate

(b) False Positive Rate

**Figure 5.11:** Performance of the likelihood-based electron identification algorithm on simulated $Z \to ee$ signal 5.11(a) and dijet background 5.11(b) events at 13 TeV, expressed in terms of the percentage of reconstructed electrons that pass the three electron identification criteria (*Loose, Medium, Tight*). Reconstructed electrons matched to true electrons in the background samples are removed. The Images reproduced from Ref. [32].

Shower shape variables computed in the EM calorimeter can be combined with hadronic information, hit location and count, track quality requirements, and track impact parameter quantities to tighten the selection purity.

Electron identification in Run II takes place via the construction of a naive likelihood-based discriminant that improves the Run I sequential cut-based approach [313]. The signal efficiency (TPR) and background efficiency (FPR) for Run II electron identification from hadronic background using the likelihood-based method are displayed in Fig. 5.11(a) and 5.11(b), respectively, in ATLAS simulated data at 13 TeV, for three electron identification operating points [32].

A new addition to the electron identification stack, known as Ringer Algorithm [371], has been integrated into the HLT trigger for faster, more powerful background rejection [372]. It consists of constructing concentric ring-shaped filters, applying them to calorimeter cells moving from the cluster barycenter outward, summing the transverse energy among all cells in the ring, and using them as discriminative features to feed into a multilayer perceptron.

Electron identification suffers from misclassification issues in complex topologies. Isolation requirements are enforced to remove topologies with high calorimeter activity in the surroundings of the electron, often characteristic of heavy-flavor decays. A variety of fixed-efficiency or fixed-cut working points have been designed to provide varying degrees of isolation at both the calorimeter and tracker level, often using $p_T$-dependent cone aperture definitions.

5.2.4  ELECTRON AND PHOTON IDENTIFICATION WITH MACHINE LEARNING

Both the sliding-window and ringer algorithms introduce the concept of investigating the local structure of energy depositions in sub-regions of the calorimeter to extract patterns that help distinguish electron and photon initiated showers. Similarly, the likelihood-based approach to electron identification makes use of input variables defined as the ratios of transverse energies recorded over different-size regions of the EM calorimeter grid, in both the transverse and longitudinal directions.

These hand-designed functions of energy depositions over nearby cells are just a tiny subset of the functions one could compute with a convolutional neural network. In fact, convolutional filters of different sizes, combined with the non-linearities of activation functions, allow the learning of arbitrary functions of the input pixel space while encoding the local information contained over regions in a grid-like topology. In an effort to generalize and modernize the current approaches to $e$-$\gamma$ identification, we have explored the use of families of convolutional networks, and investigate the performance gains by highlighting the particular regions of phase-space in which shower-shape-based classification approaches fail but convolutional neural networks succeed [33]. The aim of this work is to set competitive benchmarks on a shared dataset [373] towards the improvement of classification and regression methods for electromagnetic shower reconstruction and identification, and to justify the investigation of deep learning techniques in the `egamma` community, as natural extensions to the suite of currently deployed algorithms.

Using the dataset described in detail in Sec. 8.2.1 and publicly available in Ref. [373], the GEANT4 simulation package is used to shoot single $e^+$, $\gamma$, and $\pi^+$ particles at a slab of liquid argon interleaved with layers of lead, with longitudinal and lateral segmentation that approximate those of the ATLAS EM calorimeter. Upon interaction with the medium, the particles develop showers along their trajectories.

The task of inferring the type of incident particle from the nature of the energy depositions along the shower is investigated with a series of machine learning methods. First, a set of 20 traditional shower shapes, among those adopted by ATLAS for shower reconstruction and object identification, are computed and used as inputs to a fully connected network (FCN), to roughly approximate the performance of the likelihood-based baseline. Then, a selection of neural network architectures that operate directly on the pixel space, without the need to reconstruct engineered shower quantities, are compared to the benchmark. These include: a FCN that ingests the flattened and concatenated 1D array of pixel (or, more properly, voxel) intensities, and three different multi-stream convolutional architectures (a locally-connected network, a traditional convolutional network, and a DenseNet) that treat energy deposits across each of the longitudinal calorimeter layers as an image in 2D pixel space. In depth description of the network design and training parameters obtained after a scan over sensible options is provided in Ref. [33][*].

Convolutional networks present several advantages beyond the automatic processing of local features. Through the desirable property of parameter sharing, convolutional networks pose lower memory burdens because of their ability to "recycle" parameters in multiple locations. As a further benefit, they exhibit translational invariance to the location of features across input vectors, making them more robust to shifts in the shower configuration. On the other hand, locally-connected layers abandon location-invariant filters in favor of purposefully learned location-specific filters, which may be competitive in analyzing images that, given the physical setup, are centered and scaled to a common level across the entire dataset. Finally, densely connected convolutional networks (DenseNets) simplify the training of deep convolutional networks by devising smart connectivity patterns among layers that generalize the concept of skip-connections in ResNets, thus increasing the number of viable paths for efficient information forward flow and back propagation.

The methods are independently trained and evaluated on two classification tasks, *i.e.* the binary discrimination of $e^+$ from $\gamma$, and $e^+$ from $\pi^+$. In the former, more complex task, the methods are tasked with identifying differences in the electromagnetic showers produced by positrons and photons in a EM calorimeter, without making use of tracking information. The physical processes that determine the phenomenology of shower formation in calorimeter detectors are explained in Sec. 2.2.1.2. As noted in Sec.5.2.2 and 5.2.3, showers from $e$ and $\gamma$ leave very similar signatures in the calorimeter layers, making the separation between the two classes based on calorimetric information only extremely difficult. On the other hand, the second classification task aims at distinguishing electron showers from those generated by hadrons such as charged pions. In this case, based on topological considerations, higher rejection values are easily attainable, but, because of the high class imbalance that favors hadronic background

---

[*]Code is available at https://github.com/hep-lbdl/CaloID

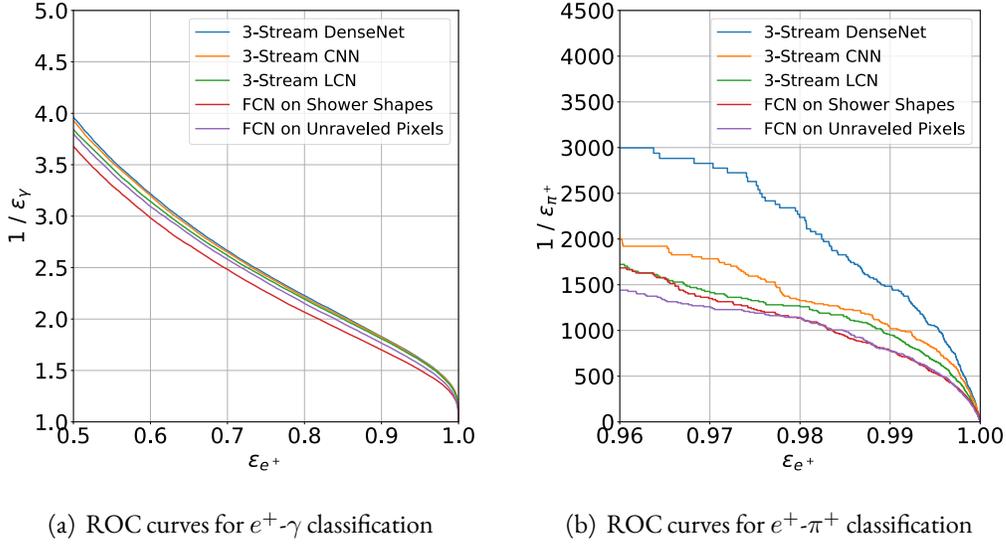

(a) ROC curves for $e^+$-$\gamma$ classification

(b) ROC curves for $e^+$-$\pi^+$ classification

**Figure 5.12:** Performance of various binary classifiers on the $e^+$-$\gamma$ and $e^+$-$\pi^+$ classification tasks defined in Ref. [33]. The three-stream DenseNet (in blue) shows higher background rejections across a broad range of meaningful, task-specific signal efficiencies. Images reproduced from Ref. [33].

events, even percent-level improvements result in significant gains when methods are employed in realistic applications.

Classification performance results are displayed using ROC curves in Fig. 5.12. In addition, Tables 5.2 and 5.3 report the relative background rejection improvements provided by the various pixel-based methods over the shower shapes baseline. These results establish the superiority, for the tasks at hand, of the DenseNet architecture designed in this study. The DenseNet outperforms other methods across all meaningful efficiency ranges that are relevant for the two tasks under investigation, while also providing higher parameter efficiency than the alternative architectures tested in this work. The rejection gains are more moderate in the $e$-$\gamma$ task because of its intrinsic complexity and because of the lack of full separability between the two classes. The $e$-$\pi$ results show more dramatic performance gains, with over 100% improvement at selected working points. It is in this latter task more so than in the former that the DenseNet is able to show impressive performance, with 3x to 10x the improvement of the convolutional network. At the same time, the experiments performed in this study remind us that applying simple, feed forward networks on the array of per-cell energies often results in similar performance levels to likelihood methods that make use of well-understood, interpretable, physics-driven features.

In addition to establishing benchmarks for object identification, we demonstrate that shower kine-matic information can be extracted by regressing the quantities of interest directly from the low-level detector representation explored in the previous task, without the necessity for clustering algorithms. A FCN on the raveled voxel intensities, similar to the one employed in the classification tasks above, is successfully trained to deduce the shower's position of incidence $(x_0, y_0)$ at the surface of the calorimeter, and its direction expressed in terms of the polar and azimuthal angles $(\theta_C, \phi_C)$. The experiments confirm that these shower properties can be recovered even from a simplified 1D array of energy deposits per volumetric unit. The performance measurements obtained in this work are displayed (for photon showers) in Fig. 5.13, which highlights the mapping between true and reconstructed shower properties, and in Table 5.4, which provides numerical benchmarks for these regression tasks in terms of mean absolute error (MAE) and root mean squared error (RMSE).

**Table 5.2:** Percentage relative increase or decrease in $\gamma$ rejection at five different $e^+$ efficiency working points compared to the baseline fully-connected network trained on shower shape variables.

|  |  | \multicolumn{5}{c}{$e^+$ efficiency} |
|---|---|---|---|---|---|---|
|  |  | 60% | 70% | 80% | 90% | 99% |
| Model | FCN on shower shapes | - | - | - | - | - |
|  | FCN on unraveled pixels | +3.7% | +3.9% | +4.1% | +3.9% | +2.1% |
|  | 3-Stream Locally-Connected | +5.3% | +5.4% | +6.1% | +6.4% | +5.2% |
|  | 3-Stream Conv Net | +5.5% | +6.8% | +6.9% | +7.3% | +6.0% |
|  | 3-Stream DenseNet | **+7.5%** | **+7.4%** | **+7.7%** | **+7.6%** | **+6.4%** |

**Table 5.3:** Percentage relative increase or decrease in $\pi^+$ rejection at five different $e^+$ efficiency working points compared to the baseline fully-connected network trained on shower shape variables.

|  |  | \multicolumn{5}{c}{$e^+$ efficiency} |
|---|---|---|---|---|---|---|
|  |  | 96% | 97% | 98% | 99% | 99.99% |
| Model | FCN on shower shapes | - | - | - | - | - |
|  | FCN on unraveled pixels | −14.4% | −7.6% | +0.76% | +0.0% | −34.6% |
|  | 3-Stream Locally-Connected | +2.3% | +4.8% | +11.9% | +22.3% | −43.7% |
|  | 3-Stream Conv Net | +20.3% | +31.0% | +17.9% | +32.4% | −6.8% |
|  | 3-Stream DenseNet | **+81.6%** | **+107.5%** | **+100.0%** | **+90.1%** | **+34.9%** |

Table 5.4: Regression evaluation metrics on a test set of 200,000k unseen $\gamma$ showers.

| Variable (units) | MAE | RMSE |
|---|---|---|
| $\phi_C$ (rad) | 0.024 | 0.030 |
| $\theta_C$ (rad) | 0.026 | 0.032 |
| $x_0$ (mm) | 6.162 | 7.876 |
| $y_0$ (mm) | 2.959 | 4.221 |

Since shower simulation tasks involving generative networks (see Sec. 8) often make use of additional loss terms to enforce constraints on conditional shower attributes, such as energy, position, and direction, it is important to demonstrate that even a simple FCN is capable of reconstructing these shower properties from the fixed-length shower representation adopted here and in later generative studies in Sec. 8.2. These regression results verify the plausibility of designing conditional generative models in which the discriminator is tasked to enforce physical constraints.

The promise of these preliminary studies will have to be confirmed in the presence of fully simulated detector effects and pile-up interactions within the context of an experiment, but, though under simplified conditions, the results clearly add to the evidence in favor of the current trend of replacing manual information summarization with automated machine learning systems that operate on high-dimensional, low-level detector information. Although the dataset lacks fidelity and realism, the performance gains warrant examination of similar machine learning methods by the LHC collaborations.

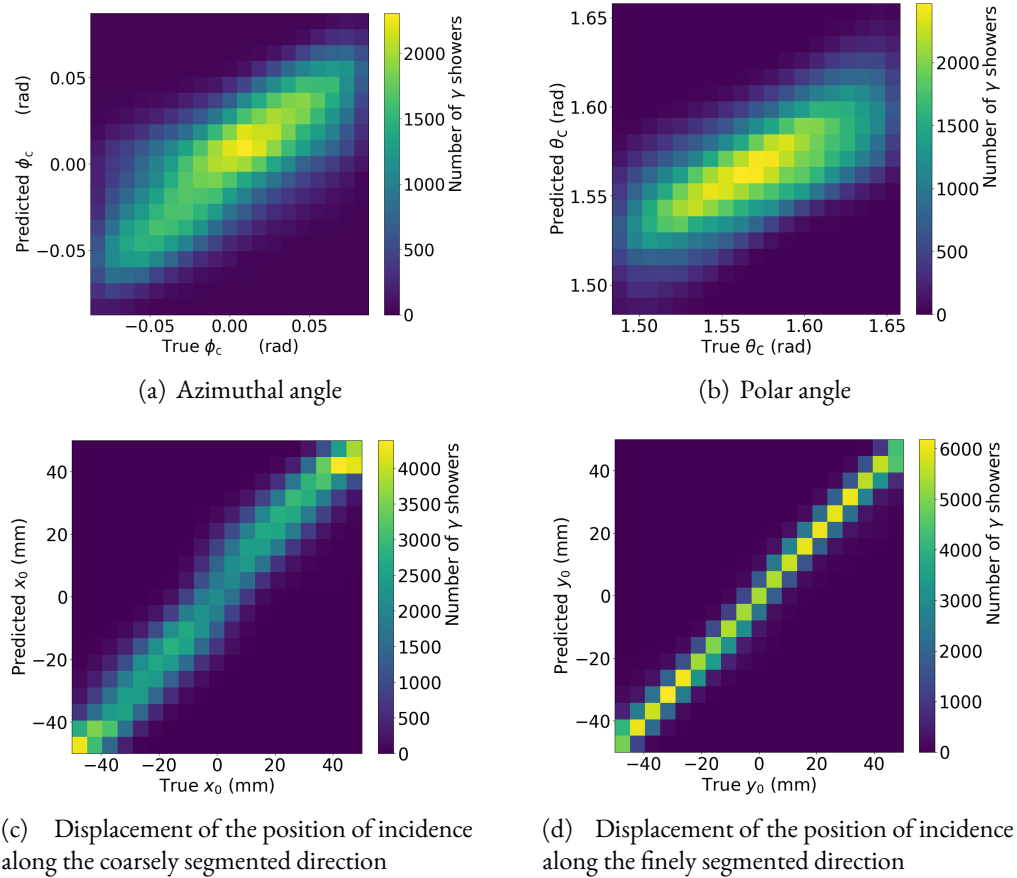

**Figure 5.13:** Regression results, evaluated on a test set of 200,000k unseen $\gamma$ showers, for the four variables that define the incident direction and position of photons upon contact with the first layer of the electromagnetic calorimeter. Images reproduced from Ref. [33].

5.2.5   Muons

Muons are important flags for electro-weak processes at the LHC, so much so that the entire muon spectrometer (MS) described in Sec. 2.2.1.3 has been dedicated to their detection and accurate property measurement. Information from the inner detectors (ID) and calorimeters contribute to the reconstruction and identification of muons.

The dedicated muon trigger chambers in the muon spectrometer identify interesting events in which muons are present withing the $\sim 2.5\mu s$ latency requirements of the L1-trigger. Muons may appear as final state particles in a broad range of events at the LHC, each with distinctive event features. For instance, reconstructing soft muons in the busy environment defined by jets is important to correctly account for the total energy of heavy-flavor jets with leptonic decays. Conversely, higher momentum muons produced in boson decays require a clean signature, which includes the passing of isolation requirements.

Energy deposits in the ID and in the MS are used for tracking purposes to reconstruct a clean trajectory, and are complemented by relatively smaller energy deposits in the calorimeter. Unlike other particles, muons often reach the muon detector, making their identification and classification a cleaner task than usual. The track reconstruction in the ID and MS are usually carried out independently before being combined.

In ATLAS, muons are divided into *calo-tagged muons*, if they leave traces in both the ID and the calorimeter, *combined muons*, if they leave traces in both the ID and the MS, *segmented muons*, if only one segment of their track is recovered in the MS in addition to the ID track, and *standalone* or *extrapolated muons*, if they are only detected in the MS. The definition of these four categories of muons is represented in the diagram in Fig. 5.14.

In the MS, track segments are built from connected hits. Pattern finding algorithms are used to search for aligned hit patterns in the individual muon chambers and form segments. These are first fitted with a straight line approximation. Then, a second fitting routine is called to connect track segments into track candidates across different detector layers. The candidates are seeded using high-quality segments from the middle layers, after which a combinatorial procedure considers segments from outer and inner layers in order of segment quality. Tracks are then refined by removing shared segments when possible, reject-ing candidates with high $\chi^2$, or refitting candidates after the removal or addition of hits. This information is then combined with measurements from the calorimeters and inner detector; at this stage, muons are categorized according to the schema above, depending on the availability of sub-detector information for combination. Quality cuts are enforced on the $\chi^2$ of the combined track and on the number of hits and holes in various tracking layers.

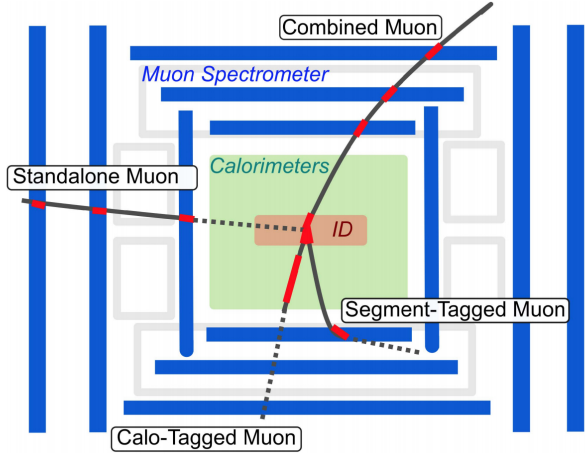

**Figure 5.14:** Type of muons reconstructed in ATLAS, based on the origin of the information used by the reconstruction algorithms.

The latest available muon identification and reconstruction performance studies in ATLAS are available in Ref. [374]. Identification proceeds by applying a series of cuts to efficiently separate prompt muons from background signatures, primarily caused by pion and kaon decays. Discriminating features are constructed to make use of the knowledge that, in background events, track measurements in the ID and MS are not expected to be compatible. For example, the $q/p$ significance $\text{sig}_{q/p} = \frac{|q/p_{\text{ ID}} - q/p_{\text{ MS}}|}{\sqrt{\sigma^2_{q/p\text{ ID}} + \sigma^2_{q/p\text{ MS}}}}$ is defined as the ratio of the absolute difference between the charge-to-momentum ratio measured in the ID and in MS to the square root of the sum of squared uncertainties, and $\rho' = \frac{|p_{T,\text{ ID}} - p_{T,\text{ MS}}|}{p_{T,\text{ combined}}}$ is defined as the ratio of the absolute difference between the track $p_T$ measured in the ID and in MS to the combined track $p_T$.

Three muon identification categories (*Loose*, *Medium*, *Tight*) are defined based on muon types and cuts on the features described above, in addition to a *high-$p_T$* category designed to explicitly improve the resolution of muons above 100 GeV.

Reconstruction efficiency is measured with a tag-and-probe method in $J/\psi \to \mu^+\mu^-$ (low $p_T$)

and $Z \to \mu^+\mu^-$ (high $p_T$) events. Efficiency scale factors are calculated to account for the difference in measured efficiency in data versus Monte Carlo [374].

Semi-muonic decays of heavy hadrons in jets result in unaccounted momentum that should be combined with that of the jet to more accurately resolve the kinematic properties of the initiating hadron. Analyses and performance groups independently apply corrections to the jet collections to account for the effects caused by these decays. These procedure include a muon finding step, in which muons that overlap with the jet (or with a smaller radius jet within a large jet) are selected, and a correction step, in which the four-momenta of all selected muons (or the highest ranked muon in terms of $p_T$ or other criteria) are added back to the jet four-momentum. Muon-in-jet corrections are observed to improve the mass resolution of systems that rely on the precise measurement of heavy flavor jet kinematics.

If the goal is *not* that of finding muons embedded within jet cones to correct the jet kinematics to account for semi-leptonic decays, then muon isolation requirements can be enforced to increase the purity of the muon selection. These can be track-based or calorimeter-based isolation cuts, in which ambient activity contributions to the transverse momentum within a cone around the muon trajectory are expressed as fractional quantities with respect to the transverse momentum of the muon.

### 5.2.6 Taus

Tau leptons, unlike the other charged leptons in the SM, have a large mass (1.777 GeV) and an extraordinarily short mean lifetime ($\mathcal{O}(10^{-13})$ seconds, which corresponds to a decay length of only 87 $\mu$m) [375]. They, therefore, decay before reaching the ATLAS detector, and are only detectable through their decay products. Taus decay hadronically approximately 65% of the time, or leptonically, into lighter leptons. Experimentally, leptonic tau decays are hard to distinguish from the production of prompt electron and muons, while the tau identification efforts focused on the hadronic decay modes face large background contributions from the many hadronic processes at the LHC. 68% of hadronic decays produce a neutral pion in association with other decay products; 72% of times the hadronic decay is single-pronged (containing one charged pion), and 22% of times it is triple-pronged (containing three charged pions). All decay modes include a fraction of invisible energy carried by neutrinos.

Hadronic tau decays manifest themselves in the ATLAS detector as hadronic showers associated to tracks in the ID. Jets that represent tau candidates are traditionally reconstructed from topoclusters with the anti-$k_t$ algorithm with radius parameter $R = 0.4$. The baseline tau selection rejects candidates that do not pass the following criteria: $p_T > 20$ GeV, $|\eta| < 2.5$ (also excluding the $1.37 < |\eta| < 1.52$ region), exactly 1 or 3 associated tracks.

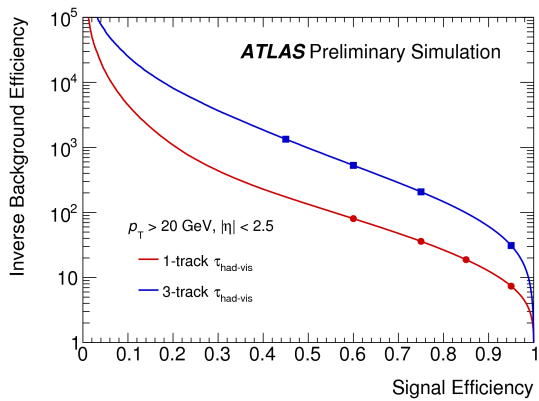

**Figure 5.15:** Performance of the boosted decision trees trained to distinguish hadronic tau decays from QCD jets, with the four selected working points (*Very Loose*, *Loose*, *Medium* and *Tight*) marked on each curve. One BDT is optimized for 1-prong decays (red), the other for 3-prong decays (blue). Signal and background events are required to be reconstructed with exactly 1 or 3 tracks. The signal events are extracted from a simulated $\gamma^* \rightarrow \tau\tau$ sample, the background events from a simulated multi-jet sample.

A set of discriminating features listed in Ref. [376] are designed as inputs to machine learning methods or cut-based selections to distinguish hadronic tau decays from QCD jets and electrons; these include calorimetric shower shape observables, track properties, and secondary vertex information. The secondary vertex associated to the tau decay vertex is selected as the one with the largest sum of momenta from high quality tracks found within $\Delta R = 0.2$ of the jet axis. Quality, isolation, and vertex-matching criteria described in Ref. [376] are imposed on the tracks used to reconstruct the tau decay. The classifier used for tau identification against dijet events from quark and gluon jets is a boosted decision tree (BDT), with two separate instances trained to target single- and triple-prong decays. Fig. 5.15 shows the ROC curves that quantify the performance of the two BDTs. Four working points (*Very Loose*, *Loose*, *Medium* and *Tight*) are independently defined for different fixed values of the true positive rate for the two BDTs. These correspond to efficiency values of 0.95, 0.75, 0.6, 0.45 for the 3-track BDT, and 0.95, 0.85, 0.75, 0.6 for the 1-track BDT.

Tau jets are also tagged as part of the flavor tagging effort discussed in Sec. 6.1. In addition, dedicated algorithms that make use of jet substructure properties have been investigated to improve the ATLAS tau identification capabilities [377].

## 5.2.7 Missing Transverse Energy

Because of conservation of momentum, the sum of momenta in the plane transverse to the beam axis should be identical before and after the collision. This implies that, if no net primordial transverse component is present in the colliding partons, the outgoing transverse components should cancel out. Any residual momentum in the transverse direction is interpreted as being complementary to an equal and opposite momentum vector carried by undetected particles, such as weakly interacting neutrinos or potential new particles such as dark matter candidates. On the other hand, the missing transverse energy signal could also be mimicked by lack of hermeticity in the detector, unaccounted for reconstruction inefficiencies, and other systematic factors.

The missing transverse energy in the $x$ and $y$ directions is calculated as the opposite of the corresponding projections of the vectorial sum of all transverse momenta of all reconstructed and fully calibrated physics objects and additional soft radiation:

$$E^{\text{miss}}_{x(y)} = -\sum_{i \in \{\text{hard objects}\}} p_{x(y),\, i} - \sum_{j \in \{\text{soft signals}\}} p_{x(y),\, j} \tag{5.1}$$

An ambiguity resolution method is required to avoid double counting (for example, of photons also reconstructed as jets).

The performance of ATLAS missing transverse momentum ($E^{\text{miss}}_T$) algorithms in 13 TeV data is available in Ref. [378].

# 6
# Jet Tagging

With the abundance of jets produced at hadron colliders such as the LHC, one of the main efforts within the ATLAS collaboration is to continue improving the performance of jet classification algorithms, in order to recover the nature of the elementary particles that initiate the cascade of nearby charged and neutral particles identified as a jet. *Tagging* refers to the task of labeling jets as originating from a specific type of particle, and distinguishing them from jets with similar signatures. From the machine learning standpoint, this amounts to a series of classification problems. Instead of taking a holistic approach to jet tagging, various separate working groups within ATLAS coordinate the development of specialized algorithms that focus on individual signals and topologies.

Jet tagging has been the proving ground for some of the more innovative machine learning techniques deployed in experiments like ATLAS and CMS, and has given rise to a young, thriving community working on the exploration and implementation of state-of-the-art learning algorithms. This is in part due to the complexity of the jet tagging discrimination tasks, the potential for improvement upon traditional algorithms, the possibility for positive impact towards the physics program, and the need for automation in light of the future increases in data rates and detector redesigns that will directly affect jet tagging.

This chapter will detail available ATLAS flavor tagging techniques (Sec. 6.1), and briefly touch on other developments in top tagging (Sec. 6.2), boson tagging (Sec. 6.3), and quark-gluon tagging (Sec. 6.4) with particular attention towards machine learning solutions.

## 6.1 Flavor Tagging

Flavor tagging amounts to identifying the flavor of the quark from which a jet originates. Within the set of particle recognition tasks at the LHC, $b$-tagging, $i.e.$ the separation of jets containing $b$-hadrons from jets initiated by lighter quark flavors, is of the utmost importance and is a fundamental tool for the ATLAS physics program. This is due to the abundance of physics processes of interest with $b$-hadrons in their final state, versus the overpowering quantity of background light-flavored jets to be rejected. While exotic new physics, hadronic top decays, and the primary decay mode for a SM Higgs boson with $m_H \sim 125$ GeV all involve the presence of $b$-jets, hadron colliders such as the LHC also produce copious amounts of background gluon, $c$, and light ($u$, $d$, and $s$ – often referred to as QCD) jets that can be mistakenly identified as $b$-jets.

Recasting it as a classification problem, $b$-tagging consists of learning a high-dimensional decision boundary among jets of different flavor, using as inputs the trajectories of charged particles in the inner detector and the energy deposits in the calorimeter. Ground truth labels are assigned based on a geometric match between reconstructed jets and simulated hadrons in Monte Carlo simulated events, according to the following logic:

```
if (ΔR(jet, B_hadron) < 0.3) & (p_{T_B hadron} > 5 GeV) then
    jet flavor ← b
else if (ΔR(jet, C_hadron) < 0.3) & (p_{T_C hadron} > 5 GeV) then
    jet flavor ← c
else if (ΔR(jet, τ_lepton) < 0.3) & (p_{T_τ lepton} > 5 GeV) then
    jet flavor ← τ
else
    jet flavor ← light
end if
```

The extended flavor labeling scheme recently introduced in ATLAS also includes double-$b$, double-$c$, and $bc$ topologies as separate classes. With $c$-jet identification having always been an afterthought, and $\tau$ leptons being more of a nuisance than a target for flavor tagging, the primary task of interest is the discrimination of $b$-jets from the light jet background. No ground truth labels are available for collected data events.

Improving $b$-tagging algorithms corresponds to increasing the number of rejected background jets at a given $b$-jet efficiency, or, in other words, to reducing the false positive rate (FPR) at a given true positive rate (TPR).

Many physical properties of $b$-quark fragmentation render the topology of $b$-jets unique. The $b$-quark hadronizes to form a $b$-hadron, and much of the $b$-quark's momentum is transfered to the hadron. Multiple diagrams containing intermediate virtual $W$ bosons contribute to the total decay rate of $b$-hadrons [2]. The decay chain of heavy hadrons usually proceeds from a $b$-hadron primarily through a $c$-hadron to a strange or lighter hadron, etc., following the heavy-flavor quark's decay chain governed by the CKM matrix (see Sec 1.1.2.1.2). Both hadronic and semi-leptonic decays are possible. The heavy-flavor hadrons have relatively long lifetimes of $\approx$ 1.5 ps that allow them to traverse $\mathcal{O}(\text{mm})$ distances within the detector before decaying. The impact parameter can be calculated as $\beta\gamma c\tau \cdot \sin(\alpha)$ for relativistic factors $\beta$ and $\gamma$, average lifetime $\tau$ and opening angle $\alpha$. The magnitude of this displacement provides a strong signature for heavy flavor jets, and makes it viable for detection and reconstruction in the ATLAS ID, yielding events characterized by one or more vertices displaced from the collision point.

The algorithms used to discriminate between $b$-, $c$- and light-flavor jets are called *taggers*. This section describes the suite of taggers currently in use in the ATLAS experiment at the LHC, which includes five low-level taggers (IP3D in Sec. 6.1.4, RNNIP in Sec. 6.1.5, SMT in Sec. 6.1.6, SV1 in Sec. 6.1.2, and JetFitter in Sec. 6.1.3), and two high-level taggers (MV2 in Sec. 6.1.7, and DL1 in Sec. 6.1.8).

The jet collection used for $b$-tagging is reconstructed using the anti-$k_t$ algorithm [332] with radius $R = 0.4$ (`AntiKt4EMTopoJets`), and the following set of selection criteria is imposed [37]:

- $p_\text{T}^\text{jet} > 20$ GeV
- $\left|\eta^\text{jet}\right| > 2.5$
- if $\left|\eta^\text{jet}\right| < 2.4$ and $p_\text{T}^\text{jet} < 60$ GeV: JVT$^\text{jet}$ > 0.59

where JVT is the output of the Jet Vertex Tagger algorithm [379] used to suppress jets from pile-up interactions. Jets also need to satisfy the overlap removal requirement designed to discard leptons reconstructed as jets: jets within a radius $R = 0.3$ of electrons and hard muons are removed.

Other jet collections are currently in use for the training and calibration of $b$-tagging algorithms, including track jets clustered with the anti-$k_t$ algorithm with radius $R = 0.2$ (`AntiKt2PV0TrackJets`), variable-radius track jets clustered with anti-$k_t$ (`AntiKtVR30Rmax4Rmin02TrackJets`), and particle-flow jets [380] clustered with anti-$k_t$ and radius $R = 0.4$ (`AntiKt4EMPFlowJets`). Track jet $b$-tagging is discussed more in depth in Sec. 6.1.9.

Sec. 6.1.1 elaborates on the selection criteria and jet association strategies applied to tracks used for flavor tagging.

### 6.1.1 Track Selection and Track-to-Jet Association

A $b$-induced jet contains tracks originating from the $b$-decay, as well as other fragmentation tracks.

After a jet-clustering algorithm has constructed the jet, another tool modifies it by adding in all tracks compatible with an angular separation requirement. Tracks are associated to the nearest jet based on a maximum allowed $\Delta R(\text{track, jet})$ distance, which decreases as a function of the jet $p_T$, to account for the higher collimation of decay products in boosted scenarios. The maximum spatial separation is parametrized as a decaying exponential function of the jet transverse momentum:

$$\Delta R(p_T) = a + e^{b+c \cdot p_T/\text{GeV}}, \tag{6.1}$$

with fit parameters $[0.239, -1.22, -0.0164]$ initially found in Run I [381]. This is equivalent to a $\Delta R$ value of 0.45 for a 20 GeV jet, and approximately 0.239 for a $\gtrsim$ 2 TeV jet. The constant term defines the value at which the function plateaus at high $p_T$.

The optimality of this functional form and parameter values has been reevaluated for Run II using `mc15` samples at 13 TeV. These $t\bar{t}$ events are simulated using POWHEG+PYTHIA. The objective of this study is to check the validity of the Run I fit for calorimeter jets, and derive a new one for the track jet collection. The ideal association cone size intuitively shrinks as a function of $p_T$ in order to reduce the contamination from fragmentation and pile-up tracks at high $p_T$, and maintain high $b$-tagging efficiency. The optimization procedure looks for charged particles at truth level, computes the angular separation of each charged particle to the nearest jet, and fits a function that describes the necessary cone aperture to include 97% of charged particles in each $p_T$ bin. The value of the 97$^\text{th}$ percentile is bootstrapped by evaluating it 5,000 times per bin, drawing 100,000 samples with replacement each time. The functional forms considered for the fit include the decaying exponential suggested in prior optimization studies, as well as a hyperbolic function, justified by the $\Delta R \sim m/p_T$ rule of thumb.

The Run I parameters are found to be approximately optimal for calorimeter jets, with the new best fit setting them to $[0.222, -1.11, -0.0198]$, which results in a slightly larger radius at very low $p_T$, and slightly narrower radius at very high $p_T$. A significantly improved fit can be achieved for track jets clustered with the anti-$k_t$ algorithm with radius parameter $R = 0.3$. In this case, the best fit parameters are found to be $[0.172, -1.02, -0.0275]$, which entails a much steeper decline, with overall narrower association cones.

Optimal values for the parameters $a$ and $b$ in the functional fit $\Delta R(p_T) = \frac{a}{p_T} \cdot \text{GeV} + b$ are set to $[5.88, 0.214]$ for $R = 0.4$ calorimeter jets, and $[2.74, 0.195]$ for $R = 0.3$ track jets. Parameters have also been estimated for the inclusion of the 98$^\text{th}$ and 99$^\text{th}$ percentile of true tracks.

Tracks are then independently selected according to specific quality criteria before being used by track-based $b$-tagging algorithms.

### 6.1.2 SECONDARY VERTEX TAGGING

The presence of a secondary vertex (SV) in a jet is a clear hint that the jet may have originated from the decay cascade of a $b$-hadron. For the purpose of identifying heavy-flavor jets, only secondary vertices produced by $b$-hadron decays are of interest, whereas displaced vertices that originate from kaon and $\Lambda$ decays, photon conversion, track crossings, or hadronic material interactions are considered background sources and ought to be removed before the formation of multi-track vertices and the calculation of the final discriminant.

The design and implementation of the SV-based $b$-tagging algorithm used in ATLAS is documented in Ref. [382] and summarized in this section. The generic Secondary Vertex Finder algorithm SVF examines the collection of tracks within a jet, and returns a list of secondary vertices with associated features that can be used downstream for jet classification. The algorithm explicitly tests for the presence of a secondary vertex in the jet by measuring the agreement of each track pair with a displaced vertex hypothesis. The properties of the reconstructed secondary vertices are then passed on to a SV-based $b$-tagging algorithm, SV1, that classifies jets based on the characteristics of their secondary vertex candidates.

The tracks used for secondary vertex-based flavor tagging must pass the following requirements [34]:

- $N_{\text{Si}}^{\text{hits}} \geq 7$ and $N_{\text{Si}}^{\text{shared}} \leq 1$

- if $|\eta| > 1.5$: $N_{\text{Si}}^{\text{hits}} \geq 8$

- $\chi^2/\text{dof} < 3$

- $d_0/\sigma_{d_0} \geq 2$ and $z_0/\sigma_{z_0} \leq 6$

To reduce the contribution of fragmentation tracks to the formation of fake vertices in high-$p_T$ jets, at most 25 tracks ordered by decreasing $p_T$ are used to reconstruct a secondary vertex (compared to an average charge particle multiplicity from $b$-hadrons of 5, as shown in Fig. 6.1).

A secondary vertex is defined as the point of intersection of several tracks that is displaced from the primary vertex. Any two tracks that intersect within track-parameter uncertainties form a two-track vertex, and several nearby two-track vertices signal the presence of an $n$-track vertex. The vertex finding algorithm attempts to form two-track vertices from each track pair, and then iteratively re-fit the set of selected tracks to obtain one single secondary vertex (Single SVF, or SSVF), or embed two-track vertices into a compatibility graph, and merge overlapping ones to obtain multiple SV candidates (MSVF). In this last case, if a track is associated with more than one vertex, the ambiguity is resolved by exclusively assigning the track to the vertex with the lowest track-to-vertex association $\chi^2$. Two-track vertices are excluded if the sum of impact parameter significances exceeds 2, if the $\chi^2$ of the track system exceeds 4.5, or if the invariant mass at the vertex is greater than 6 GeV [34].

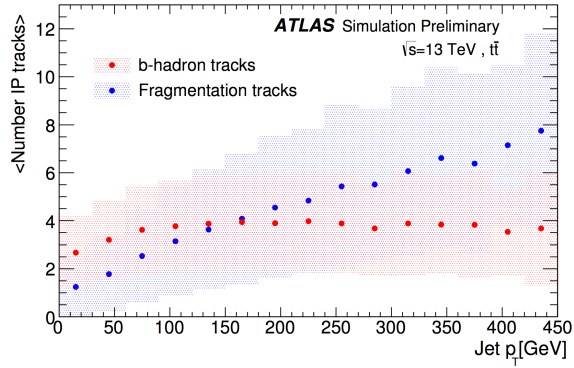

**Figure 6.1:** Average charged particle multiplicity as a function of jet transverse momentum, with tracks separated by origin: the number of tracks from fragmentation are shown in blue, those from $b$-hadron decays in red. While the number of tracks of interest remains approximately constant with $p_T$, the number of fragmentation tracks within jets becomes dominant at medium-to-high $p_T$. The shaded bands identify the per-bin RMS. The image is replicated from Ref. [34]

The finite resolution of ATLAS tracking detector prevents us from resolving all $b$ and $c$ decay vertices with full efficiency. Vertices are identified as fake or background based on kinematic and topological properties such as the reconstructed vertex mass, which identifies $K_0$ and $\Lambda_0$ decays, the number of hits in layers that precede the secondary vertex position, which flags fake secondary vertices in dense environments for removal, or the radial position of the secondary vertices, which easily singles out vertices that occur in the vicinity of a detector material layer. These identifiable sources of fake secondary vertices are immediately removed and do not enter the vertex reconstruction efficiency calculations. The percentage of jets of each flavor with reconstructed secondary vertices is plotted, as a function of jet $|\eta|$ and $p_T$, in Fig. 6.2(a) and 6.2(b). The performance degradation at high $|\eta|$ is due to the increase in the thickness of the material that particles traverse in the endcap region, with subsequent losses in tracking efficiency and track parameter resolution. At high $p_T$, the worsened performance for light flavor jets is due to the increase in number of fragmentation tracks, and track crossings in dense environments.

Viable vertices are selected based on the significance of their tracks' distances from the PV, the distance to material layers, the tracks' combined $\chi^2$, and the consistency of the hit pattern with the SV location. Secondary vertex properties are then propagated to a jet flavor tagging algorithm and SV candidates are examined for $b$-tagging purposes based on features such as the invariant mass, the jet energy fraction at the vertex, the number of two-track vertices, and the angular distance between jet axis and the axis that connects the PV and SV. Some of these highly discriminative variables that enter the multi-dimensional likelihood-based SV1 $b$-tagging algorithm are shown in Fig. 6.3. Details on the transformations that these quantities undergo before the likelihood is constructed, as well as the technical implementation information to arrive to the final discriminant are discussed in Ref. [382]. Vertices can be rejected based on the output of the secondary-vertex tagger if they fit the scenario of background jets that fake the presence of $b$-looking secondary vertices.

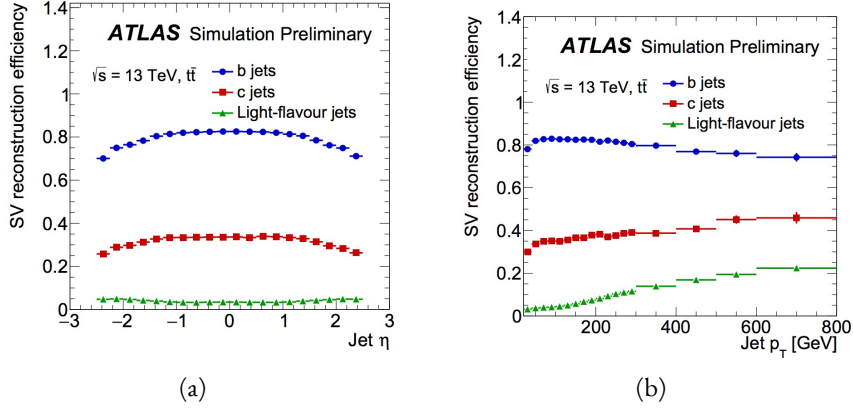

**Figure 6.2:** Percentage of jets of each flavor in which at least one secondary vertex is reconstructed. Detector resolution and track reconstruction inefficiencies contribute to the sub-optimal performance of the SV finding algorithm in $b$ and $c$ decays. On the other hand, light jets should have no real secondary vertices of the kind that passes the selection cuts designed to reject known contamination sources, so the green contributions can be interpreted as false positives for the SV finding method. SV reconstruction performance degrades at high $p_T$ and high $|\eta|$ from a maximum of $\approx 80\%$ for $b$-jets. Images reproduced from Ref. [35].

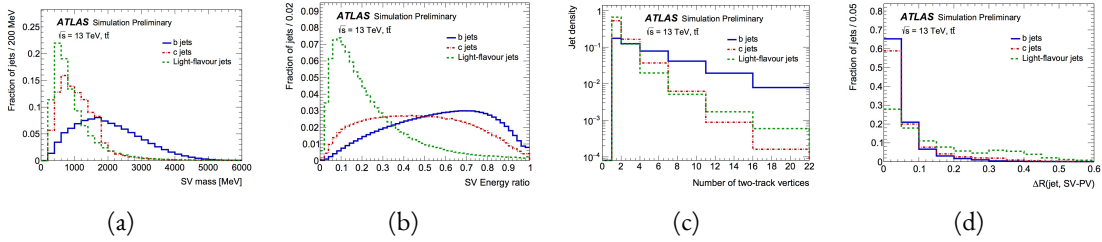

**Figure 6.3:** Distributions of a selection of inputs to the SV-based $b$-tagging algorithm SV1 for $b$, $c$, and light jets. Images reproduced from Ref. [35].

The upper bound on the achievable performance by secondary-vertex based $b$-tagging algorithms is fundamentally limited by the SV reconstruction efficiency, so the TPR cannot be greater than approximately 80% for $b$-favored jets. That is why, when visualized on a ROC curve, the performance of SV taggers terminates around the 80% $b$-efficiency mark.

### 6.1.3 JetFitter

Another tagger that exploits the unique vertex topology within jets originating from heavy flavor hadrons and extends lifetime-based tagging capabilities beyond the single vertex assumption is known as JetFitter (JF) [383]. The constraints of the single vertex hypothesis reduce the efficiency in track association whenever more real vertices are present, because of the inability to obtain a low $\chi^2$ value for the association of all tracks to a single vertex. At the same time, high quality identification of two or more separate vertices within a jet incurs in compounding inefficiencies from track and vertex reconstruction, as well as parameter resolution.

JF circumvents the issues with independent multi-vertex reconstruction by assuming that vertices that belong to the $b$-hadron decay chain approximately reside along the same line. Specifically, it makes use of a Kalman Filter that processes input track parameters and covariance matrices in order to fit the full decay topology, including secondary and tertiary vertices found along the hadron's line of flight. This opens the possibility of reconstructing complex, incomplete topologies such as single tracks associated to a vertex. The algorithm description and implementation details are outlined in Ref. [383].

The fitted topology is described in terms of discriminating variables, such as the number of reconstructed vertices, the number of two-track vertices, the track multiplicity at each vertex, the reconstructed vertex mass, the vertex energy fraction, and the impact parameter significance. These physically-motivated JF observables can be combined into a likelihood function, which computes jet compatibilities with different flavor hypotheses based on likelihood templates obtained from Monte Carlo, or can be used as inputs to higher-level taggers. In particular, including topological multi-vertex reconstruction information in taggers such as MV2 achieves significant performance improvements for both light and $c$-rejection [384].

### 6.1.4 Impact Parameter Taggers

A third low-level approach to flavor tagging exploits the difference in impact parameter properties for tracks originating along the decay chain of heavy flavor and light flavor hadrons. These lifetime-based taggers aim at discriminating $b$-topologies using the knowledge that the large mean path length for a $b$-hadron will cause the decay vertex to be significantly displaced from the PV, thus creating tracks with high impact parameters.

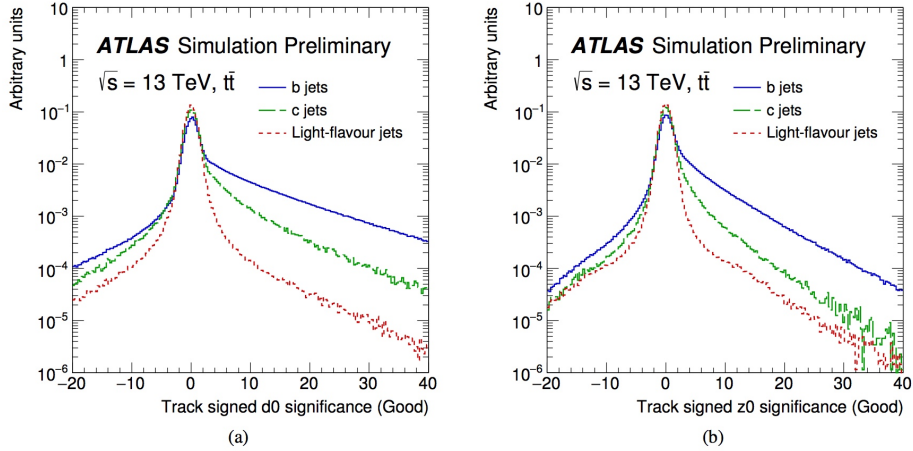

**Figure 6.4:** Significance of the transverse and longitudinal impact parameters for good tracks in jets of different flavors. The 'good' category includes 83.6% of tracks in $b$ jets, 84.8% of tracks in $c$ jets, and 86.4% of tracks in light jets [34]. The histograms are normalized to unit area for distributional shape comparison. Lighter flavor jets tend to have tracks with impact parameter significances that are more likely to be compatible with zero, *i.e.* with the PV hypothesis, while displaced tracks from $b$-hadron decays have higher impact parameter values. Images reproduced from Ref. [34].

The impact parameter is the distance of closest approach between a track and the primary vertex. It can be decomposed into the transverse component, $d_0$, *i.e.* the shortest track-to-PV distance in the $r$-$\phi$ projection plane, and the longitudinal component, $z_0$, *i.e.* the distance between the primary vertex and the projection of the track's position of closest approach onto the beam line [34]. If the jet axis needs to be extended backward beyond the PV to cross the path of a given track, the track's impact parameter will be negative; if, instead, the two lines cross in front of the PV with respect to the jet direction of flight, the track's impact parameter will be positive. The impact parameter significance is computed as the ratio of the impact parameter to its measurement uncertainty.

Tracks are reconstructed in the ATLAS detector as described in Sec 5.1.1. To ensure the quality of tracks and their readiness for physics analysis, only tracks that satisfy the following requirements are used for impact parameter-based flavor tagging [34]:

- $p_T^{\text{track}} > 1$ GeV

- $|d_0| < 1$ mm and $|z_0 \sin\theta| < 1.5$ mm

- $N_{\text{Si}}^{\text{hits}} \geq 7$ and $N_{\text{Si}}^{\text{holes}} \leq 2$ with $N_{\text{pixel}}^{\text{holes}} \leq 1$.

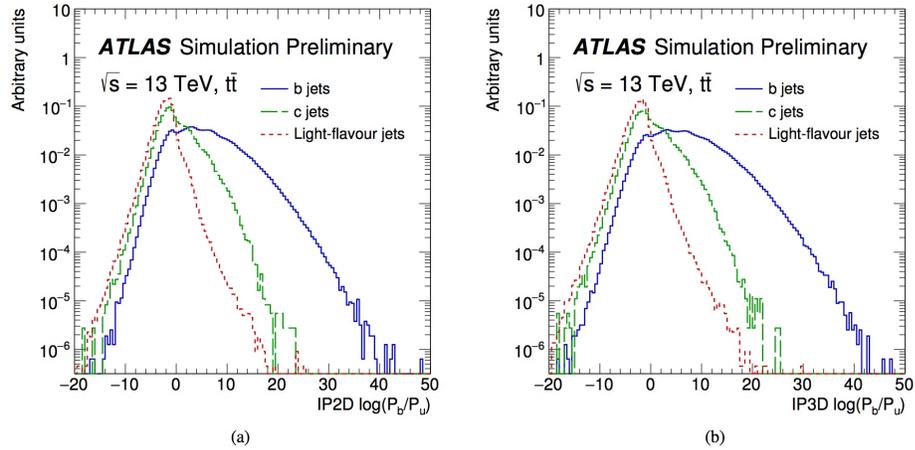

**Figure 6.5:** Log of the likelihood ratio between the $b$ and light hypotheses obtained from the construction of binned templates from Monte Carlo simulated $t\bar{t}$ samples. The left plot (a) shows the output distribution of the IP2D algorithm, while the right plot (b) shows the output for IP3D. The plot omits the lowest bin in which jets with no tracks that pass the selection are located. Images reproduced from Ref. [34].

In general, tracks originating from the $b$-hadron decay chain tend to have larger impact parameters compared to background tracks that originate from the PV, as clearly visible in Fig. 6.4. Light jets, by definition, do not contain tracks coming from in-flight decays of heavy hadron, and most tracks originate from the PV, so their impact parameter significances are typically consistent with zero; tracks generated from $b$-hadron decays away from the PV are, instead, more likely to have impact parameter significances that differ significantly from zero [34]. This reasoning provides the physical grounding for using impact parameter variables as distinctive features in a classification task. Tracks in light-flavor jets with high transverse impact parameter significance can be explained as coming from $\Lambda$, and $K_S$ decays, hadronic interactions or from photon conversions [34].

Traditional impact parameter taggers are designed to take advantage of this property and search for tracks originating away from the collision point. They score tracks by calculating a log-likelihood ratio that expresses the track's compatibility with the PV hypothesis. First, binned likelihood templates are constructed for individual track-quality categories starting from simulated distributions of transverse and longitudinal impact parameter significances. The definition of the track categories based on measured and expected hits in various inner detector layers, the categories' relative size, and the fraction of $b$-hadron versus fragmentation tracks as a function of jet $p_T$ in each category are presented in Ref. [34]. Then, test tracks are assigned a weight that corresponds to the value of the probability density function associated to each flavor hypothesis in the bin defined by its impact parameter properties.

The IP2D tagger draws its discriminative power from the distribution of the transverse impact parameter significance, while the IP3D tagger combines transverse and longitudinal impact parameter significances into 2D templates. The IP2D and IP3D discriminants are computed for a jet with $N$ tracks as the sums of per-track contributions

$$\text{LLR} = \sum_{i=1}^{N} \log \frac{p_{b_i}}{p_{u_i}} \tag{6.2}$$

where $p_b$, $p_u$ are the template PDFs for the $b$- and light flavor hypotheses obtained from Monte Carlo simulated reference distributions. The output distributions of the IP2D and IP3D discriminants for $b$, $c$, and light jets are shown in Fig. 6.5.

The formula above assumes that the tracks are independent and ignores the conditional probability terms.

### 6.1.5 RNNIP

The Recurrent Neural Network Impact Paramater (RNNIP) tagger enhances previous impact parameter-based taggers by exploiting track correlations within jets.

Within the decay of a $b$-hadron, several charged particles with large impact parameters may emerge from the secondary (or tertiary) vertex. Therefore, the tracks' impact parameters must be intrinsically correlated: in $b$-jets, if there exists a track with a large impact parameter, then the presence of a second track with large impact parameter is also likely. If, instead, there is no displaced decay, which is often times the case for light-flavor jets, then such a correlation should not exist. The different behavior between tracks from $b$ and light jets can be observed in Fig. 6.6, which clearly shows how, in the case of light jets, finding a track with high impact parameter significance is not correlated with the presence of any other high impact parameter track. Traditional impact parameter-based taggers such as IP3D disregard this correlation and assume that the properties of each track in a jet are independent of all other tracks. This clearly limits their tagging performance.

RNNIP is a Recurrent Neural Network (RNN) that uses Long-Short Term Memory units (LSTMs) [242] to process jets represented as sequences of tracks ordered by their absolute transverse impact parameter significance. In the final RNNIP setup, each track is described by a vector of the following properties: transverse and longitudinal impact parameter significances ($S_{d_0}$ and $S_{z_0}$), the fraction of the jet's trans-verse momentum carried by the track $p_T^{\text{frac}}$, the distance between the track and the jet axis $\Delta R(\text{track}, \text{jet})$, and a learned 2D embedding of the track grade, which is a categorical variable.

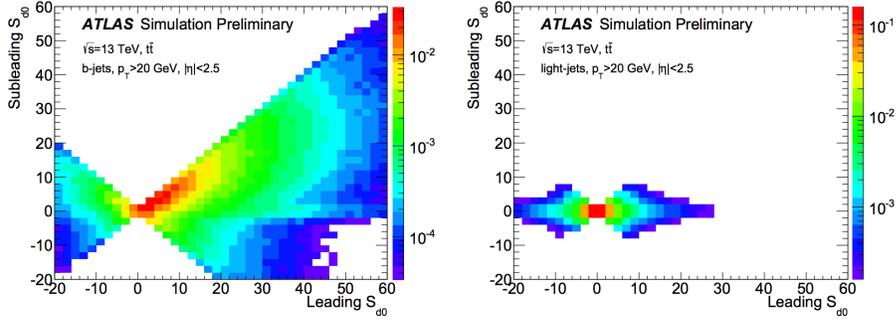

**Figure 6.6:** Two dimensional histograms highlighting the relation (left), or lack thereof (right), between the transverse impact parameter significances of the two highest $S_{d_0}$ tracks in a jet, for $b$ jets (left) and light jets (right). The presence of this type of correlation among tracks is clearly a discriminating feature between heavy and light jets, which impact parameter taggers such as IP3D ignore, while taggers like RNNIP try to exploit. The histograms are normalized to unit area. Images reproduced from Ref. [36].

The track grade parametrization, which traditionally subdivides tracks into 14 categories identified by integer labels and based on the amount of hits and holes left in the tracking detectors, is turned into a learned, two-dimensional unit vector representation, equivalent to a single continuous degree of freedom that encodes the track quality information. No meaningful pattern can be extracted from the learned position of the 14 categories around the unit circle. In addition, in practice, no significant performance improvement is observed in simulation when choosing different dimensionalities for the embedding or substituting the raw hit pattern with the embedded track grade, as shown in Fig. 6.7. However, suspicions that the possible mismodeling of the number of hits in each detector layer may affect the data-MC agreement of the classifier output are widespread enough to push towards the 2D embedding setup. Similarly, Fig. 6.8 demonstrates that ordering tracks by transverse impact parameter significance is suboptimal compared to the choice of ordering tracks by the euclidean norm computed from the transverse and longitudinal impact parameter significances.

The LSTM unit outputs a 50-dimensional hidden vector that is further processed by a fully-connect layer with softmax activation and four output nodes, corresponding to the $b$, $c$, $\tau$, and light classes. The softmax guarantees that the sum of the four outputs for each jet sum to one, so that they can be loosely interpreted as probabilities of the jet belonging to each class.

Despite the narrative that neural networks are parameter inefficient, RNNIP has approximately 60%

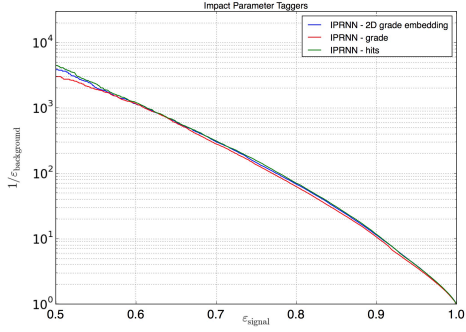
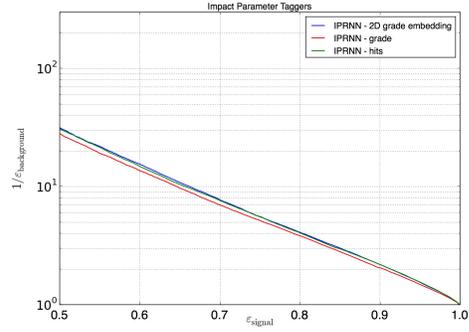

(a) Effects of track quality representations on the ROC curves for $b$ versus light classification.

(b) Effects of track quality representations on the ROC curves for $b$ versus $c$ classification.

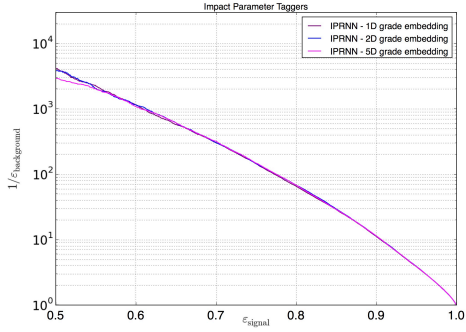
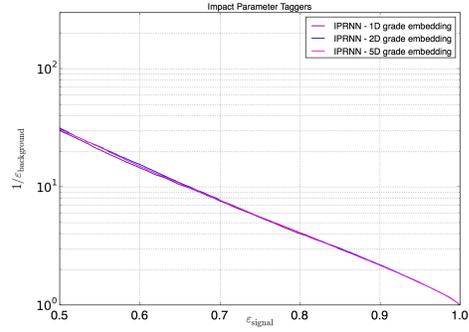

(c) Effects of the track quality embedding dimensionality on the ROC curves for $b$ versus light classification.

(d) Effects of the track quality embedding dimensionality on the ROC curves for $b$ versus light classification.

**Figure 6.7:** Performance studies to assess the impact on performance of varying the representation format and the embedding dimensionality of the track grade information. The 'track grade' representation follows the current ATLAS categorization scheme; the 'hits' representation uses the number of hits and holes in tracking detector layers as direct inputs to the neural network, instead of summarizing this information to construct the grade categories. Embedding the track grade allows the network to learn an ideal representation, starting from the ATLAS categorization as input.

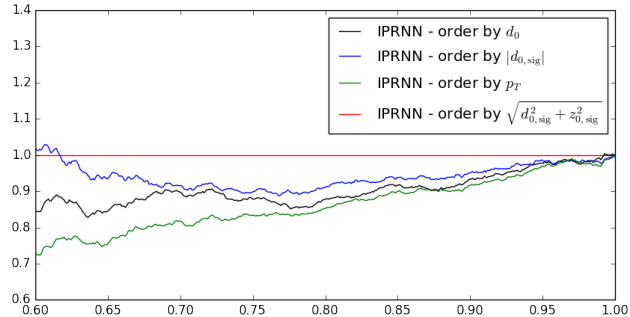

**Figure 6.8:** Ratio of ROC curves obtained from the retraining of the RNN-based impact parameter tagger using different ordering schemes for the sequence of tracks that represent a jet. Here, $d_{0,\,\text{sig}}$ and $z_{0,\,\text{sig}}$ are the transverse and longitudinal impact parameter significances, otherwise referred to in the text as $S_{d_0}$ and $S_{z_0}$. The ratio is computed with respect to the euclidean norm $\sqrt{d_{0,\,\text{sig}}^2 + z_{0,\,\text{sig}}^2}$, which appears to be superior to other track orders.

fewer parameters than the 35 × 20 (bins) × 14 (categories) × 3 (flavors considered) = 29,400 bin height values stored in the IP3D template histograms. The simple, shallow architecture of RNNIP can easily be augmented by more, or more sophisticated, layers to boost its performance, while balancing the computational efficiency requirements of the experiment.

RNNIP is a multi-class tagger trained to predict four output quantities associated with each flavor. The class assignment can be chosen using maximum a posteriori estimation, or by constructing a single discriminant (as a function of the four outputs) and defining fixed $b$-efficiency cuts on the univariate discriminant. Calibration favors the latter, so the four outputs are combined into the following quantity, where the importance of each background can be varied after training according to the specific use case:

$$\text{RNNIP}(f_c, f_\tau) = \log\left(\frac{p_b}{f_c \cdot p_c + f_\tau \cdot p_\tau + (1 - f_c - f_\tau) \cdot p_u}\right) \tag{6.3}$$

where $f_c$ and $f_\tau$ are parameters that regulate the relative importance of $c$ and $\tau$ jets in the discriminant. Given that $\tau$ reconstruction and identification is the subject of a completely different line of work in ATLAS [385, 377], and $\tau$ jets are a relatively minor source of background for $b$-jet tagging, $f_\tau$ is conventionally set to zero; $f_c$ is instead commonly set to 0.07, equivalent to the percentage of $c$-jets in the $t\bar{t}$ sample used at training time. Improved performance for $c$-discrimination tasks can be achieved, without the need to retrain the classifier, by increasing the magnitude of $f_c$. Therefore, any $b$-versus-$c$ performance plot shown in this chapter that involves the use of RNNIP or any other multi-class classifier is not truly indicative of its best achievable performance level.

The network is designed and trained using the `Keras` library [275], and weights are updated using the Adam optimizer [177][*]. The `lwtnn` package [386] is used within the ATLAS Athena framework [162] to load the trained network weights and evaluate the forward pass within the reconstruction pipeline.

The maximum number of tracks in a sequence is set to 15, although no major performance degradation is found when this threshold is increased. On the other hand, artificially decreasing the sequence length reduces the amount of useful information that the network can use, thus decreasing the overall performance. It is important to note that the maximum sequence length is tightly coupled to the choice of track selection criteria adopted. In fact, while only a negligible percentage of jets possess more than 15 tracks when these are selected according to the same criteria used for track selection in IP3D, looser track selection cuts that admit a higher number of tracks per jet would require an increase in the maximum sequence length. The distribution of number of tracks in a jet for different track selection criteria is provided in Fig. 6.9. While the desire to harmonize the track selection with the *Loose* working point defined by the Tracking CP group might provide a valid track selection alternative, applying no track quality selection at all causes the track container to be contaminated by excessive amounts of pile-up tracks.

---

[*]Code is available at https://github.com/mickypaganini/RNNIP.

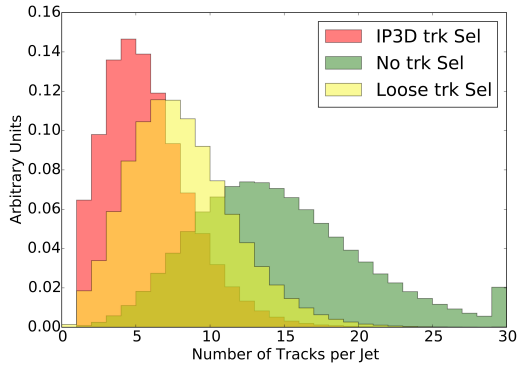

**Figure 6.9:** Distribution of the number of tracks within a jet when different track selection criteria are applied. The red distribution represents the set of cuts imposed prior to the IP3D evaluation (see Sec. 6.1.4); the yellow distribution corresponds to the *Loose* working point defined by the ATLAS Tracking CP group; and the green distribution corresponds to a minimal requirement of hits to reconstruct the track. Depending on the chosen track selection, the maximum sequence length considered by RNNIP may be varied.

Jets that contain fewer than 15 tracks are padded with entries set, by default, to a value of -999, which are then masked and ignored by the neural network. Zero-padding is considered unsafe given the non-negligible probability and physical significance of having a track described by a vector of all zero features.

Performance studies investigating the discrimination gains due to the addition of kinematic variables have been summarized in Fig. 6.10, which shows how successively adding $p_T^{\text{frac}}$ and $\Delta R(\text{track}, \text{jet})$ steadily moves the ROC curves towards more performant configurations. Fig. 6.10 also reveals how RNNIP is able to boost the performance of impact parameter tagging over likelihood-based methods even when using the same inputs as IP3D. While the curse of dimensionality affects the ability to add more discriminating variables to the likelihood-based approach, a powerful neural network-based method such as RNNIP has the advantage of being able to handle a growing set of input features without difficulty. In fact, models with additional input variables, such as the jet kinematics, the track impact parameters, and the track $\chi^2$, were successfully trained but not productionalized. In future redesigns of the $b$-tagging algorithms landscape, one could consider including vertex information in an RNNIP retraining to increase the completeness of information available to the net, and join the power of secondary vertex and impact parameter tagging.

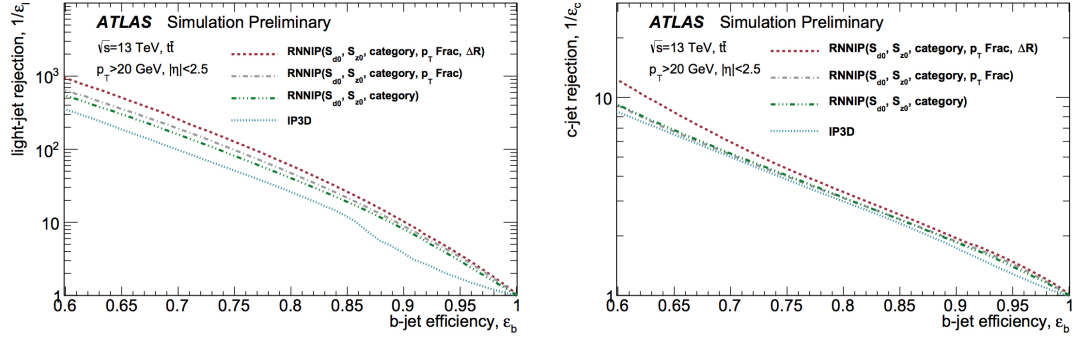

**Figure 6.10:** ROC curves for $b$ versus light classification (left) and $b$ versus $c$ classification (right) for the baseline IP3D tagger and three variations of the RNNIP method with incremental addition of input variables. Unlike the likelihood-based IP3D tagger, the RNNIP neural network can easily increase the dimensionality of the input space and make use of additional sources of information. Even the base RNNIP configuration that uses the same input variables available to IP3D provides a definite increase in performance across both tasks. Images reproduced from Ref. [36].

The comparison between the overall performance of RNNIP and that of other $b$-tagging algorithms is presented in Fig. 6.11. The plot shows the performance improvement provided by RNNIP over other track-based low-level taggers, such as IP3D, and the SV1 tagger that relies on secondary vertex reconstruction. For reference, the ROC curve for the MV2 tagger, which aggregates information from all traditional low-level taggers, is also shown on the plot. For comparison, at the 70% efficient working point, RNNIP improves the light jet rejection of IP3D by 2.5×. In particular, at 70% constant $b$-jet efficiency, RNNIP outperforms IP3D in both light and $c$-rejections across all the $p_T$ bins used to produce Fig. 6.12. The performance of both algorithms deteriorates both at low and high $p_T$ because of tracking inefficiencies and statistical considerations.

Despite RNNIP's superior performance across most dimensions, IP3D is observed to perform particularly well on topologies characterized by low track multiplicity and short heavy-hadron decay length. This is not unexpected: the naive approach of summing per-track log likelihoods is based on the assumption that each and every track in the jet provides equally relevant evidence towards assessing the jet flavor. However, as the track multiplicity increases, so does the probability of having wrongly associated a certain number of tracks to the jet. Therefore, without the introduction of a discount factor $\gamma^i$ to progressively reduce the relevance of track $i$ in the sequence, or of a weighting scheme to downweight tracks based on the goodness of their association to the jet, the IP3D discriminant will, by construction, depend on the number of tracks $N$ in each jet, favoring topologies with low track multiplicity. This level of complementarity between the two taggers, that otherwise rely on similar input information, warrants the usage of both as inputs to higher-level taggers, instead of advocating for a full replacement of IP3D by RNNIP. In fact, removing the IP3D variables from the MV2 tagger results in a 0.2× drop in light jet rejection at 77% $b$-efficiency compared to the full configuration.

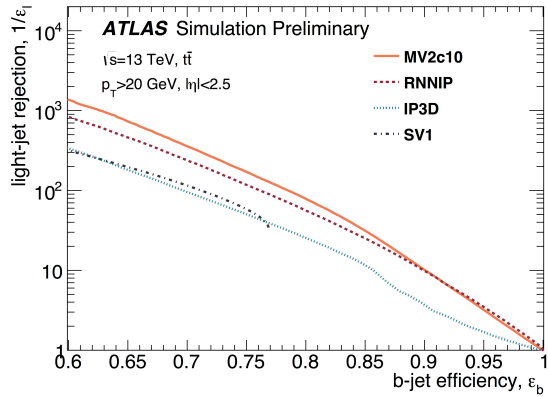

**Figure 6.11:** Light jet rejection as a function of the $b$-jet selection efficiency in simulated $t\bar{t}$ events for three low-level taggers (the traditional impact parameter tagger IP3D, the traditional secondary vertex tagger SV1, and the impact parameter neural network tagger RNNIP) and the high-level tagger MV2. Image reproduced from Ref. [36]

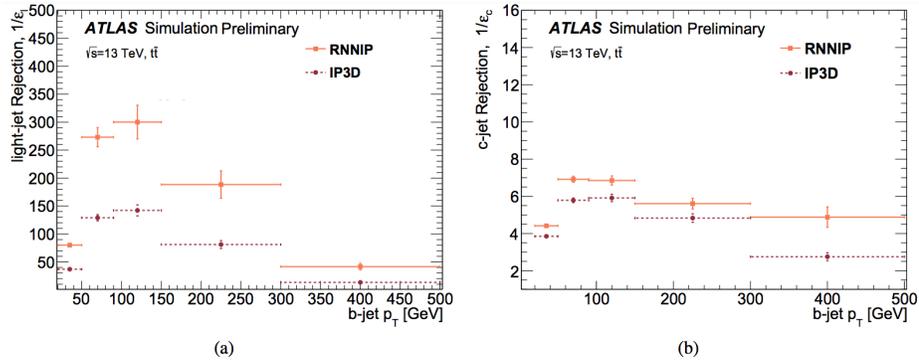

**Figure 6.12:** Performance comparison between the impact parameter taggers IP3D and RNNIP as a function of transverse momentum, at their 70% efficient working points. The left plot (a) shows the behavior of light jet rejection, while the right plot (b) shows the $c$-jet rejection. RNNIP outperforms IP3D across all $p_T$ bins in both tasks. Images reproduced from Ref. [36]

The outputs of RNNIP are then added as inputs to higher-level taggers, which significantly increases their performance, as shown later in Fig. 6.16 and 6.18, especially at high $p_T$, where the RNN excels over other complementary taggers (Fig. 6.19), and in $b$-versus-$c$ discrimination. In addition, the inclusion of RNNIP in the training of high-level taggers does not impact the agreement between data and Monte Carlo distributions, as one can verify in Fig. 6.20 and 6.21. These plots are obtained by including the RNNIP outputs into the training of both DL1 (in its DL1MuRnn version) and MV2 (in its MV2MuRnn version). The quality of the modeling of the data distribution by the Monte Carlo output, for both the $t\bar{t}$ and $Z$ sample, suggests that the performance gains observed in simulation can be expected to apply to data as well.

### 6.1.6 Soft Muon Tagger

The Soft Muon Tagger (SMT) is a recent addition to the flavor tagging suite of algorithms introduced in Run 2 [37, 387]. It aims at identifying heavy flavor jets by reconstructing the presence of soft muons from the semi-leptonic decay of the corresponding hadrons. Semi-leptonic decays occur with a branching ratio of ≈ 21%, which upper-bounds the fraction of jets that would benefit from a dedicated muon tagger. On top of being intrinsically limited by the semi-leptonic branching ratio, SMT's performance is further reduced by the inefficiencies in muon reconstruction and association to jets: in $t\bar{t}$ events, only ∼ 65% of $b$-jets containing muons are properly reconstructed as such. However, the tagger provides a useful complement to impact parameter- and vertex-based taggers, thus contributing to lower the overall false positive rate when deployed in association with orthogonal tagging information.

Muons from heavy hadron decays are called 'soft' because of their reduced transverse momentum compared to the leptons that originate from weak boson decays. In spite of this, they are characterized by non-negligible transverse momentum with respect of the jet axis – a feature that can be utilized for the reconstruction of this event topology.

The signature of muons from heavy hadron decays can be imitated by hadrons that punch through to the muon spectrometer and fake the presence of a muon ($< 0.1\%$ contamination), as well as true muons originating from the decay of lighter hadrons ($\sim 1\%$ contamination) and $W$ bosons ($\sim 1\%$ contamination). The SMT combines a group of 6 physically-motivated observables into a BDT to optimize the rejection of these background events.

In addition to the momentum component perpendicular to the jet axis ($p_T^{\text{rel}}$), other features that are found to have discriminating power are the angular distance between the muon and the jet ($\Delta R$(muon, jet)), the transverse impact parameter of the muon track with respect to the primary vertex ($d_0^{\text{PV}}$), as well as the following track quality variables:

- $\mathcal{S} = q \sum_i \frac{\Delta \phi_{\text{scat}}^i}{\sigma_{\Delta \phi_{\text{scat}}^i}}$, the muon charge-weighted sum of scattering angle significances for pairs of track-segments ending and originating at each hit in the inner detector. This variable is designed to identify kinks in the track that point to pion or kaon decays, which are expected to carry a higher significance $\mathcal{S}$.

- $\mathcal{M} = \frac{p_{\text{ID}} - p_{\text{MS}}^{\text{extr.}}}{\sigma_{E_{\text{loss}}}}$, the momentum imbalance significance, which computes the difference between the momentum measured in the ID and MS (extrapolated back to the vertex), weighted by the uncertainty in the calorimeter energy measurement.

- $\mathcal{R} = \frac{(q/p)_{\text{ID}}}{(q/p)_{\text{MS}}}$, the ratio of the charge-to-momentum ratio measured in the ID and in the MS, which is a measure of the track curvature.

The distributions of these features for $b$, $c$, and light jets are shown in Fig. 6.13.

Jets chosen for the evaluation of this method contain combined muons with $p_T > 5$ GeV and $|\eta| < 2.5$ within a radius of $\Delta R = 0.4$ from the jet axis. A cut on the transverse impact parameter ($d_0 < 4$ mm) is imposed on the muons to reduce the contribution of pile-up tracks. Only approximately 12% of $b$-jets in a typical $t\bar{t}$ sample used for flavor tagging contain a muon that passes the criteria above [37].

The distribution of the BDT output for the different jet flavors is provided in Fig. 6.14, along with the comparison of the output shape in simulated and collected event. The SMT unfortunately suffers from localized data-MC mismodeling of up to 50%, which stems from the poor agreement between the simulated and data distributions of its inputs.

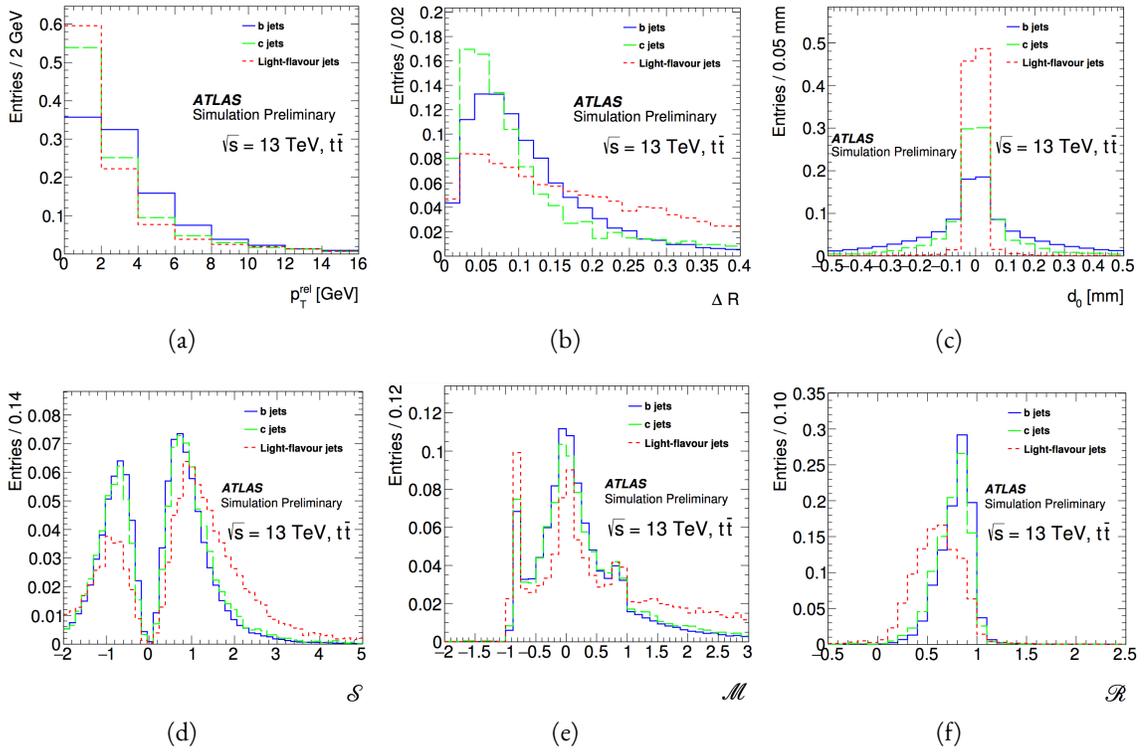

**Figure 6.13:** Distribution of the input features used by the SMT. The plots are obtained for $b$ jets (blue), $c$ jets (red) and light jets (green) from $t\bar{t}$ simulated events. The histograms are normalized to unit area for shape comparison. Images reproduced from Ref. [37].

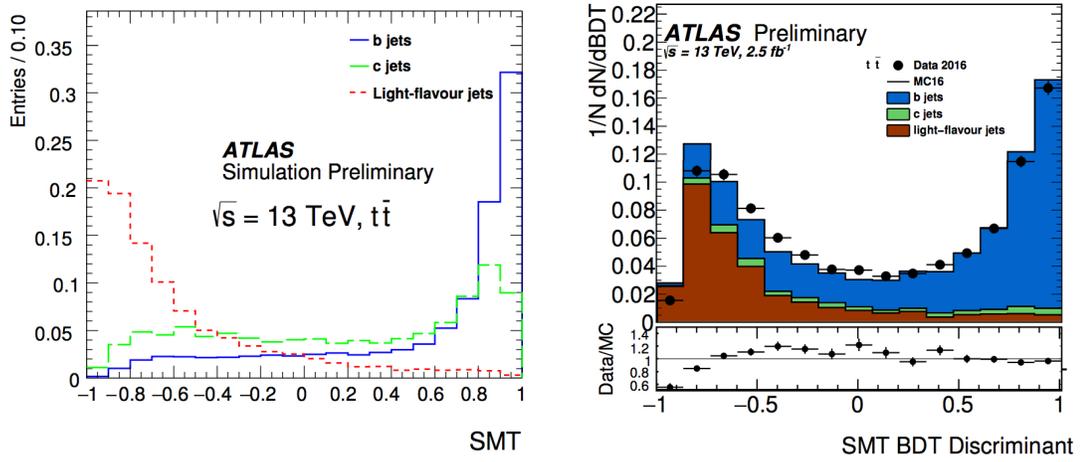

(a)   SMT BDT output distribution for the three jet flavors. The histograms are normalized to unit area for shape comparison.

(b)   Data-MC comparison for the output of the SMT BDT in 2016 $t\bar{t}$ events.

**Figure 6.14:** Individual per-flavor (6.14(a)) and stacked (6.14(b)) SMT BDT output distributions. Image reproduced from Ref. [37].

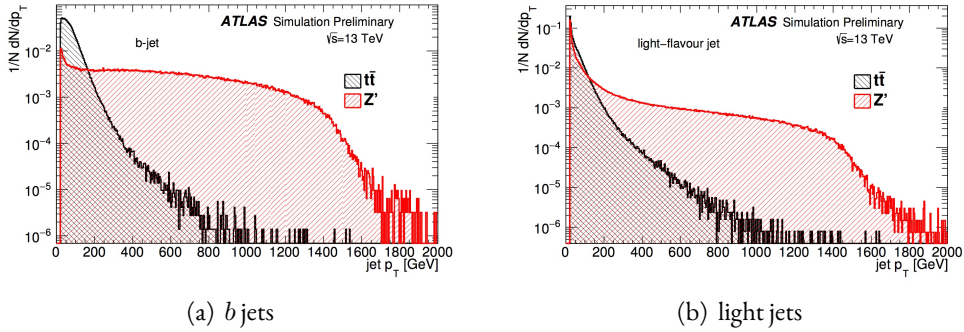

(a) $b$ jets  (b) light jets

**Figure 6.15:** Jet transverse momentum spectrum for the $t\bar{t}$ and $Z'$ events that make up the hybrid sample introduced in 2017 for $b$-tagging algorithm retraining. The $Z'$ sample significantly enhances the representation of high-$p_T$ jets to mitigate the previous inefficiencies in that regime. Image reproduced from Ref. [37].

### 6.1.7 MV2 - A Boosted Decision Tree for Flavor Tagging

All high-level $b$-tagging algorithms in ATLAS are machine learning classifiers that combine the output of low-level taggers acting on reconstructed track and vertex information. MV2 is one such high-level tagger that uses as inputs various combinations of the outputs of low-level taggers, in order to provide a final jet discriminant.

MV2 is, in fact, a family of BDT-based taggers designed and trained with the TMVA package [388] to perform binary classification tasks. The taggers that fall under the MV2 umbrella can be differentiated based on their configuration properties defined on two orthogonal axes: the classes considered for classification (two of $b$, $c$, light), and the number of input variables. In terms of the class choice, the base method is trained for $b$ versus non-$b$ discrimination, with fractional background contributions from $c$-jets set manually prior to training: MV2C00 contains no $c$-jets, the MV2C10 background is 7% $c$-jets, and the MV2C20 background is 15% $c$-jets. These solutions represent different trade-offs designed to provide a single competitive tagger for both $b$-versus-light and $b$-versus-$c$ discrimination, in the presence of different mixtures of light and $c$-jet backgrounds. In addition, two $c$-tagging specific variants exist: MV2C100 is trained for $b$-versus-$c$ classification with no light jets in the mix, while MV2CL100 is trained on $c$-versus-light with no $b$-jets in the training sample. The charm dedicated taggers also make use of 12 additional observables obtained from JetFitter reconstructed topologies with a unique SV and listed in Ref. [37].

With respect to the input variables, multiple combinations have been investigated internally, with the three main publicly documented options being:

- MV2: the baseline model that makes use of 24 input variables that include jet kinematic properties and properties computed by low-level taggers such as IP2D, IP3D, SV1, and JETFITTER. Ref. [389] documents these features, their meaning, and their distributions;

- MV2MU: an enhancement of the baseline tagger with the addition of the SMT output as an input;

- MV2MURNN: a further enhancement which adds the RNNIP outputs to the list of inputs.

The choice of preferring the output of the SMT BDT to the list of six reconstructed muon variables described in Sec. 6.1.6 is informed by preliminary MV2 performance studies.

The recently optimized MV2 hyper-parameters [34] yield a final configuration of 1000 trees with maximum tree depth of 30, and minimum node size of 0.05% of the training sample.

Prior to training, the input samples are reweighted using importance sampling to flatten the spectrum along $p_T$ and $\eta$ simultaneously. The models are trained on a hybrid sample that joins two individually simulated $t\bar{t}$ and $Z'(m = 4\text{ TeV})$ samples. The addition of the $Z'$ events contributes to increasing the number of training examples in the high-$p_T$ end of the spectrum, as shown in Fig. 6.15. Previously, the lack of high transverse momentum examples was cause for significant degradation of $b$-tagging performance in that region, which represented a severe obstacle for analyses with high-$p_T$ jets. The $Z'$ sample is decayed by setting the branching ratio to one third for $bb$, $cc$, and double-light decays, in order to evenly populate the high-$p_T$ portion of the spectrum with jets of all flavors.

The performance of the MV2C20 models trained with different input-variable configurations is separately tested on exclusive held-out sets of $t\bar{t}$ and $Z'$ examples. Four fixed cuts define working points at constant values of signal selection efficiency (60%, 70%, 77%, and 85%) computed for the $t\bar{t}$ sample, which can be calibrated and distributed to the analyses. Fig. 6.16 presents the ROC curves for $b$-versus-light and $b$-versus-$c$ discrimination in simulated $t\bar{t}$ events. Since the SMT information is designed to separate semi-leptonically decaying $b$-hadrons from light hadrons (but not from $c$-hadrons, in which muons may also be present), the MV2MU variant is not expected to contribute to improving the performance of the baseline configuration in the $b$-versus-$c$ classification task. The differential distributions for the background rejection levels at the global 77% efficient $b$-identification cutoff as functions of $p_T$ are show, for $t\bar{t}$ events, in Fig. 6.17. Similar plots for performance evaluation on $Z'$ events are provided in Fig. 6.18 and 6.19. While the inclusion of SMT is observed to boost $b$-to-light separation, the benefit of this tagger gradually weakens at high $p_T$, as observed in Fig. 6.19. On the other hand, much of the improvement provided by the addition of the RNNIP tagger to the input list is visible at high $p_T$ and in the $b$-versus-$c$ classification task. For reference, the MV2MuRnn variant outperforms the 2017 baseline MV2 configuration by a factor of $\sim 2$ in light-rejection and a factor of $\sim 1.3$ in $c$-jet rejection for 1 TeV $Z'$ jets. These types of data visualization techniques, however, fail to specify the per-bin TPR, which might differ from the global, fixed one.

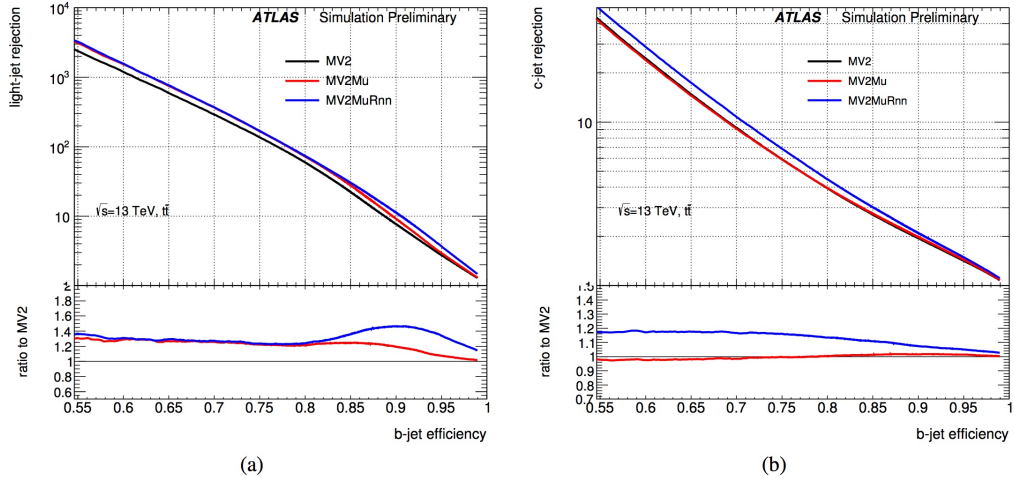

**Figure 6.16:** ROC curves showing the trade-off between TPR (or $b$-efficiency), on the $x$-axis, and light jet rejection in (a), or $c$-jet rejection in (b), on the $y$-axis, in simulated $t\bar{t}$ events for the following MV2 configurations: baseline MV2 (black), MV2Mu (red), and MV2MuRnn (blue). Images reproduced from Ref. [37].

Despite the MV2 retraining performed on the hybrid sample, a comparison between Fig. 6.16 and Fig. 6.18 immediately shows that about an order of magnitude difference between the performance of $b$-tagging algorithms on samples with different properties can be expected.

Although the introduction of the SMT presents known mismodeling issues, the output distribution of the MV2MuRnn tagger is able to achieve good levels of agreement between data and Monte Carlo, as displayed in Fig. 6.20.

### 6.1.8 DL1 - A Deep Feed-Forward Neural Network for Flavor Tagging

The family of taggers known as DL1 [390] has been developed to assess the added benefit of adopting deep learning techniques and open-source deep learning libraries for high-level flavor tagging, and to compare them to the performance obtained by traditional, powerful but cumbersome, C++ implementations of BDT-based classifiers such as MV2.

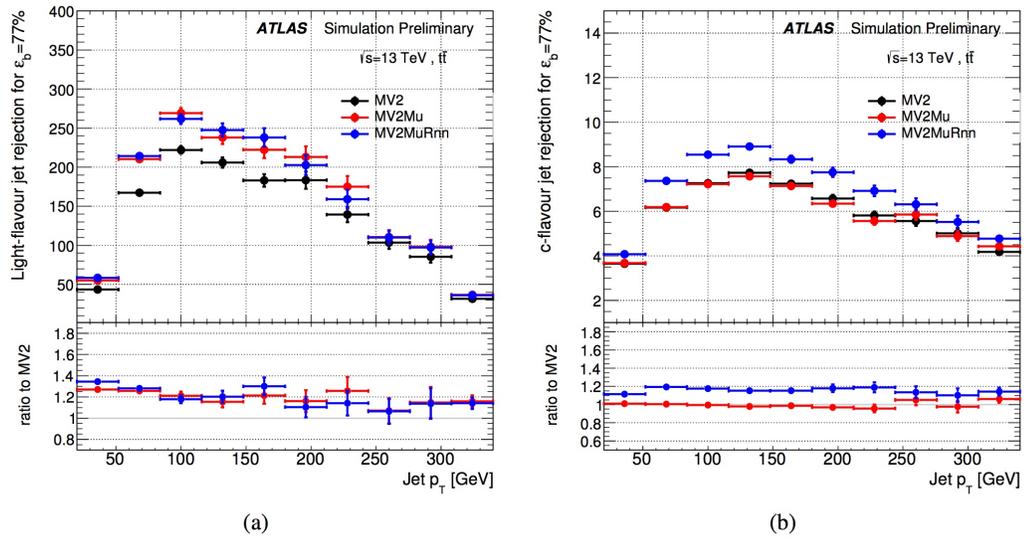

**Figure 6.17:** Light (a) and $c$-jet (b) rejection in simulated $t\bar{t}$ events at 77% $b$-jet efficiency in bins of jet $p_T$ for the following MV2 configurations: baseline MV2 (black), MV2Mu (red), and MV2MuRnn (blue). The ratio plots are drawn using the baseline MV2 configuration at the denominator. Images reproduced from Ref. [37].

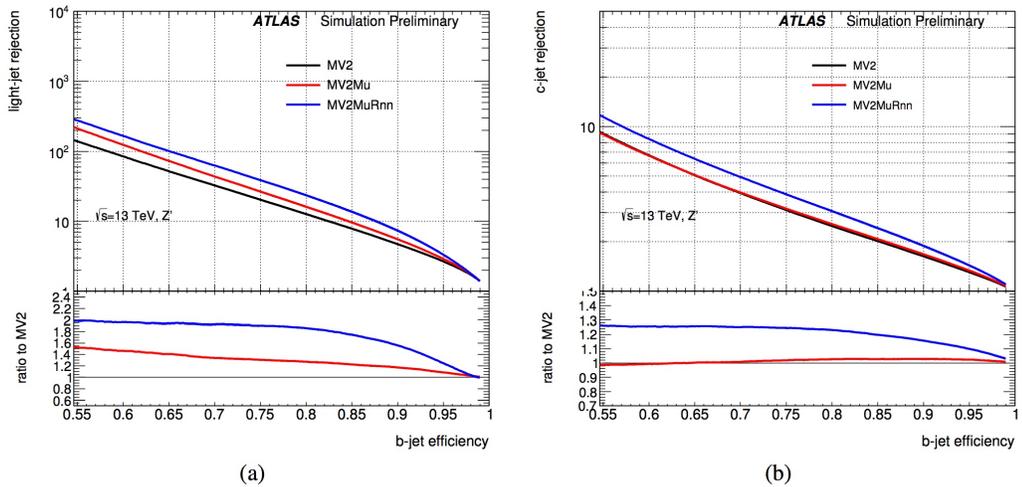

**Figure 6.18:** ROC curves showing the trade-off between TPR (or $b$-efficiency), on the $x$-axis, and light jet rejection in (a), or $c$-jet rejection in (b), on the $y$-axis, in simulated $Z'$ events for the following MV2 configurations: baseline MV2 (black), MV2Mu (red), and MV2MuRnn (blue). Images reproduced from Ref. [37].

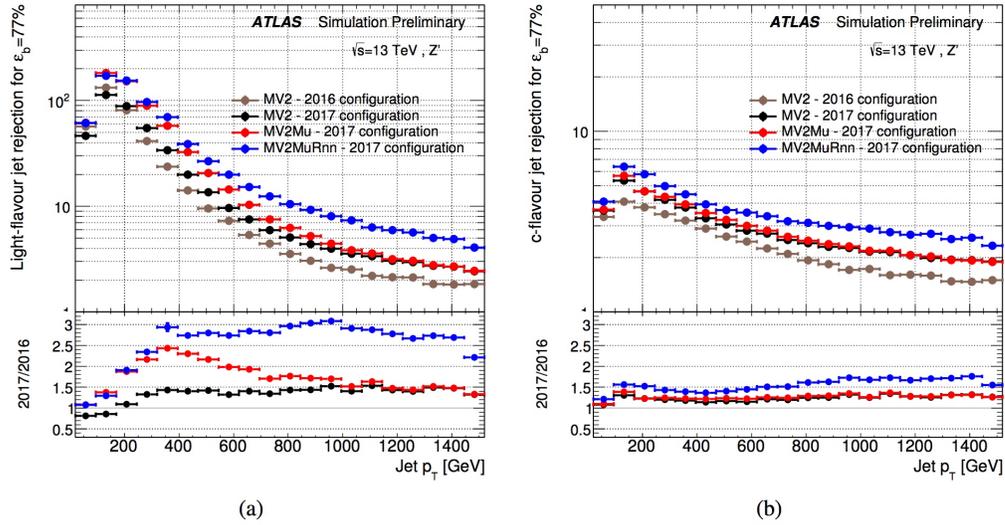

**Figure 6.19:** Light (a) and $c$-jet (b) rejection in simulated $Z'$ events at 77% $b$-jet efficiency in bins of jet $p_T$ for the following MV2 configurations: the baseline MV2 tagger trained using the 2016 software configuration and training samples (brown), the baseline MV2 tagger trained using the 2017 software configuration and training samples (black), MV2Mu (red), and MV2MuRnn (blue). The ratio plots are drawn using the 2016 MV2 configuration at the denominator. Images reproduced from Ref. [37].

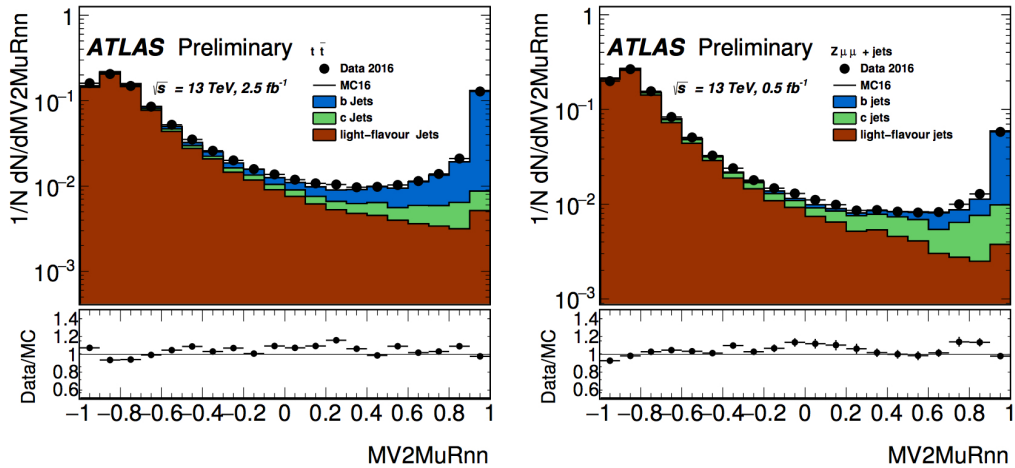

**Figure 6.20:** Data-MC comparison between the distributions of the MV2MuRnn tagger in a $t\bar{t}$-dominated sample (left) and a $Z \to \mu^+\mu^-$+jets-dominated sample (right). Images reproduced from Ref. [37].

After undergoing lengthy and detailed design, implementation, and evaluation processes, in which an extensive grid search among numerous hyper-parameters and configuration setups was performed while simultaneously monitoring the performance of often competing objectives and indicators, the configuration to which the flavor tagging community converged is an 8-layer deep stack of fully-connected, maxout [391], and batch normalization [202] layers activated by rectified linear units [226] and interleaved with dropout [201] layers for stochastic regularization at training time. The number of nodes per layer, number of layers, and other learning parameters are all tuned by this procedure. This network architecture is completed by a set of three softmax-activated output nodes that map the high-dimensional hidden representation to output probabilities for each jets to be associated with one of the three jet flavors considered for this task ($p_b, p_c, p_u$). This multi-class classifier is trained to produce highly non-linear functions of the input space by minimizing the cross-entropy criterion through the iterative update its parameters following the strategy set forth by the Adam optimizer [177]. The multi-class nature of the network provides increased flexibility over the convoluted MV2 solution of training and shipping multiple BDT instances for different $c$-jet fractions.

Similar to RNNIP, DL1 produces a final univariate discriminant that combines its three outputs, and is parameterized by fractional weights that control the relative importance of light and $c$-jet discrimination without the need for custom retrainings. The tunable function of DL1 outputs used to discriminate $b$-jets from mixtures of background light and $c$-jets can be expressed as:

$$\mathrm{DL1}(f_c) = \log\left(\frac{p_b}{f_c \cdot p_c + (1 - f_c) \cdot p_u}\right), \tag{6.4}$$

where $f_c \in [0, 1]$ measures the prevalence of $c$-jets in the background source and regulates the trade-off between light and $c$-rejection at a given $b$-tagging efficiency. Like RNNIP, DL1 can be used for $c$-tagging as well, by swapping the flavor probabilities such that the signal flavor appears at the numerator, and the background flavors appear at the denominator with the preferred fractional composition. The formula in Eq. 6.4 is used for comparisons to other $b$-tagging algorithms after setting the $c$-fraction to its natural value of 7%, which corresponds to the observed composition of the $t\bar{t}$ sample.

The network is designed and trained using the Keras library [275], and subsequently incorporated into the ATLAS reconstruction code base [162] through the use of the lwtnn package [386]. As with RNNIP, the development of the infrastructure needed to support experimentation and optimization with DL1 is based on industry-standard tools and data models such as HDF5 [392], h5py [393], pandas [394], scikit-learn [395], matplotlib [396], numpy [397], and HEP-specific converters such as rootpy [398] and root_numpy [399].

Various input sets have been tested, resulting in a final base configuration, selected for ease of comparison with MV2, that includes the same inputs as MV2, including the JetFitter observables optimized for $c$-tagging. In addition, to match the practice introduced for MV2 studies, two variants, called DL1Mu and DL1MuRnn, are trained by adding the SMT and RNNIP variables. The input distributions are reweighted to match the two-dimensional $(p_T, |\eta|)$ distribution of $b$-jets, in order to remove undesirable kinematic biases among classes that can be exploited by the classifier. In addition, sensible default values equivalent to the mean of each distribution are assigned to jets for which a low-level tagger is unable to produce a score. The input set is then augmented by binary features that flag the absence of any original tagger score [390]. All training distributions are normalized to zero mean and unit standard deviation, and equivalent transformations must be applied to the input distributions at test time.

The performance improvement provided by DL1 makes it the state-of-the-art tagger for flavor tagging in ATLAS. The relevant figure of merit for this multi-class classifier can be extracted from the iso-efficiency curves, in which the background composition is varied while imposing a threshold on the output distribution such that the percentage of signal jets that correctly pass the cut remains constant. When compared to the well-established MV2 tagger, preliminary studies show that DL1 is able to match or improve its performance [390]. For example, for a fixed working point that is 77% efficient at selecting $b$-jets while rejecting 99% of light jets, DL1 is capable of a 9% overall improvement in $c$-jet rejection over MV2, the magnitude of which varies as a function of jet $p_T$. Alternatively, for a fixed working point that is 25%(40%) efficient at selecting $c$-jets while rejecting almost 94% (75%) of $b$-jets, DL1 is capable of a 110% (65%) overall improvement in light-jet rejection over MV2, with improvement gaps still varying with $p_T$.

In addition to possible performance improvements, DL1 shares many desirable properties and advantages with RNNIP, such as higher flexibility, modularity, portability, and extensibility, which simplify and speed up the training process for future R&D campaigns. The lightweight training infrastructure affords the possibility of streamlining the processes of reproducing past results, adding extra variables, and testing new configurations, while the adoption of standardized tools lowers the barrier to entry for new contributors. The DL1 training setup is GPU enabled to take advantage of hardware acceleration when available. The framework can be extended to jointly train the RNNIP and DL1 taggers in an end-to-end fashion, or to augment them with an adversarial loss to reduce data-Monte Carlo disagreements or the taggers' dependence on kinematic or nuisance parameters [400].

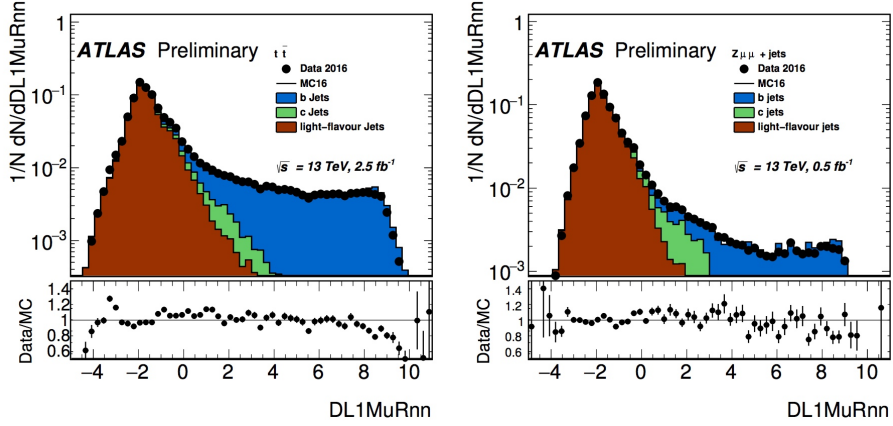

**Figure 6.21:** Contribution of all jet flavors to the distributional shape of the DL1MuRnn tagger output in simulated and collected events from $t\bar{t}$ (left) and $Z \to \mu^+\mu^-$+jets (right) samples. Images reproduced from Ref. [37].

Data-Monte Carlo agreement levels for the 2017 DL1MuRnn version of the tagger can be verified in Fig. 6.21 for $t\bar{t}$-dominated and $Z \to \mu^+\mu^-$+jets-dominated data samples collected by the ATLAS experiment at $\sqrt{s} = 13$ TeV.

### 6.1.9 Track Jet $b$-Tagging

The search for heavy resonances decaying into known SM particles has become central to the physics program of the ATLAS experiment. According to favored theories, these objects may give rise to high-$p_T$ top quarks, with subsequent $b$-hadrons appearing in boosted, dense jet environments among overlapping, collimated decay products.

Track jets provide an extension to traditional jet topologies investigated in flavor tagging, as they target the dense environments of overlapping decay products from boosted objects. They can therefore be useful to identify $b$-tagged sub-jets within larger-radius jets, but also enhance the tagging capabilities of standard calorimeter jets by infusing track-level information from the match of track jets to calorimeter jets. Track jet $b$-tagging algorithms can be optimized independently from calorimeter algorithms and later integrated to combine benefits of track-level and hadronic final state information. The importance of including designated track jet $b$-tagging methods in the ATLAS flavor tagging arsenal is discussed in Ref. [343]. Alternative techniques (not discussed here) for flavor tagging in boosted topologies include the use of variable-radius, exclusive-$k_t$, and center-of-mass sub-jet tagging [401].

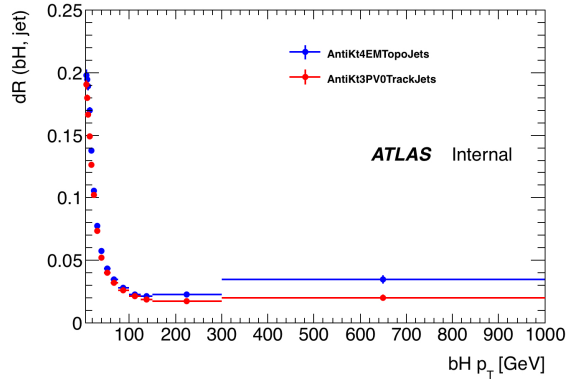

**Figure 6.22:** Angular distance $\Delta R$ between the $b$-hadron direction and the jet axis for two jet collections (calorimeter jets reconstructed from topoclusters using the anti-$k_t$ algorithm with radius $R = 0.4$, and track jets reconstructed using the anti-$k_t$ algorithm with radius $R = 0.3$), as a function of the $b$-hadron transverse momentum. This profile shows the mean angular distance in each $p_T$ bin. Track jets improve the resolutions across the entire $p_T$ spectrum. The generally degraded resolution at high $p_T$ is in part due to the poor estimate of centrality of the mean compared to the median, for skewed distributions.

Track jets present the advantage of improving the competitiveness and robustness of flavor tagging algorithms in high pile-up environments, thanks to the PV origin constraint imposed on the tracks that are used to reconstruct track jets, thus allowing for narrower radius parameters $R$ at reconstruction time. The independence of track jets from calorimeter information reduces the need for calibrations and JES corrections to account for calorimeter energy response. Furthermore, compared to calorimeter jets, track jets achieve superior direction and angular resolution with respect to the $b$-hadron line of flight, especially at high $b$-hadron $p_T$, as shown in Fig. 6.22. Capturing the directionality of the $b$-hadron correctly is important because $b$-tagging efficiency is known to scale with proximity between the hadron and the jet axis [381]. In general, however, $b$-hadron matching efficiency decreases as $p_T$ decreases, because of the $p_T$ thresholds imposed on (both calorimeter and track) jets. The issue is exacerbated for small-radius track jets that are less likely to catch enough constituents to pass the $p_T$ cut and to accurately reconstruct the hadron's kinematics from non-collimated decay products.

A caveat in the use of track jets is that they only measure the charged component of a jet, so they need corrections to account for the missing neutral component or one-to-one matching with calorimeter jets (or truth jets, in simulation) to set a common baseline for performance evaluation in terms of object kinematics.

The designated strategy of ghost-matching [344, 345, 346] calorimeter and track jets requires a resolution strategy in the case of multiple matches; if only one track jet is to be used to extract the $b$-tagging label and assign it to the calorimeter jet, then the selection of the best track jet candidate in the collection becomes an optimization problem. An optimal solution could be learned from the list of associated track jets, their properties, and those of the calorimeter jet under investigation, but a simpler selection based on the maximization of a single physics variable is often preferred and considered more intuitive. For example, the flavor of the highest $p_T$ track jet, or that of the track jet with the highest MV2 score, may be transferred to the corresponding calorimeter jet.

The best single-variable matching criterion is investigated using mc14 $t\bar{t}$ events simulated with Powheg+Pythia. LCW-calibration calorimeter jets are selected for this study if: $p_T > 20$ GeV, they pass the electron overlap removal criteria, and are separated from other jets by $\Delta R > 0.8$. The $R = 0.3, 0.4$ anti-$k_t$ track jet collections used in this application contain track jets with $p_T > 5$ GeV.

Transverse momentum and angular separation are found to be a sub-optimal matching criteria, whereas the $b$-tagging score obtained from a competitive $b$-tagging algorithm, such as MV2, is certainly preferable. However, it is observed that, among all events in which more than one track jet could be matched to the calorimeter jet, 18% (15%) of the time for $R = 0.4$ ($R = 0.3$) track jets the track jet with the highest $b$-tagging score is *not* the true $b$-track jet, due to tagging inefficiencies. The problem would naturally be exacerbated by selecting a less competitive tagger than MV2 for the track jet flavor assignment. While this track jet to calorimeter jet matching criterion is highly efficient for true $b$-calorimeter jets, which get assigned to a true $b$-track jet about 81.7% (84.6%) of times for $R = 0.4$ ($R = 0.3$) track jets, it also biases the track jet assignment to non-$b$ calorimeter jets towards the track jet with the highest compatibility with the $b$-hypothesis, thus potentially increasing the mistag rate for background jets. When looking for simultaneous effects of the matching criterion choice on the TPR and FPR, it is evident that most $b$-tagging algorithms perform similarly and could be successfully used to select the track jet from which the calorimeter jet will borrow the flavor tagging score, with the MV2 taggers providing slightly higher overall trade-off performance in light of the superior signal efficiency. These conclusions can be verified in Fig. 6.23.

An alternative to retraining track jet-specific $b$-tagging algorithms involves rescaling the track jets to the calorimeter jet kinematics and applying $b$-tagging methods that have been optimized for calorimeter jet tagging. A new linear fit to transform the track jet transverse momentum to the same range as the transverse momentum of the corresponding calorimeter jet is found to significantly differ from the previous implementation, as one can see in Fig. 6.24. This fit is obtained by only considering track jets with 30 GeV $< p_{T,\text{ trackjet}} <$ 150 GeV. With the optimized parameters, the suggested rescaling function becomes:

$$p_{T,\text{ track jet}}^{\text{uncalib. calo jet scale}} = 0.915 \cdot (p_{T,\text{ track jet}}) + 10255 \text{ MeV}, \tag{6.5}$$

compared to the values $[1.4, 8000]$ from the previous fit.

Unfortunately, neither the improved track-to-jet association (see Sec. 6.1.1) nor the improved $p_T$ rescaling result in performance improvements with respect to the tagging efficiency of calorimeter jets matched to track jets, without an appropriate retraining of the whole $b$-tagging stack specifically for track jets.

The campaign for dedicated retrainings of $b$-tagging algorithms resulted in the reoptimization of taggers such as IP3D and MV2 for various track jet collections. To avoid confounding inefficiencies that come from the track jet to calorimeter jet matching process, only track jets that are one-to-one matched with a calorimeter jet are considered in this study. This amounts to $\sim$ 92% (85%) for radius $R =$ 0.4 ($R =$ 0.2) track jets. Training took place on $\sim$ 10M jets from mc15 $t\bar{t}$ samples, while testing was performed on $\sim$ 6M jets, passing the standard $b$-tagging requirements for kinematics and isolation.

To optimize the MV2 tagger, the BDT is trained on a sample in which the fraction of $c$-jets is set to 15%, corresponding to the MV2c20 configuration. The BDT is composed of 700 trees (though a simpler method with 100 trees is explored for comparison purposes for $R =$ 0.4 track jets) and trained using the TMVA package [388]. The maximum depth of the decision tree is set to 30, the minimum number of training examples required in a leaf is 0.05% of the total training size, and the number of possible cuts investigated at each splitting is 200.

Different BDT instances are trained on the $R =$ 0.2 and $R =$ 0.4 track jet collections, resulting in collection-specific taggers. Their performances, compared to the default MV2c20 tagger trained on

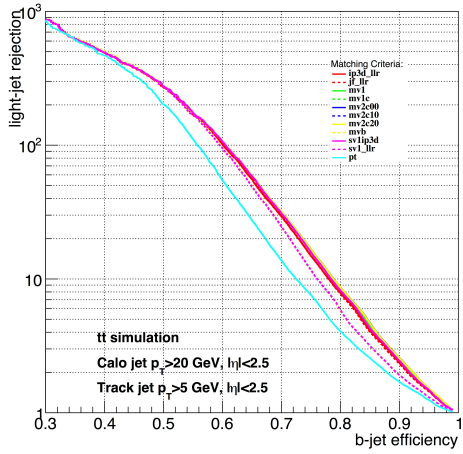
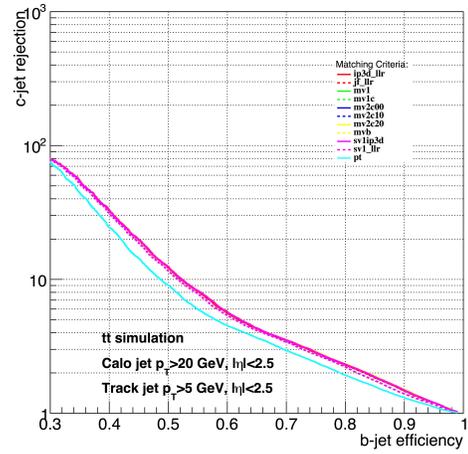

(a) ROC curve for $b$ versus light calorimeter jet discrimination, when the calorimeter jet inherits the flavor tagging score from the track jet with the highest value of one of the matching criteria.

(b) ROC curve for $b$ versus $c$ calorimeter jet discrimination, when the calorimeter jet inherits the flavor tagging score from the track jet with the highest value of one of the matching criteria.

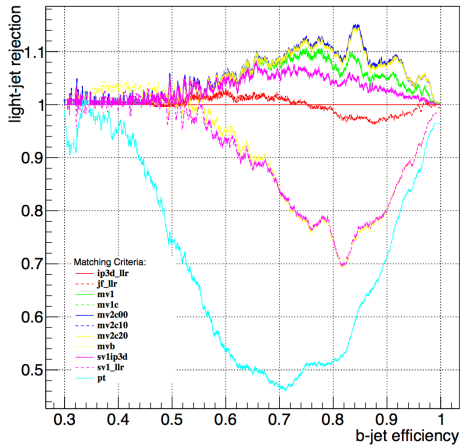
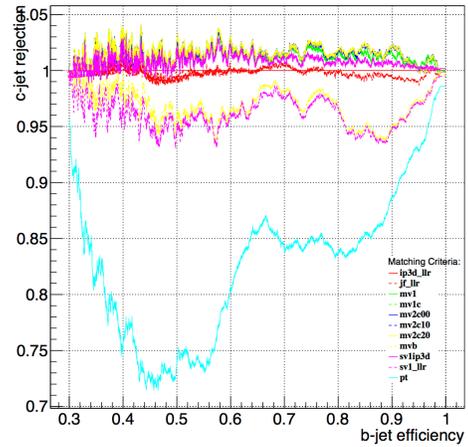

(c) Ratio of ROC curves for $b$ versus light calorimeter jet discrimination for each matching criterion to the performance obtained by selecting the track jet with the highest IP3D score as a baseline.

(d) Ratio of ROC curves for $b$ versus $c$ calorimeter jet discrimination for each matching criterion to the performance obtained by selecting the track jet with the highest IP3D score as a baseline.

**Figure 6.23:** Calorimeter tagging performance for jets with $p_T > 20$ GeV in the $|\eta| < 2.5$ region with more than one possible track jet match with $p_T > 5$ GeV in the $|\eta| < 2.5$ region. Various matching criteria are tested by letting the calorimeter jet inherit the $b$-tagging score from the track jet with the highest value of the selected matching criterion, and observing the outcome in terms of calorimeter jet tagging efficiency and background rejection. Selecting the track jet with the highest $p_T$ is found to be highly inefficient compared to using one of the $b$-tagging algorithms as a selection strategy. The MV2 taggers appear superior to others for both $c$ and light rejection tasks.

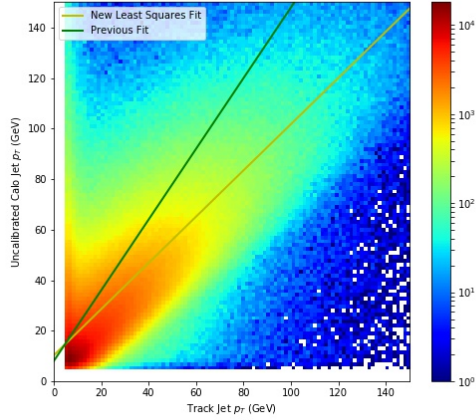

**Figure 6.24:** Two-dimensional histogram that highlights the relation between calorimeter jet transverse momentum and the transverse momentum of the matched track jet. In seeking a linear fit to rescale the track jet $p_T$ to the calorimeter jet scale, an improved set of parameters represented by the gold line is proposed as a replacement for the previous fit depicted by the green line. The relationship is clearly not exactly linear, with multiple low-$p_T$ track jets matched to very high-$p_T$ calorimeter jets, and viceversa (the extent of these issues in not fully captured by the selection $p_{T,\text{ trackjet}} < 150$ GeV chosen for this study and displayed in this plot).

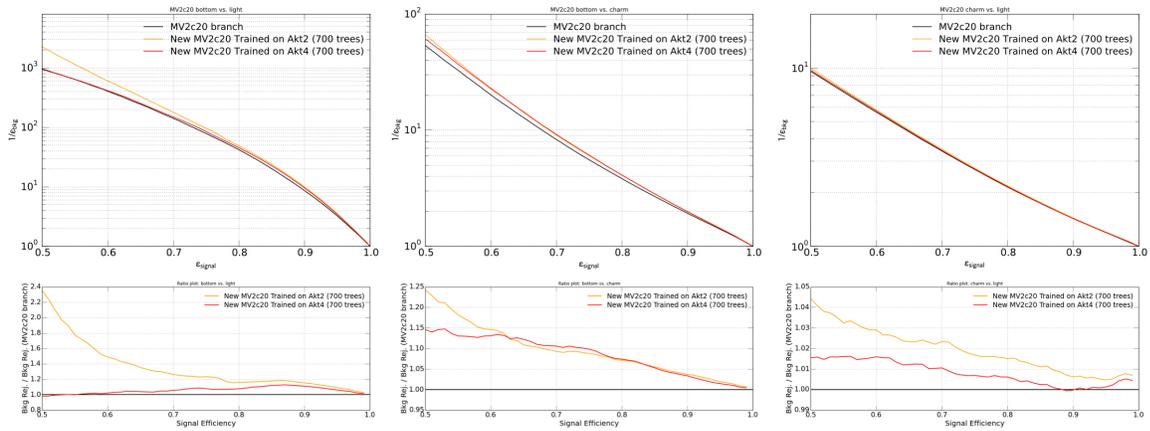

**Figure 6.25:** ROC curves for $b$-versus-light (left), $b$-versus-$c$ (center), and $c$-versus-light (right) track jet classification for MV2c20-like BDTs trained either on track jets from the $R = 0.2$ collection (orange) or $R = 0.4$ collection (red), and evaluated on a separate test set of $R = 0.2$ track jets. The two retrainings are compared to the MV2c20 tagger score obtained from the evaluation of the method originally optimized for calorimeter jets. The ratio plots quantify the relative improvement over the default MV2c20 tagger and justify the desire for collection-specific retrainings of $b$-tagging algorithms.

calorimeter jets, are evaluated on $R = 0.2$ track jets and visualized in the form of ROC curves and their ratios in Fig. 6.25. Visible improvements in rejection levels at all values of signal efficiency can be observed when applying track jet specific taggers. Even the application of a BDT trained on a different track jet collection is preferable to the default calorimeter jet MV2C20 tagger.

## 6.2 TOP TAGGING

Flavor tagging extends its range of applicability to the classification of several other topologies. Many exciting searches at the LHC look for high mass hadronic resonances resulting in collimated decay products reconstructed within the cone of large-radius jets. Among the topologies of interests are those characterized by the presence of top quark jets with high transverse momentum (*boosted*), making their correct reconstruction and identification crucial for many BSM analyses.

The recurrent task of identifying top jet events among the omnipresent QCD background of quark- and gluon-initiated jets has given rise to a well-established field of study that focuses on the joint theoretical and empirical effort of designing powerful and discriminative physics-driven observables to quantify the expected difference in jet substructure among different classes of jets. Strong theoretical knowledge of particle theory and phenomenology, coupled with the precise description of detector effects, has allowed physicists to theorize, simulate, and quantify the predictive power of expert-designed features that can be used individually or combined into high-level taggers in order to push the boundary of classification algorithms towards the information limit. An extensive list of references and an accessible review of developments of both theoretical jet-substructure quantities and applied techniques, with attention to machine learning methods, can be found in Ref. [347].

Boosted decision trees and neural networks are popular choices of learning algorithms for high-level top taggers that combine the power of theoretically motivated variables. Ref. [38] documents a recent ATLAS attempt at comparing the performance of two machine learning classifiers with a traditional baseline top tagger. In this study, hadronically decaying top quarks are obtained from the PYTHIA8 simulation of $Z' \to t\bar{t}$ samples at various resonant mass values in the range [0.4, 5] TeV. Background dijet samples are simulated in slices of $p_T$, and are later combined to produce an unphysical but balanced distribution of events across the $p_T$ spectrum in the range [0.2, 2] TeV. Event reweighting can be applied to restore the physical exponentially decaying distribution. The large-radius jets selected for this task are reconstructed from LCW-calibrated calorimeter topoclusters using the anti-$k_t$ algorithm with radius parameter $R = 1.0$. Trimming (see Sec. 5.2.1) is applied to remove jet sub-components from pile-up, using $f_\text{cut} = 5\%$. Only the leading and sub-leading sub-jet (the two highest $p_T$ jets) per event that pass $|\eta|$ and $p_T$ cuts are selected and calibrated. Reconstructed jets are $\Delta R$-matched to truth jets and particles. The events are reweighted so that the truth jets' transverse momentum distributions are uniform, and the signal jets' $p_T$ distribution matches that of the background jets.

A shared set of candidate input variables to the classifiers, listed in Ref. [38], is slimmed down by iterative variable selection studies based on empirical measurements, which results in different final input sets for the BDT and NN. A similar strategy is employed for model hyper-parameter selection. The final, optimized configurations for the BDT, constructed using the `TMVA` library, and the feed-forward neural network, implemented using `Keras`, are trained for binary discrimination and evaluated on a separate test set, using the input variables optimized for each method. Prior to testing, a jet mass cut is manually applied to favor compatibility with the traditional analysis strategy. The background rejection results, compared to the scan on the single 3-subjettiness variable $\tau_{32}$ [402], which is known to be one of the most powerful discriminating observables for top tagging, are displayed in Fig. 6.26 in bins of $p_T$, after a fixed threshold has been imposed on the classifier output corresponding to 50% TPR. All machine learning methods trained on the two different sets of input features show comparable performance and improve upon the $\tau_{32}$ cut across the entire transverse momentum spectrum.

However, contrary to traditional taggers, the BDT and NN examined in Ref. [38] significantly alter the mass distribution of background jets that pass the tagger selection, as evident in Fig. 6.27. In other words, although the mass is not used as input to the classifiers, the two methods still manage to pick up on it and make use of this highly discriminative variable through other observables' correlations with it. This makes the two classifiers very efficient at rejecting background jets that don't possess a mass compatible with the signal peak. However, this is an undesirable effect; in fact, by sculpting the background distribution to fall within the signal mass window, these two methods also reduce the opportunity of using sidebands to constrain the background shape and assess its contribution in the signal region in downstream analysis steps, thus making the background contamination almost irreducible. Adversarial techniques [400] can be adopted to reduce the unwanted classifier selection bias and dependence on this jet trait, as explained in Sec. 6.3 in the context of $W$ tagging [39].

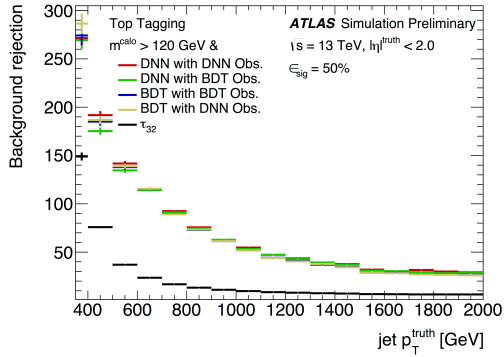 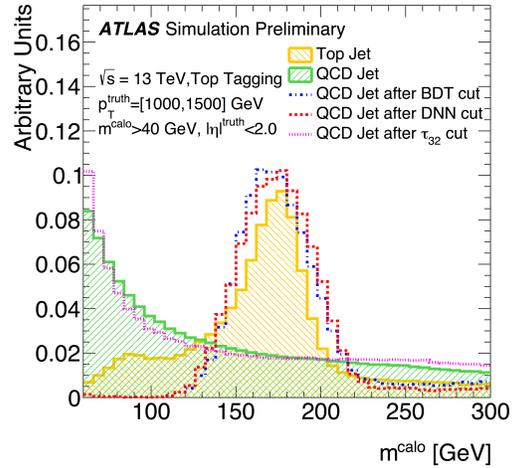

**Figure 6.26:** Background QCD jet rejection levels in bins of $p_T$ for the tagger obtained by scanning over the $\tau_{32}$ variable (black), the outputs of the BDTs trained on variables optimized for the BDT itself (blue) and for the NN (yellow), and the outputs of the NN trained on variables optimized for the NN itself (red) and for the BDT (green), after applying a cut on the discriminator output that correctly identifies 50% of the top jets. Image reproduced from Ref. [38].

**Figure 6.27:** Jet mass distribution for signal (yellow) and background (green) events prior to the application of any top tagger, as well as the modified background distributions after the application of the $\tau_{32}$ cut (pink), the BDT classifier cut (blue), and the NN classifier cut (red). All histograms are normalized to unit area for shape comparison, though the events that pass the various tagger cuts are small subsets of the original events. Image reproduced from Ref. [38].

As an alternative to training neural networks on top of a heavily engineered input space, one can design architectures that are able to process low-level representations of the radiation pattern within jet areas. Approaches such as the one documented in Ref. [403] examine the jet topology by constructing feed-forward neural networks that take as inputs the four-momenta of jet constituents. Computer vision solutions that make use of convolutional, locally-, or fully-connected layers have also appreciably increased their popularity due to their substantial performance gains and automation capabilities [359, 362].

## 6.3 Boson Tagging

Another crucial jet tagging task at the LHC addresses the identification and correct classification of jets initiated by the (possibly collimated) hadronic decay products of weak or Higgs bosons. There exist several tagging problem definitions involving bosons, given the variety of analyses and event topologies in which they appear and the respective forms of background sources. For example, one might first want to select boosted boson jets from the abundant QCD background, which can be performed by investigat-ing jet-substructure properties, and then learn to distinguish hadronically decaying $Z$ bosons from $W$ bosons, which can be carried out, for example, by building a likelihood from discriminative observables such as the jet mass, charge, and associated $b$-tagging score [404]. The identification of boosted boson large-radius jets and their separation from QCD jets has, in fact, traditionally been performed through the design and combination of physics-motivated observables that make use of the knowledge of expected differences in jet substructure and flavor composition. These variables are constructed to expose the most salient features that differentiate the radiation patterns within jets of different origins. For an in-depth review of jet-substructure techniques, see Ref. [347].

A recent ATLAS study on the competitiveness of boosted decision trees and neural networks that make use of known substructure variables for boosted $W$ tagging versus QCD is presented in Ref. [38]. The simulation, event, and object selections resemble the ones summarized in Sec. 6.2 for top tagging, with the exception that boosted $W$ bosons are obtained from the simulation of BSM $W' \to WZ \to qqqq$ events. Performance improvements over the baseline tagger provided by a scan over the single physically-motivated observable $D_2$ [405], shown in Fig. 6.28, are present yet reduced in magnitude compared to those observed in the top tagging scenario.

The undesirable mass sculpting that occurs when machine learning classifiers display strong bias along a certain dimension (in this case, the jet mass, as shown in Fig. 6.29), can be reduced by the addition of adversarial components to the taggers' training [400]. This is the subject of the ATLAS work presented in Ref. [39], which successfully applies adversarial training of neural networks for mass decorrelation in $W$ tagging. Robust background estimation is severely impacted by tagger outputs that are highly correlated with jet mass and that remove the majority of events in the sideband region. This occurs because the classifier is able to indirectly learn biases in the training distributions: specifically, the mass spectra of signal and background jets are not equally populated, as all signal jets fall in a restricted mass region centered around the nominal signal resonance mass.

Mass decorrelation then becomes a competing objective that needs to be balanced with the pure classification performance quantified by ROC curves. Striking the right compromise between the two remains a subjective task. From a quantitative standpoint, mass bias in a tagger is computed in terms of the Jensen-Shannon divergence between the normalized distribution of background jets that are selected by the classifier and those that are rejected.

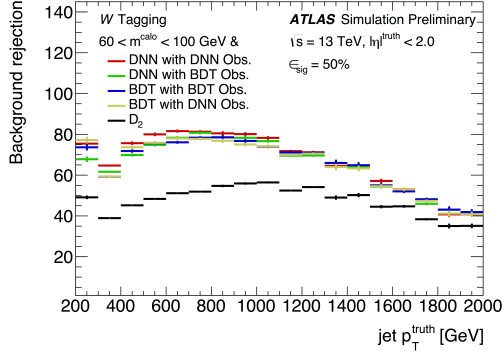 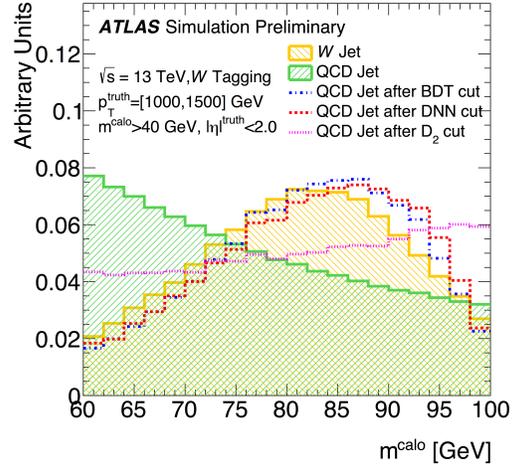

**Figure 6.28:** Background QCD jet rejection levels in bins of $p_T$ for the tagger obtained by scanning over the $D_2$ variable (black), the outputs of the BDTs trained on variables optimized for the BDT itself (blue) and for the NN (yellow), and the outputs of the NN trained on variables optimized for the NN itself (red) and for the BDT (green), after applying a cut on the discriminator output that correctly identifies 50% of the $W$ jets. Image reproduced from Ref. [38].

**Figure 6.29:** Jet mass distribution for signal (yellow) and background (green) events prior to the application of any $W$ tagger, as well as the modified background distributions after the application of the $D_2$ cut (pink), the BDT classifier cut (blue), and the NN classifier cut (red). All histograms are normalized to unit area for shape comparison, though the events that pass the various tagger cuts are small subsets of the original events. Image reproduced from Ref. [38].

The adversarial neural network setup for mass decorrelation in taggers, first implemented within this context by Ref. [258], consists of a standard classifier tasked with learning a high dimensional separation between the jet classes, and an adversary tasked with inferring the mass of a jet from its classifier output. If the adversary were to succeed, it would provide evidence for the existence of a correlation between the classifier output and mass. The joint objective is then to train a competitive classifier whose output is not telling of the mass of the input example, keeping that information inaccessible to the adversary. Formally, in the adversarial framework with a classifier network parametrized by weights $\theta_{\text{clf}}$ and an adversarial network parametrized by weights $\theta_{\text{adv}}$, the problem can be phrased as the two-player minimax game

$$\min_{\theta_{\text{clf}}} \max_{\theta_{\text{adv}}} L_{\text{clf}}(\theta_{\text{clf}}) - \lambda L_{\text{adv}}(\theta_{\text{clf}}, \theta_{\text{adv}}), \qquad (6.6)$$

where $L_{\text{clf}}$ is the binary cross-entropy classification loss, $L_{\text{adv}}$ is the mass reconstruction loss, which depends on the specific nature and implementation of the adversary, and $\lambda$ is a hyper-parameter with use case-driven value that regulates the relative importance of the two competing objectives. When trained with the adversarial component, the neural network loses competitiveness in the classification task, but gains the desirable property of preserving the original background shape, as shown in Fig. 6.30 for the tagger developed in Ref. [39]. Another visualization method, provided in Fig. 6.31, consists of verifying that the FPR becomes more uniform across jet mass thanks to the adversarial component, as opposed to growing disproportionately around certain mass values, as typical of standard taggers. Strategies for decorrelating other types of non-neural-network-based taggers are also presented in Ref. [39].

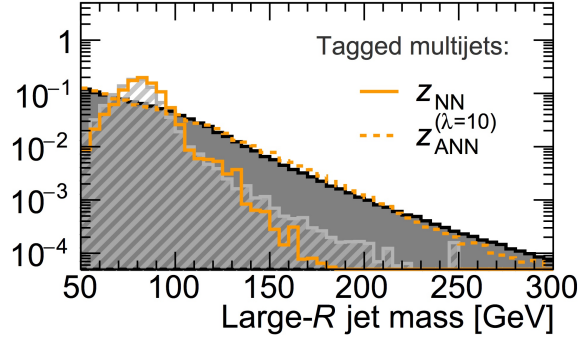

**Figure 6.30:** Normalized jet mass distribution for background jets (dark gray) and signal $W$ jets (striped light gray) before the classifier selection. A cut is then applied to the classifiers' outputs at the value corresponding to TPRs of 50%. The normalized distribution of background jet masses for jets that pass the regular neural network tagger cut is shown by the solid orange line, while the normalized distribution of background jet masses for jets that pass the adversarial neural network tagger cut is shown by the dashed orange line. Image reproduced from Ref. [39].

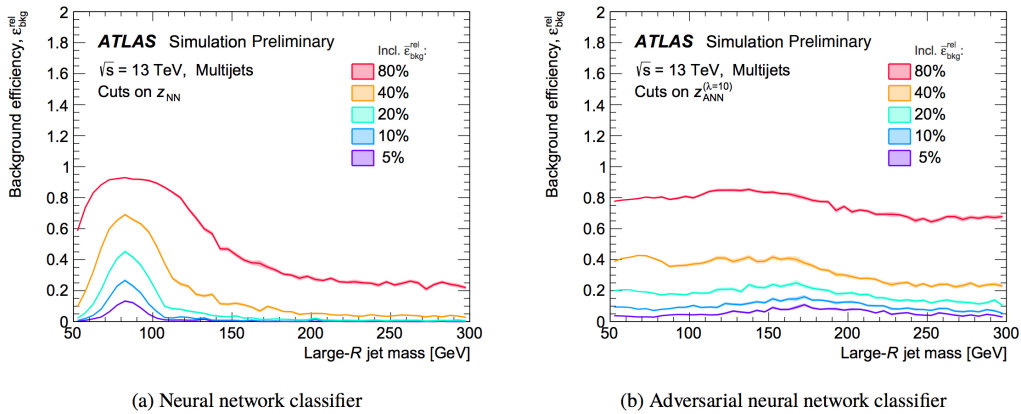

(a) Neural network classifier

(b) Adversarial neural network classifier

**Figure 6.31:** False positive rate as a function of jet mass for the regular (a) and adversarial (b) neural networks at different values of the true positive rate. Image reproduced from Ref. [39].

To complement machine learning techniques that make use of engineered substructure variables, novel deep learning techniques acting on raw energy deposits within jets have shown promise towards recovering the knowledge synthesized by the high-level physics observables and extracting even more information from a low detector-level data representation. The radiation pattern within jets can be visualized in the form of jet images (see Sec. 5.2.1.2.4), two dimensional representations of the energy deposits

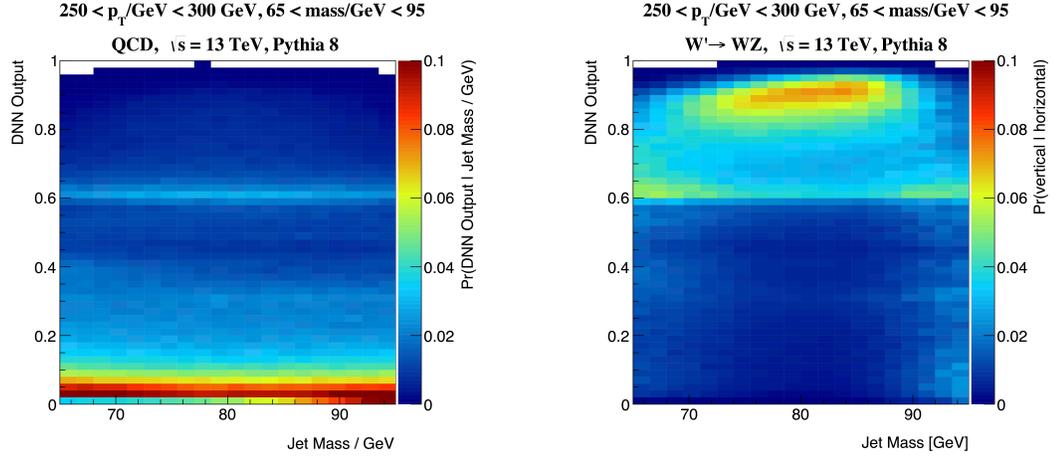

(a) Convnet output versus jet mass for QCD jets

(b) Convnet output versus jet mass for $W$ jets

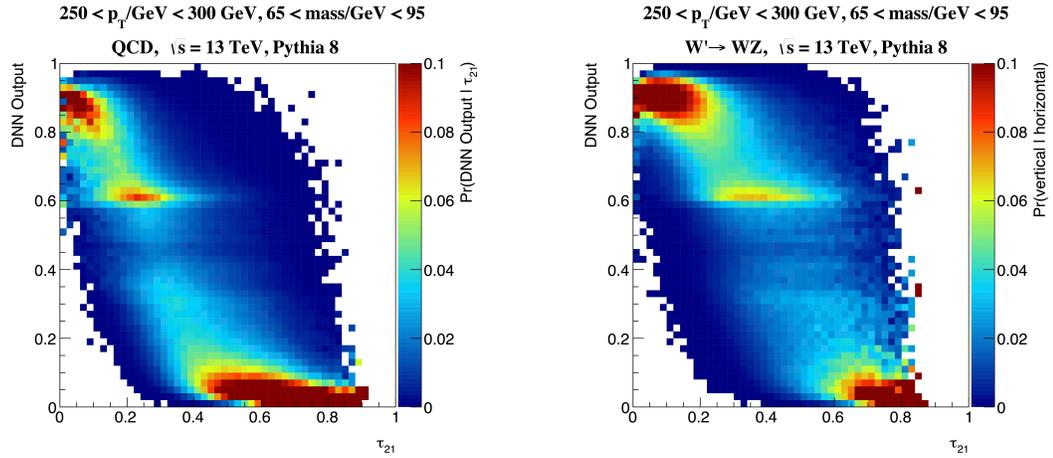

(c) Convnet output versus $\tau_{21}$ for QCD jets

(d) Convnet output versus $\tau_{21}$ for $W$ jets

**Figure 6.32:** Distribution of the convnet output in each bin of the mass and $\tau_{21}$ distributions for background and signal jets, normalized to unity in each bin of the jet property. In general, the neural network is able to assign a very low (high) score to QCD ($W$) jets regardless of their mass, with only a small visible bias for $W$ jets that fall in the signal mass window. This implies that mass is not an important variable that the network learns to use when classifying jets. On the other hand, $\tau_{21}$ strongly influences the network's decision making strategy, with low (high) $\tau_{21}$ jets often being assigned to the signal (background) class, regardless of their true nature. Images reproduced from Ref. [30].

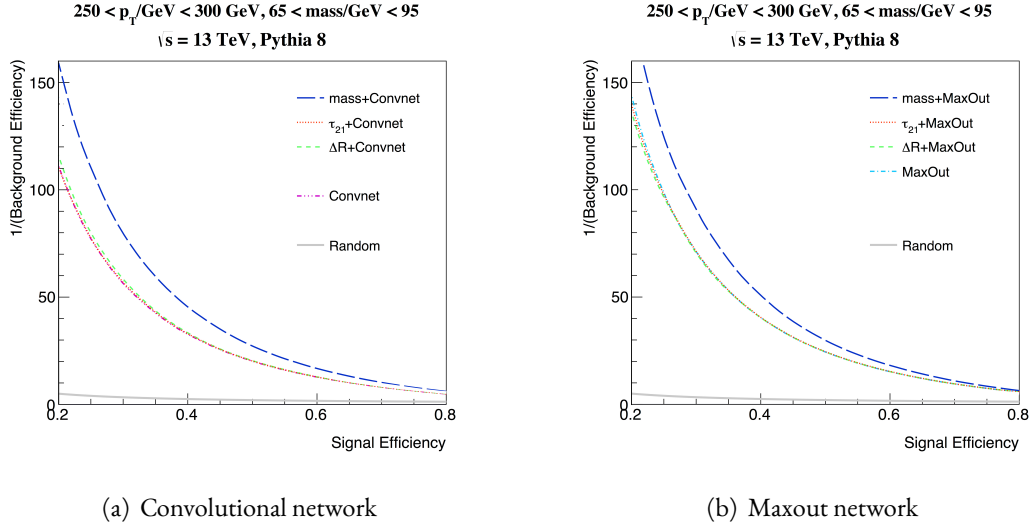

(a) Convolutional network  (b) Maxout network

**Figure 6.33:** ROC curves obtained from the 2D likelihood combination of the output of the convnet 6.33(a) or maxout networks 6.33(b) and one the three physics-driven discriminative observables (jet mass, 2-subjettiness, and $\Delta R$ between sub-jets). Only the addition of mass improves performance. Images adapted from Ref. [30].

in an $\eta$-$\phi$ grid that mimics the granularity of calorimeter detectors. Principle Components Analysis or Fisher's Linear Discriminant can be used as baseline models to build a discriminant starting from the image representation provided by jet images [356]. Fully-connected and convolutional neural networks are found to achieve higher class separation when applied to the input pixel space. They are able to build highly non-linear deep transformation of the inputs and extract more powerful hidden features that enhance the performance of boosted boson versus QCD classifiers [30].

Beyond performance, one of the many advantages of these approaches is that they make use of a natural and intuitive data format, which bypasses the need for the construction of theoretically-motivated engineered features, while leaving the possibility open for a post-facto information-theoretic validation and comparison between the neural network output or weights and known discriminative variables. In addition, within the context of the recent trend of designing mass-decorrelated taggers, the jet image classifiers in Ref. [30] are not observed to be able to fully learn the mass distribution of the input jets starting from their low-level pixel representation. This equips them with the perhaps fortuitous yet attractive property of lacking a strong mass dependence. This is corroborated by the 2D distributions in Fig. 6.32, in which the neural network's output does not exhibit any strong bias at different jet mass values. On the other hand, one can observe a clear direct correlation between the network's output and powerful substructure variables such as $\tau_{21}$. This endorses the theory that deep neural networks act-ing on low-level detector readouts could successfully be used to bypass the traditional time-consuming feature-engineering step, as they are able to automatically learn useful hidden features that encompass the ones experts would construct. Another way of substantiating these claims is to combine the neural nets' outputs with jet properties (such as mass, 2-subjettiness, or the distance between the sub-jets) in a 2D likelihood to verify whether the addition of these known discriminative variables improves upon the network's performance, or if the information they provide has already been fully recovered by the neural network. The ROC curves corresponding to these combinations can be found in Fig. 6.33. They again show that while there is no clear benefit in adding $\tau_{21}$ and $\Delta R$ information, which the networks automatically learns, the addition of jet mass outwardly improves the performance from a ROC curve standpoint, but simultaneously introduces a possibly unwanted dependence on this variable. Although, in hindsight, not extracting discriminative power from jet mass may appear as an advantage in the broader context, it begs the question of what other useful information the deep neural networks in Ref. [30] are incapable of learning, and how well they can be expected to approximate the upper bound of expected performance on this task.

For comparison, a similar task is tackled in Ref. [361], in which case the neural network acting on the input image space does develop a strong mass dependence. However, this can be partially explained by the absence of a window cut around the signal mass region in the preprocessing phase. This leads to more pronounced differences in the proportions of signal and background jets away from the signal peak region, which ultimately forces the neural network to have to learn to apply that cut itself, thus significantly impacting the shape of the distribution.

## 6.4 Quark-Gluon Tagging

The discrimination of light quark- from gluon-initiated jets is yet another active area of research, where machine learning has shown promising performance improvements.

Intuitively, the low-energy particle multiplicity in gluon jets should be higher than in quark jets, because of the values of the Casimir color factors that appear in the DGLAP splitting functions: $C_A = 3$ for gluon splitting from a gluon, and $C_F = 4/3$ for gluon emission from a quark (see Sec. 1.1.2.2.1). Experimentally, the measurable number of tracks in a jet serves as a proxy for particle multiplicity, and is consequently often employed by LHC experiments, along with other jet observables sensitive to parton fragmentation characteristics, in order to construct classifiers to distinguish quark jets from gluon jets using likelihoods or other machine learning methods [406, 407, 408, 409, 410, 411].

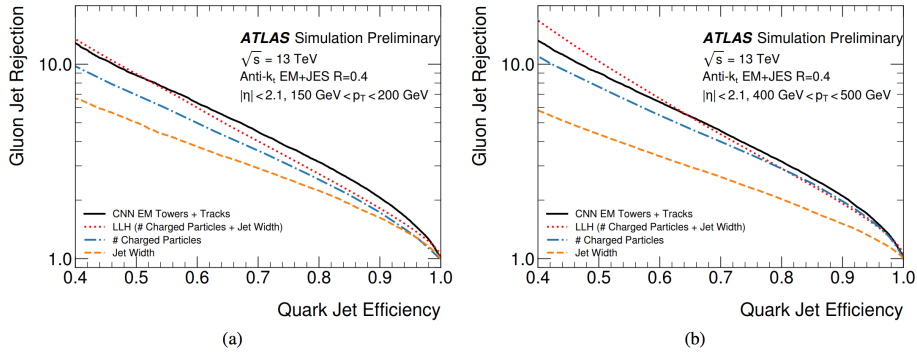

**Figure 6.34:** ROC curves representing the gluon jet rejection versus quark jet efficiency for various quark-gluon tagging methods, evaluated on jets in the $p_T$ range [150, 200] GeV (a) and [400, 500] GeV (b). Images reproduced from Ref. [40].

As in the case of top and boosted boson tagging, quark versus gluon tagging can also be performed by probing the energy depositions within jets represented as jet images. The convolutional neural network developed in Ref. [40] achieves comparable or slightly improved performance compared to standard approaches based on physics-driven observables, as shown in Fig. 6.34. A performance enhancement due to the use of convolutional networks over physics observables or simpler machine learning methods is also documents in Ref. [358], where the authors suggest augmenting the conventional jet image representation by creating an RGB-like input that separates charged particles' $p_T$, neutral particles' $p_T$, and charged particle multiplicity across the three different image channels.

Across many object identification tasks at the LHC, deep learning is emerging as a valiant substitute for traditional multi-dimensional likelihood methods, thanks to its ability of scaling to higher input dimensions and reducing the need for time-consuming, manual feature discovery. While architecture optimization seems to be the main engineering focus for the community at the moment, the recent conversation around data representation is likely to catalyze breakthroughs towards more efficient and responsible information extraction from the data collected by the ATLAS experiment.

# 7

# Search for Resonant and Enhanced Non-Resonant di-Higgs Production in the $\gamma\gamma b\bar{b}$ Channel

Measuring the physical properties of the Higgs boson may shed light on the phenomenology of the Higgs sector, with important implications for many BSM theories (see Sec. 1.2 for a theoretical introduction). Its couplings to fermions and gauge bosons have been measured with increasing confidence over the past few years, demonstrating consistency with SM predictions [412]. Continuing to probe the structure of the Higgs potential, including the terms involving its trilinear self-coupling, which is accessible through the measurement the di-Higgs production rate, is of fundamental relevance for electroweak physics and is still an outstanding task for researchers at the LHC.

The trilinear interaction term emerges from EWSB and gives rise to a vertex where a Higgs boson splits into two Higgs bosons. The strength of this interaction can be experimentally measured at the LHC by looking for signatures of the decay products of the two Higgs bosons produced at this vertex. Although the di-Higgs SM cross-section is too low to be observed at the current luminosity levels, setting limits on the di-Higgs production rate allows to constrain, at a given confidence level, the phase space for anomalous couplings and new interaction vertices with an outgoing Higgs pair induced by new physics models.

The analysis presented here, also published in Ref. [413], searches for $gg$-initiated di-Higgs events in the $\gamma\gamma b\bar{b}$ decay channel. This channel takes advantage of the clean reconstruction of the narrow mass peak in the diphoton channel, while increasing the event rate by relying on the dominant branching ratio of the second Higgs boson to $b\bar{b}$. The branching ratios for the Run I measured value of the Higgs mass $m_h$ = 125.09 GeV are reported in Ref. [66] to be: BR($h \to b\bar{b}$) = 5.809 × $10^{-1}$ $^{+0.65\%}_{-0.65\%}$ (theory) $^{+0.72\%}_{-0.74\%}$ (parametric uncertainties from the quark masses) $^{+0.77\%}_{-0.79\%}$ ($\alpha_S$), and BR($h \to \gamma\gamma$) = 2.270 × $10^{-3}$ $^{+1.73\%}_{-1.72\%}$ (theory) $^{+0.97\%}_{-0.94\%}$ (parametric uncertainties from the quark masses) $^{+0.66\%}_{-0.61\%}$ ($\alpha_S$).

The analysis focuses on resonant production via a narrow-width massive scalar $X$ in the range $m_X \in$ [260, 1000] GeV, as well as non-resonant production, which could deviate from the expected SM rate due to BSM alterations to the trilinear Higgs self-coupling $\lambda_{hhh}$.

It supersedes previous ATLAS di-Higgs searches in the $\gamma\gamma b\bar{b}$ channel (see Sec. 7.7.1 and 7.7.3) and extends the reach of the joint ATLAS and CMS multi-channel di-Higgs program (see Sec. 7.7.4).

## 7.1 Data

This analysis looks for di-Higgs events in the 36.1 fb$^{-1}$ of $pp$ collision data that were delivered to the ATLAS detector by the LHC in the 2015-2016 data-taking periods and that pass the data quality assessment checks [414, 415]. The average number of interactions per bunch crossing in the data delivered to ATLAS in that period is $\langle \mu \rangle$ = 23.7. For an introduction to data collection and storage in ATLAS, see Sec. 2.3.

Events are considered if they pass the `HLT_g35_loose_g25_loose` diphoton trigger, which identifies events with at least two reconstructed photons with transverse energy greater than 35 GeV, for the leading photon, and 25 GeV, for the sub-leading photon. This trigger is seeded by the `L1_2EM15VH` Level-1 trigger, which flags the presence of two EM clusters with transverse energy > 15 GeV passing a veto on hadronic activity. The efficiency of the HLT diphoton trigger was measured in 2012 data to be 99.50 ± 0.15% [416]. The regions most affected by trigger inefficiencies are those at high-$|\eta|$ and the ∼ 5 GeV range just above the trigger $p_T$ threshold, in which performance quickly ramps up before plateauing to the expected level [416].

## 7.2 Simulated Samples

Only the dominant $gg$ production mechanism is considered when generating signal samples for this analysis. Simulators rely on Monte Carlo methods for the production of a SM di-Higgs sample, as well as several BSM resonant $gg \to X \to hh$ samples for $m_X \in \{260, 275, 300, 325, 350, 400, 450, 500, 750, 1000\}$ GeV, in the narrow-width approximation.

The SM di-Higgs signal is generated at NLO in QCD with the FT$_{\text{approx}}$ approximation, which accounts for the full top quark mass dependence in the real radiation, but uses Higgs effective field theory (HEFT) in the virtual part, rescaled by the ratio $B_{\text{FT}}/B_{\text{HEFT}}$ of the full theory LO contribution to the HEFT ($m_t \to \infty$) LO result [66]. Events are reweighted to the full theory to account for finite top quark mass effects, which affect the strength of the interference between triangle and box-like diagrams. Further reweighting is needed to match the total cross-section calculated at NNLO in QCD. This has recently been computed to be 31.02 fb$^{+2.2\%}_{-5.0\%}$ ± 2.6% ($m_t$) ±3.0% (PDF + $\alpha_S$) for $m_h = 125.09$ GeV at $\sqrt{s} = 13$ TeV [69], but was previously known to equal 33.41 fb$^{+4.3\%}_{-6.0\%}$ ± 5.0% (theory) +2.3% ($\alpha_S$) ±2.1% (PDF) at the moment the samples were produced [417]. While the cross-section decreases from a central value of approximately 32 fb at NLO HEFT to approximately 29 fb at NLO FT$_{\text{approx}}$ and approximately 28 fb at NLO FT$_{\text{exact}}$, large QCD corrections result in large $K$ factors relating the HEFT cross-section at different orders, with a cross-section increase of ∼ 20% going to NNLO [66, 69]. The samples are generated using aMC@NLO interfaced with Herwig++ for showering, with the CTEQ6L1 and CT10ME PDF sets for matrix elements and parton showers, and the UEEE5 underlying-event tune. The generator-level output is then processed using fast detector simulation.

In addition, for the purpose of interpretation, 10 samples with modified Higgs self-coupling are produced. The values of the self-coupling parameter $\kappa_\lambda = \lambda_{hhh}/\lambda_{\text{SM}}$ considered for this study are $\{0, \pm 1, \pm 2, \pm 4, \pm 6, \pm 10\}$. They are generated at LO in QCD with MadGraph5 and rescaled by a $K$-factor of 2.283 to reweight the cross-section to the NNLO value. They are showered using Pythia8 with the A14 tune, using the NNPDF 2.3 leading order PDF set, and then passed through fast detector simulation. When $\kappa_\lambda = 0$, the Higgs self-coupling vanishes, so that only the box diagram contributes to the di-Higgs production cross-section. Removing the interference caused by the triangle diagram increases the cross-section by a factor of $\sim 2$ compared to the SM value. On the other hand, the interference is maximal around $\kappa_\lambda \approx 2$, where the cross-section tends to zero. Different self-coupling values would also manifest themselves in modified differential distributions. Analytical relations among can be derived and used to interpolate and extrapolate to other $\kappa_\lambda$ values.

The BSM resonant di-Higgs production samples are generated at approximate NLO with MADGRAPH5_AMC@NLO in the gluon-gluon fusion production channel, using a narrow-width (10 MeV) assumption for the heavy scalar $X$ decaying to two SM-like Higgs bosons with $m_h$ = 125 GeV. The samples are showered with HERWIG++ using the CTEQ6L1 PDF set and the UEEE5 tune, and later interfaced with the ATLAS fast simulation framework to approximate detector effects.

In order to extract the shape of the dominant non-resonant background distribution, $\gamma\gamma$+jets events are generated at LO with SHERPA and the CT10 PDF set in uneven bins of $m_{\gamma\gamma}$. The overall normalization factor that determines the numerical contribution of this background to the expected number of events is estimated by template matching the MC histogram to the corresponding observed events in the diphoton sideband regions, and attributing any deficiency to the contribution of jets faking photons.

Various single Higgs production mechanisms, including $ggh$, $Zh$, $t\bar{t}h$, $th$, VBF $h$, $Wh$, $b\bar{b}h$, contribute to the background of this analysis. These samples are produced using the generators and PDF sets listed in Table 7.1, and the result of the interaction with the detector is computed using full simulation.

Pile-up contributions are simulated using the A2 tune of PYTHIA 8.186 with the MSTW2008LO PDF set, and overlaid onto generated events to resemble the conditions of the 2015-2016 data-taking campaigns.

## 7.3 Object and Event Selection

A cutflow is a series of physically meaningful, sequential, univariate cuts on object or event observables that partition events into a series of regions based on whether they pass the designed selection criteria. Physics objects are only considered if they pass certain kinematic and reconstruction quality thresholds, and events might be rejected based on ensemble properties of their physics objects or on event shape variables, which separate potential signal candidates of physical processes of interest from those that are likely to originate from uninteresting background processes. Given the decay mode chosen for this analysis, the goal of the cutflow is to carve out a region of events with at least two photons and at least two jets that pass specific object selection criteria.

| Process | Generator | Showering | PDF set | $\sigma$[fb] | Order of calculation of $\sigma$ | Simulation |
|---|---|---|---|---|---|---|
| Non-resonant SM hh | MadGraph5_aMC@NLO | Herwig++ | CT10 NLO | 33.41 | NNLO+NNLL | Fast |
| Non-resonant BSM hh | MadGraph5_aMC@NLO | Pythia 8 | NNPDF 2.3 LO | - | LO | Fast |
| Resonant BSM hh | MadGraph5_aMC@NLO | Herwig++ | CT10 NLO | - | NLO | Fast |
| $\gamma\gamma$ + jets | Sherpa | Sherpa | CT10 NLO | - | LO | Fast |
| $ggh$ | Powheg-Box NNLOPS (r3080) | Pythia 8 | PDF4LHC15 | 48520 | N$^3$LO(QCD)+NLO(EW) | Full |
| VBF | Powheg-Box (r3052) | Pythia | PDF4LHC15 | 3780 | NNLO(QCD)+NLO(EW) | Full |
| $Wh$ | Powheg-Box (r3133) | Pythia | PDF4LHC15 | 1370 | NNLO(QCD)+NLO(EW) | Full |
| $q\bar{q} \to Zh$ | Powheg-Box (r3133) | Pythia 8 | PDF4LHC15 | 760 | NNLO(QCD)+NLO(EW) | Full |
| $tt\bar{h}$ | MadGraph5_aMC@NLO | Pythia 8 | NNPDF3.0 | 510 | NLO(QCD)+NLO(EW) | Full |
| $gg \to Zh$ | Powheg-Box (r3133) | Pythia 8 | PDF4LHC15 | 120 | NLO+NLL(QCD) | Full |
| $bb\bar{h}$ | MadGraph5_aMC@NLO | Pythia | CT10 NLO | 490 | NNLO(5FS)+NLO(4FS) | Full |
| $t$-channel $th$ | MadGraph5_aMC@NLO | Pythia 8 | CT10 NLO | 70 | LO(4FS) | Full |
| $W$-associated $th$ | MadGraph5_aMC@NLO | Herwig++ | CT10 NLO | 20 | NLO(5FS) | Full |

**Table 7.1:** List of parameters used for the simulation of MC samples produced for the ATLAS di-Higgs search in the $\gamma\gamma b\bar{b}$ decay channel with $36.1$ fb$^{-1}$ of $pp$ collision data from the 2015-2016 LHC runs. The top portion lists the three types of signal the analysis is tuned to explore. The dominant background is simulated in the non-resonant photons + jets sample in the middle section of the table. Single Higgs simulated samples, organized by production mode, and used in the analysis to account for this component of the background, are listed in the bottom portion of the table. The third-to-last column indicates the cross-section computed for a Higgs boson with $m_h = 125.09$ GeV at $\sqrt{s} = 13$ TeV.

The object and event selections are implemented as a cutflow analysis in the `HHyybbTool`, which is part of the larger `HGamAnalaysisFramework` shared across multiple Higgs analyses in the diphoton channel.

### 7.3.1 Object Reconstruction

Photons are reconstructed and identified from other electromagnetically interacting physical objects as explained in Sec. 5.2.2. The photons used in this analysis are reconstructed with an efficiency of $97\%$. They are classified into converted and unconverted photons, depending on the presence of a secondary vertex and tracks compatible with a $\gamma \to e^+e^-$ conversion event. Energy is associated to photon candidates by summing the calibrated EM energy deposits in the calorimeter cluster.

A neural network re-analyzes the event to select the optimal primary vertex from which the diphoton pair originated. This method selects the correct primary vertex over $85\%$ of times. Objects' kinematic variables are then translated to be expressed with respect to the newly selected origin.

Hadronic showers from the dijet system are reconstructed as jets starting from topological clus-ters [25] in the calorimeter using the FASTJET library [334]. Jets are defined as clusters of hadronic activity in nearby cells by drawing a cone of radius $\Delta R = 0.4$ built using the anti-$k_t$ algorithm [332]. Jets are calibrated to account for different responses in various calorimeter regions [418].

The kinematics of jets are corrected by adding the four momenta of all muons with $p_T > 4$ GeV found within a radius of $\Delta R = 0.4$ of the jet axis. This procedure is implemented to recover higher fidelity on the reconstructed properties of $b$-jets with leptonic decays, where the jet is accompanied by a muon. The sum of the jet and muon contributions increases the reconstruction resolution of the $b$-hadron's kinematics.

The muons considered for jet kinematics corrections must pass the medium quality criteria [374], and be reconstructed within the $|\eta| < 2.7$ acceptance region. Additionally, the muons are required to have transverse impact parameter significance $S_{d_0} < 3$ and longitudinal impact parameter $|z_0| < 0.5$ mm, both measured with respect to the PV.

Alternative muon correction strategies are evaluated but ultimately discarded based on the improvement on the resolution of the dijet and four-body mass. For example, we explored the option of only adding the contribution from the highest-$p_T$ muon, or from the closest muon to the jet axis. Regardless, the majority of jets contain no muon or at most one muon, with two muons only appearing in a small fraction of events, as shown in Fig. 7.1. Therefore, in the majority of cases, all three jet correction strategies investigated in this analysis are equivalent. More accurately reconstructing the properties of the $b$-hadron results in a more precise and more accurate reconstruction of the dijet mass, with a visible shift in the peak of the distribution towards the Higgs mass, as shown in Fig. 7.2.

Electrons are reconstructed for the explicit goal of rejecting jets that overlap with them. The medium electron ID working point and the loose isolation working point are selected. Electrons are further required to satisfy: $p_T > 10$ GeV and $|\eta| < 2.47$.

### 7.3.2 OBJECT SELECTION CUTS

All optimization studies and decisions are made based on cross-section values set to: (SM $hh$ production cross-section) $\times$ BR($hh \to \gamma\gamma b\bar{b}$) = 33.41 fb $\times 2 \times 2.270 \times 10^{-3} \times 5.809 \times 10^{-1} = 0.08811$ fb for the non-resonant analysis, and (5 pb) $\times$ BR($hh \to \gamma\gamma b\bar{b}$) = 12.89 fb for the resonant analysis, where 5 pb corresponds to the resonant cross-section exclusion point obtained in the early Run II analysis (see

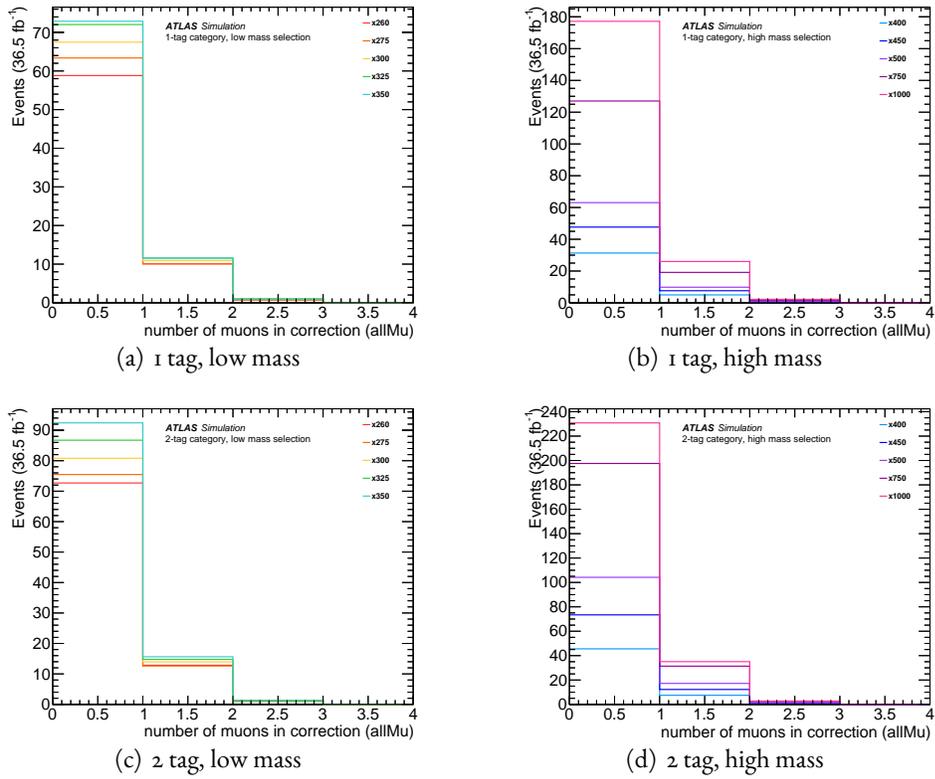

**Figure 7.1:** Number of reconstructed muons passing the selection criteria listed in the text found in each jet, for jets in events in the 1 tag category (top row) and 2 tag category (bottom row) that pass the low mass selection (left column) and the high mass selection (right column). The different colors distinguish the distributions of different resonant BSM signal samples.

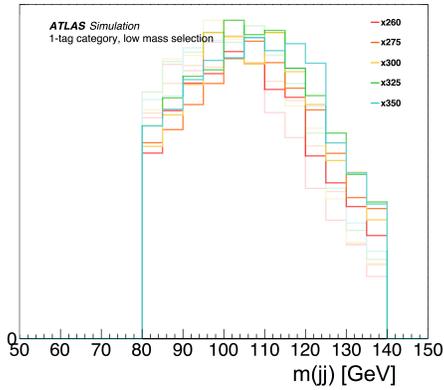
(a) 1 tag, low mass

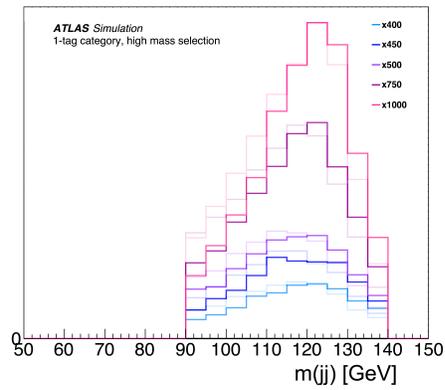
(b) 1 tag, high mass

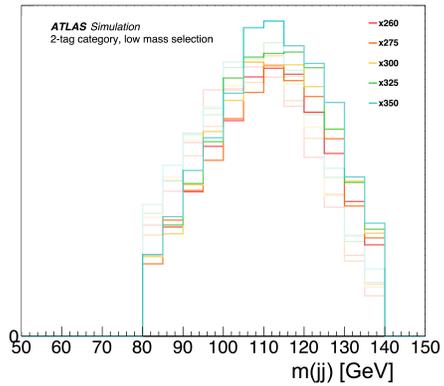
(c) 2 tag, low mass

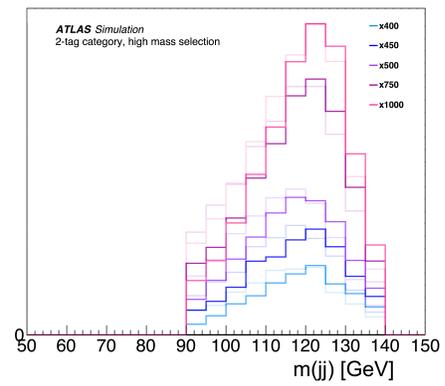
(d) 2 tag, high mass

**Figure 7.2:** Shift in dijet reconstructed mass for events in the 1 tag category (top row) and 2 tag category (bottom row) that pass the low mass selection (left column) and the high mass selection (right column). The different colors distinguish the distributions of different resonant BSM signal samples. The lighter histograms show the mass distributions pre-correction; the dark, solid histograms show the mass distributions post-correction. A clear shift toward higher reconstructed mass values, closer to 125 GeV on average, can be observed.

Sec. 7.7.3).

Photons with $p_T > 25$ GeV are selected by applying tight identification cuts to reject the majority of background events faking a photon signature. In order to ensure their quality, photon candidates are also screened for possible overlap with other objects: non-isolated photons are removed from the cutflow to decrease the contamination from misidentified jets. This selection is approximately 98% efficient. However, an orthogonal control region is created by considering events in which two photons pass loose selection criteria but not the tight identification and isolation ones. The isolation cuts look for activity in the vicinity of the photon both in the inner detector and in the calorimeter, and are parametrized as functions of the photon $p_T$. If the net transverse momentum of tracks with $p_T > 1$ GeV found within a $\Delta R$ radius of 0.2 from the photon but not associated to its conversion is greater than 5% of the photon's transverse momentum, or if the energy in the topo-clusters not associated to the photon but within a $\Delta R$ radius of 0.2 from it is greater than 6.5% of the photon's transverse momentum, then the photon candidate is removed. Only 2% of true photon candidates are erroneously removed by these cuts.

The choice of identification and isolation working points for each photon in the diphoton system is manually checked by calculating the Asimov significance obtained from the the cutflow applied to a couple of representative resonant samples, for all possible combinations of $b$-tagging efficiency working points applied to the dijet system.

In addition, an ambiguity resolution tool, shared across Higgs searches in the diphoton channel, is adopted in order to eliminate electrons that can be misidentified as photons, especially due to the similarity of their signature to that of converted photons. Further, photons falling outside of the acceptance region of the EM calorimeter ($|\eta| > 2.7$) or within the crack region ($1.37 < |\eta| < 1.52$) are removed. Finally, the cutflow selects the two highest-$p_T$ photons in each event that satisfy the requirement $p_T/m_{\gamma\gamma} > 0.35(0.25)$ for the leading (sub-leading) candidate, where $m_{\gamma\gamma}$ is the reconstructed mass of the two candidate-system.

The initial jet container returns all jets that are consistent with the diphoton primary vertex that have $p_T > 25$ GeV and Jet Vertex Tagger (JVT) [379] value $> 0.59$, if the jet falls within the kinematic region defined by $|\eta| < 2.4$ and $20$ GeV $< p_T < 60$ GeV. The JVT cut is found to retain approximately 92% of jets that originate from the diphoton vertex. Only central jets with $|\eta| < 2.5$ are considered, and an overlap-removal cut is applied to reject all jets within a $\Delta R = 0.4$ radius of a tight, isolated photon, or a $\Delta R = 0.2$ radius of an electron.

### 7.3.3 Event Selection Cuts

Two non-exclusive categories are defined, depending on the kinematic range of the signal events that they are optimized for: the *high mass* analysis, otherwise known as the tight kinematic selection, is designed to best identify SM non-resonant di-Higgs events and BSM resonant events with $m_X \geq 500$ GeV; the *low mass* analysis, otherwise known as the loose kinematic selection, is best suited to separate low mass BSM resonances from background contributions. A quick validation of the distribution of kinematic and substructure properties of high mass BSM resonances surfaces clear discriminative trends with respect to the features observed in low mass resonances: on average, higher mass $m_X$ corresponds to higher diphon, dijet, and jet $p_T$, and consequent decrease in jet-to-jet distance $\Delta R$ in the dijet system.

Both analysis workflows start with a list of events with no duplicates that belong to a run in the Good Run Lists (GRL), which encode the results of the quality checks performed on the beam and detector conditions [*]. The events must pass the diphoton trigger and detector data quality requirements.

The standard diphoton event selection is then applied: only events with a reconstructed primary vertex are considered; out of those, we select the events that contain at least 2 loose photons that pass the $e - \gamma$ ambiguity requirement; at least two photons need to satisfy the tight ID, isolation and $p_T$ requirements, otherwise the event is rejected; a loose window is placed around the Higgs mass along the $m_{\gamma\gamma}$ distribution to carve out the set of events with 105 GeV $< m_{\gamma\gamma} <$ 160 GeV. In the analysis software, these events are tagged with a positive value for the binary flag `isPassed`.

Analysis-specific cuts are subsequently applied to target the specific di-Higgs topology. The $\gamma\gamma b\bar{b}$ selection requires at least two central jets ($|\eta| < 2.5$) to reconstruct the dijet portion of the event. The retraining of the $b$-tagging algorithm known as MV2C10 released in v20.7 of the ATLAS software is evaluated on these jets to determine whether they originate from $b$-hadrons. Three more exclusive analysis categories are defined based on the number of $b$-tagged jets. If exactly 2 jets pass the $b$-tagging cut corresponding to the 70% efficient working point, the event is placed in the 2-tag signal category. If only 1 jet passes the $b$-tagging cuts, the event is placed in the 1-tag signal category. Non-$b$-tagging-based meth-ods, such as the ones described in Sec. 7.3.4, are necessary to select the second jet in the events in the 1-tag category. If no jet passes the $b$-tagging cuts, the event is placed in the 0-tag control region, and the two highest $p_T$ jets are picked for downstream operations on the dijet system. To remain orthogonal to the selection designed by the $hh \rightarrow b\bar{b}b\bar{b}$ group and avoid double counting at combination time, events with more than 2 $b$-tagged jets are not considered in this search.

---

[*]The GRL for 2015 and 2016 data reprocessed with release 21 of the ATLAS software can be found at these two links: 2015 GRL and 2016 GRL.

The efficiency values used to distinguish the selected $b$-tagging working points refer to the percentage of $b$-jets with $b$-tagging discriminant value greater than a certain threshold in the specific sample used in flavor tagging optimization studies, and does not necessarily reflect the efficiency of these cuts applied to the samples in this analysis. The flavor tagging performance group reports the following false positive rates for $c$- and light jets contamination: the 70%-efficient WP corresponds to a cut on the MV2C10 output at the valye of 0.8244273 and lets through $\approx 8\%$ of $c$-jets and $\approx 0.3\%$ of light jets; the tighter 60%-efficient WP corresponds to a cut at 0.934906 and lets through $\approx 3\%$ of $c$-jets and $\approx 0.06\%$ of light jets.

Events inherit weight multipliers from their jet tagging and JVT scale factors.

In the low mass analysis, only events in which the (sub-)leading jet $p_T > (25)40$ GeV are selected. The dijet system is further required to have reconstructed mass that falls in the interval 80 GeV $< m_{jj} <$ 140 GeV. On the other hand, in the high mass analysis, only events in which the (sub-)leading jet $p_T > (30)100$ GeV are selected. The dijet system is further required to have reconstructed mass that falls in the interval 90 GeV $< m_{jj} <$ 140 GeV.

The signal efficiency in the $m_{jj}$ window is increased by $5 - 6\%$, for both low and high mass categories, when the muon-in-jet correction described in Sec. 7.3.2 is applied. This correction only modifies the background contamination in the $m_{jj}$ window by a negligible amount because the correction does not sculpt the continuum background mass spectrum defined by the two arbitrarily paired jets.

`HGamTools` provide centralized cutflow generation code to apply the standard diphoton-based event selection described above. The additional $hh \rightarrow \gamma\gamma b\bar{b}$-specific cuts are defined and applied within the `HHyybbTool`. The cutflows for the non-resonant analysis in the 1- and 2-tag categories are provided in Table 7.2 for the events that pass the low and high mass selections. The low mass (loose) cut selection retains 10% of the non-resonant SM di-Higgs signal events in the 2-tag category, and 7.2% in the 1-tag category. On the other hand, the high mass (tight) cut selection only retains 5.8% of the non-resonant

**Table 7.2:** Cutflows for SM non-resonant $hh \to \gamma\gamma b\bar{b}$

| Cuts | 2-tag | | | | 1-tag | | | |
|---|---|---|---|---|---|---|---|---|
| | Low mass | | High mass | | Low mass | | High mass | |
| | Yield | Efficiency | Yield | Efficiency | Yield | Efficiency | Yield | Efficiency |
| All events | 3.181 | 100.0 | 3.181 | 100.0 | 3.181 | 100.0 | 3.181 | 100.0 |
| No duplicates | 3.181 | 100.0 | 3.181 | 100.0 | 3.181 | 100.0 | 3.181 | 100.0 |
| GRL | 3.181 | 100.0 | 3.181 | 3.181 | 100.0 | 100.0 | 3.181 | 100.0 |
| Diphoton trigger | 2.184 | 68.7 | 2.184 | 68.7 | 2.184 | 68.7 | 2.184 | 68.7 |
| Detector DQ | 2.184 | 68.7 | 2.184 | 68.7 | 2.184 | 68.7 | 2.184 | 68.7 |
| Primary vertex | 2.184 | 68.7 | 2.184 | 68.7 | 2.184 | 68.7 | 2.184 | 68.7 |
| 2 loose photons | 1.894 | 59.5 | 1.894 | 59.5 | 1.894 | 59.5 | 1.894 | 59.5 |
| $e$-$\gamma$ ambiguity | 1.891 | 59.5 | 1.891 | 59.5 | 1.891 | 59.5 | 1.891 | 59.5 |
| Trigger match | 1.885 | 59.3 | 1.885 | 59.3 | 1.885 | 59.3 | 1.885 | 59.3 |
| Tight $\gamma$ ID | 1.617 | 50.8 | 1.617 | 50.8 | 1.617 | 50.8 | 1.617 | 50.8 |
| $\gamma$ isolation | 1.476 | 46.4 | 1.476 | 46.4 | 1.476 | 46.4 | 1.476 | 46.4 |
| $\gamma\, p_T$ cuts | 1.362 | 42.8 | 1.362 | 42.8 | 1.362 | 42.8 | 1.362 | 42.8 |
| $m_{\gamma\gamma} \in [105,160]$ GeV | 1.362 | 42.8 | 1.362 | 42.8 | 1.362 | 42.8 | 1.362 | 42.8 |
| 2 central jets | 1.149 | 36.1 | 1.149 | 36.1 | 1.149 | 36.1 | 1.149 | 36.1 |
| $b$-tagging category | 0.373 | 11.7 | 0.373 | 11.7 | 0.490 | 15.4 | 0.492 | 15.5 |
| $b$-jet $p_T$ cuts | 0.369 | 11.6 | 0.219 | 6.9 | 0.475 | 14.9 | 0.247 | 7.8 |
| $m_{jj}$ cut | 0.318 | 10.0 | 0.183 | 5.8 | 0.229 | 7.2 | 0.127 | 4.0 |

SM di-Higgs signal events in the 2-tag category, and 4.0% in the 1-tag category.

The signal region in the resonant search is defined by a tighter constraint on the diphoton mass: 120.39 GeV $< m_{\gamma\gamma} <$ 129.79 GeV for the low mass analysis, 120.79 GeV $< m_{\gamma\gamma} <$ 129.39 GeV for the high mass analysis, corresponding to approximately 95% of the signal. The events that fall inside the previously applied $m_{\gamma\gamma}$ cut but outside this more selective one can be used for background estimates.

In addition, within the resonant analysis workflow, the reconstructed dijet mass $m_{jj}$ is rescaled to equal the Higgs mass $m_h$, in order to improve the resolution of the $m_{\gamma\gamma jj}$ distribution. The correction results in an average improvement of $\sim 60\%$ in the signal mass resolution with little modification to the background shape. For example, Fig. 7.3 shows the effect of this correction on the four-body mass spectrum for events that pass the 2-tag, low mass selection criteria for some of the signal and background samples.

The resonant analysis requires a further tight window cut on the four-object mass $m_{\gamma\gamma jj}$, the position of which is optimized to represent the interval within which at least 95% of events in each resonant signal sample fall. The ranges are set to $m_{\gamma\gamma jj} \in [245, 610]$ GeV for the low mass analysis, and $m_{\gamma\gamma jj} \in [335, 1140]$ GeV for the high mass analysis.

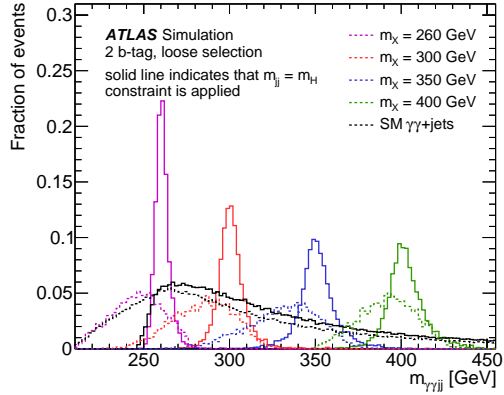

**Figure 7.3:** Effects of correcting the dijet mass to equal $m_h$ on the resolution of the four-object mass for resonant signal samples and the continuum background.

After applying the resonant search selection, between 6% and 15.4% of signal events across different mass hypotheses make it into the 2-tag category, and between 5.1% and 12.3% make it into the 1-tag category.

### 7.3.4  1-tag BDT

One of the improvements to this cycle of the analysis is the development of a simple but powerful machine learning method to select the second jet in an event in which only one jet passes the $b$-tagging selection cut. Resolving this ambiguity allows to extend the signal region to the events falling in the 1-tag category, by choosing alternative methods of selecting the dijet pair. Because of the ambiguity in how to handle the second jet selection, the 1-tag category was ignored in previous $hh \rightarrow \gamma\gamma b\bar{b}$ analysis rounds [419]. Devising a strategy to account for these events was a priority for this analysis.

This is not a straightforward task, even at truth level. To begin with, identifying a jet's origin is an ill-defined problem, given the subjective definition of a jet in terms of a clustering algorithm and its input arguments. At truth level, the correct jet pair can be selected using geometric arguments, such as calculating the $\Delta R$ distance between the dijet system and the truth-level Higgs boson, or the $\Delta R$ distance between untagged jet candidates and truth-level $b$-hadrons or $b$-quarks. A sensible strategy employed in exploratory studies is to geometrically map reconstructed jets to truth-level jets, and look for true $b$-hadron children of a Higgs boson that are located within a certain distance $\Delta R$ from the truth-level jet axes. Naïvely, in each signal sample, one would expect a total of two reconstructed jets per event to be mapped to a truth-level Higgs boson, regardless of the success or failure of the evaluation of $b$-tagging algorithms such as MV2. However, looking, for example, at a mid mass range resonant signal sample like the BSM $X \to hh \to \gamma\gamma b\bar{b}$ with $m_X = 350$ GeV, it is evident that this is not always the case. In about 40% of cases, fewer than 2 jets per signal event satisfy the jet-origin definition above, and in approximately 2% of events, more than 2 jets per event do. In the second case, at least half of the events present two jets that overlap with each other within a $\Delta R$ radius of 0.4. These are both likely to have originated from the same $b$-hadron and to geometrically overlap with it, suggesting that a possible resolution would be to merge the two jet objects. In the remaining instances, selecting the two highest-$p_T$ jets that qualify as originating from a truth $b$-hadron gives better reconstructed mass resolution and proximity to $m_H$ than merging the whole $> 2$ multi-jet system. In general, then, both $p_T$ ordering and $m_h$ matching criteria can be relevant in selecting the best possible jet pair candidate.

In the first case, instead, $< 1\%$ of truth-level jets are lost due to overlap removal and strict $p_T$ cuts. Very low transverse momentum and high $|\eta|$ truth-level bottom quarks are, on average, less likely to generate a nearby truth-level jet that satisfies the imposed requirements, and additional reconstruction inefficiencies exacerbate the problem. Therefore, even in a large percentage of simulated signal events, a correct pair of jets, in which both originate from the shower initiated by $b$-quarks from the Higgs decay, may not exist, and any forced dijet selection may lead to erroneous event reconstruction and classification.

To set up the selection of the optimal jet pair in 1-tag events as a machine learning classification problem, the following Higgs-matching labeling scheme is finally devised: if no truth-level Higgs boson decaying to a pair of $b$-quarks is found in the event, all possible jet pairings are labeled as *incorrect* (class 0); if one of the untagged jets overlaps with and is the closest jet to the Higgs boson that decays to $b\bar{b}$ at truth level (and vice versa), then the pair formed by that jet and the MV2-tagged one is labeled as *correct* (class 1), while all other jet combinations are labeled as incorrect. Within each signal event, then, at most one jet pair can be labeled as correct, while all others (plus all jet pairs in background events) are labeled as incorrect. The distributions of $\Delta R$ distances between reconstructed jet and truth-level Higgs boson are displayed in Fig. 7.4 for various resonant signal samples, along with the demarcation of the threshold for association set at $\Delta R = 0.6$.

Alternative labeling schemes are explored but eventually abandoned. These include: an exclusive truth-matching strategy between untagged reconstructed jets and truth-level $b$-quarks originating from the Higgs boson, and a $\Delta R$ match (`HadronConeExclTruthLabelID`) between untagged reconstructed jets $b$-hadrons (initially proposed to get around the absence of quark-level information in samples simulated with HERWIG++), which loses information about the Higgs origin but is in line with the labeling strategy employed by the flavor tagging group (see Sec. 6.1).

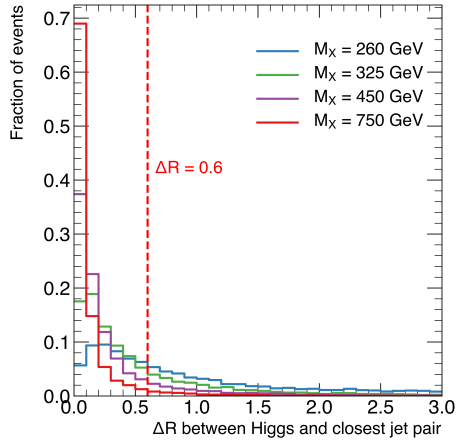

**Figure 7.4:** Distance between truth-level Higgs boson ($h \to bb$) and the closest reconstructed jet, for events in the 1-tag category across various BSM resonance samples. The association efficiency varies with signal sample as the Higgs-to-jet distance dicreases with $m_X$.

Simple, baseline suggestions for how to pick one of the untagged jets to form the dijet system were investigated in previous analysis rounds, and include selecting the highest-$p_T$ untagged jet (`pThigh`), the jet that, along with the tagged one, produces the highest-$p_T$ dijet system (`pTjbhigh`), or the reconstructed mass closest to $m_h$ (`mHmatch`).

In order to improve the efficiency in identifying the correct jet to associate to the tagged one, a boosted decision tree is trained to assign a score to each possible jet pairing in the event, based on a series of properties of the dijet system and of the individual untagged jets. These include: $|\eta_\text{j}|$ (the absolute value of the $\eta$ coordinate of the candidate jet), $|\eta_\text{jb}|$ (the absolute value of the $\eta$ coordinate of the dijet system), $\Delta\eta_\text{jb}$ (the distance along the $\eta$ coordinate between the candidate jet and the tagged jet), $m_\text{jb}$ (the reconstructed mass of the system defined by the candidate jet and the tagged jet), $p_{T,\text{j}}$ (the transverse momentum of the candidate jet), $p_{T,\text{b}}$ (the transverse momentum of the tagged jet), $m_{H,\text{ index}}$ (the ranking of the dijet system according to the 'closest reconstructed mass to the Higgs mass' approach), $p_{T,\text{ index}}$ (the ranking of the dijet system according to the 'highest transverse momentum candidate' approach), $p_{T,\text{jb, index}}$ (the ranking of the dijet system according to the 'highest transverse momentum dijet pair' approach), and *b*-tagging scores associated to candidate jets.

Since kinematic variables enter the selection in the form of inputs to the BDT, two separate models are trained to uniquely focus on the low and high mass signal regions. The distributions of the input variables for correct and incorrect jet pairs for simulated BSM resonant events that fall in the 1-tag category are plotted in Fig. 7.5 and 7.6 for the $m_X = 260$ GeV and $m_X = 1$ TeV samples, respectively, which characterize the extremes of the low and high mass regimes.

In practice, the BDT is tasked with classifying a large collection of jet pairs as either correct or incorrect, without any event-level information. In other words, the model is not informed of the fact that, in principle, if one jet pair in the event is highly likely to be the correct one, no other correct pairing should exist in that event. Therefore, even if the optimization objective is phrased as a jet pair-by-jet pair task, the final performance evaluation assesses the scores assigned by the BDT to each jet pair in each event, and selects the one with the highest BDT output score per event. It subsequently compares that choice with the actual, correct jet pair (if it exists).

The BDT is able to recover the correct jet pair out of the multiple options per event with efficiency values that range between 60% and 80% across various resonant and non-resonant signal samples. The efficiency for the selection of the correct jet pair per event is represented, for different signal samples and analysis categories, in Fig. 7.7. In addition, Fig. 7.8 displays the distribution of jet-level output predictions and the confusion matrix for the low mass BDT and competing methods evaluated on the $m_X = 260$ GeV BSM resonant sample, while Fig. 7.9 shows equivalent results for the high mass analysis, with results evaluated on the $m_X = 1$ TeV BSM resonant sample. From these distributions, it is easy to see that identifying the correct untagged jet in samples with high mass resonances is an easier task than doing so for low mass resonances.

This conclusion is further substantiated by the ROC curves and efficiency working points displayed in Fig. 7.10 and 7.11 for the $m_X = 260$ GeV and $m_X = 1$ TeV BSM resonant samples, respectively. While the BDT provides a full set of continuous decisions that can be imposed as thresholds to separate the two classes, the other methods yield binary decisions that correspond to unique points on the TPR-FPR plane. All methods perform better on the easier task of selecting the correct dijet pair in the high mass sample, compared to the low mass sample. Of the three alternative methods, `mHmatch` appears to be the most competitive one, while the BDT comfortably outperforms all of them, at least at a jet pair-by-jet pair evaluation level, *i.e.* when no event-level information is considered. At the same false positive rate levels as those achieved by the `mHmatch` method, the BDT improves the percentage of properly identified correct pairs from $40\%$ to $70\%$ in the $m_X = 260$ GeV sample, and from $80\%$ to $90\%$ in the $m_X = 1$ TeV sample. Selecting the highest-$p_T$ untagged jet is markedly suboptimal, as, for both samples, the highest $p_T$ untagged jet is more likely *not* to be the correct one than vice versa. On the other hand, selecting the untagged jet that forms the highest-$p_T$ dijet system when coupled to the tagged jet becomes a relatively competitive strategy only for high mass resonances. However, by definition, using a selection criterion that matches the constructed dijet mass to that of the Higgs biases the background samples towards the signal region.

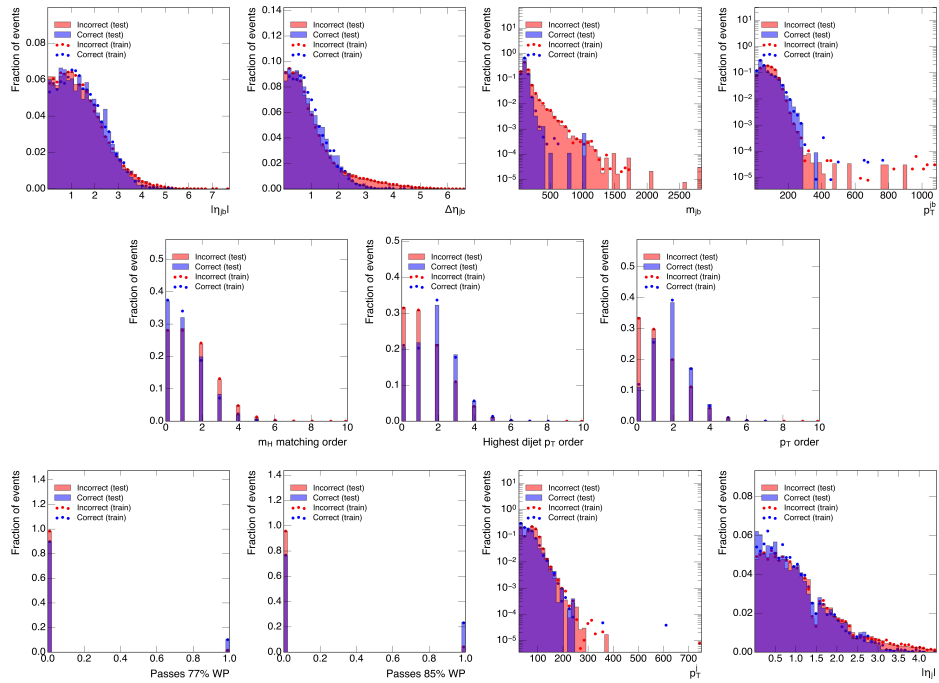

**Figure 7.5:** Distribution of BDT input variables for the $m_X = 260$ GeV BSM sample, highlighting the differences between correct and incorrect jet pairs that the model can exploit in the classification task.

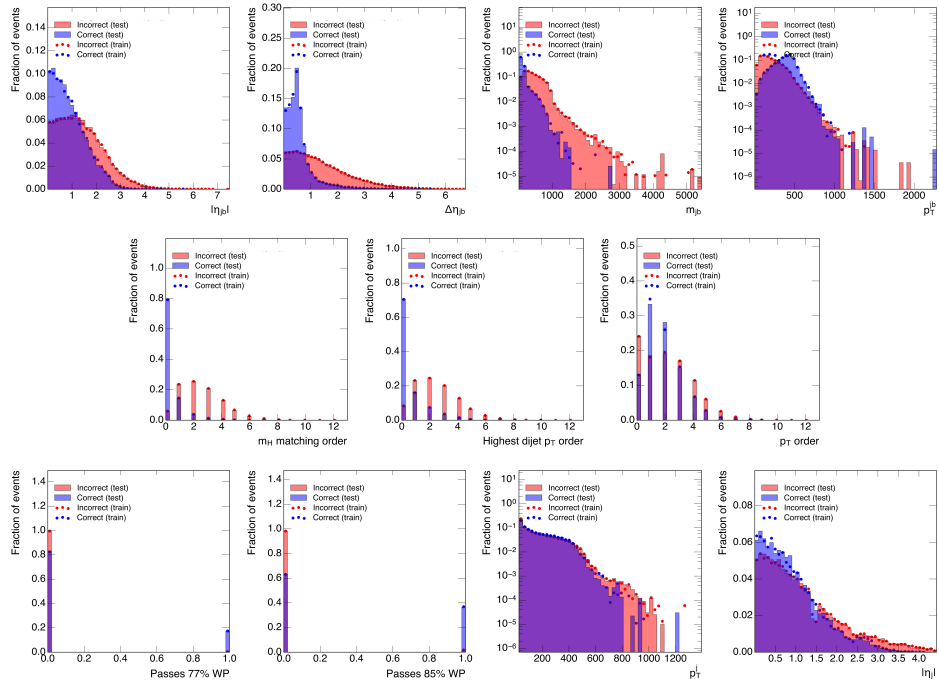

**Figure 7.6:** Distribution of BDT input variables for the $m_X = 1000$ GeV BSM sample, highlighting the differences between correct and incorrect jet pairs that the model can exploit in the classification task.
233

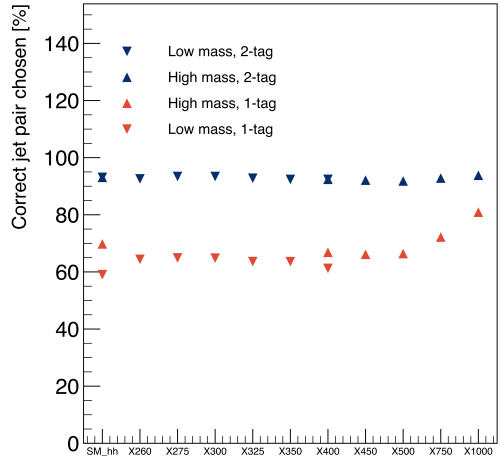

**Figure 7.7:** Jet pair selection efficiency for each signal sample in the 1- and 2-tag categories, applying the low or high mass selection, according to the sample's kinematics.

Although the ideal objective is to select the correct pair of jets that originates from the decay of a Higgs boson, in reality, the practical objective with respect to which final yields are calculated is to increase the number of signal events for which the selected jet pair (correct or not) contributes to letting the event pass the signal selection, while avoiding an increase in background contamination in the signal region. The two goals, though related, are not exactly equivalent. In fact, the empirical signal-to-background ratio (or Asimov coefficient) is computed by accounting for all different contributions to the signal region from each of the sample classes. For background events in which no Higgs boson is present, the BDT nonetheless selects the most likely jet pair to have originated from a Higgs boson, regardless of how low its probability is. If, once the downstream cuts are applied, the BDT's selection of said jet pair causes the background event to fall into the signal region, the figure of merit associated with the performance of the BDT will be negatively affected. If, on the other hand, the event is later placed in the sideband region, the BDT's performance should not be negatively impacted. A similar logic can be applied to signal events: whether a correct jet pair exists at truth level, and whether the BDT correctly identifies the correct jet pair at reconstruction level, the final figure of merit should be related to the change in the final yield in the signal region defined by downstream cuts. This is, of course, at odds with the direct objective that the BDT is instructed to optimize.

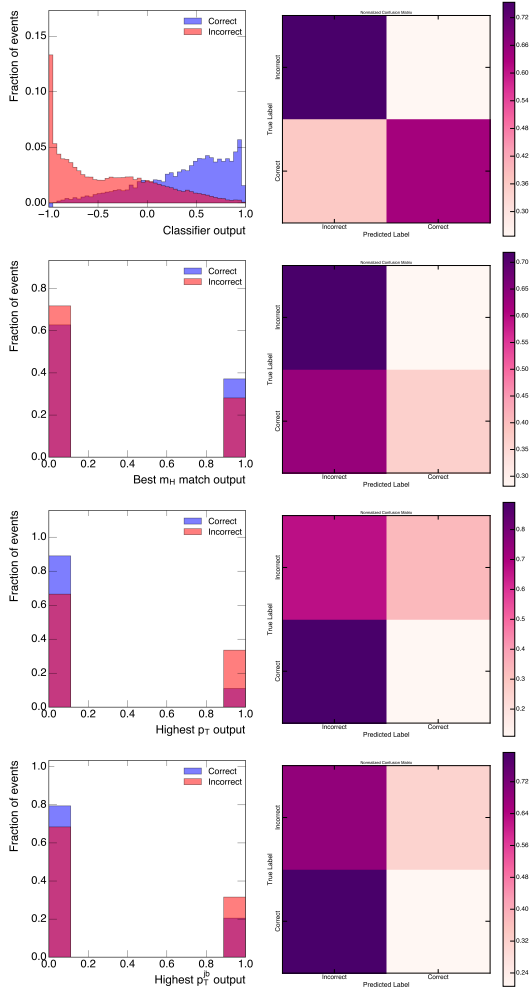

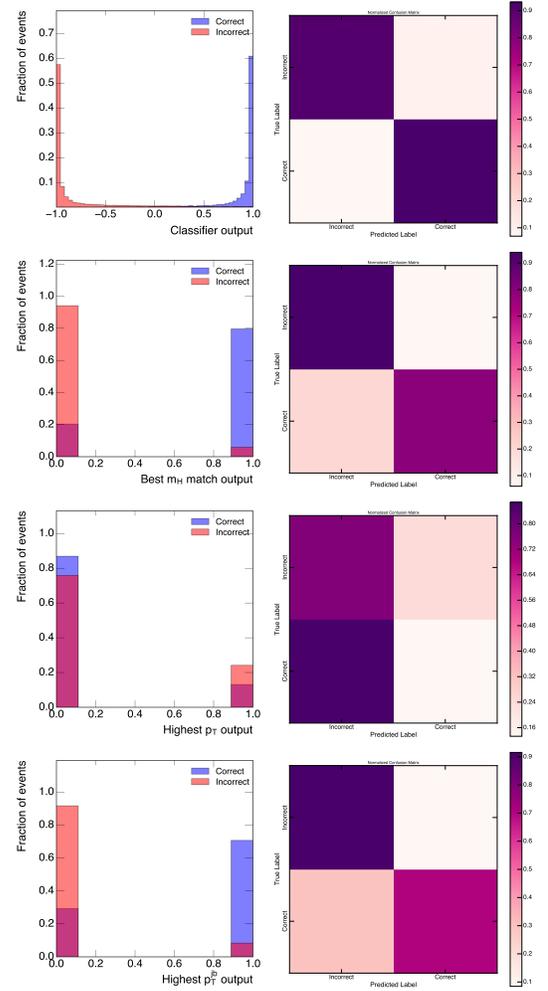

**Figure 7.8:** Classifier output distribution and confusion matrix for the low-mass BDT (row 1), `mHmatch` method (row 2), `pThigh` method (row 3), and `pTjbhigh` method (row 4), evaluated on a test set of jet pairs from events in the $m_X = 260$ GeV sample.

**Figure 7.9:** Classifier output distribution and confusion matrix for the high-mass BDT (row 1), `mHmatch` method (row 2), `pThigh` method (row 3), and `pTjbhigh` method (row 4), evaluated on a test set of jet pairs from events in the $m_X = 1$ TeV sample.

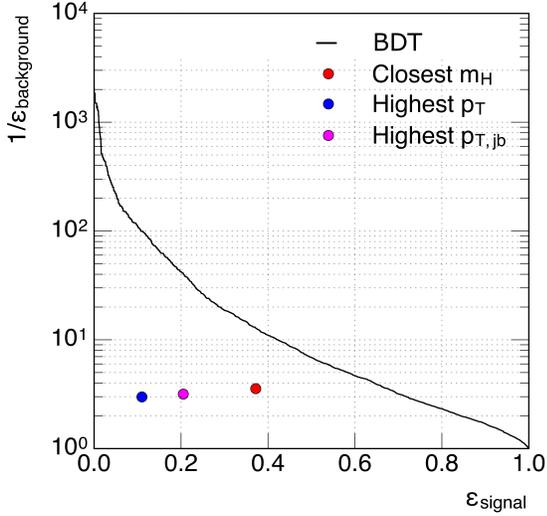 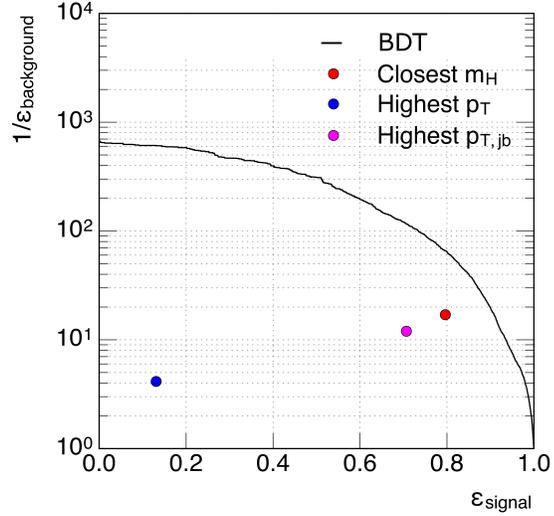

**Figure 7.10:** ROC curves showing 1/FPR versus TPR for correct/incorrect jet pair classification evaluated on a test set of jet pairs from events in the $m_X = 260$ GeV sample.

**Figure 7.11:** ROC curves showing 1/FPR versus TPR for correct/incorrect jet pair classification evaluated on a test set of jet pairs from events in the $m_X = 1$ TeV sample.

A suggested improvement consists of imposing a post-facto cut on the BDT output to set a threshold on the likelihood of the highest ranked jet pair to have originated from a Higgs boson. If no jet pair in the event passes the threshold, then the event could be automatically moved to the sideband region and any method could be employed to select the dijet system. The performance of the BDT, compared to the most promising alternative method (`mHmatch`), is quantified as a function of the BDT output in terms of the ratio of the Asimov significances obtained from the use of the two methods. However, this results in each signal sample benefiting from a different threshold cut on the BDT output to maximize its analysis significance. Employing a mass parametrization of an analysis cut would undermine the sample-independence of the analysis procedure. Still, a single, averagely optimal threshold can be selected for the low and high mass analyses independently. Fig. 7.12 shows that imposing a fixed cut on the BDT output,

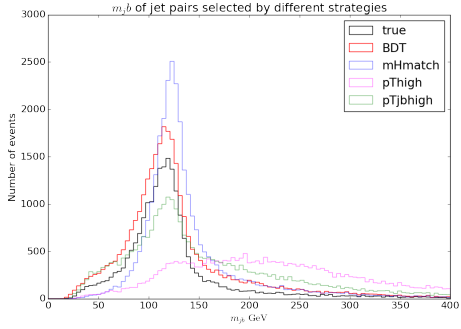 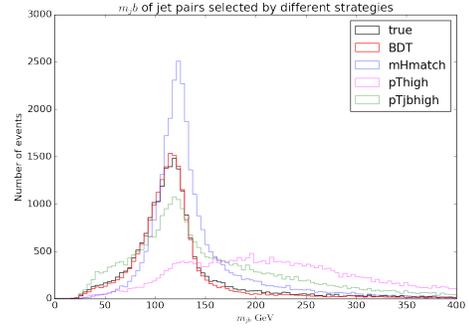

(a) No threshold       (b) Classify jet pair as correct only if BDT output $> 0.4$

**Figure 7.12:** Dijet mass spectrum for events in the 1-tag category. The distribution obtained from selecting the truth-level correct jet pair is shown in black. The dijet systems selected by the BDT form the red histograms, while the ones selected with alternative strategies are shown in blue (`mHmatch`), pink (`pThigh`), and green (`pTjbhigh`). In Fig. 7.12(a), the jet pair per event with the highest BDT score is chosen as the correct pair, without any thrsehold on its minimum value. In Fig. 7.12(b), a jet pair is selected as the correct pair only if it has the highest BDT score in the event and its BDT score is greater than 0.4. If no jet pair in an event reaches this BDT score value, all jet pairs are considered incorrect, and the event can be removed from the signal region. All distributions related to other selection methods remain unaltered.

in this case tentatively placed at a global value of 0.4 (on an output scale from -1 to 1), also provides the added value of improving the dijet mass resolution to match that of the truth-level correct dijet system. However, the reduction in signal efficiency in this already statistically limited analysis is ultimately seen as an insurmountable problem that causes the idea of imposing a threshold to the BDT output to be abandoned.

Yet another solution is to implement a different machine learning system to learn to identify the correct jet pair among the various combinations of jets in an event, with the option for the model to prefer a null solution in which no jet pair is selected. This method was investigated but not implemented in the final round of the analysis. A diagram of the proposed model, based on a recurrent neural network architecture, is presented in Fig. 7.13. In the example presented in this picture, an event with 6 candidate jet pairs is examined by the network through an LSTM module that processes the features of each jet pair in the event. The network is then tasked to output a set of outputs summing to unity that correlate to the likelihood of each jet pair (or no jet pair) to be the correct one. The location of the true, correct pair (if any) can be encoded in a one-hot vector, where the hot entry identifies the position of the correct jet pair within the sequence.

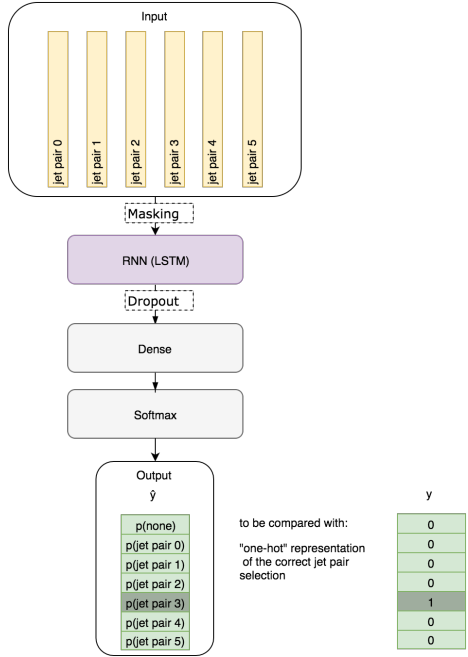

**Figure 7.13:** Schematic representation of an alternative jet selection architecture based on a recurrent neural network for the 1-tag category. This solution is not currently implemented in the $\gamma\gamma b\bar{b}$ analysis.

## 7.4 Background Estimates

The continuum background is estimated using data-driven methods, while MC simulation is used to account for the single Higgs background. The distinction is due to the observation that the Sherpa simulated continuum background does not accurately model the shape of that portion of the background observed in data.

Due to the inability to rely on accurate Monte Carlo simulation to describe processes in which jets fake photons, the $\gamma j$, $j\gamma$, and $jj$ contributions are estimated using data-driven techniques from a 2D fit to the sideband regions (2x2D method [420]). This consists of changing the photon ID and isolation working points, in order to vary the levels of jet contamination in the photon selection. This method is used to extract the fractional contribution of $\gamma\gamma$, $\gamma j$, $j\gamma$, and $jj$ to the continuum background in data, in the 1- and 2-tag categories separately. Before this method can be applied to the Sherpa sample, due to the MC shape mismodeling, an event reweighting scheme must first be derived to correct the simulated $m_{\gamma\gamma}$ shape to match that of data. To do so, the 2x2D method is applied in bins of $m_{\gamma\gamma}$ in both data and Sherpa simulated 0-tag categories to separate the various background contributions. The binned spectrum of each background source in each dataset is then fitted with an exponential function; the ratio of the analytical distribution in data to that in MC yields the per-bin correction per background type. All corrections are $< 5\%$. The effects of these corrections in the 0-tag category are shown, for each type of background source, in Fig. 7.14 for events passing the low mass selection, and in Fig. 7.15 for events passing the high mass selection.

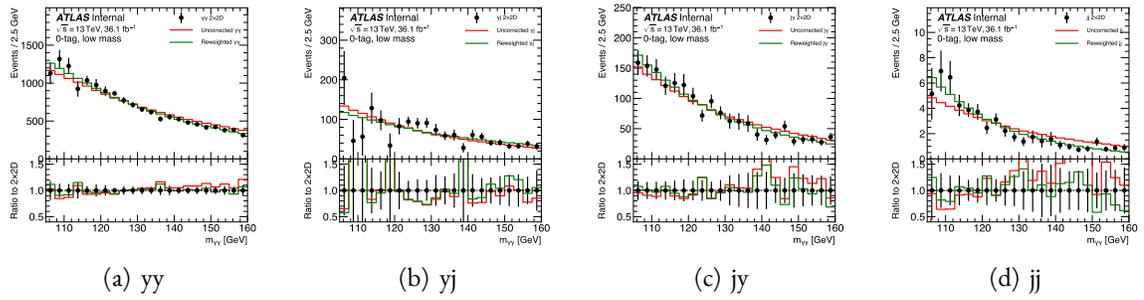

(a) γγ  (b) γj  (c) jγ  (d) jj

**Figure 7.14:** Binned $m_{\gamma\gamma}$ distributions in the 0-tag, low mass selection for the four sources that contribute to the continuum background. The plots show the shape obtained from data (black), Sherpa before correction (red) and after correction (green).

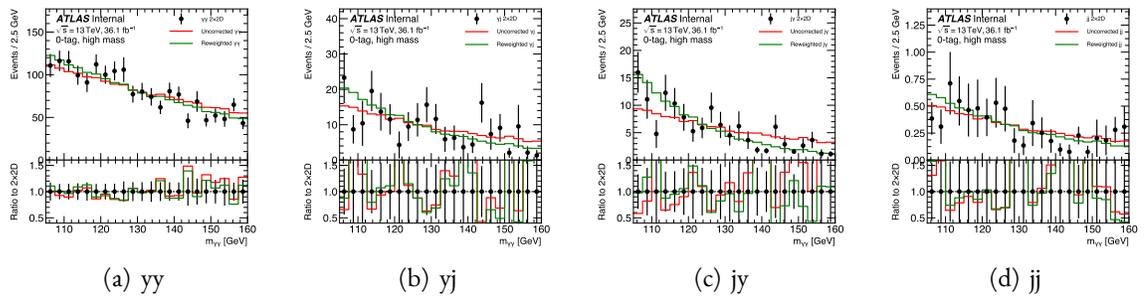

(a) γγ  (b) γj  (c) jγ  (d) jj

**Figure 7.15:** Binned $m_{\gamma\gamma}$ distributions in the 0-tag, high mass selection for the four sources that contribute to the continuum background. The plots show the shape obtained from data (black), Sherpa before correction (red) and after correction (green).

Due to limited statistics, ad-hoc per bin corrections for the 1- and 2-tag categories cannot be independently derived, so the reweighting scheme obtained from the 0-tag category is applied to these regions, using the fractional background contributions extracted from data. The total normalization factor is computed by matching the number of events in data and simulation that fall in the joint sidebands region. Since the $\gamma\gamma$ events make up between 80-90% of the continuum background, they can be further separated based on the flavor of the jets produced in association with the photons, obtained from the truth information provided by the generator.

The combined $m_{\gamma\gamma}$ and $m_{\gamma\gamma jj}$ distributions that results from the combination of the various reweighted sources that contribute to the continuum background are compared to the corresponding data distributions in Fig. 7.16 and 7.17 for events that pass the low mass analysis cutflow, and in Fig. 7.18 and 7.19 for high mass.

## 7.5 STATISTICAL ANALYSIS

If data or simulated signal and background events pass the cutflow selection defined in Sec. 7.3.3, histograms are populated to represent the distributions of discriminating variables, such as the diphoton mass in the non-resonant analysis or the four-body mass in the resonant analysis. These spectra are then fitted to obtain an analytical description. Events $x \in X$ are allocated in a signal-enriched region (SR) if they pass all selection criteria, while events $y \in Y$ that pass some but not all cuts may be allocated in statistically orthogonal control regions (CR) which can be used to provide crucial information about the signal region (in the form of constraints on nuisance parameters) which cannot be reliably extracted from simulation.

This section explains how signal and background contributions are modeled using analytical parametric functions (see Sec. 7.5.1 for modeling in the resonant analysis, and Sec. 7.5.2 for the non-resonant analysis), and how these models are then used in the final limit-setting procedure to arrive to the results summarized in Sec. 7.6.

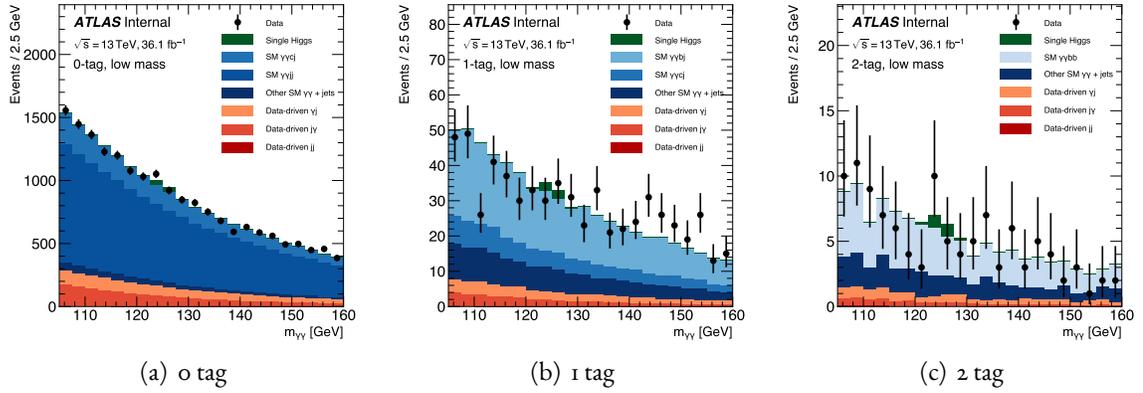

**Figure 7.16:** Data-MC comparison of the $m_{\gamma\gamma}$ distribution for events in that pass the low mass selection, in 2015+2016 data and in the various MC simulated contributions to the continuum background after reweighting and 2x2D fractional contribution estimation.

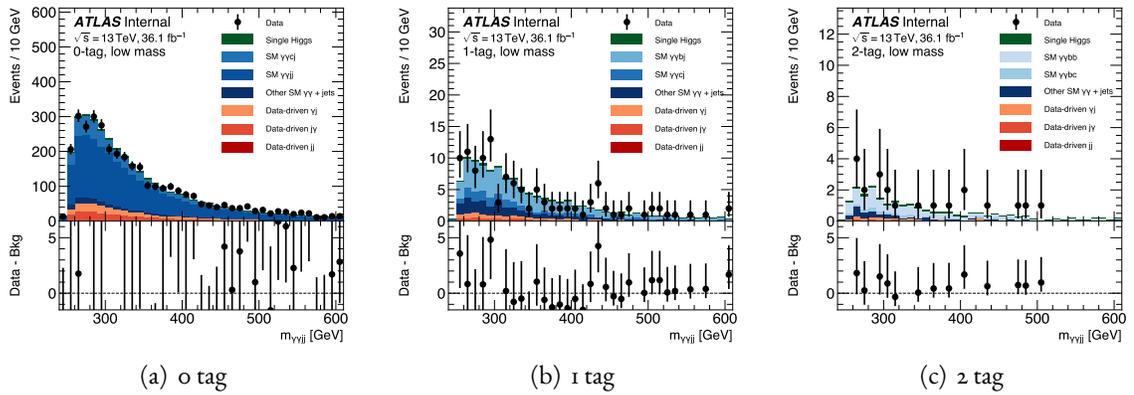

**Figure 7.17:** Data-MC comparison of the $m_{\gamma\gamma jj}$ distribution for events in that pass the low mass selection, in 2015+2016 data and in the various MC simulated contributions to the continuum background after reweighting and 2x2D fractional contribution estimation.

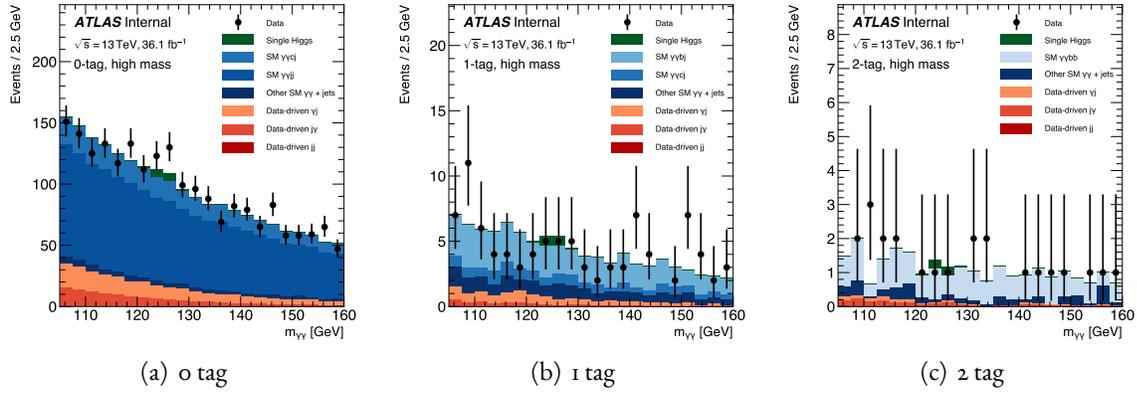

**Figure 7.18:** Data-MC comparison of the $m_{\gamma\gamma}$ distribution for events in that pass the high mass selection, in 2015+2016 data and in the various MC simulated contributions to the continuum background after reweighting and 2x2D fractional contribution estimation.

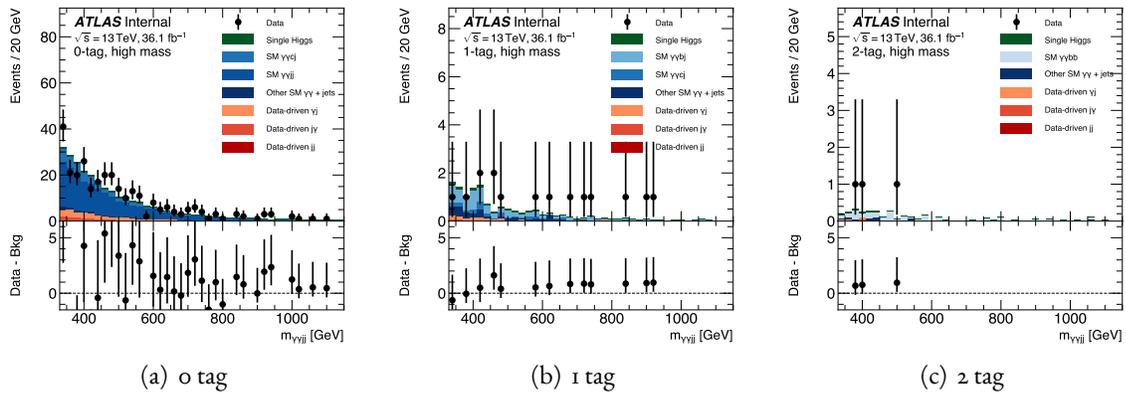

**Figure 7.19:** Data-MC comparison of the $m_{\gamma\gamma jj}$ distribution for events in that pass the high mass selection, in 2015+2016 data and in the various MC simulated contributions to the continuum background after reweighting and 2x2D fractional contribution estimation.

Given a choice of a model, the *likelihood* of a set of model parameters $\vartheta$ designates the probability of the observed data $X$ given the parameters $\vartheta$ of the model: $\mathscr{L}(\vartheta) = \prod_{x \in X} P_{\text{model}}(x; \vartheta)$. To estimate the optimal values of the model parameters, one can use maximum likelihood estimation (see Sec. 4.1.1):

$$\hat{\vartheta} = \arg\max_{\vartheta} \mathscr{L}(\vartheta) = \arg\min_{\vartheta}(-\ln \mathscr{L}(\vartheta)) \tag{7.1}$$

In an analysis, however, we want to compute not only the point estimates for the optimal value of the parameters, but also the ranges of parameter values for which the likelihood remains within a certain neighborhood of the maximum. To estimate the boundaries of this interval, it is useful to approximate the entry $V_{ij} = \text{cov}[\hat{\vartheta}_i, \hat{\vartheta}_j]$ in the covariance matrix for the MLE of two parameters $(\vartheta_i, \vartheta_j)$ as [2]:

$$(\hat{V}^{-1})_{ij} = -\left.\frac{\partial^2 \ln \mathscr{L}}{\partial \vartheta_i \partial \vartheta_j}\right|_{\hat{\vartheta}} \tag{7.2}$$

In the majority of cases, however, only a limited number of selected model parameters are physically relevant and are reported, with their confidence intervals, in the final measurement. All other quantities and terms upon which a model depends constrain the fit, and may or may not be known with 100% certainty. For example, the measurement of a signal cross-section may depend, among others, on the background cross-section, which may only be constrained by previous analyses or by measurements in control regions to within the uncertainty of those measurements. The parameters $\vartheta$ that parametrize a model can therefore be divided into parameters of interest $\mu$, which represent physically meaningful quantities we would like to estimate in this analysis, and nuisance parameters $\theta$ that symbolize all other physical or unphysical quantities whose values are allowed to fluctuate within their uncertainties. These adjustable knobs in the model enlarge the space of solutions that are available to the optimization strategy in order to reflect the uncertainty associated with the experimental procedure, by accounting for these uncertainties in secondary model parameters, in the form of quantifiable costs associated to deviations from their nominal values. Solutions that require the nuisance parameters to take on drastically different values from those provided by the constraints will be more unlikely. The way in which nuisance parameters and their constraints affect the overall likelihood of the model is quantified by extending the likelihood with extra multiplicative terms. Assuming the constraints on the nuisance parameters originate from a statistically independent set of measurements $y \in Y$:

$$\mathscr{L}(\mu, \theta) = P_x(X; \mu, \theta) P_y(Y; \theta) \tag{7.3}$$

In other words, including nuisance parameters in a model is a way of imposing a flexible prior on the value and distribution of quantities that affect the measurement of the parameters of interest.

The analysis objective then becomes that of scanning various values of the parameters of interest $\mu$ and noting whether the systematic uncertainties allow the likelihood to maintain a value compatible with its maximum through the leeway provided by the uncertainty on the nuisance parameters. Low systematic uncertainties corresponds to tight constraints on $\theta$ and steeper rates of change in the likelihood as a function of $\mu$. High systematic uncertainties, instead, provide more options for the likelihood to be maximized irrespective of the exact value of $\mu$.

For the scope of hypothesis testing of a specific value of $\mu$, we want to construct a test statistic that allows to define a region along $\mu$ in which the $p$-value for the parameter of interest is always less than a certain threshold for any choice of the nuisance parameters. The level of compatibility between the maximum of the likelihood and its value at any other point $\mu$ can be expressed in terms of the likelihood ratio of the profile likelihood to the maximum of the unconstrained likelihood itself. In the following, the hat notation indicates the maximum likelihood estimator (MLE) of a parameter, while the double hat represents the conditional MLE of a parameter, *i.e.* the value that maximizes the likelihood for a specific choice of a variable of interest in terms of which the profiled parameter can be expressed.

Since the exact MLE and the confidence intervals of the nuisance parameters are not themselves of intrinsic interest for this measurement, we seek a test statistic that is solely a function of the parameters of interest $\mu$. Approximate independence on $\theta$ can be achieved by profiling these nuisance parameters and expressing them in terms of the parameters of interest:

$$\mathscr{L}_p(\mu) = \mathscr{L}(\mu, \hat{\hat{\theta}}(\mu)) = \max_{\theta} \mathscr{L}(\mu, \theta) \tag{7.4}$$

In practice, maximizing the profile likelihood yields the same MLE $\hat{\mu}$ for the parameter of interest as maximizing the full likelihood.

Testing a hypothesis value $\mu$ for the parameter of interest can then be performed by considering the

profile likelihood ratio

$$\lambda(\mu) = \frac{\mathscr{L}(\mu, \hat{\hat{\theta}}(\mu))}{\mathscr{L}(\hat{\mu}, \hat{\theta})} \qquad (7.5)$$

which grows to 1 to represent maximum compatibility between the measurement in data and the hypothesis $\mu$, and decreases to 0 to signal maximum incompatibility.

For numerical stability, $\lambda(\mu)$ can be transformed into a related test statistic $t_\mu = -2 \ln \lambda(\mu)$ which can be directly inserted in the $p$-value calculation:

$$p_\mu = \int_{t_{\mu,\text{ obs.}}}^{\infty} f(t_\mu|\mu) dt_\mu. \qquad (7.6)$$

Asymptotically, in the limit of large sample sizes, the distribution of the test statistic $f(t_\mu)$ approaches a $\chi^2$ distribution, with well known properties related to the calculation of confidence intervals. In other circumstances, a manual scan over the range of $\mu$ can be performed by drawing Monte Carlo toy datasets and using these simulated empirical results to build the probability distribution function of the test statistic. This can then be used to construct frequentist Neyman confidence intervals, which represent the region in which the true value of the parameter of interest falls for a fraction of times corresponding to the confidence level selected for the exclusion (*coverage*).

In the case of searches in which the objective is to set upper limits on a parameter of interest $\mu$, the profile likelihood ratio can be expressed as:

$$\tilde{q}_\mu = \begin{cases} -2 \ln \frac{\mathscr{L}(\mu, \hat{\hat{\theta}}(\mu))}{\mathscr{L}(0, \hat{\hat{\theta}}(0))} & \hat{\mu} < 0 \\ -2 \ln \frac{\mathscr{L}(\mu, \hat{\hat{\theta}}(\mu))}{\mathscr{L}(\hat{\mu}, \hat{\theta}(\mu))} & 0 \leq \hat{\mu} \leq \mu \\ 0 & \hat{\mu} > \mu \end{cases} \qquad (7.7)$$

where $\hat{\hat{\theta}}(\mu))$ is the MLE of $\theta$ given a fixed $\mu$. Here, the parameter of interest, which can be taken as a signal's strength, is, by definition, a non-negative quantity. Therefore, if the MLE $\hat{\mu}$ is < 0, the most compatible value of $\mu$ is set to its lower bound, 0, in order to avoid interpreting deficits in the number of recorded events as evidence for an unphysical negative signal. On the other hand, the value of the test statistic is 0 if the data favors $\hat{\mu} > \mu$ because an even higher MLE for the parameter of interest does not represent evidence to reject the presence of a signal, and should instead point to maximum

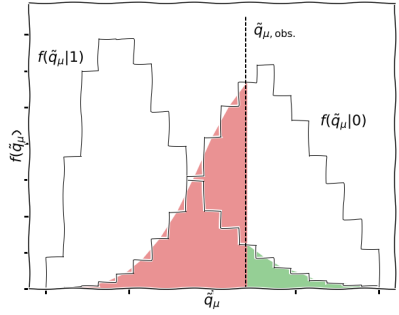

**Figure 7.20:** Probability density functions of the test statistic $\tilde{q}_\mu$ for two different values of the parameter of interest $\mu = 0, 1$. Given a hypothetical observed value $\tilde{q}_{\mu,\,\text{obs.}}$, the green and red shaded areas corresponds to the $p$-values for the two hypotheses.

incompatibility with the null hypothesis.

A diagram representing two hypothetical distributions of the test statistic $\tilde{q}_\mu$ for hypotheses $\mu = 1$ and $\mu = 0$ are displayed in Fig. 7.20, along with the visualization of the meaning of a $p$-value in the case in which a test statistic value $\tilde{q}_{\mu,\,\text{obs.}}$ is observed from data. If the green area corresponds to a $p$-value of 0.05, then the signal strength hypothesis $\mu = 1$ is excluded with 95% confidence. If, instead, $p_{\mu=1} > 0.05$, one would construct $f(\tilde{q}_\mu|\mu)$ for several signal strength hypotheses $\mu > 1$ until one finds the value of $\mu$ that the observation $\tilde{q}_{\mu,\,\text{obs.}}$ would exclude with the desired confidence level.

A refined, modified frequentist technique often used in high energy physics to exclude a signal up to a certain upper limit consists of applying a threshold on the quantity $\text{CL}_s = \frac{p_\mu}{1-p_0}$ instead of on the traditional $p$-value [421]. In practice, this rescales the probability of the signal + background hypothesis by the probability of the background-only hypothesis, ensuring that the background-only hypothesis would not be excluded in the case a low-sensitivity analysis. The creation of a conservative, conditional confidence belt results in the confidence level losing its intuitive association with coverage, but avoids spurious exclusions of parameter values beyond the parameter domain or beyond the experimental sensitivity.

In this analysis, the parameter of interest in the resonant analysis is the BSM di-Higgs production cross-section as a function of resonance mass $m_X$, while in the non-resonant analysis we aim to set limits on the SM di-Higgs production cross-section (using the high mass selection) and on the range of non-excluded values of the Higgs self-coupling $\kappa_\lambda$ (using the low mass selection). The likelihood model designed for this analysis relates the parameters of interest to all analysis decision efficiencies, theoretical and experimental systematics, detector and collider settings, and other factors that influence the mea-surements. The systematic effects caused by most sources of uncertainty in the 1- and 2-tag categories are generally correlated and implemented as shared nuisance parameters.

The statistical methodology employed in both the resonant and non-resonant analyses consists of performing a simultaneous unbinned maximum likelihood fit in the 1- and 2-tag categories that define the signal region, using parametrized functional forms optimized on simulation.

Different functional models and fitting strategies are adopted to describe the signal and background contributions in the resonant and non-resonant analyses. The fits are performed on the $m_{\gamma\gamma}$ distribution in the non-resonant analysis, and on the $m_{\gamma\gamma jj}$ distribution in the resonant analysis. The low-statistics regime encountered in this search requires the use of Monte Carlo methods to generate pseudo-datasets and extract the distributions of the test statistic $\tilde{q}_\mu$ used to set limits on the resonant di-Higgs production cross-section.

Sec. 7.5.1 and 7.5.2 discuss the modeling decisions adopted in the resonant and non-resonant analyses, respectively, while Sec. 7.5.3 quantifies the systematic uncertainties that affect the nuisance parameters in the models.

### 7.5.1 Signal and Background Modeling in the Resonant Analysis

A Gaussian function with independent exponential tails [422] (also known as `ExpGausExp`) is used to analytically model the four-object signal distribution in a mass window around the resonance mass $m_X$. This function can be expressed as:

$$P(x) = N \cdot \begin{cases} e^{k_{\text{low}}^2/2 + k_{\text{low}} x} & \text{if } x \leq -k_{\text{low}} \\ e^{-x^2/2} & \text{if } -k_{\text{low}} < x < k_{\text{high}} \\ e^{k_{\text{high}}^2/2 - k_{\text{high}} x} & \text{if } x \geq k_{\text{high}} \end{cases} \qquad (7.8)$$

where $x = (m - \mu_{\text{Gauss}})/\sigma_{\text{Gauss}}$, and $k_{\text{low}}$ and $k_{\text{high}}$ are two parameters that regulate the behavior of the function in the exponential tails.

The fit parameters are themselves expressed as functions of $m_X$, so as to factor out any mass dependence from the signal fit, and are fixed after a fit to the $X \to hh$ simulated samples.

The background is fitted separately for events passing the criteria that define the low mass and high mass cutflows. The fit is performed in the four-body $m_{\gamma\gamma jj}$ distribution for a combined simulated continuum and single Higgs background samples. The family of functions used to fit the background contribution is determined through spurious signal studies: simulated background events are fitted with several signal + background parameterizations by varying the functional form used to describe the background contribution; for each option of background fit function, the absolute value of the extracted signal (*i.e.*, the spurious signal) corresponds to the bias associated with that fit function; the functional form with the lowest measured bias (or lowest number of parameters, if more than one function results in the same bias) is selected as the preferred function to describe the background shape.

The results identify the Novosibirsk function [423] and the exponential function as the best functional forms for the low and high mass analyses, respectively, where the Novosibirsk function is defined as:

$$P(q) = N \cdot e^{-\frac{1}{2}\Lambda^2 + (\ln q)^2/\Lambda^2} \quad \text{where } q = 1 + \frac{\Lambda(m-\mu)}{\sigma} \cdot \frac{\sinh(\Lambda\sqrt{\ln 4})}{\Lambda\sqrt{\ln 4}}, \tag{7.9}$$

where $\mu$ and $\sigma$ parametrize the position and width of the peak, and $\Lambda$ describes the tail of the distribution [424].

Since the Novosibirk function is peaked and the background peak may overlap with the signal peak for low-mass resonances, extra care is necessary to constrain the fit. Once the functional forms are selected, the nominal values of their fit parameters are fixed from a fit on simulated samples, but the shape is still allowed to vary to within their statistical covariance when applied to data. The overall normalization factor representing the number of background events in the signal region is estimated via interpolation of the $m_{\gamma\gamma}$ distribution in the sidebands.

The bias in the cross-sectional measurement is assessed by drawing 10,000 pseudo-datasets from the final signal + background function for each resonance mass hypothesis, after scaling the event weights to correspond to different amounts of injected signal. These data points are refitted to extract cross-sectional estimates, and the median extracted cross-section across each set of 10,000 pseudo-datasets is compared with the injected cross-section; the difference is taken as the bias. The magnitude of the bias as a function of $m_X$ is shown, for different amounts of injected signal, in Fig. 7.21(a). To offset this bias, the following corrections are applied: for $m_X \leq 400$ GeV, the required offset grows linearly with $m_X$; for $m_X > 400$ GeV, a flat correction of 0.005 pb is applied to the final experimentally measured cross-section. The post-correction magnitude of the residual bias displayed in Fig. 7.21(b) shows further dependence on the injected signal cross-section, which is captured by adding a global $\pm$ 0.02 pb uncertainty on the correction.

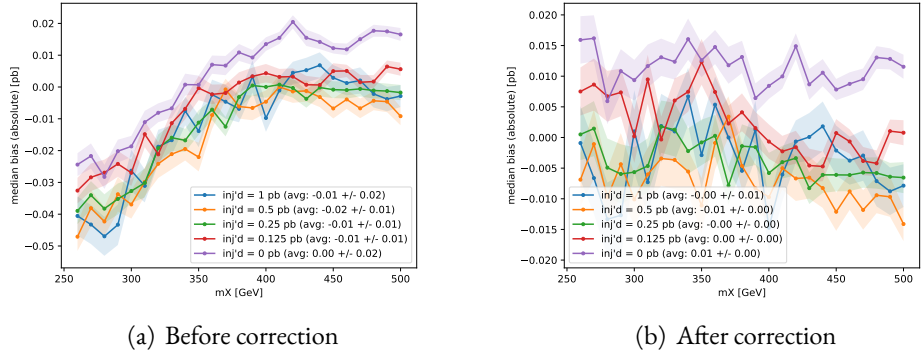

(a) Before correction

(b) After correction

**Figure 7.21:** Median bias in the extracted cross-section from 10,000 pseudo-datasets generated at different $m_X$ points for different injected signal cross-sections (obtained by reweighting events), before (Fig. 7.21(a)) and after (Fig. 7.21(b)) applying the bias offset correction described in the text.

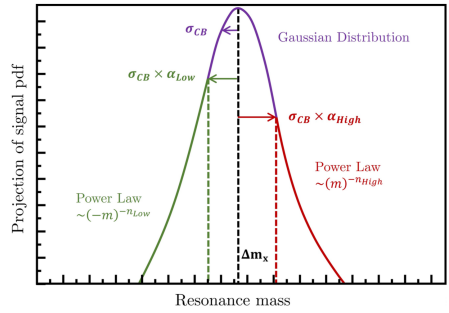

**Figure 7.22:** Representation of the double-sided Crystal Ball function and of the approximate functional behavior of each of its components. Image reproduced from Ref. [41].

### 7.5.2 SIGNAL AND BACKGROUND MODELING IN THE NON-RESONANT ANALYSIS

The $m_{\gamma\gamma}$ distribution is modeled by a double-sided Crystal Ball function [425] (Gaussian plus power-law tails) with fit parameters obtained from previously fitting simulated SM di-Higgs events. The double-sided Cristal Ball function is represented in Fig. 7.22, and its functional form is given by [426]:

$$P(x) = N \cdot \begin{cases} e^{-x^2/2} & \text{if } -\alpha_{\text{low}} \leq x \leq \alpha_{\text{high}} \\ e^{-\alpha_{\text{low}}^2/2} \left[ \frac{\alpha_{\text{low}}}{n_{\text{low}}} \left( \frac{n_{\text{low}}}{\alpha_{\text{low}}} - \alpha_{\text{low}} - x \right) \right]^{-n_{\text{low}}} & \text{if } x < -\alpha_{\text{low}} \\ e^{-\alpha_{\text{high}}^2/2} \left[ \frac{\alpha_{\text{high}}}{n_{\text{high}}} \left( \frac{n_{\text{high}}}{\alpha_{\text{high}}} - \alpha_{\text{high}} - x \right) \right]^{-n_{\text{high}}} & \text{if } x > \alpha_{\text{high}} \end{cases} \quad (7.10)$$

where $x = (m - \mu_{\text{CB}})/\sigma_{\text{CB}}$, $m$ is the independent variable, $\mu_{\text{CB}}$ and $\sigma_{\text{CB}}$ parametrize the position of the peak and the width of the Gaussian, $N$ is an overall normalization factor, $\alpha_{\text{low}}$ and $n_{\text{low}}$ parametrize the function in the low mass regime, while $\alpha_{\text{high}}$ and $n_{\text{high}}$ do so in the high mass range.

Another double-sided Crystal Ball function is used to analytically parametrized the shape of the single Higgs background, with fit parameters determined from a fit to simulated events.

Spurious signal analysis is performed to identify the lowest-bias functional form to describe the diphoton mass distribution of the continuum background. The fits are performed separately in the 1- and 2-tag categories, but the same family of fit functions is eventually selected and applied in both *b*-tagging categories. The first order exponential is chosen among the seven explored options. A data-driven approach is used for the final exponential fit to this background source.

### 7.5.3 Systematic Uncertainties

This analysis is fundamentally limited by statistical uncertainties due to the low number of events expected to pass the selection. Nonetheless, systematic uncertainties are carefully estimated and presented in this section.

The effects of various sources of systematic uncertainty on the yields in the 2- and 1-tag regions for the non-resonant and resonant analyses are summarized in Table 7.3 in terms of percentages of the nominal values of the yield in each category.

Uncertainties enter the process of cross-section extraction and limit setting in the form of constraints on the nuisance parameters included in the likelihood model.

Table 7.3: Percentage change in the nominal yields affected by the dominant sources of systematic uncertainty in the 2-tag (1-tag) categories for resonant and non-resonant analyses by sample of origin or analysis selection. The continuum background is omitted from this table because its contribution is derived through data-driven procedures instead of simulation.

| Source of systematic uncertainty | | % effect relative to nominal in the 2-tag (1-tag) category | | | |
|---|---|---|---|---|---|
| | | Non-resonant | | Resonant: BSM $hh$ signal | |
| | | SM $hh$ signal | Single $h$ bkg | Low mass | High mass |
| Luminosity | | ± 2.1  (± 2.1) | ± 2.1  (± 2.1) | ± 2.1  (± 2.1) | ± 2.1  (± 2.1) |
| Trigger | | ±0.4  (±0.4) | ±0.4  (±0.4) | ±0.4  (±0.4) | ±0.4  (±0.4) |
| Pile-up modelling | | ± 3.2  (± 1.3) | ±2.0  (±0.8) | ±4.0  (±4.2) | ±4.0  (±3.8) |
| Photon | identification | ± 2.5  (±2.4) | ± 1.7  (± 1.8) | ±2.6  (±2.6) | ± 2.5  (± 2.5) |
| | isolation | ±0.8  (±0.8) | ±0.8  (±0.8) | ±0.8  (±0.8) | ±0.9  (±0.9) |
| | energy resolution | - | - | ± 1.0  (± 1.3) | ± 1.8  (± 1.2) |
| | energy scale | - | - | ±0.9  (±3.0) | ±0.9  (±2.4) |
| Jet | energy resolution | ± 1.5  (±2.2) | ±2.9  (±6.4) | ±7.5  (± 8.5) | ±6.4  (±6.4) |
| | energy scale | ±2.9  (±2.7) | ±7.8  (± 5.6) | ±3.0  (± 3.3) | ± 2.3  (±3.4) |
| Flavour tagging | $b$-jets | ±2.4  (± 2.5) | ± 2.3  (± 1.4) | ±3.4  (±2.6) | ± 2.5  (±2.6) |
| | $c$-jets | ± 0.1  (± 1.0) | ± 1.8  (±11.6) | - | - |
| | light-jets | < 0.1  (± 5.0) | ± 1.6  (± 2.2) | - | - |
| Theory | PDF+$\alpha_S$ | ± 2.3  (± 2.3) | ± 3.1  (± 3.3) | n/a | n/a |
| | Scale | +4.3  (+4.3)  −6.0  (−6.0) | +4.9  (+ 5.3)  +7.0  (+8.0) | n/a | n/a |
| | EFT | ± 5.0  (±5.0) | n/a | n/a | n/a |

### 7.5.3.1 Theoretical Uncertainties

Theoretical uncertainties account for modeling effects and are extracted from relevant references and prior work. They amount to varying the cross-section to account for changes in the renormalization and factorizations scales, and uncertainties in the PDF and $\alpha_S$ estimates. Systematic effects due to the choice of parton showering and hadronization models are found to be negligible and are therefore disregarded in this analysis.

Uncertainties on the SM di-Higgs production cross-section, along with those on the SM Higgs branching ratios to photons and $b$-quarks, are reported in Sec. 7.2 and correspond to the values calculated in Ref. [66].

The SM single Higgs production, which proceeds through several production channels listed in Table 7.1, requires the computation of several uncertainty contributions, including the special treatment of uncertainties related to the production of Higgs bosons in association with heavy-flavor jets. The scale and PDF+$\alpha_S$ uncertainties reported in Ref. [66] are at most +20%/-24% and ±3.6%, respectively. Prior Run I work on heavy-flavor production in $t\bar{t}$ events [427] and associated production of $b$-jets with $W$ bosons [428] suggests applying 100% uncertainties on the single Higgs $gg$-fusion and $WH$ production modes.

The effects of these systematic variations on the expected yields obtained from the single and double Higgs samples in the non-resonant analysis are listed in the bottom portion of Table 7.3.

In the resonant analysis, theoretical uncertainties on the resonant samples production, as well as interference effects between the BSM signal and SM di-Higgs background are observed to have a negligible impact.

### 7.5.3.2 Experimental Uncertainties

Experimental uncertainties include object and event uncertainties that affect the expected experimental result derived from simulation, by modifying the yields in the various analysis regions, as well as the fit parameters in the analytical models. The extent of the experimental parameter variations is generally communicated in the form of centralized recommendations made by ad-hoc ATLAS performance groups.

Object-related uncertainties encode the inability to extract properties of physics object with infinite precision, or to simulate their distributions to match reality. The numerical value of each physical observable used to describe jets, photons, muons, and other objects in the simulated events considered in this analysis is inherently uncertain and cannot be taken at face value. When these variables are utilized to design sequential cuts that define signal and control regions, small fluctuations around their nominal values may cause events to move from one statistical region to another, thus affecting the analysis yields and the location and width of the analytical functions used to fit their diphoton or four-body reconstructed mass distributions.

Object uncertainties are therefore propagated into uncertainties on expected yields and other modeling parameters that enter the statistical analysis, by repeating the fits and recomputing best-fit model parameters after independently varying the values of object observables up and down within the limits of their respective uncertainties. Parameters that describe the tails of the fit functions used in the modeling process are held at their nominal values and uncertainties are not propagated to these secondary model parameters. In the resonant analysis, the effects of systematic variations are first computed independently for each $m_X$ hypothesis, and then the maximum global uncertainty is applied to the whole resonance mass range.

In the non-resonant analysis, the main systematics that contribute to the uncertainty in the peak location for both 1- and 2-tag categories in the SM $hh$ sample include the photon and jet energy scales, which, along with flavor tagging systematics, also contribute to the uncertainty in the signal width parameter. In SM single $h$ samples, flavor tagging is a prominent source of uncertainty, in addition to jet and photon energy scales and resolutions. In particular, photon energy resolution is the main source of uncertainty in the width parameter of the model across single and double Higgs samples. In the resonant analysis, the photon energy scale dominates the uncertainty on the peak's location, while the jet energy resolution strongly affects the uncertainty in the width and yield across tagging categories and $m_X$ regions. These result in uncertainties of 0.2–0.6% on the peak's location and 5–14% on the peak's width across all single and di-Higgs samples considered in the resonant and non-resonant analyses.

The spurious signal studies performed to inform the choice of background fit functions (see Sec. 7.5.2 and 7.5.1) automatically assess the uncertainty in the signal yields in the various signal regions in both resonant and non-resonant analyses. The uncertainty in the number of signal events in each category is reported in Table 7.4. The bias correction procedure discussed in Sec. 7.5.1 results in further $m_X$-dependent modification to the extracted cross-sections in the low mass analysis.

|  |  | Non-Resonant | Resonant |
|---|---|---|---|
| 1-Tag | Low Mass | -2.91 | 2.06 |
|  | High Mass | -0.25 | 0.89 |
| 2-Tag | Low Mass | -0.89 | 0.58 |
|  | High Mass | -0.63 | 0.21 |

**Table 7.4:** Spurious signal events for each analysis category which contribute to the systematic uncertainty related to the choice of fit functions used to model the $m_{\gamma\gamma}$ and $m_{\gamma\gamma jj}$ distributions. In the resonant analysis, the largest spurious signal contribution from each of the mass points included in the low and high mass ranges are used as global uncertainties in the corresponding analysis category.

Event-level uncertainties include: the uncertainty in the determination of the luminosity delivered to and collected by the ATLAS experiment (2.1% for the 2015+2016 data), the uncertainty on the diphoton trigger efficiencies (0.4%) and pile-up reweighting (up to ±4.2%).

Other negligible uncertainties include those associated with the vertex selection algorithm efficiency.

## 7.6 Results

The simultaneous unbinned fits to data in the 1- and 2-tag categories for the $m_{\gamma\gamma}$ and $m_{\gamma\gamma jj}$ distributions for the non-resonant and resonant analyses, respectively, are shown in Fig. 7.23 and 7.24 by separating the events into the exclusive *b*-tagging and non-exclusive analysis selection categories. Only the background-only fits are shown, primarily for aesthetic reasons due to the absence of signal or the presence of negative signal when adding the signal component to the fit.

The non-resonant analysis finds no statistically significant excess with respect to the SM prediction. The collected data points agree, within the uncertainties, with the background-only prediction in all signal regions. The best-fit estimate of the extracted di-Higgs cross-section is $0.04^{+0.43}_{-0.36}$ pb in the low mass selection, and $-0.21^{+0.33}_{-0.25}$ pb in the high mass selection. The resonant analysis finds the largest deviation from the background-only hypothesis around $m_X = 480$ GeV, corresponding to a local significance of $1.2\sigma$.

Therefore, limits are set on the resonant and non-resonant Higgs pair production cross-sections and Higgs self-coupling. At the $95\%$ confidence level, the observations place the upper limit on the non-resonant Higgs pair cross-section at $0.73$ pb, or $22\times$ the SM expectation, while the resonant cross-section limits are set as a function of $m_X$ in the narrow-width approximation, and range over $[0.12, 1.1]$ pb for $m_X \in [260, 1000]$ GeV. The anomalous Higgs self-coupling $\kappa_\lambda = \lambda_{hhh}/\lambda_{SM}$ is constrained to the range $[-8.2, 13.2]$ at 95% confidence.

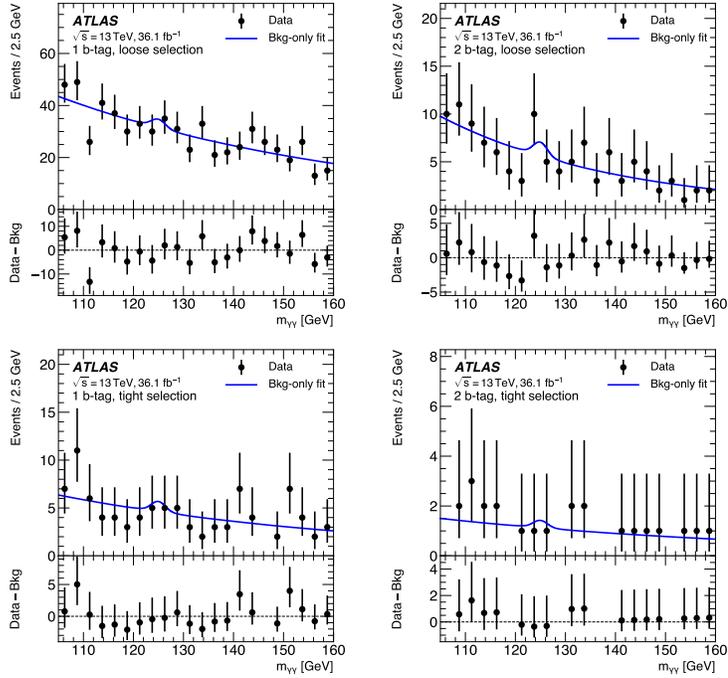

**Figure 7.23:** $m_{\gamma\gamma}$ distributions for data events in the non-resonant search that pass the loose (or low mass) selection (top row), and the tight (or high mass) selection (bottom row), divided by $b$-tagging categories (1-tag on the left, 2-tag on the right). The blue line shows the background-only fit applied to the data points, while the bottom panel shows the ratio of the data to the fit.

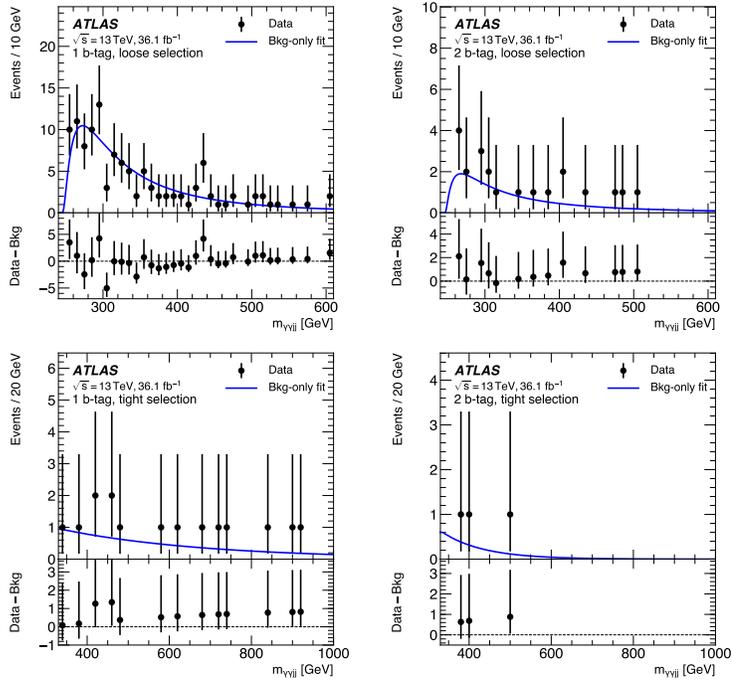

**Figure 7.24:** $m_{\gamma\gamma jj}$ distributions for data events in the resonant search that pass the loose (or low mass) selection (top row), and the tight (or high mass) selection (bottom row), divided by $b$-tagging categories (1-tag on the left, 2-tag on the right). The blue line shows the background-only fit applied to the data points, while the bottom panel shows the ratio of the data to the fit.

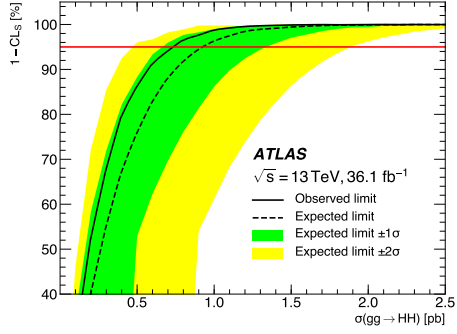 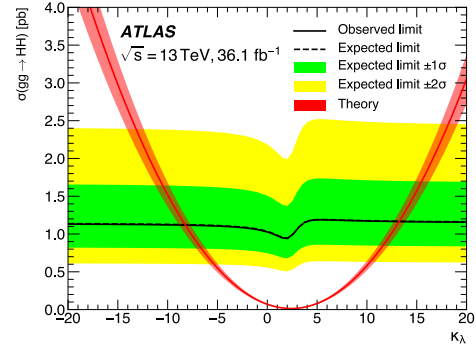

**Figure 7.25:** Predicted non-resonant di-Higgs production cross-section limits (with 1 and $2\sigma$ uncertainty bands) obtained from the fit to simulated events, plotted versus $CL_s$ confidence level, and compared to the observed exclusion limits derived from the fit to data.

**Figure 7.26:** Predicted 95% limit on the non-resonant di-Higgs production cross-section (with 1 and $2\sigma$ uncertainty bands) obtained from fits to simulated events at different values of the Higgs self-coupling constant $\kappa_\lambda$, plotted versus the tested value of $\kappa_\lambda$, and compared to the observed exclusion limits derived from the fit to data.

These limits can be compared to the predicted parameter ranges extracted before unblinding, which quantify the power of the designed analysis and result in the expected exclusion limits. To do so, the shape of the background from the sidebands in the high mass selection is extrapolated to populate the signal region. The 95% CL expected limit on the non-resonant cross-section is $0.93^{+1.4}_{-0.66}$ pb, or $28^{+40}_{-20}\times$ the SM prediction. The observed and expected limits on the non-resonant di-Higgs cross-section (and the expected 1 and $2\sigma$ bands) are plotted versus the confidence level in Fig. 7.25.

The observed and expected limits to the range of allowed values of the Higgs self-coupling parameter $\kappa_\lambda$ are derived using the low mass selection, which is kinimatically more favorable to investigate the deviations caused by the range of variations in $\kappa_\lambda$ explored in this work. In this case, instead of resorting to the production of toy datasets, the distribution of the profile-likelihood statistic is approximated using the asymptotic behavior of the function. As shown in Fig. 7.26, at 95% CL, the analysis is expected to be able to constrain the $\kappa_\lambda$ coupling to the range $[-8.3, 13.2]$.

The observed limits on the BSM resonant di-Higgs production cross-section can instead be com-pared to the predicted limits and their confidence bands shown in Fig. 7.27, which are generated by dividing the mass spectrum into a low mass and a high mass region and applying the suitable analysis selections to various $m_X$ hypothesis within the two mass regimes. The expected limits in the range $m_X \in [260, 1000]$ GeV go from 0.90 pb to 0.15 pb.

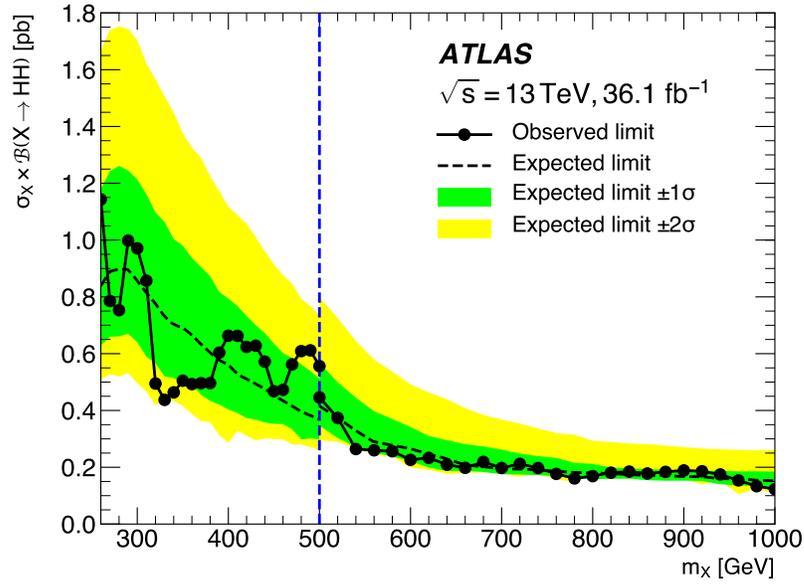

**Figure 7.27:** Predicted 95% CL resonant di-Higgs production cross-section × branching ratio limits (with 1 and 2σ uncertainty bands), plotted versus the resonance mass hypothesis, and compared to the observed exclusion limits derived from fits to data. The vertical line indicates the transition point between low and high mass regions.

## 7.7 Other di-Higgs Search Results in ATLAS and CMS

The results of this analysis can be further understood in the context of other di-Higgs searches performed in Run I and Run II by ATLAS and CMS. Both the ATLAS and the CMS collaborations have activated a number of di-Higgs analyzers, divided into groups according to Higgs decay channels ($\gamma\gamma b\bar{b}$, $b\bar{b}b\bar{b}$, $W^+W^-b\bar{b}$, $\tau^+\tau^- b\bar{b}$, $W^+W^-\gamma\gamma$, etc.). Most searches take advantage of the high branching ratio of the Higgs boson to a $b$ quark-antiquark pair to enhance their event rate. However, as in the case of single Higgs searches, this channel is plagued by copious amounts of background QCD activity. On the other side of the spectrum, channels that include a $h \to \gamma\gamma$ decay can rely on precise diphoton mass reconstruction and low background contamination. The heat maps in Fig. 7.28 show the expected number of di-Higgs events in each decay channel with 35 and 3000 fb$^{-1}$ of integrated luminosity, as well as the decay modes currently considered by Run II ATLAS analyses.

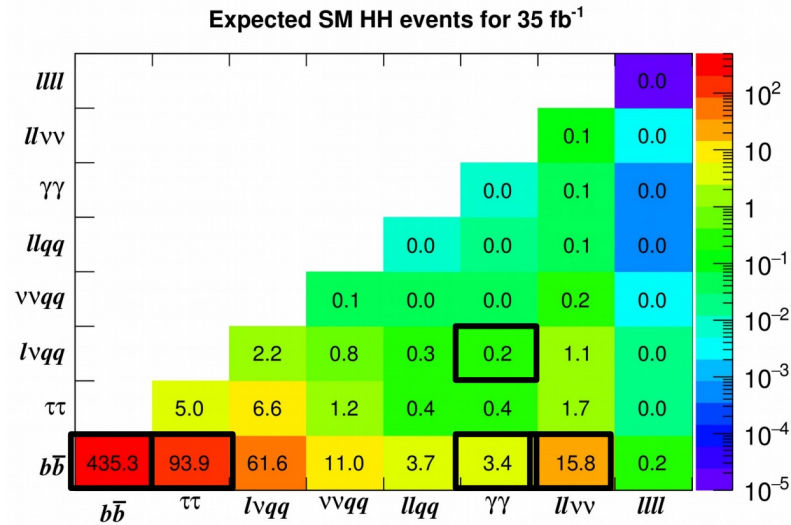

(a) Expected number of di-Higgs events in various decay channels, assuming non-resonant SM production only and SM Higgs branching ratios, at an integrated luminosity of 35 fb$^{-1}$. The black boxes indicate channels that are currently being targeted by ATLAS searches.

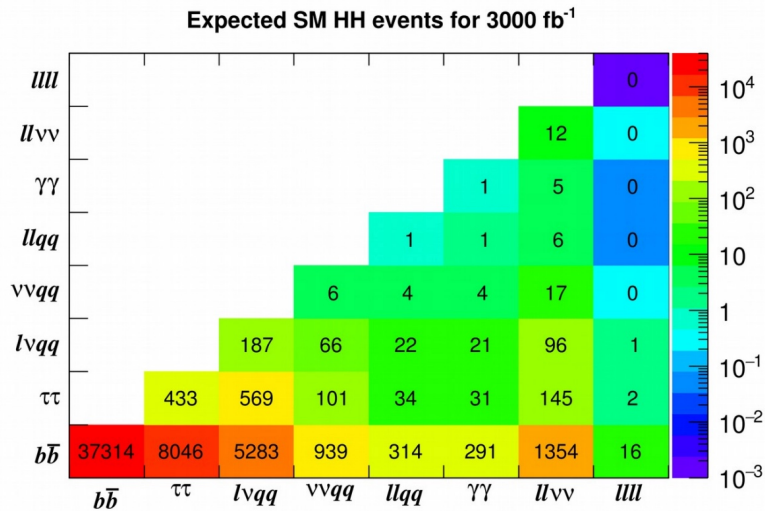

(b) Expected number of di-Higgs events in various decay channels, assuming non-resonant SM production only and SM Higgs branching ratios, at an integrated luminosity of 3000 fb$^{-1}$.

**Figure 7.28:** Standard Model branching ratios, expressed as number of expected events at the specified integrated luminosity, for a di-Higgs system in which one Higgs boson decays to the particles specified along the $x$-axis, and the second Higgs boson decays to the particles specified along the $y$-axis.

This section complements the report of the Run II $hh \to \gamma\gamma b\bar{b}$ analysis above by briefly summarizes past ATLAS di-Higgs searches – including the Run I $\gamma\gamma b\bar{b}$ search (Sec. 7.7.1), the ATLAS Run I combination (Sec. 7.7.2), and the early Run II $\gamma\gamma bb$ analysis (Sec. 7.7.3) – and other Run II efforts from both ATLAS and CMS (Sec. 7.7.4)

### 7.7.1 Run I $\gamma\gamma b\bar{b}$ ATLAS Analysis

Excitement around di-Higgs searches started during Run I, with the larger than expected di-Higgs event rate measured by the first $\gamma\gamma b\bar{b}$ ATLAS analysis [429]. The search for enhancements in the di-Higgs production was performed at $\sqrt{s}$ =8 TeV using an integrated luminosity of 20.3 ± 0.6 fb$^{-1}$, corresponding to the full dataset collected in 2012.

The results of the non-resonant search were quantified as a 2.4$\sigma$ excess from the background-only hypothesis, or as an upper limit on the non-resonant cross-section of 2.2 pb, compared to the expected analysis sensitivity that would exclude cross-sections up to 1.0 pb at 95% CL. Resonant cross-section upper limits at 95% CL were set as a function of $m_X$ in the narrow-width approximation, and range over [0.7, 3.5] pb for $m_X \in [260, 500]$ GeV. The resonant search was therefore limited to spin-0 resonances in the low mass range. The largest deviation from the background-only hypothesis was observed at $m_X = 300$ GeV, equivalent to a 3$\sigma$ local significance, which was reduced to 2.1$\sigma$ after accounting for the look-elsewhere effect [430, 431].

The results are further summarized in the following plots: Fig. 7.29 shows the excess of di-Higgs events registered in the signal region, along with the best signal + background fit to the $m_{\gamma\gamma}$ distribution applied to the data points and compared to the background-only fit; Fig. 7.30 represents the five signal data points along the four-body distribution with the best background-only fit and the shape of a 300 GeV simulated resonant sample for comparison; finally, Fig. 7.31 interprets the results in the form of 95% CL exclusion limits on the cross-section × branching ratio for the production of a spin-0 resonance at various mass values in the range between 260 and 500 GeV.

The non-resonant analysis was performed with a simultaneous fit to the $m_{\gamma\gamma}$ distribution in the ≥ 2 $b$-tag category and in the ≤ 1 $b$-tag category, using an exponential function to model the background and a Crystal Ball + Gaussian for the signal component. The resonant analysis, instead, proceeded via a cut-and-count strategy because of the low expected number of events.

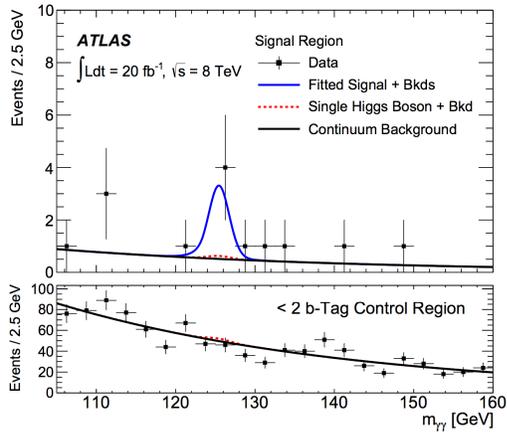
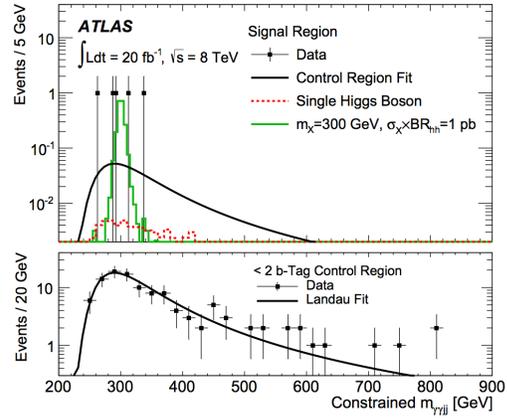

**Figure 7.29:** Diphoton mass distribution in the signal region (top) and control region (bottom) defined for the non-resonant analysis, with the best-fit functions derived for the description of the continuum background (black), the background-only hypothesis that includes the single $h$ contribution (red), and the signal + background hypothesis (blue).

**Figure 7.30:** Four-body mass distribution in the signal region (top) and control region (bottom) defined for the resonant analysis, with the best-fit functions derived for the description of data in the sidebands. The distribution of simulated events from a compatible resonance sample at $m_X = 300$ GeV is shown for comparison.

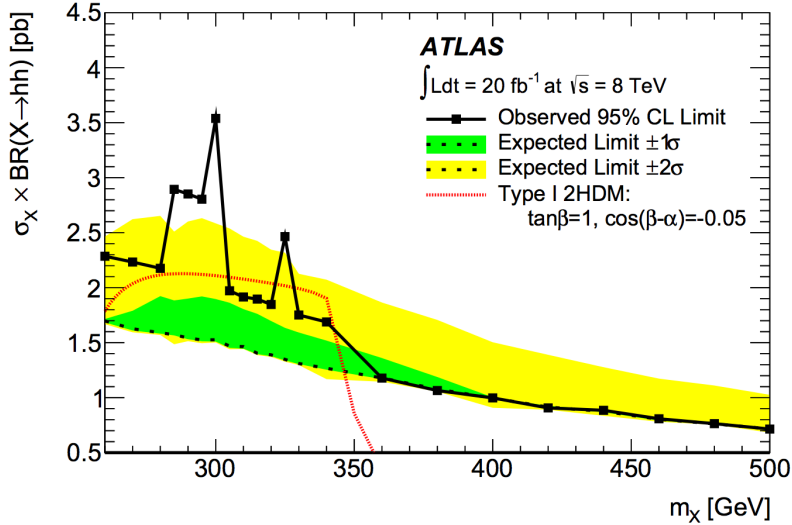

**Figure 7.31:** Predicted 95% CL resonant di-Higgs production cross-section × branching ratio limits (with 1 and $2\sigma$ uncertainty bands), plotted versus the resonance mass hypothesis, and compared to the observed exclusion limits derived from fits to data. The red line symbolizes the prediction of a type I 2HDM theory as computed in Ref. [42].

| Physical Process | Events in $|m_h - m_{\gamma\gamma}| < 2\sigma_{m_{\gamma\gamma}}$ |
|---|---|
| Continuum background | $1.3 \pm 0.5$ |
| SM $h$ | $0.17 \pm 0.04$ |
| SM $hh$ | 0.04 |
| Observed | 5 |

Table 7.5: Expected number of events in the signal region of the non-resonant analysis for the Run I di-Higgs search in the $\gamma\gamma b\bar{b}$ decay channel, and number of observed events in data.

Table 7.5 shows the expected contribution of different physical processes to number of events in the signal region for the non-resonant analysis, which is to be compared with the observation of five events in the examined data.

### 7.7.2 ATLAS Run I Combination

The Run I $\gamma\gamma b\bar{b}$ ATLAS result outlined in Sec. 7.7.1 was interpreted in the context of other di-Higgs ATLAS searches through the combination of results from the $b\bar{b}\tau^+\tau^-$, $\gamma\gamma W^+W^-$, $\gamma\gamma b\bar{b}$ [429], and $b\bar{b}b\bar{b}$ [432] channels using 20.3 fb$^{-1}$ of $pp$ collision data collected with the ATLAS detector [43].

Table 7.6 reports the expected and observed 95% CL upper limits on the non-resonant $gg \to hh$ cross-section for individual Run I ATLAS analyses, along with the combined result. The final observation had a $p$-value of 0.044 with the SM hypothesis, due to the excess reported by the $\gamma\gamma b\bar{b}$ analysis [43]. The combined analyses had the statistical power to exclude, with 95% confidence, a cross-section for this process larger than 48× the SM prediction, or larger than 0.47 pb. However, the data collected by the ATLAS detector showed upward fluctuations in the number of candidate events that led analyzers to only be able to exclude a cross-section value larger than 70× the SM prediction, or larger than 0.69 pb.

For the resonant search, several low (260-500 GeV) and high (500-1000 GeV) mass hypotheses were tested, and upper limits for each were set on the cross-section for $gg \to H$ production times the branching ratio to SM-like Higgs bosons BR($H \to hh$), assuming the SM values for the subsequent $h$ decays. The $\gamma\gamma b\bar{b}$ and $b\bar{b}\tau^+\tau^-$ analyses provided the tightest limits on the cross-section times branching ratio measurements at low $m_H$, while the $b\bar{b}b\bar{b}$ result dominated at high $m_H$. The combination showed an excess with local significance of $2.5\sigma$ at the mass value of 300 GeV [43]. The joint analyses were able to set tighter bounds on $\sigma(gg \to H) \times$ BR($H \to hh$) at high $m_H$ because of higher background suppres-sion. The observed and expected 95% CL upper limits from the combination of the results reported by di-Higgs resonant analyses across decay channels are displayed in Fig. 7.32.

| Analysis | $\gamma\gamma b\bar{b}$ | $\gamma\gamma WW^*$ | $b\bar{b}\tau^+\tau^-$ | $b\bar{b}b\bar{b}$ | Combined |
|---|---|---|---|---|---|
| | Upper limit on the cross-section [pb] | | | | |
| Expected | 1.0 | 6.7 | 1.3 | 0.62 | 0.47 |
| Observed | 2.2 | 11 | 1.6 | 0.62 | 0.69 |
| | Upper limit on the cross-section relative to the SM prediction | | | | |
| Expected | 100 | 680 | 130 | 63 | 48 |
| Observed | 220 | 1150 | 160 | 63 | 70 |

**Table 7.6:** Summary reported values for expected and observed 95% CL upper limits on the cross-section of non-resonant di-Higgs production via gluon fusion at $\sqrt{s}$ = 8 TeV from the ATLAS Run I di-Higgs analyses in four decay channels. These are expressed, in the lower portion of the table, as a ratio with respect to the SM cross-section prediction of 9.9 ± 1.3 fb. Table reproduced from Ref. [43].

These analyses were dominated by statistical uncertainties due to the low cross-section for di-Higgs production. The primary source of systematic uncertainty in both resonant and non-resonant searches was due to background modeling [43].

### 7.7.3 Early Run II $\gamma\gamma bb$ ATLAS Analysis

ATLAS repeated the search in the $\gamma\gamma b\bar{b}$ with 3.2 fb$^{-1}$ of data at $\sqrt{s}$ = 13 TeV [419].

A cut-and-count approach was used for the resonant search in the mass range $H \in [275, 400]$ GeV, while a fit on the $m_{\gamma\gamma}$ distribution was performed for the non-resonant search.

The main sources of background included a dominant continuum from photon + jets events, which was estimated using data-driven techniques, as well as SM single Higgs events, which were well-modeled in simulation. In the non-resonant analysis, the continuum was handled by applying a simultaneous fit to the $m_{\gamma\gamma}$ distribution in the control region defined by events with zero $b$-tagged jets, and in the signal region defined by the events with two $b$-tagged jets. The single Higgs background was fit with a double-sided Crystal Ball (DSCB) function. Another DSCB function was used to fit the di-Higgs signal. In the resonant analysis, the background was fitted in the 0-tag category (with the same shape applied to the 2-tag category) using an exponentially decaying function along the $m_{\gamma\gamma}$ direction, and a Landau function along the $m_{\gamma\gamma b\bar{b}}$ direction. Cuts were applied to the two-dimensional distribution defined by these two quantities so that 95% of the signal was preserved after the cuts. The background contribution in the signal region was then estimated by scaling the number of background events in the sidebands by the false positive rate caused by the cut along $m_{\gamma\gamma}$, and by the false positive rate caused by the cut along $m_{\gamma\gamma b\bar{b}}$ after the application of the first cut. In practice, interpreting the false positive rate associated with a cut as the efficiency $\epsilon^{\mathrm{background}}$ for the selection of background events:

$$N^{\mathrm{background}}_{\mathrm{signal\ region}} = N_{\mathrm{sidebands}} \epsilon^{\mathrm{background}}_{m_{\gamma\gamma}} \frac{\epsilon^{\mathrm{background}}_{m_{\gamma\gamma b\bar{b}}}}{1 - \epsilon^{\mathrm{background}}_{m_{\gamma\gamma}}}. \tag{7.11}$$

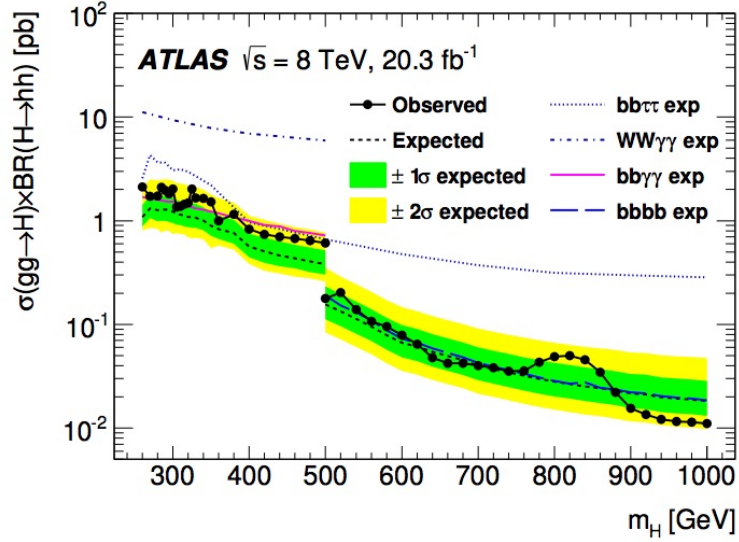

**Figure 7.32:** Observed and expected 95% CL upper limits on $\sigma(gg \to H) \times \mathrm{BR}(H \to hh)$ as a function of the heavy Higgs mass $m_H$, obtained from the combination of resonant di-Higgs searches with data collected in Run I with the ATLAS detector at a center of mass energy of 8 TeV, assuming SM branching ratios for light Higgs $h$ decays. Image reproduced from Ref [43].

The fit functions described above were integrated to estimate the false positive rate of the signal-region-defining cuts. The location of the $m_{\gamma\gamma b\bar{b}}$ cut was varied depending on the resonant mass of the signal hypothesis.

The statistical uncertainty was the dominant limitation to this analysis. For the continuum background, this was estimated from the variance of the fluctuations in the percentage of accepted background events, after repeating the cut in the diphoton mass distribution $\epsilon^{\mathrm{background}}_{m_{\gamma\gamma}}$ on a series of Monte Carlo toy distributions obtained by applying Poisson fluctuations to the bins' content in the original distribution.

The main contribution to the the systematic uncertainty was the uncertainty related to the modeling

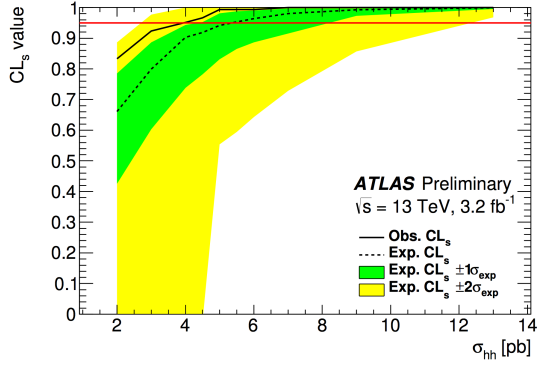
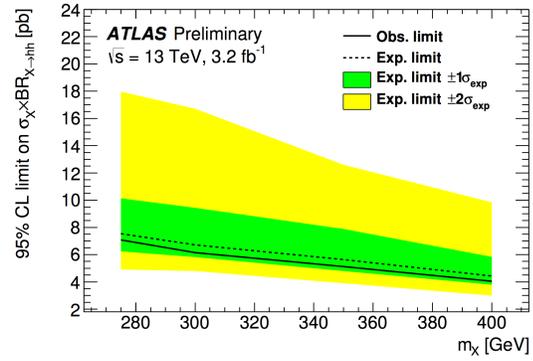

Figure 7.33: Predicted non-resonant di-Higgs production cross-section limits (with 1 and $2\sigma$ uncertainty bands) obtained from the fit to simulated events selected by the early Run II $\gamma\gamma b\bar{b}$ ATLAS analysis. The expected limits are plotted versus $CL_s$ confidence level, and compared to the observed exclusion limits derived from the fit to 2015 data at $\sqrt{s}$ = 13 TeV.

Figure 7.34: Predicted 95% CL resonant di-Higgs production cross-section $\times$ branching ratio limits (with 1 and $2\sigma$ uncertainty bands) obtained from the selection devised by the early Run II $\gamma\gamma b\bar{b}$ ATLAS analysis. The expected limits are plotted versus the resonance mass hypothesis, and compared to the observed exclusion limits derived from fits to 2015 data at $\sqrt{s}$ = 13 TeV.

of background processes, which are poorly described in Monte Carlo simulation.

The analysis reported a deficit of events in the signal region defined by the analysis cutflow compared to expectations, which corresponded to an observed upper bound on the non-resonant di-Higgs production cross-section of 3.9 pb at 95% CL, compared to expected limit of $5.4^{+2.8}_{-1.0}$ pb. The 95% CL exclusion limits for the di-Higgs production cross-section via narrow-width resonance are found to be in the range [7.0, 4.0] pb for masses in the range [275, 400] GeV, compared to the expected limits ranging between 7.5 and 4.4 pb. The results of the non-resonant and resonant analyses are condensed into the plots in Fig. 7.33 and 7.34, respectively.

### 7.7.4 OTHER RUN II DI-HIGGS RESULTS

Throughout Run II of the LHC, the ATLAS and CMS collaborations have continued analyzing the data collected by their detectors to look for di-Higgs signatures across multiple decay channels. CMS has so far published results on 13 TeV data in the following channels: $\gamma\gamma b\bar{b}$ [433, 434], $b\bar{b}l\nu l\nu$ [435], $b\bar{b}\tau^+\tau^-$ [436], and $b\bar{b}b\bar{b}$ [437, 438, 439, 440]. ATLAS has published 13 TeV results in the following channels: $\gamma\gamma b\bar{b}$ [419, 413], $b\bar{b}b\bar{b}$ [441, 442], $b\bar{b}\tau^+\tau^-$ [443], and $\gamma\gamma WW^*$ [444]. These analyses have contributed to the progressive tightening of the upper limits on the resonant and non-resonant di-Higgs production cross-sections.

Fig. 7.35, produced prior to the publication of the latest $\gamma\gamma b\bar{b}$ ATLAS analysis described earlier in this

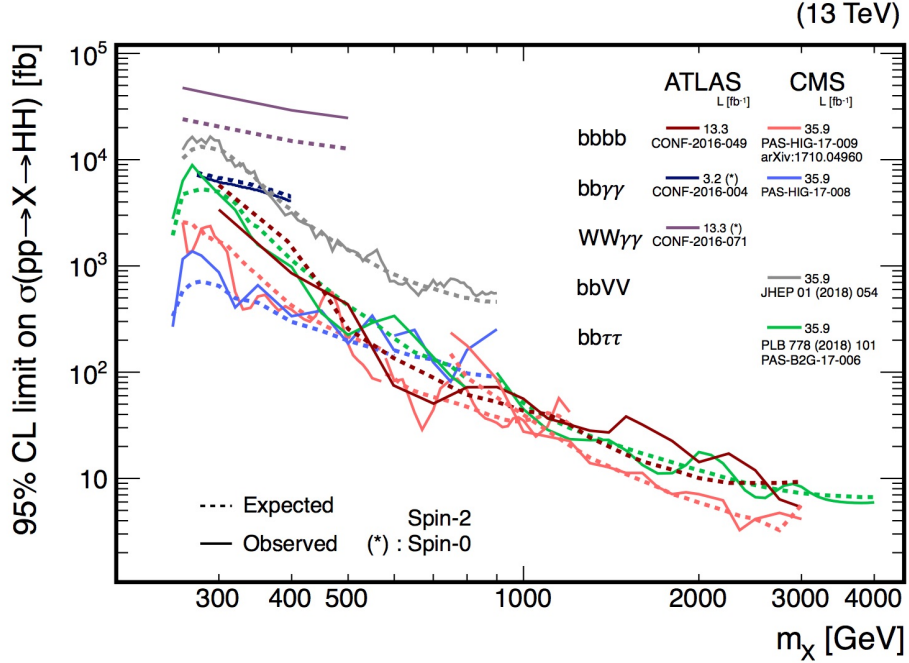

**Figure 7.35:** Expected and observed limits on the production cross-section of a heavy resonance times its branching ratio to a Higgs pair as a function of resonance mass, for various ATLAS and CMS analyses (publication ID available in the legend) performed at $\sqrt{s} = 13$ TeV. The integrated luminosity used in each analysis is also listed in the legend. Image reproduced from Ref [44].

chapter but after the early Run II iteration described in Sec. 7.7.3, shows the individual contributions of the various published ATLAS and CMS analyses on 13 TeV data in setting the upper limits to the cross-section for resonant di-Higgs production as a function of the mass of the intermediate state. The $\gamma\gamma b\bar{b}$ channel continues to dominate in the low mass region ($m_X \lesssim 500$ GeV) due to the cleanliness of the channel and the ability to reject background, while the $b\bar{b}b\bar{b}$ search dominates the higher portion of the spectrum because of the high branching ratio [44].

After ensuring the orthogonality of each analysis selection, the results obtained by each experiment's Run II di-Higgs analyses have been combined into joint statistical results [445, 446].

The plots in Fig. 7.36 show the upper limits on the non-resonant di-Higgs production cross-section, normalized to the SM cross-section, set by individual di-Higgs analyses in ATLAS and CMS using 13 TeV data. The combined results are presented in the bottom panels.

The statistical combination of CMS results using 35.9 fb$^{-1}$ of $pp$ collision data at 13 TeV sets the limit on the production cross-section to 21.8× the SM prediction, compared to the expected ability to exclude values up to 12.4×$\sigma_{SM}$, while limits on the resonant cross-section for the production of a spin-0 radion

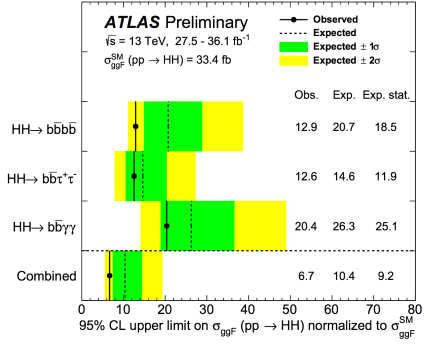
(a) ATLAS results across 3 decay channels.

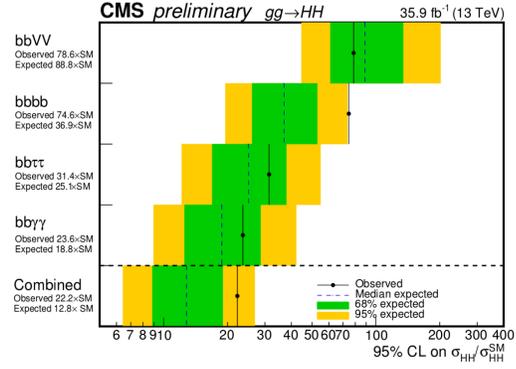
(b) CMS results across 4 decay channels with combination.

**Figure 7.36:** Observed and expected upper limits of the ratio of the di-Higgs production cross-sections to the SM di-Higgs production cross-section from ATLAS and CMS analyses at 13 TeV. The green (yellow) bands mark the extend of the expected $\pm 1(2)\sigma$ intervals.

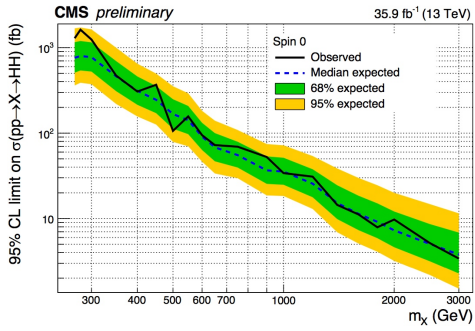
(a) Spin-0

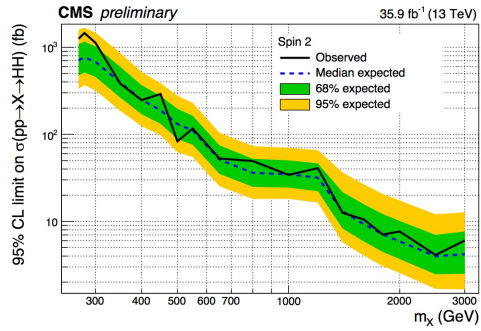
(b) Spin-2

**Figure 7.37:** 95% CL upper limits on the cross-section times branching ratio of narrow, spin-0 or spin-2, resonances decaying to pairs of SM-like Higgs bosons as a function of the resonance mass $m_X$ obtained from the combination of CMS di-Higgs analyses carried out at $\sqrt{s} = 13$ TeV with 35.9 fb$^{-1}$ of data in Run II. Images reproduced from Ref. [45].

and spin-2 graviton times the branching ratio to a Higgs pair are set, in the narrow width approximation, for $m_X$ up to 3 TeV and displayed in Fig. 7.37 [446].

The statistical combination of the $b\bar{b}b\bar{b}$, $b\bar{b}\tau^+\tau^-$, and $b\bar{b}\gamma\gamma$ ATLAS results using 36.1 fb$^{-1}$ of $pp$ collision data at 13 TeV sets the limit on the production cross-section to 0.22 pb, equivalent to 6.7× the SM prediction, compared to the expected ability to exclude values up to 0.35 pb, equivalent to 10.4×$\sigma_{SM}$ [445].

Limits on the cross-section for the production of spin-0 and spin-2 heavy resonances times the branching ratio to a Higgs pair are set, in the narrow width approximation, for resonance masses up to 1 TeV. The 95% CL limits on the heavy scalar production cross-section are plotted versus resonance mass in Fig. 7.38. Only the $b\bar{b}b\bar{b}$ and $b\bar{b}\tau^+\tau^-$ analyses contribute to the spin-2 limits.

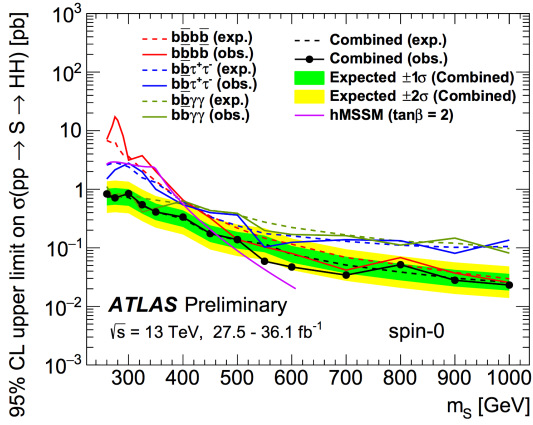
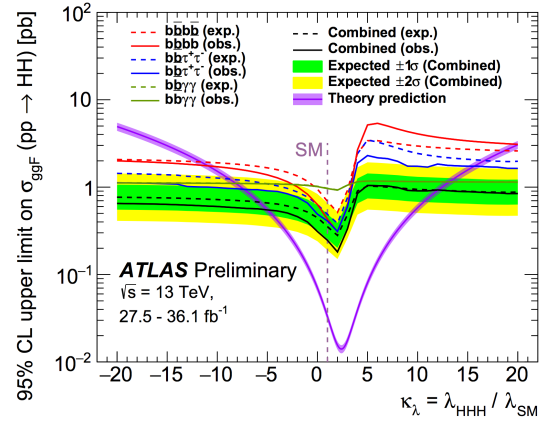

Figure 7.38: Predicted 95% CL limits on the resonant di-Higgs production cross-section from a heavy scalar (with 1 and 2$\sigma$ uncertainty bands) obtained from the combination the $b\bar{b}b\bar{b}$, $b\bar{b}\tau^+\tau^-$, and $b\bar{b}\gamma\gamma$ ATLAS analyses performed on 36.1 fb$^{-1}$ of data at $\sqrt{s} = 13$ TeV. The expected and observed limits are displayed for each individual analysis, as well as for the combined statistical analysis, and are plotted versus the resonance mass hypothesis.

Figure 7.39: Individual and combined predicted 95% limits on the non-resonant di-Higgs production cross-section (with 1 and 2$\sigma$ uncertainty bands) at different values of the Higgs self-coupling constant $\kappa_\lambda$, obtained from simulation by the $b\bar{b}b\bar{b}$, $b\bar{b}\tau^+\tau^-$, and $b\bar{b}\gamma\gamma$ ATLAS analyses performed on 36.1 fb$^{-1}$ of data at $\sqrt{s} = 13$ TeV. The expected limits are denoted by dashed lines, the observed limits by solid lines.

Combined constraints on the Higgs self-coupling parameter $\kappa_\lambda$ are depicted in Fig. 7.39 and set at 95% CL to the range $\kappa_\lambda \in [\text{-}5.0, +12.1]$, compared to the expected exclusion limits ranging between -5.8 and +12.0.

## 7.8 Prospects at the HL-LHC

The future phase of the LHC lifetime, known as HL-LHC, provides a more rosy outlook for the measurement of the di-Higgs production cross-section.

At the future center of mass energy of 14 TeV, the expected NNLL matched to NNLO $gg \to hh$ production cross-section including top quark mass effects to NLO for $m_h = 125$ GeV is 39.64 fb$^{+4.4\%}_{-6.0\%}$ (QCD scale) $\pm 2.1\%$ (PDF) $\pm 2.2\%$ ($\alpha_s$) [66]. On the other hand, the cross-section for $gg \to hhh$ at NLO QCD will remain below 100 ab.

Preliminary, generator-level studies performed by the ATLAS collaboration predict that the expected 3000 fb$^{-1}$ of data to be collected during that phase could yield an expected significance of $1.05\sigma$ in the $\gamma\gamma b\bar{b}$ decay channel, while constraining at 95% confidence level the ratio of the Higgs self-coupling to the SM prediction to between -0.8 and 7.7 [447]. In these investigations, the *MC truth* final state parti-cles are smeared to emulate the detector resolution and response, and overlaid with an expected pile-up rate of $\langle \mu \rangle = 200$.

Prospects for $pp \to hh$ cross-section measurements and anomalous Higgs self-coupling constraints in the $\gamma\gamma b\bar{b}$ channel for higher energy upgrades to the LHC (27 TeV and 100 TeV) are documented in Ref. [448].

# 8

# Generative Adversarial Networks for Fast Simulation in High Energy Physics

The simulation of scientific datasets, often used as test beds for the development and evaluation of application and domain specific machine learning algorithms, is, in various disciplines such as high energy physics, a slow and complex, yet necessary, step in many scientists' workflows. Scientific disciplines utilize computer simulations and modeling techniques to describe natural phenomena and parameterize their functional description in terms of latent factors. These may be free parameters of the theory, for which constraints from measurement may or may not exist, or theoretically computable properties, for which competing models predict variations. Simulated datasets are used to investigate how model parameters affect physical realizations and outcomes, both for well-established and hypothesized theories. At the core of each model is an analytical or estimated, theory- or data-driven probability density function. Simulation, therefore, provides a computational basis to perform a large number of experiments and to build statistical expectations for hypothesis testing, when performing physical exper-iments is prohibitively expensive or time-consuming, or when inference of the model parameters from observation is intractable. In the context of hardware development, simulation is also used during the planning phase of detector R&D to establish the effect of design choices on physics desiderata such as shower containment.

To achieve the desired resolution and confidence in parameter estimation from simulation, large production campaigns of petabytes of simulated data are necessary to guarantee the appropriate statistics availability.

Ordinary scientific simulators often rely on precise solvers to exactly compute trajectories, kinematic properties, and cross-sections for particle or ensemble interactions under various physical processes. In addition to deterministic computations, the simulation of particle interactions with matter, central to the field of high energy physics, requires the implementation of Monte Carlo techniques to model the stochastic nature of phenomena involving particles traversing detector volumes. In the context of LHC experiments like ATLAS, GEANT4 provides the computational engine to produce detailed *full simulation* of detector effects, while a suite of *fast simulation* software packages have been accompanying it for those use cases that can afford the trade-off between simulation accuracy and latency. Nonetheless, GEANT4 remains necessary to validate fast simulation outputs and serve the analyses that require the lowest possible systematic uncertainties. GEANT4 implements state-of-the-art physics models that encode the probabilistic nature of the propagation and interaction of each individual particle with portions of the detector, allowing the user to specify arbitrary detector geometries.

Of particular interest in the simulation workflow is the faithful description of shower development in calorimeters – a crucial task given the extent to which calorimetry contributes to the detection and measurement of particles in the experiments at the LHC. For a reminder on the role of calorimeters within detectors like ATLAS, and on the type of physical reactions particles undergo while traversing calorimeter volumes, refer to Sec. 2.2.1.2.

From the computational standpoint, tracking hundreds of thousands of particles as they interact with atomic nuclei and contribute to particle cascades has heavy compute requirements which limit the practical feasibility of full detector simulation. While, as of Run I, ATLAS Monte Carlo sample production altogether already occupied approximately 75% of total CPU utilization and more than 60% of the total available disk space [449], the growing demands for full simulation of large datasets in Run II (and in the future, at the HL-LHC) will continue to put a large strain on the ATLAS computing resources. As a matter of fact, the precise description of cascade formation as particles interact with material in the calorimeter regions is the most computationally demanding step in the ATLAS simulation pipeline, which itself annually requires billions of CPU hours [450, 451, 449].

Currently, the high CPU cost of production of fully-simulated GEANT4 samples also has a bearing upon disk space that needs to be reserved for the storage of these valuable datasets. The promise of a fast simulation chain is to reduce its computational footprint to the point that its required processing time becomes comparable to that of other elements in the pipeline; at that point, simulating events on the fly will free up storage space from otherwise pre-simulated and stored datasets.

Significant effort has been spent in properly understanding and modeling the complexity of electromagnetic and hadronic shower components spawned by a variety of physical processes in the EM and nuclear sectors. The details of such simulation portions are not always captured to satisfactory levels of agreement, and proposed improvements are continuously validated against experimental data.

The notion that shower development originating from different types of particles, or from particles with different energies, differs quantitatively and qualitatively in both the longitudinal and the lateral profile has suggested the development of particle identification algorithms based on the shower profile characteristics (see Sec. 5.2). Features in both integrated and differential shower profiles need to be properly modeled in simulation to maintain fidelity to the measurements performed by experiments.

While the distributional behavior of electromagnetic and hadronic showers is modeled sufficiently well by empirical laws, individual events exhibit large fluctuations from the average shower behavior, enriching the dataset with a diverse set of outcomes from similar initial conditions. This is due both to the inherent stochasticity of the physical processes involved in the shower development phase, as well as the asymmetry in detector geometry.

In current fast simulation techniques such as FASTCALOSIM, shower-by-shower stochastic fluctuations, cell occupancy levels, and radiations distributions are often parametrized in overly simplistic manners, requiring the user to specify rather unintuitive conditioning factors such as the numbers of 'packs of energy' to deposit per shower, as documented in Ref. [152]. In addition, the structure and relation of topologically connected energy deposits in hadronic showers, which represent the macroscopic units used for jet and particle reconstruction in ATLAS calorimeters, are known to be inadequately captured by fast simulation techniques, thus rendering full simulation the only employable method in the context of analyses and optimization studies that rely on precise jet substructure description [152]. For more details on simulation in ATLAS, see Sec. 3.1.

Although algorithmic advances and hardware accelerators continue to improve the computational footprint of scientific simulation programs, the sequential and causal nature of the simulation, as well as the complexity of the output space and the need for the number of simulated examples to far exceed the amount of corresponding collected data points, still warrant the exploration of novel, machine learning-based techniques to address the intrinsic limitations of classical methods. Automated, data-driven, simulation techniques proposed in this chapter vow to abandon a principled modeling approach in favor of a direct data imitation strategy.

The search for ideas and inspiration from the machine learning literature to address the imminent simulation crisis high energy physics experiments might face coincided with the parallel rise in popularity and success of deep generative models applied to other academic and industry-related audio and visual tasks. Speculation on their applicability to the physics domain burgeoned, along with the first practical attempts at bridging the two fields, thanks to the increasing resourcefulness, enterprise, and foresight of a portion of the high energy physics community, and to the ties established with machine learning collaborators. Among the proposed approaches were the use of variational autoencoders, autoregressive models, and generative adversarial networks.

After careful evaluation of pros and cons of the various architectural solutions and learning strategies, our group converged on the decision to investigate a GAN-based approach to fast calorimeter simulation. For fast prototyping, we began our exploration from the generation of jet images in a single-layer calorimeter, to then graduate to the higher-complexity task of generating electromagnetic showers in a heterogeneously segmented calorimeter with much more realistic hardware characteristics. While the former constitutes an entirely speculative first attempt at applying this technology to high energy physics simulation, the latter represents a more mature and conscious evolution of the architectures and experimental proposals developed in the earlier stage.

This chapter recounts my personal discovery journey towards opening the door to a new, exciting field of research in deep generative modeling for particle physics, and provides detailed accounts and insights into the architectural decisions and analysis interpretations that emerged from these experiments.

Specifically, material sourced from published and unpublished work has been drawn together to form two main sections that reflect the two major stages of this research campaign: Sec. 8.1 discusses the first phase of the project, focused on the generation of jet images, while Sec. 8.2 presents the more advanced application of this technology to the generation of showers in a multi-layer electromagnetic calorimeter.

## 8.1 LAGAN

The first exploratory work in the direction of arming the high energy physics community with deep generative models for simulation and sample augmentation focuses on the generation of single-channel (gray-scale) jet images, which have already proven popular in various classification applications in this field (see Sec. 5.2.1.2.4). Jet images are used as a proving ground to demonstrate the prospects of using generative adversarial networks to speed up high energy physics simulation, and to develop and test ad-hoc techniques in a controlled scenario.

This work represents the one of the first successful applications of GANs to the physical sciences. This contribution is a concrete first step in understanding both the potential upsides and limitations of these networks for computationally intensive physics applications, and lays the groundwork for understanding the viability of GANs and other generative modeling frameworks for reliable acceleration of detailed simulation of particle collision, production, and detection, that preserves physical properties across multiple energy scales.

This section is structured as follows: Sec. 8.1.1 describes the dataset used in this application, Sec. 8.1.2 outlines the characteristics of the architectures tested on this problem, and Sec. 8.1.3 provides insights into the results obtained by the models.

The content of this section is adapted from my previously published work in Ref. [31].

### 8.1.1 DATASET

This work on deep generative models for jet images is accompanied by the release of a public dataset for reproducibility purposes as well as future developments on a common benchmark problem [29].

The dataset consists of a collection of highly sparse, yet structured images of the energy patterns that originate from subatomic particles interacting with a single layer calorimeter (see Sec. 5.2.1.2.4 for an introduction to jet images and their preprocessing procedure). The nature of jet images varies significantly depending on the class of jets they represent and on the physical process from which the jets originate. Drawing inspiration from a high profile classification task, this dataset contains jet images originating from boosted $W$ bosons from $W' \rightarrow WZ$ (signal), and from generic quark and gluon jets (background). Figs. 8.1(a) and 8.1(b) show the average image from the signal and background classes. The difference between the two average images is more clearly displayed by Fig. 8.2, in which the average background image is subtracted from the average signal image, revealing a marked propensity for $W$ jets to possess a more evident two-prong structure, as well as more energetic and collimated leading sub-jet.

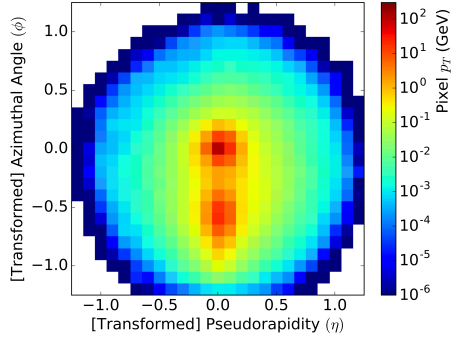 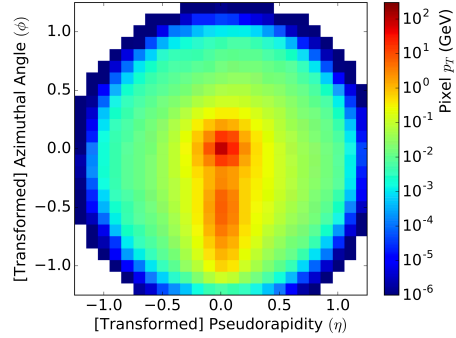

(a) Averaged signal images  (b) Averaged background images

**Figure 8.1:** These mean images are obtained by sampling 200,000 examples from the dataset, and averaging over $W$ boson jets (8.1(a)) and QCD jets (8.1(b)) independently. Signal images tend to be more concentrated and have a more pronounced two-prong structure.

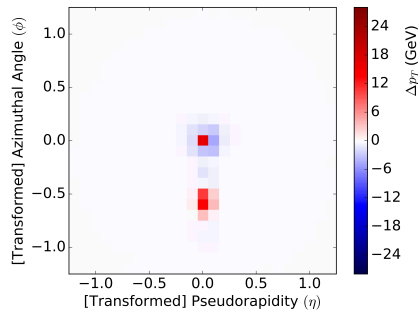

**Figure 8.2:** Per-pixel difference between the average signal (8.1(a)) and background (8.1(b)) image, which highlights the location-specific patterns that determine the nature of jet images. Red-colored pixels are more intensely activated in signal images, while blue-colored pixels are more intensely activated in background images.

Both signal and background are simulated using PYTHIA 8.219 [140, 452] at $\sqrt{s} = 14$ TeV [*]. In order to try to factor out the impact of the jet transverse momentum, the jet images are produced in the range 250 GeV $< p_T^{\text{jet}} <$ 300 GeV. Jets are clustered and trimmed using the `FastJet 3.2.1` software package [334]. A tight mass window is applied to exclude jets outside of the [60, 100] GeV jet mass range. The `scikit-image 0.12.0` [454] implementation of cubic spline rotation is used in the preprocessing phase.

The dataset focuses on the calorimeter portion of a particle detector, collapses it into a cylindrical surface surrounding the interaction point, and measures the energy of the jets that interact with it. The structure of jet images consist of a regular $0.1 \times 0.1$ grid of pixels in $\eta$-$\phi$ space, analogous to a finite-granularity calorimeter without segmentation along the depth dimension (*e.g.*, the CMS detector [92]). In reality, calorimeters such as the ATLAS one are multi-layered and have non-uniform pixel sizes with partially overlapping sections. This pioneering work only considers a one-layer calorimeter with uniform pixel size, while leaving the exploration of generative models for multi-layer detector showers to later applications (see Sec. 8.2).

The resulting image is a $25 \times 25$ gray-scale image with pixel width of $0.1$ along the $\eta$ and $\phi$ directions, and with the following characteristics: pixel intensity representing the transverse momentum of all particles incident to a given calorimeter cell; localized features; most energetic cluster (*leading sub-jet*) located at the center of the image; second most energetic cluster (*sub-leading sub-jet*) placed below it along the vertical axis; low-intensity energy deposits around the central region; high degree of sparsity ($\sim 10\%$ active pixels); pixel activations spanning multiple orders of magnitude. The principal unique feature of jet images is location dependence: if a particular sub-object is shifted or perturbed, properties of the image, including any posterior on class, can wildly shift, as most class semi-separable manifolds over the space admit a high Lipschitz constant [185, 186, 187]. Therefore, this dataset demands a generative modeling solution that preserves location-dependent physical properties.

---

[*]Generation and analysis code available at https://github.com/hep-lbdl/adversarial-jets [453]. Docker image available on Docker Hub under lukedeo/ji:latest.

Jet observables are a series of powerful, physically meaningful, non-linear, low-dimensional features of the jet that can be computed from the jet image pixel space in order to capture properties of the jet it represents and later validate a generative model's ability to reproduce the data distribution. These are theory-motivated functions $f : \mathbb{R}^{25 \times 25} \to \mathbb{R}$ whose features can easily be interpreted by domain experts. Three such one-dimensional features of a jet image $I$ are the mass $m$, transverse momentum $p_T$, and $n$-subjettiness $\tau_{21}$, which are defined as follows for an image of $N$ pixels:

$$p_T^2(I) = \left(\sum_{i=1}^{N} I_i \cos(\phi_i)\right)^2 + \left(\sum_{i=1}^{N} I_i \sin(\phi_i)\right)^2 \tag{8.1}$$

$$m^2(I) = \left(\sum_{i=1}^{N} I_i\right)^2 - p_T^2(I) - \left(\sum_{i=1}^{N} I_i \sinh(\eta_i)\right)^2 \tag{8.2}$$

$$\tau_{21}(I) = \frac{\tau_2(I)}{\tau_1(I)}, \tag{8.3}$$

where:

$$\tau_n(I) \propto \sum_{i=1}^{N} I_i \min_{1 \leq a \leq n} \left\{\sqrt{(\eta_i - \eta_a)^2 + (\phi_i - \phi_a)^2}\right\}$$

and the index $a$ runs over the number of sub-components within the jet, and $I_i$, $\eta_i$, and $\phi_i$ are the pixel intensity, pseudo-rapidity, and azimuthal angle, respectively. The quantities $\eta_a$ and $\phi_a$ are axis values determined with the one-pass $k_t$ axis selection using the winner-take-all combination scheme [455]. Intuitively, $\tau_n$ is a physically meaningful mapping that quantifies how likely it is for a given jet image to have $n$ subcluster-like structures. The distributions of $m(I)$, $p_T(I)$, and $\tau_{21}(I)$ are shown in Fig. 8.3. These quantities are highly non-linear one-dimensional manifolds of the 625-dimensional space in which jet images live.

The released dataset is composed of a total of 872,666 jet images made available in HDF5 format [392]. The archive contains the following branches:

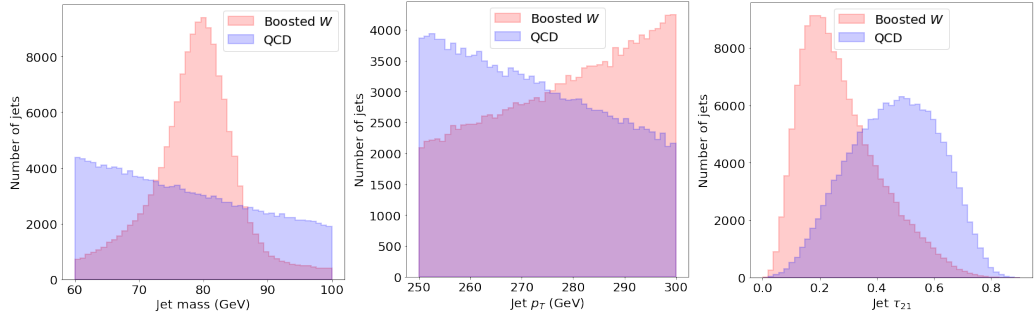

**Figure 8.3:** The distributions of image mass $m(I)$, transverse momentum $p_T(I)$, and $n$-subjettiness $\tau_{21}(I)$.

- `image`: array of dimension (872666, 25, 25), contains the pixel intensities of each 25×25 image
- `signal`: binary array to identify signal (1, *i.e.* $W$ boson) from background (0, *i.e.* QCD)
- `jet_eta`: $\eta$ coordinate of the jet
- `jet_phi`: $\phi$ coordinate of the jet
- `jet_mass`: mass of the jet
- `jet_pt`: transverse momentum of the jet
- `jet_delta_R`: distance between leading and sub-leading sub-jets if 2 sub-jets are present, otherwise set to zero
- `tau_1`, `tau_2`, `tau_3`: $n$-subjettiness substructure variables of the jet
- `tau_21`: `tau_2`/`tau_1` for each jet
- `tau_32`: `tau_3`/`tau_2` for each jet

Fig. 8.4 plots the pairwise relationships among the jet variables stored in the dataset and helps visualize this $d$-dimensional space along its various 2D projections. The plots along the diagonal show the kernel density estimation of the distribution of each jet observable, and clearly identify the class separability provided by features such as jet mass, $\Delta R$ and $\tau_{21}$. The off-diagonal scatter plots highlight, among others, the mass dependence of certain jet observables. Visualizing the high-dimensional distribution of jets in this dataset along these 2D planes also shows why machine learning classifiers operating on these jet observables would outperform iterative rectangular cuts.

Jet substructure variables are shown to be some of the most discriminative single variables for $W$ jet identification versus QCD by computing the absolute value of their linear correlation coefficients with the class label, as shown in Fig. 8.5.

### 8.1.2 Model

Encapsulating the data distribution of jet images in a generative model represents a fundamental challenge for most vanilla GAN architectures, due to the extreme levels of sparsity and unbounded nature of pixel activations [356]. In order to overcome this challenge, this work proposes a series of architectural modifications to the DCGAN [21] framework to take advantage of the pixel-level symmetry properties of jet images, while explicitly inducing location-based feature detection.

An auxiliary task is built into the system, following the ACGAN [292] formulation. In addition to the primary task where the discriminator network must learn to identify fake jet images from real ones, the discriminator is also tasked with jointly learning to classify boosted $W$ bosons (signal) and QCD (background). Jets stemming from both sources are generated using the same conditional model in an attempt to learn the class-conditional distribution.

The proposed Location-Aware Generative Adversarial Network (LAGAN) follows a set of experimentally-motivated architectural guidelines for building GANs for sparse, location-dependent data distributions. The characteristics of a LAGAN can be summarized as follows:

- **Locally-Connected Layers** - or any attentional component to attend to or learn location-specific features

- **Rectified Linear Units** in the last layer of $G$ to induce sparsity

- **Batch normalization** [202], as also recommended in [21], to help with weight initialization and gradient stability

- **Minibatch discrimination** [294], which is experimentally found to be the lynch-pin in modeling both the high dynamic range and levels of sparsity of this dataset.

A detailed diagram of the architecture is available in Fig. 8.6. The latent space is encoded into a low-dimensional vectors $z \in \mathbb{R}^{200}$, where, $z \sim \mathcal{N}(0, I)$, while the final generated outputs belong to $\mathbb{R}_{\geq 0}^{25\times 25}$. A Hadamard (element-wise matrix) product [456] is performed between $z$ and a trainable lookup-table embedding of the desired class (boosted $W$, or QCD), effectively conditioning the generation procedure [292].

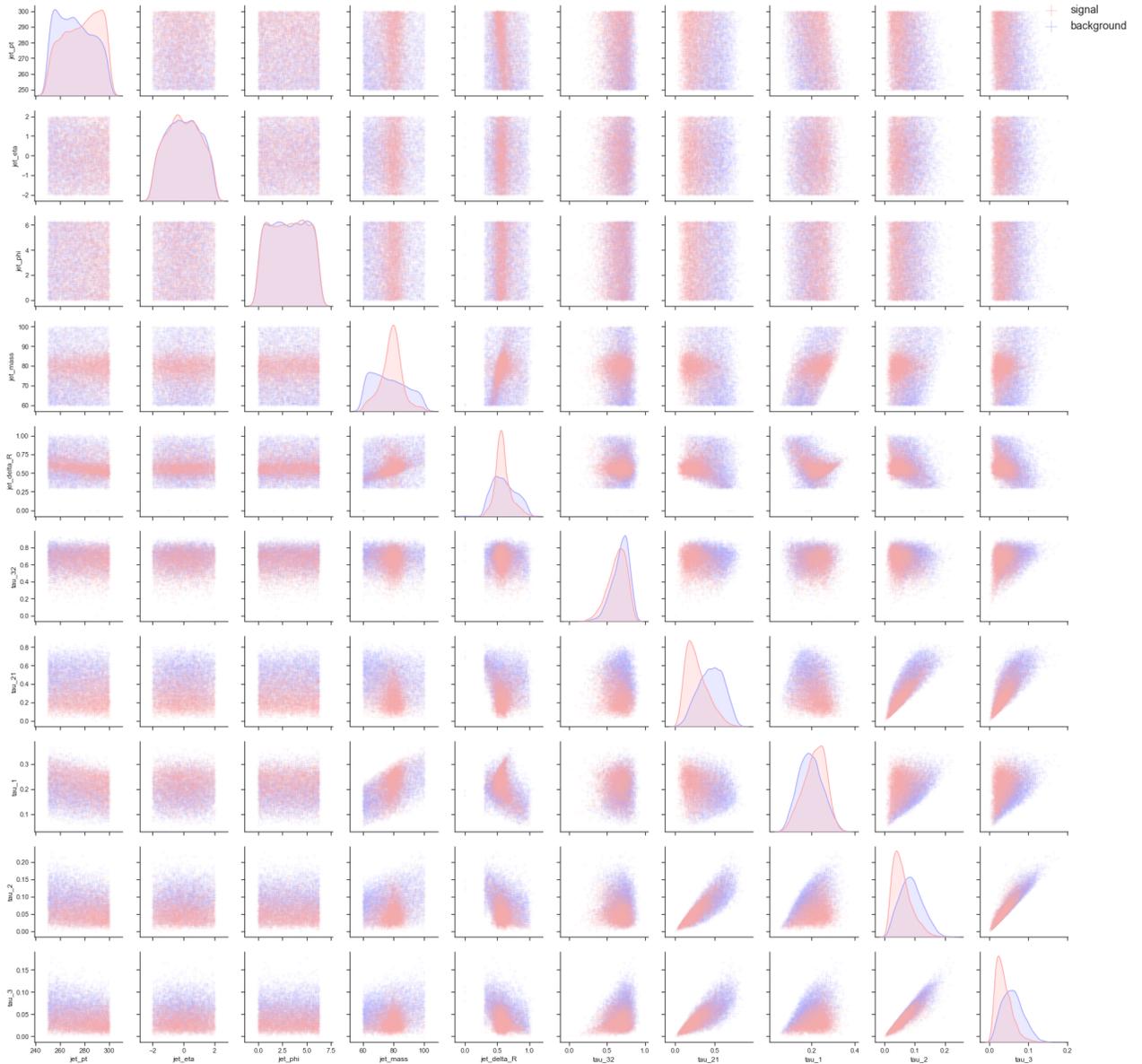

**Figure 8.4:** Univariate and bivariate analysis of the distribution of jets in the dataset. $W$ signal jets are shown in red, QCD background jets in blue. The plots along the diagonal show the distributions obtained from fitting KDE to the empirical distributions in the dataset.

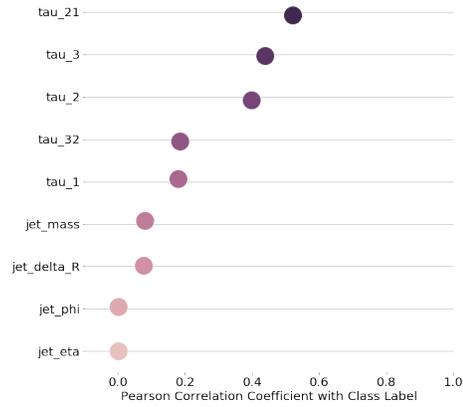

**Figure 8.5:** Absolute value of the Pearson correlation coefficient between each jet observable and the class label.

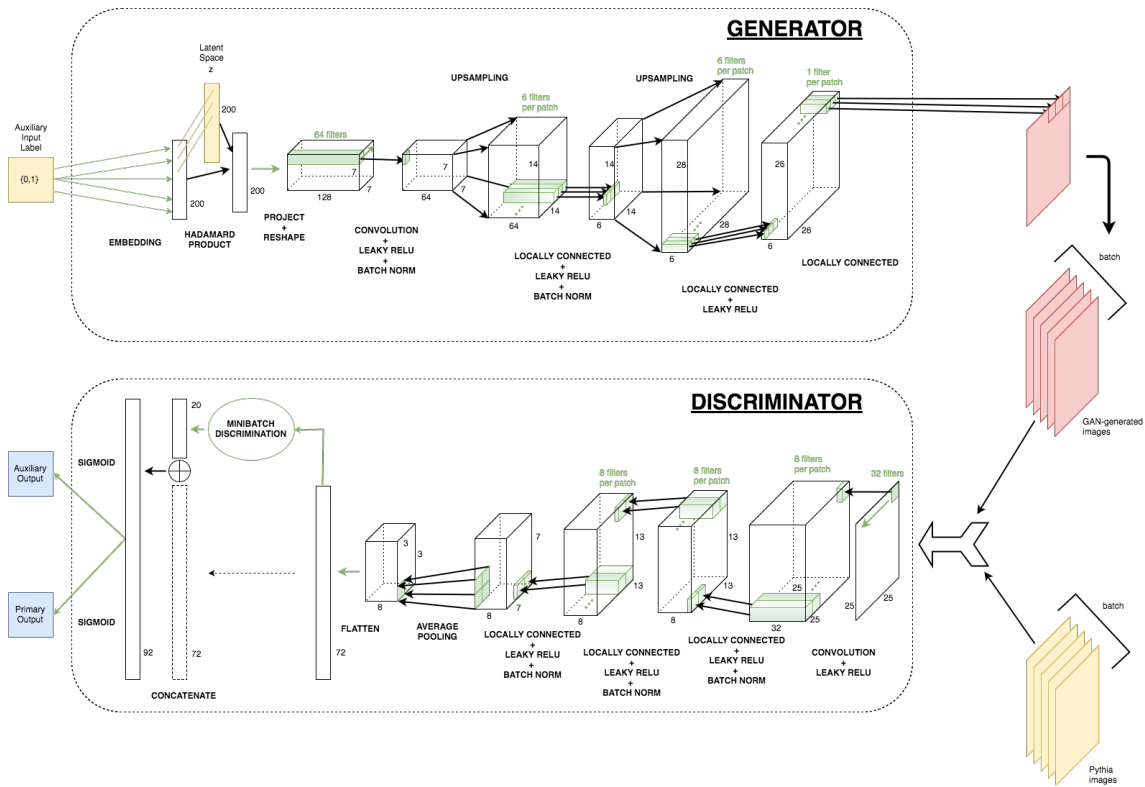

**Figure 8.6:** A schematic representation of the LAGAN architecture with the specific network structure and dimensionality choices used to learn a generative model from the dataset described in Sec. 8.1.1.

The generator $G$ consists of a `same`-bordered 2D convolution, followed by 2 `valid`-bordered 2D locally-connected layers, with 64, 6, and 6 feature maps respectively. Though locally-connected layers are rarely seen as useful components of most computer vision systems applied to natural images due to their lack of translation invariance and parameter efficiency, the location specificity of locally-connected layers allows them to outperform their convolutional counterparts on tasks related to jet images, as previously shown in Ref. [30, 361]. Visualizations of the nature of convolutional and locally-connected layers are available in Figures 8.7 and 8.8. The convolutional layer has receptive fields of size 5×5, while the ones in the two locally-connected layers have dimension 5×5 and 3×3 respectively. Sandwiched between the layers are 2×-upsampling operations and channel-wise batch normalization layers that help reduce internal covariate shift and stabilize gradients. The standard procedure of adding batch normalization is observed to be extremely important in modeling the large dynamic range encountered in this application. A ReLU-activated locally-connected layer with one feature map, a 2×2 receptive field, and no bias term, is placed on top of the last layer of $G$. The ReLU, though not commonplace in most GAN architectures due to issues with sparse gradients [286], is used to induce the desired level of sparsity in generated images. However, to remain consistent with Ref. [21, 286], Leaky Rectified Linear Units [230] are used throughout both generator and discriminator.

The discriminator $D$ consists of a `same`-bordered 2D convolutional layer with 32 $5\times 5$, unique filters, followed by 3 `valid`-bordered 2D locally-connected layers all with 8 feature maps with receptive fields of size $5\times 5$, $5\times 5$, and $3\times 3$, After each locally-connected layer, channel-wise batch normalization is applied. The last feature layer is input to a minibatch discrimination operation with twenty 10-dimensional kernels. This operation is crucial to obtain both stability and sample diversity in the face of an extremely sparse data distribution. For scientific applications, complete exploration of the data support is essential for any GAN-based system to be useful. Inasmuch as it directly encourages this behavior through batch-level statistics, minibatch discrimination has proven useful to aid in this direction. The batch-level features are then concatenated with the hidden feature layer, before being mapped using sigmoids to the outputs of both primary and auxiliary tasks.

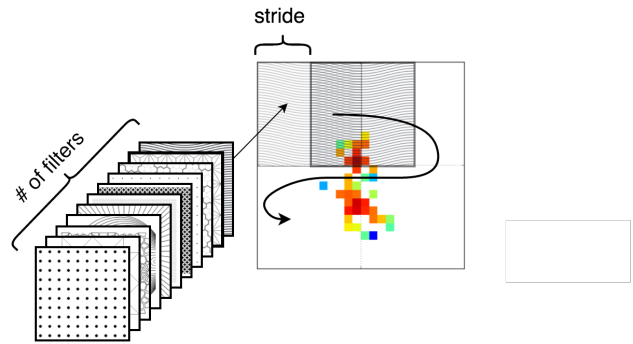

**Figure 8.7:** In the simplest (*i.e.*, all-square) case, a convolutional layer consists of $N$ filters of size $F \times F$ sliding across an $L \times L$ image with stride $S$. For a `valid` convolution, the dimensions of the output volume will be $W \times W \times N$, where $W = (L - F)/S + 1$.

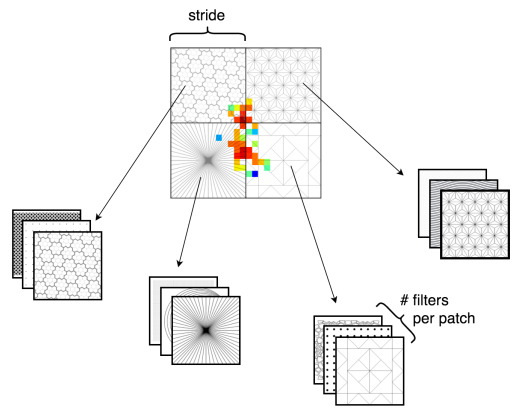

**Figure 8.8:** A locally connected layer consists of $N$ unique filters applied to *each* individual patch of the image. Each group of $N$ filters is specifically learned for one patch, and no filter is slid across the entire image. The diagram shows the edge case in which the stride $S$ is equal to the filter size $F$, but in general patches would partially overlap. A convolution, as described above, is simply a locally connected layer with a weight sharing constraint.

#### 8.1.2.1 Benchmark Models

The LAGAN solution is compared to other GAN architectures. For fairness of comparison, all models use a ReLU output for sparsity, and minibatch discrimination in the discriminator.

##### 8.1.2.1.1 Vanilla DCGAN
The proposed LAGAN design is compared to a traditional deep convolutional architecture (DCGAN) [21] with batch normalization in the generator and discriminator, and transposed convolutions in the generator. Two transposed convolutions with a stride of 2 and filter size of $5 \times 5$ are followed by a convolutional layer to ensure consistent image size. The discriminator contains four 2D convolutional blocks with filter size $5 \times 5$.

##### 8.1.2.1.2 Fully-Connected GAN
Given the importance of centralized, high energy pixel clusters (as opposed to the spread out, low intensity energy patterns) for the reconstruction of the physical properties of jet images, a GAN composed of fully-connected layers (FCGAN) is benchmarked on this dataset. The architecture resembles that of the previous models, with fully-connected layers replacing the convolutional (or locally-connected) components.

##### 8.1.2.1.3 Double-Stream Hybrid GAN
Given the complementarity of information learned by the convolutional and fully-connected GANs, a hybrid model (HYBRIDGAN) is built by merging the two architectures described above. Both generator and discriminator consist of a convolutional and fully-connected stream that are jointly learned and whose outputs are merged to produce the final output. This multi-headed model essentially employs a localization network and a dispersion network, where each stream learns a portion of the data distribution that reflects its comparative advantage.

#### 8.1.2.2 Training Procedure

All proposed models are trained end-to-end by taking alternating steps in the gradient direction for both $G$ and $D$. The dynamics of this procedure, coupled with the parametrization of GANs in terms of two non-convex players, generally causes the training to be quite unstable, without guarantees of convergence [457]. To combat this, a variety of ad-hoc procedures for controlling convergence have emerged in the field, in particular, relating to the generation of natural images. Architectural constraints and opti-mizer configurations introduced in Ref. [21] provide a well studied set of defaults and starting points for training GANs.

Many empirical improvements also help avoid mode collapse, a very common point of failure for GANs where a generator learns to generate a single element of the data distribution that is maximally confusing for the discriminator. For instance, minibatch discrimination and feature matching allow the discriminator to use batch-level features and statistics that effectively render mode collapse suboptimal. It has also been empirically shown in Ref. [291, 292, 293, 294] that adding auxiliary tasks can reduce the tendency towards mode collapse and improve convergence stability; side information is therefore often used as either additional information to the discriminator and/or generator [294, 290, 291], as a quantity to be reconstructed by the discriminator [458, 293], or both [292].

The training procedure employs the ADAM optimizer [177], utilizing the sensible parameters outlined in Ref. [21], with a batch size of 100 and for 40 epochs. All models are constructed using `Keras` [275] and `TensorFlow` [274]. Two NVIDIA Titan X Pascal Architecture GPUs are used for training.

At training time, label flipping alleviates the tendency of the GAN to produce very signal-like and very background-like samples, and attempts to prevent saturation at the label extremes. Labels are flipped according to the following scheme: when training $D$, 5% of the target labels for the primary classification output are flipped, as well as 5% of the target labels for the auxiliary classification output on batches composed solely fake images, essentially tricking $D$ into misclassifying fake versus real images in the first case, and signal versus background GAN-generated images in the second case; in addition, while training the generator, 9% of the time $G$ is required to produce signal images that the discriminator would classify as background, and vice versa. These tricks are an attempt to extend the support of the generated distribution.

This exploratory work avoids performing extensive hyper-parameter optimization or latent size analysis, as any such study is not expected to generalize to future GAN implementations for specific applications. Hyper-parameters are chosen based on settings recommended in the literature [21], and the results are found to be insensitive to small perturbations in their values.

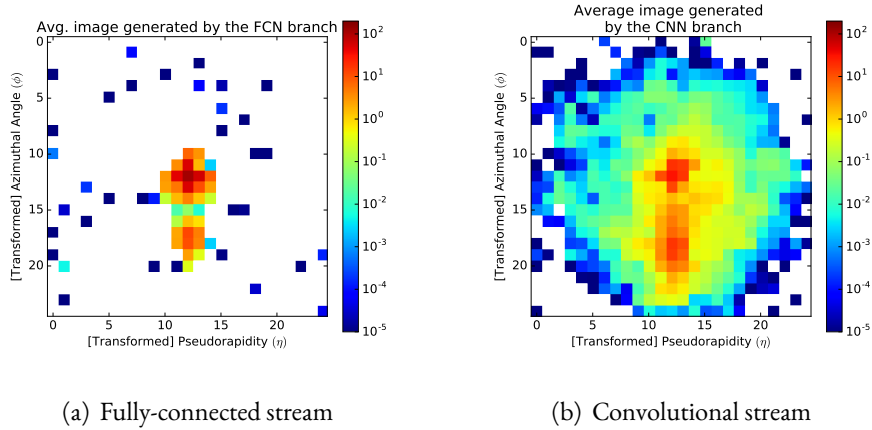

(a) Fully-connected stream    (b) Convolutional stream

**Figure 8.9:** Average image generated by the fully-connected and convolutional streams in the double-stream hybrid GAN (see Sec. 8.1.2.1.3). The fully-connected branch succeeds at reproducing the high-energy deposits in the core of the two sub-jets, but fails at capturing any of the information contained in the low-energy radiation disk around the center. The opposite can be said about the convolutional stream. Similar results are obtained when training the two branches separately to obtain the vanilla DCGAN and fully connected GAN described in Sec. 8.1.2.1.1 and 8.1.2.1.2 respectively.

### 8.1.3 Results

In preliminary investigations, experimental evidence shows the efficacy of fully-connected networks in producing the central constituents of jet images, which is reminiscent of the findings of Ref. [30] on the discriminative power of fully-connected networks applied to jet images. On the other hand, fully-convolutional systems excel at capturing the less location-specific low-energy radiation dispersion pattern. This remains true when the two types of architectures are connected in the double-stream hybrid GAN described in Sec. 8.1.2.1.3. The individual areas of competence of fully-connected and convolutional networks, already observed in individual trainings, are preserved when the two are connected as streams of a combined generator and are jointly trained. A desirable – yet not necessarily expected – feature of this compositional structure is that two streams learn to specialize and take on complementary roles at generation time. In fact, in the double-stream hybrid generator, the fully-connected and convolutional branches cooperate to produce high-fidelity radiation patterns with pixel intensities that span multiple orders of magnitude. This behavior is clearly visible in Fig. 8.9, which shows the average image produced by the fully-connected and convolutional streams of the hybrid GAN, before the two are combined.

The complementarity between the two types of architectures is also evident in the exploration of the type of information they capture about the physical traits that separate boosted $W$ from QCD images. The average signal and background images generated by the fully-connected portion of the hybrid GAN are displayed, respectively, in Fig. 8.10(a) and 8.10(c), with the difference between the two plotted on linear scale in Fig. 8.10(b). This stream produces signal images that are much more energetic than background images in the pixels that represent the core of the two sub-jets, while the background images it generates are characterized by more energetic depositions in a slightly wider area around the leading sub-jet. The fully-connected network recognized the importance of a more pronounced, dispersed radiation pattern in QCD jets, yet is unable to generate a realistic version of it. Simultaneously, the noticeable second prong that identifies the sub-leading sub-jet in boosted $W$ jets is reduced to almost inappreciable size for QCD jets. The difference image, once again, confirms the fully-connected branch's specialization and concentration on the highly-energetic portion of the image.

On the other hand, the convolutional stream is able to discover more nuanced dissimilarities between signal and background jets, which focus on the lower-energy dispersion pattern around the two leading sub-jets. This stream produces a strong background-like energy deposit in the area between the two sub-jets, as well as a more widespread and pronounced background contribution that is especially visible on the right and lower side of the image. Furthermore, it intuitively learns hidden features that map to the $\Delta R$ jet observable, which augment the signal-to-background differentiation by stressing the importance of the distance between the two principal components. The average boosted $W$ and QCD images generated by the convolutional portion of the hybrid GAN are displayed, respectively, in Fig. 8.11(a) and 8.11(c), with the difference between the two plotted on linear scale in Fig. 8.11(b).

Upon visual inspection, the contribution of the two streams to the productions of both signal and background images confirms that, while the fully-connected network is mainly responsible for learning locally-confined, highly-energetic, spiky energy depositions in the core locations of the two sub-jets, the convolutional component provides further discrimination power by filling in the pixels between the two sub-jets and concentrating on the larger, smeared, low-energy radiation pattern in the outer areas of the image.

These observations inform the choice of 2D locally-connected layers in the LAGAN architecture, in order to obtain a parameter-efficient model while retaining location-specific information. The hybrid architecture can, however, be preferable in applications in which the user wants to retain more direct

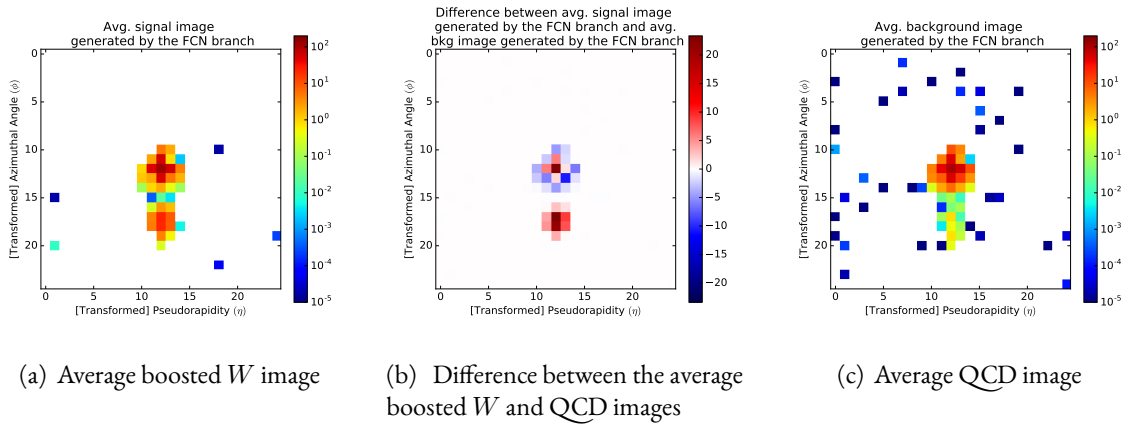

(a) Average boosted $W$ image (b) Difference between the average boosted $W$ and QCD images (c) Average QCD image

**Figure 8.10:** Average signal and background images produced by the fully-connected component of the hybrid GAN architecture.

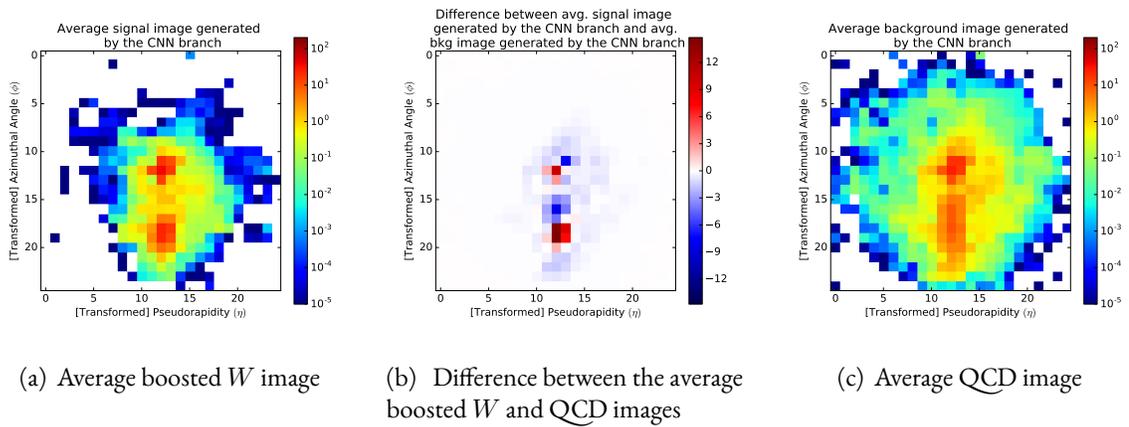

(a) Average boosted $W$ image (b) Difference between the average boosted $W$ and QCD images (c) Average QCD image

**Figure 8.11:** Average signal and background images produced by the convolutional component of the hybrid GAN architecture.

control over the relative contribution of the two streams.

The proposed candidate architectures are validated through quantitative and qualitative means on the task of generating realistic jet images. The use of physics-driven performance evaluation metrics is preferred to alternative application-agnostic evaluation methods, in order to obtain an objective quantification of each model's performance while keeping into consideration domain-specific requirements and constraints. To quantify the reconstruction of physically meaningful observables under the data and learned distributions, and to standardize generative model performance comparisons, a simple universal scoring function is introduced.

Given the true data distribution of jet images $\mathcal{D}$, and a generated distribution $\mathcal{S}$ from a particular generative model, $\mathcal{M}_\mathcal{D}(x)$ and $\mathcal{M}_\mathcal{S}(x)$ are defined as the distributions of any number of physically meaningful observables $x$ under the data and generated distributions respectively. Ideally, both conditional distributions should be well-modeled, or, more formally, for a sensibly chosen distance metric $d(\cdot, \cdot)$,

$$d\left(\mathcal{M}_\mathcal{S}(x|c), \mathcal{M}_\mathcal{D}(x|c)\right) \to 0, \ \forall c \in \mathcal{C}, \tag{8.4}$$

where $\mathcal{C}$ is the set of classes. To ensure good convergence for all classes, the performance scoring function is designed to minimize the maximum distance between the real and generated class-conditional distributions. Formally, such a minimax scoring function is defined as:

$$\sigma(\mathcal{S}, \mathcal{D}) = \max_{c \in \mathcal{C}} d(\mathcal{M}_\mathcal{D}(x|c), \mathcal{M}_\mathcal{S}(x|c)). \tag{8.5}$$

Eq. 8.5 is minimized when the generative model exactly recovers the data distribution. For this application, the similarity metric $d$ is chosen to be the Earth Mover's Distance [459].

From a quantitative standpoint, data exploration studies and prior work on the classification of jet images from boosted $W$ bosons and QCD jets [30] indicate that the combination of jet mass and $\tau_{21}$ is expected to be among the most discriminating high-level variable combinations out of the infinite space of valid manifolds. Therefore, these two features are chosen to form the reduced space used for performance evaluation using the scoring function in Eq. 8.5. The exclusion of $p_T$ is also due to its unphysical shape caused by the application of the tight window cut at sample generation time. The empirical 2D probability mass function (PMF) over $(m, \tau_{21})$ is discretized into a 40×40 equi-spaced grid contain-ing the full support, and the distance between any two points in the PMF is defined to be the euclidean distance of the respective coordinate indices. This distance is used in the internal flow optimization procedure when calculating the Earth Mover's Distance. Given the jet image dataset described in Sec. 8.1.1, two classes are used in the minimax calculation of $\sigma(\cdot, \cdot)$: jets from $W$ bosons and jets from QCD.

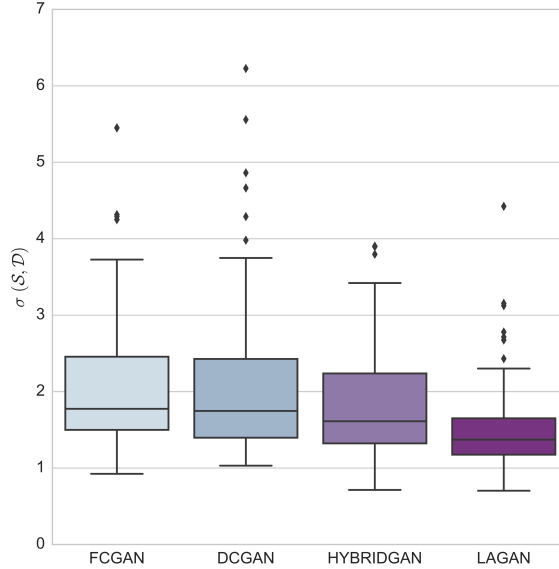

**Figure 8.12:** Boxplots quantifying the performance of the 4 benchmarked network architectures in terms of the scoring function $\sigma$ defined in Eq. 8.5. The boxes represent the interquartile range (IQR), and the line inside the box represents the media. The whiskers extend to include all points that fall withing 1.5×IQR of the upper and lower quartiles.

The LAGAN and benchmark models described in Sec. 8.1.2 are each evaluated over 150 different trainings by generating 12,000 jet images from each iteration. The corresponding jet $m$ and $\tau_{21}$ distributions are calculated using Eqs. 8.2 and 8.3, and combined into the scoring function in Eq. 8.5. The evaluation results are summarized in Fig. 8.12, which shows that the LAGAN architecture achieves the lowest median distance score, as well as the lowest spread between the first and third quartiles, confirming it as both a performant and robust architecture for this task. The jet $m$ and $\tau_{21}$ distributions used to compute the scoring function of a trained LAGAN are provided for reference in Fig. 8.13, along with the corresponding distributions calculated over 12,000 images from the reference dataset.

One of the unique advantages of developing GANs for high energy physics is the availability of these useful jet observables that can be used not only to compare competing architecture but also to assess their overall ability to mimic PYTHIA. Since the models are not directly informed of the relevance of matching these one-dimensional projections of the data distribution, there is no guarantee that these non-trivial mappings will be preserved under generation. However this property is desirable and easily verifiable. In fact, LAGAN-generated images reproduce many of the jet features present in Pythia images. For example, the signal jet mass distribution exhibits a peak at $\sim$ 80 GeV, which corresponds to the mass of the $W$ boson that generates the hadronic shower. This is a learned, emergent property of the network. Importantly, the generated GAN images are as diverse as the true Pythia images used for training – the fake images do not simply occupy a small subspace of credible images.

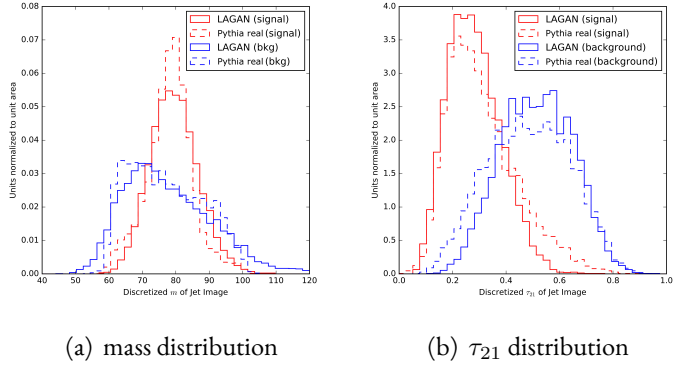

(a) mass distribution  (b) $\tau_{21}$ distribution

**Figure 8.13:** Class conditional distributions of two physically-meaningful 1D projections of the data distribution: mass (8.13(a)) and $\tau_{21}$ (8.13(b)). The dashed lines represent the true distributions from true reference images, while the solid lines are the distributions calculated on LAGAN-generated images.

After having established its superiority to alternative architectures considered in this work over the task of reproducing the Pythia-generated jet distribution, the final LAGAN candidate model is tasked with the production of 200,000 jet images, which are then compared to 200,000 Pythia images, in order to monitor the development of the training procedure (Sec. 8.1.3.1), evaluate - both quantitatively and qualitatively - the content of generated images (Sec. 8.1.3.2), explore the powerful information provided by some of the most representative images (Sec. 8.1.3.3), dig deeper into the inner workings of the generator (Sec. 8.1.3.4) and discriminator (Sec. 8.1.3.5), evaluate the generated images in the context of traditional jet classification tasks (Sec. 8.1.3.6), and briefly discuss computational efficiency of the proposed method (Sec. 8.1.3.7).

### 8.1.3.1 Training Dynamics

During training, a handle on the performance of $D$ is always directly available through the evaluation of its ability to correctly classify real versus GAN-generated images. The discriminator's performance on its primary task can be visualized in Fig. 8.14 by constructing histograms of the output of the discriminator evaluated on GAN and Pythia generated images. Perfect confusion would result in the discriminator's output being equal to $1/2$ for every image. It is worth nothing, however, that this visualization fails to provide quantifiable information on the performance achieved by $G$. In fact, a failure in the classification tasks can also be symptomatic of under-training in both $G$ and $D$ or mode collapse in $G$, and is observed in the first few epochs of training if $D$ is not pre-trained. Similarly, a slow divergence towards $D$ winning the game against $G$ cannot lead to the conclusion that the performance of $G$ is *not* improving; it might simply be doing so at a slower pace than $D$, and the type of representation shown in Fig. 8.14 does not help with the identification of when the peak performance of $G$ is obtained.

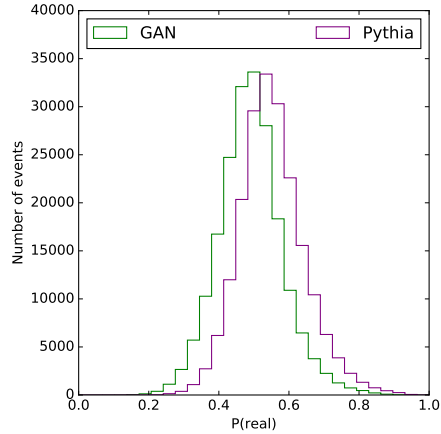

**Figure 8.14:** Histograms of the discriminator's adversarial output for generated and real images at the epoch chosen to generate the images used in this application (34[th] epoch).

Instead, in practice, convergence of generated mass and $p_T$ distributions to the corresponding real distributions from the reference dataset is used to select a stopping point for the training procedure (Fig. 8.3). This leads to the selection of the 34[th] epoch of training for the generation of the images shown in this chapter, but qualitatively similar results can be observed after only a handful of epochs. By the 34[th] epoch, the training seems to have almost reached the equilibrium point in which $D$ is unsure of how to label most samples, regardless of their true origin, without any strong dependence on the physical process that generated the image.

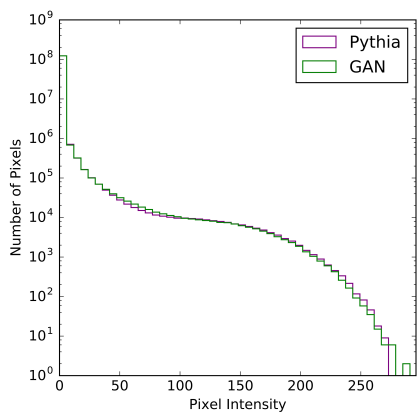

**Figure 8.15:** The distribution of the pixel intensities for both fake (GAN) and real (Pythia) jet images, aggregated over all pixels in the jet image.

#### 8.1.3.2 Image Content Quality Assessment

Quantifying the efficacy of the generator is challenging because of the dimensionality of the images and the lack of visual cues to facilitate the task. However, it is possible to visualize what aspects of the radiation pattern the network is most strongly associating with signal and background processes, and what regions of the image are harder to generate via adversarial training.

A first qualitative evaluation relies on the comparison of histograms showing the distribution of the pixel intensities, aggregated over all pixels in an image, for both generated and reference images. Intensities span a wide range of values, from the energy scale of the jet $p_T$ ($\mathcal{O}(100)$ GeV) down to the machine epsilon for double precision arithmetic. Because of inherent numerical degradation in the preprocessing steps [†], images acquire unphysically low pixel intensity values. The distribution of pixel intensities is therefore truncated at $10^{-3}$ GeV to discard all unphysical contributions. Fig. 8.15 shows this distribution for both Pythia and GAN-generated jet images. The full physical dynamic range is explored by the GAN, and the important high-$p_T$ region is faithfully reproduced.

Another useful method to understand the network's behavior is to directly investigate the images' appearance. This is inspired by the visualization techniques proposed in Ref. [30].

In accordance with the common procedure from the GAN literature, randomly selected images from the data distribution are compared to their nearest generated neighbor by euclidean distance, in order to show that the GAN has not simply memorized the training set (Fig 8.53). Although the euclidean distance usually fails to convey meaning as a similarity metric over natural images, its use in this application is warranted by the physical relevance of meaningful pixel-level shifts in determining the properties of a high energy jet. This analysis suggests that the network is generating credible, physical images across a variety of topologies without memorizing the training set. The dispersion patterns appear realistic, and, although the generative model seems inadequate in the very low intensity region, this is of little practical consequence because most of the important jet information is encoded in the higher $p_T$ radiation pattern. This check also confirms that the generative model is not just copying images from the training dataset, but is instead learning to sample from an approximation to the original data distribution.

---

[†]Bicubic spline interpolation in the rotation process causes a large number of pixels to be interpolated between their original value and zero, the most likely intensity value of neighboring cells. Though a zero-order interpolation would solve sparsity problems, we empirically determine that the loss in jet-observable resolution is not worth the sparsity preservation.

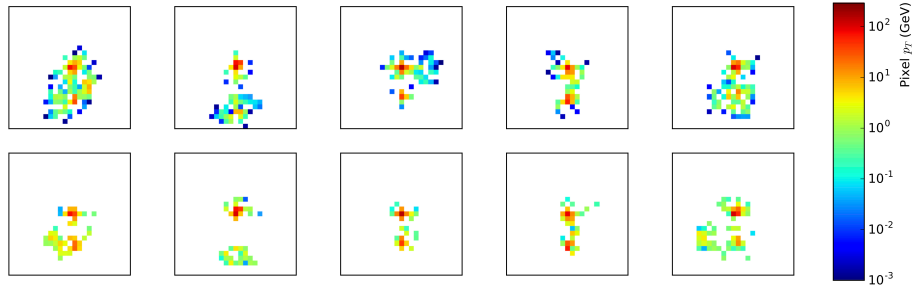

**Figure 8.16:** Randomly selected Pythia images (top row) and their nearest GAN-generated neighbor (bottom row).

The topology of jet images becomes more visible when examining averages over multiple images with specific characteristics. Figure 8.17 shows the average Pythia image, the average generated image, and their difference . On average, the generative model produces large, spread out, roughly circular dispersion patterns that approximate those of Pythia jet images. Even though the pixel regions with average activation $< 10^{-2}$ GeV are not as carefully reproduced, the generated images span multiple orders of magnitude in pixel intensity, and differ from Pythia images by only a few percent in the activation of the central regions at the core of the jet.

### 8.1.3.3 Most Representative Images

To better understand the distribution that emerges from the generative model, one can investigate the features displayed by the 500 most representative jet images of a particular kind, and average over them. This helps identifying striking and unique features that could point to pathological issues.

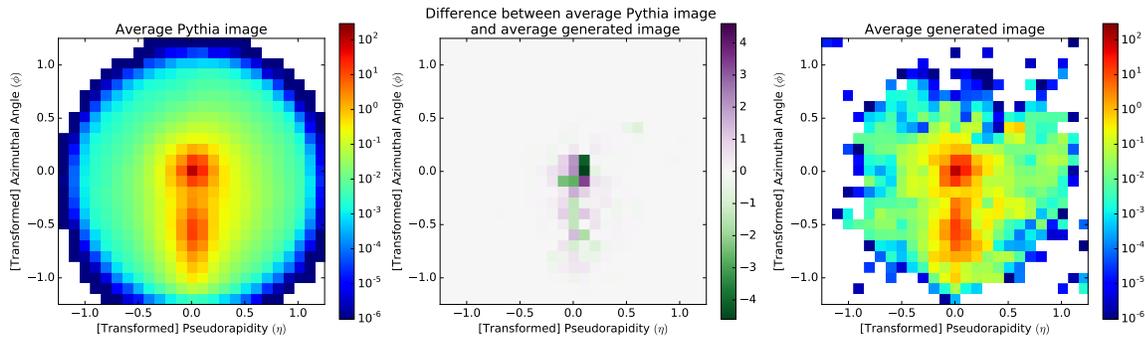

**Figure 8.17:** Average image produced by Pythia (left) and by the GAN (right), displayed on log scale, with the difference between the two (center), displayed on linear scale.

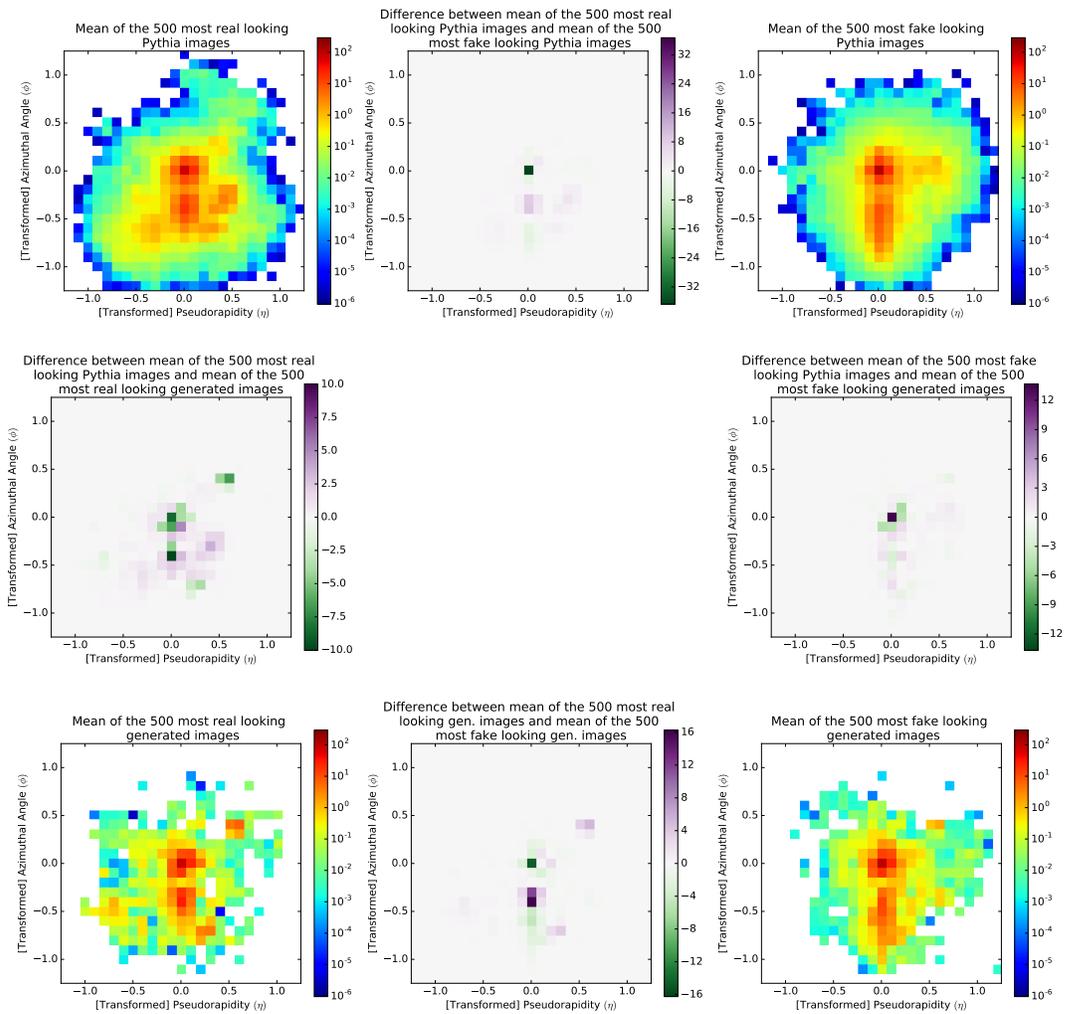

**Figure 8.18:** Comparison between the 500 most real and most fake looking images, generated by Pythia, on the left, and by the GAN, on the right.

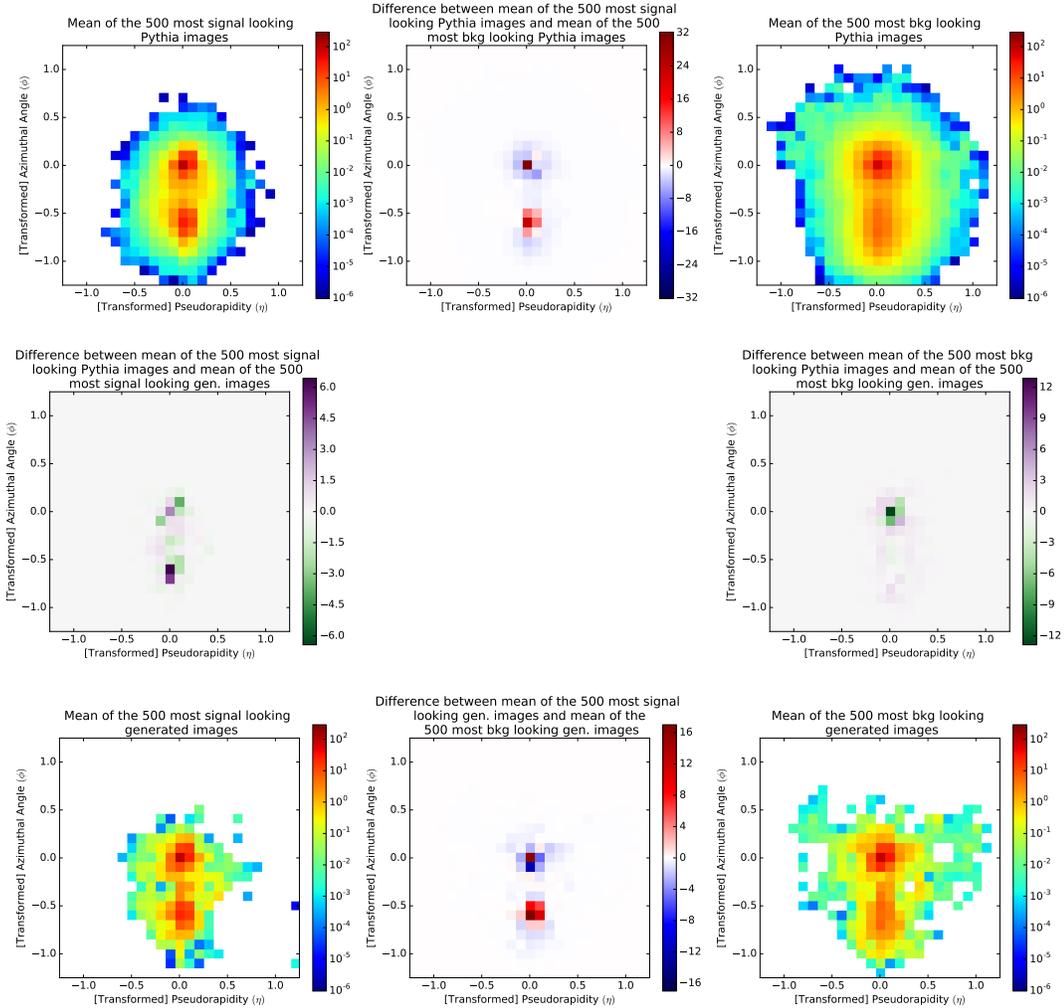

**Figure 8.19:** Comparison between the 500 most signal and most background looking images, generated by Pythia, on the left, and by the GAN, on the right.

The plots shown in Fig. 8.18 provide unique insight into what features the discriminator is learning and using in its adversarial task of distinguishing generated images from real images. Although Pythia images qualitatively differ from generated images, the metric that $D$ applies to discriminate between real and fake samples is consistent among the two. The images with the highest $P(\text{real})$ are largely asymmetric, with strong energy deposits in the bottom corners and a much closer location of the sub-leading sub-jet to the leading one. The net learns that more symmetric images, with a more uniform activation just to the right of the leading sub-jet, stronger intensity in the central pixel, and larger distance between the sub-jets, are more easily identifiable as originating from the generator, and therefore labeled as fake.

In addition, it is possible to isolate the primary information that $D$ pays attention to when classifying boosted $W$ and QCD images. The learned metric is consistently applied to both PYTHIA and GAN images, as shown in Fig. 8.19. When identifying signal images, $D$ learns to looks for more concentrated images, with well-defined two prong structure. On the other hand, the network learns that background images have a wider radiation pattern and a more fuzzy structure around the location of the second subjet.

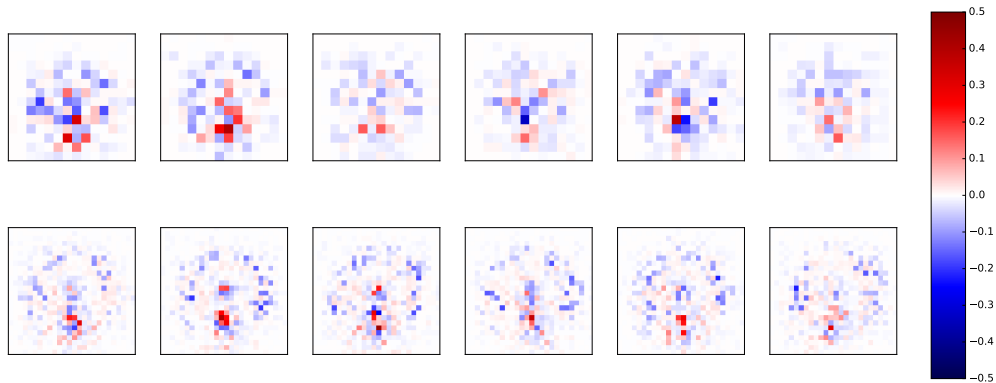

**Figure 8.20:** Pixel activations of images representing the various channels in the outputs of the two locally-connected hidden layers that form the generator, highlighting the difference in activation between the production of the average signal and average background samples. The first row represents the six $14 \times 14$ channels of the first locally-connected layer's output; the second row represents the six $26 \times 26$ channels of the second locally-connected layer's output.

#### 8.1.3.4 GENERATOR PERFORMANCE

Exploring the inner workings of the generator network allows to further understand the generation process of jet images using GANs. As outlined in Sec. 8.1.2, $G$ consists of a 2D convolution followed by 3 consecutive locally-connected layers, the last of which yields the final generated images. By peeking through the layers of the generator and following the path of a jet image in the making, it is possible to visually explore the steps that lead to the production of jet images with their desired physical characteristics.

The network is probed after each locally-connected hidden layer to investigate how the average signal and average background images develop into physically distinct classes as a function of depth in the generator network. The difference in activation between the average signal and background image for each channel in the first two locally-connected layers is independently plotted in Fig. 8.20. The output from the first locally-connected layer consists of six $14 \times 14$ images (*i.e.*, a $14 \times 14 \times 6$ volume), and the second consists of six 26 × 26 images. The red pixels are more strongly activated for signal images, while blue pixels activate more strongly in the presence of background images, which suggests that the generator is learning, from its very early stages that spread out energy depositions are needed to produce more typical background images, and that the presence of an energetic second sub-jet is emblematic of boosted $W$ jets.

### 8.1.3.5 Discriminator Performance

The discriminator network undertakes the task of hierarchical feature extraction and learning for classification. Its role as an adversary to the generator is augmented by an auxiliary classification task. With interpretability being paramount in high energy physics, various techniques are employed to try to visualize and understand the representations learned by the discriminator and their correlation with the known physical processes driving the production of jet images.

The convolutional layer at the beginning of the network provides the most significant visual aid to guide the interpretation process. In this layer, the 32 5×5 learned convolutional filters are convolved with patches of pixels by sliding them across the image. The weights of the 5×5 matrices that compose the convolutional kernels of the first hidden layer are displayed in the top panel of Fig. 8.21. Each filter is then convolved with two physically relevant images: in the middle panel, the convolution is performed with the difference between the average GAN-generated signal image and the average GAN-generated background image; in the lower panel, the convolution is performed with the difference between the average Pythia image drawn from the data distribution and the average GAN-generated image. By highlighting pixel regions of interest, these plots shed light on the highest level representation learned by the discriminator, and how this representation translates into the auxiliary and adversarial classification outputs.

Furthermore, one can compute linear correlations between the pixel intensities in the average images and each of the discriminator's outputs, and visualize the learned location-dependent discriminative information in Fig. 8.22. These plots accentuate the distribution of distinctive local features in jet images that the discriminator picks up on.

Further visual analysis of Pythia and GAN images aims at identifying what features the discriminator network $D$ uses to differentiate real and fake images. Fig. 8.23 displays the average radiation pattern in real images correctly labeled as real and fake images incorrectly labeled as real, as well as the difference between the two. Conversely, Fig. 8.24 shows the average pattern in real images incorrectly labeled as fake and fake images correctly labeled as fake, as well as their difference. In both cases, the most striking feature is that a higher intensity in the central pixel is associated with a higher probability of the image being classified as fake, while real-looking images tend to have a more diffuse radiation around the leading sub-jet, primarily concentrated in the top right area adjacent to the leading sub-jet itself.

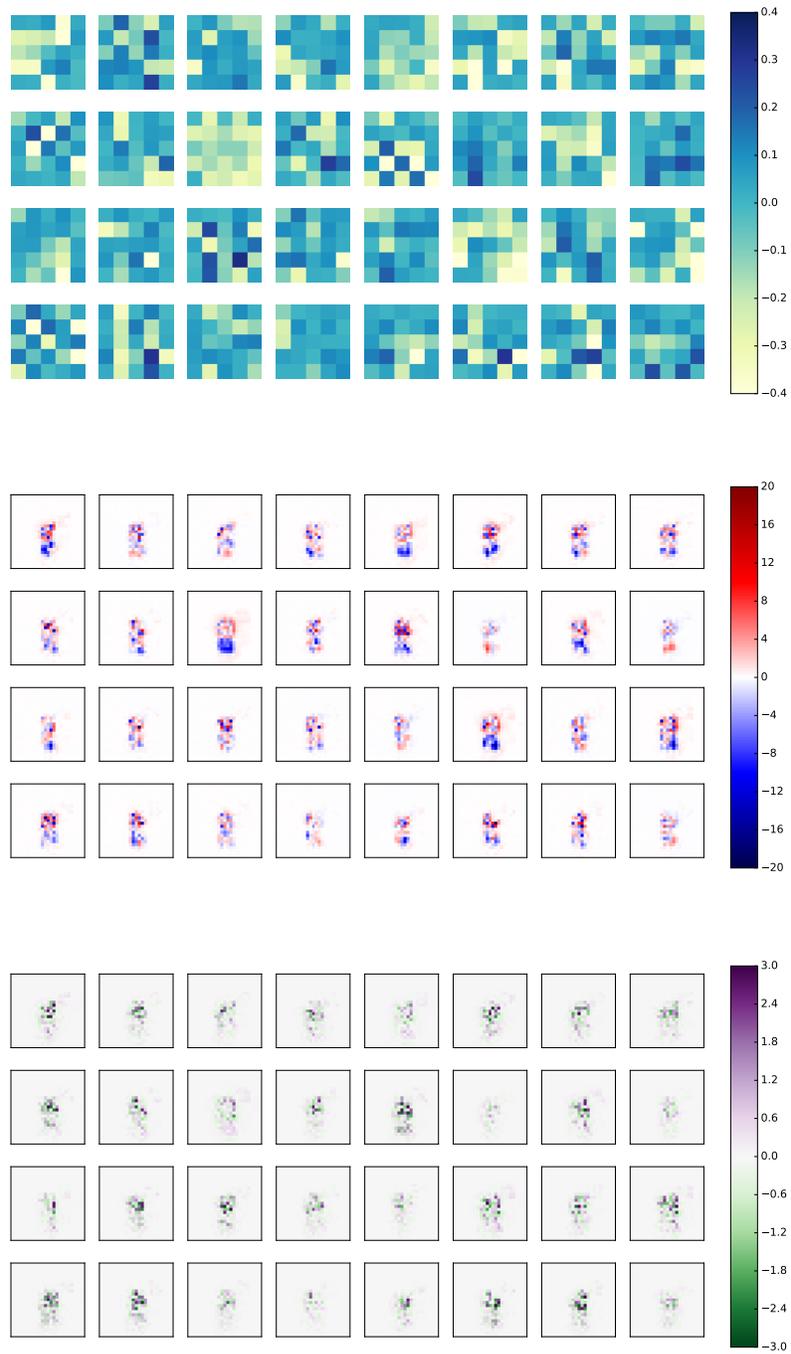

**Figure 8.21:** Convolutional filters in the first layer of the discriminator (top), their convolved version with the difference between the average signal and background generated image (center), and their convolved version with the difference between average Pythia and average generate image.

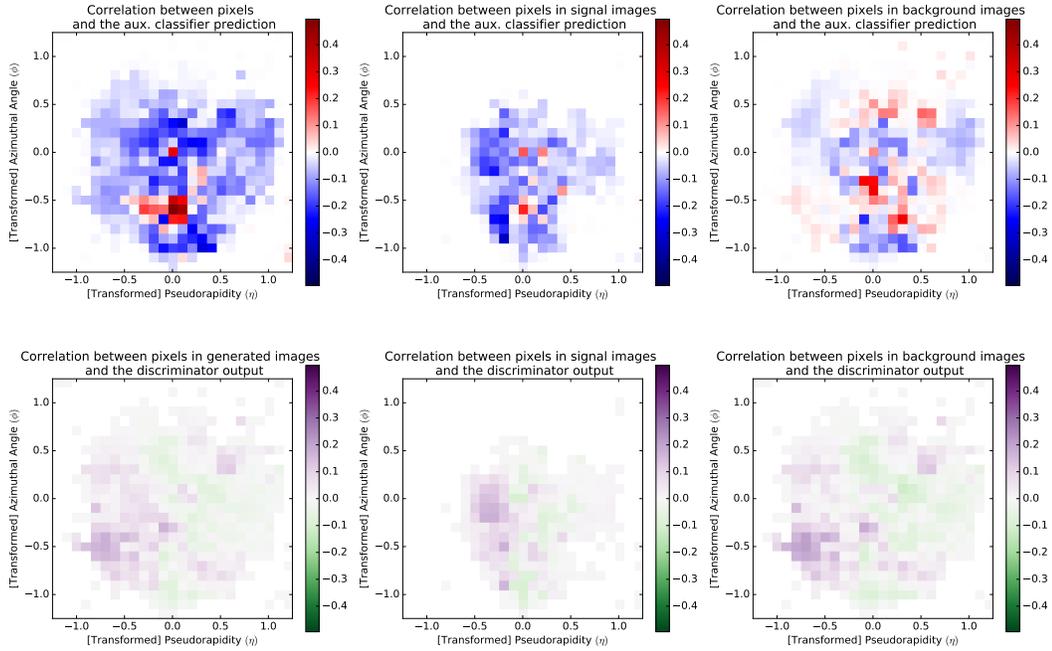

**Figure 8.22:** Per-pixel linear correlation with the discriminator auxiliary (top) and adversarial (bottom) classification output for combined signal and background images (left), signal only images (center), and background only images (right).

Another insight is that the LAGAN may not only be learning to produce samples with a diverse range of $m$, $p_T$ and $\tau_{21}$, but it may also be making use of them internally to aid the discriminator in its two classification tasks. Evidence for this hypothesis can be found by investigating the relationships between the discriminator's primary and auxiliary outputs, namely $P(\text{real})$ and $P(\text{signal})$, and the physical quantities that the generated images possess, such as mass and transverse momentum.

Fig. 8.25 shows the discriminator's ability to correctly identify the class that most generated images belong to, while exposing the dependence of the response on jet kinematic variables. The discriminator is making use of its internal representation of mass to identify signal-like images: the peak of the $m$ distribution for true signal events is located around 80 GeV, and, indeed, images with mass around that point have a higher $P(\text{signal})$ than the ones at very low or very high mass. Similarly, low $p_T$ images are more likely to be classified as background, while high $p_T$ ones have a higher probability of being categorized as signal images. This behavior is well understood from a physical standpoint and can be easily cross-checked with the $m$ and $p_T$ distribution for boosted $W$ and QCD jets displayed in Fig. 8.3. Although mass and transverse momentum influence the label assignment, $D$ is only partially relying on these quantities; there is more knowledge learned by the network that allows it, for example, to still manage to correctly classify the majority of signal and background images regardless of their $m$ and $p_T$ values.

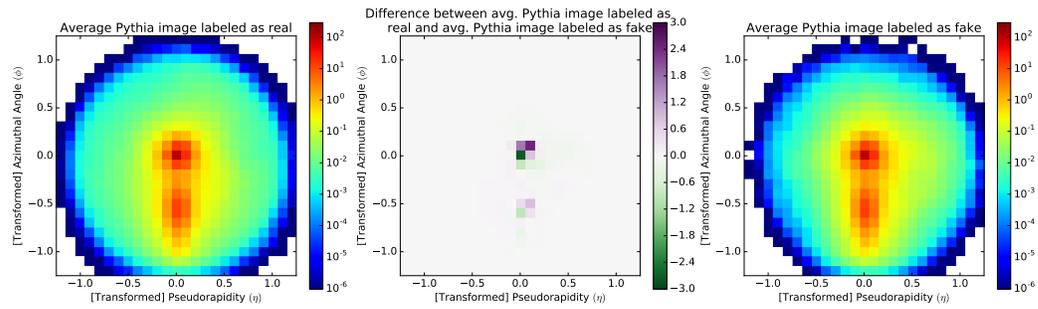

**Figure 8.23:** The average Pythia image labeled as real (left), as fake (right), and the difference between these two (middle) plotted on linear scale. All plots refer to aggregated signal and background jets.

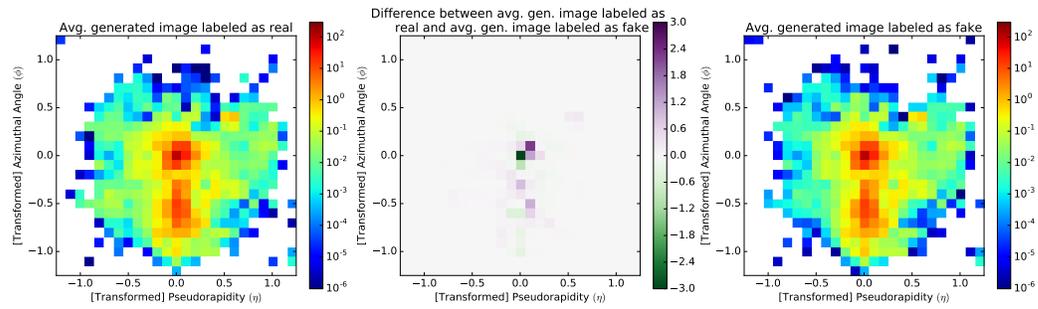

**Figure 8.24:** The average GAN-generated image labeled as real (left), as fake (right), and the difference between these two (middle) plotted on linear scale. All plots refer to aggregated signal and background jets.

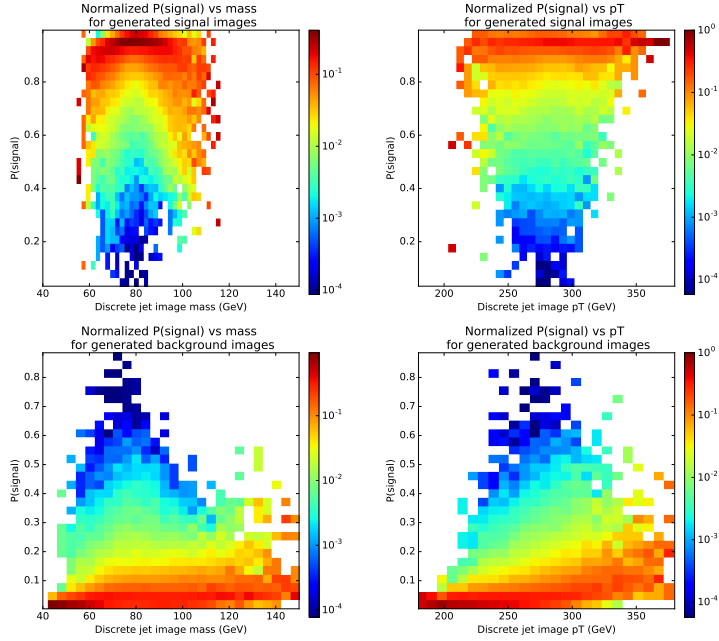

**Figure 8.25:** Auxiliary discriminator output as a function of mass, on the left, and as a function of transverse momentum, on the right. The top plots refer to signal (boosted $W$ bosons from $W'$ decay) images, while the bottom plots refer to background (QCD) images. All plots have been normalized to unity in each $m$ or $p_T$ bin, such that the $z$-axis represents the percentage of images generated in a specific $m$ or $p_T$ bin that have been assigned the value of $P(\text{signal})$ indicated on the $y$-axis.

On the other hand, Fig. 8.26 shows that the training has converged to a stable point such that $D$ outputs a $P(\text{real}) \approx 1/2$ for almost all generated images. A high level of confusion from the discriminator is not just expected – it is one of the goals of the adversarial training procedure. In addition, there is no noticeably strong $m$ or $p_T$ dependence of the output, except for background images produced with $m$ and $p_T$ values outside the ranges of the training set.

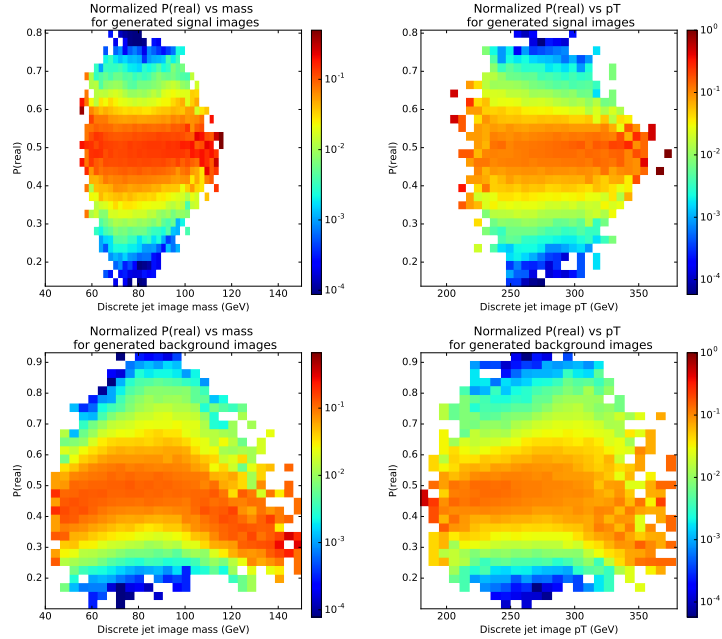

**Figure 8.26:** Discriminator output as a function of mass, on the left, and as a function of transverse momentum, on the right. The top plots refer to signal (boosted $W$ bosons from $W'$ decay) images, while the bottom plots refer to background (QCD) images. All plots have been normalized to unity in each $m$ or $p_T$ bin, such that the $z$-axis represents the percentage of images generated in a specific $m$ or $p_T$ bin that have been assigned the value of $P(\text{real})$ indicated on the $y$-axis.

#### 8.1.3.6 Classification as a GAN Performance Proxy

Another powerful, physically-motivated image quality assurance test is the preservation of pixel-level features to discriminate between $W$ and QCD jet images. For classification, the most important goal for GAN images is that the difference between signal and background be faithfully reproduced.

Fig. 8.27 shows the comparison between Pythia and LAGAN-generated mean class difference, in which the average background image is subtracted from the average signal image. Even though the stronger magnitude difference in GAN-generated images suggests that the GAN overestimates the importance of individual pixel contributions to the categorization of signal and background, the two plots generally show the correct pixel-by-pixel polarity and acceptable magnitude agreement between the true and generated datasets. Similar images are available in Fig. 8.28 and 8.29 after conditioning on the discriminator's output, which classifies images into real and fake.

A shortcoming of GANs applied to the generation of images that, like the ones in this dataset, are divided into inherently overlapping and partially indistinguishable classes, is their inability to carefully explore the gray area represented by the subspace of images that are hard to classify with the correct label. Many references [356, 30, 361] document the effort of classifying $W$ boson from QCD jet images using machine learning, and the topic is itself an active field of research in high energy physics (see Sec. 6.3). In particular, inserting this auxiliary classification task into the GAN's loss function may induce the generator to produce very distinguishably $W$-like or QCD-like jet images. The production of ambiguous images is unfavorable under the ACGAN loss formulation. Evidence of the occurrence of this phenomenon can be found by analyzing the normalized confusion matrices in Fig. 8.30 generated from the auxiliary classifier's output evaluated on Pythia and GAN-generated images. The plots show that classification appears to be easier for GAN-generated images, which $D$ correctly labels with higher success rate that Pythia images.

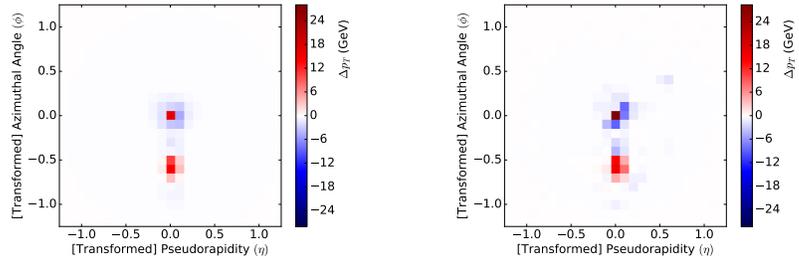

**Figure 8.27:** Difference between the average signal and the average background image produced by Pythia (left) and by the GAN (right).

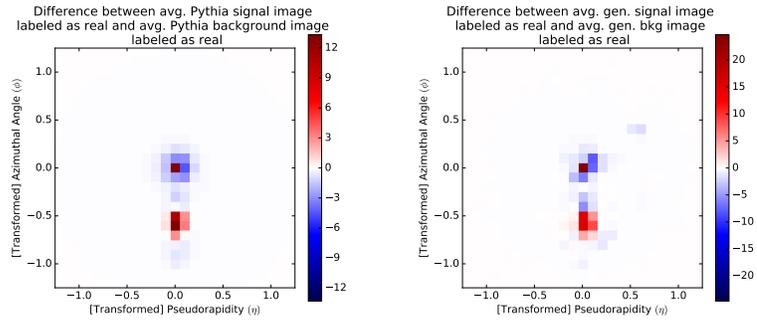

**Figure 8.28:** Difference between the average signal and the average background images labeled as real, produced by Pythia (left) and by the GAN (right).

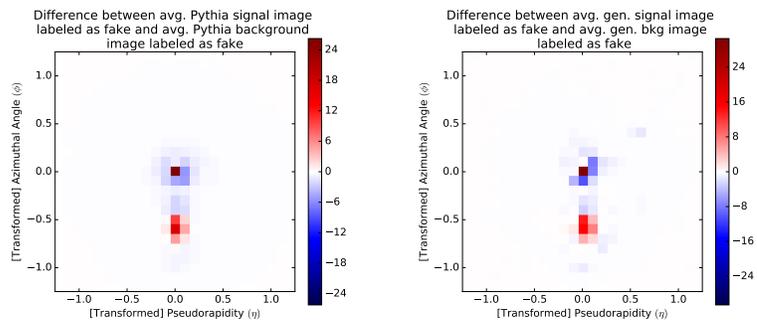

**Figure 8.29:** Difference between the average signal and the average background images labeled as fake, produced by Pythia (left) and by the GAN (right).

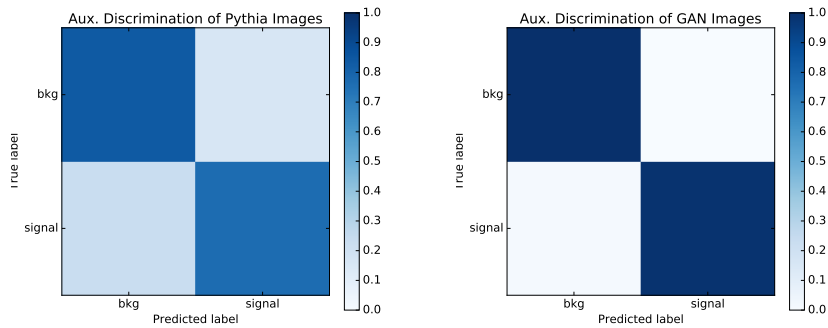

**Figure 8.30:** Normalized confusion matrices showing the percentage of signal and background images that the auxiliary classifier successfully labels. The matrices are plotted for Pythia images (left) and GAN images (right).

A consequence of the inability to produce equivocal images is that GAN-generated images currently do not represent a viable, exclusive substitute to Pythia images as a training set for a $W$-versus-QCD classifier that is to be applied for the discrimination of Pythia images. This hypothesis is checked by training a fully-connected MaxOut network [391] in the same vogue as the one in Ref. [30]. Two trainings are performed: one on a subset of Pythia images, one on GAN images. Both are evaluated on a test set composed exclusively of Pythia images. Fig. 8.31 clearly shows how, unlike Pythia images, GAN images cause the network to easily create two distinct representations for signal and background, which in turn leads to a higher misclassification rate when the model is applied to real images. With more research in this direction, coupled with better theoretical understanding, this problem can be ameliorated. Nonetheless, generated images can still be useful for data augmentation.

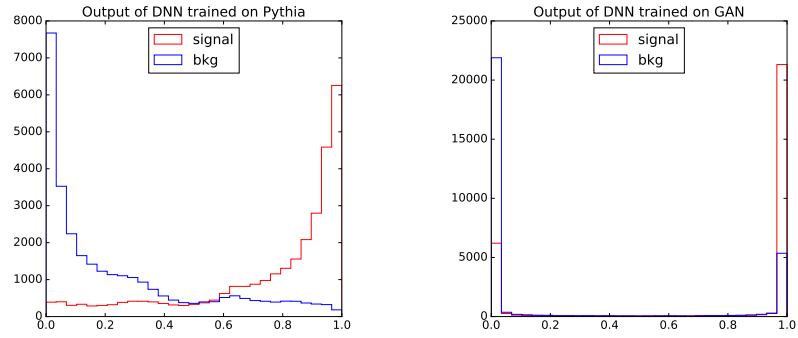

**Figure 8.31:** Output of the 2 fully-connected MaxOut nets - one trained on Pythia, one trained on GAN images - evaluated on Pythia images to discriminate boosted $W$ bosons (signal) from QCD (bkg).

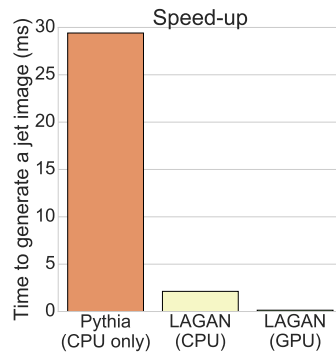

**Figure 8.32:** Time required to produce a jet image in PYTHIA-based generation, LAGAN-based generation on CPU, and LAGAN-based generation on GPU.

#### 8.1.3.7 Inference

Simulating particle collisions is extremely computationally expensive and is not inherently parallelizable, as $\mathcal{O}(\mathsf{depth}) \gg \mathcal{O}(\mathsf{width})$ for most processes. Though the production method used to generate the Pythia dataset is a significant simplification over full detector simulation, the procedure is still computationally suboptimal and a speedup may be desirable.

To evaluate the relative benefit of GAN-powered fast simulation, the event generation speed is benchmarked using Pythia on CPU and compared to the finalized LAGAN model on both CPU and GPU, as many applications of such technology in high energy physics will not have access to inference-time graphics processors. The benchmarks are performed on a `p2.xlarge` Elastic Compute Cloud (EC2) instance on Amazon Web Services (AWS) for control and reproducibility, using an Intel Xeon `E5-2686 v4` @ 2.30 GHz for all CPU tests, and an NVIDIA Tesla K80 for GPU tests. All GPU-driven generation

Table 8.1: Performance Comparison for LAGAN and Pythia-driven event generation

| Method | Hardware | Events/sec | Speed-up |
|--------|----------|------------|----------|
| Pythia | CPU | 34 | 1 |
| LAGAN | CPU | 470 | 14 |
| LAGAN | GPU | 7200 | 210 |

processes use `TensorFlow v0.12`, `CUDA 8.0`, and `cuDNN v5.1`.

GAN-based approaches can offer up to two orders of magnitude improvement over traditional event generators (see Table 8.1 and Fig. 8.32). Specifically, LAGANs provide an inference-time speedup of approximately 200× when run on GPUs, and 14× when run on CPUs, compared to Pythia-based generation. These preliminary results validate this approach as a beneficial research direction worth pursuing to solve the simulation bottleneck.

## 8.2 CaloGAN

Despite the progress achieved in the context of jet image generation by the LAGAN architecture, the levels of complexity and realism encountered in the application described in the previous section are still far from those expected in the context of developing a practical tool for faithful simulation of particle interactions with the ATLAS calorimeters.

In detectors like ATLAS, there exist two flavors of calorimeters: electromagnetic and hadronic (see Sec. 2.2.1.2). Electromagnetic calorimeters are designed to stop particles like electrons and photons, which have shallower and narrower showers compared to protons, neutrons, and charged pions. Hadronic calorimeters are thicker and deeper in order to capture penetrating radiation that forms irregular showers from nuclear interactions. In this first application of GANs to a longitudinally segmented calorimeter, the choice is to focus only on showers in electromagnetic calorimeters.

In moving from the previous application of generative modeling to jet images to a more complex and realistic task of generating electromagnetic shower developments across calorimeter volumes, the primary source of increase in the degree of difficulty comes from the introduction of a multi-layer detector geometry characterized by heterogeneous resolutions and detector cell sizes per layer, which prevent the modeling of the detector volume as an evenly segmented cuboid. Transverse segmentation is critical for particle identification and energy calibration. For example, the detailed radiation pattern can be used to distinguish prompt photons from $\pi^0 \to \gamma\gamma$, where the distance between the two photons is $\mathcal{O}(\mathrm{cm})$ for a 10 GeV $\pi^0$ at one meter from the interaction point. Pion rejection and an excellent resolution for photons in the Higgs boson $H \to \gamma\gamma$ discovery channel were driving factors for the design of the ATLAS Liquid Argon (LAr) electromagnetic calorimeter [460].

This section is structured as follows: Sec. 8.2.1 introduces the public datasets released in association with the publication of our findings in this application domain; Sec. 8.2.2 describes the neural network modules and loss formulations developed to achieve faithful generation results; and Sec. 8.2.3 evaluates the models' performance from both qualitative and quantitative standpoints.

The content of this section is adapted from the various published articles [461, 462, 463, 464, 33] that detail the progress our group has made over the years towards building a viable deep generative model for electromagnetic showers.

### 8.2.1 Dataset

As in the case of the LAGAN application, the introduction of the CaloGAN architecture is coupled to the release of a series of public datasets [373, 46] to incentivize the growth of the literature and of a community working on generative modeling for high energy physics. In the absence of similar open-source datasets, the ones introduced for this application provide a common, freely sharable reference for shower generation tasks. The vision is for future algorithmic developments in this application domain to report performance results on these established datasets for ease of comparison and insight extraction, bar the community-driven introduction of a superior dataset that surpasses it and better addresses the research objectives and constraints of the field.

These datasets contain sets of fully-simulated EM showers induced by the electromagnetic and nuclear interactions that the incident and secondary particles undergo as they propagate through the volumes of an electromagnetic calorimeter.

In general, a detector simulation begins with a list of particles with lifetimes greater than $\mathcal{O}(\mathrm{mm}/c)$. Each particle can be parametrized by its type (*e.g.*, electron, pion, etc.), its energy, and its direction, among others. The particle type determines when and how the particle interacts with the medium along its trajectory (see Sec. 2.2.1.2). Material interactions with the detector factorize: the energy deposited in a calorimeter by various particles is the sum of the energy from each shower treated independently [‡].

Table 8.2: Specifications of the calorimeter layers built with Geant4 for the CaloGAN application.

| Layer Number | Depth along $\hat{z}$ (mm) | $N_{\text{cells},x}$ | Cell width along $\hat{x}$ (mm) | $N_{\text{cells},y}$ | Cell width along $\hat{y}$ (mm) |
|---|---|---|---|---|---|
| 0 | 90 | 3 | 160 | 96 | 5 |
| 1 | 347 | 12 | 40 | 12 | 40 |
| 2 | 43 | 12 | 40 | 6 | 80 |

The ATLAS electromagnetic calorimeter serves as an inspiration for the detector used in this study. The geometry of the calorimeter portion constructed in these datasets consists of a cube of size 480 mm$^3$ at a distance $z_0 = 288$ mm from the origin and with no material in front of it. There are three instrumented layers stacked along the radial ($\hat{z}$) direction [§] with thicknesses 90 mm, 347 mm, and 43 mm. The active material is LAr and the absorber is lead. In contrast to the complex accordion geometry in the actual ATLAS calorimeter, the simplified setup adopted for the creation of these datasets (built from the GEANT4 B4 example [465]) uses flat alternating layers of lead and LAr that are 2 mm and 4 mm thick, respectively. The total energy per layer is computed by including both the active and inactive contributions. Each of the three longitudinal layers has different granularity and resolution, with pixels that are not square in the first and third layers. Instead, the cells in the first layer are 160 mm × 5 mm, the ones in the second layer are 40 mm × 40 mm, and those in the third layer are 40 mm × 80 mm. Table 8.2 summarizes the calorimeter geometry and the voxel dimensions. The detector volume is big enough to contain more than 99% of all showers in the training and test sets.

The short direction in the first layer ($\eta$) corresponds to the $pp$-beam direction in the real ATLAS experiment, while the short direction in the third layer ($\phi$) is perpendicular to $\eta$. In the native GEANT4 coordinate system (G), the $\hat{z}$-direction corresponds to the radial direction of a collider experiment (C).

---

[‡]Energy losses factorize, but detector readout does not. Due to threshold and digitization effects, the energy readout from two energy deposits in different detector elements need not be the same as the recorded energy from the two deposits in the same element. In detector simulations, these non-linear effects are treated after accounting for the material interactions and are therefore beyond the scope of the CALOGAN. It may be interesting in future work to consider an end-to-end generator that includes these effects, but it may not save a lot of time since simulation is much more costly than reconstruction.

[§]This is the direction that prompt neutral particles at $\eta$=0 would enter the calorimeter without any prior material interaction.

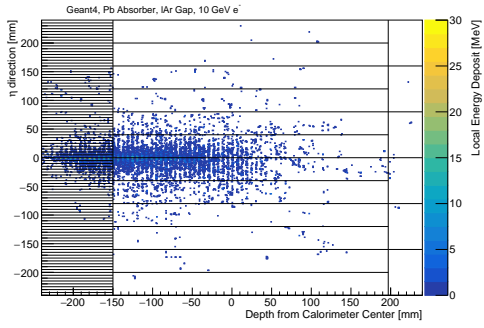
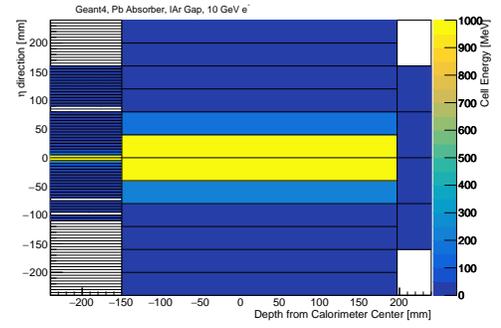

(a) Each dot represents one energy deposit from GEANT4 and the color of the dot encodes the energy. The absorber-gap structure is clearly visible, where most of the energy is lost in the absorber.

(b) Discretized version of (a), in which energy depositions are assigned to individual, discrete detector cells.

**Figure 8.33:** The electromagnetic shower from one 10 GeV electron event. The boundaries of the cells are shown along the $\eta$-$z$ plane.

The following coordinate transformations relate the two coordinate systems:

$$\hat{x}_G = \hat{y}_C; \quad \hat{y}_G = \hat{z}_C; \quad \hat{z}_G = \hat{x}_C \tag{8.6}$$

where the subscripts identify the coordinate system, $\hat{z}_C$ points along the beam line, $\hat{x}_C$ points towards the center of the LHC ring, and $\hat{y}_C$ points upwards towards the sky, as customary in ATLAS. Given the coordinate transformation in Eq. 8.6, the momenta are related by the following:

$$p_x^G = p_y^C \qquad p\sin\theta_G \cos\phi_G = p\sin\theta_C \sin\phi_C \tag{8.7}$$

$$p_y^G = p_z^C \qquad p\sin\theta_G \sin\phi_G = p\cos\theta_C \tag{8.8}$$

$$p_z^G = p_x^C \qquad p\cos\theta_G = p\sin\theta_C \cos\phi_C \tag{8.9}$$

The distribution of angles in collider coordinates can therefore be obtained from the momenta in GEANT4 coordinates:

$$\theta_C = \cos^{-1}\left(\frac{p_y^G}{p}\right) \tag{8.10}$$

$$\phi_C = \tan^{-1}\left(\frac{p_x^G}{p_z^G}\right) \tag{8.11}$$

In practice, the datasets are prepared as follows. Geant4 v10.2.0 [143] is used to generate parti-cles and simulate their interaction with the calorimeter using the FTFP_BERT physics list based on the FRITIOF [466, 467, 468, 469] and BERTINI intra-nuclear cascade models [470, 471, 472] with the standard GEANT4 electromagnetic physics package [473]. The magnetic field is turned off, removing curvature directions that would separate negatively and positively charged particles. Positrons, photons, and charged pions with uniform energies in the range between [1, 100] GeV are incident upon the calorimeter. A two-dimensional representation of a 10 GeV electron shower in the detector volume designed for these datasets is presented in Fig. 8.33, where Fig. 8.33(a) shows the exact, truth-level energy deposits from GEANT4, and Fig. 8.33(b) discretizes them according to the designed calorimeter readout geometry. A three-dimensional diagram of an electromagnetic shower after energy discretization is provided in Fig. 8.34, while Fig. 8.35 represents the corresponding series of 3 planar images in $\eta$-$\phi$ space. In these diagrams, the pixel intensity represents the sum of the energies of all particles incident to each cell ¶. Therefore, the final, fixed-length representation used to describe the showers in the datasets is a 3-image data format, in which the energy depositions in the first layer can be represented as a $3 \times 96$ image, the ones in the middle layer as a $12 \times 12$ image, and those in the last layer as a $12 \times 6$ image.

Two separate datasets are generated in order to progressively increase the difficulty of the generation task. In the first [373], particles enter the detector with a perpendicular trajectory at the impact point at the center of the detector ($\eta = 0, \phi = 0$). In the second [46], instead, both the incident angle and incident position are varied, so as to become particle properties to be used to condition the generation process. The incoming particles are shot isotropically in a circle around the $\hat{z}_G$-axis, with a maximum aperture angle $\theta_G$ of 5° (Fig. 8.36) and uniform lateral displacement of 5 cm in both $\hat{x}_G$ and $\hat{y}_G$ directions.

The first dataset is composed of three files, each containing 100,000 shower image sets originating from $e^+$, $\pi^+$, and $\gamma$ (as identified by the file names). The second public dataset is composed of 500,000 $e^+$, 500,000 $\pi^+$, and 400,000 $\gamma$ showers. The datasets are released as HDF5 files [392]. In each file, the `energy` entry specifies the true energy of the incoming particle in units of GeV. The branches named `layer_0`, `layer_1`, and `layer_2` represent the energy deposited in each cell of the three calorimeter layers, in `numpy.ndarray` format [397]. The `overflow` key maps to the amount of energy deposited outside of the calorimeter section considered in this application. The second dataset contains additional fields (x0, y0, z0, t0, px, py, pz) that store truth information that reflects the position and kinematic properties of the incident particle.

---

¶For the purposes of this study, the cell energy is the sum of the energy deposited in the lead and the argon; in practice, only the LAr energy deposits are measured. Dividing out these two components is left for future work.

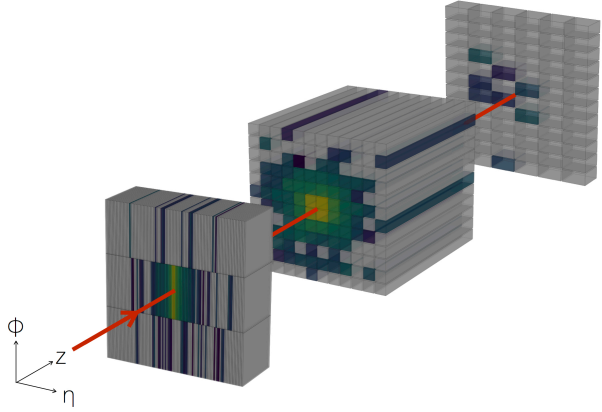

**Figure 8.34:** Three-dimensional representation of a 10 GeV $e^+$ incident perpendicular to the center of the detector. Not-to-scale separation among the longitudinal layers is added for visualization purposes.

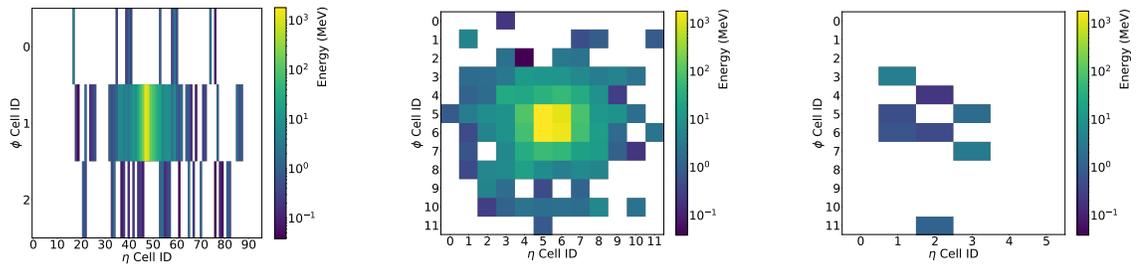

**Figure 8.35:** Two-dimensional, per-layer representation of the same shower as in Fig. 8.34.

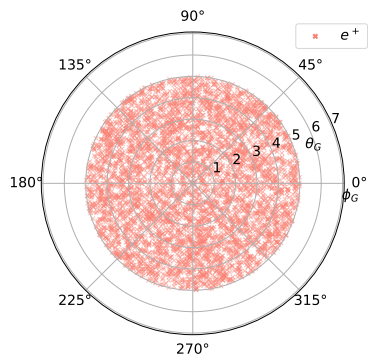

**Figure 8.36:** Isotropic distribution of 5,000 positrons from the dataset available in Ref. [46]. The concentric marks indicate the extent of the angle $\theta_G$, while the distribution is uniform across $360°$ in $\phi_G$. The $\hat{z}_G$-axis is into the page.

### 8.2.2 Model

Prior to entering the CaloGAN input layers, particle shower energies (expressed in MeV) are scaled down by a factor of $10^2$ and multiplied into the 1024-dimensional latent space vector $z \in \mathbb{R}^{1024}$. The generator $G$ then maps this 1024-dimensional vector to three gray-scale image outputs with different numbers of pixels, which represent the energy patterns collected by the three calorimeter layers as the requested particle propagates through them. The discriminator $D$ accepts the three images as inputs, along with $E$, the chosen value for the particle energy. The inputs are mapped to a binary output that classifies showers into real and fake, and a continuous output which calculates the total energy deposited in the three layers, then compares it with the requested energy $E$. Additional inputs and outputs are present in case of additional attribute conditioning (see Sec. 8.2.2.3).

Given the similarities in the datasets described in Sec. 8.1.1 and 8.2.1, the CaloGAN models designed for the generation of shower images follow the LAGAN guidelines (see Sec. 8.1.2) for how to modify a base DCGAN architecture to handle sparse, location-specific scientific data. In other words, this work builds on the prior experience and knowledge accumulated from the success of the preliminary LAGAN work, in order to start the model design from a base configuration that has previously been optimized not to suffer from some of the known complications that arise from the unique properties of calorimetric data.

To first address the generation of electromagnetic showers without variations in incident angle and position, the CaloGAN generator design combines parallel LAGAN-like processing streams, with a trainable attention mechanism that encodes the sequential connection among calorimeter layers. The LAGAN sub-modules are composed of a 2D convolutional unit followed by two locally-connected units with batch-normalization [202] layers in between. The dimensionality and granularity mismatch among the three longitudinal segmentations of the detector demand separate streams of operations with suitably sized kernels. Towards providing a readily adaptable tool, the architecture construction is simply a function of the desired output image size, so as to ensure flexibility for the application to a variety of particle types and detector configurations, to quickly obtain reasonable baselines in future R&D cycles. This desirable feature of the proposed network allows it to be trivially extended and configured for a different detector geometry by specifying the number of sampling layers and their resolutions.

Modeling the sequential nature of the relationship among the energy patterns collected by the three layers requires extra care. Drawing inspiration from Ref. [474], the chosen attention mechanism allows for dependence among layers by defining a trainable transfer functions to optimally resize and apply knowledge of the energy pattern in previous layers to the generation of the subsequent layer's readout. More specifically, the in-painting module takes as input a resized image from a previous layer, $\mathcal{I}'$, and the hypothesized image from the current layer, $\mathcal{I}$, and learns a per-pixel attention weight matrix $W$ via a weighting function $\omega(\mathcal{I}, \mathcal{I}')$ such that the pre-ReLU version of the current layer is $W \odot \mathcal{I} + (1 - W) \odot \mathcal{I}'$, where $\odot$ is the Hadamard product. This end-to-end trainable unit can utilize knowledge about the two adjacent layers to decide what information to propagate forward from the energy deposition in the anterior calorimeter layer. Alternative architectural choices, including the use of a recurrent connection, can be tested in future extensions of this work. Long Short Term Memory units, for example, are expected to be superior at retaining state across detector geometries with a high number of layers stacked along the longitudinal dimension.

Leaky Rectified Linear Units [230] are chosen as activation functions throughout the system, with the exception of the output layers of $G$, in which Rectified Linear Units [226] are preferred for the creation of sparse samples.

In the discriminator, three optimized LAGAN-like discriminator modules are employed for the immediate processing of the three input images that represent each shower. The feature space produced by each LAGAN-style stream is augmented by minibatch discrimination [294] on the hidden features, to encourage a well-examined distribution support.

In the second iteration of CaloGAN model design, aimed at generating electromagnetic showers with variable ingress position and direction, the locally-connected layers, which excel in the presence of highly location-specific features that arise across centered jet images and showers, are replaced with convolutional layers, more suitable to account for the translation invariance displayed by the showers contained in the second dataset [46]. In addition, convolutional kernels are added as blurring operations to specific detector layers for smearing effects, where weights are initialized to $\frac{1}{nm}$ for an $n \times m$ filter and are not updated during training. This represents the inductive bias we have over the shower image topology, and is empirically found to improve generation quality in light of otherwise very localized energetic energy deposits obtained over this more complex task. This explicit blur operation allows the model to learn a factored problem which marginalizes out smearing effects, and to focus on the factors in shower patterns that are due to conditional attributes.

### 8.2.2.1 Loss Formulation

In the original GAN formulation [285], a loss $\mathcal{L}_{\mathsf{adv}}$ is constructed in order to guide the learning towards equilibrium:

$$\mathcal{L}_{\mathsf{adv}} = \underbrace{\mathbb{E}_{z \sim p_z(z)}[\log(1 - D(G(z)))]}_{\text{term associated with } D \text{ classifying a sample from } G \text{ as fake}} + \underbrace{\mathbb{E}_{x \sim p_{\mathsf{data}}(x)}[\log D(x)]}_{\text{term associated with } D \text{ perceiving a real sample as real}}. \tag{8.12}$$

$G$ minimizes $\mathcal{L}_{\mathsf{adv}}$ while $D$ minimizes $-\mathcal{L}_{\mathsf{adv}}$, *i.e.*, the game is zero-sum.

However, this formulation suffers from gradient saturation when a synthetic sample $G(z)$ is seen as very fake, *i.e.*, when $D(G(z)) \approx 0$, which stagnates the learning procedure due to near-zero gradients. To overcome this, Ref. [285] proposes a non-saturating heuristic, replacing the generator objective with the minimization of the following:

$$\mathcal{L}_G = -\mathbb{E}_{z \sim p_z(z)}[\log D(G(z))] \tag{8.13}$$

This is the first term in the cost utilized in the applications described in this section; the discriminator, instead, still begins by minimizing:

$$\mathcal{L}_D = -\mathcal{L}_{\mathsf{adv}}. \tag{8.14}$$

To ensure the realism of the CALOGAN setup, an additional constraint is imposed in order to encourage the generator to produce showers originating from user-defined incoming particle energies. That is, the learned, implicit PDF $p_G$ needs to converge to the hypothetical data generating function $p_{\mathsf{data}}$ for any initial nominal energy $E_0$, *i.e.*, that $p_G(x|E = E_0) \longrightarrow p_{\mathsf{data}}(x|E = E_0)$ for all $E_0 \in [E_{\mathsf{min}}, E_{\mathsf{max}}]$, the energy range considered by the application.

The total shower energy is trivial to compute from the 3-image representation adopted in this work. The sum of intensities across all pixels gives the desired quantity, without the need to learn to extract or regress its value from each input image. A specific sub-module of the discriminator is directed to compute the sum across all input pixels to obtain the reconstructed energy $\hat{E}$. To penalize generated instances of too little or too much deposited energy, the reconstructed energy is then internally compared to the true (or requested) energy $E$, which is passed as an input to both the generator and the discriminator, by computing the mean absolute energy error expressed as $|e - e'|$, for any two energy values $e$ and $e'$. To account for inefficiencies and leakage, an additional binary feature is constructed in order to determine whether the reconstructed energy falls withing a 10 GeV well around the true energy value, to avoid penalizing events in which the energy deposited across the calorimeter volume does not exactly match the true energy. These energy-consistency features are then concatenated with the remaining hidden features automatically constructed by the discriminator to form the input to its final decision layers. The logic behind this choice is that the difference between true and reconstructed energy (besides being helpful as direct, energy-related feedback to the generator) can be used to easily discriminate real from fake images, in the case in which the generator is not producing showers around the user-requested energy. To further penalize incorrect energy generation, a new component is added to the generator loss, which consists of the mean absolute error between the requested energies $E$ and the reconstructed energies $\hat{E}$:

$$\mathcal{L}_E = \mathbb{E}_{z \sim p_z(z)}[|E - \hat{E}(G(z))|] \qquad (8.15)$$

$\mathcal{L}_E$ is a constant with respect to the discriminator's parameters, so it will not directly affect its training. This solution not only helps ensure the confinement of the generated energy to a desirable range, but also allows to encode a 'soft' physical notion of conservation of energy. In fact, this formulation discourages, but does not forbid, the deposition of more energy than requested. Unphysical results can be rejected by sampling the learned distribution until energy preservation is met.

Finally, the generator minimizes Eq. 8.16, and the discriminator minimize Eq. 8.17:

$$\mathcal{L}_G = \lambda_E \mathcal{L}_E - \mathbb{E}_{z \sim p_z(z)}[\log D(G(z))], \qquad (8.16)$$

$$\mathcal{L}_D = -\mathcal{L}_{\text{adv}}. \qquad (8.17)$$

$\lambda_E$ is manually set to a value of $0.05$ to achieve a sensible scaling towards a reasonable numerical range. A more carefully tuned trade-off parameter value can be systematically optimized to balance the adversarial and conditional losses.

### 8.2.2.2 Sparsity Considerations

In direct opposition to natural images, most scientific data is inherently sparse. As a consequence, in most scientific applications, sparsity plays an important role in determining how viable a generated sample is.

Even though traditional GANs are not directly designed for a non-dense output space, previous work on generative models for jet images (see Sec. 8.1) has proposed preliminaty solutions to address the issue. A simple approach is to utilize ReLUs [226] to directly induce sparsity in the output layers. However, though utilizing ReLU pointwise non-linearities introduces sparsity in the generated elements, since ReLUs produce sparse gradients in the generator, care must be given to the training procedure in order not to suffer from a lack of feedback.

A proposed solution to improve sparsity modeling is to design a quantity closely related to sparsity that the discriminator can compute and utilize to encourage the generator to produce samples with adequate occupancy levels. In the first iteration of model construction, the discriminator is explicitly instructed to perform a raw sparsity calculation and concatenate the result to its internal features for downstream categorization. For an image $X \in \mathbb{R}^{m \times n}$, the sparsity is calculated as:

$$\mathsf{sparsity}(X) = \frac{1}{nm} \sum_{i<m} \sum_{j<n} \mathbb{I}[X_{i,j} \neq 0]. \tag{8.18}$$

Although it allows the discriminator to learn to reject generated samples that do not match the sparsity levels of real samples, this formulation does not let any gradient signal propagate to the generator, due to the all-zero sub-derivatives of the indicator function $\mathbb{I}[\cdot]$. To ameliorate this deficiency, a *soft sparsity* formulation is introduced in later CaloGAN models as a quantity that, in the limit of tunable hyper-parameters, converges to Eq. 8.18. The quantity that the discriminator internally computes and uses as an input feature to its decision layer is formally defined as:

$$\mathsf{softsparsity}(X) = \frac{1}{nm} \left\| \left( \frac{|X|^\alpha + |X|^{1/\beta}}{|X|^\alpha + |X|^{1/\beta} + 1} \right)^{1/\gamma} \right\|_1, \tag{8.19}$$

where $\alpha, \beta, \gamma > 0$, all matrix powers and $|\cdot|$ operators are assumed to act pointwise on $X$, and $\|\cdot\|_1$ is the entry-wise 1-norm rather than the induced norm. Note that $\mathsf{softsparsity} : \mathbb{R}^{m \times n} \longrightarrow [0, 1]$, and that the limit behavior for Eq. 8.19 is consistent with Eq. 8.18, *i.e.*,

$$\lim_{\alpha, \beta, \gamma \to \infty} \mathsf{softsparsity}(X) = \frac{1}{nm} \sum_{i<m} \sum_{j<n} \mathbb{I}[X_{i,j} \neq 0]. \tag{8.20}$$

In practice, in the code [475], $\alpha = \beta = \gamma = 5$, but these parameters can be tuned.

Even with the discriminator now computing a sub-differentiable approximation to sparsity, there is no direct guarantee as to whether generated samples will display occupancy levels commensurate with the distribution of sparsity over the full data distribution. Hence, the discriminator needs to be instructed to make use of it as part of its minibatch discrimination [294] step, in order to allow the model to learn the correct distribution over batches of images.

Despite our direct intervention in encouraging the model to produce realistic image sparsity distributions, the faithful modeling of this quantity still appears as a major challenge for the CaloGAN. A future improvement could perhaps consists of decoupling the sparsity across layers into separate quantities, in order to prioritize per-layer distribution matching over the overall quantity.

### 8.2.2.3 The Conditional CaloGAN

While categorical conditioning can be taken care of using traditional methods [290, 292], continuous values that cannot be directly computed from the shower pixel representation necessitate the introduction of a simple methodology to condition on continuous characteristics.

To create a GAN-based simulator that presents a useful solution for scientific fields, it is often not sufficient to only learn $p_{\mathsf{data}}(x)$, but it is, instead, preferable to also learn to approximate $p_{\mathsf{data}}(x|\xi)$, where $\xi \in \Xi$ is a vector of conditional attributes. This is crucial for most scientific applications, in which $\{\xi\}$ is a set of attribute vectors of theoretical importance that can be conditioned on or interpolated between. Although parameter interpolation is not necessarily relevant to the shower simulation problem explored in this work, this formalism easily generalizes to fields, such as Cosmology, in which this task is paramount [476]. In that context, only a limited set of values $\{\xi\} \subseteq \Xi$ can be simulated with realistic computing resources, so that $\Xi$ can, in practice, only be sparsely sampled. Instead, the goal is to be able to directly sample from the continuous distribution over the convex hull of sampled $\{\xi\}$. This allows dense interpolation in spite of finite samples in a conditioning space.

In this application (where the objective is simply conditioning, instead of interpolating), given the second dataset with physical attributes described in Sec. 8.2.1, $\xi$ is chosen to be $\xi = (E, x_0, y_0, \theta, \phi)$, where $x_0$ and $y_0$ are the incident coordinates of the incoming particle, and $\theta$ and $\phi$ are the incident angles. A separate head of the discriminator is built into the model in order to for it to learn to regress on said quantities given the 3-image shower representation as input. For any (real or fake) input image $x$, the regression sub-module, parametrized by a subset of discriminator weights, specializes in the computation of the function $\hat{\xi}(x)$. Unlike in the energy case, the reconstruction of these conditioning attributes is not hard-coded as an analytical function for the discriminator to compute, but is itself learned. An extra term is then added to both the generator's and discriminator's losses: the discriminator adjusts its weights to learn to correctly compute the function $\hat{\xi}(x)$ to reconstruct the attributes of the reference shower images from GEANT4 $\hat{\xi}(x) \approx \xi$, while the generator is trained to produce images $G(z|\xi)$ such that the attributes reconstructed by the discriminator match those requested by the user in the form of inputs to the generative task $\hat{\xi}(G(z|\xi)) \approx \xi$. The loss terms in Eq. 8.16 and Eq. 8.17 are, therefore, augmented by conditioning on continuous variables that encode the location and kinematic properties of the incoming particle, so that both players cooperate in minimizing the attribute reconstruction error:

$$\mathcal{L}_{\hat{\xi}, D} = \sum_{i=1}^{\mathsf{dim}(\xi)} \lambda_{\xi_i} \mathbb{E}_{x \sim p_{\mathsf{data}},\, \xi_i \sim p_{\xi_i}(\xi_i|x)} \left[ |\xi_i - \hat{\xi}_i(x)| \right], \tag{8.21}$$

$$\mathcal{L}_{\hat{\xi}, G} = \sum_{i=1}^{\mathsf{dim}(\xi)} \lambda_{\xi_i} \mathbb{E}_{z \sim p_z(z),\, \xi_i \sim p_{\xi_i}(\xi_i|G(z))} \left[ |\xi_i - \hat{\xi}_i(G(z|\xi))| \right]. \tag{8.22}$$

where the index $i$ tracks each feature in conditioning space. Hyper-parameters $\lambda_{\xi_i}$ are included to allow the user to individually control the contribution to the loss of each attribute reconstruction task. In this application, they are simply set to $0.01$ until further flexibility becomes necessary.

Further specifications of the exact hyper-parameter and architectural choices as well as software ver-sioning constraints are available in the source code [475].

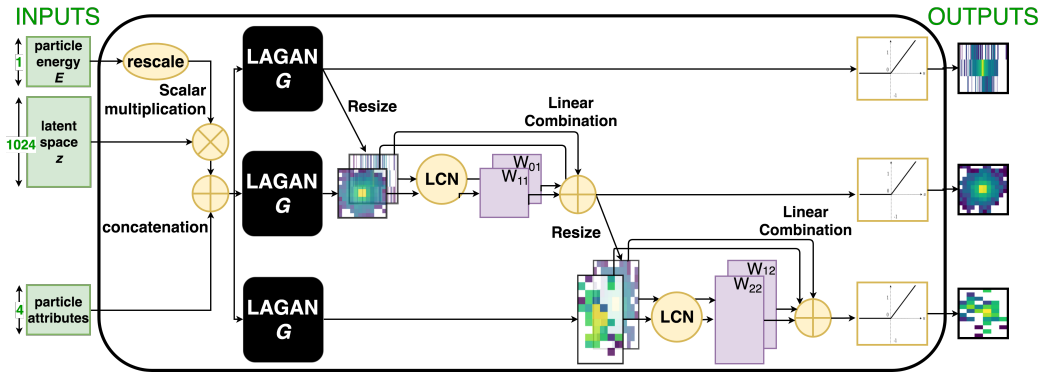

**Figure 8.37:** Composite conditional CaloGAN generator $G$, with three LAGAN-like streams connected by attentional layer-to-layer dependence.

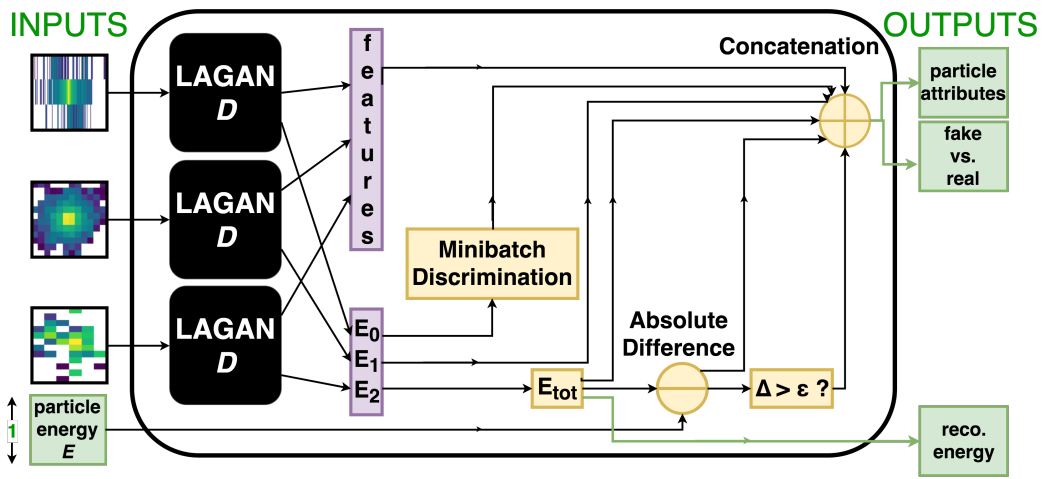

**Figure 8.38:** Composite conditional CaloGAN discriminator $D$, with three LAGAN-like streams and additional domain-specific energy calculations included in the final feature space.

The architecture diagrams for the conditional CALOGAN generator and discriminator are available in Fig. 8.37 and 8.38, respectively. They differ from the traditional, baseline CALOGAN model for the presence of input particle attributes in the generator, are their corresponding regressed attributes at the output of the discriminator.

Two additional architectural modifications are tested in order to build a particle-type conditioning system directly into the learning process. Neither the AC-GAN [292] nor the conditional GAN [290] frameworks are found to be able to handle the substantial differences among the three particle shower types, to an extent that would favor a single model solution over the training of specialized per-particle generative models. The unique features that pion showers exhibit, compared to the other particles, make it unfavorable to train a single model for multiple particle types. Perhaps, both a significantly richer model and a larger latent space could alleviate some problems associated with conditioning using the investigated approaches. Although building a fully joint model is an interesting Machine Learning challenge, the practicality and flexibility of this application may suffer from having one single model for all particle showers.

### 8.2.2.4 Training Procedure

Three individual CaloGAN models are trained, one per particle type (photons, positrons, and charged pions). Separating per-particle-type CaloGAN architectures and implementations affords many practical benefits. It is easy to imagine a situation where the life cycles surrounding models for different particle types could be very different. In addition, this allows for total independence of versioning, framework, or language.

The weights in the generator and discriminator are optimized in an alternating fashion over a set of 100,000 Geant4-simulated events for each particle type in batches of 256, using the Adam optimizer [177]. The learning rates are $2 \times 10^{-5}$ for the discriminator and $2 \times 10^{-4}$ for the generator. Outside of initial rough hyper-parameter tuning, no dedicated per-particle-type hyper-parameter optimization is performed, and the same training parameters are applied to all three networks. Significant performance improvements are expected (especially for pions) with dedicated optimizations.

Each system is trained for 50 epochs. The number of epochs is increased to 150 for the more complex conditional CaloGAN. Sixteen NVIDIA K80 graphics cards are used for initial hyper-parameter sweeps, with two Titan X Pascal Architecture cards used for final training. `Keras v2.0.3` [275] is used to construct all models, with the `TensorFlow v1.1.0` backend [274].

The sparsity and high dynamic ranges that characterize shower images require careful considerations with respect to gradient properties and batch sizes. In fact, larger batch sizes are necessary to smooth out gradients during each update step, because most parameters receive no update in a small batch. In addition, the sparsity levels can lead to a truth-bit being present in generated samples. To discourage this violation of the generator distribution being absolutely continuous with respect to $p_{\text{data}}$, label flipping is used in order create more overlap between the true and generated distributions.

### 8.2.3 Results

Similar to the strategy adopted in earlier LAGAN evaluation studies (see Sec. 8.1.3), the CaloGAN performance is analyzed in this section using application-driven methods focused on sample quality. A first qualitative assessment in Sec. 8.2.3.1 is accompanied by a quantitative evaluation based on physics-driven similarity metrics in Sec. 8.2.3.2. The choice reflects the domain-specific procedure for data-simulation comparison. The fidelity of GAN-based simulation is verified through the use of on one-dimensional statistics of the shower probability distribution and their pair plots. Visualizing and verifying the performance in higher dimensions, however, is still an open challenge. One way to probe the distributional modeling in higher dimensions is to investigate the distinguishability of different particle types in classification tasks, and study the classifiers' performance differences when trained or tested on either Geant4 or CaloGAN-generated showers (Sec. 8.2.3.3). The computational advantage due to the speed-up factors achieved by the CaloGAN is quantified in Sec. 8.2.3.4.

This section first reports results on the simpler task of generating showers with perpendicular incident direction with respect to the center of the calorimeter volume. Sec. 8.2.3.5, instead, is reserved for the performance evaluation of the conditional CaloGAN.

#### 8.2.3.1 Image Content Quality Assessment

Although the 3-image representation is less intuitively interpretable than the single jet image representation used in the LAGAN application, it is still possible to begin the exploration of the distribution of showers produced by the CaloGAN models with a visual assessment of their quality.

Given the equivalence of frame of reference for every captured particle shower, computing per-pixel averages is a physically meaningful operation that exposes ensemble properties that are not as clearly visible at the individual image level due to the inherent stochasticity of the generation process. After averaging over thousands of electromagnetic showers originating from the same kind of particle, the output of the CaloGAN generator is visualized in Figs. 8.39, 8.40, and 8.41 for cascades of $e^+$, $\gamma$, and $\pi^+$, respectively, with uniform energy between 1 GeV and 100 GeV (the range explored by the reference training dataset). On average, the systems appear to learn a complete picture of the outcome of the physical processes governing shower generation. For photon showers, for instance, the mean per-layer cell variations only show a $\sim 4\%$ and $\sim 1\%$ discrepancy in the first two layers where most energy is deposited for $e/\gamma$. Incidentally, the level of fidelity demanded of the model in this comparison, although useful for benchmarking, may be greater than what is necessary in practice, as the following steps of data readout and reconstruction are generally lossy.

Diversity and over-training concerns can be investigated by considering the nearest neighbors among the training and generated datasets. Figs. 8.42, 8.43, and 8.44 shows the shower evolution across the three calorimeter layers for five randomly selected electromagnetic showers, along with their GAN-generated nearest neighbors, for $e^+$, $\gamma$ and $\pi^+$ showers respectively. Good qualitative agreement can be observed between the true and learned distributions across all layers, without obvious signs of mode collapse. Compared to the other two particle types explored in this application, at the individual image level, charged pions clearly display a higher degree of complexity and diversity in their showers. Some $\pi^+$ deposit energy in all cells of a given layer, some only hit a handful of them. This is because charged pions undergo nuclear interactions in addition to electromagnetic interactions.

Concerns about training set memorization and sample diversity among generated samples are also addressed by reporting the distribution of euclidean distances between each image and its nearest neighbor among each particle-specific GEANT4 dataset (Fig. 8.45(a)), each particle-specific CALOGAN dataset (Fig. 8.45(b)), between each CALOGAN image and its nearest neighbor in the corresponding GEANT4 dataset (Fig. 8.45(c)), and vice versa (Fig. 8.45(d)). These histograms show that the intra-class distance between each shower and its nearest neighbor is comparable for GEANT4 reference showers and CALOGAN-generated showers. Specifically, generated images are not all similar to one another (which would point to potential mode collapse), but they are approximately as diverse as the ones presented in the GEANT4 dataset. Unlike the photon and positron ones, the pion distribution presents clear bimodal behavior, with both modes explored by the GAN.

The one failure mode identified in this study concerns the production of very collimated $\pi^+$ showers, which would create extremely sparse images with only very few non-zero pixel activations. In the reference dataset generated with the GEANT4 package, these showers appear to be very close to their nearest neighbor because of the lack of variation in their shower topologies. This subset of electromagnetic showers is not as faithfully reproduced by the CALOGAN as the rest of the distribution. In fact, by under-representing ultra-sparse $\pi^+$ showers, the CALOGAN ends up producing fewer samples with

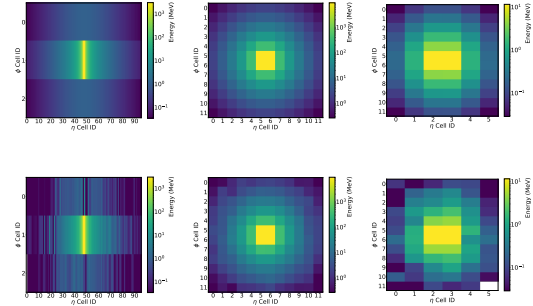

**Figure 8.39:** Average $e^+$ Geant4 shower (top), and average $e^+$ CaloGAN shower (bottom), with progressive calorimeter depth (left to right).

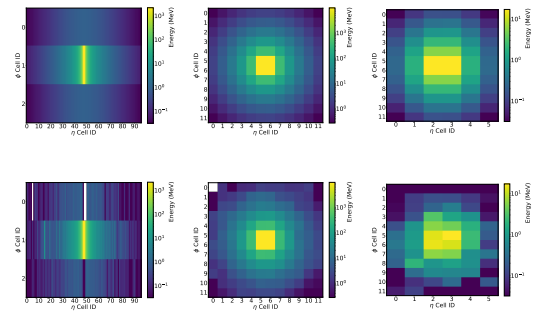

**Figure 8.40:** Average $\gamma$ Geant4 shower (top), and average $\gamma$ CaloGAN shower (bottom), with progressive calorimeter depth (left to right).

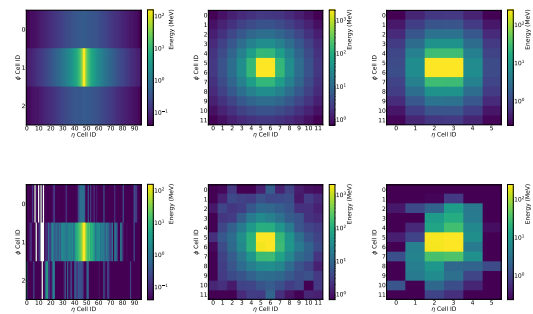

**Figure 8.41:** Average $\pi^+$ Geant4 shower (top), and average $\pi^+$ CaloGAN shower (bottom), with progressive calorimeter depth (left to right).

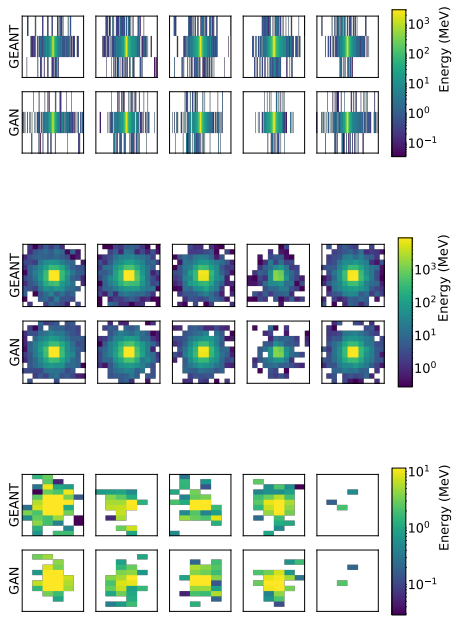

**Figure 8.42:** Five randomly selected $e^+$ showers per calorimeter layer from the training set (top) and the five nearest neighbors (by euclidean distance) from a set of CaloGAN candidates.

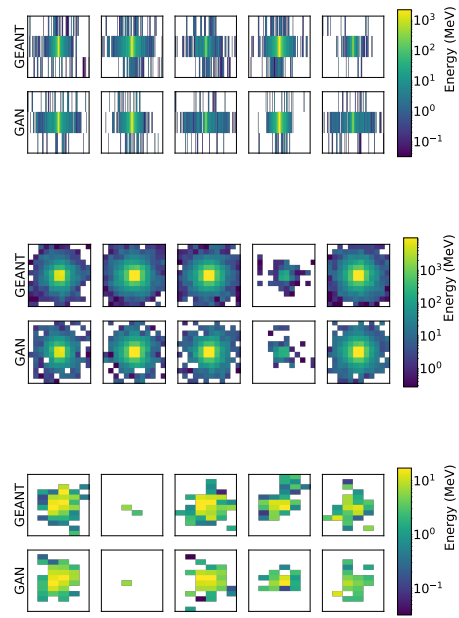

**Figure 8.43:** Five randomly selected $\gamma$ showers per calorimeter layer from the training set (top) and the five nearest neighbors (by euclidean distance) from a set of CaloGAN candidates.

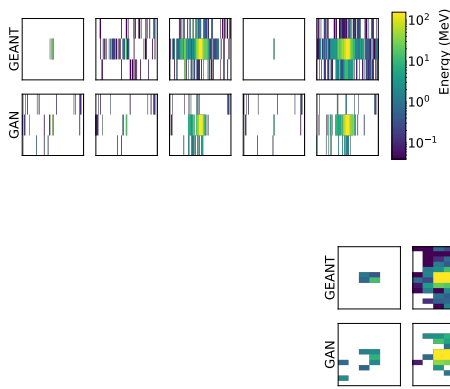

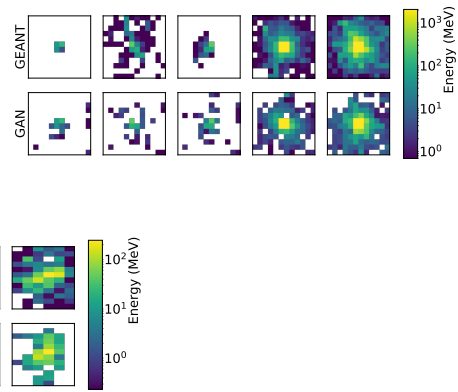

**Figure 8.44:** Five randomly selected $\pi^+$ showers per calorimeter layer from the training set (top) and the five nearest neighbors (by euclidean distance) from a set of CaloGAN candidates.

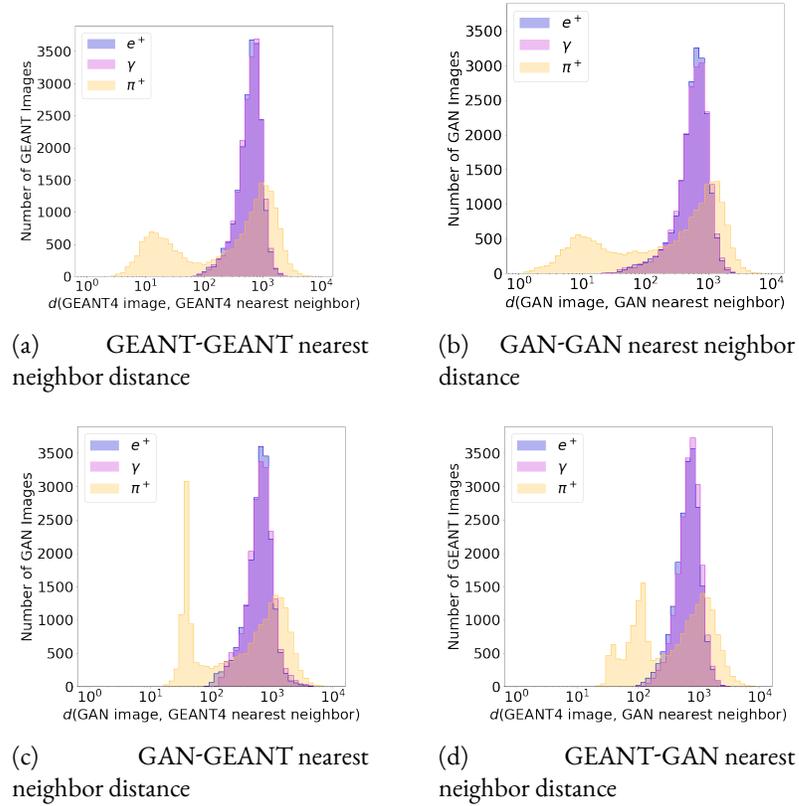

**Figure 8.45:** Euclidean distance between nearest neighbors across datasets, for the three particle types ($e^+$, $\gamma$, and $\pi^+$.)

very closely related nearest neighbors. It also causes those collimated GEANT4 $\pi^+$ images to find their nearest neighbors in the CALOGAN dataset much further away than their nearest neighbors in the GEANT4 dataset. This can be observed in Fig. 8.46(a), which shows the ratio between the euclidean distances of each GEANT4 shower to its nearest neighbors in the GEANT4 and CALOGAN dataset, and in Fig. 8.46(b), which plots GEANT4 images on a plane using as coordinates the euclidean distance to their GEANT4 (on the $\hat{x}$-axis) and to their CALOGAN (on the $\hat{y}$-axis) nearest neighbors. The pion distribution veers significantly away from the diagonal for GEANT4 images with low euclidean distance to their GEANT4 nearest neighbor. The photon and positron generated nearest neighbor to each GEANT4 shower, instead, tends to be approximately as far away from the reference image as the nearest neighbor in the GEANT4 dataset. As mentioned, due to their complexity and diversity, pion showers would probably benefit from a dedicated CALOGAN optimization and re-training, which could address some of the inefficiencies of the current model. This study is carried out using 20,000 GEANT4 and 20,000 CALOGAN-generated shower images per particle type.

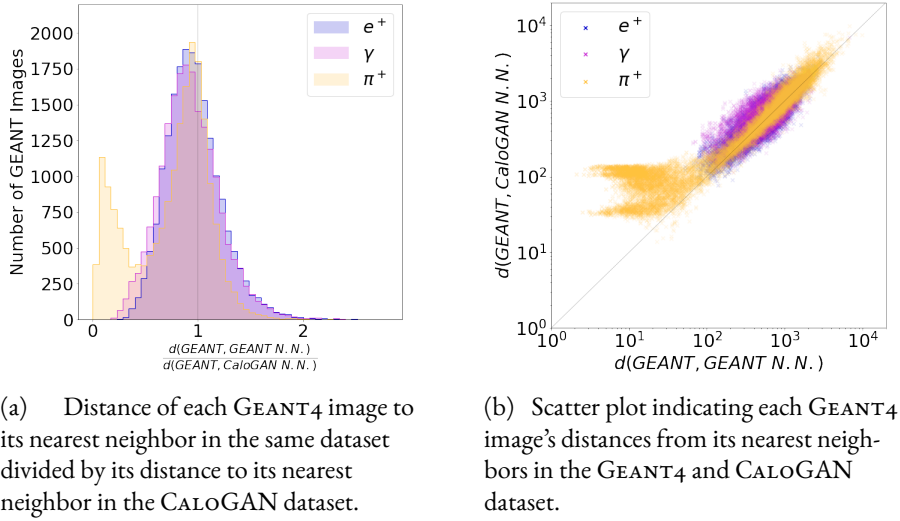

(a) Distance of each GEANT4 image to its nearest neighbor in the same dataset divided by its distance to its nearest neighbor in the CALOGAN dataset.

(b) Scatter plot indicating each GEANT4 image's distances from its nearest neighbors in the GEANT4 and CALOGAN dataset.

**Figure 8.46:** Comparisons between the distances from each Geant4 image to its nearest neighbors across two datasets: the same one from which the image originated, and the one generated by the CaloGAN.

#### 8.2.3.2 Shower Properties

Traditional photon and electron classification and energy calibration in ATLAS make use of geometric and kinematic properties of the calorimeter shower (see Sec. 5.2.2 and 5.2.3). To allow for intuitive domain-specific interpretation of the GAN's performance, these same features are used to assess the quality of the GAN-generated electromagnetic showers. These are physics-driven one-dimensional projections of the multi-dimensional, inaccessible, data distribution that can be used as further validation and introspection into the capabilities of the CALOGAN to adequately capture and model non-linear functional representations of the simulated data distribution. In fact, it is desirable for the approximation to the true data distribution learned by the generative model to match the shape of its unidimensional projections along these meaningful directions and recover the target distribution of these 1D statistics. . Table 8.3 contains the description and mathematical definition of the shower shape variables used to compare the generated and target distributions. These are defined as functions of $\mathcal{I}_i$, the vector of pixel intensities for an image in layer $i$, where $i \in \{0, 1, 2\}$.

The 1D distributions of shower shapes for the three types of particles in this study are shown in Fig. 8.47 for both GEANT4 and CALOGAN-generated showers.

The networks are never directly shown any shower shape variable besides the total energy, and the only other type of information they have access to is the pixel representation of each shower. It is there-fore encouraging to note that the CaloGAN recovers the main trends and some complex features of the Geant4 simulated data distribution for a variety of shower shapes across several orders of magnitude and for all three particle types. However, certain features of some distributions are not well-described. This is a challenge for the future and will likely require improvements to the architecture and training procedure. Longer trainings of higher capacity architectures have shown promise in rectifying some of these issues. Another possible method to try improve the matching of these features consists of explicitly including sub-differentiable versions of the shower shape variables of interest into the training process and integrating the explicit per-variable distributional matching objective into the loss function itself, thus inserting a prior on the importance for the GAN to properly model these observables. At the moment, however, it is preferable to withhold this information from the model and use it for a comprehensive validation assessment.

Table 8.3: One-dimensional observables used to assess the quality of the GAN samples

| Shower Shape Variable | Formula | Notes |
|---|---|---|
| $E_i$ | $E_i = \sum_{\text{pixels}} \mathcal{I}_i$ | Energy deposited in the $i^{th}$ layer of calorimeter |
| $E_{\text{tot}}$ | $E_{\text{tot}} = \sum_{i=0}^{2} E_i$ | Total energy deposited in the electromagnetic calorimeter |
| $f_i$ | $f_i = E_i / E_{\text{tot}}$ | Fraction of measured energy deposited in the $i^{th}$ layer of calorimeter |
| $E_{\text{ratio},i}$ | $\dfrac{\mathcal{I}_{i,(1)} - \mathcal{I}_{i,(2)}}{\mathcal{I}_{i,(1)} + \mathcal{I}_{i,(2)}}$ | Difference between the highest and second highest energy deposit in the cells of the $i^{th}$ layer, divided by the sum |
| $d$ | $d = \max\{i : \max(\mathcal{I}_i) > 0\}$ | Deepest calorimeter layer that registers non-zero energy |
| Depth-weighted total energy, $l_d$ | $l_d = \sum_{i=0}^{2} i \cdot E_i$ | Sum of the energy per layer, weighted by layer number |
| Shower Depth, $s_d$ | $s_d = l_d / E_{\text{tot}}$ | Energy-weighted depth in units of layer number |
| Shower Depth Width, $\sigma_{s_d}$ | $\sigma_{s_d} = \sqrt{\dfrac{\sum_{i=0}^{2} i^2 \cdot \mathcal{I}_i}{E_{\text{tot}}} - \left(\dfrac{\sum_{i=0}^{2} i \cdot \mathcal{I}_i}{E_{\text{tot}}}\right)^2}$ | Standard deviation of $s_d$ in units of layer number |
| $i^{\text{th}}$ Layer Lateral Width, $\sigma_i$ | $\sigma_i = \sqrt{\dfrac{\mathcal{I}_i \odot H^2}{E_i} - \left(\dfrac{\mathcal{I}_i \odot H}{E_i}\right)^2}$ | Standard deviation of the transverse energy profile per layer, in units of cell numbers |

Examining 1D statistics does not probe correlations between shower shapes or higher dimensional aspects of the probability distribution. The two-dimensional projections available in Fig. 8.48, 8.49, and 8.50 for photons, positrons, and pions, respectively, go one step further in providing useful visualizations of the true and learned distributions of electromagnetic shower images. These show the extent to which the CaloGAN learns (or does not learn) to cover the full support of the Geant4 distribution, and help identify the image topologies that can (or cannot) be faithfully generated by the CaloGAN, as well as those the CaloGAN generates that are not present in the reference Geant4 dataset. Among the hardest topologies to recover are very low-energy showers, showers with over 40% of energy deposited in the front layer of the calorimeter, and showers with less than 60% of energy deposited in the middle layer of the calorimeter. In spite of the helpfulness of these two-dimensional scatter plots, the full $n$-dimensional distribution is still not trivial to visualize.

As mentioned above, the only kinematic observable the base CaloGAN model is directly provided with is the total shower energy, which is passed to both generator and discriminator as an attribute to condition the generative process. Consequently, the trained model can be instructed to produce showers at any given energy specified by the user. It is important, then, to verify the fidelity and response of the CaloGAN model under the energy conditioning task. The loss formulation in Eq. 8.15, coupled with the uniform energy distribution in the training dataset, admits an approximately symmetric conditional output energy distribution, the width of which quantifies the expected proximity between the reconstructed shower energy and the requested value. Experimentally, the CaloGAN is tasked with the production of 70,000 $e^+$ showers at five specific point values of the incoming particle energy, four of which lie within the energy range of showers shown to the GAN at training time (1, 25, 50, and 100 GeV), while the last falls outside of that domain (150 GeV). The empirical shower energy, computed as the sum of the discretized energy deposits produced by the CaloGAN, is plotted for each of the five requests, in Fig. 8.51. The vertical lines, located at the nominal requested energy, approximately coincide with the mode of each distribution. As the shower energy increases, so does the complexity of the shower evolution process, which results in broader reconstructed energy distributions around the requested value. Thanks to the flexibility of the CaloGAN framework, requesting out-of-range attributes is allowed, though not necessarily recommended without careful validation of shower properties against similar Geant4 showers. The distribution of reconstructed energies for 150 GeV showers displays a broader width and shifted mode towards the training domain.

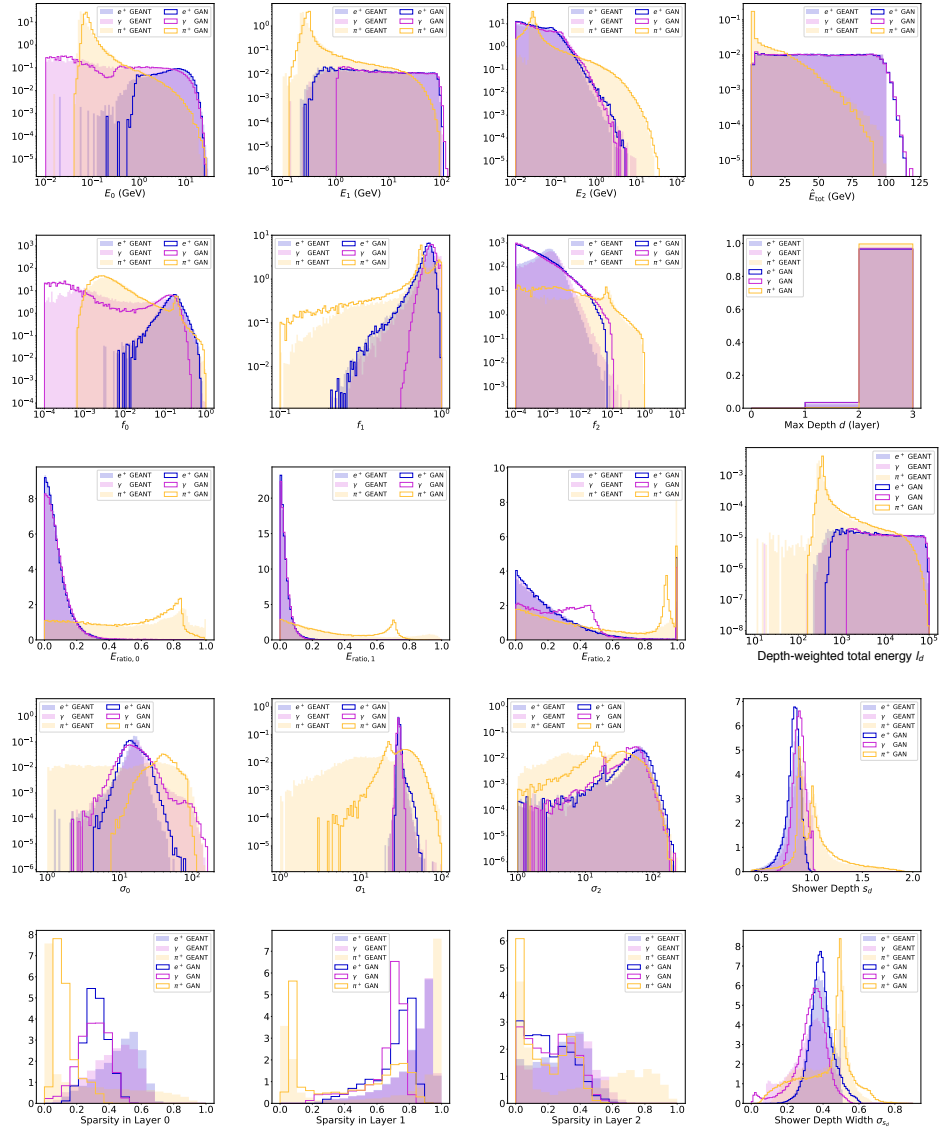

**Figure 8.47:** Comparison of shower shape variables, introduced in Table 8.3, and other variables of interest, such as the sparsity level per layer, for the Geant4 and CaloGAN datasets for $e^+$, $\gamma$ and $\pi^+$.

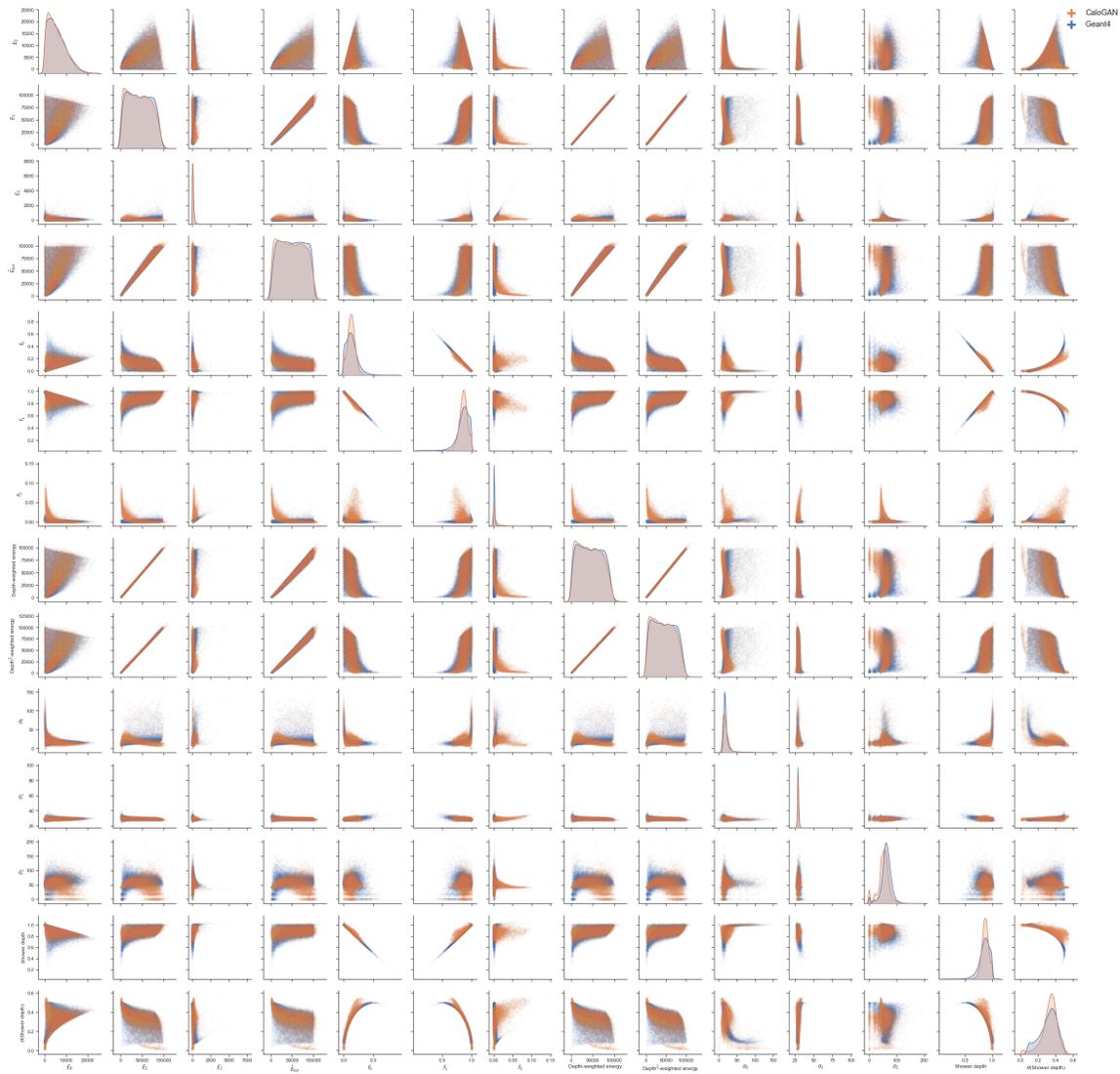

**Figure 8.48:** Pair plots of selected shower shape distributions for approximately 20,000 Geant4-generated (in blue) and 20,000 CaloGAN-generated (in orange) $\gamma$ showers. The diagonal plots show the individual variable distributions using KDE. The off-diagonal plots display pairwise scatter plots along each 2D-projection of the high-dimensional data distribution.

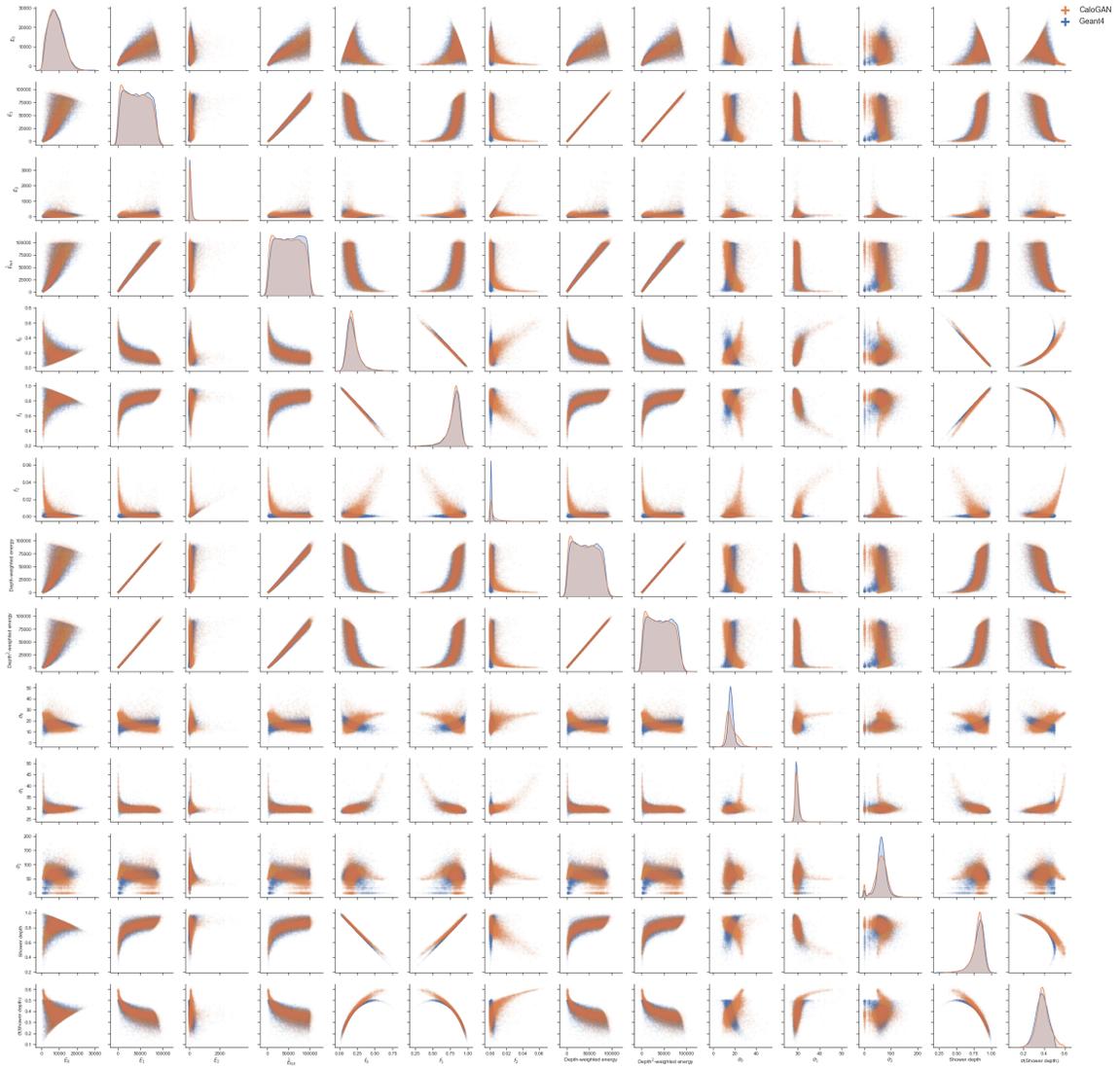

**Figure 8.49:** Pair plots of selected shower shape distributions for approximately 20,000 Geant4-generated (in blue) and 20,000 CaloGAN-generated (in orange) $e^+$ showers. The diagonal plots show the individual variable distributions using KDE. The off-diagonal plots display pairwise scatter plots along each 2D-projection of the high-dimensional data distribution.

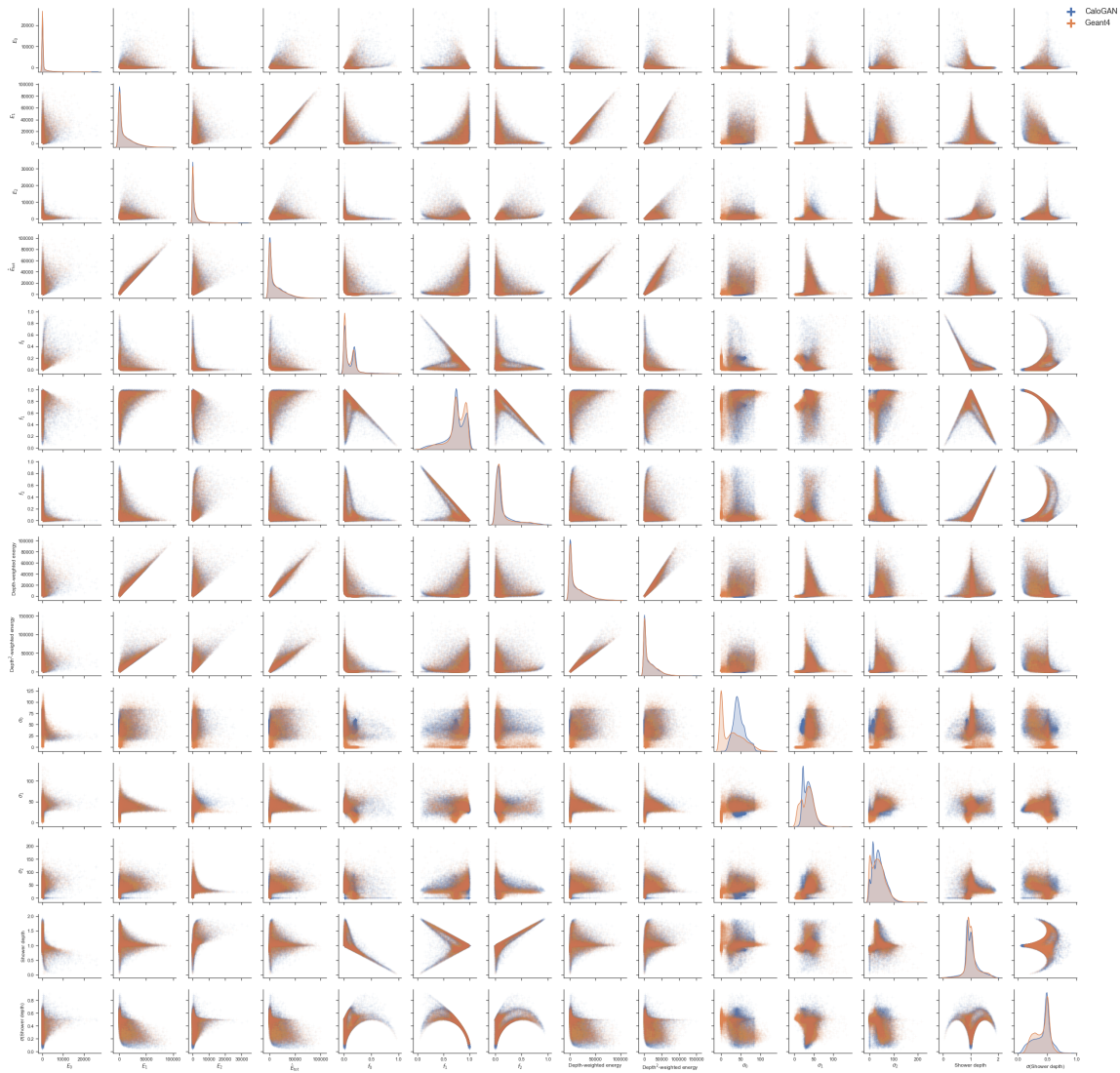

**Figure 8.50:** Pair plots of selected shower shape distributions for approximately 20,000 Geant4-generated (in blue) and 20,000 CaloGAN-generated (in orange) $\pi^+$ showers. The diagonal plots show the individual variable distributions using KDE. The off-diagonal plots display pairwise scatter plots along each 2D-projection of the high-dimensional data distribution.

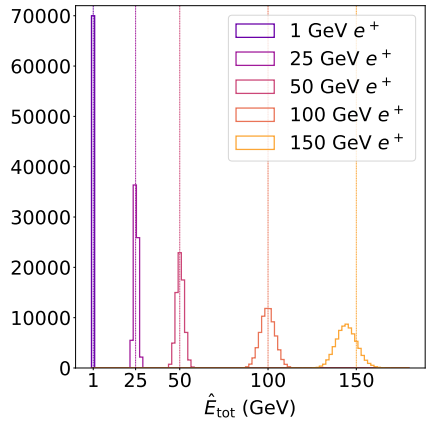

**Figure 8.51:** Post energy-conditioned empirical energy response for $e^+$ incident at 1, 25, 50, 100, and 150 GeV. Though our model is only trained on the uniform range between 1 and 100 GeV, it still admits a compelling peak at 150 GeV.

### 8.2.3.3 Classification as a GAN Performance Proxy

Although human visual perception limits the investigation of the level of agreement between the true and learned high-dimensional data distributions to checks for distributional mismatches in one or two dimensions, machine learning classifiers have the power and ability to compute complex multi-dimensional transformations of the input space, with the promise of fully examining the whole high dimensional space in which shower images live. This is, indeed, the role of the discriminator network in the adversar-ial setup.

Analogously, one way to examine the full shower space is to study the performance of classifiers in discriminating classes of showers according to the nature of the initiating particles. Specifically, transferability of classification performance from GAN-generated samples to Geant4-generated samples can be used as a proxy both for CaloGAN image quality and potential utility in a practical fast simulation setting.

In practice, ten repeated trainings of simple models (in this case, six-layer fully-connected classifiers) are performed for each $e$-$\gamma$ and $e$-$\pi$ task and for each combination of Geant4 and CaloGAN-generated showers as train/test sets, starting from the 504-dimensional input pixel space of the concatenated representation of shower energy depositions across all calorimeter layers. Table 8.4 reports the accuracies for in-domain and out-of-domain testing along with the standard deviations across the ten identical trainings, which quantify the effect of the stochasticity in the training process. The CaloGAN showers are *not* reweighted to the Geant4 distributions prior to training or testing.

The following observations and considerations can be drawn from the numerical results in Table 8.4:

- when training on Geant4, testing on the generated CaloGAN dataset yields similar results to testing on a separate Geant4 data set, which suggests that the GAN has learned most of the discriminating physics between the classes of particles. Percent-level differences in accuracy may, however, be relevant for particular applications;

- the significantly higher performance obtained on the CaloGAN-generated test set when training on a separate dataset of CaloGAN-generated images highlights a greater inter-class differentiation in the GAN synthetic dataset than originally present in the target Geant4 distribution.

This could either be due to new unphysical, class-dependent features produced by the GAN, or to the inability of the GAN to cover the entire feature space for at least one of the particle classes. It is likely that both of these contribute. To some extent, unphysical features are mitigated by the discriminator network of the GAN itself, but both physical and unphysical features that are not very useful for distinguishing real from fake samples could turn into very useful features for the two-particle classification case. Such information would therefore appear discriminative in GAN images but not in Geant4.

Using classification as a generator diagnostic is an important tool for exposing the modeling of inter-

Table 8.4: Mean and standard deviation over 10 particle classification trials using a six-layer fully-connected network with dropout. The networks are trained using a dataset from the domain specified in the first column, and tested on an independent dataset from the domain specified in the header.

$e^+$ vs. $\pi^+$

|  |  | Test on | |
|---|---|---|---|
|  |  | Geant4 | CaloGAN |
| Train on | Geant4 | 99.6% ± 0.1% | 96.5% ± 1.1% |
|  | CaloGAN | 98.2% ± 0.9% | 99.9% ± 0.2% |

$e^+$ vs. $\gamma$

|  |  | Test on | |
|---|---|---|---|
|  |  | Geant4 | CaloGAN |
| Train on | Geant4 | 66.1% ± 1.2% | 70.6% ± 2.6% |
|  | CaloGAN | 54.3% ± 0.8% | 100.0% ± 0.0% |

class shower variations. Nonetheless, while classification is a useful metric for probing the high-dimensional feature space, there are still challenges for interpreting and improving upon the outcome.

#### 8.2.3.4 Inference

In addition to the promise of high-fidelity simulation, the CaloGAN affords many orders of magnitude in computational speedups compared to traditional full simulation techniques[‖]. As a comparison, the generation times of $e^+$ with incident energy drawn uniformly between 1 GeV and 100 GeV for Geant4 and CaloGAN are benchmarked on Intel Xeon 2.6 GHz processors on nearly identical CPU compute-nodes on the PDSF distributed cluster at the National Energy Research Scientific Computing Center (NERSC), and numerical results are obtained over an average of 100 runs. CaloGAN computational performance on GPU hardware is benchmarked on an Amazon Web Service (AWS) `p2.8xlarge` instance, where a single NVIDIA K80 card is used for the purposes of benchmarking.

Table 8.5 reveals the average time required to generate a single particle shower in milliseconds. The CaloGAN framework is evaluated with different batch sizes, as different use-cases will have different demands around batching computation. Different energy values can be requested for the showers in each batch. With a batch size of 1024 on GPU, the GAN-based method admits a speedup of $\sim 100,000\times$ compared to the single-threaded Geant4 benchmark. Neither the Geant4 nor the CaloGAN exact timings should be interpreted as indicators of the best achievable performance by either framework, yet the rough order of magnitude comparison is significant and relevant.

---

[‖]Note that non-distributed training can take day(s), depending on the total number of training epochs, but it always executes in constant time regardless of the number of showers requested at generation time, so it is a fixed cost that is not relevant for the majority of the total computing budget.

**Table 8.5:** Total expected time (in milliseconds) required to generate a single shower under various algorithm-hardware combinations

| Simulator | Hardware | Batch Size | ms/shower |
|---|---|---|---|
| Geant4 | CPU | N/A | 1772 |
| CaloGAN | CPU | 1 | 13.1 |
| | | 10 | 5.11 |
| | | 128 | 2.19 |
| | | 1024 | 2.03 |
| | GPU | 1 | 14.5 |
| | | 4 | 3.68 |
| | | 128 | 0.021 |
| | | 512 | 0.014 |
| | | 1024 | 0.012 |

In addition, directly generating the deposited energy per calorimeter cell rather than simulating particle showering dynamics renders the model's time-complexity invariant to nominal energy, whereas Geant4 shower simulation runtime increases significantly with higher energy.

These preliminary results for the computational performance of GAN-based simulation models, coupled with the promising yet improvable generation quality, confirm previous results and intuitions that deep generative models may be a powerful tool worth investigating for the design of next-generation high energy physics simulation software.

#### 8.2.3.5 The Conditional CaloGAN Results

The results obtained upon evaluation of the CaloGAN version with additional conditional attributes are qualitatively similar to those described in previous sections for the base model. The task is itself only partially complicated by the presence of further attributes $\xi = (E, x_0, y_0, \theta, \phi)$ that define the motion and properties of the electromagnetically interacting primary particles, and that therefore cause changes in the aspect of the generated showers.

Given the explicit definition of a latent subspace, it is vital to verify that the model learns to translate these salient attributes into meaningful representations in pixel space. Guaranteeing the successful conditioning on these variables is key for foreseeable applications of this technology that require direct at-tribute constraint and manipulation, and explains the choice of explicitly defining key attributes within the latent space, as opposed to letting the network learn them in an unsupervised way.

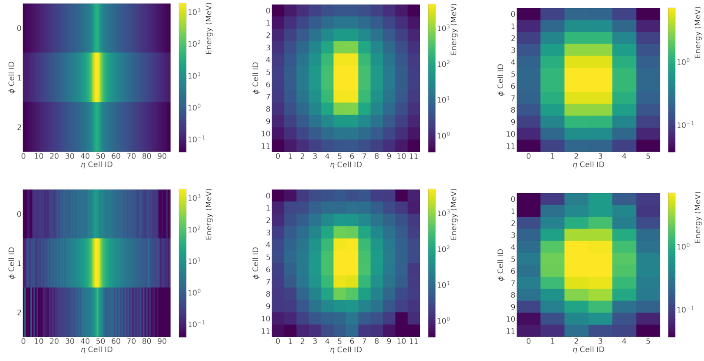

**Figure 8.52:** Average $e^+$ shower image with incoming particle position and direction variation, generated by Geant4 (top) and by the conditional CaloGAN model (bottom), proceeding, from left to right, through the layers of the segmented calorimeter.

The outcome of the conditional CaloGAN training is explored, for greater clarity, in the context of $e^+$ shower simulation. Consistent conclusions can otherwise be derived from any of the other particle types produced in this application.

As first qualitative validation, Fig. 8.52 displays the average energy deposition per calorimeter layer for $e^+$ showers obtained from Geant4 simulation and from the forward pass of the trained conditional CaloGAN. Compared to Fig. 8.39, here the spread in incoming particle position and direction are detectable by eye in the form of a larger volume over which the shower cores deposit energy across the calorimeter layers. This increased energy dispersion around the center of the detector encodes the difference between the dataset used to train the conditional CaloGAN and the original one. The qualitative comparison between the first and second row in Fig. 8.52 shows high levels of visual agreement between GAN-generated and Geant4-generated showers, which removes the concern that the model might have learned to generate showers only in one preferential position or direction.

In addition to aggregate image pattern validation, nearest GAN neighbors are retrieved for seven Geant4 images and used to validate that the conditional CaloGAN model does not simply memorize shower patterns from the reference dataset, and that the full space of displacements (both angular and positional) are explored. At an individual image level, the model produces convincing energy deposition patterns, as shown in Fig. 8.53. No sign of memorization of the training dataset is visible. In addition, positional variation (observed by noticing energy centroid deviations from the center of the calorimeter

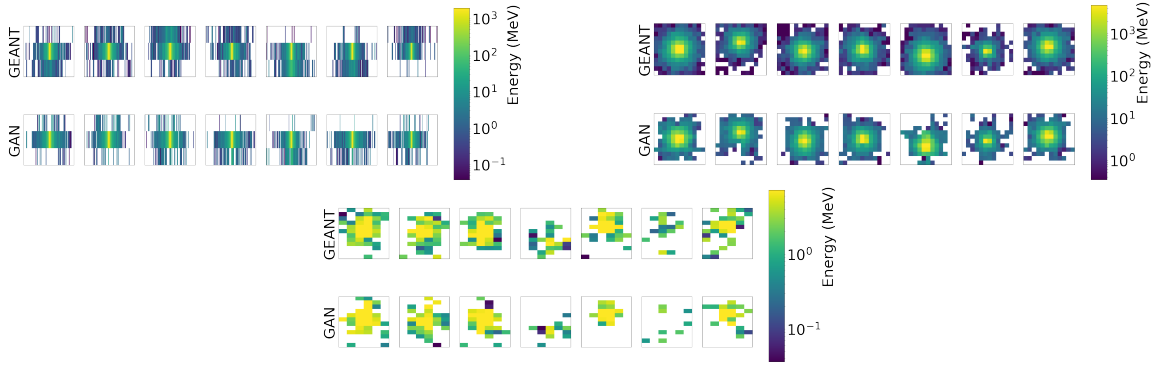

**Figure 8.53:** Nearest GAN-generated neighbors (bottom rows) for seven random Geant4-generated $e^+$ showers (top rows) for the front, middle, and last layer of the calorimeter.

image) appears to be well explored by the GAN.

To further verify the conditional CALOGAN model's ability to condition on physical attributes, the latent space for each conditioning variable is traversed, showing what the model learns about how each conditioning factor affects the outcome of the shower generation process. To illustrate the model's internal representation, incident energy, $x_0$, and $\theta$ manifolds are explored at regular intervals along the range of values provided during the training. The position and angle are defined with respect to GEANT natural coordinates. In practice, this simply consists of varying the entries in the $\xi$ vector to take on the desired values. When traversing the space along a single direction $i$, only the entry $\xi_i$ is gradually modified, while all others (including those in the random latent vector $z$) are held fixed.

Fig. 8.54 shows the manifold traversal along the incident energy direction, from low energy on the left to high energy on the right. For ease of visual interpretation, each column set of three images represents an average over ten single showers. Increases in the nominal incident energy are correctly mapped by the GAN to increases in the overall energy deposited across layers. Again, it is important to remember that while energy is varied, all other latent factors, including the batch of ten latent vectors $z$, are held constant while moving from left to right across the plot.

Similarly, the effects on shower topology of modulating other latent codes can be scrutinized using the methods described above. As a second example, the latent $x_0$ direction is traversed, and its resulting impact on generated showers is shown in Fig. 8.55. As $x_0$ increases, shower position shifts downward, which is consistent with the definition of the coordinates used in the dataset and described in Sec. 8.2.1.

Finally, Fig. 8.56 illustrates the average shower variation as a function of the angle $\theta$ of incidence, by highlighting the difference between the middle point in interpolation space and each point along the $\theta$ traversal. As $\theta$ increases, the width of the energy distribution decreases and the showers become significantly more centralized. Areas colored in blue indicate that less energy is deposited in that particular section of the image compared to the $\theta = 0$ average image. As a reminder, Eq. 8.9 relates the change in angle $\theta$ to the change in the magnitude of the momentum component in the GEANT $\hat{z}$-direction. In other words, at constant total momentum magnitude $p$, as $\theta$ increases, $p_z$ decreases as $p\cos(\theta)$, thus redistributing the energy depositions across a wider region in the $x$-$y$ plane, and losing its more collimated shower aspect.

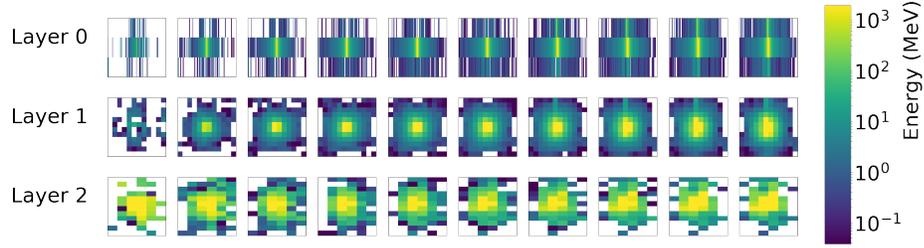

**Figure 8.54:** Interpolation across the physical range of incident energy as a conditioning latent factor for $e^+$ showers, with energy increasing from 1 GeV to 100 GeV from left to right. Each point in the interpolation is an average of 10 showers, with each point along the traversal built from an identical set of latent prior vectors $z$.

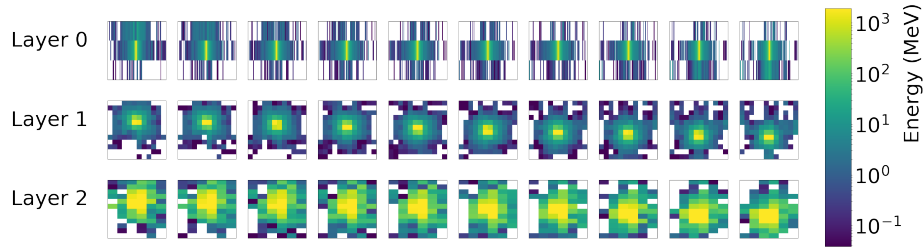

**Figure 8.55:** Interpolation across the physical range of $x_0$ as a conditioning latent factor for $e^+$ showers. Note in the $\hat{x}$-axis in the ATLAS coordinate system, represents the vertical direction in this dataset. Each point in the interpolation is an average of 10 showers, with each point along the traversal built from an identical set of latent prior vectors $z$.

Finally, in accordance with the priorities of the field, the individual one-dimensional shower shape matching ability of the conditional CALOGAN is verified by comparing the distributions obtained from GAN-generated showers from those extracted from the reference GEANT4 dataset. The histograms provided in Fig. 8.57 allow to draw substantially similar conclusions to those that apply to the base CALOGAN model in Sec. 8.2.3.2.

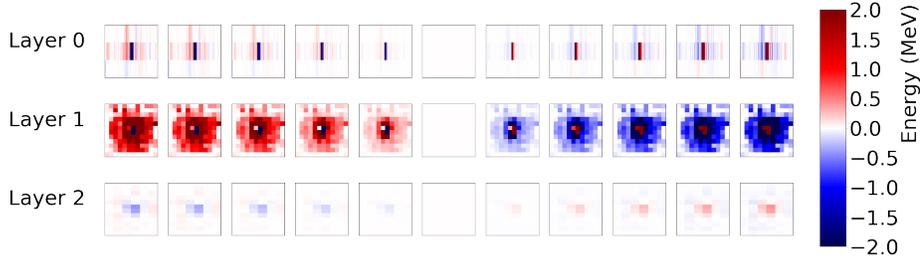

**Figure 8.56:** Interpolation across the physical range of $\theta$ as a conditioning latent factor for $e^+$ showers, with $\theta$ increasing from left to right. Each point in the interpolation is an average of 10 showers subtracted from the middle point along the interpolation path, with each point along the traversal built from an identical set latent prior vectors $z$.

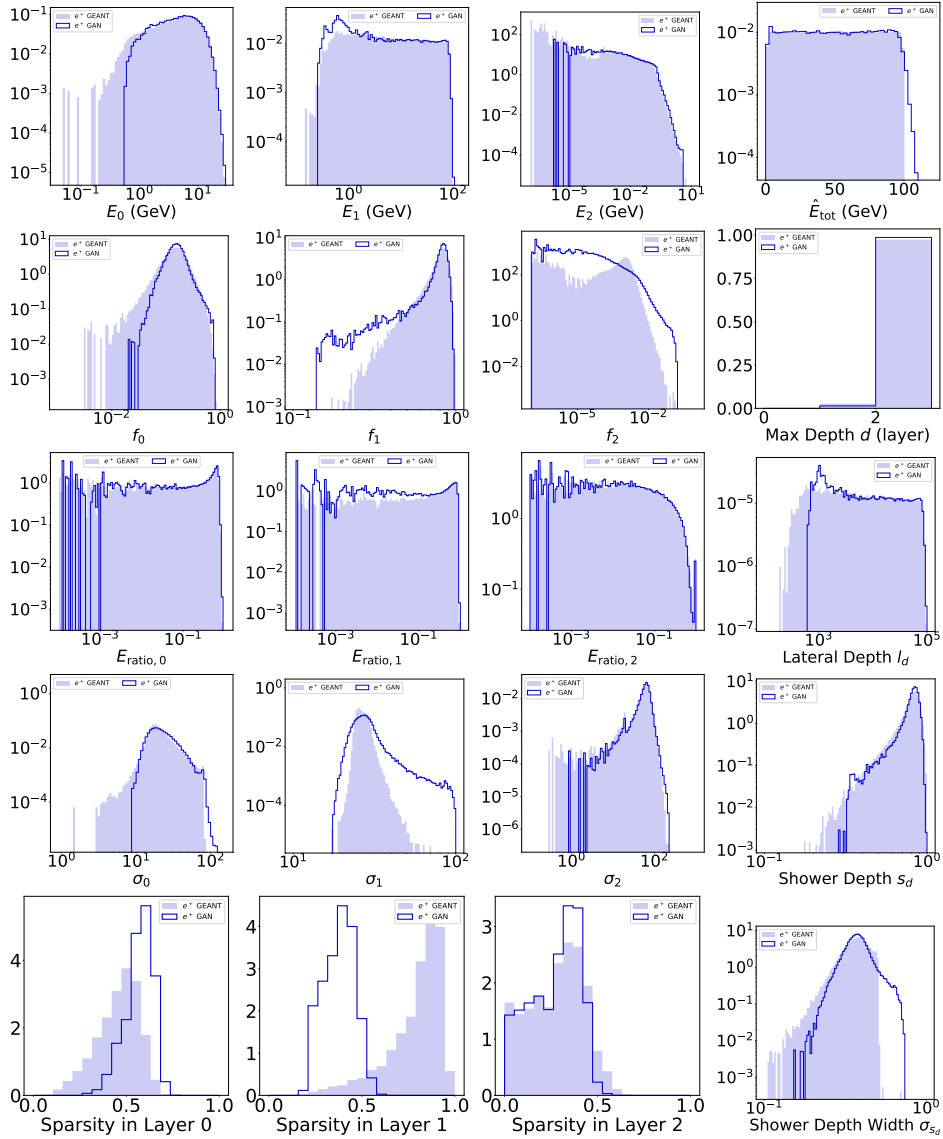

**Figure 8.57:** Comparison of shower shape variables, introduced in Table 8.3, and other variables of interest, such as the sparsity level per layer, for the $e^+$ Geant4 dataset with variable incident angle and position, and the corresponding conditional CaloGAN-generated dataset.

# 9
# Conclusion

This thesis presented a collection of research projects that attest my involvement in the effort of bridging the machine learning and high energy physics communities, with the purpose of maximizing the discovery potential of the ATLAS experiment at the LHC.

Chapter 1 provided an introduction to the Standard Model of particle physics, introducing the set of elementary particles it describes and the way in which they arise, from a mathematical standpoint, as excitations of the fields governed by the physical laws that can be expressed in the language of Quantum Field Theory.

Chapter 2 described the experimental setup of the ATLAS detector at the LHC, and explained the physical processes that drove the design of sub-detector components for the optimal measurement of the properties of different particles.

In light of the nature and focus of the contributions presented in later chapters, Ch. 3 expanded on the details of the ATLAS computing strategy and infrastructure, in particular with respect to the traditional software packages adopted for event generation and simulation.

Chapter 4 aimed to provide a functional introduction to common concepts (and misconceptions) in machine learning and statistical learning theory in an unostentatious and accessible way, for the primary benefit of present and future colleagues approaching the field of machine learning for the first time. Despite the initial grand ambition, practical considerations prevented this chapter from being anywhere near as exhaustive and encyclopedic as I once wished. Nonetheless, I do hope but do not expect that somebody will find utility in its content.

Chapter 5 discussed the many techniques and algorithms adopted by ATLAS to reconstruct primary physical entities, such as tracks and vertices, and to combine them into more abstract physics objects, such as jets, photons, etc. It further introduced a series of alternative jet representations that unlock the possibility of employing the most performant machine learning algorithms for downstream processing. It also presented a concrete example of how the use of state-of-the-art machine learning models, such as densely connected convolutional neural networks, resulted in significant increases in electron and photon identification efficiencies.

Chapter 6 outlined recent developments in the domain of jet tagging in ATLAS, with strong emphasis on machine-learning-driven solutions to which I personally contributed. The chapter primarily focused on flavor tagging applications and on the improvements provided by the development of deep learning taggers such as RNNIP and DL1.

Chapter 7 was entirely dedicated to the search for Higgs pairs at the LHC, with particular attention to the analysis carried out in Run II in the $\gamma\gamma b\bar{b}$ decay channel using 36.1 fb$^{-1}$ of data collected by the ATLAS detector in 2015 and 2016, which was marked, among others, by the introduction of a jet-selection BDT to recover untagged jets in events in which standard $b$-tagging algorithms had failed to identify the two candidate jets from the Higgs decay.

Finally, Ch. 8 related the advances achieved thanks to the exploration and utilization of deep generative models, such as generative adversarial networks, towards the fast and accurate production of simulated physics samples, in light of current and future computational demands of traditional detector simulation packages.

The common thread of this work has been the identification of inefficiencies, either from the computational or the performance standpoint, in some of the long-established algorithms adopted by the ATLAS collaboration, and their rectification through the formulation of machine learning solutions that would be attentive to the needs and desiderata of the field.

## 9.1 Future Directions

Shifts in priorities [477, 478, 479] and funding [480, 481, 482] have made it possible for more and more students to see their technical contributions on the software side be recognized and valued on similar footing as hardware projects. Focused research groups in the ATLAS collaboration have begun implementing the more exploratory solutions proposed in the work discussed in this thesis and expanding upon them for future deployment in the realistic scenarios of demanding simulation environments [483]. In addition, many more ATLAS results (as well as those from other collaborations) that make use of machine learning techniques have necessarily not found room in this thesis, which, as a personal and, therefore, biased collection of research endeavors, does not aim to be a comprehensive snapshot of the progress made by the field to this day. Furthermore, currently unpublished internal projects within the collaboration have already pushed the boundaries of knowledge and particle recognition efficiency beyond the state of the art publicly available at the time of this writing, and open-source preprints continuously update the achievements of the research community working at the intersection of high energy physics and machine learning.

In the coming years, high energy physics is likely to see the adoption of machine learning methods continue to grow on all fronts of analysis, simulation, trigger, particle reconstruction and identification. While the work presented here marks a clear evolution from the state of the field approximately five years ago, many research projects that would undeniably benefit from the capabilities of artificial intelligence remain unexplored. The development of more flexible and powerful deep learning libraries, and their advisable integration into HEP experiments' code bases, will certainly empower researchers to design and tune models that fully satisfy the rigorous requirements of the field from the theoretical, hardware, and interpretability points of view.

As is unchanged since the beginning of my graduate education, designing, developing, and deploying machine learning solutions for high energy physics experiments remains an exciting prospect for researchers who want to innovate and transform the analysis methods adopted in science. With the eager and enthusiastic contribution of the next generations of graduate students and the enlightened vision of research leads, ATLAS – and CERN, in general – have the potential to become powerhouses on the machine learning research landscape. The constraints imposed by the application domain and the pre-eminence of interpretation and generalization, coupled with the diversity of personal and professional backgrounds found at CERN, blend to create, at least in theory, an academic environment ripe for the development of compelling and innovative machine learning methods.

Homogenizing data formats, tools, and jargons is only the first necessary step to be able to source talent and solutions from adjacent fields and create cross-disciplinary collaborations for the benefit of science. A cultural shift towards openness and recognition of the validity of software engineering and computer science as full-fledged fields of research has occurred in the past years and will hopefully continue to mature beyond my time in the field.

# References

[1] L. Lönnblad, C. Peterson, and T. Rögnvaldsson, *Using neural networks to identify jets*, Nuclear Physics B **349** no. 3, (1991) 675 – 702, <http://www.sciencedirect.com/science/article/pii/055032139190392B>.

[2] Particle Data Group Collaboration, M. Tanabashi et al., *Review of Particle Physics*, Phys. Rev. B **98** (2018).

[3] CTEQ Collaboration, J. Botts, J. G. Morfin, J. F. Owens, J.-w. Qiu, W.-K. Tung, and H. Weerts, *CTEQ parton distributions and flavor dependence of sea quarks*, Phys. Lett. **B304** (1993) 159–166, <arXiv:hep-ph/9303255 [hep-ph]>.

[4] M. E. Peskin and D. V. Schroeder, *An introduction to quantum field theory*. Westview, Boulder, CO, 1995. <https://cds.cern.ch/record/257493>. Includes exercises.

[5] R. Boughezal et al., *Color singlet production at NNLO in MCFM*, Eur. Phys. J. **C77** no. 1, (2017) 7, <arXiv:1605.08011 [hep-ph]>.

[6] J. M. Campbell, R. K. Ellis, and W. T. Giele, *A Multi-Threaded Version of MCFM*, Eur. Phys. J. **C75** no. 6, (2015) 246, <arXiv:1503.06182 [physics.comp-ph]>.

[7] C. De Melis, *The CERN accelerator complex. Complexe des accélérateurs du CERN*, <https://cds.cern.ch/record/2119882>, General Photo.

[8] *LHC Guide*, tech. rep., Mar, 2017. <https://cds.cern.ch/record/2255762>.

[9] *The accelerator complex*, <http://cds.cern.ch/record/1997193>, Video.

[10] O. S. Brüning et al., *LHC Design Report*. CERN Yellow Reports: Monographs. CERN, Geneva, 2004. <https://cds.cern.ch/record/782076>.

[11] J. Pequenao, *Computer generated image of the whole ATLAS detector*, Mar, 2008.

[12] ATLAS Collaboration, M. Aaboud et al., *Study of the material of the ATLAS inner detector for Run 2 of the LHC. Study of the material of the ATLAS inner detector for Run 2 of the LHC*, JINST **12** no. CERN-EP-2017-081, (2017) P12009. 71 p, <https://cds.cern.ch/record/2273894>.

[13] K. Potamianos, *The upgraded Pixel detector and the commissioning of the Inner Detector tracking of the ATLAS experiment for Run-2 at the Large Hadron Collider*, PoS **EPS-HEP2015** (2015) 261, <arXiv:1608.07850 [physics.ins-det]>.

[14] ATLAS Collaboration, G. Aad et al., *Performance of the ATLAS muon trigger in pp collisions at $\sqrt{s} = 8$ TeV*, Eur. Phys. J. **C75** (2015) 120, arXiv:1408.3179 [hep-ex].

[15] M. Aaboud et al., *Performance of the ATLAS trigger system in 2015*, Eur. Phys. J. **C77** no. 5, (2017) 317, https://doi.org/10.1140/epjc/s10052-017-4852-3.

[16] ATLAS Collaboration,, *Trigger Menu in 2017*, Tech. Rep. ATL-DAQ-PUB-2018-002, CERN, Geneva, Jun, 2018. http://cds.cern.ch/record/2625986.

[17] T. Gleisberg, S. Höche, F. Krauss, M. Schönherr, S. Schumann, F. Siegert, and J. Winter, *Event generation with SHERPA 1.1*, JHEP **2** (2009) 007, arXiv:0811.4622 [hep-ph].

[18] N. S. Keskar, D. Mudigere, J. Nocedal, M. Smelyanskiy, and P. T. P. Tang, *On Large-Batch Training for Deep Learning: Generalization Gap and Sharp Minima*, arXiv:1609.04836.

[19] K. McGuinness, *Deep Learning for Computer Vision: Generative models and adversarial training*, 2016. https://bit.ly/2HbhszQ.

[20] I. Goodfellow, *NIPS 2016 Tutorial: Generative Adversarial Networks*, arXiv:1701.00160 [cs.LG].

[21] A. Radford, L. Metz, and S. Chintala, *Unsupervised Representation Learning with Deep Convolutional Generative Adversarial Networks*, arXiv:1511.06434 [cs.LG].

[22] ATLAS Collaboration,, *ATLAS event at 13 TeV - First stable beam, 3 June 2015 - run: 266904, evt: 25884805*, General Photo, Jun, 2015.

[23] *Early Inner Detector Tracking Performance in the 2015 data at $\sqrt{s} = 13$ TeV*, Tech. Rep. ATL-PHYS-PUB-2015-051, CERN, Geneva, Dec, 2015. https://cds.cern.ch/record/2110140.

[24] *Vertex Reconstruction Performance of the ATLAS Detector at $\sqrt{s} = 13$ TeV*, Tech. Rep. ATL-PHYS-PUB-2015-026, CERN, Geneva, Jul, 2015. http://cds.cern.ch/record/2037717.

[25] ATLAS Collaboration, G. Aad et al., *Topological cell clustering in the ATLAS calorimeters and its performance in LHC Run 1*, Eur. Phys. J. **C77** (2017) 490, arXiv:1603.02934 [hep-ex].

[26] G. P. Salam and G. Soyez, *A Practical Seedless Infrared-Safe Cone jet algorithm*, JHEP **05** (2007) 086, arXiv:0704.0292 [hep-ph].

[27] R. Atkin, *Review of jet reconstruction algorithms*, Journal of Physics: Conference Series **645** no. 1, (2015) 012008, http://stacks.iop.org/1742-6596/645/i=1/a=012008.

[28] T. Cheng, *Recursive Neural Networks in Quark/Gluon Tagging*, Comput Softw Big Sci **2** no. 1, (2018) 3.

[29] B. Nachman, L. de Oliveira, and M. Paganini, *Pythia Generated Jet Images for Location Aware Generative Adversarial Network Training*, Feb, 2017. Data set, DOI: 10.17632/4r4v785rgx.1.

[30] L. de Oliveira, M. Kagan, L. Mackey, B. Nachman, and A. Schwartzman, *Jet-images — deep learning edition*, JHEP **07** (2016) 069, arXiv:1511.05190 [hep-ph].

[31] L. de Oliveira, M. Paganini, and B. Nachman, *Learning Particle Physics by Example: Location-Aware Generative Adversarial Networks for Physics Synthesis*, Comput. Softw. Big Sci. **1** no. 1, (2017) 4, arXiv:1701.05927 [stat.ML].

[32] ATLAS Collaboration,, *Electron efficiency measurements with the ATLAS detector using the 2015 LHC proton-proton collision data*, Tech. Rep. ATLAS-CONF-2016-024, CERN, Geneva, Jun, 2016. http://cds.cern.ch/record/2157687.

[33] L. de Oliveira, B. Nachman, and M. Paganini, *Electromagnetic Showers Beyond Shower Shapes*, arXiv:1806.05667 [hep-ex].

[34] ATLAS Collaboration,, *Optimisation of the ATLAS b-tagging performance for the 2016 LHC Run*, Tech. Rep. ATL-PHYS-PUB-2016-012, CERN, Geneva, Jun, 2016. http://cds.cern.ch/record/2160731.

[35] ATLAS Collaboration, S. Heer, *The secondary vertex finding algorithm with the ATLAS detector*, Tech. Rep. ATL-PHYS-PROC-2017-195, CERN, Geneva, Oct, 2017. https://cds.cern.ch/record/2287604.

[36] ATLAS Collaboration,, *Identification of Jets Containing b-Hadrons with Recurrent Neural Networks at the ATLAS Experiment*, Tech. Rep. ATL-PHYS-PUB-2017-003, CERN, Geneva, Mar, 2017. https://cds.cern.ch/record/2255226.

[37] ATLAS Collaboration,, *Optimisation and performance studies of the ATLAS b-tagging algorithms for the 2017-18 LHC run*, Tech. Rep. ATL-PHYS-PUB-2017-013, CERN, Geneva, Jul, 2017. https://cds.cern.ch/record/2273281.

[38] ATLAS Collaboration,, *Identification of Hadronically-Decaying W Bosons and Top Quarks Using High-Level Features as Input to Boosted Decision Trees and Deep Neural Networks in ATLAS at $\sqrt{s}$ = 13 TeV*, Tech. Rep. ATL-PHYS-PUB-2017-004, CERN, Geneva, Apr, 2017. https://cds.cern.ch/record/2259646.

[39] ATLAS Collaboration,, *Performance of mass-decorrelated jet substructure observables for hadronic two-body decay tagging in ATLAS*, Tech. Rep. ATL-PHYS-PUB-2018-014, CERN, Geneva, Jul, 2018. http://cds.cern.ch/record/2630973.

[40] ATLAS Collaboration,, *Quark versus Gluon Jet Tagging Using Jet Images with the ATLAS Detector*, Tech. Rep. ATL-PHYS-PUB-2017-017, CERN, Geneva, Jul, 2017. https://cds.cern.ch/record/2275641.

[41] R. G. Reed and B. Mellado, *The search for new physics in the diphoton decay channel and the upgrade of the Tile-Calorimeter electronics of the ATLAS detector.*, Feb, 2017. https://cds.cern.ch/record/2271844. Presented 24 Apr 2017.

[42] R. Harlander, M. Mühlleitner, J. Rathsman, M. Spira, and O. Stål, *Interim recommendations for the evaluation of Higgs production cross sections and branching ratios at the LHC in the Two-Higgs-Doublet Model*, arXiv:1312.5571 [hep-ph].

[43] G. Aad et al., *Searches for Higgs boson pair production in the $hh \to bb\tau\tau$, $\gamma\gamma WW^*$, $\gamma\gamma bb$, $bbbb$ channels with the ATLAS detector*, Phys. Rev. **D92** no. 9, (2015) 092004, arXiv:1509.04670 [hep-ex].

[44] ATLAS, CMS and Higgs collaborations Collaboration, L. Cadamuro, *Higgs properties and decays, searches for high mass Higgs boson and di-Higgs production*, Tech. Rep. CMS-CR-2018-061, CERN, Geneva, Jun, 2018. https://cds.cern.ch/record/2627473.

[45] CMS Collaboration, D. Majumder, *Searches for pair-production of Higgs bosons using the CMS detector at the LHC*, Slides presented at the XXXIX international conference on high energy physics, seoul, south korea, 2018. URL: https://indico.cern.ch/event/686555/contributions/2971066/attachments/1682141/2702869/DMajumder_ICHEP2018.pdf. Last visited on 2018/07/13.

[46] M. Paganini, L. de Oliveira, and B. Nachman, *Electromagnetic Calorimeter Shower Images with Variable Incidence Angle and Position*, Aug, 2017. Data set, DOI: 10.17632/5fnxs6b557.2.

[47] J. Friedman, T. Hastie, and R. Tibshirani, *The elements of statistical learning*, vol. 1. Springer series in statistics New York, NY, USA:, 2001.

[48] I. Goodfellow, Y. Bengio, and A. Courville, *Deep Learning*. MIT Press, 2016. http://www.deeplearningbook.org.

[49] J. Hauptman, *Particle physics experiments at high energy colliders.* 2011. http://eu.wiley.com/WileyCDA/WileyTitle/productCd-3527408258.html.

[50] J. S. Schwinger, *A Theory of the Fundamental Interactions*, Annals Phys. **2** (1957) 407–434.

[51] E. C. G. Sudarshan and R. E. Marshak, *THE NATURE OF THE FOUR-FERMION INTERACTION*, Current Science **63** no. 2, (1992) 65–75, http://www.jstor.org/stable/24095422.

[52] C.-N. Yang and R. L. Mills, *Conservation of Isotopic Spin and Isotopic Gauge Invariance*, Phys. Rev. **96** (1954) 191–195, [,150(1954)].

[53] S. L. Glashow, *Partial-symmetries of weak interactions*, Nuclear Physics **22** no. 4, (1961) 579 – 588, http://www.sciencedirect.com/science/article/pii/0029558261904692.

[54] I. van Vulpen and I. Angelozzi, *The Standard Model Higgs Boson, Part of the Lecture Particle Physics II, UvA Particle Physics Master 2013-2014*, 2018. URL: https://www.nikhef.nl/~ivov/HiggsLectureNote.pdf. Last visited on 2018/06/05.

[55] F. Englert and R. Brout, *Broken Symmetry and the Mass of Gauge Vector Mesons*, Phys. Rev. Lett. **13** (1964) 321–323, https://link.aps.org/doi/10.1103/PhysRevLett.13.321.

[56] P. W. Higgs, *Broken Symmetries and the Masses of Gauge Bosons*, Phys. Rev. Lett. **13** (1964) 508–509, https://link.aps.org/doi/10.1103/PhysRevLett.13.508.

[57] G. S. Guralnik, C. R. Hagen, and T. W. B. Kibble, *Global Conservation Laws and Massless Particles*, Phys. Rev. Lett. **13** (1964) 585–587, https://link.aps.org/doi/10.1103/PhysRevLett.13.585.

[58] M. J. D. Hamilton, *The Higgs boson for mathematicians. Lecture notes on gauge theory and symmetry breaking*, arXiv:1512.02632 [math.DG].

[59] ATLAS Collaboration, G. Aad et al., *Observation of a new particle in the search for the Standard Model Higgs boson with the ATLAS detector at the LHC*, Phys. Lett. **B716** (2012) 1–29, arXiv:1207.7214 [hep-ex].

[60] CMS Collaboration, S. Chatrchyan et al., *Observation of a new boson at a mass of 125 GeV with the CMS experiment at the LHC*, Phys. Lett. **B716** (2012) 30–61, arXiv:1207.7235 [hep-ex].

[61] P. Soding, B. Wiik, G. Wolf, and S. L. Wu, *The First evidence for three jet events in e+ e- collisions at PETRA: First direct observation of the gluon*, pp. , 3–14. 1996.

[62] J. Ellis, M. K. Gaillard, and G. G. Ross, *Search for gluons in e+e− annihilation*, Nuclear Physics B **111** no. 2, (1976) 253 – 271, http://www.sciencedirect.com/science/article/pii/0550321376905423.

[63] S. Bethke, *Determination of the QCD coupling $\alpha_s$*, J. Phys. **G26** (2000) R27, arXiv:hep-ex/0004021 [hep-ex].

[64] J. Xiangdong, *Physics 741 Lecture Notes: Introduction to QCD and the Standard Model*, 2018. URL: https://www.physics.umd.edu/courses/Phys741/xji/chapter1.pdf. Last visited on 2018/04/29.

[65] G. Altarelli, *QCD evolution equations for parton densities*, Scholarpedia **4** no. 1, (2009) 7124, revision #91681.

[66] D. de Florian et al., *Handbook of LHC Higgs Cross Sections: 4. Deciphering the Nature of the Higgs Sector*, arXiv:1610.07922 [hep-ph].

[67] J. Baglio et al., *The measurement of the Higgs self-coupling at the LHC: theoretical status*, JHEP **2013** no. 4, (2013) 1–40, http://dx.doi.org/10.1007/JHEP04(2013)151.

[68] S. Dawson, S. Dittmaier, and M. Spira, *Neutral Higgs-boson pair production at hadron colliders: QCD corrections*, Phys. Rev. **D58** no. 11, (1998) 115012, hep-ph/9805244.

[69] G. M. et al., *Higgs boson pair production at NNLO with top quark mass effects*, JHEP **5** (2018) 59, arXiv:1803.02463 [hep-ph].

[70] M. J. Dolan, C. Englert, and M. Spannowsky, *New physics in LHC Higgs boson pair production*, Phys. Rev. D **87** (2013) 055002, http://link.aps.org/doi/10.1103/PhysRevD.87.055002.

[71] C. Englert et al., *Precision Measurements of Higgs Couplings: Implications for New Physics Scales*, J. Phys. **G41** (2014) 113001, arXiv:1403.7191 [hep-ph].

[72] R. Contino et al., *Anomalous couplings in double Higgs production*, JHEP **2012** no. 8, (2012) 1–21, http://dx.doi.org/10.1007/JHEP08(2012)154.

[73] I. F. Ginzburg, M. Krawczyk, and P. Osland, *Resolving SM like scenarios via Higgs boson production at a photon collider. 1. 2HDM versus SM*, arXiv:hep-ph/0101208 [hep-ph].

[74] I. F. Ginzburg, M. Krawczyk, and P. Osland, *Potential of photon collider in resolving SM like scenarios*, Nucl. Instrum. Meth. **A472** (2001) 149–154, arXiv:hep-ph/0101229 [hep-ph].

[75] I. F. Ginzburg, M. Krawczyk, and P. Osland, *Two Higgs doublet models with CP violation*, pp. , 703–706. 2002. arXiv:hep-ph/0211371 [hep-ph].

[76] I. F. Ginzburg and M. Krawczyk, *Symmetries of two Higgs doublet model and CP violation*, Phys. Rev. **D72** (2005) 115013, arXiv:hep-ph/0408011 [hep-ph].

[77] N. Craig, J. Galloway, and S. Thomas, *Searching for Signs of the Second Higgs Doublet*, arXiv:1305.2424 [hep-ph].

[78] A. Djouadi, *The Higgs sector of supersymmetric theories and the implications for high-energy colliders*, Eur. Phys. J. **C59** (2009) 389–426, arXiv:0810.2439 [hep-ph].

[79] I. J. Aitchison, *Supersymmetry and the MSSM: An Elementary Introduction*, arXiv:hep-ph/0505105 [hep-ph].

[80] M. Rauch, *Determination of Higgs-boson couplings (SFitter)*, arXiv:1203.6826 [hep-ph].

[81] L. Randall and R. Sundrum, *A Large mass hierarchy from a small extra dimension*, Phys. Rev. Lett. **83** (1999) 3370–3373, arXiv:hep-ph/9905221 [hep-ph].

[82] Y. Tang, *Implications of LHC searches for massive graviton*, JHEP **2012** no. 8, (2012) 1–15, http://dx.doi.org/10.1007/JHEP08(2012)078.

[83] G. F. Giudice, C. Grojean, A. Pomarol, and R. Rattazzi, *The strongly-interacting light Higgs*, JHEP **6** (2007) 045, `hep-ph/0703164`.

[84] G. D. Kribs and A. Martin, *Enhanced di-Higgs production through light colored scalars*, Phys. Rev. D **86** no. 9, (2012) 095023, `arXiv:1207.4496 [hep-ph]`.

[85] E. Lopienska, *Distribution of all CERN Users by Nationality on 24 January 2018*, https://cds.cern.ch/record/2302064, General Photo.

[86] A. W. Chao and M. Tigner, eds., *Handbook of accelerator physics and engineering*. 1999.

[87] *LEP design report*. CERN, Geneva, 1984. http://cds.cern.ch/record/102083.

[88] *Radiofrequency cavities*, http://cds.cern.ch/record/1997424.

[89] *HL-LHC Preliminary Design Report: Deliverable: D1.5*, https://cds.cern.ch/record/1972604.

[90] L. Evans and P. Bryant, *LHC Machine*, JINST **3** no. 08, (2008) S08001, http://stacks.iop.org/1748-0221/3/i=08/a=S08001.

[91] ATLAS Collaboration, *The ATLAS Experiment at the CERN Large Hadron Collider*, JINST **3** no. 08, (2008) S08003, http://stacks.iop.org/1748-0221/3/i=08/a=S08003.

[92] CMS Collaboration, S. Chatrchyan et al., *The CMS Experiment at the CERN LHC*, JINST **3** (2008) S08004.

[93] LHCb Collaboration, A. A. Alves, Jr. et al., *The LHCb Detector at the LHC*, JINST **3** (2008) S08005.

[94] ALICE Collaboration, K. Aamodt et al., *The ALICE experiment at the CERN LHC*, JINST **3** (2008) S08002.

[95] LHCf Collaboration, O. Adriani et al., *The LHCf detector at the CERN Large Hadron Collider*, JINST **3** (2008) S08006.

[96] TOTEM Collaboration, G. Anelli et al., *The TOTEM experiment at the CERN Large Hadron Collider*, JINST **3** (2008) S08007.

[97] MoEDAL Collaboration, J. Pinfold et al., *Technical Design Report of the MoEDAL Experiment*,.

[98] ATLAS Collaboration, M. Aaboud et al., *Luminosity determination in pp collisions at $\sqrt{s} = 8$ TeV using the ATLAS detector at the LHC*, Eur. Phys. J. **C76** no. 12, (2016) 653, `arXiv:1608.03953 [hep-ex]`.

[99] ATLAS Collaboration, G. Aad et al., *Improved luminosity determination in pp collisions at sqrt(s) = 7 TeV using the ATLAS detector at the LHC*, Eur. Phys. J. **C73** no. 8, (2013) 2518, `arXiv:1302.4393 [hep-ex]`.

[100] *LHC Page 1*, 2018. URL: https://op-webtools.web.cern.ch/vistar/vistars.php. Last visited on 2018/06/17.

[101] ATLAS Collaboration, M. Aaboud et al., *Measurement of the Inelastic Proton-Proton Cross Section at $\sqrt{s} = 13$ TeV with the ATLAS Detector at the LHC*, Phys. Rev. Lett. **117** no. 18, (2016) 182002, arXiv:1606.02625 [hep-ex].

[102] ATLAS Collaboration, *ATLAS: technical proposal for a general-purpose pp experiment at the Large Hadron Collider at CERN*. LHC Tech. Proposal. CERN, Geneva, 1994. https://cds.cern.ch/record/290968.

[103] H. H. J. ten Kate, *ATLAS superconducting toroids and solenoid*, IEEE Trans. Appl. Supercond. **15** no. 2 pt.2, (2005) 1267–1270, http://cds.cern.ch/record/912244.

[104] ATLAS Collaboration, M. Capeans et al., *ATLAS Insertable B-Layer Technical Design Report*, Tech. Rep. CERN-LHCC-2010-013. ATLAS-TDR-19, Sep, 2010. https://cds.cern.ch/record/1291633.

[105] ATLAS Collaboration, *Performance of the ATLAS Track Reconstruction Algorithms in Dense Environments in LHC Run 2*, arXiv:1704.07983 [hep-ex].

[106] A. Vogel, *ATLAS Transition Radiation Tracker (TRT): Straw Tube Gaseous Detectors at High Rates*, Tech. Rep. ATL-INDET-PROC-2013-005, CERN, Geneva, Apr, 2013. https://cds.cern.ch/record/1537991.

[107] R. Wigmans, *Calorimetry: Energy Measurement in Particle Physics*. International Series of Monographs on Physics. OUP Oxford, 2017. https://books.google.com/books?id=vJc4DwAAQBAJ.

[108] B. Aubert et al., *lR&D proposal: liquid argon calorimetry with LHC-performance specifications*, Tech. Rep. CERN-DRDC-90-31. DRDC-P-5, CERN, Geneva, 1990. https://cds.cern.ch/record/292608.

[109] ATLAS Collaboration, A. M. Henriques Correia, *The ATLAS Tile Calorimeter*, Tech. Rep. ATL-TILECAL-PROC-2015-002, CERN, Geneva, Mar, 2015. https://cds.cern.ch/record/2004868.

[110] P. Puzo, *ATLAS calorimetry*, NIMA **494** no. 1, (2002) 340 – 345, http://www.sciencedirect.com/science/article/pii/S0168900202014900, Proceedings of the 8th International Conference on Instrumentatio n for Colliding Beam Physics.

[111] A. Hrynevich, *Performance of the ATLAS Tile Calorimeter*, JINST **12** no. 06, (2017) C06021, http://stacks.iop.org/1748-0221/12/i=06/a=C06021.

[112] C. W. Fabjan and F. Gianotti, *Calorimetry for particle physics*, Rev. Mod. Phys. **75** (2003) 1243–1286, https://link.aps.org/doi/10.1103/RevModPhys.75.1243.

[113] Y. Zolnierowski, *The ATLAS liquid-argon electromagnetic calorimeter*, NIMA **384** no. 1, (1996) 230 – 236, http://www.sciencedirect.com/science/article/pii/S0168900296009497, BEAUTY '96.

[114] ATLAS Collaboration,, *ATLAS muon spectrometer: Technical Design Report*. Technical Design Report ATLAS. CERN, Geneva, 1997. https://cds.cern.ch/record/331068.

[115] G. Cattani and the RPC group, *The Resistive Plate Chambers of the ATLAS experiment: performance studies*, Journal of Physics: Conference Series **280** no. 1, (2011) 012001, http://stacks.iop.org/1742-6596/280/i=1/a=012001.

[116] E. Etzion et al., *The Certification of ATLAS Thin Gap Chambers Produced in Israel and China*, arXiv:physics/0411136.

[117] ATLAS Collaboration, A. Airapetian et al., *ATLAS detector and physics performance: Technical Design Report, 1*. Technical Design Report ATLAS. CERN, Geneva, 1999. https://cds.cern.ch/record/391176.

[118] F. Bauer et al., *Construction and Test of the Precision Drift Chambers for the ATLAS Muon Spectrometer*, IEEE Trans. Nucl. Sci. **48** (2001) 302–307, arXiv:1604.02259 [physics.ins-det].

[119] ATLAS Outreach, *ATLAS Fact Sheet: To raise awareness of the ATLAS detector and collaboration on the LHC*, 2010.

[120] A. Ruiz-Martinez and A. Collaboration, *The Run-2 ATLAS Trigger System*, Tech. Rep. ATL-DAQ-PROC-2016-003, CERN, Geneva, Feb, 2016. https://cds.cern.ch/record/2133909.

[121] M. Shochet et al., *Fast TracKer (FTK) Technical Design Report*, Tech. Rep. CERN-LHCC-2013-007. ATLAS-TDR-021, Jun, 2013. https://cds.cern.ch/record/1552953. ATLAS Fast Tracker Technical Design Report.

[122] ATLAS Collaboration, T. Iizawa, *The ATLAS Fast Tracker system*, Tech. Rep. ATL-DAQ-PROC-2017-036, CERN, Geneva, Oct, 2017. https://cds.cern.ch/record/2289580.

[123] ATLAS Collaboration, R. M. Bianchi et al., *Event visualization in ATLAS*, Journal of Physics: Conference Series **898** no. 7, (2017) 072014, http://stacks.iop.org/1742-6596/898/i=7/a=072014.

[124] J. Sánchez, A. F. Casaní, and S. G. de la Hoz, *Distributed Data Collection for the ATLAS EventIndex*, Journal of Physics: Conference Series **664** no. 4, (2015) 042046, http://stacks.iop.org/1742-6596/664/i=4/a=042046.

[125] J. Catmore, *The ATLAS data processing chain: from collisions to papers*, 2016. https://indico.cern.ch/event/472469/contributions/1982677/attachments/1220934/1785823/intro_slides.pdf.

[126] M. Gaillard and S. Pandolfi, *CERN Data Centre passes the 200-petabyte milestone*, http://cds.cern.ch/record/2276551.

[127] A. Dorigo, P. Elmer, F. Furano, and A. Hanushevsky, *XROOTD - A highly scalable architecture for data access*, http://citeseerx.ist.psu.edu/viewdoc/download?doi=10.1.1.127.9281&rep=rep1&type=pdf.

[128] L. Chen, E. A. Rundensteiner, and S. Wang, *XCache: a semantic caching system for XML queries*, pp. , 618–618, ACM. 2002.

[129] ATLAS Collaboration, V. Garonne et al., *Rucio – The next generation of large scale distributed system for ATLAS Data Management*, Journal of Physics: Conference Series **513** no. 4, (2014) 042021, http://stacks.iop.org/1742-6596/513/i=4/a=042021.

[130] G. Aad et al., *The ATLAS Simulation Infrastructure*, Eur. Phys. J. **70** no. 3, (2010) 823–874, https://doi.org/10.1140/epjc/s10052-010-1429-9.

[131] M. R. Whalley, D. Bourilkov, and R. C. Group, *The Les Houches accord PDFs (LHAPDF) and LHAGLUE*, pp. , 575–581. 2005. arXiv:hep-ph/0508110 [hep-ph].

[132] A. Buckley et al., *LHAPDF6: parton density access in the LHC precision era*, Eur. Phys. J. **C75** (2015) 132, arXiv:1412.7420 [hep-ph].

[133] S. Dulat et al., *New parton distribution functions from a global analysis of quantum chromodynamics*, Phys. Rev. **D93** no. 3, (2016) 033006, arXiv:1506.07443 [hep-ph].

[134] R. D. Ball et al., *Parton distributions for the LHC run II*, JHEP **4** (2015) 40, arXiv:1410.8849 [hep-ph].

[135] J. R. Andersen et al., *Les Houches 2017: Physics at TeV Colliders Standard Model Working Group Report*, in *10th Les Houches Workshop on Physics at TeV Colliders (PhysTeV 2017) Les Houches, France, June 5-23, 2017*. 2018. arXiv:1803.07977 [hep-ph].

[136] A. Buckley et al., *General-purpose event generators for LHC physics*, Phys. Rept. **504** (2011) 145–233, arXiv:1101.2599 [hep-ph].

[137] J. Alwall et al., *The automated computation of tree-level and next-to-leading order differential cross sections, and their matching to parton shower simulations*, JHEP **7** (2014) 79, arXiv:1405.0301 [hep-ph].

[138] B. Andersson, G. Gustafson, G. Ingelman, and T. Sjostrand, *Parton Fragmentation and String Dynamics*, Phys. Rept. **97** (1983) 31–145.

[139] I. Bird et al., *Update of the Computing Models of the WLCG and the LHC Experiments,*.

[140] T. Sjöstrand, S. Ask, J. R. Christiansen, R. Corke, N. Desai, P. Ilten, S. Mrenna, S. Prestel, C. O. Rasmussen, and P. Z. Skands, *An introduction to PYTHIA 8.2*, Computer Physics Communications **191** (2015) 159 – 177, http://www.sciencedirect.com/science/article/pii/S0010465515000442.

[141] W. Lukas, *Fast Simulation for ATLAS: Atlfast-II and ISF*, Tech. Rep. ATL-SOFT-PROC-2012-065, CERN, Geneva, Jun, 2012. https://cds.cern.ch/record/1458503.

[142] A. E. Kiryunin et al., *GEANT4 physics evaluation with testbeam data of the ATLAS hadronic end-cap calorimeter*, Journal of Physics: Conference Series **160** no. 1, (2009) 012075, http://stacks.iop.org/1742-6596/160/i=1/a=012075.

[143] GEANT4 Collaboration, S. Agostinelli et al., *GEANT4: A Simulation toolkit*, Nucl. Instrum. Meth. **A506** (2003) 250–303.

[144] A. Dotti, *On Static Vs Dynamic*, https://www.dropbox.com/s/uapsf9rplruff0u/Static%20Vs%20Shared.pdf?dl=0.

[145] A. Dotti, *Geant4 Optimization Opportunities*, 2017. https://indico.cern.ch/event/621867/contributions/2511638/attachments/1427281/2193227/ATLAS_SCWeek_G4_Optimizations.pdf.

[146] J. D. Chapman, *Approaches to speed up Geant4 Simulation in ATLAS*, Tech. Rep. ATL-COM-SOFT-2018-005, CERN, Geneva, Mar, 2018. https://cds.cern.ch/record/2310377.

[147] ATLAS Collaboration, M. Beckingham et al., *The simulation principle and performance of the ATLAS fast calorimeter simulation FastCaloSim*, tech. rep., CERN.

[148] ATLAS Collaboration,, *The new Fast Calorimeter Simulation in ATLAS*, Tech. Rep. ATL-SOFT-PUB-2018-002, CERN, Geneva, Jul, 2018. https://cds.cern.ch/record/2630434.

[149] ATLAS Collaboration, A. Hasib, *The new ATLAS Fast Calorimeter Simulation*, https://cds.cern.ch/record/2273474.

[150] ATLAS Collaboration, P. Jacka, *The new ATLAS Fast Calorimeter Simulation*, https://cds.cern.ch/record/2622459.

[151] ATLAS Collaboration, K. Gasnikova and A. Glazov, *Frozen-shower simulation of electromagnetic showers in the ATLAS forward calorimeter*, https://cds.cern.ch/record/2220762.

[152] *Performance of the Fast ATLAS Tracking Simulation (FATRAS) and the ATLAS Fast Calorimeter Simulation (FastCaloSim) with single particles*, Tech. Rep. ATL-SOFT-PUB-2014-001, CERN, Geneva, Mar, 2014. https://cds.cern.ch/record/1669341.

[153] S. Hamilton et al., *The ATLAS Fast Track Simulation Project (FATRAS)*, https://cds.cern.ch/record/1302995.

[154] ATLAS Collaboration, A. Basalaev and Z. Marshall, *The Fast Simulation Chain for ATLAS*, Tech. Rep. ATL-SOFT-PROC-2017-022. 4, CERN, Geneva, Jan, 2017. https://cds.cern.ch/record/2243045.

[155] *ATLAS Athena Guide*, 2018. URL: https://atlassoftwaredocs.web.cern.ch/athena/. Last visited on 2018/07/03.

[156] *CLHEP - A Class Library for High Energy Physics*, 2018. URL: http://cern.ch/clhep. Last visited on 2018/07/02.

[157] *Eigen - a C++ template library for linear algebra: matrices, vectors, numerical solvers, and related algorithms.*, 2018. URL: http://eigen.tuxfamily.org/. Last visited on 2018/07/02.

[158] ATLAS Collaboration, N. Styles et al., *Developments in the ATLAS Tracking Software ahead of LHC Run 2*, Journal of Physics: Conference Series **608** no. 1, (2015) 012047, http://stacks.iop.org/1742-6596/608/i=1/a=012047.

[159] ATLAS Collaboration, C. Leggett et al., *AthenaMT: Upgrading the ATLAS Software Framework for the Many-Core World with Multi-Threading*, https://cds.cern.ch/record/2222298.

[160] C. Leggett et al., *AthenaMT: upgrading the ATLAS software framework for the many-core world with multi-threading,*.

[161] G. Amadio et al., *The GeantV project: preparing the future of simulation*, Journal of Physics: Conference Series **664** no. 7, (2015) 072006, http://stacks.iop.org/1742-6596/664/i=7/a=072006.

[162] ATLAS Collaboration, G. Duckeck, D. Barberis, R. Hawkings, R. Jones, N. McCubbin, G. Poulard, D. Quarrie, T. Wenaus, and E. Obreshkov, *ATLAS computing: Technical design report,*.

[163] *WLCG Structure*, 2018. URL: http://wlcg-public.web.cern.ch/structure. Last visited on 2018/06/22.

[164] ATLAS Collaboration, F. H. Barreiro Megino, *The Future of Distributed Computing Systems in ATLAS: Boldly Venturing Beyond Grids*, http://cds.cern.ch/record/2627546.

[165] ATLAS Collaboration, J. Chudoba, *Exploitation of heterogeneous resources for ATLAS Computing*, http://cds.cern.ch/record/2624252.

[166] A. P. Dempster, N. M. Laird, and D. B. Rubin, *Maximum Likelihood from Incomplete Data via the EM Algorithm*, Journal of the Royal Statistical Society. Series B (Methodological) **39** no. 1, (1977) 1–38, http://www.jstor.org/stable/2984875.

[167] H. Zhang, M. Cisse, Y. N. Dauphin, and D. Lopez-Paz, *mixup: Beyond Empirical Risk Minimization*, arXiv:1710.09412 [cs.LG].

[168] C. Szegedy et al., *Intriguing properties of neural networks*, arXiv:1312.6199 [cs.CV].

[169] L. Bottou, F. E. Curtis, and J. Nocedal, *Optimization Methods for Large-Scale Machine Learning*, arXiv:1606.04838 [stat.ML].

[170] S. L. Smith and Q. V. Le, *A Bayesian Perspective on Generalization and Stochastic Gradient Descent*, arXiv:1710.06451 [cs.LG].

[171] S. L. Smith, P.-J. Kindermans, C. Ying, and Q. V. Le, *Don't Decay the Learning Rate, Increase the Batch Size*, arXiv:1711.00489 [cs.LG].

[172] P. Goyal et al., *Accurate, Large Minibatch SGD: Training ImageNet in 1 Hour*, arXiv:1706.02677 [cs.CV].

[173] S. De, A. K. Yadav, D. W. Jacobs, and T. Goldstein, *Big Batch SGD: Automated Inference using Adaptive Batch Sizes*, arXiv:1610.05792 [cs.LG].

[174] L. Balles, J. Romero, and P. Hennig, *Coupling Adaptive Batch Sizes with Learning Rates*, arXiv:1612.05086 [cs.LG].

[175] R. A. Jacobs, *Increased rates of convergence through learning rate adaptation*, Neural Networks **1** no. 4, (1988) 295 – 307, http://www.sciencedirect.com/science/article/pii/0893608088900032.

[176] G. Hinton, N. Srivastava, and K. Swersky, *Neural Networks for Machine Learning*, February, 2014. http://www.cs.toronto.edu/~tijmen/csc321/slides/lecture_slides_lec6.pdf. Lecture Notes.

[177] D. P. Kingma and J. Ba, *Adam: A Method for Stochastic Optimization*, arXiv:1412.6980 [cs.LG].

[178] T. Dozat, *Incorporating nesterov momentum into adam*, http://cs229.stanford.edu/proj2015/054_report.pdf.

[179] J. Duchi, E. Hazan, and Y. Singer, *Adaptive subgradient methods for online learning and stochastic optimization*, Journal of Machine Learning Research **12** no. Jul, (2011) 2121–2159.

[180] M. D. Zeiler, *ADADELTA: An Adaptive Learning Rate Method*, `arXiv:1212.5701 [cs.LG]`.

[181] P. Chaudhari, A. Choromanska, S. Soatto, Y. LeCun, C. Baldassi, C. Borgs, J. T. Chayes, L. Sagun, and R. Zecchina, *Entropy-SGD: Biasing Gradient Descent Into Wide Valleys*, `arXiv:1611.01838 [cs.LG]`.

[182] Ç. Gülçehre and Y. Bengio, *ADASECANT: Robust Adaptive Secant Method for Stochastic Gradient*, `arXiv:1412.7419 [cs.LG]`.

[183] C. Darken and J. E. Moody, *Note on learning rate schedules for stochastic optimization*, In proceedings of Neural Information Processing Systems.

[184] A. C. Wilson, R. Roelofs, M. Stern, N. Srebro, and B. Recht, *The Marginal Value of Adaptive Gradient Methods in Machine Learning*, ArXiv e-prints (2017), `arXiv:1705.08292 [stat.ML]`.

[185] R. Lipschitz, *De explicatione per series trigonometricas instituenda functionum unius variabilis arbitrariarum, et praecipue earum, quae per variabilis spatium finitum valorum maximourm et minimorum numerum habent infinitum,*, Journal für die reine und angewandte Mathematik **63** (1864) 296–308, `http://eudml.org/doc/147922`.

[186] M. Hazewinkel, *Encyclopaedia of Mathematics.* No. v. 5 in Encyclopaedia of Mathematics. Springer Netherlands, 2013. `https://books.google.com/books?id=GRLtCAAAQBAJ`.

[187] J. Heinonen, *Lectures on Lipschitz analysis.* No. 100. University of Jyväskylä, 2005. `http://www.math.jyu.fi/research/reports/rep100.pdf`.

[188] H. Robbins and S. Monro, *A stochastic approximation method*, pp. , 102–109. Springer, 1985.

[189] Y. E. Nesterov, *A method for solving the convex programming problem with convergence rate $\mathcal{O}(1/k^2)$*, Dokl. Akad. Nauk SSSR **269** (1983) 543–547, `https://ci.nii.ac.jp/naid/10029946121/en/`.

[190] D. C. Liu and J. Nocedal, *On the limited memory BFGS method for large scale optimization*, Mathematical programming **45** no. 1-3, (1989) 503–528.

[191] Y. Dauphin, R. Pascanu, Ç. Gülçehre, K. Cho, S. Ganguli, and Y. Bengio, *Identifying and attacking the saddle point problem in high-dimensional non-convex optimization*, `arXiv:1406.2572`, `http://arxiv.org/abs/1406.2572`.

[192] H. J. Kelley, *Gradient theory of optimal flight paths*, Ars Journal **30** no. 10, (1960) 947–954.

[193] A. E. Bryson, *A gradient method for optimizing multi-stage allocation processes*, in *Proc. Harvard Univ. Symposium on digital computers and their applications*, vol. 72. 1961.

[194] S. Dreyfus, *The numerical solution of variational problems*, Journal of Mathematical Analysis and Applications **5** no. 1, (1962) 30–45.

[195] S. Linnainmaa, *The representation of the cumulative rounding error of an algorithm as a Taylor expansion of the local rounding errors*, Master's Thesis (in Finnish), Univ. Helsinki (1970) 6–7.

[196] P. J. Werbos, *Applications of advances in nonlinear sensitivity analysis*, pp. , 762–770. Springer, 1982.

[197] D. E. Rumelhart, G. E. Hinton, and R. J. Williams, *Learning representations by back-propagating errors*, Nature **323** no. 6088, (1986) 533.

[198] A. G. Baydin, B. A. Pearlmutter, and A. A. Radul, *Automatic differentiation in machine learning: a survey*, arXiv:1502.05767 [cs.SC].

[199] Y. Yoshida and T. Miyato, *Spectral Norm Regularization for Improving the Generalizability of Deep Learning*, arXiv:1705.10941 [stat.ML].

[200] T. Miyato, T. Kataoka, M. Koyama, and Y. Yoshida, *Spectral Normalization for Generative Adversarial Networks*, arXiv:1802.05957 [cs.LG].

[201] N. Srivastava et al., *Dropout: A Simple Way to Prevent Neural Networks from Overfitting*, J. Mach. Learn. Res. **15** no. 1, (2014) 1929–1958.

[202] S. Ioffe and C. Szegedy, *Batch Normalization: Accelerating Deep Network Training by Reducing Internal Covariate Shift*, in *Proceedings of the 32nd International Conference on International Conference on Machine Learning - Volume 37*, pp. , 448–456. JMLR.org, 2015.

[203] R. Caruana, S. Lawrence, and C. L. Giles, *Overfitting in neural nets: Backpropagation, conjugate gradient, and early stopping*, In proceedings of Neural Information Processing Systems.

[204] S. Merity, *The NUGGET Non-Linear Piecewise Activation*, February, 2018. http://smerity.com/arxiv.org/abs/1804.404/1804.404.html.

[205] G. E. Hinton, N. Srivastava, A. Krizhevsky, I. Sutskever, and R. R. Salakhutdinov, *Improving neural networks by preventing co-adaptation of feature detectors*, arXiv:1207.0580 [cs.NE].

[206] P. Baldi and P. J. Sadowski, *Understanding dropout*, In proceedings of Neural Information Processing Systems.

[207] Y. Gal and Z. Ghahramani, *Dropout as a Bayesian Approximation: Representing Model Uncertainty in Deep Learning*, arXiv:1506.02142 [stat.ML].

[208] M. D. Zeiler and R. Fergus, *Stochastic Pooling for Regularization of Deep Convolutional Neural Networks*, arXiv:1301.3557 [cs.LG].

[209] S. Semeniuta, A. Severyn, and E. Barth, *Recurrent Dropout without Memory Loss*, `arXiv:1603.05118 [cs.CL]`.

[210] L. Wan, M. Zeiler, S. Zhang, Y. Le Cun, and R. Fergus, *Regularization of neural networks using dropconnect*, pp. , 1058–1066. 2013.

[211] S. Wang and C. Manning, *Fast dropout training*, pp. , 118–126. 2013.

[212] D. Warde-Farley, I. J. Goodfellow, A. Courville, and Y. Bengio, *An empirical analysis of dropout in piecewise linear networks*, `arXiv:1312.6197 [stat.ML]`.

[213] Z. Ghahramani, *Unsupervised Learning*. Springer Berlin Heidelberg, Berlin, Heidelberg, 2004.

[214] M. E. Celebi and K. Aydin, *Unsupervised Learning Algorithms*. Springer, 2016.

[215] A. Valassi, *ROC curves, AUC's and alternatives in HEP event selection and in other domains*, January, 2018. https://indico.cern.ch/event/679765/contributions/2814562/attachments/1590383/2516547/20180126-ROC-AV-IML_v008_final.pdf.

[216] R. H. R. Hahnloser, R. Sarpeshkar, M. A. Mahowald, R. J. Douglas, and H. S. Seung, *Digital selection and analogue amplification coexist in a cortex-inspired silicon circuit*, Nature **405** (2000) 947–951.

[217] X. Glorot, A. Bordes, and Y. Bengio, *Deep Sparse Rectifier Neural Networks*, in *AISTATS*, p. , 275. 2011.

[218] J. Bang-Jensen and G. Z. Gutin, *Digraphs: Theory, Algorithms and Applications*. Springer Publishing Company, Incorporated, 2nd ed., 2008.

[219] G. Cybenko, *Approximation by superpositions of a sigmoidal function*, Mathematics of control, signals and systems **2** no. 4, (1989) 303–314.

[220] Y. Bengio, P. Simard, and P. Frasconi, *Learning long-term dependencies with gradient descent is difficult*, IEEE Transactions on Neural Networks **5** no. 2, (1994) 157–166.

[221] A. Choromanska, M. Henaff, M. Mathieu, G. B. Arous, and Y. LeCun, *The Loss Surface of Multilayer Networks*, `arXiv:1412.0233 [cs.LG]`.

[222] R. Vidal, J. Bruna, R. Giryes, and S. Soatto, *Mathematics of Deep Learning*, `arXiv:1712.04741 [cs.LG]`.

[223] B. D. Haeffele and R. Vidal, *Global Optimality in Tensor Factorization, Deep Learning, and Beyond*, `arXiv:1506.07540 [cs.NA]`.

[224] B. D. Haeffele and R. Vidal, *Global optimality in neural network training*, pp. , 7331–7339. July, 2017.

[225] K. Kawaguchi, *Deep Learning without Poor Local Minima*, `arXiv:1605.07110 [stat.ML]`.

[226] V. Nair and G. E. Hinton, *Rectified Linear Units Improve Restricted Boltzmann Machines*, in *Proceedings of the 27th International Conference on International Conference on Machine Learning*, pp. , 807–814. Omnipress, USA, 2010.

[227] A. Krizhevsky, I. Sutskever, and G. E. Hinton, *Imagenet classification with deep convolutional neural networks*, In proceedings of Neural Information Processing Systems.

[228] G. E. Dahl, T. N. Sainath, and G. E. Hinton, *Improving deep neural networks for LVCSR using rectified linear units and dropout,* pp. , 8609–8613. May, 2013.

[229] B. Xu, N. Wang, T. Chen, and M. Li, *Empirical Evaluation of Rectified Activations in Convolutional Network*, `arXiv:1505.00853 [cs.LG]`.

[230] A. L. Maas, A. Y. Hannun, and A. Y. Ng, *Rectifier nonlinearities improve neural network acoustic models*, in *Proc. ICML*, vol. 30. 2013.

[231] D. Scherer, A. Müller, and S. Behnke, *Evaluation of pooling operations in convolutional architectures for object recognition*, pp. , 92–101. Springer, 2010.

[232] D. C. Ciresan, U. Meier, J. Masci, and J. Schmidhuber, *Flexible, high performance convolutional neural networks for image classification,*.

[233] M. Ranzato, F. J. Huang, Y. Boureau, and Y. LeCun, *Unsupervised Learning of Invariant Feature Hierarchies with Applications to Object Recognition,* pp. , 1–8. June, 2007.

[234] A. J. Bray and D. S. Dean, *Statistics of Critical Points of Gaussian Fields on Large-Dimensional Spaces*, Phys. Rev. Lett. **98** (2007) 150201, `https://link.aps.org/doi/10.1103/PhysRevLett.98.150201`.

[235] S. Hochreiter and J. Schmidhuber, *Simplifying neural nets by discovering flat minima*, In proceedings of Neural Information Processing Systems.

[236] S. Hochreiter and J. Schmidhuber, *Flat minima*, Neural Computation **9** no. 1, (1997) 1–42.

[237] K. He, X. Zhang, S. Ren, and J. Sun, *Deep residual learning for image recognition*, pp. , 770–778. 2016.

[238] H. Li, Z. Xu, G. Taylor, C. Studer, and T. Goldstein, *Visualizing the Loss Landscape of Neural Nets*, `arXiv:1712.09913 [cs.LG]`.

[239] G. Montavon, G. B. Orr, and K.-R. Müller, *Neural Networks: Tricks of the Trade,*.

[240] W. S. McCulloch and W. Pitts, *A logical calculus of the ideas immanent in nervous activity*, The bulletin of mathematical biophysics **5** no. 4, (1943) 115–133.

[241] R. K. Srivastava, K. Greff, and J. Schmidhuber, *Highway Networks*, arXiv:1505.00387 [cs.LG].

[242] S. Hochreiter and J. Schmidhuber, *Long short-term memory*, Neural computation **9** no. 8, (1997) 1735–1780.

[243] Y. N. Dauphin, A. Fan, M. Auli, and D. Grangier, *Language modeling with gated convolutional networks*, arXiv:1612.08083 [cs.CL].

[244] S. Hochreiter, Y. Bengio, and P. Frasconi, *Gradient Flow in Recurrent Nets: the Difficulty of Learning Long-Term Dependencies*, in *Field Guide to Dynamical Recurrent Networks*, J. Kolen and S. Kremer, eds. IEEE Press, 2001.

[245] G. Klambauer, T. Unterthiner, A. Mayr, and S. Hochreiter, *Self-Normalizing Neural Networks*, 2017. In proceedings of Neural Information Processing Systems.

[246] K. He, X. Zhang, S. Ren, and J. Sun, *Delving Deep into Rectifiers: Surpassing Human-Level Performance on ImageNet Classification*, arXiv:1502.01852 [cs.CV].

[247] P. W. Battaglia et al., *Relational inductive biases, deep learning, and graph networks*, arXiv:1806.01261 [cs.LG].

[248] D. Bahdanau, J. Chorowski, D. Serdyuk, P. Brakel, and Y. Bengio, *End-to-end attention-based large vocabulary speech recognition*, 2016 IEEE International Conference on Acoustics, Speech and Signal Processing (ICASSP) (2016) 4945–4949.

[249] X. Glorot and Y. Bengio, *Understanding the difficulty of training deep feedforward neural networks*, in *In Proceedings of the International Conference on Artificial Intelligence and Statistics (AISTATS'10). Society for Artificial Intelligence and Statistics.* 2010.

[250] A. M. Saxe, J. L. McClelland, and S. Ganguli, *Exact solutions to the nonlinear dynamics of learning in deep linear neural networks*, arXiv:1312.6120 [cs.NE].

[251] H. Shimodaira, *Improving predictive inference under covariate shift by weighting the log-likelihood function*, Journal of statistical planning and inference **90** no. 2, (2000) 227–244.

[252] S. Santurkar, D. Tsipras, A. Ilyas, and A. Madry, *How Does Batch Normalization Help Optimization?*, arXiv:1805.11604 [stat.ML].

[253] R. Pascanu, T. Mikolov, and Y. Bengio, *On the difficulty of training Recurrent Neural Networks*, arXiv:1211.5063 [cs.LG].

[254] Y. Bengio, N. Boulanger-Lewandowski, and R. Pascanu, *Advances in Optimizing Recurrent Networks*, arXiv:1212.0901 [cs.LG].

[255] P. E. Utgoff, *Shift of Bias for Inductive Concept Learning*. PhD thesis, New Brunswick, NJ, USA, 1984. AAI8507161.

[256] D. Haussler, *Quantifying inductive bias: AI learning algorithms and Valiant's learning framework*, Artificial intelligence **36** no. 2, (1988) 177–221.

[257] G. Louppe, K. Cho, C. Becot, and K. Cranmer, *QCD-Aware Recursive Neural Networks for Jet Physics*, arXiv:1702.00748 [hep-ph].

[258] C. Shimmin, P. Sadowski, P. Baldi, E. Weik, D. Whiteson, E. Goul, and A. Søgaard, *Decorrelated Jet Substructure Tagging using Adversarial Neural Networks*, Phys. Rev. **D96** no. 7, (2017) 074034, arXiv:1703.03507 [hep-ex].

[259] G. Louppe, M. Kagan, and K. Cranmer, *Learning to pivot with adversarial networks*, In proceedings of Neural Information Processing Systems.

[260] M. M. Bronstein, J. Bruna, Y. LeCun, A. Szlam, and P. Vandergheynst, *Geometric Deep Learning: Going beyond Euclidean data*, IEEE Signal Processing Magazine **34** (2017) 18–42, arXiv:1611.08097 [cs.CV].

[261] S. Thrun and L. Pratt, *Learning To Learn*. Kluwer Academic Publishers, November, 1997.

[262] S. Hochreiter, A. S. Younger, and P. R. Conwell, *Learning To Learn Using Gradient Descent*, pp. , 87–94. Springer, 2001.

[263] B. Zoph and Q. V. Le, *Neural Architecture Search with Reinforcement Learning*, arXiv:1611.01578 [cs.LG].

[264] J. Andreas, M. Rohrbach, T. Darrell, and D. Klein, *Learning to Compose Neural Networks for Question Answering*, arXiv:1601.01705 [cs.CL].

[265] M. Andrychowicz, M. Denil, S. G. Colmenarejo, M. W. Hoffman, D. Pfau, T. Schaul, and N. de Freitas, *Learning to learn by gradient descent by gradient descent*, arXiv:1606.04474 [cs.NE].

[266] T. Elsken, J. Hendrik Metzen, and F. Hutter, *Neural Architecture Search: A Survey*, arXiv:1808.05377 [stat.ML].

[267] C. Finn, P. Abbeel, and S. Levine, *Model-Agnostic Meta-Learning for Fast Adaptation of Deep Networks*, pp. , 1126–1135. 2017.

[268] V. N. Vapnik and A. Y. Chervonenkis, *On the uniform convergence of relative frequencies of events to their probabilities*, pp. , 11–30. Springer, 2015.

[269] P. L. Bartlett, N. Harvey, C. Liaw, and A. Mehrabian, *Nearly-tight VC-dimension and pseudodimension bounds for piecewise linear neural networks*, arXiv:1703.02930 [cs.LG].

[270] J. Hestness, S. Narang, N. Ardalani, G. Diamos, H. Jun, H. Kianinejad, M. M. A. Patwary, Y. Yang, and Y. Zhou, *Deep Learning Scaling is Predictable, Empirically*, arXiv:1712.00409 [cs.LG].

[271] R. Collobert, S. Bengio, and J. Mariéthoz, *Torch: a modular machine learning software library*, tech. rep., Idiap, 2002.

[272] J. Bergstra, O. Breuleux, F. Bastien, P. Lamblin, R. Pascanu, G. Desjardins, J. Turian, D. Warde-Farley, and Y. Bengio, *Theano: A CPU and GPU math compiler in Python*, in *Proc. 9th Python in Science Conf*, vol. 1. 2010.

[273] S. Tokui, K. Oono, S. Hido, and J. Clayton, *Chainer: a next-generation open source framework for deep learning*, in *Proceedings of workshop on machine learning systems (LearningSys) in the twenty-ninth annual conference on neural information processing systems (NIPS)*, pp. , 1–6. 2015.

[274] A. Martin et al., *TensorFlow: Large-Scale Machine Learning on Heterogeneous Systems*, 2015. http://tensorflow.org/. Software available from https://www.tensorflow.org/.

[275] F. Chollet et al., *Keras*, https://keras.io, 2015.

[276] A. Paszke, S. Gross, S. Chintala, G. Chanan, E. Yang, Z. DeVito, Z. Lin, A. Desmaison, L. Antiga, and A. Lerer, *Automatic differentiation in PyTorch*, in *NIPS-W*. 2017.

[277] L. B. Rall and G. F. Corliss, *An Introduction to Automatic Differentiation*, pp. , 1–17. SIAM, Philadelphia, PA, 1996.

[278] M. C. Bartholomew-Biggs, S. Brown, B. Christianson, and L. C. W. Dixon, *Automatic Differentiation of Algorithms*, Journal of Computational and Applied Mathematics **124** (2000) 171–190.

[279] A. Griewank, *A Mathematical View of Automatic Differentiation*, in *Acta Numerica*, pp. , 321–398. Cambridge University Press, 2003.

[280] Y. Lecun, L. Bottou, Y. Bengio, and P. Haffner, *Gradient-based learning applied to document recognition*, Proceedings of the IEEE **86** no. 11, (1998) 2278–2324.

[281] R. Raina, A. Madhavan, and A. Y. Ng, *Large-scale Deep Unsupervised Learning Using Graphics Processors*, in *Proceedings of the 26th Annual International Conference on Machine Learning*, pp. , 873–880. ACM, New York, NY, USA, 2009. http://doi.acm.org/10.1145/1553374.1553486.

[282] S. Arora, R. Ge, Y. Liang, T. Ma, and Y. Zhang, *Generalization and Equilibrium in Generative Adversarial Nets (GANs)*, pp. , 224–232. 2017. arXiv:1703.00573 [cs.LG].

[283] A. van den Oord, N. Kalchbrenner, L. Espeholt, k. kavukcuoglu, O. Vinyals, and A. Graves, *Conditional Image Generation with PixelCNN Decoders*, pp. , 4790–4798. Curran Associates, Inc., 2016. http://papers.nips.cc/paper/6527-conditional-image-generation-with-pixelcnn-decoders.pdf.

[284] D. P. Kingma and M. Welling, *Auto-Encoding Variational Bayes*, arXiv:1312.6114.

[285] I. J. Goodfellow, J. Pouget-Abadie, M. Mirza, B. Xu, D. Warde-Farley, S. Ozair, A. Courville, and Y. Bengio, *Generative Adversarial Networks*, `arXiv:1406.2661 [stat.ML]`.

[286] S. Chintala, E. Denton, M. Arjovsky, and M. Mathieu, *How to train a GAN?*, NIPS, Workshop on Generative Adversarial Networks (2016), `https://github.com/soumith/ganhacks`.

[287] S. Liu, O. Bousquet, and K. Chaudhuri, *Approximation and Convergence Properties of Generative Adversarial Learning*, pp. , 5545–5553. Curran Associates, Inc., 2017. `http://papers.nips.cc/paper/7138-approximation-and-convergence-properties-of-generative-adversarial-learning.pdf`.

[288] V. Nagarajan and J. Z. Kolter, *Gradient descent GAN optimization is locally stable*, pp. , 5585–5595. Curran Associates, Inc., 2017. `http://papers.nips.cc/paper/7142-gradient-descent-gan-optimization-is-locally-stable.pdf`.

[289] V. Dumoulin and F. Visin, *A guide to convolution arithmetic for deep learning*, `arXiv:1603.07285 [stat.ML]`.

[290] M. Mirza and S. Osindero, *Conditional Generative Adversarial Nets*, `arXiv:1411.1784 [cs.LG]`.

[291] S. E. Reed, Z. Akata, S. Mohan, S. Tenka, B. Schiele, and H. Lee, *Learning What and Where to Draw*, `arXiv:1610.02454 [cs.CV]`.

[292] A. Odena, C. Olah, and J. Shlens, *Conditional Image Synthesis With Auxiliary Classifier GANs*, `arXiv:1610.09585 [stat.ML]`.

[293] X. Chen, Y. Duan, R. Houthooft, J. Schulman, I. Sutskever, and P. Abbeel, *InfoGAN: Interpretable Representation Learning by Information Maximizing Generative Adversarial Nets*, `arXiv:1606.03657 [cs.LG]`.

[294] T. Salimans, I. Goodfellow, W. Zaremba, V. Cheung, A. Radford, and X. Chen, *Improved Techniques for Training GANs*, `arXiv:1606.03498 [cs.LG]`.

[295] C. Szegedy, V. Vanhoucke, S. Ioffe, J. Shlens, and Z. Wojna, *Rethinking the Inception Architecture for Computer Vision*, `arXiv:1512.00567 [cs.CV]`.

[296] E. L. Denton, S. Chintala, A. Szlam, and R. Fergus, *Deep Generative Image Models using a Laplacian Pyramid of Adversarial Networks*, `arXiv:1506.05751 [cs.CV]`.

[297] T. Karras, T. Aila, S. Laine, and J. Lehtinen, *Progressive Growing of GANs for Improved Quality, Stability, and Variation*, `arXiv:1710.10196 [cs.NE]`.

[298] N. Kodali, J. Abernethy, J. Hays, and Z. Kira, *On convergence and stability of GANs*,.

[299] M. Heusel, H. Ramsauer, T. Unterthiner, B. Nessler, and S. Hochreiter, *GANs trained by a two time-scale update rule converge to a local nash equilibrium*, In proceedings of Neural Information Processing Systems.

[300] S. M. Ali and S. D. Silvey, *A general class of coefficients of divergence of one distribution from another.*, J. R. Stat. Soc., Ser. B **28** (1966) 131–142.

[301] F. Liese and I. Vajda, *On Divergences and Informations in Statistics and Information Theory*, IEEE Transactions on Information Theory **52** no. 10, (2006) 4394–4412.

[302] S. Nowozin, B. Cseke, and R. Tomioka, *f-GAN: Training Generative Neural Samplers using Variational Divergence Minimization*, arXiv:1606.00709 [stat.ML].

[303] M. Arjovsky and L. Bottou, *Towards principled methods for training generative adversarial networks*, arXiv:1701.04862 [stat.ML].

[304] M. Arjovsky, S. Chintala, and L. Bottou, *Wasserstein GAN*, arXiv:1701.07875 [stat.ML].

[305] I. Gulrajani, F. Ahmed, M. Arjovsky, V. Dumoulin, and A. Courville, *Improved Training of Wasserstein GANs*, arXiv:1704.00028 [cs.LG].

[306] ATLAS Collaboration, G. Ripellino, *The alignment of the ATLAS Inner Detector in Run 2*, Tech. Rep. ATL-INDET-PROC-2016-003, CERN, Geneva, Sep, 2016. https://cds.cern.ch/record/2213441.

[307] ATLAS Collaboration, *dE/dx measurement in the ATLAS Pixel Detector and its use for particle identification*, Tech. Rep. ATLAS-CONF-2011-016, CERN, Geneva, Mar, 2011. https://cds.cern.ch/record/1336519.

[308] G. Aad et al., *ATLAS pixel detector electronics and sensors*, JINST **3** no. 07, (2008) P07007, http://stacks.iop.org/1748-0221/3/i=07/a=P07007.

[309] ATLAS Collaboration, *Training and validation of the ATLAS pixel clustering neural networks*, Tech. Rep. ATL-PHYS-PUB-2018-002, CERN, Geneva, Mar, 2018. https://cds.cern.ch/record/2309474.

[310] ATLAS Collaboration, G. Aad et al., *A neural network clustering algorithm for the ATLAS silicon pixel detector*, JINST **9** (2014) P09009, arXiv:1406.7690 [hep-ex].

[311] ATLAS Collaboration, B. P. Nachman and W. P. McCormack, *Splitting Strip Detector Clusters in Dense Environments*, https://cds.cern.ch/record/2311048.

[312] ATLAS Collaboration, *Improved electron reconstruction in ATLAS using the Gaussian Sum Filter-based model for bremsstrahlung*, Tech. Rep. ATLAS-CONF-2012-047, CERN, Geneva, May, 2012. https://cds.cern.ch/record/1449796.

[313] ATLAS Collaboration, S. D. Jones, *The ATLAS Electron and Photon Trigger*, Tech. Rep. ATL-DAQ-PROC-2018-014, CERN, Geneva, Jul, 2018. https://cds.cern.ch/record/2630597.

[314] *The Expected Performance of the ATLAS Inner Detector*, Tech. Rep. ATL-PHYS-PUB-2009-002. ATL-COM-PHYS-2008-105, CERN, Geneva, Aug, 2008. https://cds.cern.ch/record/1118445.

[315] *Track Reconstruction Performance of the ATLAS Inner Detector at $\sqrt{s} = 13$ TeV*, Tech. Rep. ATL-PHYS-PUB-2015-018, CERN, Geneva, Jul, 2015. https://cds.cern.ch/record/2037683.

[316] ATLAS Collaboration,, *Modelling of Track Reconstruction Inside Jets with the 2016 ATLAS $\sqrt{s} = 13$ TeV pp dataset*, Tech. Rep. ATL-PHYS-PUB-2017-016, CERN, Geneva, Jul, 2017. https://cds.cern.ch/record/2275639.

[317] *The Optimization of ATLAS Track Reconstruction in Dense Environments*, Tech. Rep. ATL-PHYS-PUB-2015-006, CERN, Geneva, Mar, 2015. https://cds.cern.ch/record/2002609.

[318] *Kaggle Competition: TrackML Particle Tracking Challenge*, https://www.kaggle.com/c/trackml-particle-identification/data, 2018. Accessed: 2018-05-17.

[319] ATLAS Collaboration,, *Performance of primary vertex reconstruction in proton-proton collisions at $\sqrt{s} =7$ TeV in the ATLAS experiment*, Tech. Rep. ATLAS-CONF-2010-069, CERN, Geneva, Jul, 2010. http://cds.cern.ch/record/1281344.

[320] *An imaging algorithm for vertex reconstruction for ATLAS Run-2*, Tech. Rep. ATL-PHYS-PUB-2015-008, CERN, Geneva, Apr, 2015. http://cds.cern.ch/record/2008700.

[321] V. M. Cairo, *Improvements to ATLAS Track Reconstruction for Run II*, Tech. Rep. ATL-PHYS-PROC-2015-037, CERN, Geneva, Jun, 2015. https://cds.cern.ch/record/2026476.

[322] S. Hageböck and E. von Toerne, *Medical Imaging Inspired Vertex Reconstruction at LHC*, Journal of Physics: Conference Series **396** no. 2, (2012) 022021, http://stacks.iop.org/1742-6596/396/i=2/a=022021.

[323] A. Tsaris et al., *The HEP.TrkX Project: Deep Learning for Particle Tracking*, in *Proceedings of the 18th International Workshop on Advanced Computing and Analysis Techniques in Physics Research*. 2017. https://indico.cern.ch/event/567550/papers/2629737/files/6146-hepTrkX_acat.pdf.

[324] W. Lampl et al., *Calorimeter Clustering Algorithms: Description and Performance*, Tech. Rep. ATL-LARG-PUB-2008-002. ATL-COM-LARG-2008-003, CERN, Geneva, Apr, 2008. https://cds.cern.ch/record/1099735.

[325] ATLAS Collaboration, N. Anjos, *The ATLAS Jet Trigger for LHC Run 2*, Tech. Rep. ATL-DAQ-PROC-2015-034, CERN, Geneva, Oct, 2015. https://cds.cern.ch/record/2057590.

[326] ATLAS Collaboration, S. Schramm, *Triggering on hadronic signatures: developments for 2017 + 2018*, https://cds.cern.ch/record/2630348.

[327] D. Krohn, J. Thaler, and L.-T. Wang, *Jet Trimming*, JHEP **02** (2010) 084, arXiv:0912.1342 [hep-ph].

[328] S. Catani, Y. L. Dokshitzer, M. H. Seymour, and B. R. Webber, *Longitudinally invariant $K_t$ clustering algorithms for hadron hadron collisions*, Nucl. Phys. **B406** (1993) 187–224.

[329] S. D. Ellis and D. E. Soper, *Successive combination jet algorithm for hadron collisions*, Phys. Rev. **D48** (1993) 3160–3166, arXiv:hep-ph/9305266 [hep-ph].

[330] Y. L. Dokshitzer, G. D. Leder, S. Moretti, and B. R. Webber, *Better jet clustering algorithms*, JHEP **08** (1997) 001, arXiv:hep-ph/9707323 [hep-ph].

[331] M. Wobisch and T. Wengler, *Hadronization corrections to jet cross-sections in deep inelastic scattering*, pp. , 270–279. 1998. arXiv:hep-ph/9907280 [hep-ph].

[332] M. Cacciari, G. P. Salam, and G. Soyez, *The Anti-k(t) jet clustering algorithm*, JHEP **04** (2008) 063, arXiv:0802.1189 [hep-ph].

[333] D. Krohn, J. Thaler, and L.-T. Wang, *Jets with Variable R*, JHEP **06** (2009) 059, arXiv:0903.0392 [hep-ph].

[334] M. Cacciari, G. P. Salam, and G. Soyez, *FastJet User Manual*, Eur. Phys. J. **C72** (2012) 1896, arXiv:1111.6097 [hep-ph].

[335] M. Cacciari, G. P. Salam, and G. Soyez, *SoftKiller, a particle-level pileup removal method*, Eur. Phys. J. **C75** no. 2, (2015) 59, arXiv:1407.0408 [hep-ph].

[336] S. D. Ellis, C. K. Vermilion, and J. R. Walsh, *Recombination Algorithms and Jet Substructure: Pruning as a Tool for Heavy Particle Searches*, Phys. Rev. **D81** (2010) 094023, arXiv:0912.0033 [hep-ph].

[337] M. Dasgupta, A. Fregoso, S. Marzani, and G. P. Salam, *Towards an understanding of jet substructure*, JHEP **09** (2013) 029, arXiv:1307.0007 [hep-ph].

[338] D. Bertolini, P. Harris, M. Low, and N. Tran, *Pileup Per Particle Identification*, JHEP **10** (2014) 059, arXiv:1407.6013 [hep-ph].

[339] A. J. Larkoski, S. Marzani, G. Soyez, and J. Thaler, *Soft Drop*, JHEP **05** (2014) 146, arXiv:1402.2657 [hep-ph].

[340] B. Nachman, P. Nef, A. Schwartzman, M. Swiatlowski, and C. Wanotayaroj, *Jets from Jets: Re-clustering as a tool for large radius jet reconstruction and grooming at the LHC*, JHEP **02** (2015) 075, arXiv:1407.2922 [hep-ph].

[341] P. T. Komiske, E. M. Metodiev, B. Nachman, and M. D. Schwartz, *Pileup Mitigation with Machine Learning (PUMML)*, JHEP **12** (2017) 051, arXiv:1707.08600 [hep-ph].

[342] F. A. Dreyer, L. Necib, G. Soyez, and J. Thaler, *Recursive Soft Drop*, JHEP **06** (2018) 093, arXiv:1804.03657 [hep-ph].

[343] *Flavor Tagging with Track Jets in Boosted Topologies with the ATLAS Detector*, Tech. Rep. ATL-PHYS-PUB-2014-013, CERN, Geneva, Aug, 2014. http://cds.cern.ch/record/1750681.

[344] M. Cacciari and G. P. Salam, *Pileup subtraction using jet areas*, Phys. Lett. **B659** (2008) 119–126, arXiv:0707.1378 [hep-ph].

[345] M. Cacciari, G. P. Salam, and G. Soyez, *The Catchment Area of Jets*, JHEP **04** (2008) 005, arXiv:0802.1188 [hep-ph].

[346] ATLAS Collaboration, G. Aad et al., *Performance of jet substructure techniques for large-R jets in proton-proton collisions at $\sqrt{s}$ = 7 TeV using the ATLAS detector*, JHEP **09** (2013) 076, arXiv:1306.4945 [hep-ex].

[347] A. J. Larkoski, I. Moult, and B. Nachman, *Jet Substructure at the Large Hadron Collider: A Review of Recent Advances in Theory and Machine Learning*, arXiv:1709.04464 [hep-ph].

[348] L. Mackey, B. Nachman, A. Schwartzman, and C. Stansbury, *Fuzzy Jets*, JHEP **06** (2016) 010, arXiv:1509.02216 [hep-ph].

[349] M. Zaheer, S. Kottur, S. Ravanbakhsh, B. Póczos, R. Salakhutdinov, and A. J. Smola, *Deep Sets*, arXiv:1703.06114 [cs.LG].

[350] I. Henrion et al., *Neural Message Passing for Jet Physics*, in *2017 Deep Learning for Physical Sciences Workshop*. 2017. https://dl4physicalsciences.github.io/files/nips_dlps_2017_29.pdf.

[351] M. H. Seymour, *Searches for new particles using cone and cluster jet algorithms: A Comparative study*, Z. Phys. **C62** (1994) 127–138.

[352] R. K. Böck, W. Krischer, A. Gheorghe, L. Levinson, and Z. Natkaniec, *A commercial image processing system considered for triggering in future LHC experiments*, Nucl. Instrum. Methods Phys. Res., A **356** (1995) 304–308, https://cds.cern.ch/record/301336.

[353] R. K. Böck, *Techniques of image processing in high-energy physics*, <https://cds.cern.ch/record/323781>.

[354] M. A. Thomson, *The Use of maximum entropy in electromagnetic calorimeter event reconstruction*, Nucl. Instrum. Methods Phys. Res., A **382** no. CERN-PPE-96-055, (1996) 553–560. 14 p, <https://cds.cern.ch/record/302717>.

[355] J. Shelton, pp. 303–340. 2013. `arXiv:1302.0260 [hep-ph]`. Proceedings, Theoretical Advanced Study Institute in Elementary Particle Physics: Searching for New Physics at Small and Large Scales (TASI 2012): Boulder, Colorado, June 4-29, 2012.

[356] J. Cogan, M. Kagan, E. Strauss, and A. Schwarztman, *Jet-Images: Computer Vision Inspired Techniques for Jet Tagging*, JHEP **02** (2015) 118, `arXiv:1407.5675 [hep-ph]`.

[357] M. A. KAGAN and B. NACHMAN, *Machine learning, computer vision, and probabilistic models in jet physics. Data Science @ LHC 2015 Workshop*, <https://cds.cern.ch/record/2069153>.

[358] P. T. Komiske, E. M. Metodiev, and M. D. Schwartz, *Deep learning in color: towards automated quark/gluon jet discrimination*, JHEP **01** (2017) 110, `arXiv:1612.01551 [hep-ph]`.

[359] L. G. Almeida, M. Backović, M. Cliche, S. J. Lee, and M. Perelstein, *Playing Tag with ANN: Boosted Top Identification with Pattern Recognition*, JHEP **07** (2015) 086, `arXiv:1501.05968 [hep-ph]`.

[360] J. Barnard, E. N. Dawe, M. J. Dolan, and N. Rajcic, *Parton Shower Uncertainties in Jet Substructure Analyses with Deep Neural Networks*, Phys. Rev. **D95** no. 1, (2017) 014018, `arXiv:1609.00607 [hep-ph]`.

[361] P. Baldi, K. Bauer, C. Eng, P. Sadowski, and D. Whiteson, *Jet Substructure Classification in High-Energy Physics with Deep Neural Networks*, Phys. Rev. **D93** no. 9, (2016) 094034, `arXiv:1603.09349 [hep-ex]`.

[362] G. Kasieczka, T. Plehn, M. Russell, and T. Schell, *Deep-learning Top Taggers or The End of QCD?*, JHEP **05** (2017) 006, `arXiv:1701.08784 [hep-ph]`.

[363] A. Butter, G. Kasieczka, T. Plehn, and M. Russell, *Deep-learned Top Tagging with a Lorentz Layer*, `arXiv:1707.08966 [hep-ph]`.

[364] F. A. Dreyer, G. P. Salam, and G. Soyez, *The Lund Jet Plane*, `arXiv:1807.04758 [hep-ph]`.

[365] W. Bhimji, S. A. Farrell, T. Kurth, M. Paganini, Prabhat, and E. Racah, *Deep Neural Networks for Physics Analysis on low-level whole-detector data at the LHC*, in *18th International Workshop on Advanced Computing and Analysis Techniques in Physics Research (ACAT 2017) Seattle, WA, USA, August 21-25, 2017*. 2017. `arXiv:1711.03573 [hep-ex]`.

[366] J. Guo, J. Li, T. Li, F. Xu, and W. Zhang, *Deep learning for the R-parity violating supersymmetry searches at the LHC*, arXiv:1805.10730 [hep-ph].

[367] T. Q. Nguyen et al., *Topology classification with deep learning to improve real-time event selection at the LHC*, arXiv:1807.00083 [hep-ex].

[368] K. Mistry et al., *Data-MC shower shape comparisons: Supporting documentation for the Photon identification in 2015 and 2016 ATLAS data*, Tech. Rep. ATL-COM-PHYS-2017-363, CERN, Geneva, Apr, 2017. https://cds.cern.ch/record/2258832.

[369] ATLAS Collaboration,, *Electron and photon reconstruction and performance in ATLAS using a dynamical, topological cell clustering-based approach*, Tech. Rep. ATL-PHYS-PUB-2017-022, CERN, Geneva, Dec, 2017. https://cds.cern.ch/record/2298955.

[370] ATLAS Collaboration,, *Electron and photon energy calibration with the ATLAS detector using data collected in 2015 at $\sqrt{s} = 13$ TeV*, Tech. Rep. ATL-PHYS-PUB-2016-015, CERN, Geneva, Aug, 2016. https://cds.cern.ch/record/2203514.

[371] T. Ciodaro, D. Deva, J. M. de Seixas, and D. Damazio, *Online particle detection with Neural Networks based on topological calorimetry information*, Journal of Physics: Conference Series **368** no. 1, (2012) 012030, http://stacks.iop.org/1742-6596/368/i=1/a=012030.

[372] ATLAS Collaboration, J. V. Da Fonseca Pinto, W. Spolidoro Freund, and J. Seixas, *An Ensemble of Neural Networks for Online Electron Filtering at the ATLAS Experiment.*, https://cds.cern.ch/record/2624036.

[373] B. Nachman, L. de Oliveira, and M. Paganini, *Electromagnetic Calorimeter Shower Images*, Data set, DOI: 10.17632/pvn3xc3wy5.1.

[374] ATLAS Collaboration, *Muon reconstruction performance of the ATLAS detector in proton–proton collision data at $\sqrt{s}$=13 TeV*, arXiv:1603.05598 [hep-ex].

[375] Particle Data Group Collaboration, J. Beringer et al., *Review of Particle Physics*, Phys. Rev. D **86** (2012) 010001.

[376] ATLAS Collaboration,, *Measurement of the tau lepton reconstruction and identification performance in the ATLAS experiment using pp collisions at $\sqrt{s} = 13$ TeV*, Tech. Rep. ATLAS-CONF-2017-029, CERN, Geneva, May, 2017. https://cds.cern.ch/record/2261772.

[377] ATLAS Collaboration, G. Aad et al., *Reconstruction of hadronic decay products of tau leptons with the ATLAS experiment*, Eur. Phys. J. **C76** no. 5, (2016) 295, arXiv:1512.05955 [hep-ex].

[378] ATLAS Collaboration, M. Aaboud et al., *Performance of missing transverse momentum reconstruction with the ATLAS detector using proton-proton collisions at $\sqrt{s} = 13$ TeV*, arXiv:1802.08168 [hep-ex].

[379] *Tagging and suppression of pileup jets with the ATLAS detector*, Tech. Rep. ATLAS-CONF-2014-018, CERN, Geneva, May, 2014. https://cds.cern.ch/record/1700870.

[380] ATLAS Collaboration, M. Aaboud et al., *Jet reconstruction and performance using particle flow with the ATLAS Detector*, Eur. Phys. J. **C77** no. 7, (2017) 466, arXiv:1703.10485 [hep-ex].

[381] *b-tagging in dense environments*, Tech. Rep. ATL-PHYS-PUB-2014-014, CERN, Geneva, Aug, 2014. https://cds.cern.ch/record/1750682.

[382] ATLAS Collaboration,, *Secondary vertex finding for jet flavour identification with the ATLAS detector*, Tech. Rep. ATL-PHYS-PUB-2017-011, CERN, Geneva, Jun, 2017. https://cds.cern.ch/record/2270366.

[383] G. Piacquadio and C. Weiser, *A new inclusive secondary vertex algorithm for b-jet tagging in ATLAS*, Journal of Physics: Conference Series **119** no. 3, (2008) 032032, http://stacks.iop.org/1742-6596/119/i=3/a=032032.

[384] ATLAS Collaboration, G. Gilles, *Topological b-hadron decay reconstruction and application for heavy-flavour jet tagging in ATLAS*, https://cds.cern.ch/record/2275033.

[385] ATLAS Collaboration, G. Aad et al., *Identification and energy calibration of hadronically decaying tau leptons with the ATLAS experiment in pp collisions at $\sqrt{s}$=8 TeV*, Eur. Phys. J. **C75** no. 7, (2015) 303, arXiv:1412.7086 [hep-ex].

[386] D. H. Guest et al., lwtnn/lwtnn: *Version 2.6,*.

[387] ATLAS Collaboration, A. Sciandra, *Development of a new Soft Muon Tagger for the identification of b-jets in ATLAS*, Tech. Rep. ATL-PHYS-PROC-2017-190, CERN, Geneva, Oct, 2017. https://cds.cern.ch/record/2287545.

[388] A. Hoecker et al., *TMVA - Toolkit for Multivariate Data Analysis*, arXiv:physics/0703039.

[389] *Expected performance of the ATLAS b-tagging algorithms in Run-2*, Tech. Rep. ATL-PHYS-PUB-2015-022, CERN, Geneva, Jul, 2015. http://cds.cern.ch/record/2037697.

[390] ATLAS Collaboration, M. C. Lanfermann, *Deep Learning in Flavour Tagging at the ATLAS experiment*, Tech. Rep. ATL-PHYS-PROC-2017-191, CERN, Geneva, Oct, 2017. https://cds.cern.ch/record/2287551.

[391] I. J. Goodfellow, D. Warde-Farley, M. Mirza, A. Courville, and Y. Bengio, *Maxout Networks*, in *Proceedings of the 30th International Conference on International Conference on Machine Learning - Volume 28*, pp. , III–1319–III–1327. JMLR.org, 2013.

[392] The HDF Group, *Hierarchical Data Format, version 5*, 1997-2018. http://www.hdfgroup.org/HDF5/.

[393] A. Collette, *Python and HDF5*. O'Reilly, 2013.

[394] W. McKinney, *Data Structures for Statistical Computing in Python*, pp. , 51 – 56. 2010.

[395] F. Pedregosa et al., *Scikit-learn: Machine Learning in Python*, Journal of Machine Learning Research **12** (2011) 2825–2830.

[396] J. D. Hunter, *Matplotlib: A 2D graphics environment*, Computing In Science & Engineering **9** no. 3, (2007) 90–95.

[397] S. van der Walt, S. C. Colbert, and G. Varoquaux, *The NumPy array: a structure for efficient numerical computation*, arXiv:1102.1523 [cs.MS].

[398] N. Dawe, P. Waller, E. K. Friis, et al., *rootpy: 0.8.0*, Jun, 2015. DOI: 10.5281/zenodo.18897.

[399] N. Dawe, P. Ongmongkolkul, and G. Stark, *root_numpy: The interface between ROOT and NumPy*, The Journal of Open Source Software **2** (2017).

[400] Y. Ganin et al., *Domain-Adversarial Training of Neural Networks*, arXiv:1505.07818 [stat.ML].

[401] ATLAS Collaboration,, *Variable Radius, Exclusive-$k_T$, and Center-of-Mass Subjet Reconstruction for Higgs($\to b\bar{b}$) Tagging in ATLAS*, Tech. Rep. ATL-PHYS-PUB-2017-010, CERN, Geneva, Jun, 2017. https://cds.cern.ch/record/2268678.

[402] J. Thaler and K. Van Tilburg, *Identifying Boosted Objects with N-subjettiness*, JHEP **03** (2011) 015, arXiv:1011.2268 [hep-ph].

[403] J. Pearkes, W. Fedorko, A. Lister, and C. Gay, *Jet Constituents for Deep Neural Network Based Top Quark Tagging*, arXiv:1704.02124 [hep-ex].

[404] ATLAS Collaboration, G. Aad et al., *A new method to distinguish hadronically decaying boosted Z bosons from W bosons using the ATLAS detector*, Eur. Phys. J. **C76** no. 5, (2016) 238, arXiv:1509.04939 [hep-ex].

[405] A. J. Larkoski, I. Moult, and D. Neill, *Analytic Boosted Boson Discrimination*, JHEP **05** (2016) 117, arXiv:1507.03018 [hep-ph].

[406] ATLAS Collaboration, G. Aad et al., *Light-quark and gluon jet discrimination in pp collisions at $\sqrt{s} = 7$ TeV with the ATLAS detector*, Eur. Phys. J. **C74** no. 8, (2014) 3023, arXiv:1405.6583 [hep-ex].

[407] ATLAS Collaboration,, *Discrimination of Light Quark and Gluon Jets in pp collisions at $\sqrt{s} = 8$ TeV with the ATLAS Detector*, Tech. Rep. ATLAS-CONF-2016-034, CERN, Geneva, Jul, 2016. http://cds.cern.ch/record/2200202.

[408] ATLAS Collaboration„ *Quark versus Gluon Jet Tagging Using Charged Particle Multiplicity with the ATLAS Detector*, Tech. Rep. ATL-PHYS-PUB-2017-009, CERN, Geneva, May, 2017. https://cds.cern.ch/record/2263679.

[409] ATLAS Collaboration„ *Impact of Pile-up on Jet Constituent Multiplicity in ATLAS*, Tech. Rep. ATL-PHYS-PUB-2018-011, CERN, Geneva, Jul, 2018. http://cds.cern.ch/record/2630603.

[410] CMS Collaboration„ *Performance of quark/gluon discrimination in 8 TeV pp data*, Tech. Rep. CMS-PAS-JME-13-002, CERN, Geneva, 2013. https://cds.cern.ch/record/1599732.

[411] CMS Collaboration„ *Performance of quark/gluon discrimination in 13 TeV data*, https://cds.cern.ch/record/2234117.

[412] ATLAS Collaboration„ *Combined measurements of Higgs boson production and decay using up to 80 fb$^{-1}$ of proton–proton collision data at $\sqrt{s} = 13$ TeV collected with the ATLAS experiment*, Tech. Rep. ATLAS-CONF-2018-031, CERN, Geneva, Jul, 2018. https://cds.cern.ch/record/2629412.

[413] ATLAS Collaboration, M. Aaboud et al., *Search for Higgs boson pair production in the $\gamma\gamma b\bar{b}$ final state with 13 TeV pp collision data collected by the ATLAS experiment*, arXiv:1807.04873 [hep-ex].

[414] T. Golling, H. S. Hayward, P. Onyisi, H. J. Stelzer, and P. Waller, *The ATLAS Data Quality Defect Database System*, Tech. Rep. ATL-COM-DAPR-2011-010, CERN, Geneva, Jul, 2011. https://cds.cern.ch/record/1370528.

[415] ATLAS Collaboration, G. Aad et al., *Data Quality from the Detector Control System at the ATLAS Experiment*, Tech. Rep. ATL-SOFT-PROC-2010-002, CERN, Geneva, Mar, 2010. http://cds.cern.ch/record/1248213.

[416] G. Pásztor, *The Upgrade of the ATLAS Electron and Photon Triggers towards LHC Run 2 and their Performance*, arXiv:1511.00334 [hep-ex].

[417] S. Borowka et al., *Higgs Boson Pair Production in Gluon Fusion at Next-to-Leading Order with Full Top-Quark Mass Dependence*, Phys. Rev. Lett. **117** no. 1, (2016) 012001, arXiv:1604.06447 [hep-ph], [Erratum: Phys. Rev. Lett.117,no.7,079901(2016)].

[418] ATLAS Collaboration, M. Aaboud et al., *Jet energy scale measurements and their systematic uncertainties in proton-proton collisions at $\sqrt{s} = 13$ TeV with the ATLAS detector*, Phys. Rev. **D96** no. 7, (2017) 072002, arXiv:1703.09665 [hep-ex].

[419] ATLAS Collaboration„ *Search for Higgs boson pair production in the $b\bar{b}\gamma\gamma$ final state using pp collision data at $\sqrt{s} = 13$ TeV with the ATLAS detector*,.

[420] L. Carminati et al., *Measurement of the isolated di-photon cross section in 4.9 fb-1 of pp collisions at sqrt(s) = 7 TeV with the ATLAS detector*, Tech. Rep. ATL-COM-PHYS-2012-592, CERN, Geneva, May, 2012. https://cds.cern.ch/record/1450063.

[421] A. L. Read, *Presentation of search results: The CL(s) technique*, J. Phys. **G28** (2002) 2693–2704, [,11(2002)].

[422] S. Das, *A simple alternative to the Crystal Ball function*, arXiv:1603.08591 [hep-ex].

[423] *A detailed test of the CsI(Tl) calorimeter for BELLE with photon beams of energy between 20 MeV and 5.4 GeV*, NIMA **441** no. 3, (2000) 401 – 426, http://www.sciencedirect.com/science/article/pii/S0168900299009924.

[424] A. Bevan and F. Wilson, *AFit User Guide*, tech. rep., 2010. http://pprc.qmul.ac.uk/~bevan/afit/afit.pdf.

[425] M. Oreglia, *A Study of the Reactions $\psi' \to \gamma\gamma\psi$*. PhD thesis, SLAC, 1980. http://www-public.slac.stanford.edu/sciDoc/docMeta.aspx?slacPubNumber=slac-r-236.html.

[426] ATLAS Collaboration, M. Aaboud et al., *Search for resonances in diphoton events at $\sqrt{s}$=13 TeV with the ATLAS detector*, JHEP **09** (2016) 001, arXiv:1606.03833 [hep-ex].

[427] ATLAS Collaboration, G. Aad et al., *Study of heavy-flavor quarks produced in association with top-quark pairs at $\sqrt{s} = 7$ TeV using the ATLAS detector*, Phys. Rev. **D89** no. 7, (2014) 072012, arXiv:1304.6386 [hep-ex].

[428] ATLAS Collaboration, G. Aad et al., *Measurement of the cross-section for W boson production in association with b-jets in pp collisions at $\sqrt{s} = 7$ TeV with the ATLAS detector*, JHEP **06** (2013) 084, arXiv:1302.2929 [hep-ex].

[429] G. Aad et al., *Search for Higgs Boson Pair Production in the $\gamma\gamma b\bar{b}$ Final State Using pp Collision Data at $\sqrt{s}$=8 TeV from the ATLAS Detector*, Phys. Rev. Lett. **114** no. 8, (2015) 081802, arXiv:1406.5053 [hep-ex].

[430] E. Gross and O. Vitells, *Trial factors for the look elsewhere effect in high energy physics*, Eur. Phys. J. **70** no. 1, (2010) 525–530.

[431] L. Lyons, *Open statistical issues in Particle Physics*, Ann. Appl. Stat. **2** no. 3, (2008) 887–915.

[432] ATLAS Collaboration, G. Aad et al., *Search for Higgs boson pair production in the $b\bar{b}b\bar{b}$ final state from pp collisions at $\sqrt{s} = 8$ TeV with the ATLAS detector*, Eur. Phys. J. **C75** no. 9, (2015) 412, arXiv:1506.00285 [hep-ex].

[433] CMS Collaboration,, *Search for Higgs boson pair production in the final state containing two photons and two bottom quarks in proton-proton collisions at $\sqrt{s} = 13$ TeV*, Tech. Rep. CMS-PAS-HIG-17-008, CERN, Geneva, 2017. https://cds.cern.ch/record/2273383.

[434] CMS Collaboration, A. M. Sirunyan et al., *Search for Higgs boson pair production in the $\gamma\gamma b\bar{b}$ final state in pp collisions at $\sqrt{s} = 13$ TeV*, arXiv:1806.00408 [hep-ex].

[435] CMS Collaboration, A. M. Sirunyan et al., *Search for resonant and nonresonant Higgs boson pair production in the $b\bar{b}\ell\nu\ell\nu$ final state in proton-proton collisions at $\sqrt{s} = 13$ TeV*, JHEP **01** (2018) 054, arXiv:1708.04188 [hep-ex].

[436] CMS Collaboration, A. M. Sirunyan et al., *Search for Higgs boson pair production in events with two bottom quarks and two tau leptons in proton–proton collisions at $\sqrt{s} = 13$ TeV*, Phys. Lett. **B778** (2018) 101–127, arXiv:1707.02909 [hep-ex].

[437] CMS Collaboration, *Search for resonant pair production of Higgs bosons decaying to bottom quark-antiquark pairs in proton-proton collisions at 13 TeV*, Tech. Rep. CMS-PAS-HIG-17-009, CERN, Geneva, 2017. https://cds.cern.ch/record/2292044.

[438] CMS Collaboration, A. M. Sirunyan et al., *Search for a massive resonance decaying to a pair of Higgs bosons in the four b quark final state in proton-proton collisions at $\sqrt{s} = 13$ TeV*, Phys. Lett. **B781** (2018) 244–269, arXiv:1710.04960 [hep-ex].

[439] CMS Collaboration, *Search for non-resonant pair production of Higgs bosons in the $b\bar{b}b\bar{b}$ final state with 13 TeV CMS data*, Tech. Rep. CMS-PAS-HIG-16-026, CERN, Geneva, 2016. https://cds.cern.ch/record/2209572.

[440] CMS Collaboration, A. M. Sirunyan et al., *Search for production of Higgs boson pairs in the four b quark final state using large-area jets in proton-proton collisions at $\sqrt{s} = 13$ TeV*, arXiv:1808.01473 [hep-ex].

[441] ATLAS Collaboration, M. Aaboud et al., *Search for pair production of Higgs bosons in the $b\bar{b}b\bar{b}$ final state using proton–proton collisions at $\sqrt{s} = 13$ TeV with the ATLAS detector*, Phys. Rev. **D94** no. 5, (2016) 052002, arXiv:1606.04782 [hep-ex].

[442] ATLAS Collaboration, M. Aaboud et al., *Search for pair production of Higgs bosons in the $b\bar{b}b\bar{b}$ final state using proton-proton collisions at $\sqrt{s} = 13$ TeV with the ATLAS detector*, arXiv:1804.06174 [hep-ex].

[443] ATLAS Collaboration, M. Aaboud et al., *A search for resonant and non-resonant Higgs boson pair production in the $b\bar{b}\tau^+\tau^-$ decay channel in pp collisions at $\sqrt{s} = 13$ TeV with the ATLAS detector*, Submitted to: Phys. Rev. Lett. (2018), arXiv:1808.00336 [hep-ex].

[444] ATLAS Collaboration, M. Aaboud et al., *Search for Higgs boson pair production in the $\gamma\gamma WW^*$ channel using pp collision data recorded at $\sqrt{s} = 13$ TeV with the ATLAS detector*, Submitted to: Eur. Phys. J. (2018), arXiv:1807.08567 [hep-ex].

[445] ATLAS Collaboration, *Combination of searches for Higgs boson pairs in pp collisions at 13 TeV with the ATLAS experiment.*, Tech. Rep. ATLAS-CONF-2018-043, CERN, Geneva, Sep, 2018. https://cds.cern.ch/record/2638212.

[446] CMS Collaboration,, *Combination of searches for Higgs boson pair production in proton-proton collisions at $\sqrt{s} = 13$ TeV*, Tech. Rep. CMS-PAS-HIG-17-030, CERN, Geneva, 2018. https://cds.cern.ch/record/2628486.

[447] ATLAS Collaboration,, *Study of the double Higgs production channel $H(\to b\bar{b})H(\to \gamma\gamma)$ with the ATLAS experiment at the HL-LHC*, Tech. Rep. ATL-PHYS-PUB-2017-001, CERN, Geneva, Jan, 2017. http://cds.cern.ch/record/2243387.

[448] D. Gonçalves, T. Han, F. Kling, T. Plehn, and M. Takeuchi, *Higgs boson pair production at future hadron colliders: From kinematics to dynamics*, Phys. Rev. **D97** no. 11, (2018) 113004, arXiv:1802.04319 [hep-ph].

[449] C. Bozzi, *LHCb Computing Resource usage in 2014 (II)*, Tech. Rep. LHCb-PUB-2015-004. CERN-LHCb-PUB-2015-004, CERN, Geneva, Jan, 2015. https://cds.cern.ch/record/1984010.

[450] J. Flynn, *Computing Resources Scrutiny Group Report*, Tech. Rep. CERN-RRB-2015-014, CERN, Geneva, Mar, 2015. https://cds.cern.ch/record/2002240.

[451] E. Karavakis, J. Andreeva, S. Campana, S. Gayazov, S. Jezequel, P. Saiz, L. Sargsyan, J. Schovancova, I. Ueda, and the ATLAS Collaboration, *Common Accounting System for Monitoring the ATLAS Distributed Computing Resources*, Journal of Physics: Conference Series **513** no. 6, (2014) 062024, http://stacks.iop.org/1742-6596/513/i=6/a=062024.

[452] T. Sjostrand, S. Mrenna, and P. Z. Skands, *PYTHIA 6.4 Physics and Manual*, JHEP **0605** (2006) 026, arXiv:hep-ph/0603175 [hep-ph].

[453] L. de Oliveira and M. Paganini, *lukedeo/adversarial-jets: Initial Release*, Mar., 2017. https://doi.org/10.5281/zenodo.400708.

[454] S. van der Walt, J. L. Schönberger, J. Nunez-Iglesias, F. Boulogne, J. D. Warner, N. Yager, E. Gouillart, T. Yu, and the scikit-image contributors, *scikit-image: image processing in Python*, PeerJ **2** (2014) e453.

[455] A. J. Larkoski, D. Neill, and J. Thaler, *Jet Shapes with the Broadening Axis*, JHEP **04** (2014) 017, arXiv:1401.2158 [hep-ph].

[456] G. P. Styan, *Hadamard products and multivariate statistical analysis*, Linear Algebra and its Applications **6** (1973) 217 – 240.

[457] I. J. Goodfellow, *On distinguishability criteria for estimating generative models*, arXiv:1412.6515 [stat.ML].

[458] A. Odena, *Semi-Supervised Learning with Generative Adversarial Networks*, arXiv:1606.01583 [stat.ML].

[459] Y. Rubner, C. Tomasi, and L. J. Guibas, *The Earth Mover's Distance As a Metric for Image Retrieval*, Int. J. Comput. Vision **40** no. 2, (2000) 99–121, http://dx.doi.org/10.1023/A:1026543900054.

[460] ATLAS Collaboration,, *ATLAS liquid-argon calorimeter: Technical Design Report*. Technical Design Report ATLAS. CERN, Geneva, 1996. https://cds.cern.ch/record/331061.

[461] M. Paganini, L. de Oliveira, and B. Nachman, *Accelerating Science with Generative Adversarial Networks: An Application to 3D Particle Showers in Multilayer Calorimeters*, Phys. Rev. Lett. **120** no. 4, (2018) 042003, arXiv:1705.02355 [hep-ex].

[462] M. Paganini, L. de Oliveira, and B. Nachman, *CaloGAN : Simulating 3D high energy particle showers in multilayer electromagnetic calorimeters with generative adversarial networks*, Phys. Rev. **D97** no. 1, (2018) 014021, arXiv:1712.10321 [hep-ex].

[463] L. de Oliveira, M. Paganini, and N. Benjamin, *Tips and Tricks for Training GANs with Physics Constraints*, in *2017 Deep Learning for Physical Sciences Workshop*. 2017. https://dl4physicalsciences.github.io/files/nips_dlps_2017_26.pdf.

[464] L. de Oliveira, M. Paganini, and B. Nachman, *Controlling Physical Attributes in GAN-Accelerated Simulation of Electromagnetic Calorimeters*, in *18th International Workshop on Advanced Computing and Analysis Techniques in Physics Research (ACAT 2017) Seattle, WA, USA, August 21-25, 2017*. 2017. arXiv:1711.08813 [hep-ex].

[465] *Geant4 Example B4*, http://geant4-userdoc.web.cern.ch/geant4-userdoc/Doxygen/examples_doc/html/ExampleB4.html.

[466] B. Andersson, G. Gustafson, and B. Nilsson-Almqvist, *A model for low-pT hadronic reactions with generalizations to hadron-nucleus and nucleus-nucleus collisions*, Nuclear Physics B **281** no. 1, (1987) 289 – 309.

[467] B. Andersson, A. Tai, and B.-H. Sa, *Final state interactions in the (nuclear) FRITIOF string interaction scenario*, Zeitschrift für Physik C Particles and Fields **70** no. 3, (1996) 499–506.

[468] B. Nilsson-Almqvist and E. Stenlund, *Interactions Between Hadrons and Nuclei: The Lund Monte Carlo, Fritiof Version 1.6*, Comput. Phys. Commun. **43** (1987) 387.

[469] B. Ganhuyag and V. Uzhinsky, *Modified FRITIOF code: Negative charged particle production in high energy nucleus nucleus interactions*, Czech. J. Phys. **47** (1997) 913–918.

[470] M. P. Guthrie, R. G. Alsmiller, and H. W. Bertini, *Calculation of the capture of negative pions in light elements and comparison with experiments pertaining to cancer radiotherapy*, Nucl. Instrum. Meth. **66** (1968) 29–36.

[471] H. W. Bertini and M. P. Guthrie, *News item results from medium-energy intranuclear-cascade calculation*, Nucl. Phys. **A169** (1971) 670–672.

[472] V. A. Karmanov, *Light Front Wave Function of Relativistic Composite System in Explicitly Solvable Model*, Nucl. Phys. **B166** (1980) 378–398.

[473] H. Burkhardt, V. M. Grichine, P. Gumplinger, V. N. Ivanchenko, R. P. Kokoulin, M. Maire, and L. Urban, *Geant4 standard electromagnetic package for HEP applications*, in *IEEE Symposium Conference Record Nuclear Science 2004.*, pp. , 1907–1910 Vol. 3. Oct, 2004.

[474] H. Zhang, T. Xu, H. Li, S. Zhang, X. Huang, X. Wang, and D. Metaxas, *StackGAN: Text to Photo-realistic Image Synthesis with Stacked Generative Adversarial Networks*, `arXiv:1612.03242 [cs.CV]`.

[475] M. Paganini, L. de Oliveira, and bnachman, *hep-lbdl/CaloGAN: CaloGAN generation, training, and analysis code*, May, 2017. `https://doi.org/10.5281/zenodo.584155`.

[476] M. Mustafa, D. Bard, W. Bhimji, Z. Lukić, R. Al-Rfou, and J. Kratochvil, *Creating Virtual Universes Using Generative Adversarial Networks*, `arXiv:1706.02390 [astro-ph.IM]`.

[477] A. A. Alves, Jr et al., *A Roadmap for HEP Software and Computing R&D for the 2020s*, `arXiv:1712.06982 [physics.comp-ph]`.

[478] K. Albertsson et al., *Machine Learning in High Energy Physics Community White Paper*, `arXiv:1807.02876 [physics.comp-ph]`.

[479] HEP Software Foundation Collaboration, J. Apostolakis et al., *HEP Software Foundation Community White Paper Working Group - Detector Simulation*, `arXiv:1803.04165 [physics.comp-ph]`.

[480] P. Elmer, M. Neubauer, and M. D. Sokoloff, *Strategic Plan for a Scientific Software Innovation Institute (S2I2) for High Energy Physics*, `arXiv:1712.06592 [physics.comp-ph]`.

[481] National Science Foundation Collaboration, J. Chamot et al., *New institute to address massive data demands from upgraded Large Hadron Collider*, 09, 2018. `https://www.nsf.gov/news/news_summ.jsp?org=NSF&cntn_id=296456`.

[482] National Science Foundation Collaboration, D. S. Katz et al., *Implementation of NSF CIF21 Software Vision (SW-Vision)*, `https://www.nsf.gov/funding/pgm_summ.jsp?pims_id=504817`.

[483] ATLAS Collaboration,, *Deep generative models for fast shower simulation in ATLAS*, Tech. Rep. ATL-SOFT-PUB-2018-001, CERN, Geneva, Jul, 2018. `https://cds.cern.ch/record/2630433`.